# DECENTRALIZED THERMAL CONTROL OF BUILDING SYSTEMS

BY

VIKAS CHANDAN

DISSERTATION

Submitted in partial fulfillment of the requirements
for the degree of Doctor of Philosophy in Mechanical Engineering
in the Graduate College of the
University of Illinois at Urbana-Champaign, 2013

Urbana, Illinois

Doctoral Committee:

Professor Andrew Alleyne, Chair
Professor Geir Dullerud
Professor Dušan Stipanović
Dr. John Seem

# Abstract

Energy requirements for heating and cooling of buildings constitute a major fraction of end use energy consumed. Therefore, it is important to provide the occupant comfort requirements in buildings in an energy efficient manner. However, buildings are large scale complex systems, susceptible to sensor, actuator or communication network failures in their thermal control infrastructure, that can affect their performance in terms of occupant comfort and energy efficiency. The degree of decentralization in the control architecture determines a fundamental tradeoff between performance and robustness. This thesis studies the problem of thermal control of buildings from the perspective of partitioning them into clusters for decentralized control, to balance underlying performance and robustness requirements. Measures of deviation in performance and robustness between centralized and decentralized architectures in the Model Predictive Control framework are derived. Appropriate clustering algorithms are then proposed to determine decentralized control architectures which provide a satisfactory trade-off between the underlying performance and robustness objectives. Two different partitioning methodologies – the CLF-MCS method and the OLF-FPM method – are developed and compared. The problem of decentralized control design based on the architectures obtained using these methodologies is also considered. It entails the use of decentralized extended state observers to address the issue of unavailability of unknown states and disturbances in the system. The potential use of the proposed control architecture selection and decentralized control design methodologies is demonstrated in simulation on a real world multi-zone building.

*To my parents.*

# Acknowledgments

This thesis would be incomplete without expressing my gratitude towards all the people who have directly or indirectly helped me in the course of my graduate study at the University of Illinois. It starts with Prof. Andrew Alleyne, my research advisor. I am extremely grateful to him for providing me the opportunity to be a part of the Alleyne Research Group (ARG) since January 2008. Through the course of the last five years within the ARG, I had the opportunity to work on some challenging and practical research problems in the area of control and modeling of energy systems. This thesis is a concatenation of those research efforts. I am very obliged to Prof. Alleyne for his time and efforts in directing my research despite his busy schedule. Throughout these years, he has helped me metamorphose from an undergraduate newcomer to the field of research to a serious graduate student capable of taking ownership of research directions and progress. Besides being an adviser to my research, I am also thankful to him for the role he has played in several other aspects of my overall professional development. In particular, his emphasis on leadership, collaboration and communication has helped me throughout my stint as a graduate student and I am confident that these skills would prove beneficial in the course of my future professional career. In addition to these, I am also thankful to him for helping me make a few critical career decisions, keeping in mind my future professional goals.

Next, I would like to thank the members of my PhD preliminary and defense examination committees – Prof. Geir Dullerud, Prof. Dusan Stipanovic, Prof. Prashant Mehta and Dr. John Seem – for their time and effort in evaluating my research and providing me valuable feedback. I am especially grateful to Dr. John Seem for the discussions sessions that I had with him which helped me in improving the quality and the practical appeal of my work. I would also like to express my gratitude to all the professors at Illinois who provided me the opportunity to learn through interactions both within and outside the classroom. The coursework that I undertook in the area of controls introduced me to several important concepts and tools that I found useful for my research at various points in time. These courses have also enabled me to a build fundamental understanding of controls which I am confident would help me through the course of my research career. I would also like to acknowledge the technical assistance provided by Prasad Mukka at Siemens Corporate Research, Princeton, NJ in the development of the EnergyPlus model that was used in Chapter 6 for the validation and verification of the tools presented in this thesis.

Lastly, I would like to thank all my colleagues within ARG – both past and present – with whom I had the opportunity to work with and interact very closely. In particular, I would like to thank Neera, Tim, Dave, Kira, Bin, Nanjun, Yangmin and Erick with whom I had the privilege to interact most closely during my time at the ARG. I would also like to thank some other members of the group – Brian, Scott, Serena, Doug, Gina, Brandon, Rich, Tom, Justin, Joey, Megan, Kasper and Matt – with whom I shared shorter but valuable stints. Finally, I would like to conclude by thanking my family members whose love and support was critical in keeping me motivated and strong throughout my journey towards a PhD.

# Table of Contents

# List of Tables

# List of Figures

# Chapter 1

# Introduction

## 1.1 Motivation and background

In recent times, there has been an increased emphasis, both nationally and internationally on the importance of efficient utilization of energy [1, 2]. It has primarily been driven by concerns related to environmental, economic and sustainability aspects associated with energy. The impact of climate change and the rapid depletion of non-renewable natural resources is more visible today than at any time in the past. Together with a greater emphasis or renewable and non-polluting sources of energy, efficient use of energy can help to mitigate these effects. There are also substantial economic benefits associated with lesser energy consumption in the form of reduced costs both for energy suppliers and consumers. Reduction in energy demand leads to lower utility expenses at the consumers' end. Similarly, it translates into lesser energy supply and therefore reduced capital and operating costs at the suppliers' end.

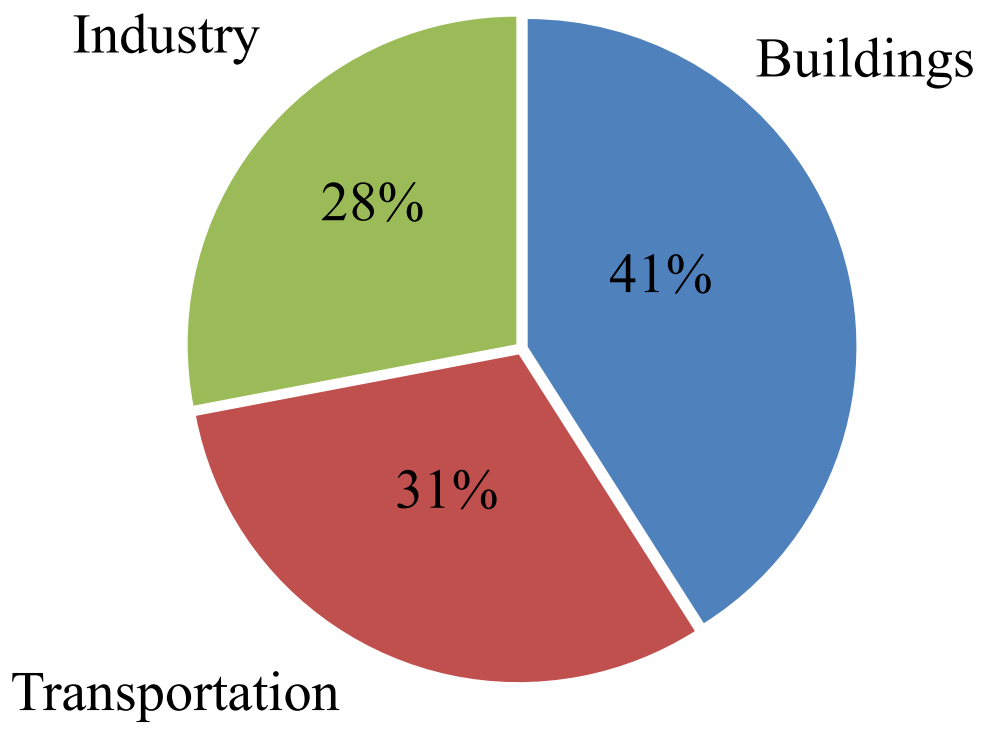

**Figure 1.1 Energy consumption by end use [3]**

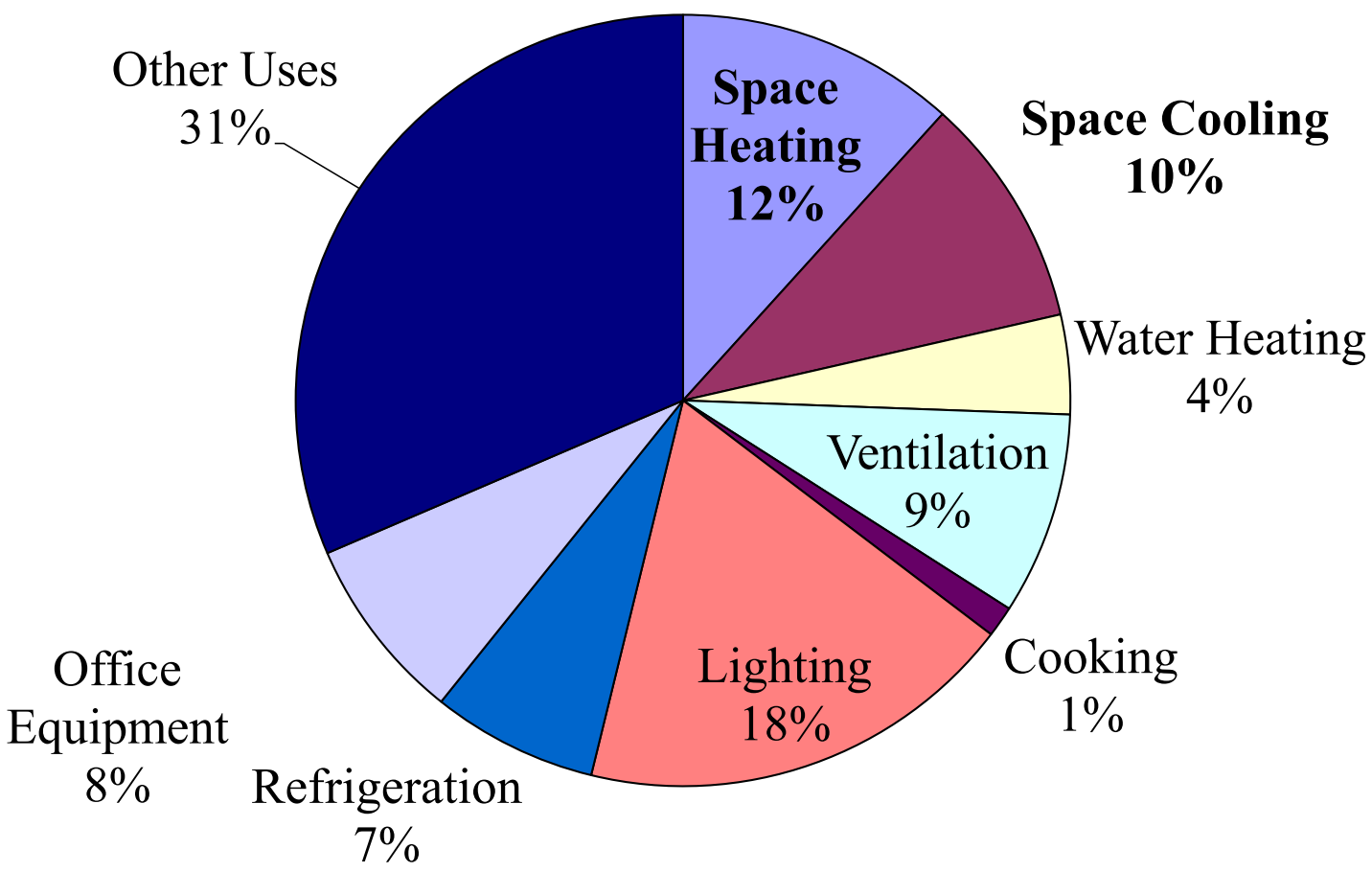

**Figure 1.2  End-use energy consumption in commercial buildings [4]**

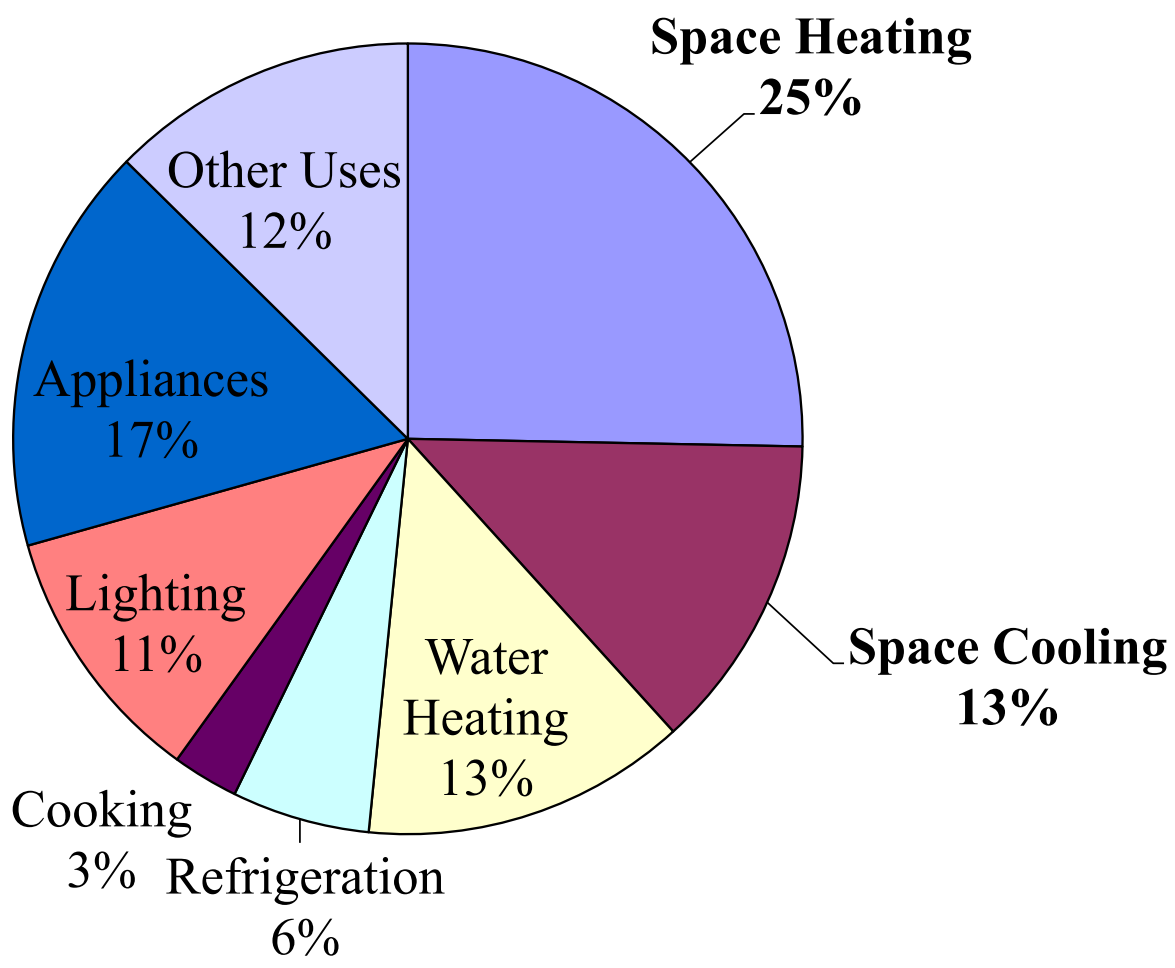

**Figure 1.3 End use energy consumption in residential buildings [5]**

The importance of efficient energy usage motivates a strong emphasis on sectors which account for a large fraction of energy consumption, in order to create a meaningful impact on the global energy and emissions scenario. Figure 1.1 shows sector-wise statistics on end use energy consumption in the United States. The buildings sector is important because it accounts for more than 40% of the total energy consumption and a similar share of greenhouse gas emissions in the United States [3]. In most buildings, more than one-third of the energy usage is attributed to space heating and cooling (Figure 1.2 and Figure 1.3). Therefore, improvements related to energy efficiency in building thermal management can significantly impact the utilization, costs and environmental sustainability aspects of the overall energy consumption.

The problem of efficient thermal management in buildings is inherently multidisciplinary and presents diverse opportunities for several different areas of technology such as design, architecture, alternative energy, modeling and control. In this regard, the opportunities offered by the field of controls are particularly important for existing buildings where re-modeling and design retrofits may be infeasible due to engineering or economic reasons. Strong arguments for energy efficiency in the existing buildings stock [6] have recently been made, therefore underlining the usefulness of controls in achieving such goals. Hence, control of the building heating, ventilation and air-conditioning (HVAC) systems for energy efficient operation has received considerable attention [7, 8, 9].

The underlying control objectives in the context of building thermal management are manifold. The primary objective is to achieve the thermal demands corresponding to the various zones in the building, which are specified by desired levels of temperature, humidity and other indexes of occupant thermal comfort. As discussed above, another important objective is to achieve these thermal demands in an energy efficient manner, which can lead to reduced energy consumption, equipment operating costs and emissions. It is also desired to satisfy the thermal demands robustly, meaning that the control design and architecture should ensure resilience against failures such as thermostat malfunction. Lastly, the controllers should preferably be easily tunable and scalable when going from one building to another.

Two key aspects associated with the control of large scale complex systems such as buildings are control architecture and control algorithm. This thesis intends to analyze the role of these aspects and address them appropriately with the aim of achieving the objectives in building thermal control mentioned above. The outcome is a set of modeling and control tools that appear promising when subjected to detailed simulation studies to examine their efficacy in meeting these objectives. The next section describes the specific objectives of this thesis in more detail.

## 1.2 Research objectives

The primary objective of this research, as mentioned in the previous section is to aid the development of novel and promising modeling and control tools capable of satisfactorily addressing the underlying objectives in the thermal control of buildings. The specific research objectives pursued in this thesis are described below.

### 1.2.1 Control architecture selection

From a systems engineering perspective, buildings are multi-time scale, complex systems with multiple states, inputs, outputs and disturbances. For such systems, the closed loop performance is affected by the choice of the control architecture. In theory, a centralized controller (see illustration in Figure 1.4) using complete information of the system dynamics, and having access to building-wide sensory data could control the building optimally, i.e. satisfy the thermal comfort requirements in the various zones of the building with the least energy consumption. In this framework, control decisions for the entire plant are made by a single controller.

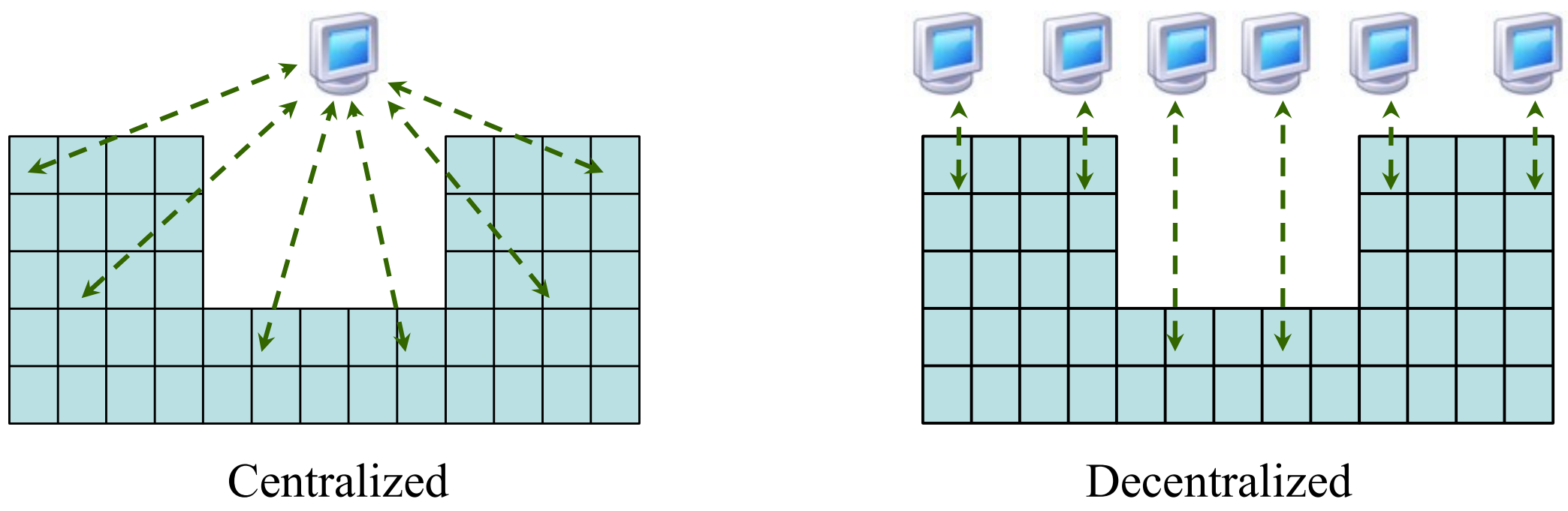

**Figure 1.4 Illustration of centralized and decentralized architectures for thermal control of a multi-zone building**

However, a key limitation of centralized decision making is potentially inferior robustness to sensor and communication network failures. A faulty reading by one sensor can affect the control decisions communicated to all actuators, thus distributing the effect of a local failure plant-wide [10, 11]. In the context of building thermal control, it implies that many of the conditioned zones will be affected until the fault is detected and diagnosed.

Due to these robustness concerns, decentralized decision making (see illustration in Figure 1.4) may be preferable for such large scale systems [12, 13]. It is more resilient to sensor and communication network faults because of its ability to contain them locally. Other benefits of decentralization include flexibility in operation, and simplified design and tuning. A decentralized control architecture consists of multiple disjoint control clusters, where each cluster determines only a subset of the plant-wide control inputs. The clusters do not communicate, i.e. decisions for the control inputs within a cluster are independent of any other cluster. Thus, failures originating in one cluster are prevented from affecting other clusters. It is therefore clear that with smaller clusters, the effect of such failures is more localized.

Although decentralization has merits from a robustness perspective, control decisions in a centralized architecture are better informed than in a decentralized architecture because the latter disregards any inter-cluster interactions in the decision making process. Therefore, it is expected that a decentralized control scheme yields suboptimal performance with respect to a centralized scheme and furthermore, the performance deterioration increases with the extent of decentralization. Hence, the `degree of decentralization' results in a trade-off between optimality and robustness.

Decentralized control has been applied to a wide variety of applications such as coordination of multi-robot systems [14, 15], control of satellite formations [16], and control of automated manufacturing systems [17, 18]. However, the decentralized control architecture is chosen in such a way that each controller caters to an individual physical unit of the overall system such as a robot, a satellite or a single machine. Decentralization in the context of building thermal control has also been studied previously [19-23]. However, similar to the applications previously mentioned, the most common architecture is a multi-agent scheme where each control agent is matched to a single zone in the building. This choice results in the smallest possible cluster size, which is beneficial from a robustness point of view. What is desired is a systematic decentralization procedure that can quantify the specific trade-offs under consideration that exist in a control design context. In this thesis we seek to address this need in the specific context of building thermal control, by developing methodologies to determine appropriate decentralized control architectures, which provide a satisfactory trade-off between optimality and robustness objectives.

### 1.2.2 Decentralized control design

The design of controllers based on decentralized architectures obtained using the methodologies developed in this thesis is another important research objective. A key issue in control design for thermal control of buildings is the unavailability of accurate and reliable information about certain aspects relevant to the thermal dynamics. In particular, thermostats installed in buildings only measure the zone air temperatures which are associated with the thermal comfort of occupants. Therefore, temperatures of walls which also participate in the building's thermal dynamics are usually not known. Additionally, thermal contribution from factors such occupants, lighting, appliances and radiation are difficult to quantify and predict accurately, resulting in potentially large uncertainties in the description of thermal dynamics inside a building.

We observe that existing literature in the area of building thermal control seek to address these issues for a particular building by (a) using data-driven or parameter identification type of modeling approaches [24], (b) describing the dynamics only in terms of zone temperature states [8], (c) adding additional sensors for prediction of unknown states or thermal loads [25, 26], and (d) using high-fidelity models such as EnergyPlus [27] for prediction of states and loads which are otherwise not known [28]. In this thesis, we aim to explore control design methodologies which can be applied to a general class of buildings without the need to add additional sensors or develop potentially expensive high-fidelity models. We also seek to address other challenges associated with control design for thermal control of buildings such as potentially large dimension of the state-space for the underlying thermal dynamics [29] and the presence of constraints originating from practical considerations.

### 1.2.3 Evaluation of proposed tools and methodologies

It is important to verify the performance of the modeling and control tools developed in this work to achieve the above mentioned objectives. For this purpose, a realistic testing environment – either experimental or simulated – needs to be developed. Therefore, in the absence of experimental facilities, an important objective of this research is to develop a simulated test environment and employ it to validate the usefulness of the tools proposed for control architecture selection and decentralized control design.

## 1.3 Literature survey

### 1.3.1 Resources for building systems research

Some important resources providing information on the statistics, challenges and past and current efforts related to energy management of buildings are as follows.

#### 1.3.1.1 ASHRAE

The American Society of Heating, Refrigeration and Air-Conditioning Engineers (ASHRAE) [30], founded in 1894 is an international organization of engineers, industrialists, scientists and researchers associated with the HVAC field. A few ASHRAE publications cater specially to building systems such as High Performing Buildings (a quarterly magazine presenting case studies on exemplary buildings designed for sustainability), Building Information Modeling Guide (available for free online) and the Load Calculation Applications Manual. In addition to these, the ASHRAE Journal, a monthly magazine, often features articles which focus on issues and technologies related to the design, operation and control of building HVAC systems. The society also publishes four handbooks related to the field (Fundamentals, HVAC Systems and Equipment, HVAC Applications and Refrigeration) which are periodically updated. These provide detailed technical descriptions of various HVAC components, together with general and component specific physical and modeling insights. ASHRAE also releases standards and guidelines to aid the design, selection and operation of HVAC systems.

#### 1.3.1.2 Energy Information Administration

The Energy Information Administration (EIA) [31], created by the US Congress in 1977 is an independent statistical agency within the US Department of Energy. The following articles are periodically published by the EIA and made available online, which contain both overall and sector-wise statistics regarding the national and international energy supply and demand:

1. Short Term Energy Outlook: Energy projections for the next 18 months, updated monthly.
2. Annual Energy Outlook: Projection and analysis of US energy supply, demand, and prices through 2030 based on EIA's National Energy Modeling System.
3. International Energy Outlook: Assessment of the outlook for international energy markets through 2030.
4. Monthly Energy Review: Statistics on monthly and annual US national energy consumption going back approximately 30 years, broken down by source.
5. Annual Energy Review: Primary report of historical annual energy statistics.

The statistics are presented sector-wise and at various levels of detail. For the building sector, both heating and cooling data is made available based on geographical region, building type and building features.

#### 1.3.1.3 Europe's Energy Portal

Europe's Energy Portal [32] is an independently run commercial organization located within the European Union (EU). It features articles presenting statistics, issues, and technological and policy initiatives concerning emissions and energy in Europe. It also publishes EU directives related to energy and the environment. Detailed country-wise and sector-wise data, news and analysis are also provided.

#### 1.3.1.4 Other resources

Some other general resources that provide background information and updates on activities related to the building energy area are as follows:

1. USGBC [33]: The U.S. Green Building Council (USGBC), founded in 1993, is a nonprofit trade organization that promotes sustainability in how buildings are designed, built and operated. USGBC provides online resources related to energy efficiency in buildings systems including technical information, statistics, and case studies in the form of articles, webcasts, videos and presentations.
2. Facilitiesnet [34]: This is an online portal containing articles related to building technologies

and building management strategies. It also includes some case studies and links to other resources on energy efficient design and operation of buildings and data centers.

3. Building Technologies Program [35]: The Building Technologies Program (BTP) is funded by the US Department of Energy to promote research and technology development to reduce commercial and residential building energy usage. The program's website features resources such as guidelines for best practices and also links to other agencies and online information repositories.
4. ENERGY STAR [36]: It is a joint program of the U.S. Environmental Protection Agency and the US Department of Energy. It provides online resources such as strategies and guidelines for the design of energy efficient buildings and plants.
5. The Green Grid [37]: The Green Grid is a consortium of IT companies and professionals seeking to improve energy efficiency especially in data centers. Its website contains articles, survey findings, forum discussions and news updates.

### 1.3.2 Modeling and simulation of building thermal dynamics

A summary of papers and other resources on control oriented modeling and simulation of building thermal dynamics is presented here. Lumped parameter methods constitute the most common approach employed in literature to model the thermal interactions inside a building for control design. This is because other potentially more accurate characterizations, such as the use of partial differential equations to represent conductive and convective heat transfers, would require computationally intensive, finite-element solution methodologies, involving high dimensional state vectors. This limits their suitability for use in a control design procedure for a complex, interconnected system such as a building.

One of the first attempts at developing lumped parameter dynamic models of buildings was considered in [38]. A first order representation of the wall thermal dynamics based on construction properties was proposed, using the concept of "accessibility factors". Zones were also represented as first order systems and were connected to walls through resistances, hence resulting in a resistance capacitance (RC) network representation of the building thermal dynamics. More details on the underlying framework, also known as the 2R1C framework are provided in Section 3.3 of this thesis. This framework was experimentally validated by [39]. A higher order lumped parameter framework, which represents each wall as a combination of three capacitors and two resistors (also known as the 3R2C framework) was proposed in [40]. In this paper, the resistances and capacitances were obtained by applying a model reduction procedure involving nonlinear constrained optimization on a higher order model. More details on the 3R2C framework are provided in Section 5.2 of this thesis. A related work is presented in [41], where a genetic algorithm based parameter identification methodology is proposed to obtain resistances and capacitances from experimental data to construct 3R2C representation of walls. A recent improvement on 3R2C modeling has been claimed in [42], which proposes a rule based methodology involving the concept of "dominant layer model" to compute resistances and capacitances. The method was applied to a real construction to demonstate improvements in accuracy over previously proposed 3R2C modeling approaches. The 3R2C framework has been further investigated in [29] to propose an aggregation based model reduction methodology which was shown to provide sufficient accuracy even after a large reduction in model order. An advantage of this method over other model reduction methods, as claimed by the authors is that the reduced order models retain the 3R2C framework.

Other types of lumped parameter modeling methods have also been inverstigated in literature, besides 2R1C and 3R2C. Grey-box modeling methodologies were investigated in [24, 43]. A semiparametric regression analysis was proposed in [24] to estimate unknown parameters and thermal loads in a grey-box model for building thermal dynamics. The methodology was used in conjunction with model predicive control to show reduction in energy consumption on an experimental test-bed. An Unscented Kalman Filter based approach was proposed in [43] to estimate the parameters of a grey-box model of building thermal dynamics. The approach was validated using EnergyPlus simulation data. Black-box system identification methods were proposed in [44, 28] to identify lumped parameter models from high-fidelity EnergyPlus models. Subspace identification methods were used in [44] and the identified models were experimentally validated. In [28], balanced model reduction was employed to reduced the order of the identified black-model for design of a model predictive controller. A model reduction method was proposed in [45] for non-linear models of building thermal dynamics. This method exploited the structure of the non-linear models and was shown to retain sufficient accuracy when compared to the full order model.

The US Department of Energy provides a list [46] of simulation tools that are available for free of cost and can be used to simulate the thermal dynamics in builings. Most of these tools use static or slowly-sampled modeling paradigms, which limits their use for control design or analysis. Therefore, they are primarily intended to provide the ability to test and improve the design of building construction and HVAC systems. Still, some of these tools can be used for control design and anlysis and are compared in Table 1.1.

**Table 1.1 An overview of control-oriented building simulation programs [46]**

| Program | Level of state resolution | Level of time resolution | Software platform |
|---|---|---|---|
| EnergyPlus [27] | Zone and wall temperatures | Upto 1 minute | Fortran compiler with text based input and output interfaces |
| ESP-r [47] | Zone and wall temperatures | Upto 1 minute | C/Fortran |
| HAMLAB [48] | Zone and wall temperatures | < 1 minute allowed | MATLAB/SIMULINK |
| BuildingSim [49] | Zone and wall temperatures | Upto 1 minute | Java |
| SMILE [50] | Zone and wall temperatures | < 1 minute allowed | C and Python |

Among the programs listed above, EnergyPlus is a popularly used modeling environment which used detailed models to simulate the thermal dynamics of a building. The interested reader is directed to Chapter 6 and online tutorials provided in [27] for more information on EnergyPlus. Analysis of other energy simulation programs is beyond the scope of this thesis.

### 1.3.3 Thermal control of buildings

In this section, we provide a survey of literature on thermal control of buildings. Model Predictive Control (MPC) has been applied extensively to this area because of its ability to handle large scale, constrained optimal control problems. Furthermore, the computational complexity concerns which are typically associated with MPC are mitigated because of the slow evolution of thermal dynamics in buildings. For a background and technical details on MPC, the reader is directed to [51, 52].

MPC was used in [53, 54] to determine optimal zonal set-points and charging and discharging strategies for thermal energy storage, so as to optimize the energy usage at the building level. In particular, a detailed investigation of weather forecasting accuracy on the closed loop performance was undertaken. Field results from experimental investigation of the proposed MPC framework were reported. A stochastic MPC strategy for building climate control was proposed in [33] that takes into account weather predictions and comfort constraints. Nonlinear models with stochastic uncertainty were used for control design. The control strategy was experimentally investigated under various weather conditions and occupancy scenarios.

A quadratic MPC framework was employed in [55, 25] on a heating system for an experimental multi-zone building, and improvement in energy consumption over a baseline PID scheme was demonstrated. Learning based MPC was applied in [24] to demonstrate improvements in energy consumption over baseline control strategies on an experimental building. The proposed approach uses statistical techniques to learn the unmodeled dynamics and therefore improve model accuracy.

A min-max MPC framework with shrinking horizon lengths and pre-cooling was proposed in [28] to minimize the energy cost associated with building thermal management. The methodology was applied in simulation on an EnergyPlus model of a building to demonstrate improvements over baseline strategies. The potential of occupancy information to reduce energy consumption in buildings was investigated in [26]. Decentralized MPC strategies using current and predicted occupancy information were implemented at the zone level. Simulation results on a real world building model were presented to show improvements in energy efficiency. A supervisory MPC scheme was explored in [56] to minimize the electrical utility cost in buildings. A special emphasis was placed on the difficulties in optimization due to the demand charge component of the utility cost. The proposed framework was implemented on an EnergyPlus model of a building using MATLAB-EnergyPlus co-simulation approaches, and energy savings with respect to baseline control strategies were reported.

Distributed MPC for thermal control of multi-zone buildings was studied in [57] to address the computational challenges associated with centralized MPC. A multi-agent control architecture involving coordination among agents was proposed, where each control agent uses Sequential Quadratic Programming, proximal minimization and dual decomposition to handle nonlinearities in the optimization framework. Simulation results were presented to demonstrate the improvements in energy efficiency over a baseline control strategy. Distributed MPC using coordination among agents was also investigated in [58]. The proposed methodology was based on linear ARX models (auto-regressive models with external inputs) and quadratic objective functions. Improvements in energy efficiency and computational complexity over a baseline controller and a centralized controller respectively were reported in simulation.

Robust MPC methodologies to address the issue of uncertainty in the thermal dynamics of buildings were investigated in [59]. Closed loop and open loop formulations of robust MPC were compared in simulation and it was concluded that the former outperforms the latter in terms of energy efficiency and robustness to disturbances. Lastly, very few studies in literature have focused on non MPC based strategies for thermal control of buildings. For example, a mean field decentralized control approach using a game-theoretic framework was proposed in [60] to address the complexity of centralized control. Optimal control strategies were derived based on the Hamilton-Jacobi-Bellman (HJB) principle and implemented in simulation to demonstrate energy savings over a baseline PID strategy.

## 1.4 Outline of the thesis

The remainder of this thesis is organized as follows. Chapter 2 provides a physical description of building systems with a special focus on variable air volume HVAC systems. Chapter 3 describes centralized and decentralized control architectures for the thermal control of buildings. These details are then used in Chapter 4 for the development of appropriate tools to enable control architecture decisions that balance the optimality and robustness requirements in the thermal control of buildings. Chapter 5 presents a control design framework for decentralized architectures obtained using the methodologies developed in Chapter 4. A simulated real world building example is studied in Chapter 6 to demonstrate the applicability of the control architecture selection and control design tools developed in this thesis. Lastly, the conclusions and research contributions from this work are presented in Chapter 7.

# Chapter 2

# Physical Details of Building Systems

## 2.1 Introduction

This chapter presents the physical details of building systems. The important control systems in most modern buildings are briefly described. This is followed by a detailed discussion on popularly used building thermal management systems known as variable air volume systems.

## 2.2 Control systems in a building

Most modern buildings consist of various control systems to meet the requirements of occupants such as thermal comfort, lighting, power and security, which are together referred to as a building management system. In this section, we describe the main features of systems used for thermal comfort, lighting and security systems. For more details the reader is directed to [61, 62].

### 2.2.1 Thermal systems

Thermal management in buildings is accomplished using heating, ventilation and air-conditioning (HVAC) systems. HVAC systems used in modern buildings perform the following key functions:

1. Production of thermal energy by conversion from other forms such as mechanical, electrical, chemical etc.
2. Distribution of thermal energy to conditioned spaces in buildings.
3. Control and monitoring through sensors, controllers and actuators.

HVAC systems in buildings range from small scale window units to large scale district heating and cooling systems. For the building examples considered later in this work, we assume that thermal management is provide by use variable air volume (VAV) systems which are t employed in most medium and large scale buildings. More details on VAV systems are provided in Section 2.3. Discussion of other types of HVAC systems is beyond the scope of this thesis but the interested reader is directed to [63] for more details.

### 2.2.2 Lighting systems

Most modern buildings employ lighting control systems which provide several benefits over individual switching, such as reduced energy consumption, synchronization of lighting levels with activities, longer bulb life and reduced carbon emission footprints. A lighting control system is usually centralized and is implemented using an embedded processor or an industrial computer unit. It is typically based on rule based program logic which uses if-then-else constructs and/or logical operators to determine lighting levels (on/off states and intensity of lighting) at various locations in the building. The rules are based on one or a combination of the following factors:

1. Schedules based on the time of the day.
2. Schedules based on the day of the week (weekday/weekend) or season of the year (winter/summer)
3. Occupancy based lighting schedules
4. Daylight based lighting schedules (daylight harvesting)
5. Rules for special events such as social occasions or holidays
6. Alarm triggers, e.g. "all lights on" in case of suspected intrusion.

An illustration of a rule based lighting control system is shown in Figure 2.1. The interested reader is directed to the online resource [65] for more information on lighting control systems such as equipment, architecture and protocols.

### 2.2.3 Security systems

Security systems are provided in residential and commercial buildings to prevent, alert or take remedial actions against undesired events such as intrusions, fire, excessive heat, flooding and carbon monoxide risks. Modern security systems employ multiple sensors at various locations in a building such as infrared/ultrasonic intrusion detectors, glass break detectors, video surveillance systems e.g. security cameras, and smoke, heat and/or carbon monoxide detectors. The trigger/alarm signals from each sensor is transmitted to one or more control units through wires or wireless means.

Depending upon the type of alarm, its location in the building, time of day, and other factors, the control units can automatically initiate various actions such as raising an alarm over the public announcement system, or calling an ambulance service, fire department or police department immediately. They may also be programmed to first call the property manager to verify if the alarm is genuine. The security control system can also trigger other systems such as the lighting control system to illuminate the entire building to facilitate evacuation, if necessary. A schematic of the various constituents of an automated security system is shown in Figure 2.2. The interested reader is directed to [67] for detailed information.

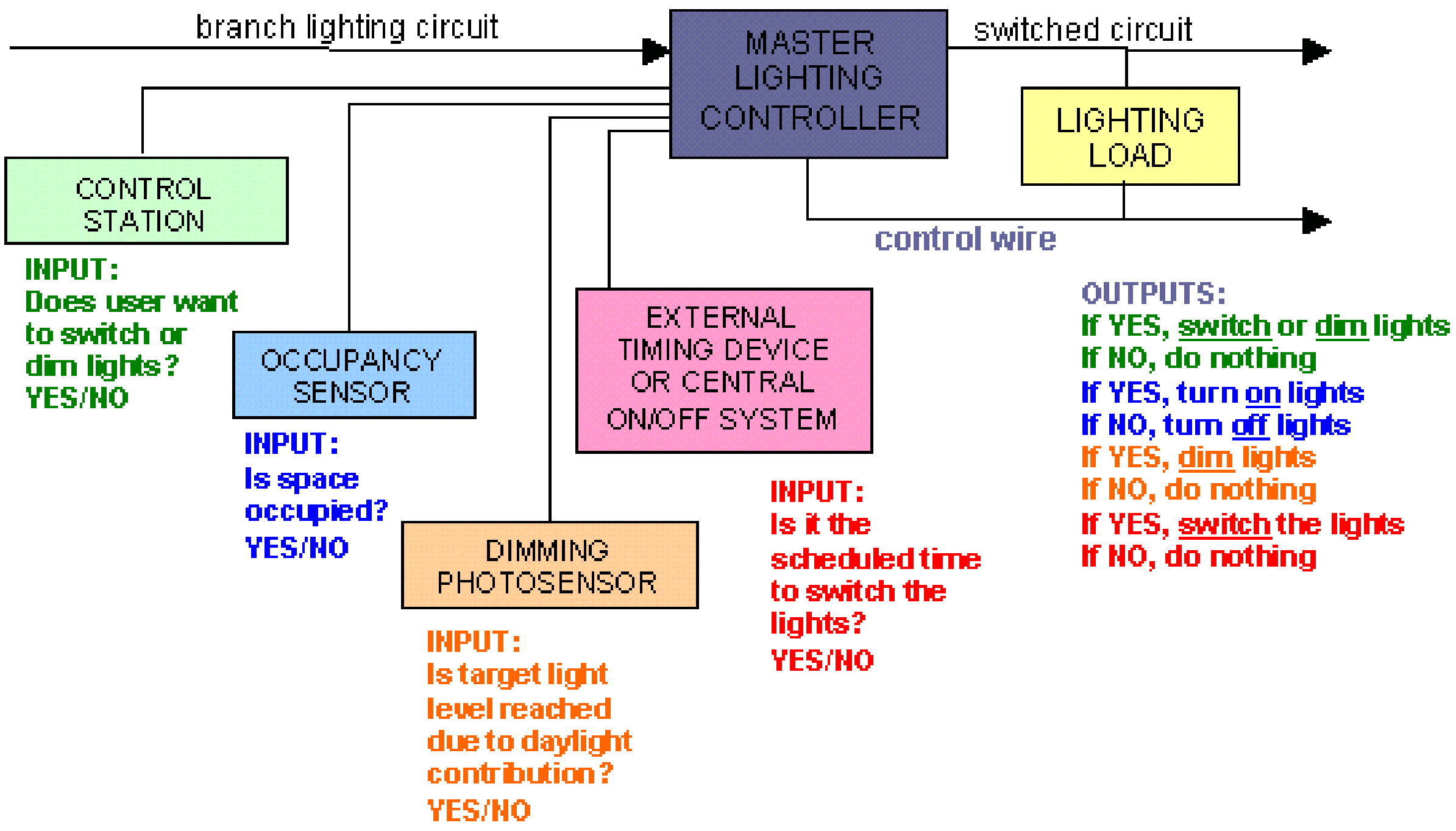

**Figure 2.1 Illustration of a rule based lighting control system (Source [64])**

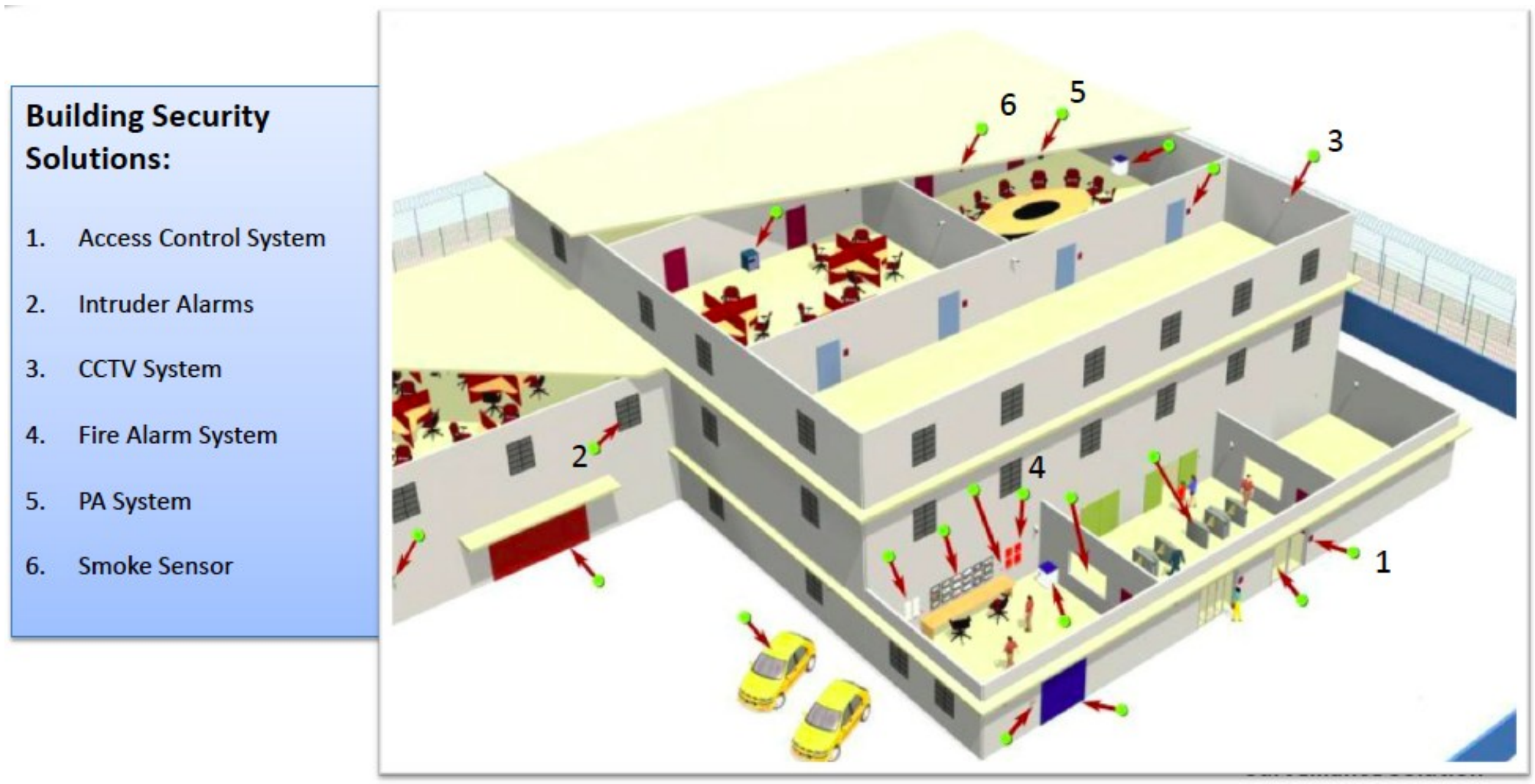

**Figure 2.2 Illustration of components in an automated building security system (Source: [66])**

## 2.3 VAV systems

A variable air volume (VAV) system is a type of HVAC system typically used for air-conditioning medium and large scale buildings. Physical details of VAV systems are described in this section.

### 2.3.1 Architecture

The architecture of a VAV system is illustrated in Figure 2.3. It consists of an air-handling unit (AHU), VAV terminal units, ducts and air terminals. A building can have one or more VAV systems depending on its size and layout. The AHU (Figure 2.4) recirculates the return air from the section of the building conditioned by it, which is then mixed with outside air. The mixing ratio is controlled using dampers. A fan is then used to transport the mixed air through a bank of cooling coils to cool the air and also reduce its humidity. If necessary, the air can also be heated and humidified through heating coils provided at the exit of the AHU. The conditioned air is then circulated to the terminal VAV units through ducts.

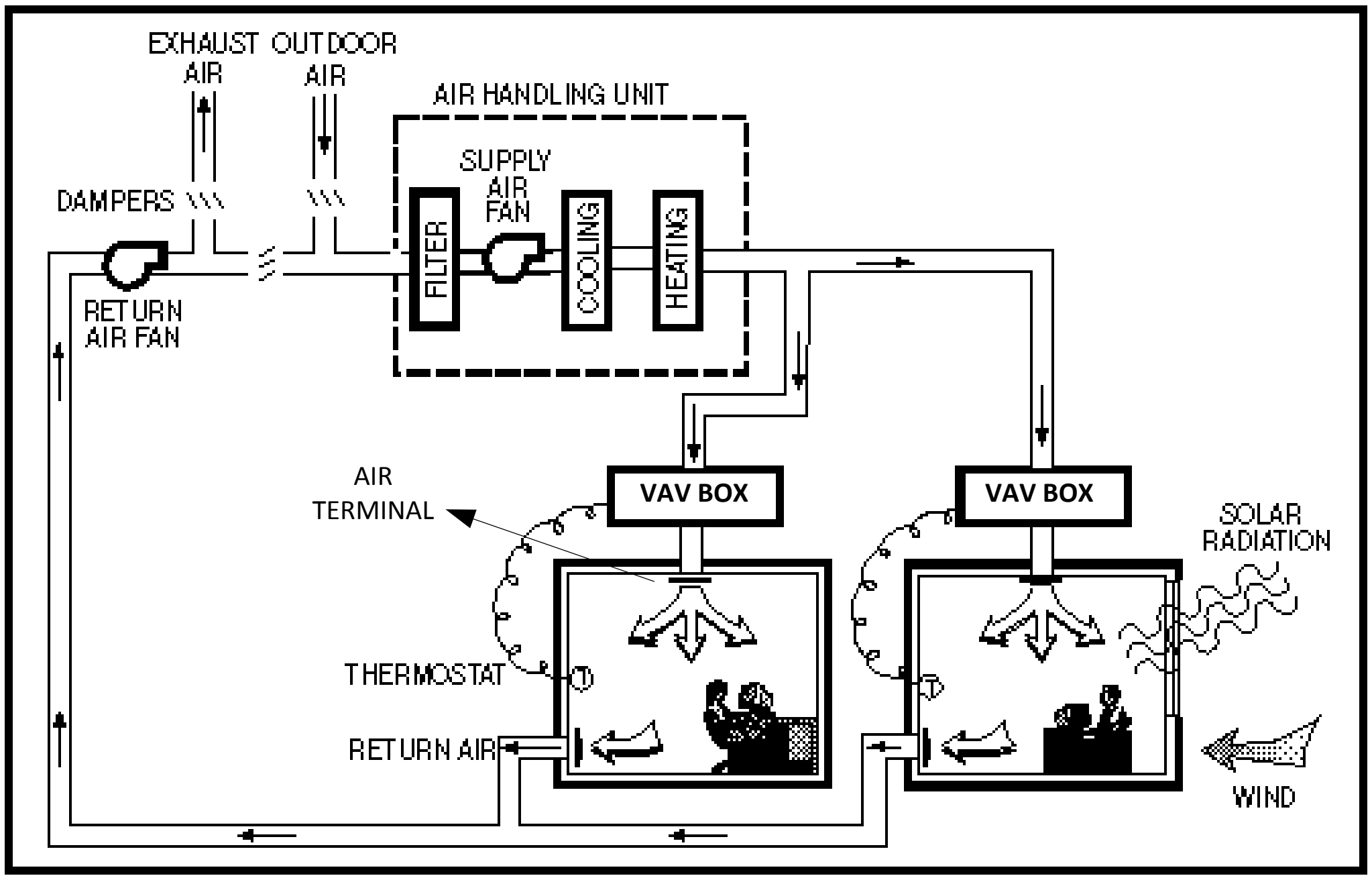

**Figure 2.3 Architecture of a VAV system (source: [68]). Note that dampers are not shown in the air handling unit.**

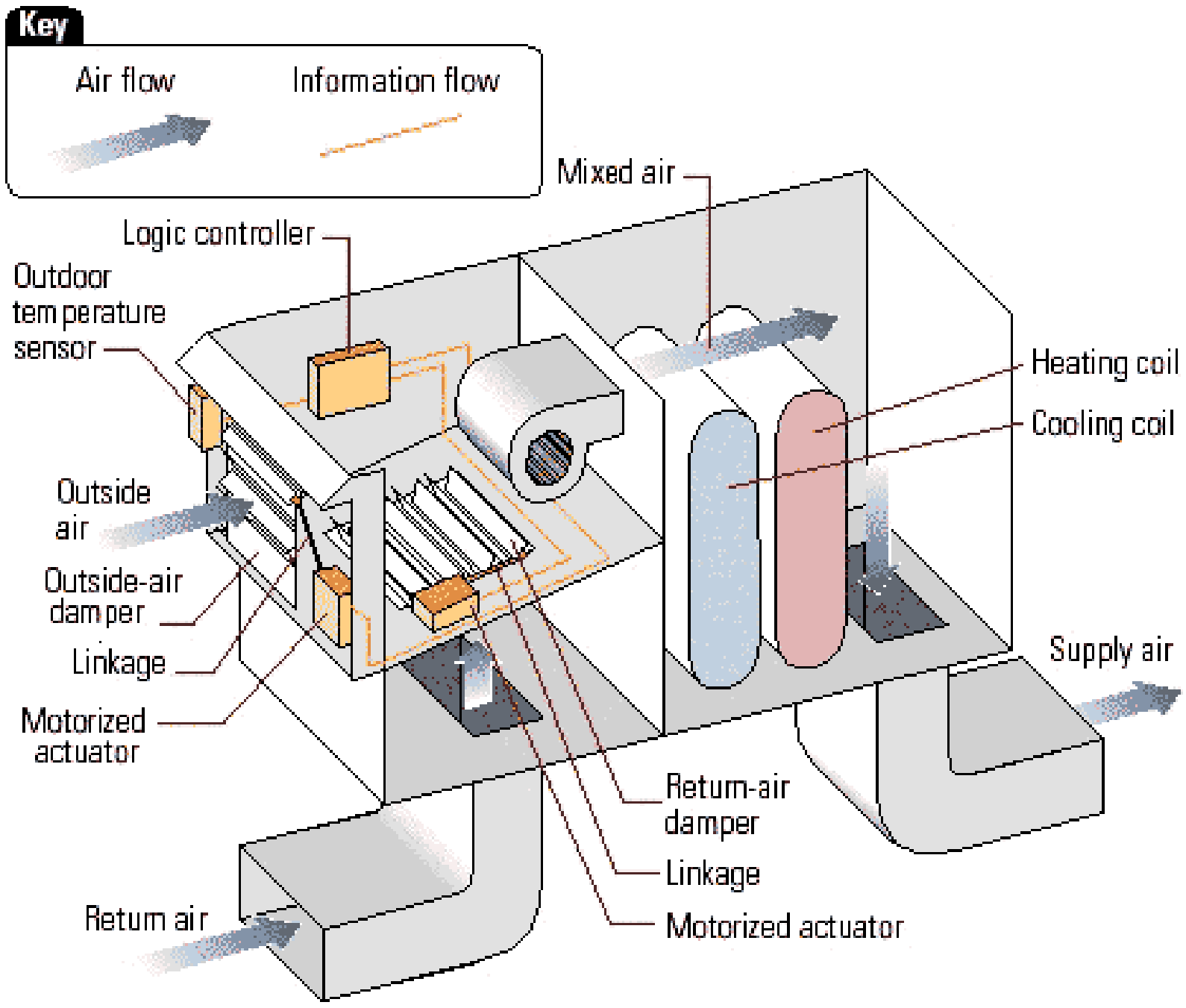

**Figure 2.4 Illustration of an air-handling unit (source: [69])**

The VAV terminal unit (Figure 2.5), also known as a VAV box is a zone level flow control device. A VAV box has two actuators – an air damper and a reheat coil. It is connected to a local or a central control system which typically seeks to achieve a specified set-point temperature in the zone by using the damper to regulate the mass flow rate of air supplied to the zone via air terminals. In the event that the zone temperature is lower than set-point and the mass flow rate of air cannot be reduced further, reheat coils are used to heat the air supplied to the zone.

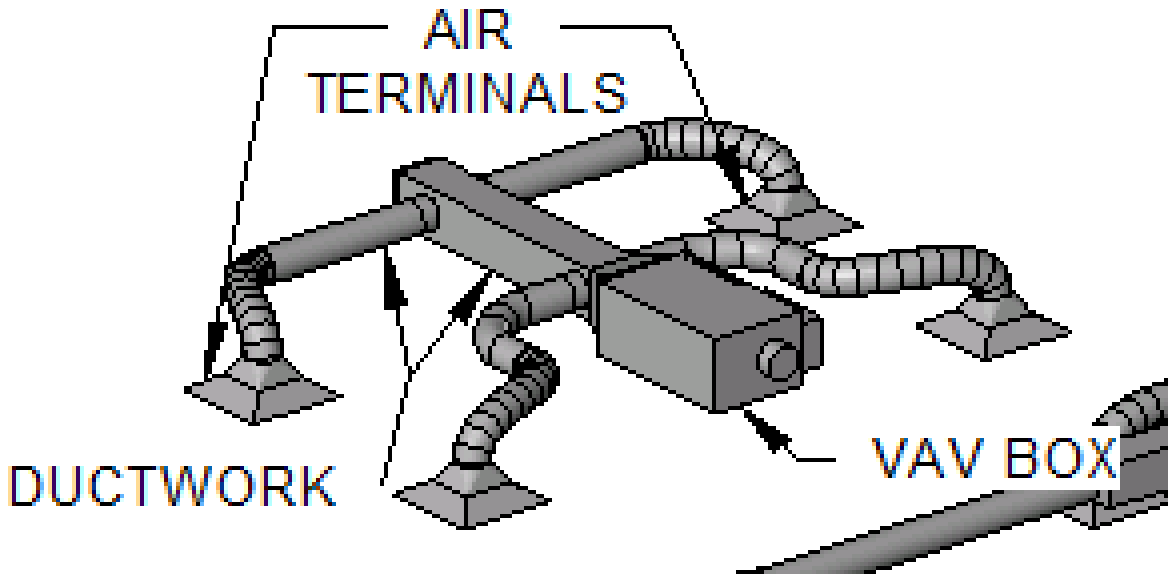

**Figure 2.5 Illustration of a VAV terminal unit (damper and reheat coils are not shown, source: Wikipedia)**

The source of cooling in the cooling coils in an AHU is usually chilled water provided by a chiller unit installed in the building or a district cooling system to which the building is connected. Similarly, the source of heating in the heating coils in an AHU and the reheat coils in a VAV box is usually hot water or steam generated by a local heating unit in the building (e.g. a boiler, air furnace or geothermal pump) or provided by a district heating system. Such water or steam based heating and cooling systems are also known as hydronic systems. The reader is directed to Chapter 2 of [70] for a detailed discussion of such systems.

In addition to the heating and cooling roles, a VAV system also provides ventilation to the conditioned spaces due to continuous circulation of air. Therefore, a separate ventilation unit is not required. The interested reader can find more details on VAV systems in the handbook [71].

### 2.3.2 Sensing and actuation

Thermostats (Figure 2.6) installed in the zones in a building measure the zone temperatures which are used by the controllers to manipulate the dampers and/or the reheat coil power as mentioned earlier. Therefore, from a thermal control perspective, the sensors correspond to the thermostats and the actuators correspond to the dampers and reheat coils in the VAV boxes. It should be noted that the temperature of air provided by the AHU is fixed (usually around 13 $^{0}$C).

The differential pressure down the duct changes as a result of changes in the damper positions in the VAV boxes, and is measured using a pressure sensor. Therefore, a fan controller is employed to change the fan speed via a variable frequency drive to regulate the differential static pressure down the duct around a specified set-point.

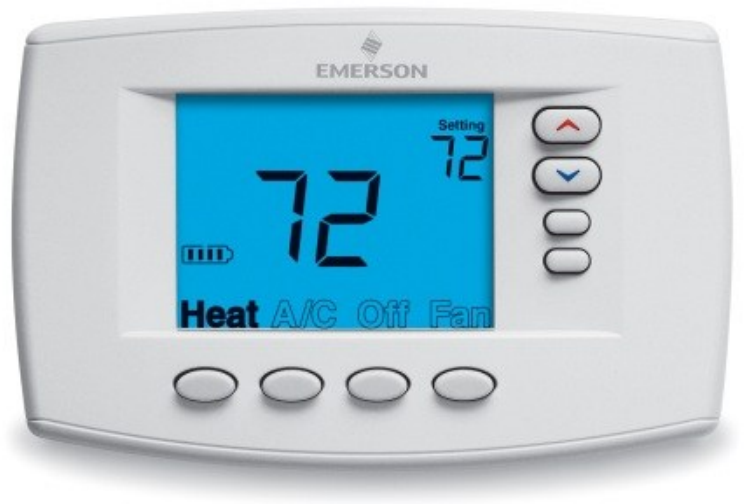

**Figure 2.6 Illustration of a thermostat (source: [72])**

### 2.3.3 Thermal control

The zones in a building are subjected to thermal loads originating from sources such as occupants, appliances, lighting, solar radiation and ambient. The primary purpose of thermal control is to offset these loads by manipulating the mass flow rate and temperature of supply air from the air terminals of the VAV system so as to maintain the zones at specified set-point temperatures. As described in Section 2.3.2, the controllers use thermostats as sensors and the dampers and reheat coils in the VAV boxes as actuators. Historically, pneumatic control was employed, but direct digital control systems (Figure 2.7) have become popular in recent times. The control architecture can be centralized (one single control agent for all VAV boxes) or decentralized (a different control agent for each VAV box) and is usually Proportional Integral Derivative (PID) [74].

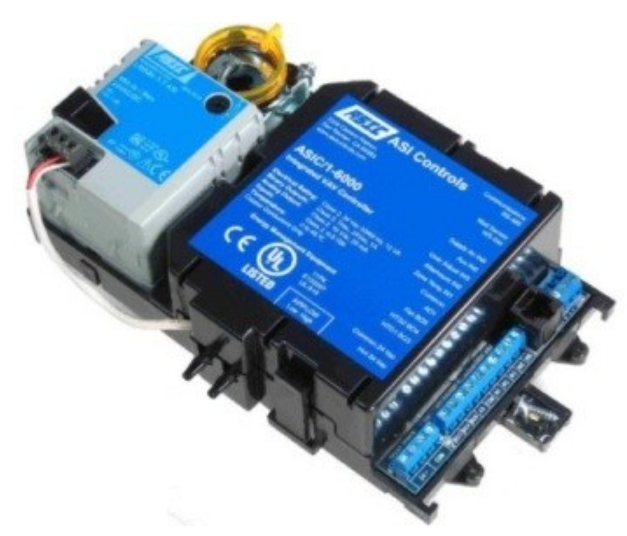

**Figure 2.7 Illustration of a digital VAV controller (Source: [73])**

# Chapter 3

# Centralized and Decentralized Control Frameworks

Centralized and decentralized Model Predictive Control (MPC) frameworks for the problem of building thermal control are presented in this chapter. These frameworks form the basis for the development of control architecture selection methodologies presented in Chapter 4. The list of common symbols used in this chapter is shown in Table 3.1.

**Table 3.1: Nomenclature of common symbols in Chapter 3**

| Symbol | Description |
|---|---|
| $S_z$ | Set of all zones in the building |
| $S_{zi}$ | $i^{th}$ cluster in a *p-partition* |
| $N_w$ | Number of walls in the building |
| $N_{wi}$ | Number of walls in cluster $S_{zi}$ |
| $N_z$ | Number of zones in the building |
| $N_{zi}$ | Number of zones in cluster $S_{zi}$ |
| $\mathbf{T_w}(k)$ | Vector of wall temperatures at time $k$ |
| $\mathbf{T_{wi}}(k)$ | Vector of wall temperatures within $i^{th}$ cluster at time $k$ |
| $\mathbf{T_z}(k)$ | Vector of zone temperatures at time $k$ |
| $\mathbf{T_{zi}}(k)$ | Vector of zone temperatures within $i^{th}$ cluster at time $k$ |
| $\mathbf{u}(k)$ | Vector of control inputs at time $k$ |
| $\mathbf{u_i}(k)$ | Vector of control inputs for $i^{th}$ cluster at time $k$ |
| $T_a(k)$ | Ambient temperature at time $k$ |
| $T_g(k)$ | Ground temperature at time $k$ |
| $\mathbf{d_w}(k)$ | Vector of unmodeled thermal loads acting on walls at time $k$ |
| $\mathbf{d_z}(k)$ | Vector of unmodeled thermal loads acting on zones at time $k$ |
| $\mathbf{d_{wi}}(k)$ | Vector of unmodeled thermal loads in $i^{th}$ cluster's walls at time $k$ |
| $\mathbf{d_{zi}}(k)$ | Vector of unmodeled thermal loads in $i^{th}$ cluster's zones at time $k$ |
| $\mathbf{T_{z,ref}}(k)$ | Vector of zone temperature set-points at time $k$ |
| $\mathbf{T_{zi,ref}}(k)$ | Vector of $i^{th}$ cluster zone temperature set-points at time $k$ |
| $\boldsymbol{\alpha}$ | Vector of weights on cost objective |
| $\boldsymbol{\alpha_i}$ | Vector of weights on $i^{th}$ cluster's cost objective |
| $\boldsymbol{\beta}$ | Vector of weights on performance objective |
| $\boldsymbol{\beta_i}$ | Vector of weights on $i^{th}$ cluster's performance objective |
| $N$ | Number of samples in the control and prediction horizon |
| $T_s$ | Sample time for discretization of thermal dynamics |
| $x(k+l\|k)$ | Predicted value of quantity $x$, after $l$ time steps in future, given $x(k)$ |
| $\mathbf{I}_N$ | Identity matrix of dimension $N \times N$ |
| $\mathbf{0}_{N,M}$ | Zero matrix of dimension $N \times M$ |

## 3.1 Thermal control of buildings

### 3.1.1 Control objectives

In this section, we revisit the objectives in the thermal control of buildings described in Secion 1.1. The primary objective in the thermal control of buildings is to provide desired levels of occupant comfort in their air conditioned sections. An important part of the occupant comfort requirements is to achieve desired temperature set-points that are prescribed manually by the users or auto-programmed by the Building Automation System (BAS). Depending on the specific requirments of occupants, activity levels, etc., the set-points can vary from one zone to another at any given time, as well as for the same zone at different times in the day. Another important control objective is to reduce the operating cost or power required by the heating, ventilation and air-conditioning (HVAC) systems while seeking to provide the occupant comfort requirements. This is motivated by the need to operate buildings efficienctly, as described in Chapter 1.

From a controls perspective, a building is a complex multi-input, multi-output (MIMO) system employing multiple sensors and actuators to meet the air-conditioning requirements as discussed in Chapter 2. Such a system can be susceptible to failures originating in the sensors, actuators or the commnucation infrastructure which integrates them with various elements of the control network. It is desired that any such failure should have a limited effect on the satisfaction of building-wide occupant comfort requirements before the fault is detected and diagnosed. Therefore, resilience to such failures is another important objective that should be considered in the thermal control of buildings. Also, from an implementation perspective, it is desired that the control framework be scalable across buildings, irrespetive of their size or layout.

### 3.1.2 Control aspects

As mentioned in Chapter 1, two important aspects that need to be considered while designing controllers to achieve the afore-mentioned objectives are (i) control architecture, and (ii) control methodology. These aspects are explained below.

#### 3.1.2.1 Control architecture

The existence of a tradeoff between optimality and robustness with respect to the degree of decentralization of the control architecture was discussed in Section 1.2.1. To further explain the demerits of centralized control, we consider a simulation case study performed on a 6 room water-cooled building shown in Figure 3.1. The underlying control objective is to achieve prescribed set-point temperatures ($25^{o}$ C) in all the rooms. A centralized control scheme based on Model Predictive Control (MPC) and a decentralized control scheme (at the room level) based on single input single output (SISO), proprtional-integral (PI) control were implemented. Details of the plant and the controllers with the relevant codes and models are provided in the media accompanying this thesis. The desired objective of temperature regulation was met by both these controllers under normal circumstances. However, the performance when a fault was introduced in the sensor in the atrium (room 1) is shown in Figure 3.2. It was observed that with centralized control, sensor failure in room 1 significantly affected the performance in the other rooms of the building (Figure 3.2 (a)). With decentralized control, however, the effect was limited to room 1, where the fault originated (Figure 3.2 (b)).

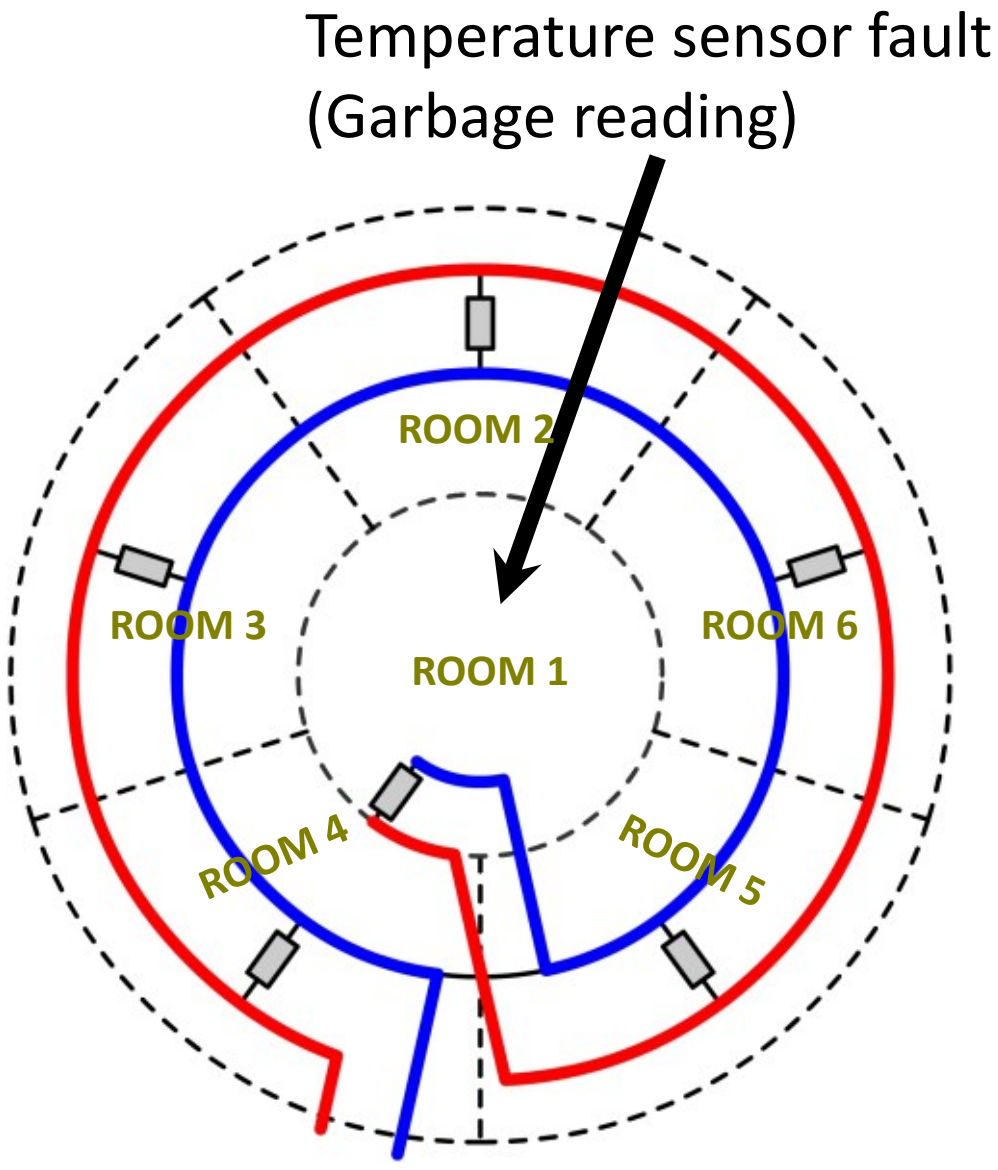

**Figure 3.1: Six zone building used in the case study (return and supply water lines for chilled water loop also shown).**

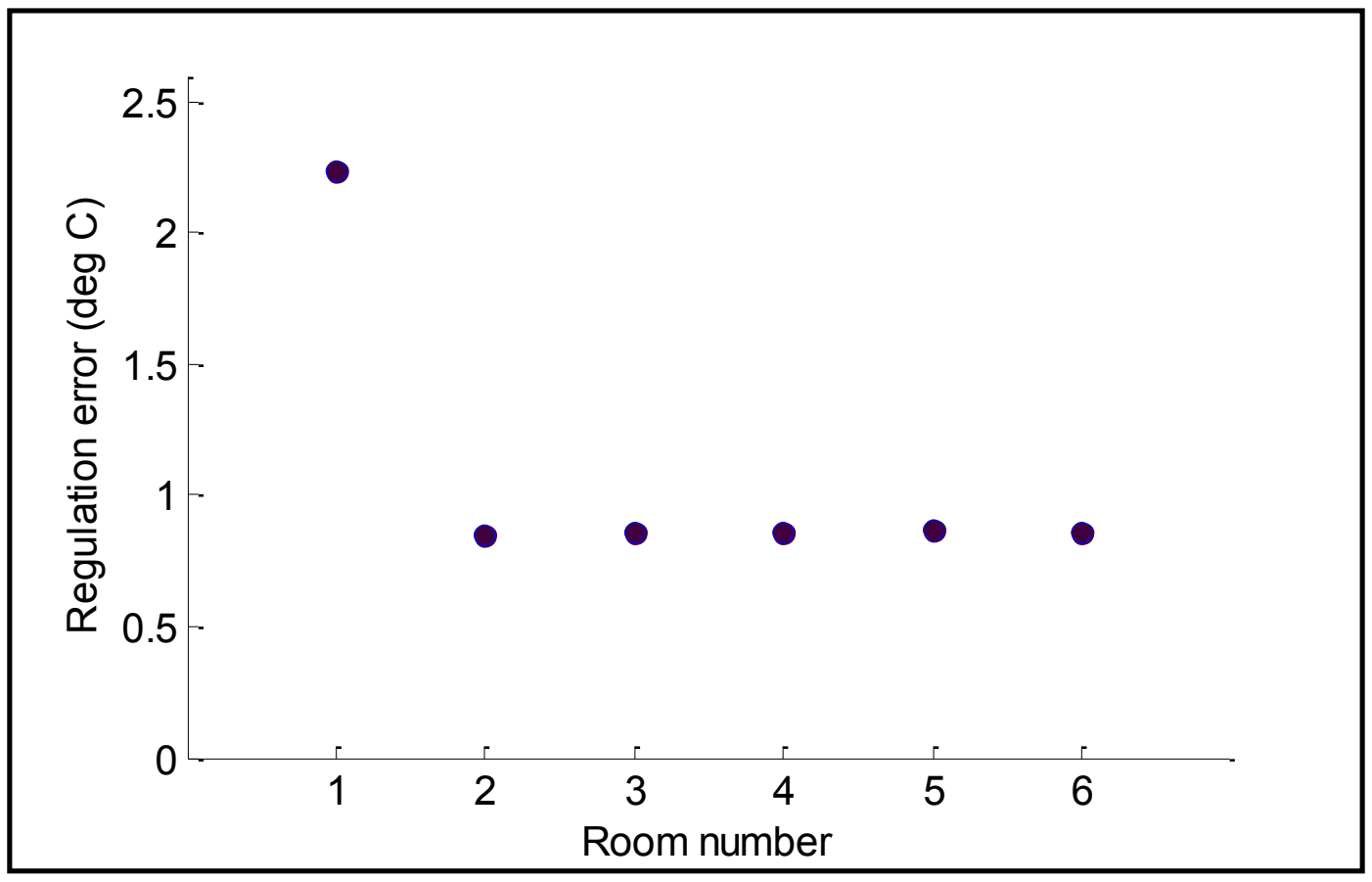

**(a) Regulation errors under centralized control**

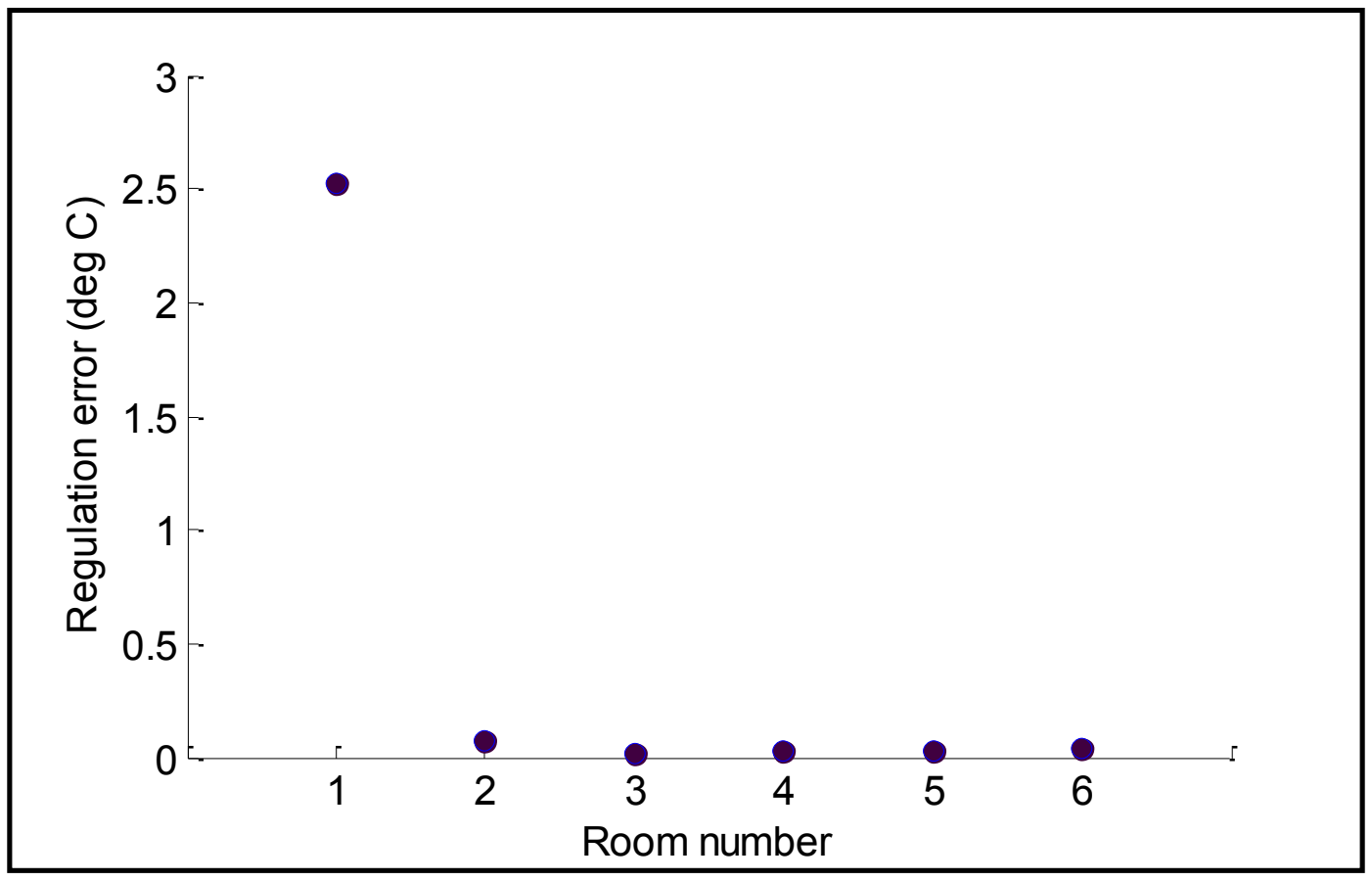

**(b) Regulation errors under decentralized control**

**Figure 3.2: Effect of sensor failure in room 1 of building shown in Fig. 3.1**

Therefore, the control architecture plays an important role in achieving the control objectives listed in section 3.1.1. Besides centralized and decentralized, other configurations such as hierarchical or overlapping architectures [75] can also be used for control. However, the scope of this thesis is limited to the study of centralized and decentralized control architectures only, because of the natural tradeoff between optimality and robustness associated with them as explained above and in Section 1.2.1.

### 3.1.2.2 Control methodology

Various control methodologies have been implemented in literature for the thermal control of buildings, ranging from model-free approaches such as on-off or PID [76,77] to model-based optimal control methods such as Linear Quadratic Regulation (LQR) [78] and model-predictive control (MPC) [8,79]. Among these approaches, MPC has been of considerable interest recently [8,79,80,81], (also see references in Section 1.3.3) because of its proven effectiveness in handling large scale, constrained, optimal control problems. Therefore, the control methodology used in this work is based on an MPC framework.

The centralized and decentralized MPC frameworks presented in this chapter are developed to serve as a basis for the development of appropriate procedures for control architecture selection in Chapter 4. However, appropriate modifications to these frameworks are necessary for their implementation. These modifications will be discussed in chapter 5.

## 3.2 Preliminaries

A few preliminaries are required before a formal description of the centralized and decentralized control frameworks can be presented.

**Definition 3.1 (Zones)**: A zone in a building is defined as a cluster of rooms for which the thermal demands are met using a common actuator. The set of all zones is denoted by $S_z$, which has $N_z$ elements.

**Definition 3.2. (*p-partition*)**: A *p-partition* (or simply a partition) of the building is defined as any set of $p$ non-empty and non-overlapping subsets of $S_z$ that cover all of $S_z$, where $p \in \{1,2,\dots,N_z\}$. The elements which constitute a *p-partition* are called its *clusters*, denoted by $S_{zi}$, where $i = 1,2,\dots,p$. The number of elements in $S_{zi}$ is denoted by $N_{zi}$. The above properties can be formally stated as:

1. $S_{zi} \neq \emptyset \quad for\ \ all\ \ i \in \{1,2,\dots,p\}$,

2. $\bigcup_{i=1}^{p} S_{zi} = S_z$, and

3. $S_{zi} \cap S_{zj} = \emptyset \quad for\ \ all\ \ i,j \in \{1,2,\dots,p\}\ \ and\ \ i \neq j$.

As an illustration of these definitions, consider a simple 3-zone building shown in Figure 3.3. The set of zones, $S_z$ for this example is $\{1, 2, 3\}$. It has exactly three *2-partitions* which are $\{\{1,2\},\{3\}\}$, $\{\{1\},\{2,3\}\}$ and $\{\{1,3\},\{2\}\}$. Furthermore, the only *1-partition* and *3-partition* of

$S_z$ are $\{\{1,2,3\}\}$ and $\{\{1\},\{2\},\{3\}\}$ respectively.

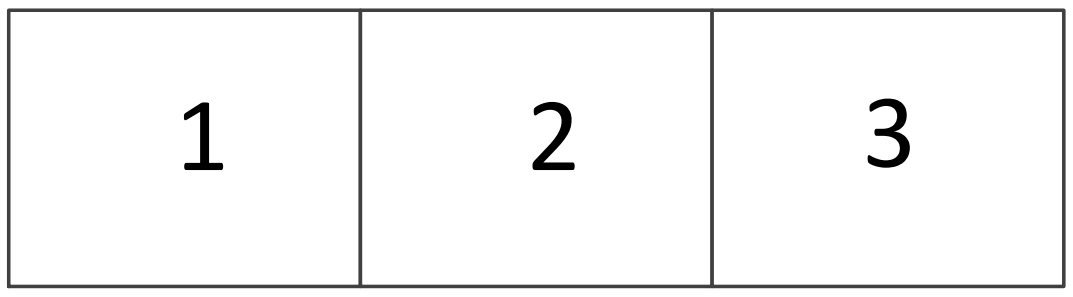

**Figure 3.3: An example 3-zone building (bottom surface of each zone faces ground – all other external surfaces are exposed to ambient)**

## 3.3 Centralized control framework

### 3.3.1 Architecture

The proposed centralized MPC architecture for the thermal control of buildings, illustrated in Figure 3.4 consists of a single control agent which determines the plant-wide control inputs, based on the feedback of building-wide sensory data (zone and wall temperatures), and appropriate forecasts of disturbances such as ambient temperature, ground temperatures and thermal loads acting on the building's walls and zones. The control decisions are arrived at using a discrete time MPC approach, where an objective function is minimized using a centralized system model that serves the purpose of constraints in the optimization. The control inputs represent the rates of energy transfer – posititve for heating and negative for cooling – provided to the zones by the HVAC system.

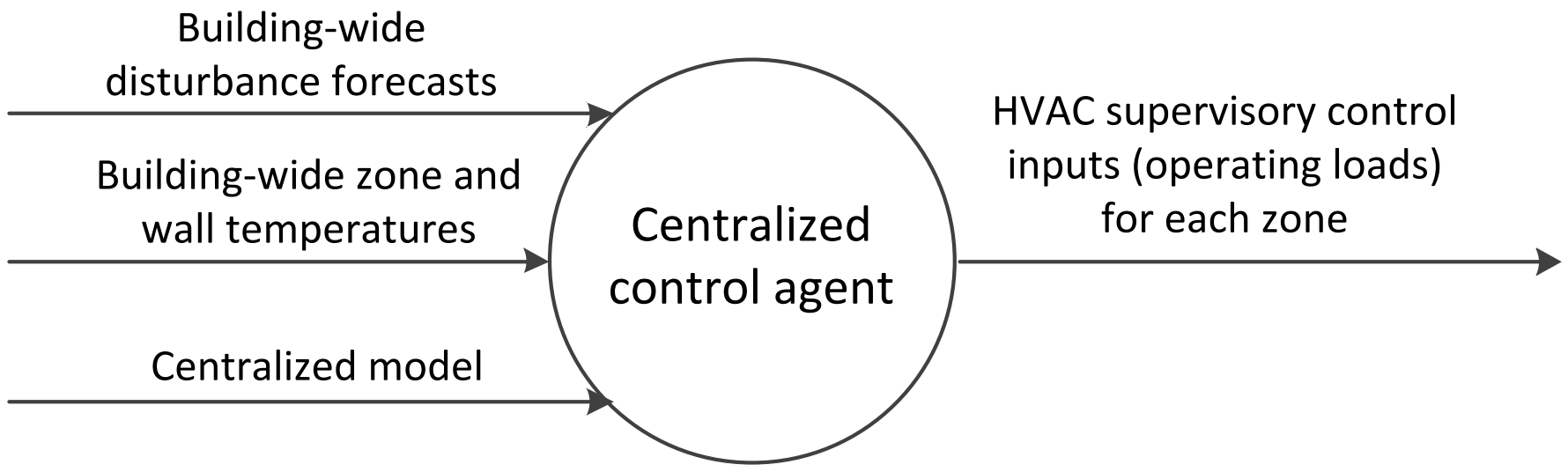

**Figure 3.4: Schematic of centralized MPC architecture**

### 3.3.2 Objective function

As explained in Section 3.1.1, a primary objective in the thermal control of buildings is to achieve desired temperature set-points for the various zones. A secondary control objective is to reduce the HVAC operating cost or energy consumption associated with meeting the specified temperature set-points. A weighted sum of these objectives is used to construct an overall objective function, as shown in (3.1), for optimization over a finite time-horizon in a discrete time setting. The size of the time-horizon, measured in terms of the number of samples, is denoted by $N$. The first term, $J^{c,perf}$ represents the temperature set-point regulation objective across all zones whereas the second term, $J^{c,cost}$ represents either the HVAC operating cost or power consumption. It should be noted that the power consumed by the HVAC system in conditioning a zone depends upon the absolute value of the energy transfer irrespective of its sign (heating or cooling). Therefore, a quadratic function of these energy transfers is used in $J^{c,cost}$. Further, the choice of quadratic functions to represent the constituent terms in the objective function imparts strict convexity, which is a desired property in static optimization problems [82]. To render the framework less restrictive, the weights $\boldsymbol{\alpha}$ and $\boldsymbol{\beta}$, in $J^{c,perf}$ and $J^{c,cost}$ respectively, are specified as vector quantities which allow the flexibility of assigning different weights for different zones. The notation used is defined in the nomenclature (Table 3.1) and is consistent with standard practice in MPC literature [12].

$$J^{c} = J^{c,perf} + J^{c,cost} \tag{3.1}$$

With,

$$J^{c,perf} = \sum_{j=1}^{N} \left(\mathbf{T_z}(k+j|k) - \mathbf{T_{z,ref}}(k)\right)^T diag(\boldsymbol{\beta}) \left(\mathbf{T_z}(k+j|k) - \mathbf{T_{z,ref}}(k)\right) \tag{3.2}$$

$$J^{c,cost} = \sum_{j=0}^{N-1} \mathbf{u}(k+j|k)^T diag(\boldsymbol{\alpha})\ \mathbf{u}(k+j|k) \tag{3.3}$$

### 3.3.3 Model

An appropriate model is required to characterize the effect of the control variables on the feedback variables of interest at each time step in the optimization. The zone temperatures are dynamically interconnected by heat flow occurring through internal walls. The ambient temperature, $T_a$ and ground temperature, $T_g$ also affect the thermal behavior in the zones through the external walls (building envelope), and can be treated as disturbances in the context of the overall system dynamics. A s reviewed in Section 1.3.2, a simple approach for modeling these thermal interactions, which is widely used in literature is to treat the building as a lumped resistive-capacitive (RC) network. Other potentially more accurate characterizations, such as the use of partial differential equations to represent conductive and convective heat transfers, would require computationally intensive, finite-element solution methodologies, involving high dimensional state vectors. This limits their suitability for use in a control design or analysis procedure for a complex, interconnected system such as a building.

The model used in this work is based on [38], revisited in [40], where the walls and zones are represented by a capacitor each, with capacitance equal to the corresponding thermal mass (Figure 3.5). The system states are the (lumped) temperatures of the walls and zones. The control inputs correspond to the energy transfer rates in $kW$ (heating or cooling) that the HVAC system provides to the zones. In a variable air volume (VAV) air-conditioning system, these can be modulated by adjusting dampers in the VAV boxes to set air flow rates or by manipulating the supply air temperature provided by the Air Handling Units (AHU) [79]. The heat transfer between a wall and any of its adjacent zones or the ambient/ground (in case of external walls) is characterized by a resistor, with resistance set to the inverse of the corresponding heat transfer coefficient. Various other factors also affect the thermal dynamics, such as heat flows contributed by occupants, lights, appliances, direct or indirect solar radiation, and thermal infiltration. In this work, these factors are not modeled separately and only their lumped contribution to each zone and wall is represented using thermal disturbance vectors $\mathbf{d}_z$ and $\mathbf{d}_\mathbf{w}$ having units of $kW$. This is because as seen in Chapter 4, these disturbances do not affect the control architecture selection methodologies.

The resulting linear, discrete time, state space model for the building thermal dynamics using the afore-mentioned assumptions is as shown.

$$\begin{bmatrix} \mathbf{T_w} \\ \mathbf{T_z} \end{bmatrix}(k+1) = \underbrace{\begin{bmatrix} \mathbf{A_{w,w}} & \mathbf{A_{w,z}} \\ \mathbf{A_{z,w}} & \mathbf{A_{z,z}} \end{bmatrix}}_{\mathbf{A}} \begin{bmatrix} \mathbf{T_w} \\ \mathbf{T_z} \end{bmatrix}(k) + \begin{bmatrix} \mathbf{0} \\ \mathbf{B_z} \end{bmatrix} \mathbf{u}(k) + \begin{bmatrix} \mathbf{B_{w,a}} & \mathbf{B_{wg}} & \mathbf{B_{w,dw}} & \mathbf{0} \\ \mathbf{0} & \mathbf{0} & \mathbf{0} & \mathbf{B_{z,dz}} \end{bmatrix} \begin{bmatrix} T_a \\ T_g \\ \mathbf{d_w} \\ \mathbf{d_z} \end{bmatrix}(k) \quad (3.4)$$

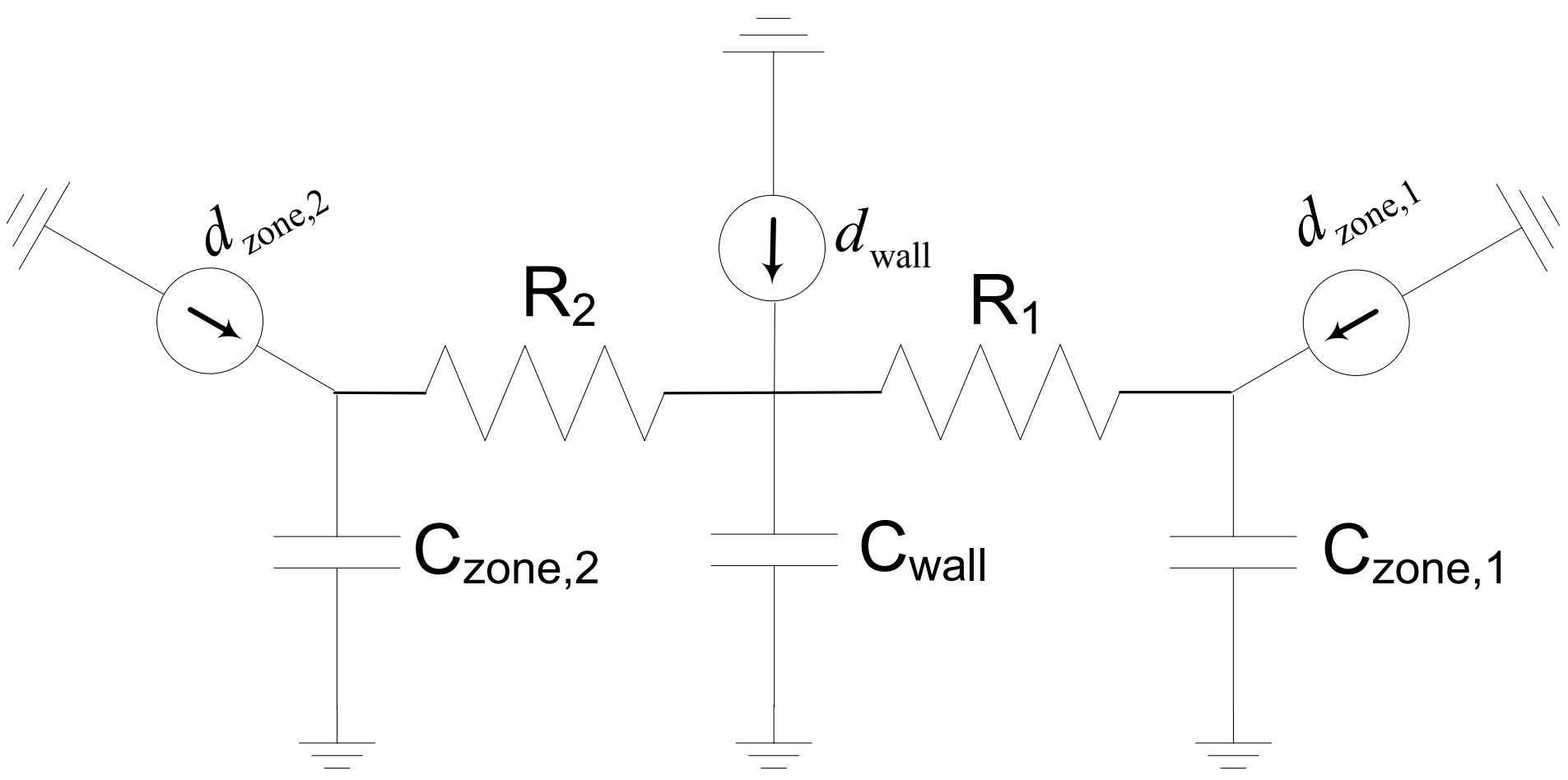

**Figure 3.5 Building block for RC network model of thermal interactions. Figure represents an internal wall flanked by zones on either side of it. The unmodeled thermal loads are also shown via current sources.**

Here, the state transition matrix, $\mathbf{A}$ is partitioned into sub-matrices $\mathbf{A_{w,w}}$, $\mathbf{A_{w,z}}$, $\mathbf{A_{z,w}}$ and $\mathbf{A_{z,z}}$. These sub-matrices, together with the other matrices $\mathbf{B_z}$, $\mathbf{B_{w,a}}$, $\mathbf{B_{w,g}}$, $\mathbf{B_{w,dw}}$ and $\mathbf{B_{z,dz}}$ appearing in the model can be obtained for any general building from a knowledge of the underlying resistance and capacitance values in the RC network via the procedure described in Algorithm 3.1. In this algorithm, a weighted graph is used to represent the resistances in the network. A Laplacian matrix is then constructed, which represents the net flow of energy into each node in the graph, thus allowing the application of the First Law of Thermodynamics (conservation of energy) at the nodes.

**Algorithm 3.1. Generation of state space model for building thermal dynamics from a RC network**

*STEP 1:* A weighted graph is created with nodes for each of the walls, the zones, the ambient and ground (see Figure 3.6). While numbering the nodes, those representing walls are numbered first, followed by the zones, the ambient and lastly the ground. Each wall node is connected by undirected edges to the two nodes to which it is thermally connected. This results in nodes representing external walls to be connected to a zone node and the ambient/ground temperature node. Similarly, the internals walls are connected to a pair of zone nodes. The weight of each edge is set to be the inverse of the corresponding thermal resistance between the two nodes it connects. The resulting weighted graph is denoted by $G = (V, E)$ along with the weight function $w: E \mapsto \mathbb{R}^{+}$. Here, $V$ and $E$ are the sets of vertices and edges, respectively, in graph G. We also define capacitance matrices, $\mathbf{C_w}$ and $\mathbf{C_z}$ which are diagonal matrices of the thermal capacitances associated with the walls and the zones respectively. The diagonal entries in these matrices are entered in the order of the corresponding node numbers in $G$.

*STEP 2:* The Laplacian matrix of $G$, denoted by $\mathbf{L_G}$ is then obtained as:

$$\mathbf{L_G} = \mathbf{D_G} - \mathbf{A_G} \tag{3.5}$$

Where,

$$\mathbf{A_G}(i,j) = \begin{cases} w(i,j) & \text{if}(i,j) \in E \\ 0 & \text{otherwise} \end{cases}$$

$$\mathbf{D_G}(i,i) = \sum_j A_G(i,j)$$

We extract a square sub-matrix from $\mathbf{L_G}$ which corresponds to its first $N_w + N_z$ rows and columns, and denote the result by $\mathbf{L_{Gx}}$. Next, a column vector denoted by $\mathbf{L_{Ga}}$ is extracted which corresponds to the first $N_w$ rows and the $(N_w + N_z + 1)^{th}$ column of $\mathbf{L_G}$. Similarly, another column vector denoted by $\mathbf{L_{Gg}}$ is extracted which corresponds to the first $N_w$ rows and the last column of $\mathbf{L_G}$.

*STEP 3:* The following matrices are now defined:

$$\mathbf{A_{cont}} = \begin{bmatrix} \mathbf{C_w}^{-1} & \mathbf{0} \\ \mathbf{0} & \mathbf{C_z}^{-1} \end{bmatrix} \mathbf{L_{Gx}} \tag{3.6}$$

$$\mathbf{B_{a,cont}} = \mathbf{C_w}^{-1} \mathbf{L_{Ga}} \tag{3.7}$$

$$\mathbf{B_{g,cont}} = \mathbf{C_w}^{-1} \mathbf{L_{Gg}} \tag{3.8}$$

$$\mathbf{B}_{\mathbf{dz,cont}} = \mathbf{B}_{\mathbf{z,cont}} = \mathbf{C_z}^{-1} \tag{3.9}$$

$$\mathbf{B}_{\mathbf{dw,cont}} = \mathbf{C_w}^{-1} \tag{3.10}$$

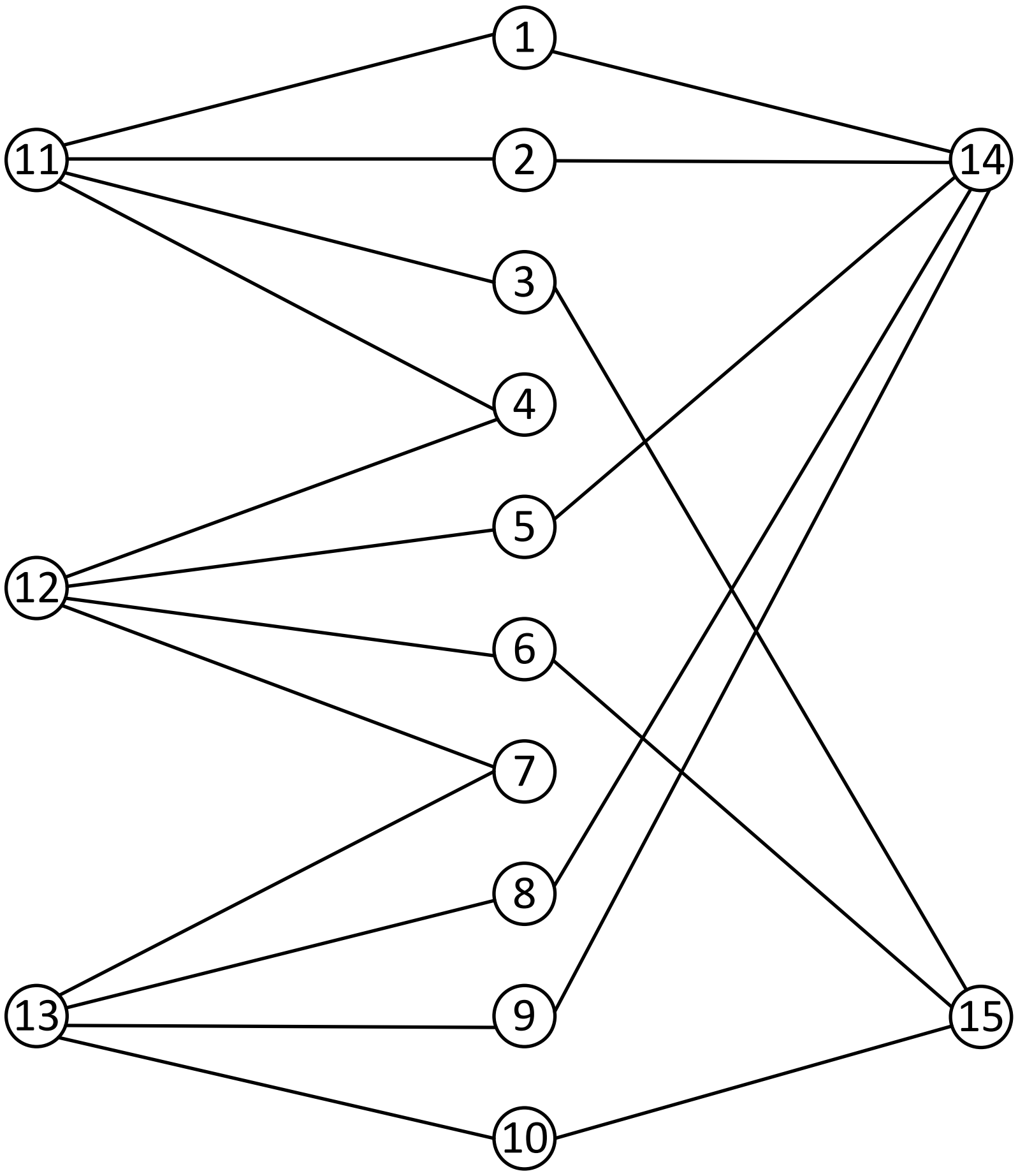

**Figure 3.6: Graph representation for the 3-zone building in Fig. 3.3. Nodes 1-10 represent walls, 11-13 are zones, 14 corresponds to ambient and 15 corresponds to ground.**

*STEP 4:* The continuous time model for the building thermal dynamics is obtained as shown in ( 3.11 ).

$$\frac{d}{dt}\begin{bmatrix}\mathbf{T_w}\\ \mathbf{T_z}\end{bmatrix} = \mathbf{A}_{\mathbf{cont}}\begin{bmatrix}\mathbf{T_w}\\ \mathbf{T_z}\end{bmatrix} + \begin{bmatrix}\mathbf{0}\\ \mathbf{B}_{\mathbf{z,cont}}\end{bmatrix}\mathbf{u} + \begin{bmatrix}\mathbf{B}_{\mathbf{a,cont}} & \mathbf{B}_{\mathbf{g,cont}} & \mathbf{B}_{\mathbf{dw,cont}} & \mathbf{0}\\ \mathbf{0} & \mathbf{0} & \mathbf{0} & \mathbf{B}_{\mathbf{dz,cont}}\end{bmatrix}\begin{bmatrix}T_a\\ T_g\\ \mathbf{d_w}\\ \mathbf{d_z}\end{bmatrix} \tag{3.11}$$

The model (3.4) is obtained by the discretization of ( 3.11 ) using an appropriately chosen sample time $T_s$. In the examples used in this thesis, the discretization is performed in MATLAB using the 'zero order hold' method. It should be noted that $\mathbf{A_{ww}}$, $\mathbf{A_{zz}}$, $\mathbf{B_z}$, $\mathbf{B_{z,dz}}$ and $\mathbf{B_{w,dw}}$ are diagonal matrices as a result of the construction procedure described above.

In most modern buildings, the zone temperature measurements are available using thermostats. However, wall temperatures measurements may not be available. We consider the estimation of wall temperatures in Chapter 5 when designing the centralized and decentralized controllers. For the purposes of control architecture selection however, we assume that full state measurement is available. This implies that at each time instant $k$, the wall temperatures $\mathbf{T_w}(k)$ and zone temperatures $\mathbf{T_z}(k)$ are known.

### 3.3.4 Conversion to Quadratic Program

Over the selected prediction horizon, the model given by ( 3.4) is used to predict the future states $\{\mathbf{T_w}(k+l|k)\}_{l=1}^{l=N}$ and $\{\mathbf{T_z}(k+l|k)\}_{l=1}^{l=N}$ in terms of any chosen current and future control inputs $\{\mathbf{u}(k+l|k)\}_{l=0}^{l=N-1}$, current state measurements $\mathbf{T_w}(k), \mathbf{T_z}(k)$, forecasted values of unmodeled thermal loads $\{\mathbf{d_w}(k+l)\}_{l=0}^{l=N-1}$ and $\{\mathbf{d_z}(k+l)\}_{l=0}^{l=N-1}$, ambient temperatures $\{T_a(k+l)\}_{l=0}^{l=N-1}$ and ground temperatures $\{T_g(k+l)\}_{l=0}^{l=N-1}$.

Let $\mathbf{x} = \begin{bmatrix} \mathbf{T_w} \\ \mathbf{T_z} \end{bmatrix}$, $\mathbf{A} = \begin{bmatrix} \mathbf{A_{w,w}} & \mathbf{A_{w,z}} \\ \mathbf{A_{z,w}} & \mathbf{A_{z,z}} \end{bmatrix}$, $\mathbf{B_u} = \begin{bmatrix} \mathbf{0}_{N_w \times N_z} \\ \mathbf{B_z} \end{bmatrix}$, $\mathbf{B_a} = \begin{bmatrix} \mathbf{B_{w,a}} \\ \mathbf{0}_{N_z \times 1} \end{bmatrix}$, $\mathbf{B_g} = \begin{bmatrix} \mathbf{B_{w,g}} \\ \mathbf{0}_{N_z \times 1} \end{bmatrix}$, $\mathbf{B_{dw}} = \begin{bmatrix} \mathbf{B_{w,dw}} \\ \mathbf{0}_{N_z \times N_w} \end{bmatrix}$ and $\mathbf{B_{dz}} = \begin{bmatrix} \mathbf{0}_{N_w \times N_z} \\ \mathbf{B_{z,dz}} \end{bmatrix}$. Using ( 3.4) we obtain

$$\mathbf{x}(k+1|k) = \mathbf{A}\mathbf{x}(k) + \mathbf{B_u}\mathbf{u}(k) + \mathbf{B_a}\mathbf{T_a}(k) + \mathbf{B_g}\mathbf{T_g}(k) + \mathbf{B_{dw}}\mathbf{d_w}(k) + \mathbf{B_{dz}}\mathbf{d_z}(k)$$

$$\begin{aligned} \mathbf{x}(k+2|k) &= \mathbf{A}\mathbf{x}(k+1|k) + \mathbf{B_u}\mathbf{u}(k+1|k) + \mathbf{B_a}\mathbf{T_a}(k+1|k) + \mathbf{B_g}\mathbf{T_g}(k+1|k) \\ &\quad + \mathbf{B_{dw}}\mathbf{d_w}(k+1|k) + \mathbf{B_{dz}}\mathbf{d_z}(k+1|k) \\ &= \mathbf{A}^2 x(k) + \mathbf{A}\mathbf{B_u}\mathrm{u(k)} + \mathbf{B_u}\mathbf{u}(k+1|k) + \mathbf{A}\mathbf{B_a}\mathbf{T_a}(k) + \mathbf{B_a}\mathbf{T_a}(k+1|k) \\ &\quad + \mathbf{A}\mathbf{B_g}\mathbf{T_g}(k) + \mathbf{B_g}\mathbf{T_g}(k+1|k) + \mathbf{A}\mathbf{B_{dw}}\mathbf{d_w}(k) + \mathbf{B_{dw}}\mathbf{d_w}(k+1|k) + \mathbf{A}\mathbf{B_{dz}} d_z(k) \\ &\quad + \mathbf{B_{dz}}\mathbf{d_z}(k+1|k) \end{aligned}$$

$$\vdots$$

$$\mathbf{x}(k+N|k) = \mathbf{A}^N\mathbf{x}(k) + \sum_{i=0}^{N-1} \mathbf{A}^{N-1-i}[\mathbf{B_u u}(k+i|k) + \mathbf{B_a T_a}(k+i|k) + \mathbf{B_g T_g}(k+i|k) + \mathbf{B_{dw} d_w}(k+i|k) + \mathbf{B_{dz} d_z}(k+i|k)]$$

The above equations can be succintly written as

$$\bar{\mathbf{x}} = \mathbf{Mx}(k) + \mathbf{S_u}\bar{\mathbf{u}} + \mathbf{S_a}\bar{\mathbf{T}}_\mathbf{a} + \mathbf{S_g}\bar{\mathbf{T}}_\mathbf{g} + \mathbf{S_{dw}}\bar{\mathbf{d}}_\mathbf{w} + \mathbf{S_{dz}}\bar{\mathbf{d}}_\mathbf{z}. \qquad (3.12)$$

Here,

$$\bar{\mathbf{x}} = \begin{bmatrix} \mathbf{x}(k+1|k) \\ \mathbf{x}(k+2|k) \\ \vdots \\ \mathbf{x}(k+N|k) \end{bmatrix}, \bar{\mathbf{T}}_\mathbf{a} = \begin{bmatrix} \mathbf{T_a}(k) \\ \mathbf{T_a}(k+1|k) \\ \vdots \\ \mathbf{T_a}(k+N-1|k) \end{bmatrix}, \bar{\mathbf{T}}_\mathbf{g} = \begin{bmatrix} \mathbf{T_g}(k) \\ \mathbf{T_g}(k+1|k) \\ \vdots \\ \mathbf{T_g}(k+N-1|k) \end{bmatrix},$$

$$\bar{\mathbf{u}} = \begin{bmatrix} \mathbf{u}(k) \\ \mathbf{u}(k+1|k) \\ \vdots \\ \mathbf{u}(k+N-1|k) \end{bmatrix} \bar{\mathbf{d}}_\mathbf{w} = \begin{bmatrix} \mathbf{d_w}(k) \\ \mathbf{d_w}(k+1|k) \\ \vdots \\ \mathbf{d_w}(k+N-1|k) \end{bmatrix}, \bar{\mathbf{d}}_\mathbf{z} = \begin{bmatrix} \mathbf{d_z}(k) \\ \mathbf{d_z}(k+1|k) \\ \vdots \\ \mathbf{d_z}(k+N-1|k) \end{bmatrix}, \mathbf{M} = \begin{bmatrix} \mathbf{A} \\ \mathbf{A}^2 \\ \vdots \\ \mathbf{A}^N \end{bmatrix},$$

$$\mathbf{S_u} = \begin{bmatrix} \mathbf{B_u} & \mathbf{0} & \cdots & \mathbf{0} \\ \mathbf{AB_u} & \ddots & \ddots & \vdots \\ \vdots & \ddots & \mathbf{B_u} & \mathbf{0} \\ \mathbf{A}^{\mathbf{N-1}}\mathbf{B_u} & \cdots & \mathbf{AB_u} & \mathbf{B_u} \end{bmatrix}, \mathbf{S_a} = \begin{bmatrix} \mathbf{B_a} & \mathbf{0} & \cdots & \mathbf{0} \\ \mathbf{AB_a} & \ddots & \ddots & \vdots \\ \vdots & \ddots & \mathbf{B_a} & \mathbf{0} \\ \mathbf{A}^{N-1}\mathbf{B_a} & \cdots & \mathbf{AB_a} & \mathbf{B_a} \end{bmatrix},$$

$$\mathbf{S_g} = \begin{bmatrix} \mathbf{B_g} & \mathbf{0} & \cdots & \mathbf{0} \\ \mathbf{AB_g} & \ddots & \ddots & \vdots \\ \vdots & \ddots & \mathbf{B_g} & \mathbf{0} \\ \mathbf{A}^{N-1}\mathbf{B_g} & \cdots & \mathbf{AB_g} & \mathbf{B_g} \end{bmatrix}, \mathbf{S_{dw}} = \begin{bmatrix} \mathbf{B_{dw}} & \mathbf{0} & \cdots & \mathbf{0} \\ \mathbf{AB_{dw}} & \ddots & \ddots & \vdots \\ \vdots & \ddots & \mathbf{B_{dw}} & \mathbf{0} \\ \mathbf{A}^{N-1}\mathbf{B_{dw}} & \cdots & \mathbf{AB_{dw}} & \mathbf{B_{dw}} \end{bmatrix}, \text{ and}$$

$$\mathbf{S_{dz}} = \begin{bmatrix} \mathbf{B_{dz}} & \mathbf{0} & \cdots & \mathbf{0} \\ \mathbf{AB_{dz}} & \ddots & \ddots & \vdots \\ \vdots & \ddots & \mathbf{B_{dz}} & \mathbf{0} \\ \mathbf{A}^{N-1}\mathbf{B_{dz}} & \cdots & \mathbf{AB_{dz}} & \mathbf{B_{dz}} \end{bmatrix}.$$

From (3.1) $J^c$ can be re-stated as

$$J^c = \left(\bar{\mathbf{T}}_\mathbf{z} - \bar{\mathbf{T}}_\mathbf{z,ref}\right)^T \mathbf{Q_1}\left(\bar{\mathbf{T}}_\mathbf{z} - \bar{\mathbf{T}}_\mathbf{z,ref}\right) + \bar{\mathbf{u}}^T\mathbf{Q_2}\bar{\mathbf{u}}. \qquad (3.13)$$

Where,

$$\overline{\mathbf{T}}_\mathbf{z} = \begin{bmatrix} \mathbf{T_z}(k+1|k) \\ \mathbf{T_z}(k+2|k) \\ \vdots \\ \mathbf{T_z}(k+N|k) \end{bmatrix}, \overline{\mathbf{T}}_\mathbf{z,ref} = \begin{bmatrix} \mathbf{T_{z,ref}}(k) \\ \mathbf{T_{z,ref}}(k) \\ \vdots \\ \mathbf{T_{z,ref}}(k) \end{bmatrix}_{N\times N_z},$$

$$\mathbf{Q_1} = \begin{bmatrix} diag(\boldsymbol{\beta}) & \mathbf{0} & \dots & \mathbf{0} \\ \mathbf{0} & diag(\boldsymbol{\beta}) & \dots & \mathbf{0} \\ \vdots & \vdots & \ddots & \vdots \\ \mathbf{0} & \mathbf{0} & \dots & diag(\boldsymbol{\beta}) \end{bmatrix}_{(N_z.N)\times(N_z.N)},$$

and $\mathbf{Q_2} = \begin{bmatrix} diag(\boldsymbol{\alpha}) & \mathbf{0} & \dots & \mathbf{0} \\ \mathbf{0} & diag(\boldsymbol{\alpha}) & \dots & \mathbf{0} \\ \vdots & \vdots & \ddots & \vdots \\ \mathbf{0} & \mathbf{0} & \dots & diag(\boldsymbol{\alpha}) \end{bmatrix}_{(N_z.N)\times(N_z.N)}.$

Let $\mathbf{C} = [\mathbf{0}_{N_z\times N_w} \quad \mathbf{I}_{N_z}]$ and $\bar{\mathbf{C}} = \begin{bmatrix} \mathbf{C} & \mathbf{0} & \dots & \mathbf{0} \\ \mathbf{0} & \mathbf{C} & \dots & \mathbf{0} \\ \vdots & \vdots & \ddots & \vdots \\ \mathbf{0} & \mathbf{0} & \dots & \mathbf{C} \end{bmatrix}_{(N_z.N)\times((N_w+N_z).N)}.$

Clearly, $$\overline{\mathbf{T}}_\mathbf{z} = \bar{\mathbf{C}}\bar{\mathbf{x}}. \qquad (3.14)$$

Using (3.12), (3.14) can be written as

$$\overline{\mathbf{T}}_\mathbf{z} = \bar{\mathbf{C}}\left(\mathbf{Mx}(k) + \mathbf{S_u}\bar{\mathbf{u}} + \mathbf{S_a}\overline{\mathbf{T}}_\mathbf{a} + \mathbf{S_g}\overline{\mathbf{T}}_\mathbf{g} + \mathbf{S_{dw}}\bar{\mathbf{d}}_\mathbf{w} + \mathbf{S_{dz}}\bar{\mathbf{d}}_\mathbf{z}\right). \qquad (3.15)$$

Substituting $\overline{\mathbf{T}}_\mathbf{z}$ from (3.15) in (3.13) and ignoring the term $\left(\mathbf{Mx}(k) + \mathbf{S_a}\overline{\mathbf{T}}_\mathbf{a} + \mathbf{S_g}\overline{\mathbf{T}}_\mathbf{g} + \mathbf{S_{dw}}\bar{\mathbf{d}}_\mathbf{w} + \mathbf{S_{dz}}\bar{\mathbf{d}}_\mathbf{z}\right)^T (\bar{\mathbf{C}})^T \mathbf{Q_1}\bar{\mathbf{C}}\left(\mathbf{Mx}(k) + \mathbf{S_a}\overline{\mathbf{T}}_\mathbf{a} + \mathbf{S_g}\overline{\mathbf{T}}_\mathbf{g} + \mathbf{S_{dw}}\bar{\mathbf{d}}_\mathbf{w} + \mathbf{S_{dz}}\bar{\mathbf{d}}_\mathbf{z}\right)$ which is independent of $\bar{\mathbf{u}}$ and therefore does not affect the optimization, we obtain

$$J^c = \bar{\mathbf{u}}^T\mathbf{H_c}\bar{\mathbf{u}} + \mathbf{f}_\mathbf{c}^T\bar{\mathbf{u}} \qquad (3.16)$$

where, $$\mathbf{H_c} = \mathbf{S}_\mathbf{u}^T\bar{\mathbf{C}}^T\mathbf{Q_1}\bar{\mathbf{C}}\mathbf{S_u} + \mathbf{Q_2}, \qquad (3.17)$$

$$\mathbf{f_c} = 2\mathbf{S}_\mathbf{u}^T\bar{\mathbf{C}}^T\mathbf{Q_1}\left[\bar{\mathbf{C}}\left(\mathbf{Mx(k)} + \mathbf{S_a}\overline{\mathbf{T}}_\mathbf{a} + \mathbf{S_g}\overline{\mathbf{T}}_\mathbf{g} + \mathbf{S_{dw}}\bar{\mathbf{d}}_\mathbf{w} + \mathbf{S_{dz}}\bar{\mathbf{d}}_\mathbf{z}\right) - \overline{\mathbf{T}}_\mathbf{z,ref}\right]. \qquad (3.18)$$

Therefore, the optimization problem corresponding to centralized MPC can be re-stated as the Quadratic Program (QP):

$$\bar{\mathbf{u}}_\mathbf{c}^* = \arg\min_{\bar{\mathbf{u}}} \; g_c(\bar{\mathbf{u}}). \qquad (3.19)$$

where, $$g_c(\bar{\mathbf{u}}) = \bar{\mathbf{u}}^T\mathbf{H_c}\bar{\mathbf{u}} + \mathbf{f}_\mathbf{c}^T\bar{\mathbf{u}}. \qquad (3.20)$$

## 3.4 Decentralized control framework

### 3.4.1 Architecture

We consider any general *p-partition* of a building (definition 3.2) with constitutive clusters $S_{zi}$, where $i \in 1,2,\ldots,p$. A multi-agent MPC scheme is considered which is decentralized with respect to these clusters. In this architecture (Figure 3.7), each agent determines the control inputs for the corresponding cluster, based on the temperature measurements of zones and walls constituting the cluster, and appropriate forecasts of disturbances such as ambient temperature, ground temperature and thermal loads acting on the walls and zones constituting the cluster. The control decisions are still arrived at using a discrete time MPC approach, but the objective function and the model used for prediction are local to the cluster corresponding to a particular control agent.

### 3.4.2 Objective function

The objective function for the $i^{th}$ control agent, $J_i^{dc}$ is obtained in (3.21) by extracting only those terms in $J^{c,perf}$ and $J^{c,cost}$ which correspond to the zones in the cluster $S_{zi}$.

$$J_i^{dc} = J_i^{dc,perf} + J_i^{dc,cost} \tag{3.21}$$

where,

$$J_i^{dc,perf} = \sum_{j=1}^{N} \left(\mathbf{T_{zi}}(k+j|k) - \mathbf{T_{zi,ref}}(k)\right)^T diag(\boldsymbol{\beta}_\mathbf{i}) \left(\mathbf{T_{zi}}(k+j|k) - \mathbf{T_{zi,ref}}(k)\right), \tag{3.22}$$

$$J_i^{dc,cost} = \sum_{j=0}^{N-1} \mathbf{u_i}(k+j|k)^T diag(\boldsymbol{\alpha}_\mathbf{i}) \ \mathbf{u_i}(k+j|k). \tag{3.23}$$

In the above equations, $\boldsymbol{\alpha}_\mathbf{i}$ and $\boldsymbol{\beta}_\mathbf{i}$ are vectors obtained from $\boldsymbol{\alpha}$ and $\boldsymbol{\beta}$ respectively by extracting entries corresponding to the zones in the $i^{th}$ cluster.

### 3.4.3 Model

Similar to the centralized MPC framework, an appropriate model is required to characterize the effect of the cluster level control variables, i.e. the thermal energy transferred to each zone in a cluster by the HVAC system, on the cluster-level state variables of interest. This relationship is obtained by first recognizing the states ($\mathbf{T_{wi}}$ and $\mathbf{T_{zi}}$), the control inputs, $\mathbf{u_i}$ and the disturbances, $\mathbf{d_{wi}}$ and $\mathbf{d_{zi}}$ that are associated with the walls and zones constituting the cluster, and then characterizing their interdependencies by extracting suitable sub-matrices from the full-order state space matrices in (3.4). Here, $\mathbf{T_{wi}}$ is the vector of temperatures of all walls which are adjacent, in terms of the graph $G$ described in Algorithm 1, to the zones constituting the $i^{th}$ cluster. The corresponding model for the $i^{th}$ cluster can be expressed in the form shown in (3.24), which uses the fact that $\mathbf{A_{w,w}}$, $\mathbf{A_{z,z}}$, $\mathbf{B_{w,dw}}$ and $\mathbf{B_{z,dz}}$ in (3.4) are diagonal matrices. The last term in the right hand side of (3.24 ) represents the influence that the zone temperatures in other clusters have on the dynamics of the $i^{th}$ cluster.

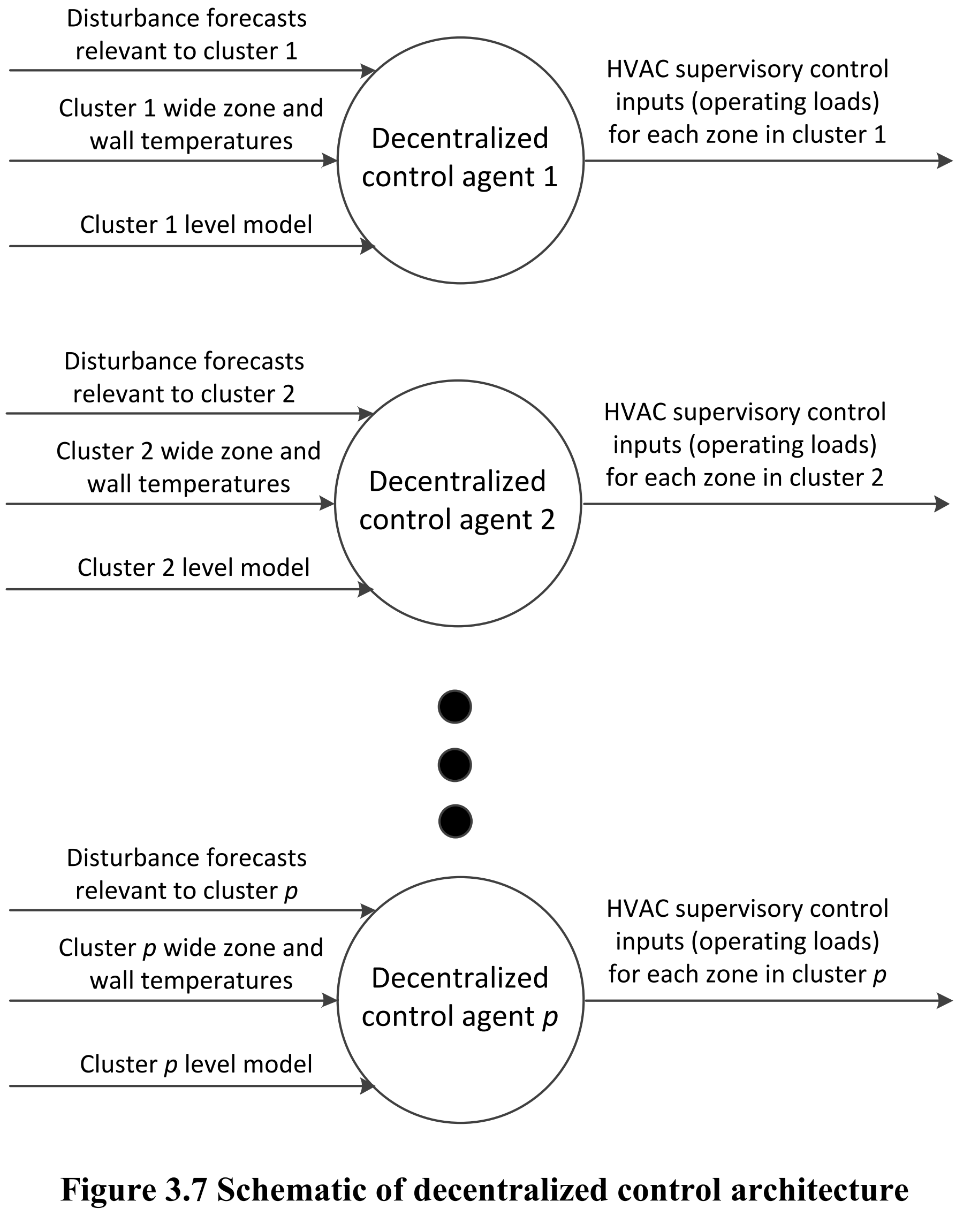

**Figure 3.7 Schematic of decentralized control architecture**

$$\begin{bmatrix}\mathbf{T_{wi}}\\ \mathbf{T_{zi}}\end{bmatrix}(k+1) = \underbrace{\begin{bmatrix}\mathbf{A_{wi,wi}} & \mathbf{A_{wi,zi}}\\ \mathbf{A_{zi,wi}} & \mathbf{A_{zi,zi}}\end{bmatrix}}_{\mathbf{A}}\begin{bmatrix}\mathbf{T_{wi}}\\ \mathbf{T_{zi}}\end{bmatrix}(k) + \begin{bmatrix}\mathbf{0}\\ \mathbf{B_{zi}}\end{bmatrix}\mathbf{u}_i(k)$$

$$+\begin{bmatrix}\mathbf{B_{wi,a}} & \mathbf{B_{wi,g}} & \mathbf{B_{wi,dw}} & \mathbf{0}\\ \mathbf{0} & \mathbf{0} & \mathbf{0} & \mathbf{B_{zi,dz}}\end{bmatrix}\begin{bmatrix}T_a\\ T_g\\ \mathbf{d_{wi}}\\ \mathbf{d_{zi}}\end{bmatrix}(k) + \sum\nolimits_{j\in\{1,2,\ldots,p\},j\neq i}\begin{bmatrix}\mathbf{B_{wi,zj}}\\ \mathbf{0}\end{bmatrix}\mathbf{T_{zj}}(k) \quad (3.24)$$

Imposition of a control architecture that is decentralized with respect to the clusters implies that the $i^{th}$ control agent does not have access to the sensory data from the other $p-1$ clusters. Therefore, the $\mathbf{T_{zj}}(k)$, for each $j \in \{1,2,\ldots,p\}, j \neq i$ appearing in (19) must be replaced with an appropriate guess or estimate, $\hat{\mathbf{T}}_{\mathbf{zj}}(k)$. For example, if the operating temperature range of the building zones is known, say $[15^0C, 25^0C]$, the estimates can be heuristically chosen values which lie in this range. A specific choice for these estimates is the set-point temperatures for the corresponding zones, assuming that the controllers are able to satisfactorily regulate the temperatures around these set-points. Therefore, the appropriate model for use by the $i^{th}$ control agent is given by

$$\begin{bmatrix}\mathbf{T_{wi}}\\ \mathbf{T_{zi}}\end{bmatrix}(k+1) = \underbrace{\begin{bmatrix}\mathbf{A_{wi,wi}} & \mathbf{A_{wi,zi}}\\ \mathbf{A_{zi,wi}} & \mathbf{A_{zi,zi}}\end{bmatrix}}_{\mathbf{A}}\begin{bmatrix}\mathbf{T_{wi}}\\ \mathbf{T_{zi}}\end{bmatrix}(k) + \begin{bmatrix}\mathbf{0}\\ \mathbf{B_{zi}}\end{bmatrix}\mathbf{u}_i(k)$$

$$+\begin{bmatrix}\mathbf{B_{wi,a}} & \mathbf{B_{wi,g}} & \mathbf{B_{wi,dw}} & \mathbf{0}\\ \mathbf{0} & \mathbf{0} & \mathbf{0} & \mathbf{B_{zi,dz}}\end{bmatrix}\begin{bmatrix}T_a\\ T_g\\ \mathbf{d_{wi}}\\ \mathbf{d_{zi}}\end{bmatrix}(k) + \sum\nolimits_{j\in\{1,2,\ldots,p\},j\neq i}\begin{bmatrix}\mathbf{B_{wi,zj}}\\ \mathbf{0}\end{bmatrix}\hat{\mathbf{T}}_{\mathbf{zj}}(k). \quad (3.25)$$

### 3.4.4 Conversion to Quadratic Program

The optimization of the cost function $J_i^{dc}$ for the $i^{th}$ control agent can be converted to a QP by proceeding similarly as in the case of centralized MPC. Over the selected prediction horizon, the model given by (3.25) is used to predict the future states $\{\mathbf{T_{wi}}(k+l|k)\}_{l=1}^{l=N}$ and $\{\mathbf{T_{zi}}(k+l|k)\}_{l=1}^{l=N}$ in terms of any chosen current and future control inputs $\{\mathbf{u}_\mathrm{i}(k+l|k)\}_{l=0}^{l=N-1}$, current state measurements $\mathbf{T_{wi}}(k), \mathbf{T_{zi}}(k)$, forecasted values of unmodeled thermal loads

$\{\mathbf{d_{wi}}(k+l)\}_{l=0}^{l=N-1}$ and $\{\mathbf{d_{zi}}(k+l)\}_{l=0}^{l=N-1}$, ambient temperatures $\{T_a(k+l)\}_{l=0}^{l=N-1}$, ground temperatures $\{T_g(k+l)\}_{l=0}^{l=N-1}$, and state estimates from other clusters $\{\widehat{\mathbf{T}}_{\mathbf{zj}}(k+l)\}_{l=0}^{l=N-1}$ (where $j \in \{1,2,\dots,p\}, j \neq i$).

Let $\mathbf{x}_{\mathrm{i}} = \begin{bmatrix}\mathbf{T_{wi}}\\ \mathbf{T_{zi}}\end{bmatrix}$, $\mathbf{A_i} = \begin{bmatrix}\mathbf{A_{wi,wi}} & \mathbf{A_{wi,zi}}\\ \mathbf{A_{zi,wi}} & \mathbf{A_{zi,zi}}\end{bmatrix}$, $\mathbf{B_{ui}} = \begin{bmatrix}\mathbf{0}_{N_{wi}\times N_{zi}}\\ \mathbf{B}_{\mathbf{z}i}\end{bmatrix}$, $\mathbf{B_{ai}} = \begin{bmatrix}\mathbf{B_{wi,a}}\\ \mathbf{0}_{N_{zi}\times 1}\end{bmatrix}$, $\mathbf{B_{gi}} = \begin{bmatrix}\mathbf{B_{wi,g}}\\ \mathbf{0}_{N_{zi}\times 1}\end{bmatrix}$, $\mathbf{B_{dw,i}} = \begin{bmatrix}\mathbf{B_{wi,dw}}\\ \mathbf{0}_{N_{zi}\times N_{wi}}\end{bmatrix}$, $\mathbf{B_{dz,i}} = \begin{bmatrix}\mathbf{0}_{N_{wi}\times N_{zi}}\\ \mathbf{B_{zi,dz}}\end{bmatrix}$ and $\mathbf{B_{ij}} = \begin{bmatrix}\mathbf{B_{wi,zj}}\\ \mathbf{0}_{N_{zi}\times N_{zj}}\end{bmatrix}$. Using (3.25), we obtain

$$\mathbf{x}_{\mathrm{i}}(k+1|k) = \mathbf{A_i}\mathbf{x}_{\mathrm{i}}(k) + \mathbf{B_{ui}}\mathbf{u}_{\mathrm{i}}(k) + \mathbf{B_{ai}}\mathbf{T_a}(k) + \mathbf{B_{gi}}\mathbf{T_g}(k) + \mathbf{B_{dw,i}}\mathbf{d_{wi}}(k) + \mathbf{B_{dz,i}}\mathbf{d_{zi}}(k) + \sum_{j\neq i}\mathbf{B_{ij}}\widehat{\mathbf{T}}_{\mathbf{zj}}(k)$$

$$\begin{aligned}\mathbf{x}_{\mathrm{i}}(k+2|k) &= \mathbf{A_i}\mathbf{x_i}(k+1|k) + \mathbf{B_{ui}}\mathbf{u_i}(k+1|k) + \mathbf{B_{ai}}\mathbf{T_a}(k+1|k) + \mathbf{B_{gi}}\mathbf{T_g}(k+1|k) \\ &\quad + \mathbf{B_{dw,i}}\mathbf{d_{wi}}(k+1|k) + \mathbf{B_{dz,i}}\mathbf{d}_{\mathbf{z}i}(k+1|k) + \sum_{j\neq i}\mathbf{B_{ij}}\widehat{\mathbf{T}}_{\mathbf{zj}}(k+1|\mathrm{k}) \\ &= \mathbf{A_i}^2\mathbf{x_i}(k) + \mathbf{A_i}\mathbf{B_{ui}}\mathbf{u_i}(\mathrm{k}) + \mathbf{B_{ui}}\mathbf{u_i}(k+1|k) + \mathbf{A_i}\mathbf{B_{ai}}\mathbf{T_a}(k) + \mathbf{B_{ai}}\mathbf{T_a}(k+1|k) \\ &\quad + \mathbf{A_i}\mathbf{B_{gi}}\mathbf{T_g}(k) + \mathbf{B_{gi}}\mathbf{T_g}(k+1|k) + \mathbf{A_i}\mathbf{B_{dw,i}}\mathbf{d_{wi}}(k) + \mathbf{B_{dw,i}}\mathbf{d_{w,i}}(k+1|k) \\ &\quad + \mathbf{A_i}\mathbf{B_{dz,i}}\mathbf{d_{zi}}(k) + \mathbf{B_{dz,i}}\mathbf{d_{zi}}(k+1|k) + \sum_{j\neq i}\left(\mathbf{A_i}\mathbf{B_{ij}}\widehat{\mathbf{T}}_{\mathbf{zj}}(\mathrm{k}) + \mathbf{B_{ij}}\widehat{\mathbf{T}}_{\mathbf{zj}}(\mathrm{k}+1|\mathrm{k})\right)\end{aligned}$$

$$\vdots$$

$$\begin{aligned}\mathbf{x}_{\mathrm{i}}(k+N|k) &= \mathbf{A_i}^N\mathbf{x}_i(k) + \sum_{i=0}^{N-1}\mathbf{A_i}^{N-1-i}[\mathbf{B_{ui}}\mathbf{u_i}(k+i|k) + \mathbf{B_{ai}}\mathbf{T_a}(k+i|k) + \mathbf{B_{gi}}\mathbf{T_g}(k+i|k) \\ &\quad + \mathbf{B_{dw,i}}\mathbf{d_{wi}}(k+i|k) + \mathbf{B_{dz,i}}\mathbf{d}_{\mathbf{z}i}(k+i|k) + \sum_{j\neq i}\mathbf{B_{ij}}\widehat{\mathbf{T}}_{\mathbf{zj}}(\mathrm{k}+\mathrm{i}|\mathrm{k})]\end{aligned}$$

The above equations can be succintly written as

$$\bar{\mathbf{x}}_{\mathrm{i}} = \mathbf{M_i}\mathbf{x}_{\mathrm{i}}(k) + \mathbf{S_{ui}}\bar{\mathbf{u}}_{\mathrm{i}} + \mathbf{S_{ai}}\bar{\mathbf{T}}_{\mathbf{a}} + \mathbf{S_{gi}}\bar{\mathbf{T}}_{\mathbf{g}} + \mathbf{S_{dw,i}}\bar{\mathbf{d}}_{\mathbf{wi}} + \mathbf{S_{dz,i}}\bar{\mathbf{d}}_{\mathbf{zi}} + \textstyle\sum_{j\neq i}\mathbf{S_{ij}}\bar{\widehat{\mathbf{T}}}_{\mathbf{zj}} \qquad (3.26)$$

Here,

$$\bar{\mathbf{x}}_{\mathrm{i}} = \begin{bmatrix} \mathbf{x}_{\mathrm{i}}(k+1|k) \\ \mathbf{x}_{\mathrm{i}}(k+2|k) \\ \vdots \\ \mathbf{x}_{\mathrm{i}}(k+N|k) \end{bmatrix}, \bar{\mathbf{T}}_{\mathbf{a}} = \begin{bmatrix} \mathbf{T}_{\mathbf{a}}(k) \\ \mathbf{T}_{\mathbf{a}}(k+1|k) \\ \vdots \\ \mathbf{T}_{\mathbf{a}}(k+N-1|k) \end{bmatrix}, \bar{\mathbf{T}}_{\mathbf{g}} = \begin{bmatrix} \mathbf{T}_{\mathbf{g}}(k) \\ \mathbf{T}_{\mathbf{g}}(k+1|k) \\ \vdots \\ \mathbf{T}_{\mathbf{g}}(k+N-1|k) \end{bmatrix},$$

$$\bar{\mathbf{d}}_{\mathbf{wi}} = \begin{bmatrix} \mathbf{d}_{\mathbf{wi}}(k) \\ \mathbf{d}_{\mathbf{wi}}(k+1|k) \\ \vdots \\ \mathbf{d}_{\mathbf{wi}}(k+N-1|k) \end{bmatrix}, \bar{\mathbf{d}}_{\mathbf{zi}} = \begin{bmatrix} \mathbf{d}_{\mathbf{zi}}(k) \\ \mathbf{d}_{\mathbf{z}i}(k+1|k) \\ \vdots \\ \mathbf{d}_{\mathbf{zi}}(k+N-1|k) \end{bmatrix}, \bar{\hat{\mathbf{T}}}_{\mathbf{zj}} = \begin{bmatrix} \hat{\mathbf{T}}_{\mathbf{zj}}(k) \\ \hat{\mathbf{T}}_{\mathbf{zj}}(k+1|k) \\ \vdots \\ \hat{\mathbf{T}}_{\mathbf{zj}}(k+N-1|k) \end{bmatrix}$$

$$\mathbf{M}_{\mathrm{i}} = \begin{bmatrix} \mathbf{A}_{\mathrm{i}} \\ \mathbf{A}_{\mathrm{i}}^{2} \\ \vdots \\ \mathbf{A}_{\mathrm{i}}^{N} \end{bmatrix}, \mathbf{S}_{\mathbf{ui}} = \begin{bmatrix} \mathbf{B}_{\mathbf{ui}} & \mathbf{0} & \dots & \mathbf{0} \\ \mathbf{A}_{\mathrm{i}}\mathbf{B}_{\mathbf{ui}} & \ddots & \ddots & \vdots \\ \vdots & \ddots & \mathbf{B}_{\mathbf{ui}} & \mathbf{0} \\ \mathbf{A}_{\mathrm{i}}^{N-1}\mathbf{B}_{\mathbf{ui}} & \dots & \mathbf{A}_{\mathrm{i}}\mathbf{B}_{\mathbf{ui}} & \mathbf{B}_{\mathbf{ui}} \end{bmatrix}, \mathbf{S}_{\mathbf{ai}} = \begin{bmatrix} \mathbf{B}_{\mathbf{ai}} & \mathbf{0} & \dots & \mathbf{0} \\ \mathbf{A}_{\mathrm{i}}\mathbf{B}_{\mathbf{ai}} & \ddots & \ddots & \vdots \\ \vdots & \ddots & \mathbf{B}_{\mathbf{ai}} & \mathbf{0} \\ \mathbf{A}_{\mathrm{i}}^{N-1}\mathbf{B}_{\mathbf{ai}} & \dots & \mathbf{A}_{\mathrm{i}}\mathbf{B}_{\mathbf{ai}} & \mathbf{B}_{\mathbf{ai}} \end{bmatrix},$$

$$\mathbf{S}_{\mathbf{gi}} = \begin{bmatrix} \mathbf{B}_{\mathbf{gi}} & \mathbf{0} & \dots & \mathbf{0} \\ \mathbf{A}_{\mathrm{i}}\mathbf{B}_{\mathbf{gi}} & \ddots & \ddots & \vdots \\ \vdots & \ddots & \mathbf{B}_{\mathbf{gi}} & \mathbf{0} \\ \mathbf{A}_{\mathrm{i}}^{N-1}\mathbf{B}_{\mathbf{gi}} & \dots & \mathbf{A}_{\mathrm{i}}\mathbf{B}_{\mathbf{gi}} & \mathbf{B}_{\mathbf{gi}} \end{bmatrix}, \mathbf{S}_{\mathbf{dw,i}} = \begin{bmatrix} \mathbf{B}_{\mathbf{dw,i}} & \mathbf{0} & \dots & \mathbf{0} \\ \mathbf{A}_{\mathrm{i}}\mathbf{B}_{\mathbf{dw,i}} & \ddots & \ddots & \vdots \\ \vdots & \ddots & \mathbf{B}_{\mathbf{dw,i}} & \mathbf{0} \\ \mathbf{A}_{\mathrm{i}}^{N-1}\mathbf{B}_{\mathbf{dw,i}} & \dots & \mathbf{A}_{\mathrm{i}}\mathbf{B}_{\mathbf{dw,i}} & \mathbf{B}_{\mathbf{dw,i}} \end{bmatrix},$$

$$\mathbf{S}_{\mathbf{dz,i}} = \begin{bmatrix} \mathbf{B}_{\mathbf{dz,i}} & \mathbf{0} & \dots & \mathbf{0} \\ \mathbf{A}_{\mathrm{i}}\mathbf{B}_{\mathbf{dz,i}} & \ddots & \ddots & \vdots \\ \vdots & \ddots & \mathbf{B}_{\mathbf{dz,i}} & \mathbf{0} \\ \mathbf{A}_{\mathrm{i}}^{N-1}\mathbf{B}_{\mathbf{dz,i}} & \dots & \mathbf{A}_{\mathrm{i}}\mathbf{B}_{\mathbf{dz,i}} & \mathbf{B}_{\mathbf{dz,i}} \end{bmatrix}, \text{and } \mathbf{S}_{\mathbf{ij}} = \begin{bmatrix} \mathbf{B}_{\mathbf{ij}} & \mathbf{0} & \dots & \mathbf{0} \\ \mathbf{A}_{\mathrm{i}}\mathbf{B}_{\mathbf{ij}} & \ddots & \ddots & \vdots \\ \vdots & \ddots & \mathbf{B}_{\mathbf{ij}} & \mathbf{0} \\ \mathbf{A}_{\mathrm{i}}^{N-1}\mathbf{B}_{\mathbf{ij}} & \dots & \mathbf{A}_{\mathrm{i}}\mathbf{B}_{\mathbf{ij}} & \mathbf{B}_{\mathbf{ij}} \end{bmatrix}$$

From ( 3.21 ), $J_i^{dc}$ can be re-stated as

$$J_i^{dc} = \left(\bar{\mathbf{T}}_{\mathbf{zi}} - \bar{\mathbf{T}}_{\mathbf{zi,ref}}\right)^T \mathbf{Q}_{\mathbf{1,i}}\left(\bar{\mathbf{T}}_{\mathbf{zi}} - \bar{\mathbf{T}}_{\mathbf{zi,ref}}\right) + \bar{\mathbf{u}}_{\mathbf{i}}^T \mathbf{Q}_{\mathbf{2,i}} \bar{\mathbf{u}}_{\mathbf{i}}, \tag{3.27}$$

where,

$$\bar{\mathbf{T}}_{\mathbf{zi}} = \begin{bmatrix} \mathbf{T}_{\mathbf{z}i}(k+1|k) \\ \mathbf{T}_{\mathbf{zi}}(k+2|k) \\ \vdots \\ \mathbf{T}_{\mathbf{zi}}(k+N|k) \end{bmatrix}, \bar{\mathbf{T}}_{\mathbf{zi,ref}} = \begin{bmatrix} \mathbf{T}_{\mathbf{zi,ref}}(k) \\ \mathbf{T}_{\mathbf{zi,ref}}(k) \\ \vdots \\ \mathbf{T}_{\mathbf{zi,ref}}(k) \end{bmatrix}_{N \times N_{zi}},$$

$$\mathbf{Q}_{\mathbf{1,i}} = \begin{bmatrix} diag(\boldsymbol{\beta}_i) & \mathbf{0} & \dots & \mathbf{0} \\ \mathbf{0} & diag(\boldsymbol{\beta}_{\mathbf{i}}) & \dots & \mathbf{0} \\ \vdots & \vdots & \ddots & \vdots \\ \mathbf{0} & \mathbf{0} & \dots & diag(\boldsymbol{\beta}_i) \end{bmatrix}_{(N_{zi}.N) \times (N_{zi}.N)},$$

and $\mathbf{Q_{2,i}} = \begin{bmatrix} diag(\boldsymbol{\alpha}_i) & \mathbf{0} & \dots & \mathbf{0} \\ \mathbf{0} & diag(\boldsymbol{\alpha}_i) & \dots & \mathbf{0} \\ \vdots & \vdots & \ddots & \vdots \\ \mathbf{0} & \mathbf{0} & \dots & diag(\boldsymbol{\alpha}_\mathbf{i}) \end{bmatrix}_{(N_{zi}.N)\times(N_{zi}.N)}$ .

Let $\mathbf{C}_\mathrm{i} = [\mathbf{0}_{N_{zi}\times N_{wi}} \quad \mathbf{I}_{N_{zi}}]$ and $\bar{\mathbf{C}}_\mathrm{i} = \begin{bmatrix} \mathbf{C_i} & \mathbf{0} & \dots & \mathbf{0} \\ \mathbf{0} & \mathbf{C_i} & \dots & \mathbf{0} \\ \vdots & \vdots & \ddots & \vdots \\ \mathbf{0} & \mathbf{0} & \dots & \mathbf{C_i} \end{bmatrix}_{(N_{zi}.N)\times((N_{wi}+N_{zi}).N)}$ .

$$\text{Clearly, } \overline{\mathbf{T}}_\mathbf{zi} = \bar{\mathbf{C}}_\mathbf{i}\bar{\mathbf{x}}_\mathbf{i} \,. \tag{3.28}$$

Using ( 3.26 ), ( 3.28 ) can be written as

$$\overline{\mathbf{T}}_\mathbf{zi} = \bar{\mathbf{C}}_\mathbf{i}\Big(\mathbf{M_i x}_\mathrm{i}(k) + \mathbf{S_{ui}}\bar{\mathbf{u}}_\mathbf{i} + \mathbf{S_{ai}}\overline{\mathbf{T}}_\mathbf{a} + \mathbf{S_{gi}}\overline{\mathbf{T}}_\mathbf{g} + \mathbf{S_{dw,i}}\bar{\mathbf{d}}_\mathbf{wi} + \mathbf{S_{dz,i}}\bar{\mathbf{d}}_\mathbf{zi} + \textstyle\sum_{j\neq i} \mathbf{S_{ij}}\hat{\overline{\mathbf{T}}}_\mathbf{zj}\Big)\,. \tag{3.29}$$

Substituting $\overline{\mathbf{T}}_\mathbf{zi}$ from (3.29) in (3.27) and ignoring the terms which are independent of $\bar{\mathbf{u}}_\mathbf{i}$ and therefore do not affect the optimization, we obtain

$$J_i^{\,dc} = \bar{\mathbf{u}}_\mathbf{i}^{\,T}\mathbf{H}_\mathbf{dc,i}\bar{\mathbf{u}}_\mathrm{i} + \mathbf{f}_\mathbf{dc,i}^T\bar{\mathbf{u}}_\mathbf{i} \tag{3.30}$$

where,
$$\mathbf{H}_\mathbf{dc,i} = \mathbf{S}_\mathbf{ui}^T\bar{\mathbf{C}}_\mathrm{i}^{\,T}\mathbf{Q}_\mathbf{1,i}\bar{\mathbf{C}}_\mathrm{i}\mathbf{S}_\mathbf{ui} + \mathbf{Q}_{\mathbf{2},\mathrm{i}}, \tag{3.31}$$

$\mathbf{f}_\mathbf{dc,i} =$

$$2\mathbf{S}_\mathbf{ui}^T\bar{\mathbf{C}}_\mathrm{i}^{\,T}\mathbf{Q}_\mathbf{1,i}\left[\bar{\mathbf{C}}_\mathrm{i}\Big(\mathbf{M_i x}_\mathrm{i}(\mathbf{k}) + \mathbf{S_{ai}}\overline{\mathbf{T}}_\mathbf{a} + \mathbf{S_{gi}}\overline{\mathbf{T}}_\mathbf{g} + \mathbf{S_{dw,i}}\bar{\mathbf{d}}_\mathbf{wi} + \mathbf{S_{dz,i}}\bar{\mathbf{d}}_\mathbf{z} + \textstyle\sum_{j\neq i} \mathbf{S_{ij}}\hat{\overline{\mathbf{T}}}_\mathbf{zj}\Big) - \overline{\mathbf{T}}_\mathbf{zi,ref}\right]. \tag{3.32}$$

Therefore, the optimization problem corresponding to decentralized MPC can be re-stated as the Quadratic Program (QP):

$$\bar{\mathbf{u}}_\mathbf{dc,i}^* = \arg\min_{\bar{\mathbf{u}}_\mathrm{i}} \; g_{dc,i}(\bar{\mathbf{u}}_\mathrm{i}). \tag{3.33}$$

where,
$$g_{dc,i}(\bar{\mathbf{u}}_\mathbf{i}) = \bar{\mathbf{u}}_\mathbf{i}^{\,T}\mathbf{H}_\mathbf{dc,i}\bar{\mathbf{u}}_\mathbf{i} + \mathbf{f}_\mathbf{dc,i}^T\bar{\mathbf{u}}_\mathbf{i}. \tag{3.34}$$

# Chapter 4
# Control Architecture Selection for Building Thermal Control

## 4.1 Introduction

The role of control architecture in achieving the thermal control objectives was described in sections 1.2.1 and 3.1.2 In particular, the fundamental tradeoff between optimality and robustness with regard to the 'degree' of decentralization was examined. The focus of this chapter is on the development of appropriate decentralized control architecture selection methodologies which consider the afore-mentioned tradeoff between optimality and robustness. Two different approaches – CLF-MCS method and OLF-FPM method – are presented and demonstrated using examples. The reader is directed to sections 3.3.1 and 3.4.1 for details on centralized and decentralized frameworks, which form the basis for these methodologies.

## 4.2 CLF-MCS approach

### 4.2.1 Overview

The CLF-MCS approach is a procedure for partitioning a building based on Coupling Loss factor (CLF) and Mean Cluster Size (MCS) as optimality and robustness metrics respectively. A divisive clustering procedure employs these metrics to create a family of partitions using combinatorial optimization. These partitions are then analyzed to select those which provide satisfactory trade-offs between optimality and robustness. A MINCUT approximation for the underlying combinatorial optimization problem is also presented to address computational concerns. The methodology is demonstrated using simulated examples.

## 4.2.2 Coupling between inputs and clusters of inputs

The concepts of coupling between inputs and clusters of inputs are presented here as preliminaries used in the development of the Coupling Loss Factor as an optimality metric.

### 4.2.2.1 Coupling between inputs

We consider the centralized MPC framework described in Section 3.3. For a prediction horizon of length $N$ samples, we introduce a vector $\bar{\mathbf{v}}$ defined at any time instant $k$ as

$$\bar{\mathbf{v}} = \begin{bmatrix} \bar{\mathbf{v}}_1 \\ \bar{\mathbf{v}}_2 \\ \vdots \\ \bar{\mathbf{v}}_{N_z} \end{bmatrix} \tag{4.1}$$

where,
$$\bar{\mathbf{v}}_i = \begin{bmatrix} u_i(k|k) \\ u_i(k+1|k) \\ \vdots \\ u_i(k+N-1|k) \end{bmatrix} \text{ for } i \in \{1,2,\dots,N_z\}. \tag{4.2}$$

Here, $u_i$ denotes the $i^{th}$ component of the vector of control inputs $\mathbf{u}$. The vector $\bar{\mathbf{u}}$ was defined in (3.12) as

$$\bar{\mathbf{u}} = \begin{bmatrix} \mathbf{u}(k|k) \\ \mathbf{u}(k+1|k) \\ \vdots \\ \mathbf{u}(k+N-1|k) \end{bmatrix} \tag{4.3}$$

Comparing (4.1) and (4.3), we observe that the entries of $\bar{\mathbf{v}}$ form a permutation of the entries of $\bar{\mathbf{u}}$. Therefore, $\bar{\mathbf{v}}$ can be expressed as shown in (4.4), where $\mathbf{P_{vu}}$ is a permutation matrix.

$$\bar{\mathbf{v}} = \mathbf{P_{vu}}\bar{\mathbf{u}}\,. \tag{4.4}$$

Using (4.4) in (3.20), the objective function for centralized MPC can be written as

$$g_c(\bar{\mathbf{u}}) = g_c'(\bar{\mathbf{v}}) = \bar{\mathbf{v}}^T\bar{\mathbf{H}}\bar{\mathbf{v}} + \bar{\mathbf{f}}^{\mathrm{T}}\bar{\mathbf{v}} \tag{4.5}$$

Where,
$$\bar{\mathbf{H}} = \mathbf{P_{vu}}\mathbf{H_c}\mathbf{P_{vu}^{-1}}, \tag{4.6}$$

and,
$$\bar{\mathbf{f}} = \mathbf{P_{vu}}\mathbf{f_c}^T. \tag{4.7}$$

The quadratic part, $\bar{\mathbf{v}}^T\bar{\mathbf{H}}\bar{\mathbf{v}}$, can be written in the expanded form as:

$$\begin{bmatrix}\bar{\mathbf{v}}_1\\ \bar{\mathbf{v}}_2\\ \vdots\\ \bar{\mathbf{v}}_{N_z}\end{bmatrix}^T \begin{bmatrix}\bar{\mathbf{H}}_{1,1} & \bar{\mathbf{H}}_{1,2} & \cdots & \bar{\mathbf{H}}_{1,N_z}\\ \bar{\mathbf{H}}_{2,1} & \bar{\mathbf{H}}_{2,2} & \cdots & \bar{\mathbf{H}}_{2,N_z}\\ \vdots & \vdots & \ddots & \vdots\\ \bar{\mathbf{H}}_{N_z,1} & \bar{\mathbf{H}}_{N_z,2} & \cdots & \bar{\mathbf{H}}_{N_z,N_z}\end{bmatrix}\begin{bmatrix}\bar{\mathbf{v}}_1\\ \bar{\mathbf{v}}_2\\ \vdots\\ \bar{\mathbf{v}}_{N_z}\end{bmatrix}$$

Each off-diagonal term, $\bar{\mathbf{H}}_{i,j} \in \mathbb{R}^{N\times N}, i \neq j$ represents the coupling between $u_i$ and $u_j$ in $g_c(\bar{\mathbf{u}})$. Therefore, we use $\left\|\bar{\mathbf{H}}_{i,j}\right\|_2$ as a measure of coupling[1] between $u_i$ and $u_j$ and extend this to define coupling between between a pairs of input clusters in Section 4.2.2.2.

#### 4.2.2.2 Coupling between clusters of inputs

Consider a p-partition of the building from definition 3.2. Consider any two clusters $S_{zi}$ and $S_{zj}$ from this partition. The coupling matrix between these clusters, $\bar{\mathbf{H}}_{S_{zi},S_{zj}}$ is defined as

$$\bar{\mathbf{H}}_{S_{zi},S_{zj}} = \begin{pmatrix}\bar{\mathbf{H}}_{p_1,q_1} & \bar{\mathbf{H}}_{p_1,q_2} & . & . & .\\ \bar{\mathbf{H}}_{p_2,q_1} & \bar{\mathbf{H}}_{p_2,q_2} & . & . & .\\ . & . & . & . & .\\ . & . & . & . & .\\ . & . & . & . & .\end{pmatrix}, \qquad (4.8)$$

where, $p_1, p_2 \ldots \in S_{zi}$ and $q_1, q_2 \ldots \in S_{zj}$.

The coupling $\mathcal{C}(S_{zi}, S_{z,j})$ between $S_{zi}$ and $S_{z,j}$ is then defined as

$$\mathcal{C}(S_{zi}, S_{zj}) = \left\|\bar{\mathbf{H}}_{S_{zi},S_{zj}}\right\|_2. \qquad (4.9)$$

### 4.2.3 Overview of divisive clustering approach

The CLF-MCS clustering procedure is carried out in a divisive sequence as illustrated in Figure 4.1. The input to each stage is a set of *parent clusters*, and the output is a set of *child clusters*. The child clusters are obtained from the parent clusters via combinatorial analysis. The input to the first stage is the root cluster containing all the control inputs, which represents the completely centralized case. The output of the last stage is a set where each control input is a cluster by itself and hence represents a fully decentralized architecture. For any intermediate stage $\text{STG}_i$, the input (set of parent clusters) is the same as the output (set of child clusters) of the previous stage $\text{STG}_{i-1}$. Two metrics representing optimality and robustness are computed for each stage. A plot of one metric versus the other is then used to identify the stage which results in a satisfactory tradeoff between robustness and optimality.

[1] It is important to scale the system first so that coupling metrics corresponding to different pairs of input channels can be compared with one another. For a discussion on scaling see [83].

### 4.2.4 Optimality and robustness metrics

Two dimensionless metrics - Coupling Loss Factor (CLF) and Mean Cluster Size (MCS) are computed for each partitioning stage.

#### 4.2.4.1 Coupling Loss Factor

The CLF for stage $\text{STG}_i$ is a normalized measure of the inter-cluster coupling among its child clusters that are denoted by $S_{z,j}^{i}$, where $j = 1,2,\ldots,n_i$. Here, $n_i$ is the total number of such child clusters. First, we introduce the coupling loss vector $\boldsymbol{\mu}_i$ for this stage $\text{STG}_i$ as the vector of the couplings $\boldsymbol{\mathcal{C}}\left(S_{z,m}^{i}, S_{z,n}^{i}\right)$ for each pair of child clusters, $\left(S_{z,m}^{i}, S_{z,n}^{i}\right)$ with $m \neq n$. More formally,

$$\boldsymbol{\mu}_i = \begin{pmatrix}\mu_{i,1} & \mu_{i,2} & \cdots & \mu_{i,n_i-1}\end{pmatrix}^T \tag{4.10}$$

where, $\mu_{i,l} = \left[\boldsymbol{\mathcal{C}}\left(S_{z,l}^{i}, S_{z,l+1}^{i}\right) \;\; \boldsymbol{\mathcal{C}}\left(S_{z,l}^{i}, S_{z,l+2}^{i}\right) \;\; \ldots \;\; \boldsymbol{\mathcal{C}}\left(S_{z,l}^{i}, S_{z,n_i}^{i}\right)\right]$ for all $l \in \{1,2,\ldots n_i\}$.

The CLF for stage $\text{STG}_i$, $\text{CLF}_i$ is then defined as

$$\text{CLF}_i = \frac{\|\boldsymbol{\mu}_i\|_2}{\|\overline{\mathbf{H}}\|_2}. \tag{4.11}$$

Here $\overline{\mathbf{H}}$ is the matrix defined in 4.6. The CLF for the parent partition to stage 1, which represents the fully centralized scenario, is clearly zero. $\text{CLF}_i$ measures the coupling that gets ignored if the system were partitioned according to the child clusters of stage $\text{STG}_i$. Therefore, it is desired to partition the system such that the corresponding CLF is small so that the resulting deviation in optimality from centralized control is small.

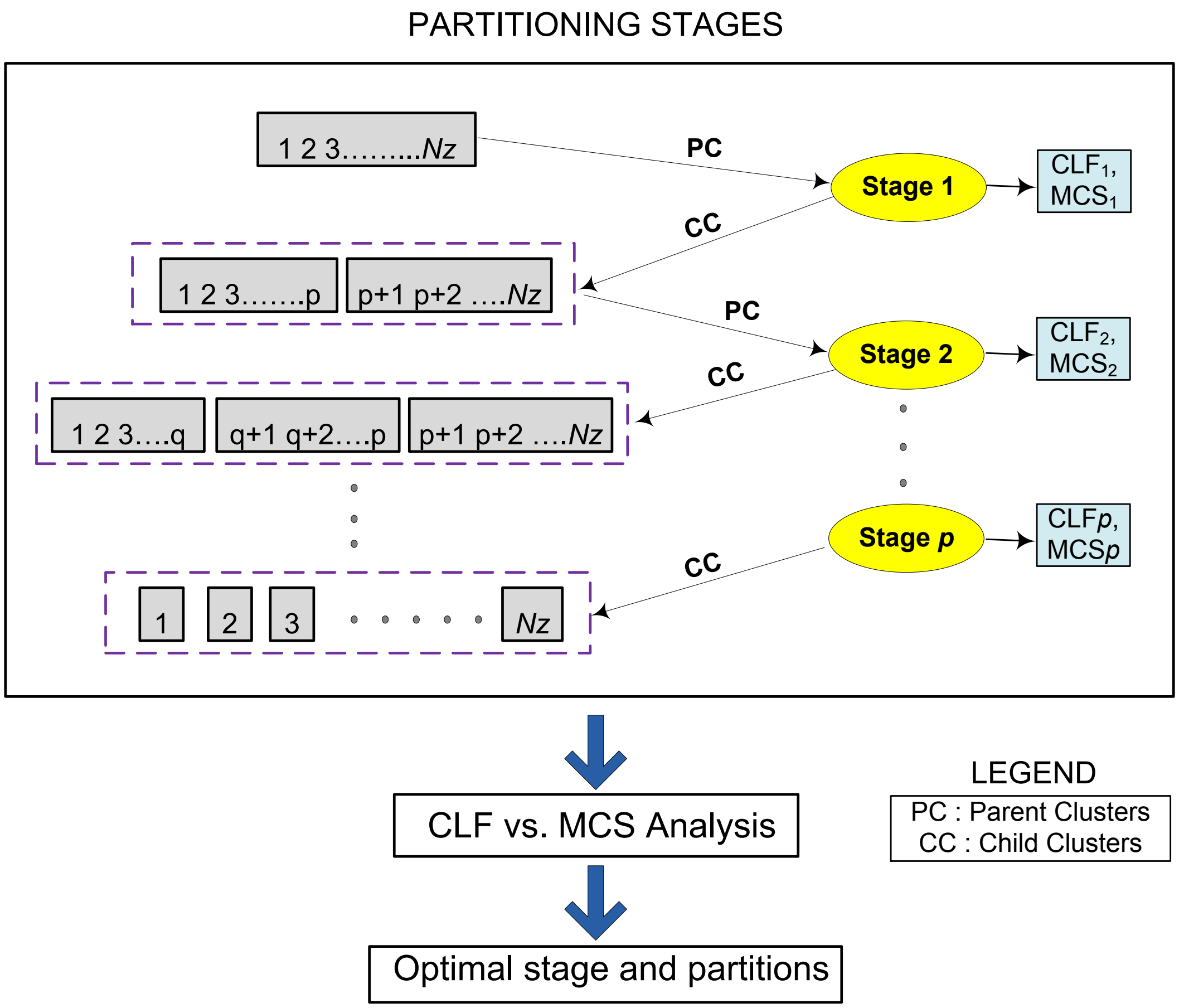

**Figure 4.1 Overview of the CLF-MCS clustering procedure**

### 4.2.4.2 Mean Cluster Size

We use $\lambda_j^i$ to denote the number of elements in child cluster $S_{z,j}^i$. $\text{MCS}_i$ for stage $\text{STG}_i$ is defined as the average number of zones per child cluster normalized with respect to the total number of zones.

$$\text{MCS}_i = \frac{\sum_{j=1}^{n_i} \lambda_j^i}{n_i N_z} = \frac{1}{n_i}. \qquad (4.12)$$

It is clear that $\text{MCS}_i \in (0,1]$. In a decentralized control architecture, the effect of a sensor or communication related fault is confined to the cluster where it originates. Therefore, the MCS is an indicator of robustness to such faults - a small value indicates that the number of clusters is large and thus the effect of failures is less widespread. Hence, it is desired to partition the building such that the corresponding MCS is small.

### 4.2.5 Stage level combinatorial optimization

As shown in Figure 4.1, for each stage, the input is a set of parent clusters and the output is a set of child clusters. The objective of the stage level optimization is to appropriately split the parent clusters to obtain corresponding child clusters. This process is based on a combinatorial optimization procedure which is explained below and illustrated in Figure 4.2.

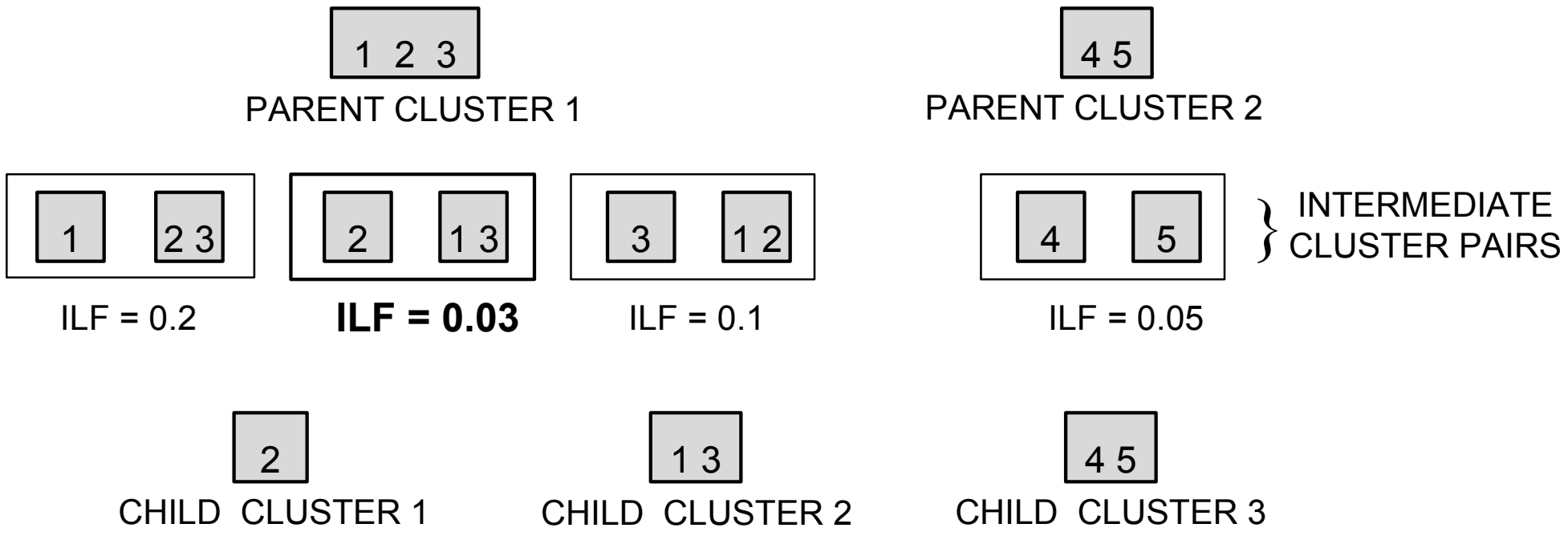

**Figure 4.2 Schematic of combinatorial optimization process for any given stage.**

Since the parent clusters for any stage $\mathrm{STG}_i$ are the child clusters from its preceding stage $\mathrm{STG}_{i-1}$, they are given by $S_{z,j}^{i-1}$ where $j = 1,2,\ldots n_{i-1}$. An *intermediate cluster pair* for any parent cluster is defined as a set of two non-empty clusters obtained by splitting that parent cluster. Therefore, the number of intermediate cluster pairs, $n_{z,j,int}^{i}$ obtained from the parent cluster $S_{z,j}^{i-1}$ is given by by the *Stirling number of the second kind* [84], $S(\lambda_j^{i-1},2)$ expressed as

$$n_{z,j,int}^{i} = S(\lambda_j^{i-1},2) = 2^{\lambda_j^{i-1}-1} - 1. \tag{4.13}$$

The **Intermediate Loss Factor** (ILF) is defined for each intermediate cluster pair $\{S_{z,j,int,l}^{i}, S_{z,j,int,l}^{*i}\}$ of $S_{z,j}^{i-1}$, indexed by $l$ as

$$\mathrm{ILF}_{z,j,int,l}^{i} = \frac{\boldsymbol{c}\left(S_{z,j,int,l}^{i}, S_{z,j,int,l}^{*i}\right)}{\boldsymbol{c}\left(S_{z,j}^{i-1}, S_{z,j}^{i-1}\right)} \tag{4.14}$$

where, $l = 1,2,\ldots n_{z,j,int}^{i}$, and $j = 1,2,\ldots n_{i-1}$.

The underlying optimization problem for the $i^{th}$ stage, $\mathrm{STG}_i$ is to find the parent cluster (indicated by $j^*,\ j^* \in \{1,2,\ldots n_{i-1}\}$) and its corresponding intermediate cluster pair (indicated by $l^*, l^* \in \{1,2,\ldots n^i_{z,j,int}\}$ ) which yield the smallest ILF. This is stated as

$$\{j^*, l^*\} = \underset{\{j,l\}}{\arg\min}\quad \mathrm{ILF}^i_{z,j,int,l} \tag{4.15}$$

The optimal parent cluster is then split to create the optimal intermediate cluster pair, whereas the other parent clusters are retained. In other words, the cluster $S^{i-1}_{z,j^*}$ is partitioned into the clusters $S^i_{z,j^*,int,l^*}$ and $S^{*i}_{z,j^*,int,l^*}$. The result is a set of child clusters having one more cluster than the set of parent clusters.

The ILF defined in (4.14) measures the 'amount of coupling' information lost when a parent cluster is split into two child clusters, normalized with respect to the coupling originally present in the parent. Therefore, the above optimization involves determination of the optimal split in the sense that such a split results in the smallest loss of coupling information among all possible splits.

### 4.2.6 MINCUT approximation

The exponential computational complexity, characterized by (4.13), of the combinatorial optimization motivates the development of a more tractable approach for the minimization problem (4.15). In what follows, for simplicity, we denote the size $\lambda^{i-1}_j$ of the parent cluster $S^{i-1}_{z,j}$ by $n$. The elements of $S^{i-1}_{z,j}$ are accordingly denoted by $p_r$, where $r = 1,2,\ldots,n$.

A matrix $\overline{\mathbf{H}}_j$ is constructed for the $j^{th}$ parent cluster $S^{i-1}_{z,j}$ from its elements in a manner analogous to the construction of the coupling matrix in (4.8)

$$\overline{\mathbf{H}}_j = \begin{pmatrix} \overline{\mathbf{H}}_{p_1,p_1} & \overline{\mathbf{H}}_{p_1,p_2} & \cdots & \overline{\mathbf{H}}_{p_1,p_n} \\ \overline{\mathbf{H}}_{p_2,p_1} & \overline{\mathbf{H}}_{p_2,p_2} & \cdots & \overline{\mathbf{H}}_{p_2,p_n} \\ \vdots & \vdots & \ddots & \vdots \\ \overline{\mathbf{H}}_{p_n,p_1} & \overline{\mathbf{H}}_{p_n,p_2} & \cdots & \overline{\mathbf{H}}_{p_n,p_n} \end{pmatrix}. \tag{4.16}$$

For any given intermediate cluster pair, $\{S^i_{z,j,int,l}, S^{*i}_{z,j,int,l}\}$ a matrix, $\overline{\mathbf{H}}^l_j$ can be obtained from $\overline{\mathbf{H}}_j$ by setting to zero all blocks which correspond to elements in one intermediate cluster only. More precisely,

$$\bar{\mathbf{H}}_j^l = \begin{pmatrix} \theta_{p_1,p_1}\bar{\mathbf{H}}_{p_1,p_1} & \theta_{p_1,p_2}\bar{\mathbf{H}}_{p_1,p_2} & \dots & \theta_{p_1,p_n}\bar{\mathbf{H}}_{p_1,p_n} \\ \theta_{p_2,p_1}\bar{\mathbf{H}}_{p_2,p_1} & \theta_{p_2,p_2}\bar{\mathbf{H}}_{p_2,p_2} & \dots & \theta_{p_2,p_n}\bar{\mathbf{H}}_{p_2,p_n} \\ \vdots & \vdots & \ddots & \vdots \\ \theta_{p_n,p_1}\bar{\mathbf{H}}_{p_n,p_1} & \theta_{p_n,p_1}\bar{\mathbf{H}}_{p_n,p_2} & \dots & \theta_{p_n,p_n}\bar{\mathbf{H}}_{p_n,p_n} \end{pmatrix}, \quad (4.17)$$

where, $$\theta_{p_r,p_s} = \begin{cases} 0 & \text{if } p_r, p_s \in S_{z,j,int,l}^{i} \text{ or } p_r, p_s \in S_{z,j,int,l}^{*i} \\ 1 & \text{otherwise.} \end{cases} \quad (4.18)$$

Using the above definitions, $\text{ILF}_{z,j,int,l}^{i}$ defined in (4.14) can be expressed as

$$\text{ILF}_{z,j,int,l}^{i} = \frac{\|\bar{\mathbf{H}}_j^l\|_2}{\|\bar{\mathbf{H}}_j\|_2}. \quad (4.19)$$

From the above expression, the problem of minimizing $\text{ILF}_{z,j,int,l}^{i}$ over intermediate cluster pairs indexed by $l$, for a particular parent, denoted by a fixed $j$, corresponds to the minimization of $\|\bar{\mathbf{H}}_j^l\|_2$ over $l$. Assuming that $\bar{\mathbf{H}}_j^l$ is sufficiently sparse, we now approximate $\|\bar{\mathbf{H}}_j^l\|_2$ by the 2-norm of the vector $\mathbf{v}_j^l$ consisting of the elements of $\bar{\mathbf{H}}_j^l$.

To make this procedure more formal, we introduce a binary vector $\mathbf{x} \in \mathbb{R}^n$ whose elements, $\mathbf{x}_r$ ($r \in \{1,2,\dots,n\}$) are defined as follows:

$$\mathbf{x}_r = \begin{cases} 1 & \text{if } p_r \in S_{z,j,int,l}^{i} \\ -1 & \text{if } p_r \in S_{z,j,int,l}^{*i} \end{cases} \quad (4.20)$$

$\left(\|\mathbf{v}_j^l\|_2\right)^2$ can be expressed as

$$\left(\|\mathbf{v}_j^l\|_2\right)^2 = \frac{\mathbf{z}^{\mathrm{T}}\mathbf{Q}^{\mathrm{T}}(\bar{\mathbf{H}}_j)^{.2}\mathbf{Q}\mathbf{z} - \mathbf{x}^T\mathbf{Q}^T(\bar{\mathbf{H}}_j)^{.2}\mathbf{Q}\mathbf{x}}{2} \quad (4.21)$$

Here, $(\bar{\mathbf{H}}_j)^{.2}$ denotes the matrix obtained by taking element-wise square of $\bar{\mathbf{H}}_j$. In other words, $(\bar{\mathbf{H}}_j)^{.2}$ is the Hadamard product of $\bar{\mathbf{H}}_j$ with itself. $\mathbf{Q}$ and $\mathbf{z}$ are defined below.

$$\mathbf{Q} = \begin{pmatrix} \mathbf{e}_N & \mathbf{0}_{N\times1} & \dots & \mathbf{0}_{N\times1} \\ \mathbf{0}_{N\times1} & \mathbf{e}_N & \dots & \mathbf{0}_{N\times1} \\ \vdots & \vdots & \ddots & \vdots \\ \mathbf{0}_{N\times1} & \mathbf{0}_{N\times1} & \dots & \mathbf{e}_N \end{pmatrix}_{N.n\times n} \quad \text{where, } \mathbf{e}_N = \begin{bmatrix} 1 \\ 1 \\ \vdots \\ 1 \end{bmatrix}_{N\times1} \quad \text{and} \quad \mathbf{z} = \begin{bmatrix} 1 \\ 1 \\ \vdots \\ 1 \end{bmatrix}_{n\times1}$$

Hence, the problem of minimizing $\text{ILF}_{z,j,int,l}^{i}$ over $l$ for a particular parent $j$ can be approximated by the following Boolean maximization, which represents a MINCUT problem in graph theory:

$$\begin{aligned} &\text{maximize} && \mathbf{x}^T \mathbf{Q}^T (\overline{\mathbf{H}}_j)^{.2} \mathbf{Q}\mathbf{x} \\ &\text{subject to} && \mathbf{x}_r \in \{1, -1\} \text{ for all } r \in \{1,2,\dots,n\}. \end{aligned}$$

The above maximization can be performed using numerical techniques such as [85] available for solving the MINCUT problem. In this way, for each parent $j$, the minimum ILF can be found and compared across all parents to solve the original minimization problem (4.15).

### 4.2.7 Optimal partition selection

Since it is desired to have both CLF and MCS small, this problem is analogous to dual objective optimization in a pareto-optimal setting [82]. Motivated by this, the optimal partition is obtained from a plot of $\text{CLF}_i$ versus $\text{MCS}_i$ illustrated in Figure 4.3. The optimal partition should be a knee point. Therefore, navigating along the curve about that point in either direction would result in a large increase in one metric but only a relatively small decrease in the other metric. The plot must be studied in the ascending order of the partitioning stages (right to left) for knee points. For instance, if the first knee is not 'sufficiently' sharp, then the second knee (if any) should be studied.

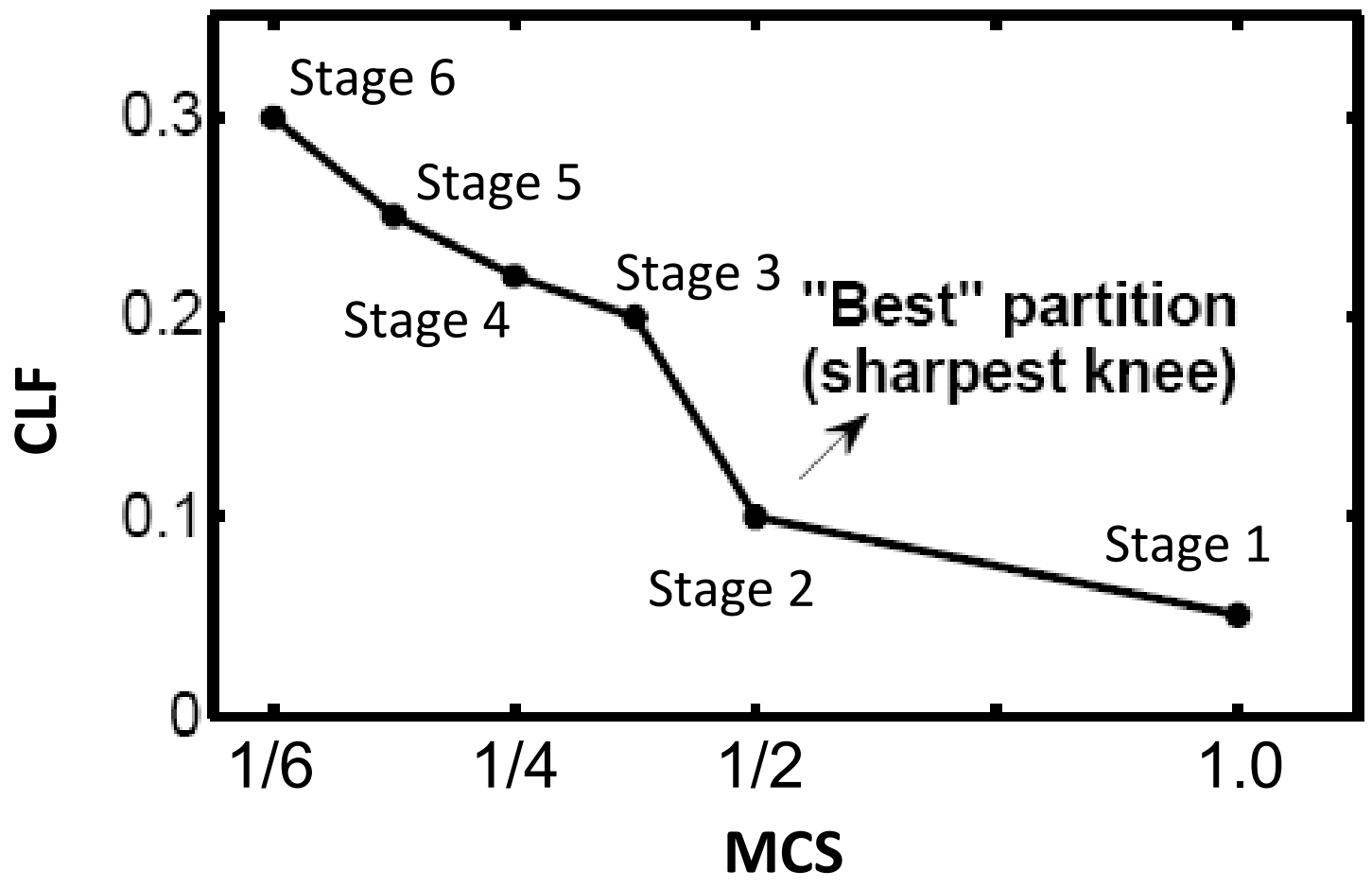

**Figure 4.3 Illustrative example of CLF vs. MCS plot.**

## 4.2.8 Twelve zone building example

The layout of the building used in this example is shown in Figure 4.4. It consists of 3 floors, with a total of 12 rooms of equal dimensions ($5 \times 5 \times 5\ m^3$) numbered as shown. The building is assumed to be surrounded by ambient on all sides. The walls were modeled as RC circuits (Figure 3.5) based on the accessibility factor method described in [38], and the zones were modeled as isolated capacitors. Each room has 6 walls – 4 side walls, 1 ceiling and 1 floor. The construction details are presented in Table 4.1, from which the resistances and capacitances for the walls were computed as shown in Table 4.2. Note that in case of external walls, $R_2$ denotes the resistance between the wall and ambient. The zonal thermal capacities were assumed to be 250 kJ/K based on air at 25 C and $10^5$ Pa. An overall system model of the form (3.4) was obtained by constructing an RC network using these details, and then applying Algorithm 3.1 where discretization was performed using the zero-order-hold method with step size of 10 minutes. This choice was justified based on the fact that it was close to one-tenth of the smallest time constant in the model. The Hessian Matrix $\overline{\mathbf{H}}$ was then created using (4.6) with prediction horizon, $N = 12$ samples (2 hours) and the weights $\boldsymbol{\alpha} = 0.1\mathbf{e}_{12}$ and $\boldsymbol{\beta} = \mathbf{e}_{12}$ (see nomenclature).

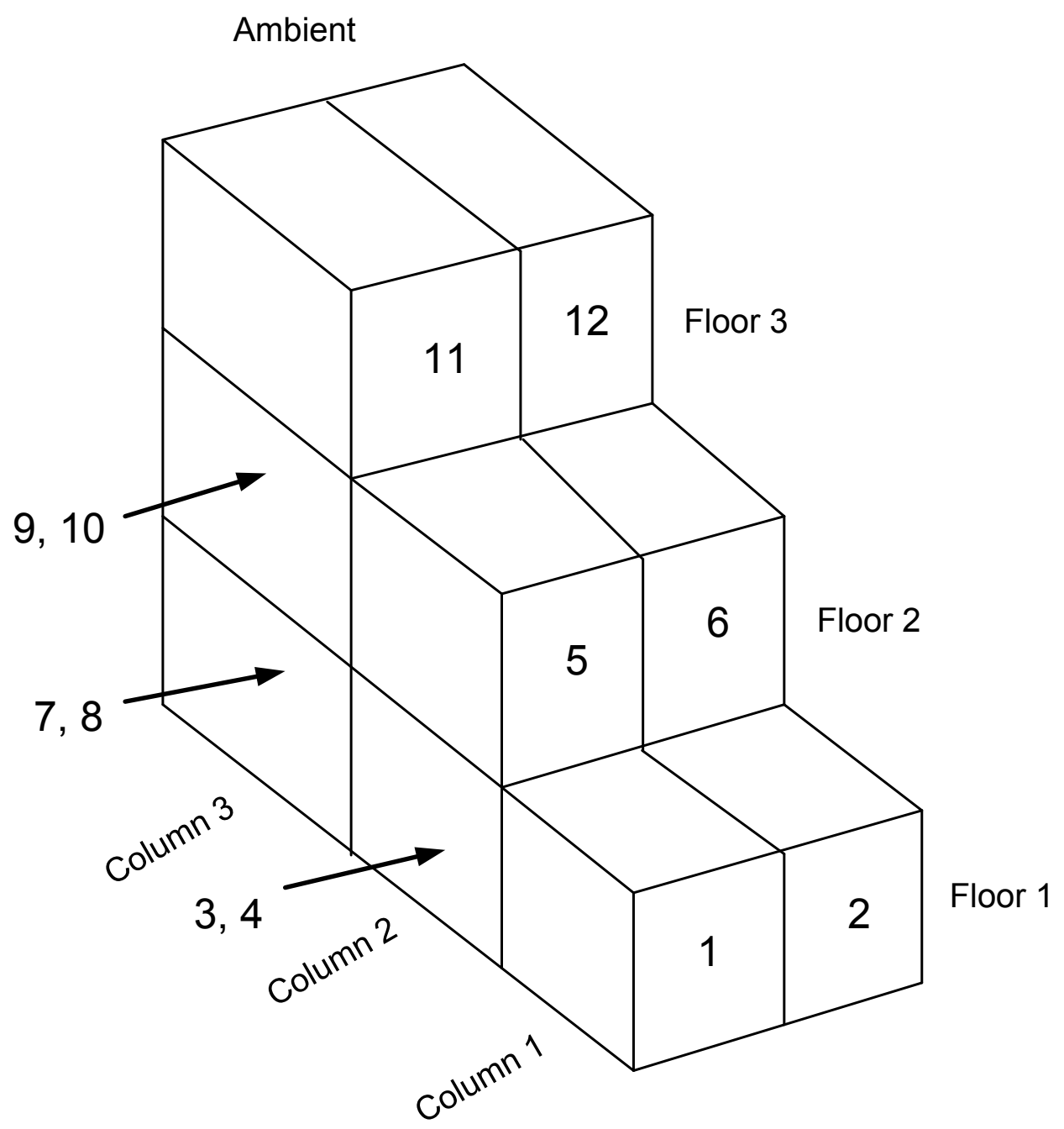

**Figure 4.4 12-zone test building architecture (Zones are numbered)**

**Table 4.1 Construction properties used for the 12-zone building model [40]**

| Element | Layering | Thickness (m) |
|---|---|---|
| External walls | Brick | 0.122 |
| | Insulation | 0.050 |
| | C-Block | 0.112 |
| | Plaster | 0.013 |
| Internal walls | Plaster | 0.013 |
| | C-Block | 0.122 |
| | Plaster | 0.013 |

**Table 4.2 Resistances and capacitances of walls (Refer to Figure 3.5)**

| Type of wall | $C_{wall}$ (kJ/kg) | $R_1$ (K/kW) | $R_2$(K/kW)* |
|---|---|---|---|
| Horizontal External | 8329.15 | 29.99 | 81.08 |
| Vertical External | 8329.15 | 36.84 | 82.00 |
| Horizontal Internal | 4660.00 | 21.32 | 21.32 |
| Vertical Internal | 4660.00 | 21.32 | 21.32 |

As seen in Table 4.2, due to symmetry, the resistances corresponding to both horizontal and vertical internal walls have the same value $R_{nom} = 21.32$ K/kW. In the case study presented, the resistances of the horizontal internal walls (floor separators) are multiplied by a factor of $\rho_1$ to introduce thermal anisotropy. Similarly, the resistances of the *column separating* vertical internal walls (e.g. between zones 1 and 3, 2 and 4 etc.) are scaled by $\rho_2$, and those of the *symmetrically splitting* vertical internal walls (e.g. between zones 1 and 2, 5 and 6, etc.) by $\rho_3$.

The clustering methodology presented in Figure 4.1 was applied for specific values of $\rho_i$ and the results have been summarized in Table 4.3. The relevant codes are provided in Appendix A. The combinatorial optimization in each stage was performed by comparing all possible intermediate cluster pairs. Evaluation of the MINCUT approximation is considered in a different example presented in section 4.2.9. The corresponding CLF vs. MCS plots are shown in Figure 4.5 to Figure 4.8. Important observations are as follows:

1. For the nominal case (case 1, Figure 4.5), a knee point is not obvious. Therefore, stage 3 was chosen to be the optimal partition where both CLF and MCS are satisfactorily small.

2. For cases 2, 3 and 4 the knee points identified for optimal partitioning (Figure 4.6 to Figure 4.8) are stages 3, 3 and 2 respectively. The corresponding clusters obtained (Table 4.3) can be explained on the basis of physical intuition. In case 2, the internal horizontal walls are more insulating than the vertical walls, therefore the building must be partitioned horizontally (along floors). In cases 3 and 4 this is reversed and a subset of vertical walls becomes more insulating. Therefore optimal clustering corresponds to partitions along such walls.

3. Figure 4.6 to Figure 4.8 indicate that clustering stages downstream of stage 3 are not optimal. This can be explained because these stages cause partitions along the low-insulation walls after separation along all high insulation walls has already been completed, therefore causing larger relative losses in coupling.

**Table 4.3 Summary of results for 12-zone test building**

| Case | $\rho_1$ | $\rho_2$ | $\rho_3$ | optimal partition |
|---|---|---|---|---|
| 1 | 1 | 1 | 1 | {1,2,3,4,7,8}, {5,6,9,10}, {11,12} |
| 2 | 3 | 1 | 1 | {1,2,3,4,7,8}, {5,6,9,10},{11,12} |
| 3 | 1 | 3 | 1 | {1,2}, {3,4,5,6}, {7,8,9,10,11,12} |
| 4 | 1 | 1 | 3 | {1,2,5,7,9}, {2,4,6,8,10} |

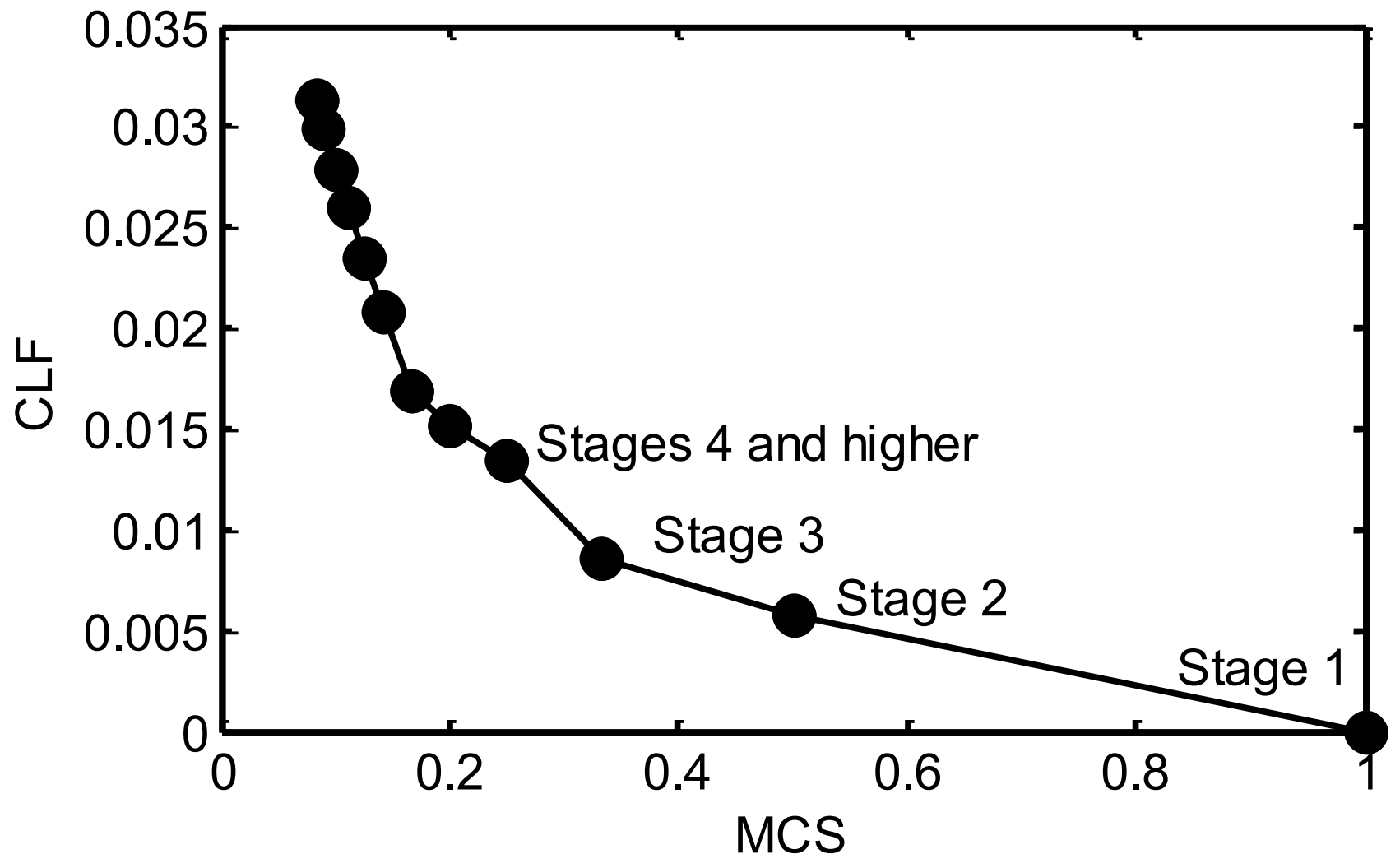

**Figure 4.5 CLF vs. MCS plot for case 1 in Table 4.3**

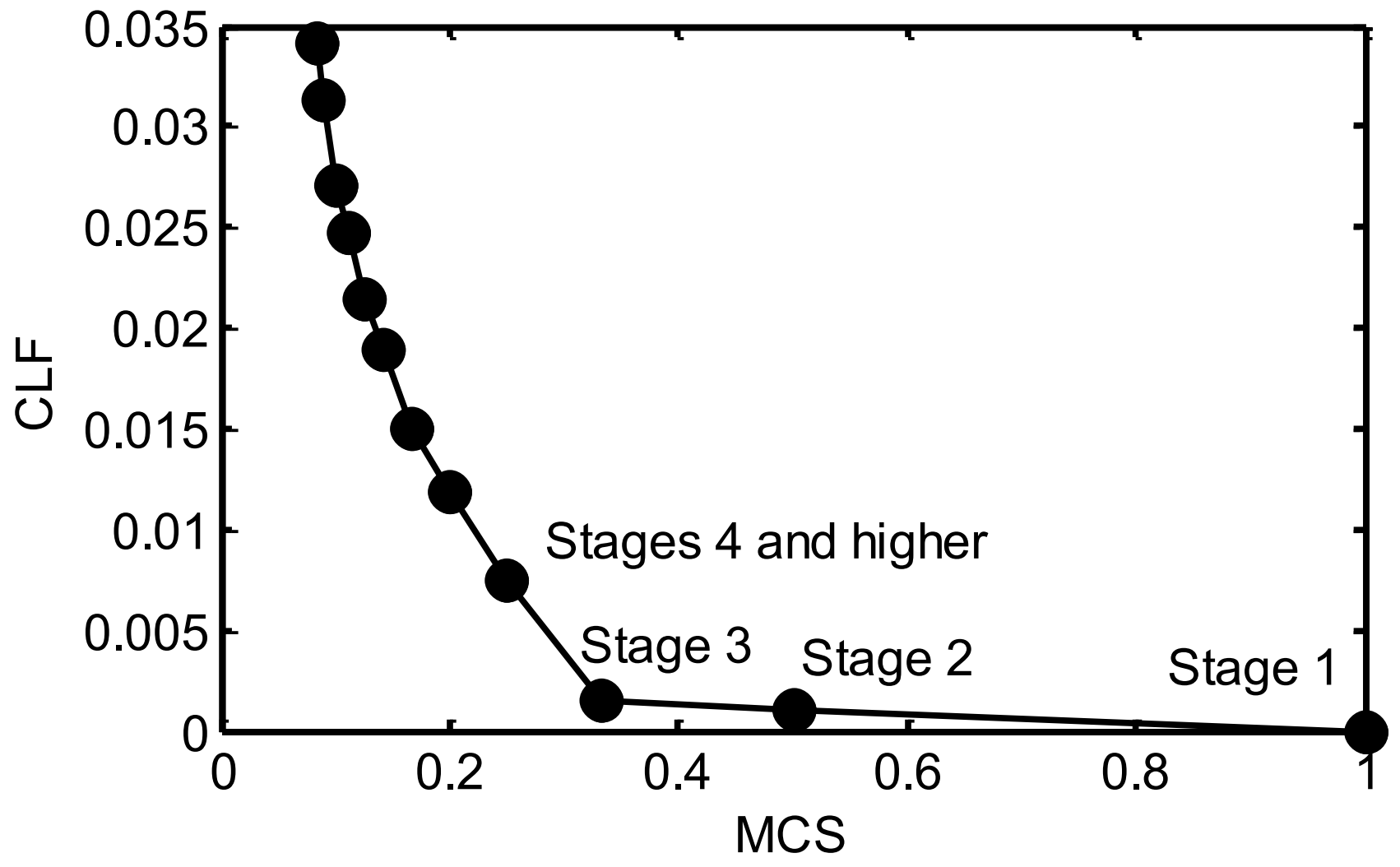

**Figure 4.6 CLF vs. MCS plot for case 2 in Table 4.3**

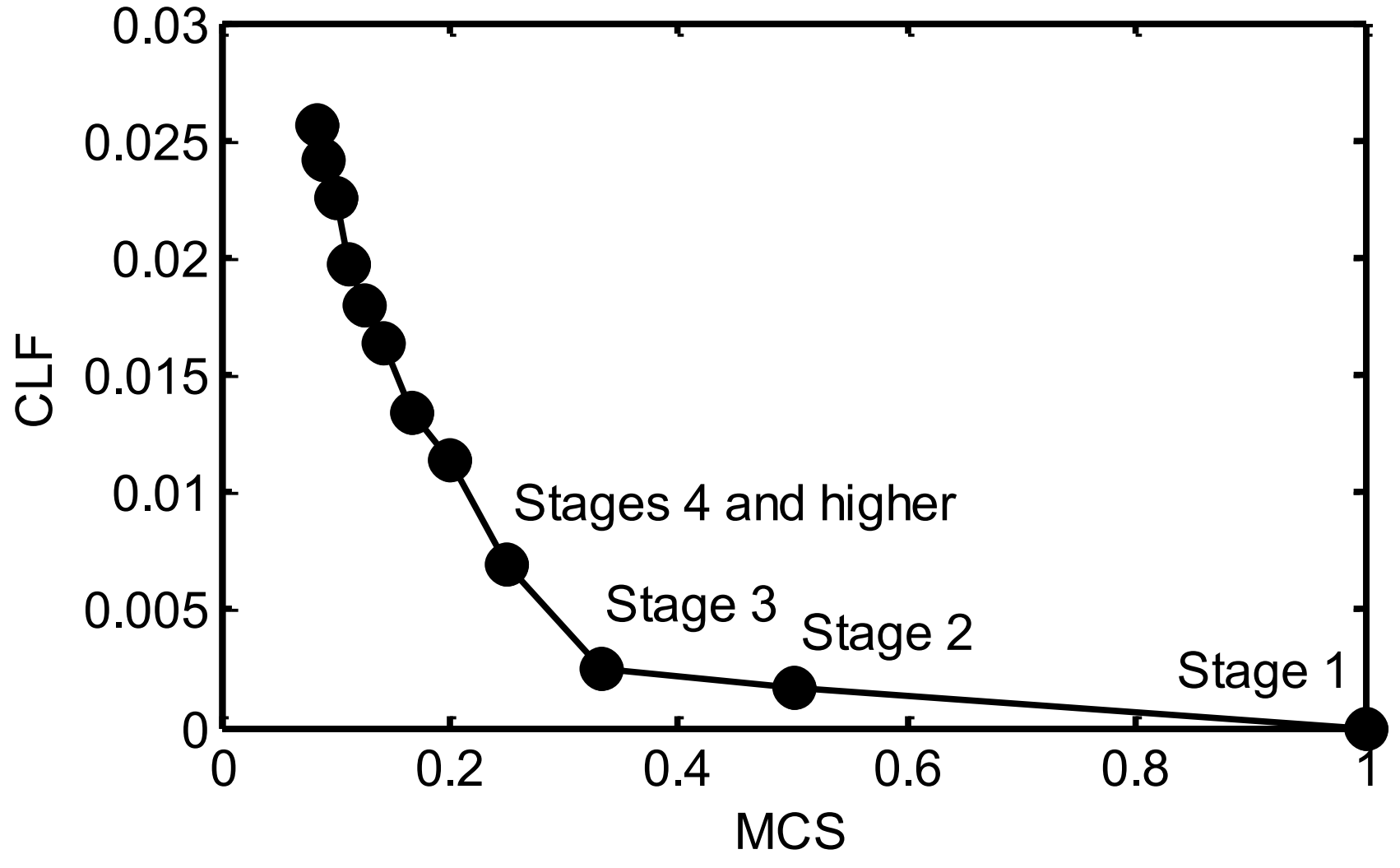

**Figure 4.7 CLF vs. MCS plot for case 3 in Table 4.3**

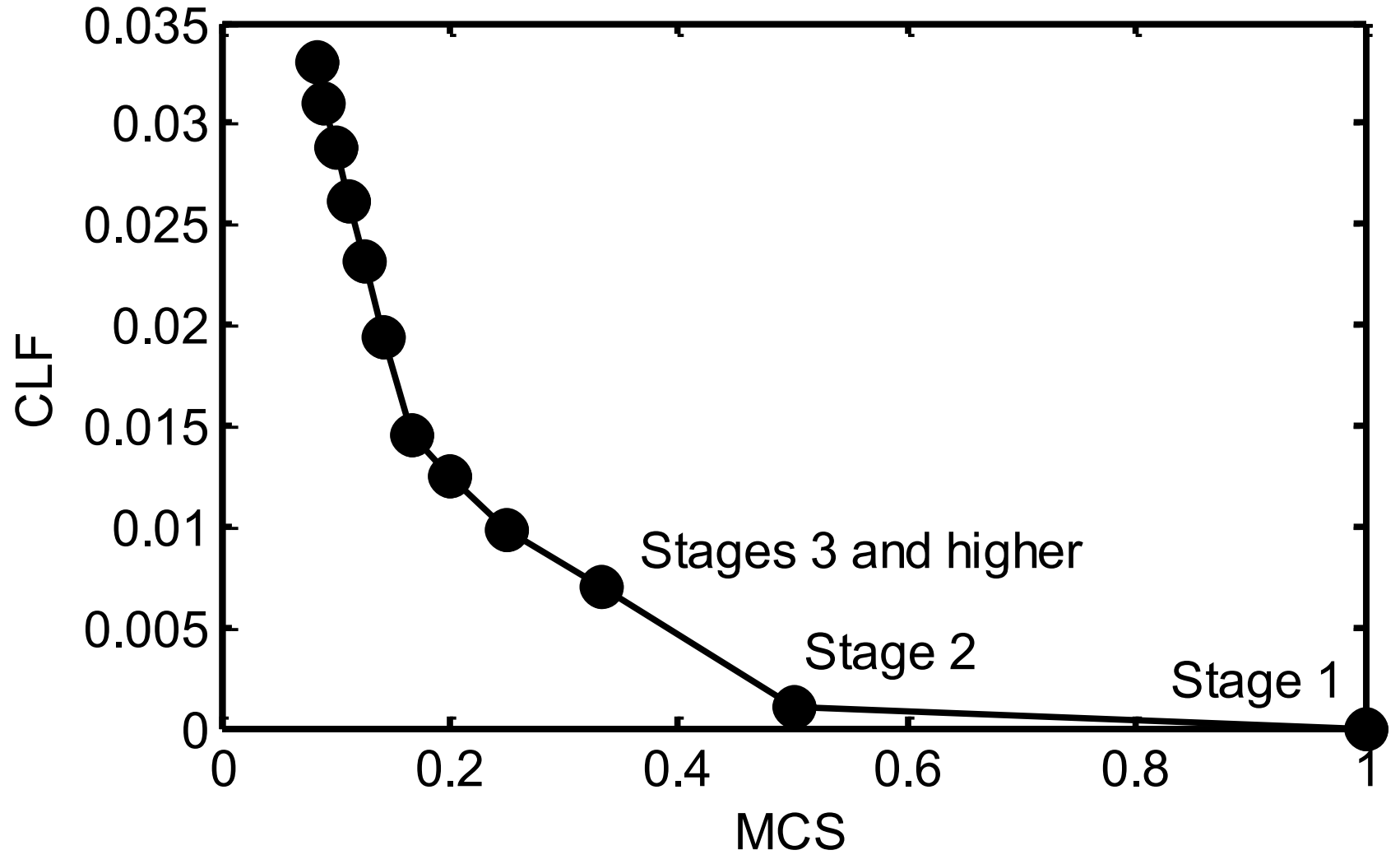

**Figure 4.8 CLF vs. MCS plot for case 4 in Table 4.3**

## 4.2.9 Nine zone building example

The layout of the building used in this example is shown in Figure 4.9. It consists of 3 floors, with a total of 9 rooms of equal dimensions ($5 \times 5 \times 5\ m^3$) numbered as shown. Similar to the previous example, the walls were modeled as RC circuits (Figure 3.5) and the zones were modeled as isolated capacitors. Each room has 6 walls - 4 side walls, 1 ceiling and 1 floor. The construction details and the resulting values of capacitances and resistances computed from them are same as for the 12-zone example, as presented in Table 4.1 and Table 4.2. The zonal thermal capacities were assumed to be 250 kJ/K based on air at 25 C and $10^5$ Pa. The dicrete-time system model for this system was then obtained in the same way as for the 12-zone building. The Hessian Matrix, $\bar{\mathbf{H}}$ was then created using (4.6) with prediction horizon, $N = 24$ samples (4 hours) and the weights $\boldsymbol{\alpha} = 0.1\mathbf{e}_9$ and $\boldsymbol{\beta} = \mathbf{e}_9$ (see nomenclature).

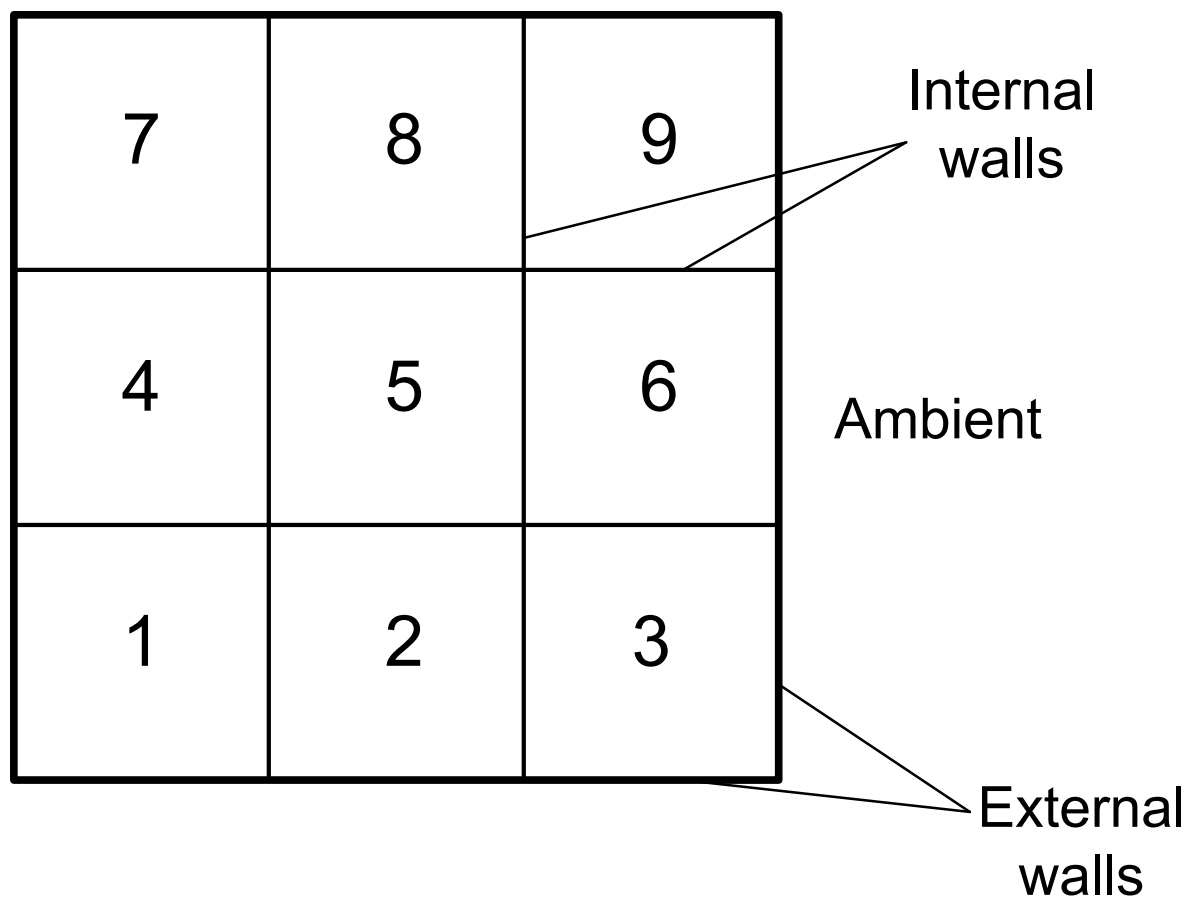

**Figure 4.9 Layout of 9-zone building (side view)**

The resistance values for all horizontal and vertical internal walls were found to be the same (see Table 4.2) as expected due to symmetry. We denote this value by $R_{nom} = 21.32$ K/kW. In the case study presented, the resistances of the horizontal internal walls are scaled up by a factor of $\rho > 0$, i.e. set to $\rho R_{nom}$. Correspondingly, the resistances of the vertical internal walls are scaled down by a factor of $\rho$, i.e. set to $R_{nom}/\rho$. Therefore, the ratio of horizontal wall resistances and vertical wall resistances is augmented by $\rho^2$.

The clustering methodology presented in Figure 4.1 was applied for various values of the factor $\rho$. The relvant codes are provided in the Appendix B. The optimization in each stage was performed using the combinatorial approach (involving all possible intermediate cluster pairs) as well as the MINCUT approximation presented in section 4.2.7. The CLF vs. MCS plots for some selected values of $\rho$ are shown in Figure 4.10 to Figure 4.12. Key observations are as follows.

1. The CLF vs. MCS plots using the MINCUT procedure and the combinatorial procedure exactly coincide in Figure 4.11 and Figure 4.12. However, they differ in Figure 4.10. This suggests that the MINCUT approximation to the combinatorial optimization problem can be potentially accurate in asymmetric situations. In general, it trades accuracy for computational simplicity as indicated by a run-time of 0.29 seconds when compared to 4.92 seconds for the combinatorial procedure[2].

2. For the nominal case ($\rho = 1$), a knee point is not immediately obvious in Figure

[2]Values are for the case $\rho = 1$, implemented on a 2.0 GHz, 960 MB, AMD Athlon machine

4.10. Therefore, stage 3 was chosen to be the optimal partition where both CLF and MCS are satisfactorily small.

3. When $\rho > 1$ (Figure 4.11 and Figure 4.12), the optimal partition is provided by stage 3 since it corresponds to a ‘sharp’ knee. It corresponds to the clusters $\{1,2,3\}$, $\{4,5,6\}$ and $\{7,8,9\}$. This can be justified on the basis of physical intuition. When $\rho > 1$, the horizontal walls are more insulating than the vertical walls, therefore the building must be sliced horizontally.

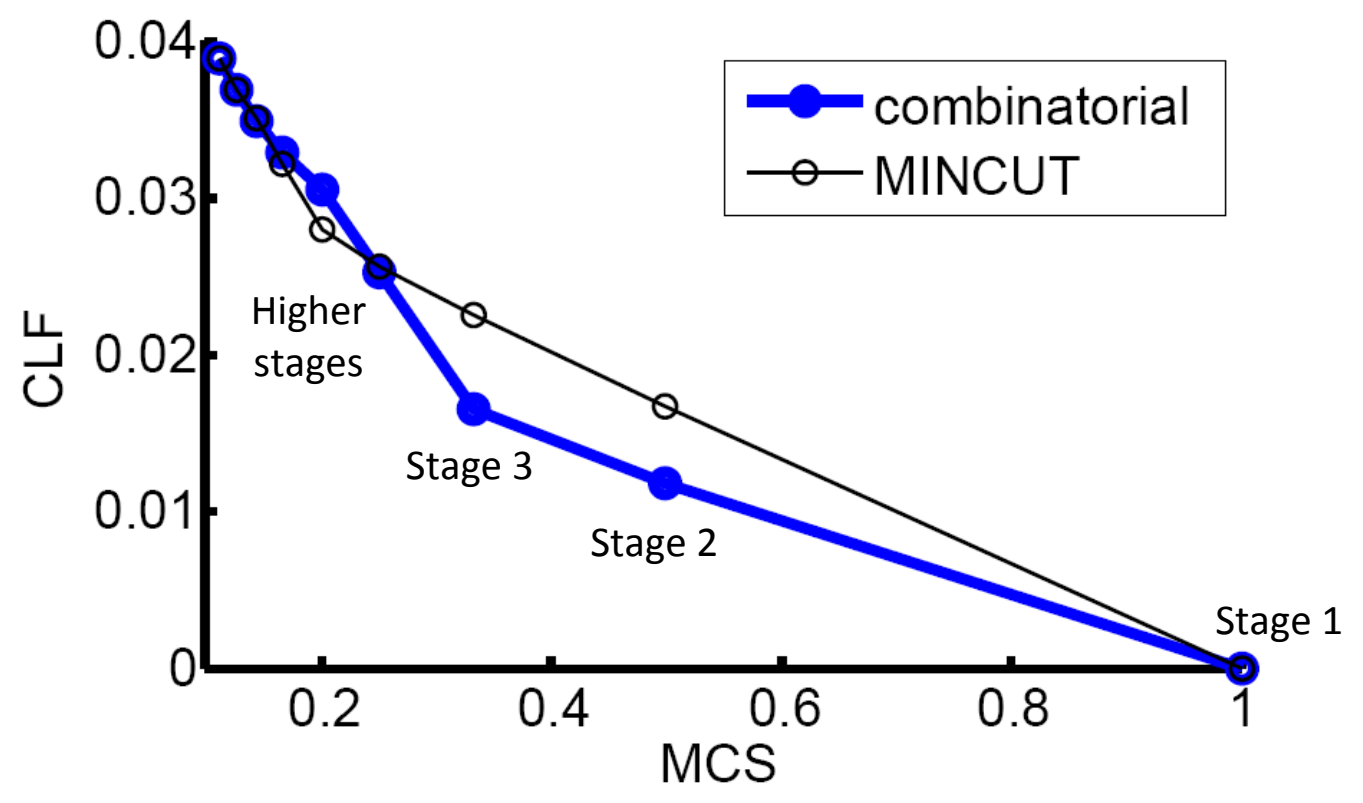

**Figure 4.10 CLF vs. MCS plot for nine-zone building with $\rho = 1$**

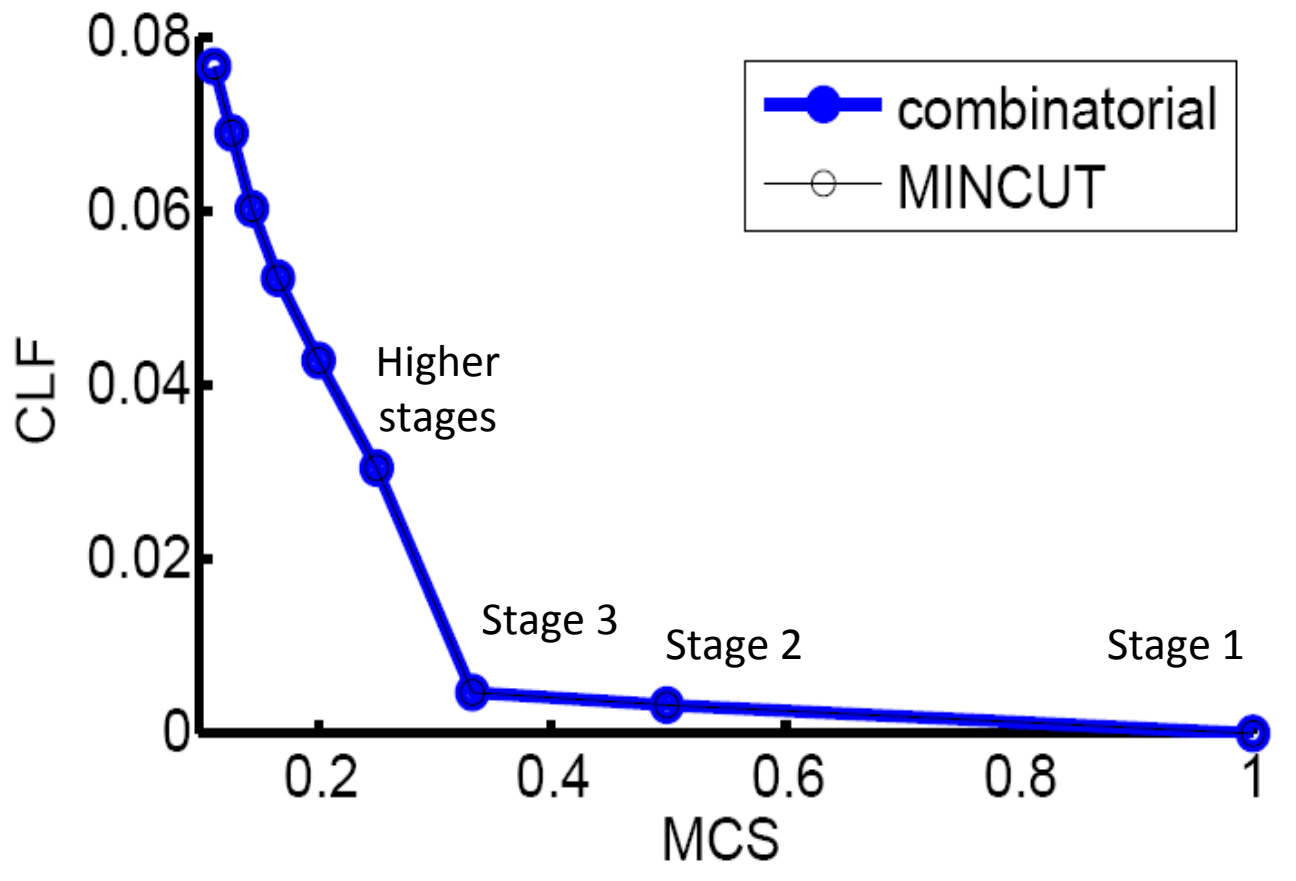

**Figure 4.11 CLF vs. MCS plot for nine-zone building with $\rho = 2$**

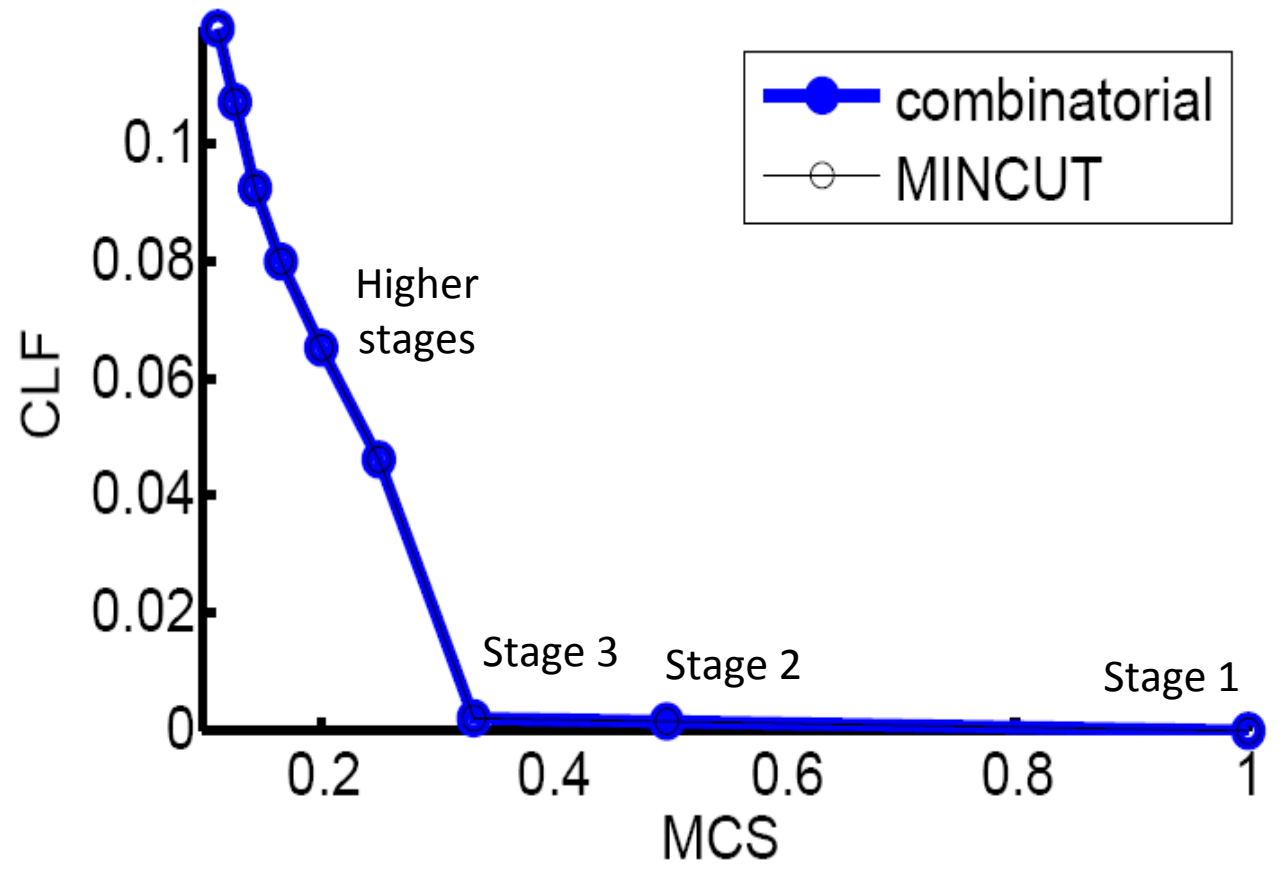

**Figure 4.12 CLF vs. MCS plot for nine-zone building with $\rho = 3$**

### 4.2.10 Remarks

The examples in Sections 4.2.8 and 4.2.9 demonstrate the capability of the CLF-MCS approach in determining architectures for decentralized control of a building that are consistent with physical intuition. In addition, Section 4.2.9 verifies that the proposed MINCUT approximation to address the high computational complexity of combinatorial optimization results in only a small loss in accuracy. However, the optimality and robustness metrics were chosen in a heuristic manner. The CLF defined in (4.11) does not directly quantify the loss in performance (optimality) in going from a centralized architecture to a decentralized architecture. Similarly, the MCS defined in (4.12) does not directly represent the true effect of fault propagation in the event of a sensor, actuator or communication infrastructure related failure in a building. Also, the divisive partitioning procedure in Figure 4.1 was found to have exponential computational complexity due to (4.13). The MINCUT approach which was proposed to address this issue is based on the assumption that $\bar{\mathbf{H}}_j^l$ is sufficiently sparse (see section 4.2.7). However, the desired level of sparsity is not quantified.

The above limitations of the CLF-MCS approach motivate the development of an approach which uses analytically derived metrics, and is computationally tractable with quantifiable computational complexity. The OLF-FPM approach presented in the next section satisfies these requirements and is proposed as an alternative to the CLF-MCS approach.

## 4.3 OLF-FPM method

### 4.3.1 Overview

The OLF-FPM approach is a procedure for partitioning a building based on Optimality Loss factor (OLF) and Fault Propagation Metric (FPM) as optimality and robustness metrics respectively. An agglomerative clustering procedure employs these metrics to create a family of partitions. Similar to the CLF-MCS approach, these partitions are then analyzed to select those which provide satisfactory trade-offs between optimality and robustness. The partitioning methodology is demonstrated using simulated examples. The optimality analysis of centralized and decentralized MPC frameworks and their comparison are presented first as necessary prerequisites for the development of OLF. The nomenclature in Chapter 3 is used several times in the remainder of Section 4.3. Additional nomenclature used is listed in Table 4.4

**Table 4.4: Nomenclature of additional symbols used (besides Table 3.1)**

| Symbol | Description |
|---|---|
| $\mathbf{e}_n$ | Vector of size $n \times 1$ with all entries 1 |
| $T_{z,i}$ | Temperature of $i^{th}$ zone in building |
| $C_{z,i}$ | Capacitance of $i^{th}$ zone in building |
| $d_{z,i}$ | Unmodeled thermal disturbance acting on $i^{th}$ zone in building |
| $d_{w,i}$ | Unmodeled thermal disturbance acting on $i^{th}$ wall in building |

### 4.3.2 Optimality analysis for centralized MPC

Assuming that $\boldsymbol{\alpha}$ and $\boldsymbol{\beta}$ in (3.2) and (3.3) are component-wise positive vectors, the objective function $J^c$ in (3.1) is strictly convex in the $(2N_z + N_w)N$ dimensional space of real variables consisting of components of $\{\mathbf{u}(k+l|k)\}_{l=0}^{l=N-1}$, $\{\mathbf{T_z}(k+l|k)\}_{l=1}^{l=N}$ and $\{\mathbf{T_w}(k+l|k)\}_{l=1}^{l=N}$. The linear constraint (3.4) represents hyperplanes in this space. Therefore, the optimization problem corresponding to centralized MPC is strictly convex with a unique global minimum [82]. In terms of the unconstrained re-statement (3.19), this implies that $g_c(.)$ is a strictly convex function of $\overline{\mathbf{u}}$, with a positive definite invertible Hessian Matrix, $\mathbf{H_c}$ in (3.20). The fact that $\mathbf{H_c}$ is positive definite can also be verified from (3.17) by noting that $\mathbf{Q_1}$ and $\mathbf{Q_2}$ are positive definite matrices when $\boldsymbol{\alpha}$ and $\boldsymbol{\beta}$ are component-wise positive vectors. The closed form expression for the unique minimizer, $\overline{\mathbf{u}}_\mathbf{c}^*$ of $g_c(.)$ in (3.19) is given by

$$\overline{\mathbf{u}}_\mathbf{c}^* = -\frac{1}{2}\mathbf{H}_\mathbf{c}^{-1}\mathbf{f}_\mathbf{c}$$

$$= -\mathbf{H}_\mathbf{c}^{-1}\mathbf{S}_\mathbf{u}^T\overline{\mathbf{C}}^T\mathbf{Q_1}\left[\overline{\mathbf{C}}\left(\mathbf{Mx(k)} + \mathbf{S_a}\overline{\mathbf{T}}_\mathbf{a} + \mathbf{S_g}\overline{\mathbf{T}}_\mathbf{g} + \mathbf{S_{dw}}\bar{\mathbf{d}}_\mathbf{w} + \mathbf{S_{dz}}\bar{\mathbf{d}}_\mathbf{z}\right) - \overline{\mathbf{T}}_\mathbf{z,ref}\right]. \quad (4.22)$$

A transformed form $\overline{\mathbf{v}}_\mathbf{c}^*$ is defined for $\overline{\mathbf{u}}_\mathbf{c}^*$ which has the structure given by (4.1). Using (4.4), $\overline{\mathbf{v}}_\mathbf{c}^*$ can be expressed as

$$\overline{\mathbf{v}}_\mathbf{c}^* = \mathbf{P_{vu}}\overline{\mathbf{u}}_\mathbf{c}^* \quad (4.23)$$

Substituting (4.23) in (4.22), we obtain

$$\overline{\mathbf{v}}_\mathbf{c}^* = -\mathbf{P_{vu}}\mathbf{H}_\mathbf{c}^{-1}\mathbf{S}_\mathbf{u}^T\overline{\mathbf{C}}^T\mathbf{Q_1}\left[\overline{\mathbf{C}}\left(\mathbf{Mx(k)} + \mathbf{S_a}\overline{\mathbf{T}}_\mathbf{a} + \mathbf{S_g}\overline{\mathbf{T}}_\mathbf{g} + \mathbf{S_{dw}}\bar{\mathbf{d}}_\mathbf{w} + \mathbf{S_{dz}}\bar{\mathbf{d}}_\mathbf{z}\right) - \overline{\mathbf{T}}_\mathbf{z,ref}\right]. \quad (4.24)$$

Equation (4.24) can be written as

$$\overline{\mathbf{v}}_\mathbf{c}^* = -\frac{1}{2}\overline{\mathbf{H}}_\mathbf{c}^{-1}\left[\mathbf{K}_1^\mathbf{c}\mathbf{T_z}(k) + \mathbf{K}_2^\mathbf{c}\mathbf{T_w}(k) + \mathbf{K}_3^\mathbf{c}\bar{\mathbf{d}}_\mathbf{w} + \mathbf{K}_4^\mathbf{c}\bar{\mathbf{d}}_\mathbf{z} + \mathbf{K}_5^\mathbf{c}\overline{\mathbf{T}}_\mathbf{a} + \mathbf{K}_6^\mathbf{c}\overline{\mathbf{T}}_\mathbf{g} + \mathbf{K}_7^\mathbf{c}\overline{\mathbf{T}}_\mathbf{z,ref}(k)\right]. \quad (4.25)$$

where,

$$\overline{\mathbf{H}}_c = \mathbf{H}_c\mathbf{P}_\mathbf{vu}^{-1}, \quad (4.26)$$

$$\mathbf{K}_1^\mathbf{c} = 2\mathbf{S}_\mathbf{u}^T\overline{\mathbf{C}}^T\mathbf{Q_1}\overline{\mathbf{C}}\begin{bmatrix}\boldsymbol{I}_{N_w}\\ \mathbf{0}_{N_z\times N_w}\end{bmatrix}, \quad (4.27)$$

$$\mathbf{K}_2^\mathbf{c} = 2\mathbf{S}_\mathbf{u}^T\overline{\mathbf{C}}^T\mathbf{Q_1}\overline{\mathbf{C}}\begin{bmatrix}\mathbf{0}_{N_w\times N_z}\\ \boldsymbol{I}_{N_z}\end{bmatrix}, \quad (4.28)$$

$$\mathbf{K}_3^\mathbf{c} = 2\mathbf{S}_\mathbf{u}^T\overline{\mathbf{C}}^T\mathbf{Q_1}\mathbf{S_{dw}}, \quad (4.29)$$

$$\mathbf{K}_4^\mathbf{c} = 2\mathbf{S}_\mathbf{u}^T\overline{\mathbf{C}}^T\mathbf{Q_1}\mathbf{S_{dz}}, \quad (4.30)$$

$$\mathbf{K}_5^\mathbf{c} = 2\mathbf{S}_\mathbf{u}^T\overline{\mathbf{C}}^T\mathbf{Q_1}\mathbf{S_a}, \quad (4.31)$$

$$\mathbf{K}_6^\mathbf{c} = 2\mathbf{S}_\mathbf{u}^T\overline{\mathbf{C}}^T\mathbf{Q_1}\mathbf{S_g}, \quad (4.32)$$

$$\mathbf{K}_7^\mathbf{c} = -2\mathbf{S}_\mathbf{u}^T\overline{\mathbf{C}}^T\mathbf{Q_1}. \quad (4.33)$$

### 4.3.3 Extraction matrices for decentralized MPC

Let $\mathbf{x} \in \mathbb{R}^{N_z}$ be a vector, the $l^{th}$ component, $\mathbf{x}^l$ of which is the value of an appropriate physical quantity associated with the $l^{th}$ zone in the building, where $l \in \{1,2,\ldots,N_z\}$. In the present context, $\mathbf{x}$ could represent a vector of temperatures, control inputs or unmodeled thermal loads for the various zones. Next, consider a general p-partition of the building consisting of clusters $S_{zi}$, where $i \in \{1,2,\ldots,p\}$. For each $i \in \{1,2,\ldots,p\}$, let $\mathbf{x}_i \in \mathbb{R}^{N_{zi}}$ be a vector such that its $r^{th}$ component, $\mathbf{x}_i^r$, where $r \in \{1,2,\ldots,N_{zi}\}$ is the value of the afore-mentioned physical quantity for the $r^{th}$ zone in the cluster $S_{zi}$. Therefore, the set of elements of $\mathbf{x}_i$ is a subset of the elements of $\mathbf{x}$. However, a sequential stack of $\mathbf{x}_i$ given by $\left(\mathbf{x}_1{}^T\ \mathbf{x}_2{}^T\ \ldots\ \mathbf{x}_p{}^T\right)^T$ does not necessarily produce $\mathbf{x}$. This is also true in the context of vectors $\mathbf{y} \in \mathbb{R}^{N_w}$ and $\mathbf{y}_i \in \mathbb{R}^{N_{wi}}$, which are analogues of $\mathbf{x}$ and $\mathbf{x}_i$ respectively, but are defined for walls instead of zones. Here, we introduce the concept of *extraction matrices* to enable an accurate representation of the mathematical relationship between the cluster-level vectors and the overall system level vectors, and extend it to include the case when these vectors are lifted in time. This concept is required for the optimality analysis of decentralized MPC that follows in the next section.

**Definition 4.1 (Cluster extraction matrix)**: Let $\mathbf{s_z} \in \mathbb{N}^{N_z}$ and $\mathbf{s}_{\mathbf{z}i} \in \mathbb{N}^{N_{zi}}$ be vector representations of the elements of the sets $S_z$ and $S_{zi}$ respectively, where $i \in \{1,2,\ldots,p\}$. The *cluster extraction matrix*, $\mathbf{P}_i \in \mathbb{Z}^{N_{zi}\times N_z}$ for the $i^{th}$ cluster, $S_{zi}$ in the p-partition is defined as the Boolean matrix with exactly one 1 in each row which satisfies

$$\mathbf{s}_{\mathbf{z}i} = \mathbf{P}_i\mathbf{s}_\mathbf{z}. \tag{4.34}$$

**Definition 4.2 (Lifted cluster extraction matrix)**: The *lifted cluster extraction matrix*, $\overline{\mathbf{P}}_i \in \mathbb{Z}^{N.N_{zi}\times N.N_z}$ for the $i^{th}$ cluster, $S_{zi}$ in the p-partitionis defined as the boolean matrix which is obtained from $\mathbf{P}_i$ by replacing all scalar ones with $\mathbf{I}_N$ and all scalar zeros with $\mathbf{0}_{N\times N}$.

**Definition 4.3 (Overall extraction matrix)**: The *overall extraction matrix*, $\mathbf{P} \in \mathbb{Z}^{N_z\times N_z}$ is defined in (4.35) by stacking the matrices $\mathbf{P}_i$, $i \in \{1,2,\ldots,p\}$ along their columns.

$$\mathbf{P} = \left(\mathbf{P}_1^T\ \mathbf{P}_2^T\ \ldots\ \mathbf{P}_p^T\right)^T. \tag{4.35}$$

**Definition 4.4 (Overall lifted extraction matrix)**: The *overall lifted extraction matrix*, $\overline{\mathbf{P}} \in$

$\mathbb{Z}^{N.N_z \times N.N_z}$ is defined in (4.36) by stacking the matrices $\bar{\mathbf{P}}_i$, $i \in \{1,2,\ldots,p\}$ along their columns.

$$\bar{\mathbf{P}} = \left(\bar{\mathbf{P}}_1^T \ \bar{\mathbf{P}}_2^T \ . \ . \ . \ \bar{\mathbf{P}}_p^T\right)^T. \tag{4.36}$$

**Definition 4.5 (Wall extraction matrix)**: Let $\mathbf{s_w} \in \mathbb{N}^{N_w}$ be a vector representing the set of walls in the building. Similarly, let $\mathbf{s_{w}}_i \in \mathbb{N}^{N_{wi}}$ be a vector representing the set of walls in the cluster $S_{zi}$. The *wall extraction matrix*, $\mathbf{R}_i \in \mathbb{Z}^{N_{wi} \times N_w}$ for the $i^{th}$ cluster, $S_{zi}$ in the p-partition is defined as the Boolean matrix with exactly one 1 in each row which satisfies

$$\mathbf{s}_{\mathbf{w},i} = \mathbf{R}_i \mathbf{s_w}. \tag{4.37}$$

**Definition 4.6. (Overall wall extraction matrix)**: The *overall wall extraction matrix*, $\mathbf{R} \in \mathbb{Z}^{(\sum_{i=1}^{p} N_{w,i}) \times N_w}$ is defined in (4.38) by stacking the matrices $\mathbf{R}_i$, $i \in \{1,2,\ldots,p\}$ along their columns.

$$\mathbf{R} = \left(\mathbf{R}_1^T \ \mathbf{R}_2^T \ . \ . \ . \ \mathbf{R}_p^T\right)^T. \tag{4.38}$$

**Definition 4.7 (Lifted wall extraction matrix)**: The *lifted wall extraction matrix*, $\bar{\mathbf{R}}_i \in \mathbb{Z}^{N.N_{wi} \times \mathrm{N}.N_w}$ for the $i^{th}$ cluster, $S_{zi}$ in the p-partition is defined as the Boolean matrix which is obtained from $\mathbf{R}_i$ by replacing all scalar ones with $\mathbf{I}_N$ and all scalar zeros with $\mathbf{0}_{N\times N}$.

**Definition 4.8 (Overall lifted wall extraction matrix)**: The *overall lifted wall extraction matrix*, $\bar{\mathbf{R}} \in \mathbb{Z}^{(\sum_{i=1}^{p} N.N_{w,i}) \times N.N_w}$ is defined in (4.39) by stacking the matrices $\bar{\mathbf{R}}_i$, $i \in \{1,2,\ldots,p\}$ along their columns.

$$\bar{\mathbf{R}} = \left(\bar{\mathbf{R}}_1^T \ \bar{\mathbf{R}}_2^T \ . \ . \ . \ \bar{\mathbf{R}}_p^T\right)^T. \tag{4.39}$$

As an example, consider the 3-zone building in Figure 3.3. We demonstrate the above definitions using the 2-partition of this building $\{\{1,3\},\{2\}\}$. Here, $S_z = \{1,2,3\}$, $S_{z1} = \{1,3\}$ and $S_{z2} = \{2\}$. Therefore, $\mathbf{s_z} = [1 \quad 2 \quad 3]^T$, $\mathbf{s_{z1}} = [1 \quad 3]^T$ and $\mathbf{s_{z2}} = 2$.

The cluster extraction matrices are given by:

$$\mathbf{P}_1 = \begin{bmatrix} 1 & 0 & 0 \\ 0 & 0 & 1 \end{bmatrix}, \ \mathbf{P}_2 = [0 \quad 1 \quad 0].$$

The lifted cluster extraction matrices are given by:

$$\bar{\mathbf{P}}_1 = \begin{bmatrix} \mathbf{I}_N & \mathbf{0}_{N\times N} & \mathbf{0}_{N\times N} \\ \mathbf{0}_{N\times N} & \mathbf{0}_{N\times N} & \mathbf{I}_N \end{bmatrix}, \ \bar{\mathbf{P}}_2 = [\mathbf{0}_{N\times N} \quad \mathbf{I}_N \quad \mathbf{0}_{N\times N}].$$

The overall extraction matrix and overall lifted extraction matrix are then given by:

$$\mathbf{P} = \begin{bmatrix} 1 & 0 & 0 \\ 0 & 0 & 1 \\ 0 & 1 & 0 \end{bmatrix}, \overline{\mathbf{P}} = \begin{bmatrix} \mathbf{I}_N & \mathbf{0}_{N\times N} & \mathbf{0}_{N\times N} \\ \mathbf{0}_{N\times N} & \mathbf{0}_{N\times N} & \mathbf{I}_N \\ \mathbf{0}_{N\times N} & \mathbf{I}_N & \mathbf{0}_{N\times N} \end{bmatrix}.$$

Using the graph shown in Figure 3.6, the vector of walls in the building is given by $\mathbf{s_w} = [1 \quad 2 \quad 3 \quad 4 \quad 5 \quad 6 \quad 7 \quad 8 \quad 9 \quad 10]^T$, $\mathbf{s_{w1}} = [1 \quad 2 \quad 3 \quad 4 \quad 7 \quad 8 \quad 9 \quad 10]^T$ and $\mathbf{s_{w2}} = [4 \quad 5 \quad 6 \quad 7]^T$.

Hence, the wall extraction matrices are given by:

$$\mathbf{R}_1 = \begin{bmatrix} 1&0&0&0&0&0&0&0&0&0 \\ 0&1&0&0&0&0&0&0&0&0 \\ 0&0&1&0&0&0&0&0&0&0 \\ 0&0&0&1&0&0&0&0&0&0 \\ 0&0&0&0&0&0&1&0&0&0 \\ 0&0&0&0&0&0&0&1&0&0 \\ 0&0&0&0&0&0&0&0&1&0 \\ 0&0&0&0&0&0&0&0&0&1 \end{bmatrix}, \mathbf{R}_2 = \begin{bmatrix} 0&0&0&1&0&0&0&0&0&0 \\ 0&0&0&0&1&0&0&0&0&0 \\ 0&0&0&0&0&1&0&0&0&0 \\ 0&0&0&0&0&0&1&0&0&0 \end{bmatrix}.$$

The overall wall extraction matrix is given by:

$$\mathbf{R} = \begin{bmatrix} 1&0&0&0&0&0&0&0&0&0 \\ 0&1&0&0&0&0&0&0&0&0 \\ 0&0&1&0&0&0&0&0&0&0 \\ 0&0&0&1&0&0&0&0&0&0 \\ 0&0&0&0&0&0&1&0&0&0 \\ 0&0&0&0&0&0&0&1&0&0 \\ 0&0&0&0&0&0&0&0&1&0 \\ 0&0&0&0&0&0&0&0&0&1 \\ 0&0&0&1&0&0&0&0&0&0 \\ 0&0&0&0&1&0&0&0&0&0 \\ 0&0&0&0&0&1&0&0&0&0 \\ 0&0&0&0&0&0&1&0&0&0 \end{bmatrix}.$$

The *lifted wall extraction matrices* $\overline{\mathbf{R}}_1$ and $\overline{\mathbf{R}}_2$ are obtained from $\mathbf{R}_1$ and $\mathbf{R}_2$ respectively by replacing all scalar ones with $\mathbf{I}_N$ and all scalar zeros with $\mathbf{0}_{N\times N}$. The *overall lifted wall extraction matrix* $\overline{\mathbf{R}}$ is obtained from $\mathbf{R}$ in a similar manner.

We state some easily verifiable properties of the extraction matrices in terms of $\mathbf{x}$, $\mathbf{x}_i$, $\mathbf{y}$

and $\mathbf{y}_i$ defined earlier, and their time-lifted analogues[3] $\bar{\mathbf{x}}$, $\bar{\mathbf{x}}_i$, $\bar{\mathbf{y}}$ and $\bar{\mathbf{y}}_i$ defined in (4.40) to (4.43).

$$\bar{\mathbf{x}} = (\bar{\mathbf{x}}^1 \ \bar{\mathbf{x}}^2 \ . \ . \ . \ \bar{\mathbf{x}}^{N_z})^T \text{ where, } \bar{\mathbf{x}}^l = [\mathbf{x}^l(k|k) \ \mathbf{x}^l(k+1|k) \ . \ . \ . \ \mathbf{x}^l(k+\mathrm{N}-1|k)], \quad (4.40)$$

$$\bar{\mathbf{x}}_i = \left(\bar{\mathbf{x}}_i^1 \ \bar{\mathbf{x}}_i^2 \ . \ . \ . \ \bar{\mathbf{x}}_i^{N_{zi}}\right)^T \text{ where, } \bar{\mathbf{x}}_i^r = [\mathbf{x}_i^r(k|k) \ \mathbf{x}_i^r(k+1|k) \ . \ . \ . \ \mathbf{x}_i^r(k+\mathrm{N}-1|k)], \quad (4.41)$$

$$\bar{\mathbf{y}} = (\bar{\mathbf{y}}^1 \ \bar{\mathbf{y}}^2 \ . \ . \ . \ \bar{\mathbf{y}}^{N_w})^T \text{ where, } \bar{\mathbf{y}}^l = [\mathbf{y}^l(k|k) \ \mathbf{y}^l(k+1|k) \ . \ . \ . \ \mathbf{y}^l(k+\mathrm{N}-1|k)], \quad (4.42)$$

$$\bar{\mathbf{y}}_i = \left(\bar{\mathbf{y}}_i^1 \ \bar{\mathbf{y}}_i^2 \ . \ . \ . \ \bar{\mathbf{y}}_i^{N_{zi}}\right)^T \text{ where, } \bar{\mathbf{y}}_i^r = [\mathbf{y}_i^r(k|k) \ \mathbf{y}_i^r(k+1|k) \ . \ . \ . \ \mathbf{y}_i^r(k+\mathrm{N}-1|k)]. \quad (4.43)$$

Properties of extraction matrices

1. $\mathbf{x}_i = \mathbf{P}_i\mathbf{x}$.
2. $\bar{\mathbf{x}}_i = \bar{\mathbf{P}}_i\bar{\mathbf{x}}$.
3. $\left(\mathbf{x}_1{}^T \ \mathbf{x}_2{}^T \ . \ . \ . \ \mathbf{x}_p{}^T\right)^T = \mathbf{P}\mathbf{x}$.
4. $\left(\bar{\mathbf{x}}_1{}^T \ \bar{\mathbf{x}}_2{}^T \ . \ . \ . \ \bar{\mathbf{x}}_p{}^T\right)^T = \bar{\mathbf{P}}\bar{\mathbf{x}}$.
5. Both $\mathbf{P}$ and $\bar{\mathbf{P}}$ are invertible permutation matrices.
6. $\mathbf{y}_i = \mathbf{R}_i\mathbf{y}$.
7. $\bar{\mathbf{y}}_i = \bar{\mathbf{R}}_i\bar{\mathbf{y}}$.
8. $\left(\mathbf{y}_1{}^T \ \mathbf{y}_2{}^T \ . \ . \ . \ \mathbf{y}_p{}^T\right)^T = \mathbf{R}\mathbf{y}$.
9. $\left(\bar{\mathbf{y}}_1{}^T \ \bar{\mathbf{y}}_2{}^T \ . \ . \ . \ \bar{\mathbf{y}}_p{}^T\right)^T = \bar{\mathbf{R}}\bar{\mathbf{y}}$.

### 4.3.4 Optimality analysis for decentralized MPC

The components of the vectors $\boldsymbol{\alpha}_i$ and $\boldsymbol{\beta}_i$ in (3.22) and (3.23) are positive if $\boldsymbol{\alpha}$ and $\boldsymbol{\beta}$ in (3.2) and (3.3) are component-wise positive. Therefore, as observed for the centralized MPC optimization problem, the optimization problem for the $i^{th}$ control agent in the decentralized control framework (3.34) is also strictly convex. The closed form expression for the unique minimizer $\bar{\mathbf{u}}^*_{\mathbf{dc,i}}$ of $g_{dc,i}(.)$ in (3.33) is given by

$$-2\mathbf{H}_{\mathbf{dc,i}}\bar{\mathbf{u}}^*_{\mathbf{dc,i}} = \mathbf{f}_{\mathbf{dc,i}}$$

$$= 2\mathbf{S}_{\mathbf{ui}}^T\bar{\mathbf{C}}_{\mathrm{i}}{}^T\mathbf{Q}_{\mathbf{1,i}}\left[\bar{\mathbf{C}}_{\mathrm{i}}\left(\mathbf{M}_{\mathbf{i}}\mathbf{x}_{\mathrm{i}}(\mathbf{k}) + \mathbf{S}_{\mathbf{ai}}\bar{\mathbf{T}}_{\mathbf{a}} + \mathbf{S}_{\mathbf{gi}}\bar{\mathbf{T}}_{\mathbf{g}} + \mathbf{S}_{\mathbf{dw,i}}\bar{\mathbf{d}}_{\mathbf{wi}} + \mathbf{S}_{\mathbf{dz,i}}\bar{\mathbf{d}}_{\mathbf{z}} + \textstyle\sum_{j\neq i}\mathbf{S}_{\mathbf{ij}}\bar{\bar{\mathbf{T}}}_{\mathbf{zj}}\right) - \bar{\mathbf{T}}_{\mathbf{zi,ref}}\right] \quad (4.44)$$

---

[3]If $x$ is used to represent the zone temperatures, $\bar{\mathbf{x}}^l$ in (32) should instead be defined as $\bar{\mathbf{x}}^l = (\mathbf{x}^l(k+1|k) \ \mathbf{x}^l(k+2|k) \ . \ . \ . \ x^l(k+N|k))$. The definition of $\bar{\mathbf{x}}_i^r$ in (( 3.26 )) should be similarly modified.

The vector $\bar{\mathbf{u}}^*_{\mathbf{dc,i}}$ is a lifted vector given by

$$\bar{\mathbf{u}}^*_{\mathbf{dc,i}} = \begin{bmatrix} \mathbf{u}^*_{\mathbf{dc,i}}(k|k) \\ \mathbf{u}^*_{\mathbf{dc,i}}(k+1|k) \\ \vdots \\ \mathbf{u}^*_{\mathbf{dc,i}}(k+N-1|k) \end{bmatrix}, \text{ where } \mathbf{u}^*_{\mathbf{dc,i}}(k+l|k) \in \mathbb{R}^{N_{zi}} \text{ for all } l \in \{0,1,\ldots,N-1\}. \quad (4.45)$$

We now introduce a vector $\bar{\mathbf{v}}^*_{\mathbf{dc,i}}$ whose elements are permutations of the elements of $\bar{\mathbf{u}}^*_{\mathbf{dc,i}}$ as shown

$$\bar{\mathbf{v}}^*_{\mathbf{dc,i}} = \begin{bmatrix} \bar{\mathbf{v}}^{*1}_{\mathbf{dc,i}} \\ \bar{\mathbf{v}}^{*2}_{\mathbf{dc,i}} \\ \vdots \\ \bar{\mathbf{v}}^{*N_{zi}}_{\mathbf{dc,i}} \end{bmatrix}, \quad (4.46)$$

where,

$$\bar{\mathbf{v}}^{*r}_{\mathbf{dc,i}} = \begin{bmatrix} u^{*r}_{dc,i}(k|k) \\ u^{*r}_{dc,i}(k+1|k) \\ \vdots \\ u^{*r}_{dc,i}(k+N-1|k) \end{bmatrix} \text{ for } r \in \{1,2,\ldots,N_{zi}\}. \quad (4.47)$$

Here, $u^{*r}_{dc,i}$ denotes the $r^{th}$ component of the vector of control inputs $\mathbf{u}^*_{\mathbf{dc,i}}$. Therefore, the elements of $\bar{\mathbf{v}}^*_{\mathbf{dc,i}}$ and $\bar{\mathbf{u}}^*_{\mathbf{dc,i}}$ are permutations of each other. Let $\mathbf{P_{vu,i}}$ be a permutation matrix such that

$$\bar{\mathbf{v}}^*_{\mathbf{dc,i}} = \mathbf{P_{vu,i}} \bar{\mathbf{u}}^*_{\mathbf{dc,i}} \,. \quad (4.48)$$

Substituting (4.48) in (4.44), we obtain

$$-2\mathbf{H_{dc,i}} \mathbf{P^{-1}_{vu,i}} \bar{\mathbf{v}}^*_{\mathbf{dc,i}} =$$

$$2\mathbf{S}^T_{\mathbf{ui}} \bar{\mathbf{C}}_{\mathrm{i}}{}^T \mathbf{Q_{1,i}} \left[ \bar{\mathbf{C}}_{\mathrm{i}} \left( \mathbf{M_i x_i(k)} + \mathbf{S_{ai}} \bar{\mathbf{T}}_{\mathbf{a}} + \mathbf{S_{gi}} \bar{\mathbf{T}}_{\mathbf{g}} + \mathbf{S_{dw,i}} \bar{\mathbf{d}}_{\mathbf{wi}} + \mathbf{S_{dz,i}} \bar{\mathbf{d}}_{\mathbf{z}} + \textstyle\sum_{j \neq i} \mathbf{S_{ij}} \bar{\bar{\mathbf{T}}}_{\mathbf{zj}} \right) - \bar{\mathbf{T}}_{\mathbf{zi,ref}} \right] \quad (4.49)$$

Equation (4.49) can be written as

$$-2\bar{\mathbf{H}}_{\mathbf{dc,i}} \bar{\mathbf{v}}^*_{\mathbf{dc,i}} = \mathbf{K}^{\mathbf{dc,i}}_{\mathbf{1}} \mathbf{T_{wi}}(k) + \mathbf{K}^{\mathbf{dc,i}}_{\mathbf{2}} \mathbf{T_{zi}}(k) + \mathbf{K}^{\mathbf{dc,i}}_{\mathbf{3}} \bar{\mathbf{d}}_{\mathbf{wi}} + \mathbf{K}^{\mathbf{dc,i}}_{\mathbf{4}} \bar{\mathbf{d}}_{\mathbf{zi}} + \mathbf{K}^{\mathbf{dc,i}}_{\mathbf{5}} \bar{\mathbf{T}}_{\mathbf{a}} + \mathbf{K}^{\mathbf{dc,i}}_{\mathbf{6}} \bar{\mathbf{T}}_{\mathbf{g}} + \mathbf{K}^{\mathbf{dc,i}}_{\mathbf{7}} \bar{\mathbf{T}}_{\mathbf{zi,ref}}$$

$$+ \textstyle\sum_{j \neq i} \mathbf{Q_{ij}} \bar{\bar{\mathbf{T}}}_{\mathbf{zj}} \quad (4.50)$$

where,

$$\bar{\mathbf{H}}_{\mathbf{dc,i}} = \mathbf{H_{dc,i}} \mathbf{P^{-1}_{vu,i}}, \quad (4.51)$$

$$\mathbf{K_1^{dc,i}} = 2\mathbf{S}_{\mathbf{ui}}^T \bar{\mathbf{C}}_\mathrm{i}^{\ T} \mathbf{Q_{1,i}} \bar{\mathbf{C}}_\mathrm{i} \begin{bmatrix} \boldsymbol{I}_{N_{wi}} \\ \mathbf{0}_{N_{zi} \times N_{wi}} \end{bmatrix}, \tag{4.52}$$

$$\mathbf{K_2^{dc,i}} = 2\mathbf{S}_{\mathbf{ui}}^T \bar{\mathbf{C}}_\mathrm{i}^{\ T} \mathbf{Q_{1,i}} \bar{\mathbf{C}}_\mathrm{i} \begin{bmatrix} \mathbf{0}_{N_{wi} \times N_{zi}} \\ \boldsymbol{I}_{N_{zi}} \end{bmatrix}, \tag{4.53}$$

$$\mathbf{K_3^{dc,i}} = 2\mathbf{S}_{\mathbf{ui}}^T \bar{\mathbf{C}}_\mathrm{i}^{\ T} \mathbf{Q_{1,i}} \mathbf{S_{dw,i}}, \tag{4.54}$$

$$\mathbf{K_4^{dc,i}} = 2\mathbf{S}_{\mathbf{ui}}^T \bar{\mathbf{C}}_\mathrm{i}^{\ T} \mathbf{Q_{1,i}} \mathbf{S_{dz,i}}, \tag{4.55}$$

$$\mathbf{K_5^{dc,i}} = 2\mathbf{S}_{\mathbf{ui}}^T \bar{\mathbf{C}}_\mathrm{i}^{\ T} \mathbf{Q_{1,i}} \mathbf{S_{ai}}, \tag{4.56}$$

$$\mathbf{K_6^{dc,i}} = 2\mathbf{S}_{\mathbf{ui}}^T \bar{\mathbf{C}}_\mathrm{i}^{\ T} \mathbf{Q_{1,i}} \mathbf{S_{gi}}, \tag{4.57}$$

$$\mathbf{K_7^{dc,i}} = -2\mathbf{S}_{\mathbf{ui}}^T \bar{\mathbf{C}}_\mathrm{i}^{\ T} \mathbf{Q_{1,i}}, \tag{4.58}$$

$$\mathbf{Q_{ij}} = 2\mathbf{S}_{\mathbf{ui}}^T \bar{\mathbf{C}}_\mathrm{i}^{\ T} \mathbf{Q_{1,i}} \mathbf{S_{ij}}. \tag{4.59}$$

### 4.3.5 Comparison of centralized and decentralized MPC

We now present a quantitative comparison of the centralized and decentralized MPC control methodologies which forms the basis of the OLF metric presented in the next section. Let $\bar{\hat{\mathbf{T}}}_\mathbf{z}^\mathbf{r}$ be a vector of temperature estimates for zone $r$ lifted in time, defined as

$$\bar{\hat{\mathbf{T}}}_\mathbf{z}^\mathbf{r} = \left(\hat{\mathbf{T}}_\mathbf{z}^\mathbf{r}(k|k)\ \hat{\mathbf{T}}_\mathbf{z}^\mathbf{r}(k+1|k)\ .\ .\ .\ \hat{\mathbf{T}}_\mathbf{z}^\mathbf{r}(k+N-1|k)\right)^T. \tag{4.60}$$

Here, $\hat{\mathbf{T}}_\mathbf{z}^\mathbf{r}(k+l|k)$ at any time instant $k$ denotes the projected estimate of the temperature of zone $r$, $l$ time steps ahead in future where $l \in \{0,1,\ldots,N-1\}$. The overall lifted vector of estimates $\bar{\hat{\mathbf{T}}}_\mathbf{z}$ is then defined as the sequential stack of $\bar{\hat{\mathbf{T}}}_\mathbf{z}^\mathbf{r}$ given by (35)

$$\bar{\hat{\mathbf{T}}}_\mathbf{z} = \left[\left(\bar{\hat{\mathbf{T}}}_\mathbf{z}^1\right)^T \left(\bar{\hat{\mathbf{T}}}_\mathbf{z}^2\right)^T\ .\ .\ .\ \left(\bar{\hat{\mathbf{T}}}_\mathbf{z}^{N_z}\right)^T\right]^T \tag{4.61}$$

The vector $\bar{\hat{\mathbf{T}}}_\mathbf{zj}$ appearing in (4.50) and defined in Section 3.4.4 can be generated by selecting appropriate components of $\bar{\hat{\mathbf{T}}}_\mathbf{z}$. It can then be used to compute $\bar{\mathbf{v}}_\mathbf{dc,i}^*$ using (4.50). Using the definition (4.46) of $\bar{\mathbf{v}}_\mathbf{dc,i}^*$, the values $\{u_{dc,i}^{*r}(k+l|k)\}_{l=0}^{l=N-1}$ in (4.47) can be obtained for each zone $r$ in the cluster $S_{zi}$. We observe that each pair $(i,r)$ maps to a zone $m$ in the building where $i \in \{1,2,\ldots p\}$, $r \in \{1,2,\ldots,N_{zi}\}$ and $m \in \{1,2,\ldots,N_z\}$. For each pair $(i,r)$ we denote $u_{dc,i}^{*r}$ by $u_{dc}^{*m}$. In this way, repeating this procedure for each cluster $S_{zi}$, where $i \in \{1,2,\ldots p\}$, we obtain the set of values $\{u_{dc}^{*n}(k+l|k)\}_{l=0}^{l=N-1}$ for each zone $n \in \{1,2,\ldots,N_z\}$ in the building. This is because the union of clusters $S_{zi}$ decomposes the entire building (see definition 3.2)

For each $n \in \{1,2,\ldots,N_z\}$, we represent the values $\{u_{dc}^{*n}(k+l|k)\}_{l=0}^{l=N-1}$ in a succint form by defining the vector

$$\bar{\mathbf{v}}_{\mathbf{dc}}^{*\mathbf{n}} = \begin{bmatrix} u_{dc}^{*n}(k|k) \\ u_{dc}^{*n}(k+1|k) \\ \vdots \\ u_{dc}^{*n}(k+N-1|k) \end{bmatrix}. \qquad (4.62)$$

Next, an overall lifted vector of control inputs, $\bar{\mathbf{v}}_{\mathbf{dc}}^{*}$ for the decentralized case is constructed by the sequential stack of $\bar{\mathbf{v}}_{\mathbf{dc}}^{*\mathbf{n}}$ given by

$$\bar{\mathbf{v}}_{\mathbf{dc}}^{*} = \left[ (\bar{\mathbf{v}}_{\mathbf{dc}}^{*1})^T \ (\bar{\mathbf{v}}_{\mathbf{dc}}^{*2})^T \ . \ . \ . \ \left(\bar{\mathbf{v}}_{\mathbf{dc}}^{*N_z}\right)^T \right]^T \qquad (4.63)$$

Note that $\bar{\mathbf{v}}_{\mathbf{dc}}^{*}$ constructed above and $\bar{\mathbf{v}}_{\mathbf{c}}^{*}$ defined in (4.23) are comparable vectors in the sense that entries in identical locations in these vectors are associated with the same zone in the building and the same time instant in the prediction horizon. This fact is easily verifiable.

**Theorem 4.1 (Centralized-Decentralized Equivalence)** *Let $\bar{\bar{\mathbf{T}}}_{\mathbf{z}}^{*} \in \mathbb{R}^{N.N_z}$ be a solution of the linear equation (4.64) in $y$. If the overall lifted vector of zone temperature estimates $\bar{\bar{\mathbf{T}}}_{\mathbf{z}}$ (defined in (4.61)) is set to $\bar{\bar{\mathbf{T}}}_{\mathbf{z}}^{*}$, then the ensuing overall lifted vector of control inputs, $\bar{\mathbf{v}}_{\mathbf{dc}}^{*}$ (defined in (4.63)) computed by the decentralized multi-agent MPC control architecture satisfies $\bar{\mathbf{v}}_{\mathbf{dc}}^{*} = \bar{\mathbf{v}}_{\mathbf{c}}^{*}$.*

$$(\mathbf{Q_{dc}\bar{P}})\mathbf{y} = \mathbf{K_1 T_w}(k) + \mathbf{K_2 T_z}(k) + \mathbf{K_3 \bar{d}_w} + \mathbf{K_4 \bar{d}_z} + \mathbf{K_5 \bar{T}_a} + \mathbf{K_6 \bar{T}_6} + \mathbf{K_7 \bar{T}_{z,ref}} \qquad (4.64)$$

where,

$$\mathbf{K_1} = \mathbf{\bar{H}_{dc}\bar{P}\bar{H}_c^{-1}K_1^c} - \mathbf{K_1^{dc}R}, \qquad (4.65)$$

$$\mathbf{K_2} = \mathbf{\bar{H}_{dc}\bar{P}\bar{H}_c^{-1}K_2^c} - \mathbf{K_2^{dc}P}, \qquad (4.66)$$

$$\mathbf{K_3} = \mathbf{\bar{H}_{dc}\bar{P}H_c^{-1}K_3^c} - \mathbf{K_3^{dc}\bar{R}}, \qquad (4.67)$$

$$\mathbf{K_4} = \mathbf{\bar{H}_{dc}\bar{P}H_c^{-1}K_4^c} - \mathbf{K_4^{dc}\bar{P}}, \qquad (4.68)$$

$$\mathbf{K_5} = \mathbf{H_{dc}\bar{P}H_c^{-1}K_5^c} - \mathbf{K_5^{dc}}, \tag{4.69}$$

$$\mathbf{K_6} = \mathbf{H_{dc}\bar{P}H_c^{-1}K_6^c} - \mathbf{K_6^{dc}}, \tag{4.70}$$

$$\mathbf{K_7} = \mathbf{H_{dc}\bar{P}H_c^{-1}K_7^c} - \mathbf{K_7^{dc}\bar{P}}, \tag{4.71}$$

$$\mathbf{Q_{dc}} = \begin{pmatrix} \mathbf{0} & \mathbf{Q_{1,2}} & . & . & \mathbf{Q_{1,p}} \\ \mathbf{Q_{2,1}} & \mathbf{0} & . & . & \mathbf{Q_{2,p}} \\ . & . & . & . & . \\ . & . & . & . & . \\ \mathbf{Q_{p,1}} & \mathbf{Q_{p,2}} & . & . & \mathbf{0} \end{pmatrix}. \tag{4.72}$$

In the above equations,

$$\mathbf{\bar{H}_{dc}} = \begin{pmatrix} \mathbf{\bar{H}_{dc,1}} & & & \\ & \mathbf{\bar{H}_{dc,2}} & & \\ & & \ddots & \\ & & & \mathbf{\bar{H}_{dc,p}} \end{pmatrix}, \tag{4.73}$$

$$\mathbf{K}_m^{\mathbf{dc}} = \begin{pmatrix} \mathbf{K}_m^{\mathbf{dc,1}} & & & \\ & \mathbf{K}_m^{\mathbf{dc,2}} & & \\ & & \ddots & \\ & & & \mathbf{K}_m^{\mathbf{dc,p}} \end{pmatrix} \text{ for } m \in \{1,2,3,4,7\}, \tag{4.74}$$

$$\mathbf{K}_m^{\mathbf{dc}} = \left[\left(\mathbf{K}_m^{\mathbf{dc,1}}\right)^T \quad \left(\mathbf{K}_m^{\mathbf{dc,2}}\right)^T \cdots \left(\mathbf{K}_m^{\mathbf{dc,p}}\right)^T\right]^T \text{ for } m \in \{5,6\}. \tag{4.75}$$

*Proof:* Using properties 1, 2, 6 and 7 of the extraction matrices (Section 4.3.3), (4.50) can equivalently be restated in the form shown below

$$-2\mathbf{\bar{H}_{dc,i}\bar{P}_i\bar{v}_{dc}^*} = \mathbf{K_1^{dc,i}R_iT_w}(k) + \mathbf{K_2^{dc,i}P_iT_z}(k) + \mathbf{K_3^{dc,i}\bar{R}_i\bar{d}_w} + \mathbf{K_4^{dc,i}P_i\bar{d}_z} + \mathbf{K_5^{dc,i}\bar{T}_a}$$
$$+\mathbf{K_6^{dc,i}\bar{T}_g} + \mathbf{K_7^{dc,i}\bar{P}_i\bar{T}_{z,ref}} + \textstyle\sum_{j\neq i}\mathbf{Q_{ij}\bar{P}_j\bar{\bar{T}}_z} \tag{4.76}$$

Using the definitions of $\mathbf{P}$, $\mathbf{\bar{P}}$, $\mathbf{R}$ and $\mathbf{\bar{R}}$ from (4.35), (4.36), (4.38) and (4.39) respectively, the combined form of (4.76) resulting from the concatenation over all clusters $(i = 1,2,\ldots,p)$, is expressed as (4.76), where $\mathbf{Q_{dc}}$, $\mathbf{H_{dc}}$ and $\mathbf{K}_m^{\mathbf{dc}}$ $(m \in \{1,2,\ldots,7\})$ are as defined in (4.73) to ( 4.75).

$$-2\mathbf{\bar{H}_{dc}\bar{P}\bar{v}_{dc}^*} = \mathbf{K_1^{dc}RT_w}(k) + \mathbf{K_2^{dc}PT_z}(k) + \mathbf{K_3^{dc}\bar{R}\bar{d}_w} + \mathbf{K_4^{dc}P\bar{d}_z} + \mathbf{K_5^{dc}\bar{T}_a} + \mathbf{K_6^{dc}\bar{T}_g} + \mathbf{K_7^{dc}\bar{P}\bar{T}_{z,ref}}$$

$$+\ \mathbf{Q}_{\mathbf{dc}}\bar{\mathbf{P}}\bar{\bar{\mathbf{T}}}_{\mathbf{z}} \tag{4.76}$$

Comparing (4.25) and (4.76), $\bar{\mathbf{v}}_{\mathbf{dc}}^{*} = \bar{\mathbf{v}}_{\mathbf{c}}^{*}$ if $\hat{\bar{\mathbf{T}}}_{\mathbf{z}}$ is such that $(\mathbf{Q}_{\mathbf{dc}}\bar{\mathbf{P}})\bar{\bar{\mathbf{T}}}_{\mathbf{z}} = \mathbf{K}_1\mathbf{T}_{\mathbf{w}}(k) + \mathbf{K}_2\mathbf{T}_{\mathbf{z}}(k) + \mathbf{K}_3\bar{\mathbf{d}}_{\mathbf{w}} + \mathbf{K}_4\bar{\mathbf{d}}_{\mathbf{z}} + \mathbf{K}_5\bar{\mathbf{T}}_{\mathbf{a}} + \mathbf{K}_6\bar{\mathbf{T}}_6 + \mathbf{K}_7\bar{\mathbf{T}}_{\mathbf{z,ref}}$ with $\mathbf{K}_m$, $m \in \{1,2,\dots,7\}$ defined in (4.65) to (4.71). This completes the proof of Theorem 4.1. □

**Remarks.** It should be noted that the centralized-decentralized equivalence condition stated above was derived based on the assumption that, apart from the underlying dynamical models, no other constraints are imposed on the centralized and decentralized optimization problems.

**Corollary 4.1** *If the overall lifted vector of zone temperature estimates,* $\hat{\bar{\mathbf{T}}}_{\mathbf{z}}$ *is chosen to be different from* $\hat{\bar{\mathbf{T}}}_{\mathbf{z}}^{*}$*, the overall lifted vector of optimal control inputs for the corresponding decentralized controller differs from that for centralized control by an amount that is linearly dependent on the deviation of* $\hat{\bar{\mathbf{T}}}_{\mathbf{z}}$ *from* $\hat{\bar{\mathbf{T}}}_{\mathbf{z}}^{*}$*. More precisely,*

$$\bar{\mathbf{v}}_{\mathbf{c}}^{*} - \mathbf{v}_{\mathbf{dc}}^{*} = -\frac{1}{2}\bar{\mathbf{P}}^{-1}\bar{\mathbf{H}}_{\mathbf{dc}}^{-1}\mathbf{Q}_{\mathbf{dc}}\bar{\mathbf{P}}\left(\hat{\bar{\mathbf{T}}}_{\mathbf{z}}^{*} - \hat{\bar{\mathbf{T}}}_{\mathbf{z}}\right) \tag{4.77}$$

*Proof:* The application of Theorem 4.1 to (4.76) provides the following alternative expression for $\bar{\mathbf{v}}_{\mathbf{c}}^{*}$, which was originally given by (4.25):

$$-2\bar{\mathbf{H}}_{\mathbf{dc}}\bar{\mathbf{P}}\bar{\mathbf{v}}_{\mathbf{c}}^{*} = \mathbf{K}_1^{\mathbf{dc}}\mathbf{R}\mathbf{T}_{\mathbf{w}}(k) + \mathbf{K}_2^{\mathbf{dc}}\mathbf{P}\mathbf{T}_{\mathbf{z}}(k) + \mathbf{K}_3^{\mathbf{dc}}\bar{\mathbf{R}}\bar{\mathbf{d}}_{\mathbf{w}} + \mathbf{K}_4^{\mathbf{dc}}\mathbf{P}\bar{\mathbf{d}}_{\mathbf{z}} + \mathbf{K}_5^{\mathbf{dc}}\bar{\mathbf{T}}_{\mathbf{a}} + \mathbf{K}_6^{\mathbf{dc}}\bar{\mathbf{T}}_{\mathbf{g}} + \mathbf{K}_7^{\mathbf{dc}}\bar{\mathbf{P}}\bar{\mathbf{T}}_{\mathbf{z,ref}}$$

$$+\ \mathbf{Q}_{\mathbf{dc}}\bar{\mathbf{P}}\hat{\bar{\mathbf{T}}}_{\mathbf{z}}^{*}. \tag{4.78}$$

Subtraction of (4.76) from (4.78) leads to (4.77). □

**Remarks.**

1. The family of solutions to (4.64) can be described by the set $\{\mathbf{y}_0 + \bar{\mathbf{P}}^{-1}\mathbf{x} : \mathbf{x} \in Ker\,(\mathbf{Q}_{\mathbf{dc}})\}$, where $\mathbf{x}$ is any particular solution of (4.64). To satisfy the conditions of Theorem 4.1, $\hat{\bar{\mathbf{T}}}_{\mathbf{z}}^{*}$ can be chosen as any element from this set. In the particular case where $\mathbf{Q}_{\mathbf{dc}}$ is invertible, $\hat{\bar{\mathbf{T}}}_{\mathbf{z}}^{*}$ has a unique closed form expression given by $\hat{\bar{\mathbf{T}}}_{\mathbf{z}}^{*} = \bar{\mathbf{P}}^{-1}\mathbf{Q}_{\mathbf{dc}}^{-1}\big(\mathbf{K}_1\mathbf{T}_{\mathbf{w}}(k) + \mathbf{K}_2\mathbf{T}_{\mathbf{z}}(k) + \mathbf{K}_3\bar{\mathbf{d}}_{\mathbf{w}} + \mathbf{K}_4\bar{\mathbf{d}}_{\mathbf{z}} + \mathbf{K}_5\bar{\mathbf{T}}_{\mathbf{a}} + \mathbf{K}_6\bar{\mathbf{T}}_6 + \mathbf{K}_7\bar{\mathbf{T}}_{\mathbf{z,ref}}\big)$.
2. Since $\bar{\mathbf{H}}_{\mathbf{dc,i}}$, $i \in \{1,2,\dots,p\}$ is full rank due to strict convexity of the decentralized MPC optimization problem, the matrix $\bar{\mathbf{H}}_{\mathbf{dc}}$ defined in (4.73) is also full rank. Hence, $\mathbf{H}_{\mathbf{dc}}^{-1}$ exists for use in (4.77).

3. The centralized-decentralized equivalence (Theorem 4.1) provides a sufficient condition for which the solutions of the centralized and decentralized (corresponding to any p-partition) MPC problems match. The temperature estimates $\hat{\bar{\bar{\mathbf{T}}}}_{\mathbf{z}}^{*}$, that the multi-agent decentralized controllers would require in such a situation, depend on system wide sensory data $\mathbf{T}_{\mathbf{z}}(k)$ and $\mathbf{T}_{\mathbf{w}}(k)$, as expressed by (4.64). Since each decentralized controller has access to only certain temperature measurements $\mathbf{T}_{\mathbf{zi}}(k)$ and $\mathbf{T}_{\mathbf{wi}}(k)$, the centralized-decentralized equivalence of Theorem 4.1, cannot be achieved in practice. However, this condition can be used to quantify the difference between the centralized and decentralized solutions as stated in Corollary 4.1.

### 4.3.6 Optimality metric

For any p-partition of the building, where $p \in \{1,2,\ldots,N_z\}$, we define an appropriate scalar metric to quantify the optimality associated with a multi-agent MPC controller that is decentralized with respect to the clusters constituting the partition. The overall lifted vector of temperature estimates $\hat{\bar{\bar{\mathbf{T}}}}_{\mathbf{z}}$ defined in (4.61) which is required for implementing the decentralized MPC controller is, at best, chosen heuristically. The centralized-decentralized equivalence (Theorem 4.1) establishes a theoretical best value of this estimate, $\hat{\bar{\bar{\mathbf{T}}}}_{\mathbf{z}}^{*}$, which if used, results in matching of the centralized and decentralized control inputs. However, an arbitrary choice of this estimate results in a deviation of the decentralized control inputs $\mathbf{v}_{\mathbf{dc}}^{*}$ from the centralized control inputs $\bar{\mathbf{v}}_{\mathbf{c}}^{*}$ which is quantified by (4.77). This deviation translates into a deviation of the centralized objective function from its optimal value. To quantify it we proceed as follows.

Using the transformation 4.4, the centralized objective function $g_c(\bar{\mathbf{u}})$ given by (3.19) was expressed as a function $g_c'(\bar{\mathbf{v}})$ shown in (4.5). An alternative expression for $g_c'(\bar{\mathbf{v}})$ is

$$g_c'(\bar{\mathbf{v}}) = g_c'(\bar{\mathbf{v}}_{\mathbf{c}}^{*}) + (\bar{\mathbf{v}} - \bar{\mathbf{v}}_{\mathbf{c}}^{*})^T \bar{\mathbf{H}} (\bar{\mathbf{v}} - \bar{\mathbf{v}}_{\mathbf{c}}^{*}). \qquad (4.79)$$

Using Corollary 4.1, we obtain

$$\begin{aligned} g_c'(\bar{\mathbf{v}}_{\mathbf{c}}^{*}) - g_c'(\bar{\mathbf{v}}_{\mathbf{dc}}^{*}) &= (\bar{\mathbf{v}}_{\mathbf{dc}}^{*} - \bar{\mathbf{v}}_{\mathbf{c}}^{*})^T \bar{\mathbf{H}} (\bar{\mathbf{v}}_{\mathbf{dc}}^{*} - \bar{\mathbf{v}}_{\mathbf{c}}^{*}) \\ &= \frac{1}{4}\left(\hat{\bar{\bar{\mathbf{T}}}}_{\mathbf{z}} - \hat{\bar{\bar{\mathbf{T}}}}_{\mathbf{z}}^{*}\right)^T \bar{\mathbf{P}}^T \mathbf{Q}_{\mathbf{dc}}^T \bar{\mathbf{H}}_{\mathbf{dc}}^{-1} \bar{\mathbf{P}} \bar{\mathbf{H}} \bar{\mathbf{P}}^T \bar{\mathbf{H}}_{\mathbf{dc}}^{-1} \mathbf{Q}_{\mathbf{dc}} \bar{\mathbf{P}} \left(\hat{\bar{\bar{\mathbf{T}}}}_{\mathbf{z}} - \hat{\bar{\bar{\mathbf{T}}}}_{\mathbf{z}}^{*}\right). \end{aligned} \qquad (4.80)$$

Here, we have used the fact that $\bar{\mathbf{H}}_{\mathbf{dc}}$ is a symmetric matrix and $\bar{\mathbf{P}}$ is a permutation matrix.

It is desired that the above deviation be small, so that the decentralized controller can provide performance which is close to that provided by centralized control. An illustration of this deviation is shown in Figure 4.13 for a hypothetical scalar case.

Using the inequality shown in (4.81), it can be concluded that the quantity $\lambda_{max}(\bar{\mathbf{P}}^T \mathbf{Q}_{\mathbf{dc}}^T \bar{\mathbf{H}}_{\mathbf{dc}}^{-1} \bar{\mathbf{P}} \bar{\mathbf{H}} \bar{\mathbf{P}}^T \bar{\mathbf{H}}_{\mathbf{dc}}^{-1} \mathbf{Q}_{\mathbf{dc}} \bar{\mathbf{P}})$ characterizes an upper bound on the above deviation in the objective function which is independent of the deviation of $\bar{\bar{\mathbf{T}}}_{\mathbf{z}}$ from $\bar{\bar{\mathbf{T}}}_{\mathbf{z}}^*$. Therefore, it represents the 'loss' in optimality in going from a centralized control architecture to a decentralized control architecture. We denote this quantity using the term **Optimality Loss Factor** (OLF) as shown in (4.82).

$$\left(\bar{\bar{\mathbf{T}}}_{\mathbf{z}} - \bar{\bar{\mathbf{T}}}_{\mathbf{z}}^*\right)^T \bar{\mathbf{P}}^T \mathbf{Q}_{\mathbf{dc}}^T \bar{\mathbf{H}}_{\mathbf{dc}}^{-1} \bar{\mathbf{P}} \bar{\mathbf{H}} \bar{\mathbf{P}}^T \bar{\mathbf{H}}_{\mathbf{dc}}^{-1} \mathbf{Q}_{\mathbf{dc}} \bar{\mathbf{P}} \left(\bar{\bar{\mathbf{T}}}_{\mathbf{z}} - \bar{\bar{\mathbf{T}}}_{\mathbf{z}}^*\right) \leq$$

$$\lambda_{max}(\bar{\mathbf{P}}^T \mathbf{Q}_{\mathbf{dc}}^T \bar{\mathbf{H}}_{\mathbf{dc}}^{-1} \bar{\mathbf{P}} \bar{\mathbf{H}} \bar{\mathbf{P}}^T \bar{\mathbf{H}}_{\mathbf{dc}}^{-1} \mathbf{Q}_{\mathbf{dc}} \bar{\mathbf{P}}) \left\| \bar{\bar{\mathbf{T}}}_{\mathbf{z}} - \bar{\bar{\mathbf{T}}}_{\mathbf{z}}^* \right\|_2^2. \quad (4.81)$$

$$\text{OLF} = \lambda_{max}(\bar{\mathbf{P}}^T \mathbf{Q}_{\mathbf{dc}}^T \bar{\mathbf{H}}_{\mathbf{dc}}^{-1} \bar{\mathbf{P}} \bar{\mathbf{H}} \bar{\mathbf{P}}^T \bar{\mathbf{H}}_{\mathbf{dc}}^{-1} \mathbf{Q}_{\mathbf{dc}} \bar{\mathbf{P}}) = \sigma_{max}^2(\bar{\mathbf{H}}^{-1} \bar{\mathbf{P}}^T \bar{\mathbf{H}}_{\mathbf{dc}}^{-1} \mathbf{Q}_{\mathbf{dc}} \bar{\mathbf{P}}). \quad (4.82)$$

The OLF is used as the appropriate optimality metric which must be minimized in the choice of partitions for decentralized control.

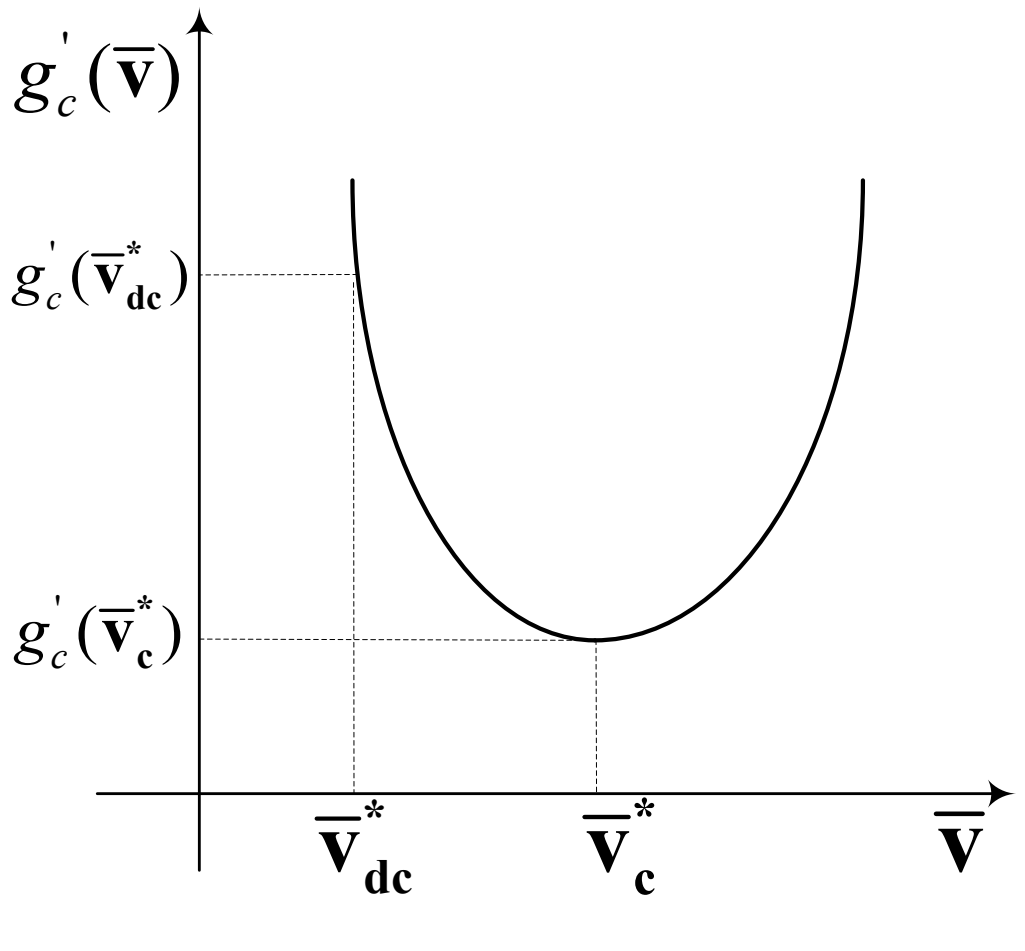

**Figure 4.13 Scalar illustration of deviation in performance between centralized and decentralized MPC when $\bar{\bar{\mathbf{T}}}_{\mathbf{z}}^* \neq \bar{\bar{\mathbf{T}}}_{\mathbf{z}}$**

### 4.3.7 Robustness metric

For any *p-partition* of the building, where $p \in \{1,2,\ldots,N_z\}$, we define an appropriate scalar metric to quantify the robustness associated with a multi-agent MPC controller that is decentralized with respect to the clusters constituting the partition. In such an architecture, the effect of a sensor or communication related fault on the resulting control inputs, $\bar{\mathbf{v}}^*_{\mathbf{dc}}$, will be confined to the cluster where the fault originates because the control agents do not communicate. Therefore, the average number of affected zones in the event of a failure is an indicator of robustness in the sense that a small value ensures that the effect of failures is less widespread.

To quantify this concept, we consider the event that a failure has occurred in one of the zones in the building. For simplicity, we assume that the probability of failure occurring in any particular zone of the building is uniform across all zones and therefore equals $1/N_z$. We define the **Fault Propagation Metric** (FPM) $\in (0,1]$ as the expected value of the aggregated thermal capacity of all affected zones in case of above failure event, normalized with respect to the total thermal capacity of zones in the building. An expression for the FPM is derived as shown

$$\text{FPM} = \frac{\text{Expected value of aggregated thermal capacity of all affected zones}}{\text{Total thermal capacity of all zones in the building}}$$

$$= \frac{\sum_{m=1}^{N_z} (\text{Probability of failure in zone } m \times \text{Net thermal capacity of zones in cluster containing zone } m)}{\text{Total thermal capacity of all zones in the building}}$$

$$= \frac{\sum_{i=1}^{p} (\text{Probability of failure in each zone} \times \text{Number of zones in cluster } i \times \text{Net thermal capacity of zones in cluster } i)}{\text{Total thermal capacity of all zones in the building}}$$

$$= \frac{\sum_{i=1}^{p} \frac{1}{N_z} N_{zi} C_{zi}}{N_z} = \frac{1}{N_z C_z} \sum_{i=1}^{p} N_{zi} C_{zi} \tag{4.83}$$

As an example, FPM for all possible partitions of the 3-zone building shown in Figure 3.3 are calculated in Table 4.5. For simplicity, the thermal capacity of each zone is set to unity.

**Table 4.5 FPM computation for 3-zone building in Figure 3.3**

| $p$ | p-partitions | FPM |
|---|---|---|
| 1 | {1,2,3} | 1 |
| 2 | {1,2},{3} OR {1,3},{2} OR {3,2},{1} | $\frac{(2\times2)+(1\times1)}{(3\times3)}=\frac{5}{9}$ |
| 3 | {1},{2},{3} | $\frac{(1\times1)+(1\times1)+(1\times1)}{3\times3}=\frac{3}{9}$ |

It should be noted that in the case of a building, which exhibits heterogeneity in the sense that its thermal zones are sized differently, the total volume of space affected is a more appropriate characterization of the effect of the failure event than just the number of affected thermal zones. Since the thermal capacity of a zone is closely related to its volume, the FPM defined above is an indicator of the effect of failure in terms of the volume of affected space. Therefore, the FPM is used as an appropriate robustness metric which must be minimized in the choice of partitions for decentralized control.

### 4.3.8 Optimal partitioning problem and complexity analysis

This section presents a formal definition of the partitioning problem for the decentralized thermal control of a building and analyzes the underlying computational complexity.

**Definition 4.9 (Optimal *p-partition*)**: An *optimal p-partition* is defined as one with the smallest OLF among all possible *p-partitions* of $S_z$, for a fixed $p \in \{1,2,\ldots,N_z\}$.

**Definition 4.10. (Optimal partitioning problem)**: The *optimal partitioning problem* is to determine a family of *optimal p-partitions*, one for each $p \in \{1,2,\ldots,N_z\}$.

For a given $p$, the number of *p-partitions* of the set $S_z$ is given by the *Stirling number of the second kind*, $S(N_z, p)$ [84], as expressed in (4.84). Therefore, the total number of partitions to be considered to solve the optimal partitioning problem is given by the sum of the Stirling numbers over $p$, which is also defined as the *Bell number* [84], denoted by $B_{N_z}$ and is expressed in (4.85 ).

$$S(N_z, p) = \frac{1}{N_z!} \sum_{j=1}^{p} (-1)^j \binom{p}{j} (p-j)^{N_z} \quad (4.84)$$

$$B_{N_z} = \sum_{p=1}^{N_z} S(N_z, p) \quad (4.85)$$

The Bell number grows exponentially[4] with $N_z$. This implies that if all possible partitions were to be considered, the optimal partitioning problem becomes intractable as the number of zones in the building increases. This motivates the development of a less computationally complex methodology for optimal partitioning using only a small subset of all partitions of $S_z$.

### 4.3.9 Agglomerative clustering

Agglomerative or `bottom-up' clustering [86] is a hierarchical methodology used in a variety of applications such as data-mining and bio-informatics [87, 88] to form clusters of objects. It starts with individual objects that are progressively grouped together into larger clusters until the root cluster containing all the objects is reached. This is typically done using a greedy approach which groups the two 'closest' clusters together at each step, based on a suitable distance function metric between clusters.

We adopt the agglomerative clustering approach using the OLF as a distance function, to address the optimal partitioning problem in the context of decentralized building thermal control.

**Algorithm 4.1. Agglomerative Clustering for Partitioning a Building (See Figure 4.14 for illustration)**

*STEP 1:* Define the initial *parent partition* as the unique $N_z$-partition of $S_z$, which consists of $N_z$ clusters, each having exactly one zone.

*STEP 2:* Agglomerate any two clusters in the parent partition to create a *child partition*. In this way, find all child partitions of the parent partition. Compute the OLF for each such child partition.

*STEP 3:* Among all child partitions found above, determine one with smallest OLF. Set the new parent partition to be this child partition. In case of multiple child partitions having the smallest OLF, select any one of them.

*STEP 4:* Repeat steps 2 and 3, until the parent partition becomes the unique 1-partition of $S_z$,

---

[4]The first few Bell numbers are given by 1, 1, 2, 5, 15, 52, 203, 877, 4140, 21147, 115975,...

which consists of exactly one cluster that contains all zones.

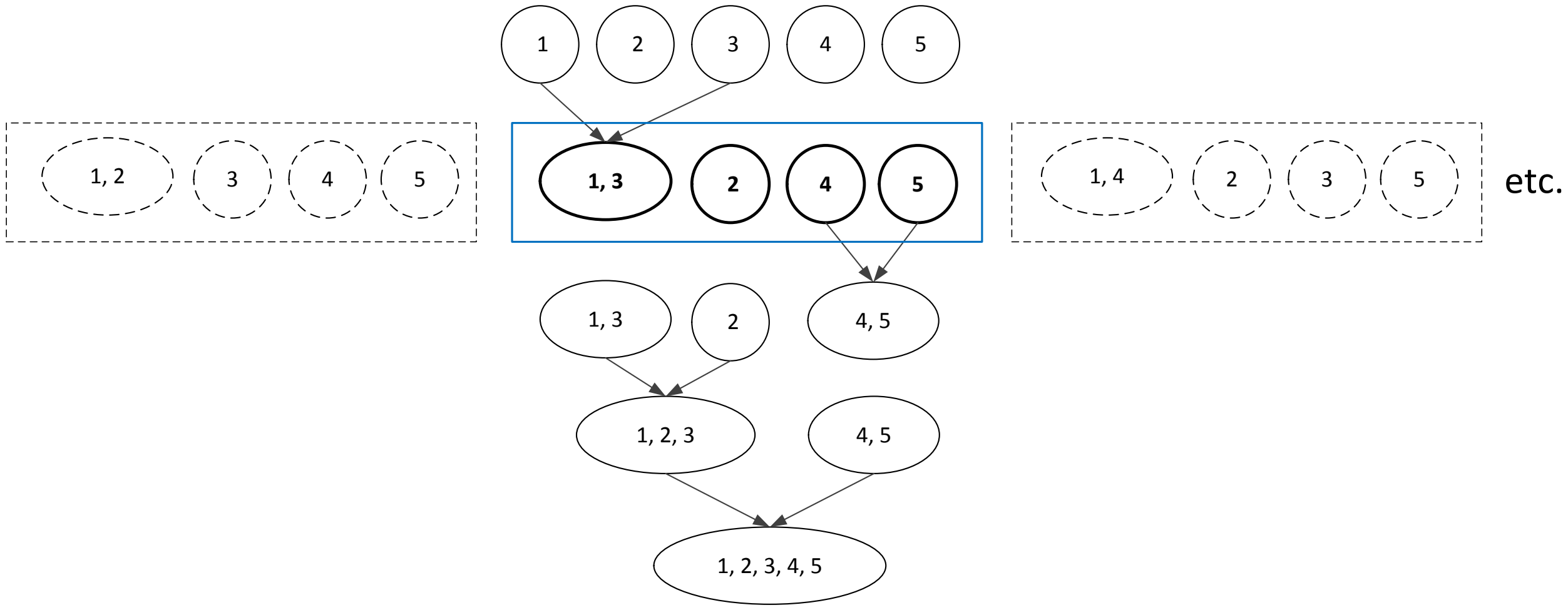

**Figure 4.14 Illustration of agglomerative clustering for a building with 5 zones**

**Remarks.**

1. The *parent partition* has $N_z$ clusters initially. At the end of each iteration, the number of clusters decreases exactly by 1. Therefore the *parent partition* obtained at the end of the $i^{th}$ iteration is a *p-partition*, with $p = N_z - i + 1$.

2. The *p-partitions* generated via these iterations are not necessarily optimal *p-partitions* (definition 4.9). They represent a guess for an optimal *p-partition*, obtained from the previous iteration in a greedy manner.

3. If the size of the parent partition in any iteration of Algorithm 4.1 is $N^{parent}$, the number of child partitions $N^{child}$ that are created using the agglomeration described in Step 2 of the algorithm is given by:

$$N^{child} = \binom{N^{parent}}{2} \qquad (4.86)$$

Noting that $N^{parent}$ starts from $N_z$ and decreases by 1 in each iteration, the total number of child partitions, $N_{total}^{child}$ considered in one run of the algorithm is $O\left({N_z}^3\right)$ as computed below.

$$N_{total}^{child} = \sum_1^{N_z} \binom{N^{parent}}{2} = \sum_1^{N_z} \frac{N^{parent}(N^{parent}-1)}{2} = \frac{N_z(N_z+1)(N_z-1)}{6} \qquad (4.87)$$

Hence a significant computational benefit is achieved when compared to the complexity associated with considering all possible partitions (Figure 4.15).

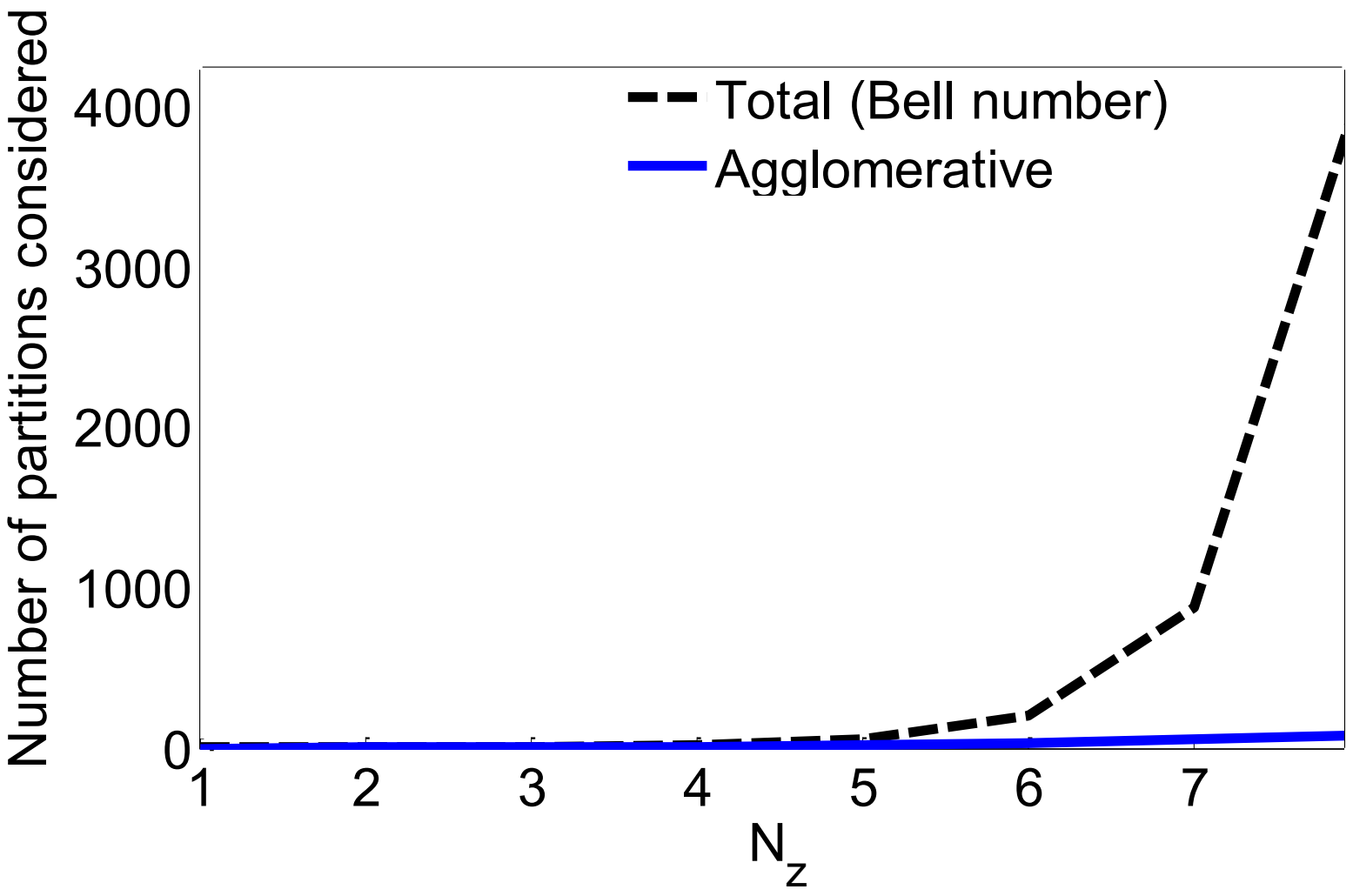

**Figure 4.15 Computational complexity comparison of partitioning approaches**

Similar to the CLF-MCS method, the results of Algorithm 4.1 can be presented on an *optimality-robustness trade-off curve*, as notionally illustrated in Figure 4.16, where the OLF and the FPM values of the resulting parent partitions from each iteration are plotted. This represents a multi-objective optimization framework, where the goal is to simultaneously minimize both the OLF and the FPM. The rightmost and leftmost points on this curve correspond to the two extremes of a completely centralized architecture (a single cluster) and a completely decentralized architecture ($N_z$ clusters) respectively. This curve serves as a useful design tool. It can be used to compare various partitions and make a decision on the appropriate intermediate architecture between these two extremes that results in a satisfactory trade-off between optimality and robustness objectives. Some heuristic guidelines, similar to the CLF-MCS approach, are presented below based on visual inspection.

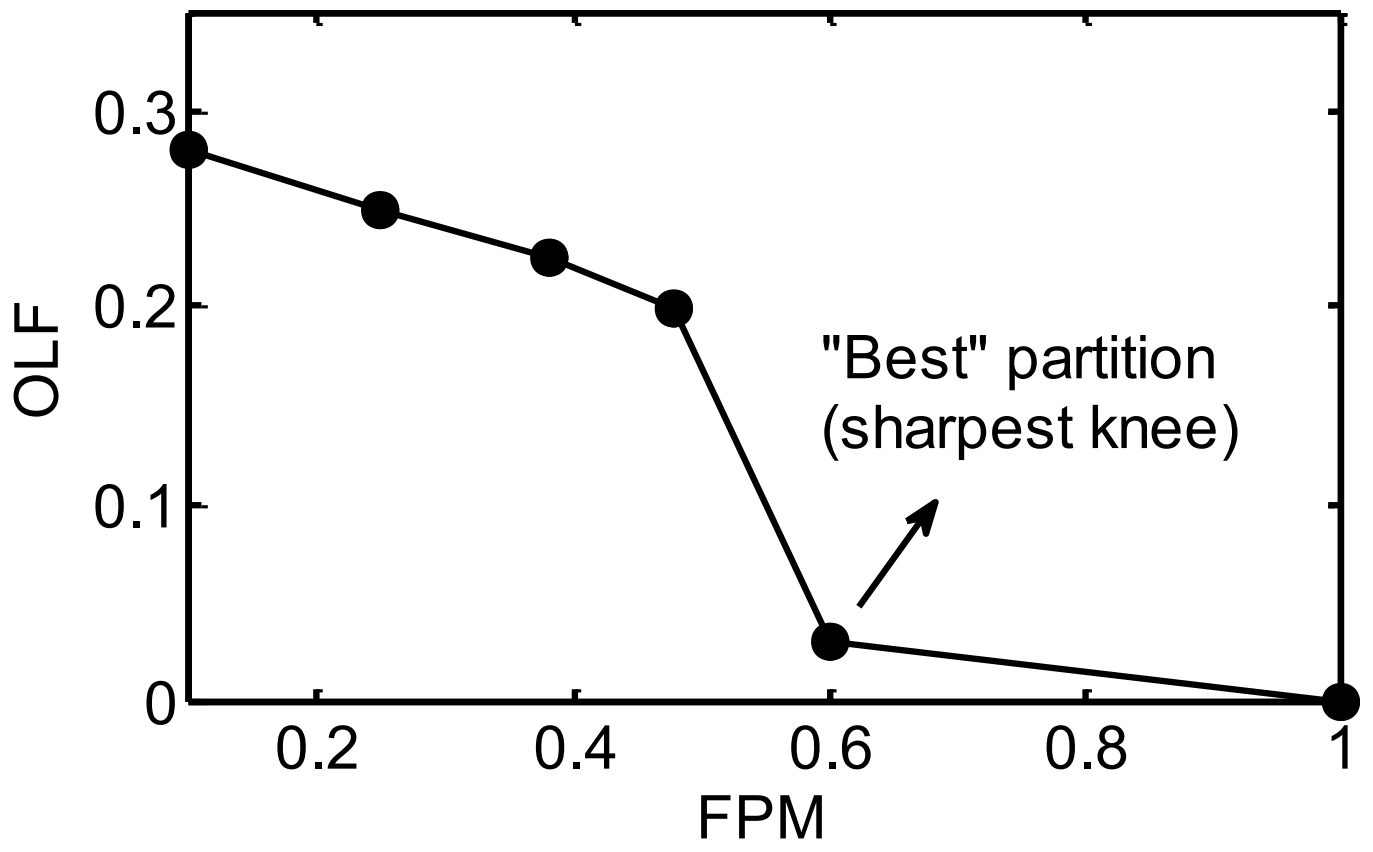

**Figure 4.16 Illustration of optimality-robustness trade-off curve for OLF-FPM approach**

Starting from the rightmost point, and proceeding left on the optimality-robustness trade-off curve, the partitions which correspond to a 'knee point' on this curve should be explored. At such points, navigation in either direction would result in a large increase in one metric but only a relatively small decrease in the other metric. Therefore, these points reflect the attainment of a satisfactory balance between the optimality and robustness objectives, and the corresponding partition of the building should preferably be used for decentralized control. In the event of multiple knee points in the curve, the sharpest among them may be considered. Also, those knee points which are more centrally located should be preferred over others. The following sections provide some examples to further explain this process.

### 4.3.10 Nine zone building example

We revisit the 9-zone building described in section 4.2.9. Its layout is shown in Figure 4.9, and the construction properties, wall resistances and capacitances are shown in Table 4.1 and Table 4.2. Similar to Section 4.2.9, the capacitance for each zone was assumed to be 250 kJ/kg. The contribution of occupants and objects to the zone capacitances was ignored for simplicity.

With the above modeling assumptions, the building is rendered thermally symmetric, meaning that the resistances offered by all internal walls are the same (Table 4.2). To introduce anisotropy, we artificially decrease the thermal resistances associated with the vertical internal walls by a factor of 3 from their originally computed values. Furthermore, we multiply each wall resistance by a factor of 0.06 to increase the coupling among zones. An overall system model of the form (3.4) was obtained by constructing an RC network using these details, and then applying Algorithm 3.1 where discretization was performed using the zero-order-hold method with step size of 10 minutes.

To partition the building, the exact solution to the optimal partitioning problem was obtained for this example by considering all possible p-partitions, for each $p \in \{1,2,\ldots,9\}$ as described in Section 4.3.8. Next, the agglomerative clustering approach (Algorithm 4.1) was applied, which provided suitable guesses for optimal p-partitions. OLF computations were based on the MPC parameters $N = 5$ samples, $\alpha = 0.1\mathbf{e}_9$ and $\beta = \mathbf{e}_9$. All relevant codes are provided in Appendix C. The resulting partitions from the agglomerative approach and the exact solution approach are compared in Figure 4.17 and Table 4.6. The corresponding optimality-robustness trade-off curves are shown in Figure 4.18. The following observations can be made from these results:

1. As seen in Figure 4.17, the p-partitions obtained using the agglomerative approach have OLF values close (in many cases identical) to optimal *p-partitions*, for all $p \in \{1,2,\ldots,9\}$.

2. Visual inspection of Figure 4.18 for knee-points suggests that the partition which offers the `best' trade-off between the OLF (optimality) and FPM (robustness) objectives is an optimal *3-partition*, which from the Table 4.6 corresponds to $\{\{1,2,3\},\{4,5,6\},\{7,8,9\}\}$. This would partition the building along its floors and is physically consistent with the thermal anisotropy that was introduced by causing horizontal walls (floors and ceilings) to be more insulated than vertical walls.

3. Another knee-point is observed in Figure 4.18 which corresponds to $p = 6$. This can possibly be explained on the basis of the partitions obtained in Table 3 for the agglomerative approach. Navigation from $p = 7$ to $p = 6$ results in fusion of the zones 8 and 9 from separate clusters into a single cluster. However, these zones have a potentially significant dynamical coupling through the relatively less insulated common vertical wall between them. This is likely to cause a large decrease in OLF for a comparatively small increase in FPM while going from $p = 7$ to $p = 6$ resulting in the knee point seen in Figure 4.18.

4. Finding the exact solution to the optimal partitioning problem required the consideration of 4140 partitions, whereas only 240 partitions were analyzed by the agglomerative approach.

From the above observations, it can be concluded that agglomeration is able to provide sufficiently accurate results with significantly less computational effort when compared to the exact solution approach involving the analysis of all partitions.

**Table 4.6 Optimal partitions vs. partitions using agglomeration for 9-zone building**

| $p$ | Optimal p-partition | | p-partition from agglomeration | |
|---|---|---|---|---|
| | Clusters | OLF ($\times 10^{-3}$) | Clusters | OLF ($\times 10^{-3}$) |
| 9 | {1}{2}{3}{4}{5}{6}{7}{8}{9} | 4.273 | {1}{2}{3}{4}{5}{6}{7}{8}{9} | 4.273 |
| 8 | {1}{2}{3}{4,5}{6}{7}{8}{9} | 4.053 | {1}{2}{3}{4,5}{6}{7}{8}{9} | 4.053 |
| 7 | {1,2}{3}{4}{5}{6}{7}{8,9} | 3.873 | {1}{2,3}{4,5}{6}{7}{8}{9} | 3.977 |
| 6 | {1}{2,3}{4,5}{6}{7}{8,9} | 2.088 | {1}{2,3}{4,5}{6}{7}{8,9} | 2.088 |
| 5 | {1}{2,3}{4,5,6}{7,8}{9} | 2.006 | {1}{2,3}{4,5}{6}{7,8,9} | 2.023 |
| 4 | {1,2,3}{4}{5,6}{7,8,9} | 1.936 | {1,2,3}{4,5}{6}{7,8,9} | 1.936 |
| 3 | {1,2,3}{4,5,6}{7,8,9} | 5.118 | {1,2,3}{4,5,6}{7,8,9} | 5.118 |
| 2 | {1,2,3,4,5,6}{7,8,9} | 2.559 | {1,2,3,4,5,6}{7,8,9} | 2.559 |
| 1 | {1,2,3,4,5,6,7,8,9} | 0 | {1,2,3,4,5,6,7,8,9} | 0 |

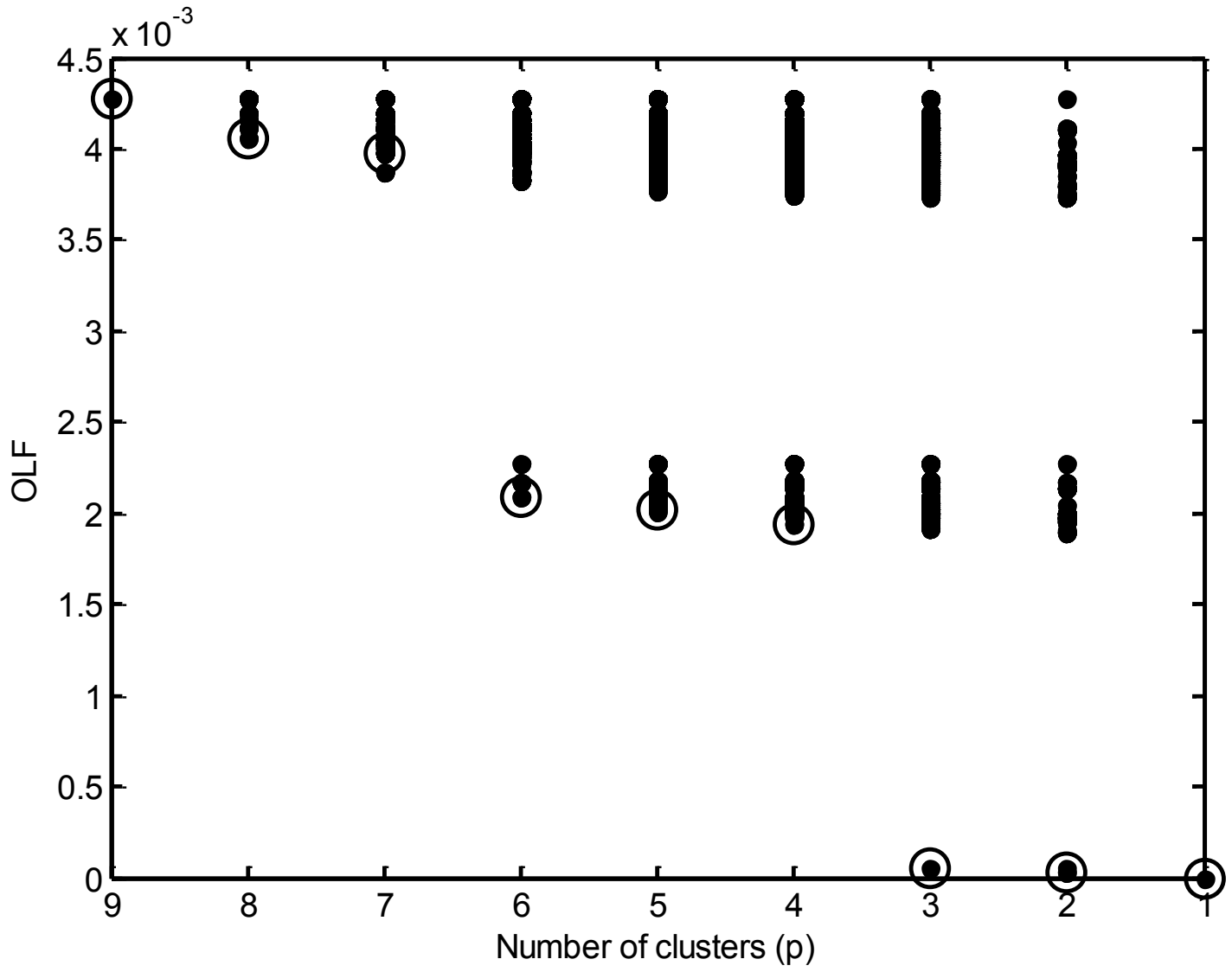

**Figure 4.17 OLF comparison of all p-partitions (solid circles) of the 9-zone building with agglomerative partitions (empty circles). The agglomerative clustering progresses from left to right, starting with the most decentralized partition ($p = 9$).**

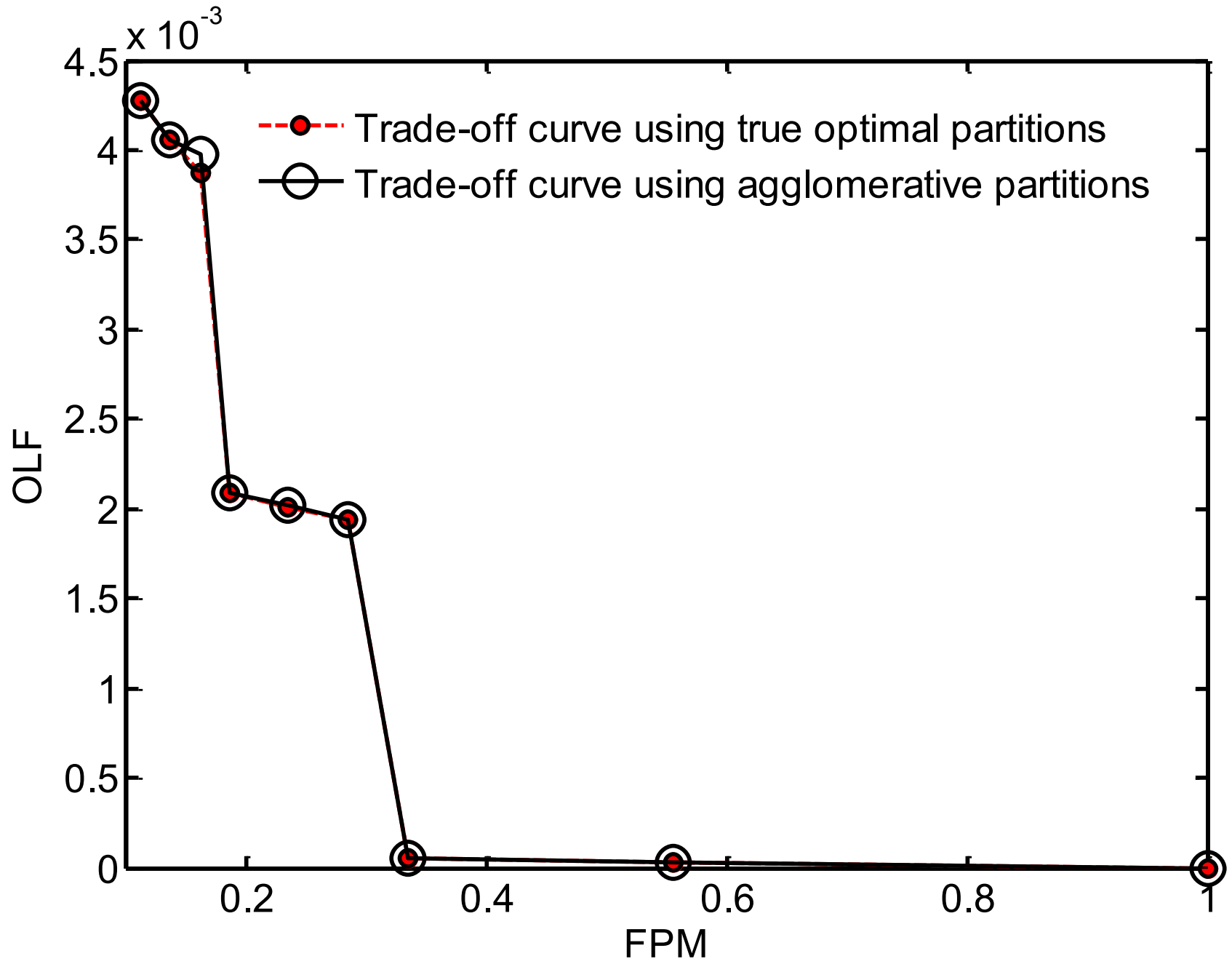

**Figure 4.18 Optimality robustness trade-off *curves* for 9-zone building using true optimal partitions and agglomerative partitions**

### 4.3.11 Eleven zone circular building

We now consider a single-story circular building with 11 zones including a central atrium (zone 1), as shown in Figure 4.19, which can be thought of as a small office building. It is surrounded by ambient on all sides. Hallways, shown shaded, are provided to facilitate the movement of people inside the building. For the purposes of modeling, the building has 27 external walls and an equal number of internal walls. The walls are assumed to have construction properties as shown in Table 4.1. The R and C parameters shown in Table 4.7 and Table 4.8 were computed in a manner similar to the 9-zone building, with the hallways modeled as resistors with a high value of resistance calculated using the thermal conduction and convection properties of air. For simplicity, the accessibility factors [38] for computation of wall resistances were assumed to be 0.5 each. Similar to the 9-zone building in the previous section, an overall system model of the form (3.4) was obtained by constructing an RC network using these details, and then applying Algorithm 3.1 where discretization was performed using the zero-order-hold method with step size of 10 minutes.

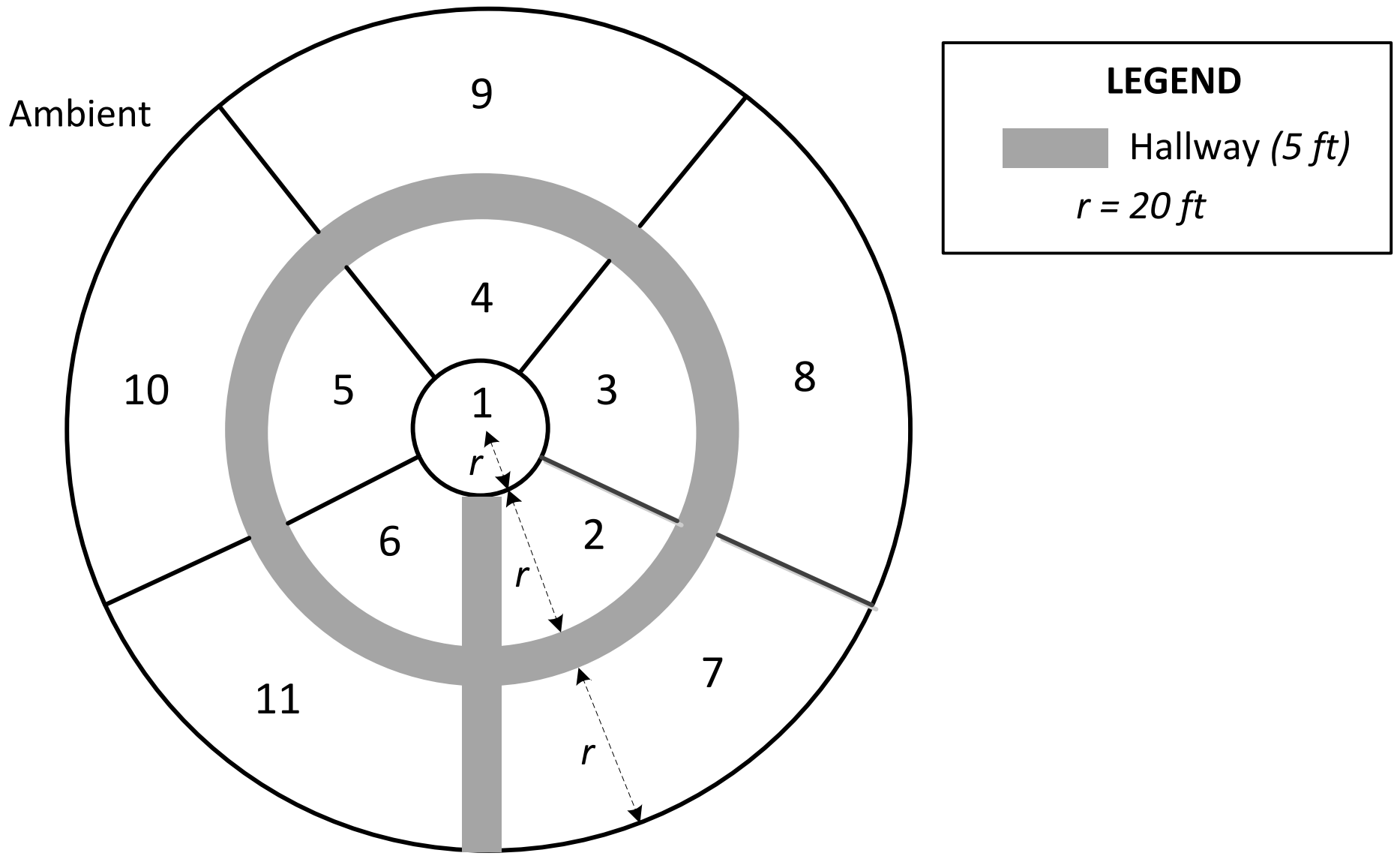

**Figure 4.19 Layout of 11-zone circular building (plan view with building height = 15 ft)**

**Table 4.7 Resistance and capacitance values for walls of 11-zone building (see Figure 3.5)**

| Type of wall | $C_{wall}$ (kJ/kg) | $R_1$ (K/kW) | $R_2$(K/kW) |
|---|---|---|---|
| Ceiling and floor for zone 1 | $4.27 \times 10^4$ | 9.178 | 9.178 |
| Ceilings and floors for zones 2-6 | $2.56 \times 10^4$ | 15.297 | 15.297 |
| Ceilings and floors for zones 7-11 | $3.36 \times 10^4$ | 8.344 | 8.344 |
| Vertical walls separating ambient from each of zones 7 – 11 | $4.17 \times 10^4$ | 9.414 | 9.414 |
| Vertical walls between zone 1 and each of zones 2 – 6 | $6.53 \times 10^3$ | 9.160 | 9.160 |
| Vertical walls separating zone pairs (2,3), (3,4), (4,5) and (5,6) | $5.20 \times 10^3$ | 11.515 | 11.515 |
| Vertical wall (hallway separation) between zones 2 and 6 | $1.04 \times 10^4$ | 1151.50 | 1151.50 |
| Vertical walls (hallway) separating zone pairs (2,7), (3,8), (4,9), (5,10), (6,11) | $2.77 \times 10^4$ | 431.25 | 431.25 |
| Vertical walls separating zone pairs (7,8), (8,9), (9,10) and (10,11) | $5.20 \times 10^3$ | 11.515 | 11.517 |
| Vertical wall (hallway separation) between zones 7 and 11 | $1.04 \times 10^4$ | 1151.50 | 1151.50 |

**Table 4.8 Zone capacitances for 11-zone building**

| Zone number | Capacitance (kJ/kg) |
|---|---|
| 1 | 643.72 |
| 2 – 6 | 386.23 |
| 7 – 11 | 708.09 |

An intuitive method to partition the building is to split it along the thermally insulating circular ring of hallway, resulting in two clusters: $\{1,2,3,4,5,6\}$ and $\{7,8,9,10,11\}$. However, it is not obvious how to further partition these clusters into smaller clusters. The partitions resulting from the application of the agglomerative clustering algorithm are shown in Table 4.9, where parameters used to compute the OLF values are $N = 5$, $\alpha = 0.1\mathbf{e}_{11}$ and $\beta = \mathbf{e}_{11}$. Relevant codes are provided in Appendix D. The corresponding optimality-robustness trade-off curve is shown in Figure 4.20. A visual inspection of this curve indicates the presence of two knee points, as labeled in the figure. Following observations are made:

1. Knee 1 corresponds to the intuitive partition $\{\{1,2,3,4,5,6\},\{7,8,9,10,11\}\}$ which was noted above. However, the unacceptably high FPM associated with this partition indicates inferior robustness to faults, motivating further partitioning of these clusters.

2. The OLF values for both knees 1 and 2 are close to zero. However, the FPM associated with knee 2 is only about 60% of the FPM for knee 1. Therefore, knee 2 provides a better trade-off between optimality and robustness than knee 1.

3. Strong dynamic coupling is expected to exist among zones 1 to 6, primarily due to the atrium (zone 1) which is connected to each of the zones 2 to 6. Therefore to ensure small deviation from optimality, the building should be partitioned such that these zones are contained in the same cluster. This is verified from the clusters constituting the partition for knee 2 ($p = 4$ in Table 4.9).

The above observations can also be explained by considering a scalarized framework, which is a widely used approach for multi-objective optimization [82]. For the multi-objective problem of minimizing OLF and FPM, we define a single scalar objective function $J_{partition}$ as shown in (4.88), where $\lambda$ is a parameter $\in [0,1]$ which can be adjusted to influence the relative weights on the optimality and robustness objectives. Here $\mathrm{OLF}_{\max}$ is the value of OLF for the most decentralized partition $\{\{1\},\{2\},\{3\},\{4\},\{5\},\{6\},\{7\},\{8\},\{9\},\{10\},\{11\}\}$, whereas $\mathrm{FPM}_{\max}$ is the value of the FPM for the most centralized partition, i.e. $\{1,2,3,4,5,6,7,8,9,10,11\}$.

$$J_{\text{partition}}(p) = \lambda \frac{\mathrm{OLF}\,(p)}{\mathrm{OLF}_{\max}} + (1-\lambda)\frac{\mathrm{FPM}\,(p)}{\mathrm{FPM}_{\max}} \qquad (4.88)$$

Figure 4.21 is a plot of $J_{\text{partition}}$ for the various p-partitions produced in Table 4.9 by the application of the agglomerative clustering approach. For $\lambda = 0.5$, it is observed that of all the partitions, the *4-partition* (knee 2 in Figure 4.20) corresponds to the global minimum, resulting in the smallest value of $J_{\text{partition}}$. However, increasing the weight on optimality by changing $\lambda$ to 0.85 causes the *2-partition* (knee 1 in Figure 4.20) to provide the global minimum. This is consistent with the analysis presented above. However since a suitable value of $\lambda$ is not obvious to decide, we prefer to use the optimality-robustness trade-off curve instead as the appropriate tool for the analysis of the partitions provided by agglomerative clustering.

**Table 4.9 Partitions using agglomeration for 11-zone building**

| ***p*** | **p-partition from agglomeration** |
|---|---|
| 11 | {1}{2}{3}{4}{5}{6}{7}{8}{9}{10}{11} |
| 10 | {1,4}{2}{3}{5}{6}{7}{8}{9}{10}{11} |
| 9 | {1,3,4}{2}{5}{6}{7}{8}{9}{10}{11} |
| 8 | {1,3,4,5}{2}{6}{7}{8}{9}{10}{11} |
| 7 | {1,2,3,4,5}{6}{7}{8}{9}{10}{11} |
| 6 | {1,2,3,4,5,6}{7}{8}{9}{10}{11} |
| 5 | {1,2,3,4,5,6}{7}{8}{9,10}{11} |
| 4 | {1,2,3,4,5,6}{7,8}{9,10}{11} |
| 3 | {1,2,3,4,5,6}{7,8}{9,10,11} |
| 2 | {1,2,3,4,5,6}{7,8,9,10,11} |
| 1 | {1,2,3,4,5,6,7,8,9,10,11} |

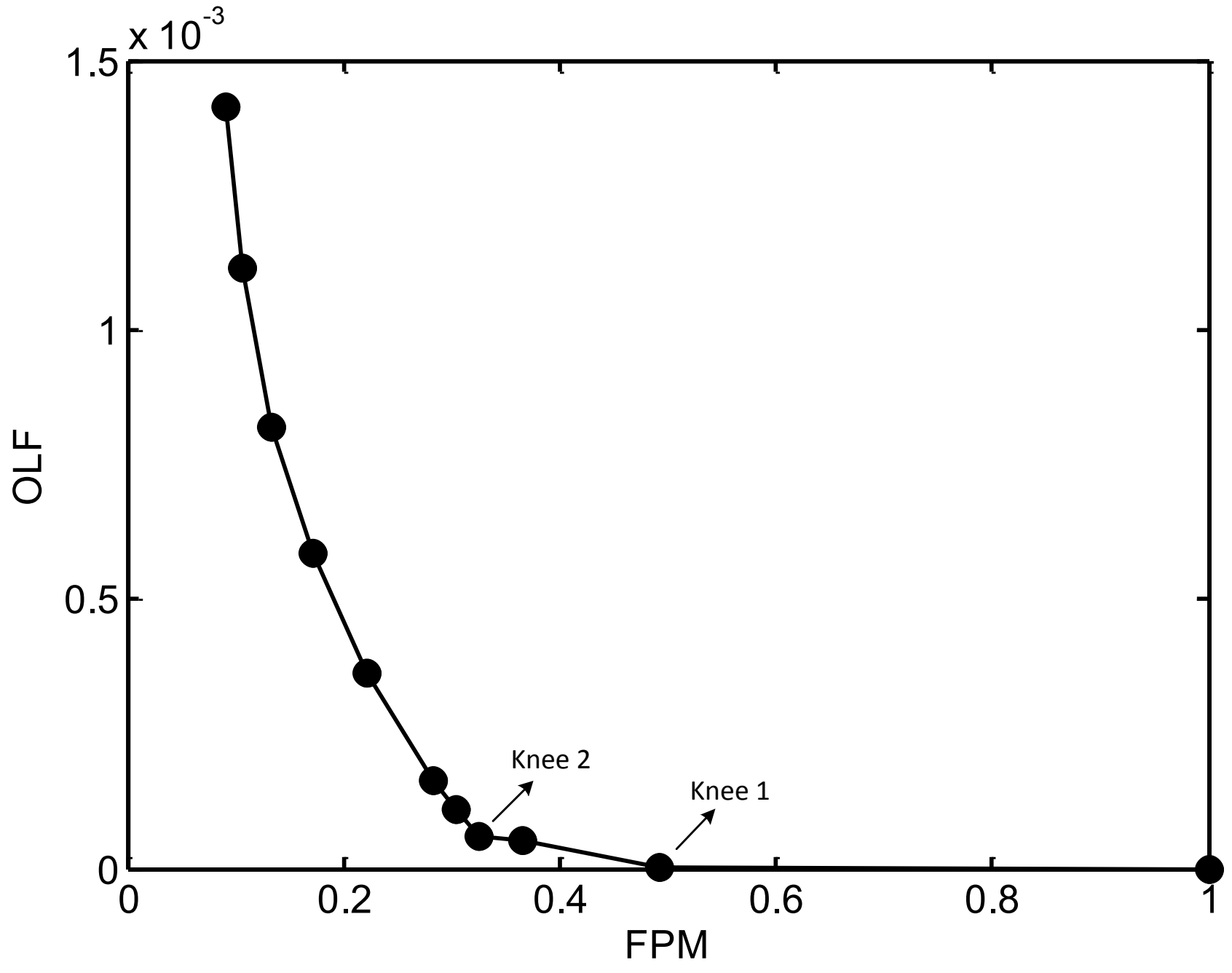

**Figure 4.20 Optimality robustness trade-off curve for 11-zone building**

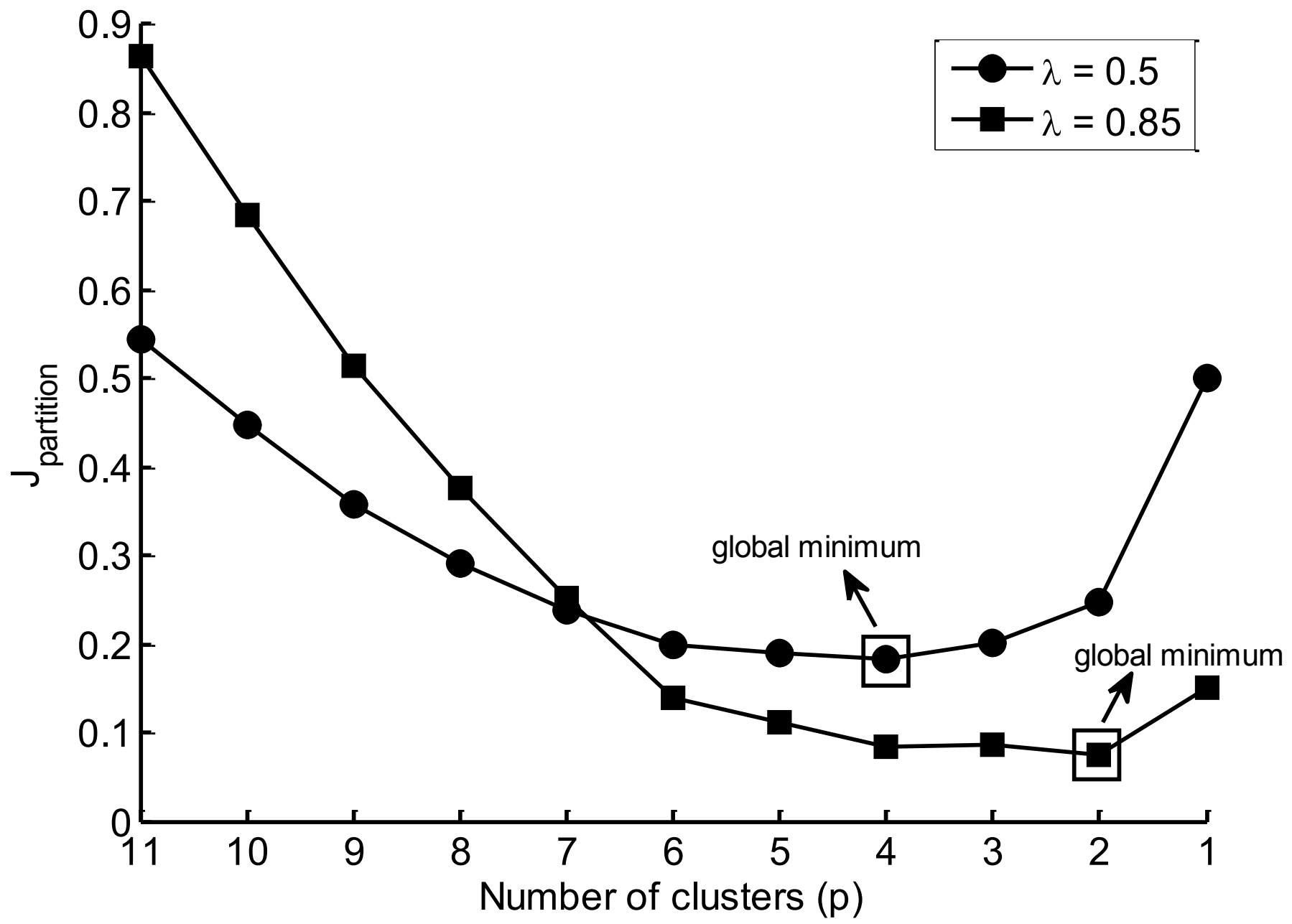

**Figure 4.21 Plot of $J_{\text{partition}}$ for partitions from agglomeration for 11-zone building**

A validation of the above findings is provided by observations from open loop simulations of the building thermal model. The building was partitioned into clusters corresponding to knee 2 in Figure 4.20. Cluster-level models, as discussed in Section 3.4.3, were obtained from the centralized model by decoupling the system along the boundaries (physical walls) of each cluster. An estimate of $20^o$ C was used to represent the temperature of the zones outside any cluster, i.e each element of the vector $\mathbf{T_{zj}}$ in (3.24) is 20. Simulation results obtained for a period of 10 days are shown in Figure 4.22, where $T_{c,mean}$ is the capacity-weighted mean temperature of the zones, defined as

$$T_{c,mean}(k) = \frac{1}{C_z} \sum_{i=1}^{N_z} C_{z,i} T_{z,i}(k). \qquad (4.89)$$

The relevant MATLAB codes are provided in Appendix D and the SIMULINK models are included in the media accompanying this thesis. In defining $T_{c,mean}$, the contribution of each zone's temperature is weighted by its thermal capacity. Since the thermal capacity of a zone is closely related to its volume, $T_{c,mean}$ indicates an effective temperature for the building as a whole. The response corresponding to the fully decentralized partition, i.e. $\{\{1\},\{2\},\{3\},\{4\},\{5\},\{6\},\{7\},\{8\},\{9\},\{10\},\{11\}\}$ is also plotted. In these simulations, the ambient temperature $T_a$ and the unmodeled thermal disturbances $d_{z,i}$ (for all $i \in \{1,2,\dots,11\}$) are assumed to be sinusoids with a 24 hour time-period as shown in Figure 4.23. The thermal disturbance $d_{w,i}$ is assumed to be 0 for all $i \in \{1,2,\dots,11\}$ for simplicity. From Figure 4.22 it is observed that the error in predicting $T_{c,mean}$ is less than 5% for the knee 2 partition, whereas, the fully decentralized partition results in a maximum error of about 20% over the simulation time window. Here, errors are evaluated with respect to $T_{c,mean}$ predicted by the centralized case. This observation verifies that by partitioning the system using knee 2, the corresponding cluster-level decentralized models do not result in significant loss in inter-cluster thermal coupling, when compared to the centralized system model.

The 11-zone building example considered above demonstrates the benefit of using OLF-FPM partitioning approach, as opposed to physical intuition, which may be absent or can only provide limited insight. For example, the partition $\{\{1,2,3,4,5,6\},\{7,8,9,10,11\}\}$ arrived at using the intuition of separation along the thermally insulating hallways generates a knee in Figure 4.20. However, as the analysis presented above clearly demonstrates, it is not the most appropriate choice.

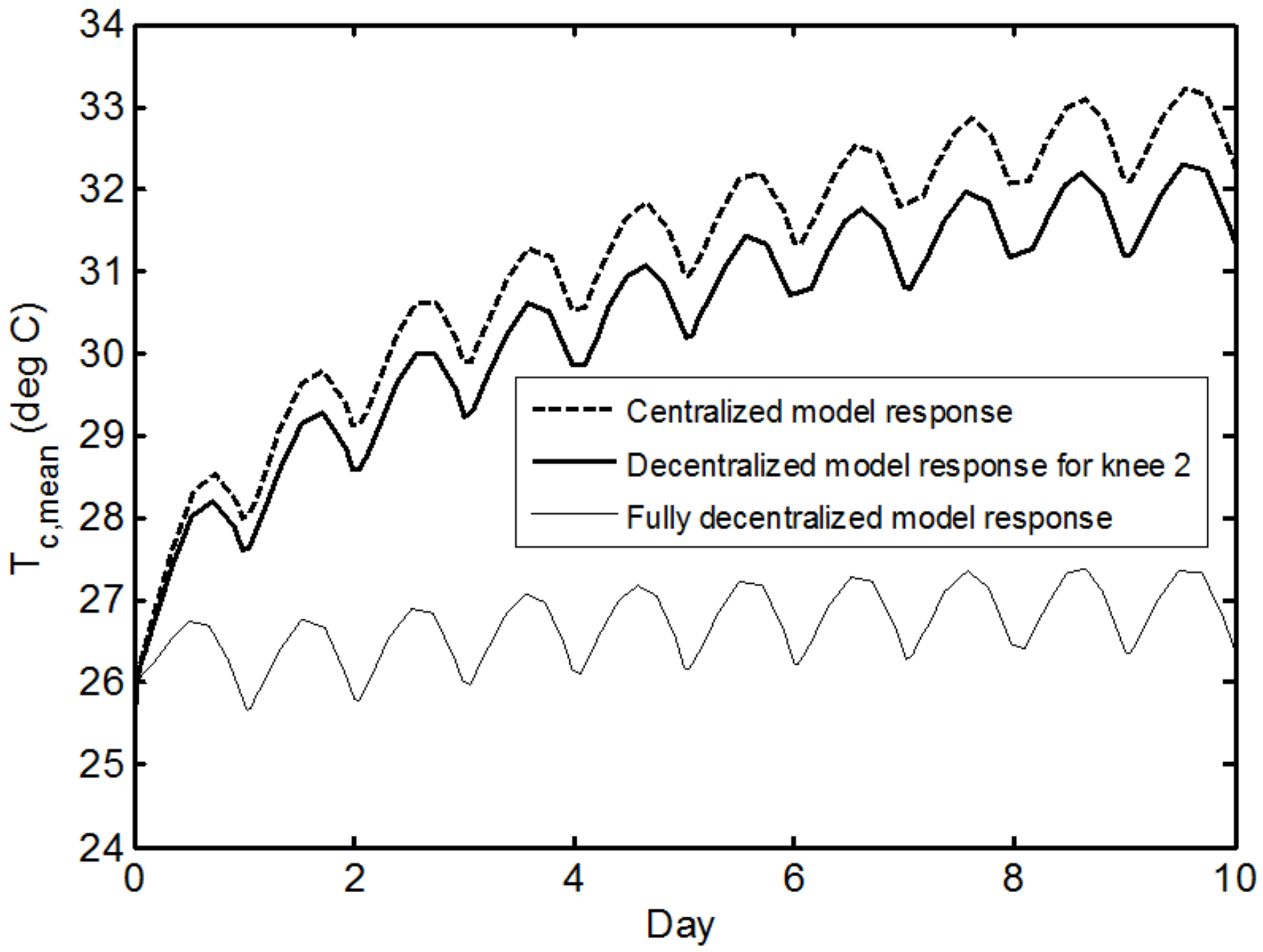

**Figure 4.22 Open loop response analysis for partitions of 11-zone building**

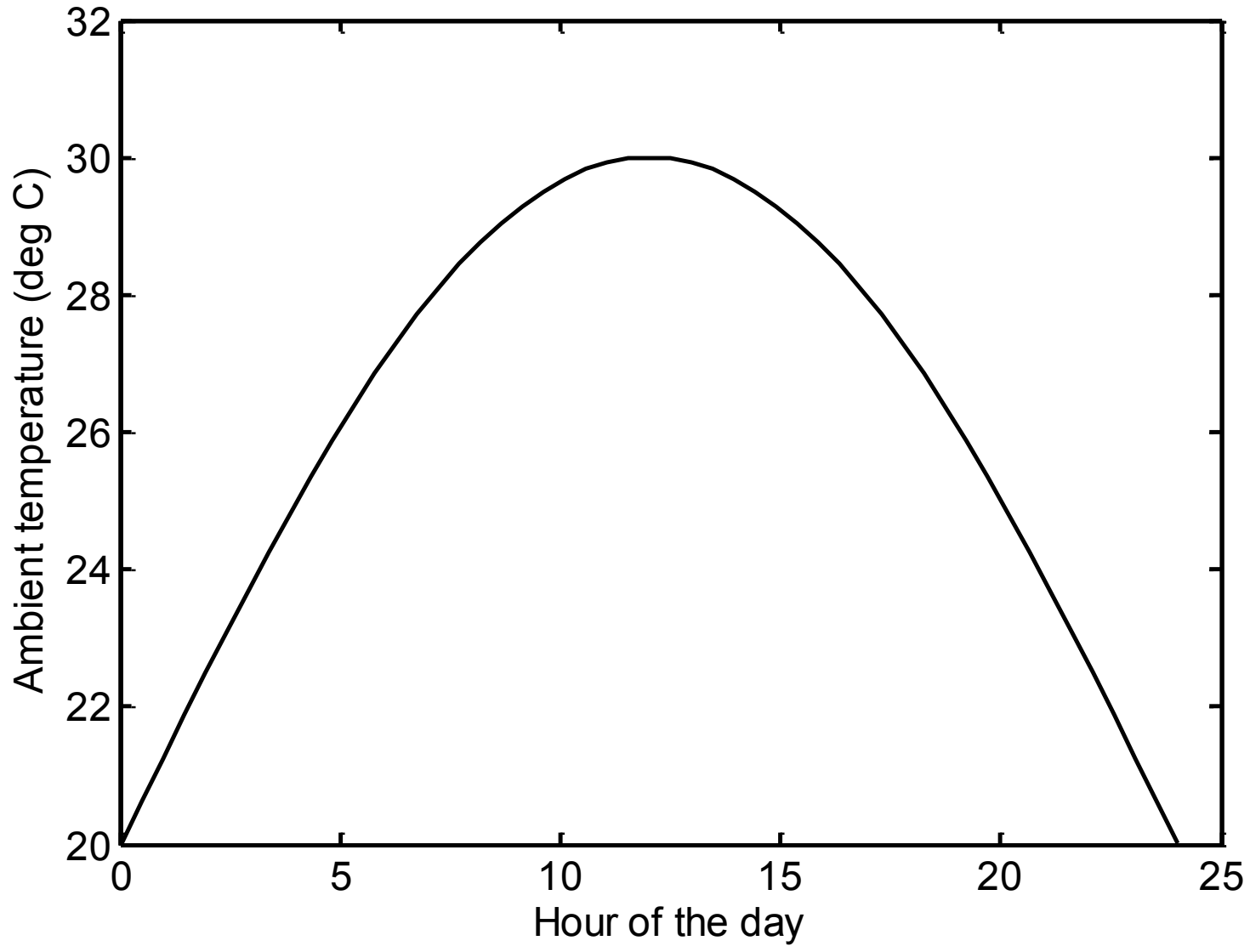

(a) Ambient temperature profile

**Figure 4.23 (cont. on next page)**

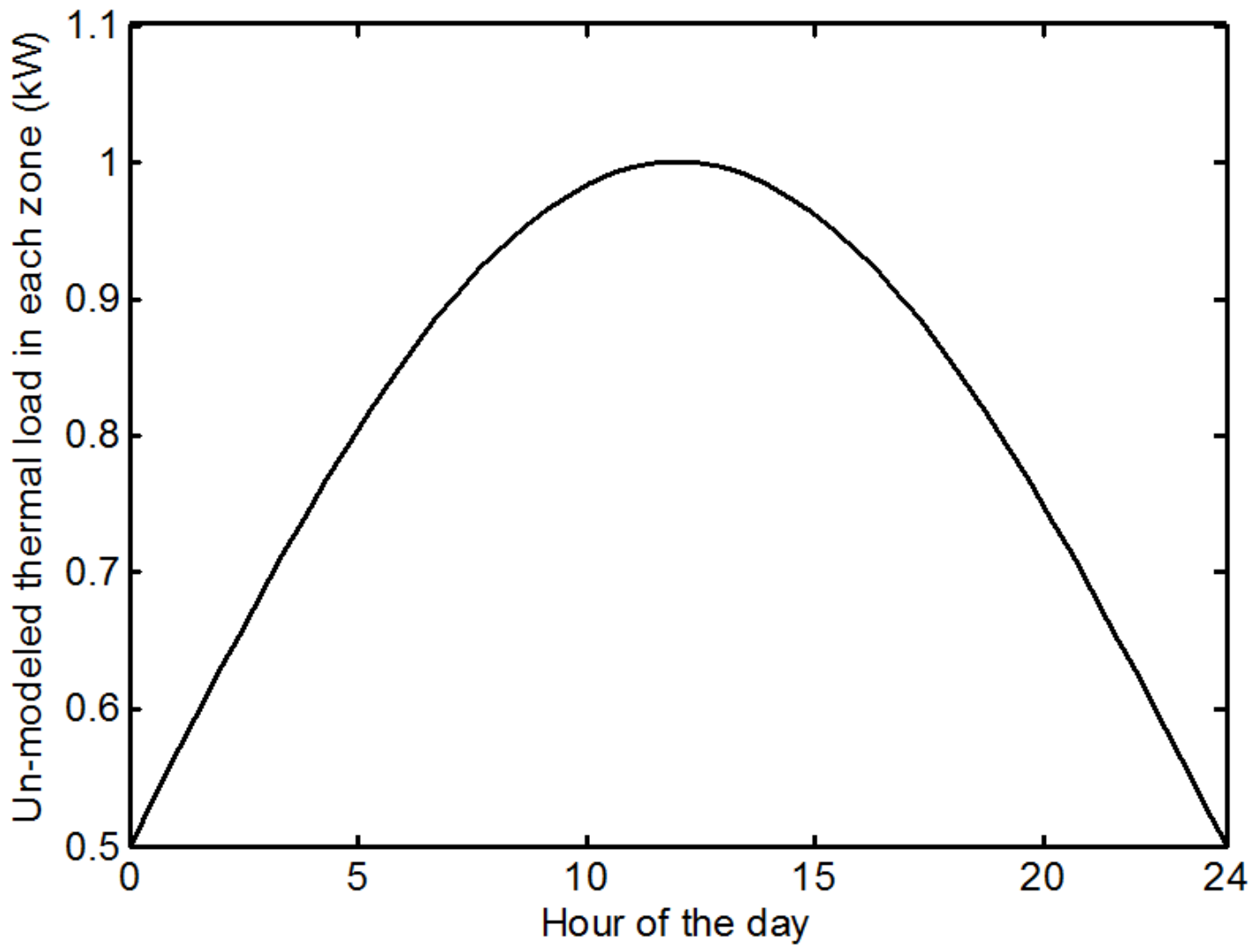

(b) Unmodeled thermal load profile for each zone

**Figure 4.23 Sinusoidal disturbance profiles used in open loop analysis of 11-zone building**

### 4.3.12 Remarks

The main advantage of the OLF-FPM approach over the CLF-MCS approach is that the optimality and robustness metrics are analytically derived and directly correlated with the notions of optimality and robustness. In particular, OLF is related to an upper bound on the deviation in performance between centralized control and decentralized control. Similarly, FPM by definition represents the expected value of the % volume of building affected in the event of failures. On the other hand, a quantifiable relationship between CLF and optimality or MCS and robustness is not available. However the CLF-MCS approach is useful because it provides an initial framework for partitioning which is improved upon by the OLF-FPM method. In the remainder of this work, only the OLF-FPM method is considered for making control architecture decisions because of its above stated benefits over the CLF-MCS method.

# Chapter 5
# Control Design with Optimal Architecture

## 5.1 Introduction

Chapter 4 presented two different approaches for the determination of decentralized control architectures which balance the underlying optimality and robustness requirements in the thermal control of buildings. However, the design of decentralized controllers based on the control architectures determined using these methodologies was not considered. This chapter seeks to address this requirement by focusing on the design aspects of decentralized control for a given architecture. Two key challenges are envisioned in this regard: all system states are not measurable and there are several unmodeled disturbances. These practical concerns are addressed in this chapter by first developing reduced order model representations for the thermal dynamics then using decentralized extended state observers to simultaneously estimate the unknown states and disturbances. This is followed by control design based on the Model Predictive Control (MPC) framework, since MPC has been extensively applied in the building systems control literature because of its proven effectiveness in handling large-scale constrained optimal control problems (see discussion in section 3.1.2.2).

The decentralized control design methodology presented in this chapter differs from Chapter 3 in three aspects. Firstly, the model used for control design is based on a three resistor – two capacitor (3R2C) framework to represent the thermal dynamics of walls, which is a more accurate representation than the two resistor – one capacitor (2R1C) framework that was used in Chapter 3. Secondly, the controllers are output-feedback as opposed to the state-feedback framework presented in chapter 3, considering the fact that typically only zone temperature measurements are available in a building via the thermostats. Lastly, the control design considers certain practical constraints which were ignored in Chapter 3 for simplicity. The nomenclature used in this chapter is shown in Table 5.1.

## 5.2 Cluster level modeling

In this section, we consider a p-partition of the building (definition 3.2) and develop models to describe the thermal dynamics associated with the clusters which constitute the partition. These models are used for the design of decentralized observers and controllers as described in subsequent sections.

**Table 5.1 Nomenclature of common symbols in Chapter 5**

| Symbol | Description |
|---|---|
| $N_w$ | Number of walls surfaces in the building |
| $N_z$ | Number of zones in the building |
| $N_{zi}$ | Number of zones in $i^{th}$ cluster |
| $\mathbf{T_w}$ | Vector of wall surface temperatures |
| $\mathbf{T_w^i}$ | Vector of wall surface temperatures in $i^{th}$ cluster |
| $T_{w,k}^i$ | Temperature of $k^{th}$ wall surface in $i^{th}$ cluster |
| $\mathbf{T_z}$ | Vector of zone temperatures |
| $\mathbf{T_z^i}$ | Vector of zone temperatures in $i^{th}$ cluster |
| $\mathbf{T_{z,ref}^i}$ | Vector of zone temperature set-points in $i^{th}$ cluster |
| $\mathbf{T_{z,meas}^i}$ | Vector of zone temperature measurements in $i^{th}$ cluster |
| $T_a$ | Ambient temperature |
| $T_g$ | Temperature of ground below building |
| $T_{z,n}^i$ | Temperature of $n^{th}$ zone in $i^{th}$ cluster |
| $T_{supp}$ | Temperature of cold air supplied to each zone |
| $\mathbf{d_w}$ | Vector of unknown thermal loads acting on wall surfaces |
| $\mathbf{d_z}$ | Vector of unknown thermal loads acting on zones |
| $\mathbf{d_w^i}$ | Vector of unknown thermal loads acting on wall surfaces in $i^{th}$ cluster |
| $\mathbf{d_z^i}$ | Vector of unknown thermal loads acting on zones in $i^{th}$ cluster |
| $\mathbf{\bar{d}_w^i}$ | Vector of aggregated thermal loads acting on wall surfaces in reduced order model for $i^{th}$ cluster |
| $\mathbf{\bar{d}_w^i}$ | Vector of aggregated thermal loads acting on zones in reduced order model for $i^{th}$ cluster |
| $C_{w,k}^i$ | Thermal capacitance associated with $k^{th}$ wall surface in $i^{th}$ cluster |
| $C_{z,n}^i$ | Thermal capacitance associated with $n^{th}$ zone in $i^{th}$ cluster |
| $c_{pa}$ | Specific heat capacity of air |
| $x(k+l\|k)$ | Projected value of quantity $x$ after $l$ time steps in future, given $x(k)$. Note $x(k\|k) = x(k)$. |
| $\mathbf{u}$ | Vector of control inputs |
| $\mathbf{u^i}$ | Vector of control inputs in $i^{th}$ cluster |
| $u_n^i$ | Control input for $n^{th}$ zone in $i^{th}$ cluster |
| $\dot{m}_{max,n}^i$ | Maximum air mass flow rate for $n^{th}$ zone in $i^{th}$ cluster |
| $\dot{Q}_{RH-max,n}^i$ | Maximum reheating available for $n^{th}$ zone in $i^{th}$ cluster |
| $\mathbf{0}_{m\times n}$ | Zero matrix of dimension $m \times n$ |
| $\mathbf{I}_n$ | Identity matrix of dimension $m \times n$ |

### 5.2.1 3R2C modeling framework

We consider a 3R2C modeling paradigm [40], where each wall is represented by a set of 3 resistances and 2 capacitances (Figure 5.1), whereas, each room is treated as a single capacitance. This is an improvement over the 2R1C paradigm considered in chapter 3 for control architecture selection. In Figure 5.1, $C_{w-in,i}$ and $C_{w-out,i}$ are capacitances for the $i^{th}$wall, the states associated with them being the temperatures of its two surfaces. For internal walls (flanked by zones on both sides), the designation of "in" and "out" for the wall surfaces is arbitrary. However, for external walls (facing the ambient or ground on one side and a zone on the other), by convention "out" refers to the surface which faces the ambient/ground and "in" refers to the surface facing the zone. $R_{w-in,i}$ and $R_{w-out,i}$ in Figure 5.1 are thermal resistances between the surfaces of wall $i$ and the appropriate elements (zones, ambient or ground) which they thermally interact with. $R_{ww,i}$ represents the resistance between inner and outer wall surfaces. Similar to the 2R1C framework, each zone in the building is represented by a lumped capacitance.

The states of the system are the wall surface temperatures and zone temperatures. The control inputs represent the rates of energy transfer – positive for heating, and negative for cooling — provided to the zones by the HVAC system. The ambient and ground temperatures are considered as measured disturbances. Various other factors also affect the thermal dynamics which are treated as unknown disturbances. These include long-wave and short-wave radiation heat transfers affecting the walls, and thermal loads from occupants, appliances and lighting which affect the zones. In this work, these factors are not modeled separately and only their lumped thermal contribution to each zone and wall is represented using disturbance vectors $\mathbf{d_z}$ and $\mathbf{d_w}$.

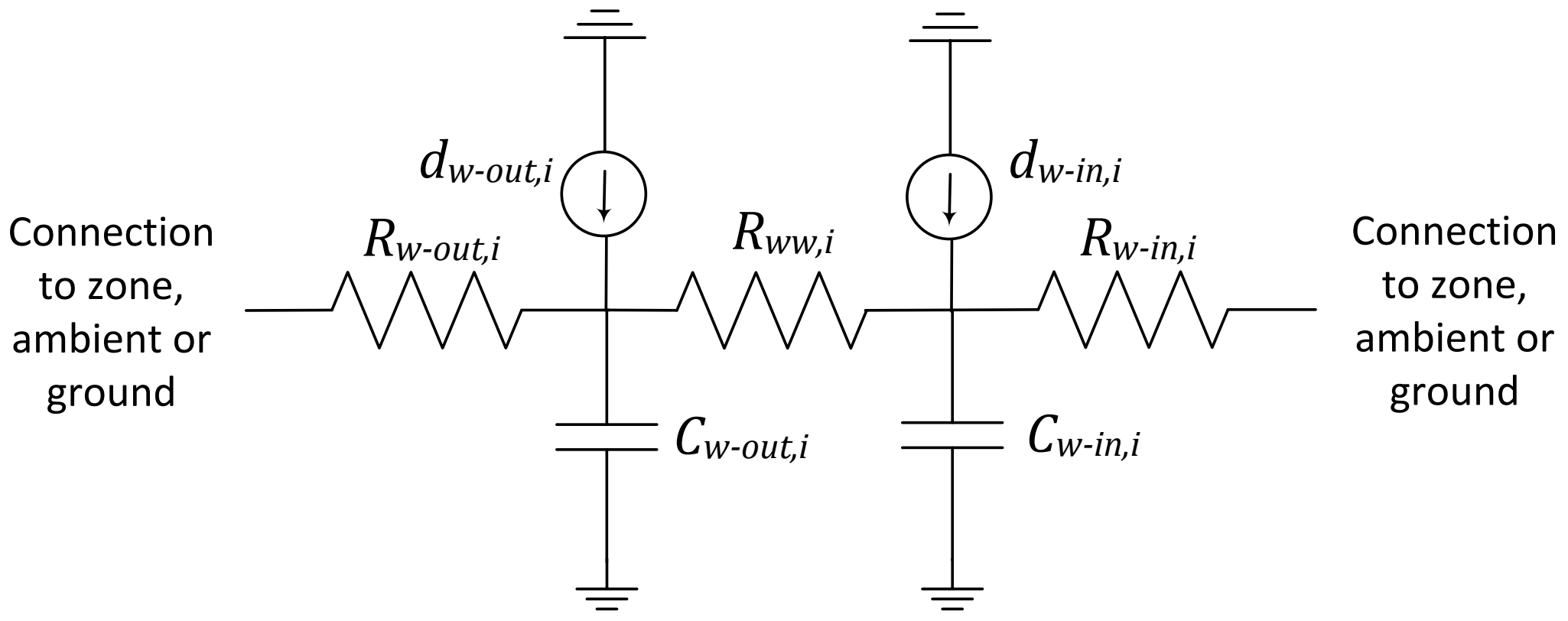

**Figure 5.1 Schematic of 3R2C modeling paradigm for wall *i*. Un-modeled thermal loads acting on the wall's surfaces are also shown.**

The ensuing linear state space model is of the form (5.1), with reference to the notation described in the nomenclature (Table 5.1). It is obtained by applying Algorithm 5.1 presented below on a RC network for the building constructed using the framework presented above. Note that Algorithm 5.1 is an extension of Algorithm 3.1 to the 3R2C case.

$$\frac{d}{dt}\begin{bmatrix}\mathbf{T_w}\\ \mathbf{T_z}\end{bmatrix} = \begin{bmatrix}\mathbf{A_{ww}} & \mathbf{A_{wz}}\\ \mathbf{A_{zw}} & \mathbf{A_{zz}}\end{bmatrix}\begin{bmatrix}\mathbf{T_w}\\ \mathbf{T_z}\end{bmatrix} + \begin{bmatrix}\mathbf{0}\\ \mathbf{B_z}\end{bmatrix}\mathbf{u} + \begin{bmatrix}\mathbf{B_a} & \mathbf{B_g} & \mathbf{B_{dw}} & \mathbf{0}\\ \mathbf{0} & \mathbf{0} & \mathbf{0} & \mathbf{B_{dz}}\end{bmatrix}\begin{bmatrix}T_a\\ T_g\\ \mathbf{d_w}\\ \mathbf{d_z}\end{bmatrix}. \tag{5.1}$$

**Algorithm 5.1. Generation of state space model for building thermal dynamics using 3R2C framework**

*STEP 1:* A weighted graph is created with nodes for each of the wall surfaces, the zones, the ambient and ground. While numbering the nodes, those representing wall surfaces are numbered first, followed by the zones, the ambient and lastly the ground. Each wall surface node is connected by undirected edges to the two nodes (zones, ambient or ground) to which it is connected via resistances in the RC network. Additionally, the inside and outside surfaces of each wall are also connetcted by an edge. The weight of each edge is set to be the inverse of the corresponding thermal resistance between the two nodes it connects. The resulting weighted graph is denoted by $G = (V, E)$ along with the weight function, $w: E \mapsto \mathbb{R}^+$. We also define capacitance matrices, $\mathbf{C_w}$ and $\mathbf{C_z}$ which are diagonal matrices of the thermal capacitances associated with the wall surfaces and the zones respectively. The diagonal entries in these matrices are entered in the order of the corresponding node numbers in $G$.

*STEP 2:* The Laplacian matrix of $G$, denoted by $\mathbf{L_G}$ is then obtained as:

$$\mathbf{L_G} = \mathbf{D_G} - \mathbf{A_G}. \tag{5.2}$$

where,

$$\mathbf{A_G}(i,j) = \begin{cases} w(i,j) & \text{if}(i,j) \in E \\ 0 & \text{otherwise} \end{cases}$$

$$\mathbf{D_G}(i,i) = \sum_j A_G(i,j)\,.$$

Note that $\mathbf{D_G}$ is a diagonal matrix. We extract the following sub-matrices from $\mathbf{L_G}$:

1. $\mathbf{L_{ww}}$ is the square sub-matrix of $\mathbf{L_G}$ which corresponds to its first $N_w$ rows and first $N_w$ columns.
2. $\mathbf{L_{wz}}$ is the sub-matrix of $\mathbf{L_G}$ which corresponds to its first $N_w$ rows and the columns $N_w + 1,\ N_w + 2, \dots, N_w + N_z$.
3. $\mathbf{L_{zw}}$ is the sub-matrix of $\mathbf{L_G}$ which corresponds to the rows $N_w + 1,\ N_w + 2, \dots, N_w + N_z$ and the first $N_w$ columns.
4. $\mathbf{L_{zz}}$ is the sub-matrix of $\mathbf{L_G}$ which corresponds to the rows $N_w + 1,\ N_w + 2, \dots, N_w + N_z$ and the columns $N_w + 1,\ N_w + 2, \dots, N_w + N_z$.
5. $\mathbf{L_a}$ is a column vector which corresponds to the first $N_w$ rows and the $(N_w + N_z + 1)^{th}$ column of $\mathbf{L_G}$
6. $\mathbf{L_g}$ is a column vector which corresponds to the first $N_w$ rows and the $(N_w + N_z + 2)^{th}$ column of $\mathbf{L_G}$

*STEP 3:* The matrices appearing in (5.1) are then obtained as follows:

$$\mathbf{A_{ww}} = -\mathbf{C_w}^{-1}\mathbf{L_{ww}}, \quad (5.3)$$

$$\mathbf{A_{wz}} = -\mathbf{C_w}^{-1}\mathbf{L_{wz}}, \quad (5.4)$$

$$\mathbf{A_{zw}} = -\mathbf{C_z}^{-1}\mathbf{L_{zw}}, \quad (5.5)$$

$$\mathbf{A_{zz}} = -\mathbf{C_z}^{-1}\mathbf{L_{zz}}, \quad (5.6)$$

$$\mathbf{B_a} = -\mathbf{C_w}^{-1}\mathbf{L_a}, \quad (5.7)$$

$$\mathbf{B_g} = -\mathbf{C_w}^{-1}\mathbf{L_g}, \quad (5.8)$$

$$\mathbf{B_{dw}} = \mathbf{C_w}^{-1}, \quad (5.9)$$

$$\mathbf{B_{dz}} = \mathbf{C_z}^{-1}. \quad (5.10)$$

### 5.2.2 Full order cluster level model

We consider a p-partition of the building. The model for the $i^{th}$ ($i \in \{1,2,\dots,p\}$) cluster is obtained from (5.1) by extracting the dynamics of the walls and the zones constituting the cluster as shown below (see nomenclature)

$$\frac{d}{dt}\begin{bmatrix}\mathbf{T_w^i}\\ \mathbf{T_z^i}\end{bmatrix} = \begin{bmatrix}\mathbf{A_{ww}^i} & \mathbf{A_{wz}^i}\\ \mathbf{A_{zw}^i} & \mathbf{A_{zz}^i}\end{bmatrix}\begin{bmatrix}\mathbf{T_w^i}\\ \mathbf{T_z^i}\end{bmatrix} + \begin{bmatrix}\mathbf{0}\\ \mathbf{B_z^i}\end{bmatrix}\mathbf{u^i} + \begin{bmatrix}\mathbf{B_a^i} & \mathbf{B_g^i} & \mathbf{B_{dw}^i} & \mathbf{0}\\ \mathbf{0} & \mathbf{0} & \mathbf{0} & \mathbf{B_{dz}^i}\end{bmatrix}\begin{bmatrix}T_a\\ T_g\\ \mathbf{d_w^i}\\ \mathbf{d_z^i}\end{bmatrix} + \sum_{j\neq i}\begin{bmatrix}\mathbf{B_{w,z}^{ij}}\\ \mathbf{0}\end{bmatrix}\mathbf{\hat{T}_z^j}\,. \quad (5.11)$$

The dynamics of the $i^{th}$ cluster depends on the zone temperatures in the other clusters $\mathbf{T_z^j}$, where $j \neq i$. However, imposition of a control architecture that is decentralized with respect to the clusters implies that the $i^{th}$ control agent does not have access to the sensory data from the other $p-1$ clusters. Therefore, as mentioned in chapter 3, $\mathbf{T_z^j}$ for each $j \neq i$ must be replaced with an appropriate guess or estimate $\mathbf{\hat{T}_z^j}$ as shown in ( 5.11).

The states $\mathbf{T_w^i}$, and the disturbances $\mathbf{d_w^i}$ and $\mathbf{d_z^i}$ are unknown and therefore ( 5.11) cannot be directly used for control design. This concern is addressed in the remainder of this section.

### 5.2.3 Model order reduction

For a building with $N_z$ zones, the number of states in the model (5.1) is of the order of $7N_z$ [29]. This motivates the development of reduced order models, which can be used for control design at the overall building level (centralized architecture) or the cluster level (decentralized architecture). An aggregation based methodology was proposed in [29] for the development of reduced order models of building thermal dynamics. A particular advantage is that the reduced order models can also be represented by RC networks. In this paper, we adopt the methodology in [29] to seek an observable reduced order model representation which allows extended state observers to estimate the unmeasured states and disturbances.

Algorithm 5.2, presented below, outlines steps for model reduction via aggregation of states, which eventually lead to an observable representation in Section 5.2.4. The resulting state space model for the $i^{th}$ cluster obtained from the algorithm consists of the states $\bar{T}^i_{w-in,m}$, $\bar{T}^i_{w-out,m}$ and $T^i_{z,m}$, where $m \in \{1,2,\dots,N_{zi}\}$ and is given by

$$\frac{d}{dt}\underbrace{\begin{bmatrix}\mathbf{\bar{T}^i_{w-in}}\\ \mathbf{\bar{T}^i_{w-out}}\\ \mathbf{T^i_z}\end{bmatrix}}_{\mathbf{x^i}} = \underbrace{\begin{bmatrix}\mathbf{\bar{A}^i_{w-in,w-in}} & \mathbf{\bar{A}^i_{w-in,w-out}} & \mathbf{\bar{A}^i_{w-in,z}}\\ \mathbf{\bar{A}^i_{w-out,w-in}} & \mathbf{\bar{A}^i_{w-out,w-out}} & \mathbf{0}\\ \mathbf{\bar{A}^i_{z,w-in}} & \mathbf{0} & \mathbf{\bar{A}^i_{z,z}}\end{bmatrix}}_{\mathbf{\bar{A}^i}}\begin{bmatrix}\mathbf{\bar{T}^i_{w-in}}\\ \mathbf{\bar{T}^i_{w-out}}\\ \mathbf{T^i_z}\end{bmatrix} + \underbrace{\begin{bmatrix}\mathbf{0}\\ \mathbf{0}\\ \mathbf{\bar{B}^i_{z,u}}\end{bmatrix}}_{\mathbf{\bar{B}^i_u}}\mathbf{u^i}$$

$$+\begin{bmatrix} \underbrace{\begin{matrix}\mathbf{0}\\ \overline{\mathbf{B}}^{\mathbf{i}}_{\mathbf{w-out,a}}\\ \mathbf{0}\end{matrix}}_{\overline{\mathbf{B}}^{\mathbf{i}}_{\mathbf{a}}} & \underbrace{\begin{matrix}\mathbf{0}\\ \overline{\mathbf{B}}^{\mathbf{i}}_{\mathbf{w-out,g}}\\ \mathbf{0}\end{matrix}}_{\overline{\mathbf{B}}^{\mathbf{i}}_{\mathbf{g}}} & \underbrace{\begin{matrix}\overline{\mathbf{B}}^{\mathbf{i}}_{\mathbf{w-in,dw}}\\ \overline{\mathbf{B}}^{\mathbf{i}}_{\mathbf{w-out,dw}}\\ \mathbf{0}\end{matrix}}_{\overline{\mathbf{B}}^{\mathbf{i}}_{\mathbf{dw}}} & \underbrace{\begin{matrix}\mathbf{0}\\ \mathbf{0}\\ \overline{\mathbf{B}}^{\mathbf{i}}_{\mathbf{z,dz}}\end{matrix}}_{\overline{\mathbf{B}}^{\mathbf{i}}_{\mathbf{dz}}} \end{bmatrix} \begin{bmatrix} T_a \\ T_g \\ \overline{\mathbf{d}}^{\mathbf{i}}_{\mathbf{w}} \\ \overline{\mathbf{d}}^{\mathbf{i}}_{\mathbf{z}} \end{bmatrix} + \sum_{j\neq i} \underbrace{\begin{bmatrix} \mathbf{0} \\ \overline{\mathbf{B}}^{\mathbf{ij}}_{\mathbf{w-out,z}} \\ \mathbf{0} \end{bmatrix}}_{\overline{\mathbf{B}}^{\mathbf{ij}}_{\mathbf{z}}} \widehat{\mathbf{T}}^{\mathbf{j}}_{\mathbf{z}}. \quad (5.12)$$

Here,

$$\overline{\mathbf{T}}^{\mathbf{i}}_{\mathbf{w-in}} = \begin{bmatrix} \overline{T}^{i}_{w-in,1} & \overline{T}^{i}_{w-in,2} & \cdots & \overline{T}^{i}_{w-in,n^{i}_{w-in}} \end{bmatrix}^{T}, \quad (5.13)$$

$$\overline{\mathbf{T}}^{\mathbf{i}}_{\mathbf{w-out}} = \begin{bmatrix} \overline{T}^{i}_{w-out,1} & \overline{T}^{i}_{w-out,2} & \cdots & \overline{T}^{i}_{w-out,n^{i}_{w-out}} \end{bmatrix}^{T}. \quad (5.14)$$

In the above equations, $n^{i}_{w-in}$ and $n^{i}_{w-out}$ are defined in step 1of Algorithm 5.2.

**Algorithm 5.2. Model reduction for the $i^{th}$ cluster via aggregation of states**

*STEP 1:* For each zone $m \in \{1,2,\dots,N_{zi}\}$ in the cluster, all wall surfaces belonging to the walls that enclose it are identified (note that a wall has two surfaces, each represented by a capacitance in the 3R2C framework). Of these, we denote those surfaces that directly face the zone by the set $\mathcal{N}^{i}_{m}$. Among the remaining surfaces which do not face the zone $m$ directly, those which also do not face another zone in the cluster directly are identified. The set of these surfaces is denoted by $\overline{\mathcal{N}}^{i}_{m}$. Note that $\overline{\mathcal{N}}^{i}_{m}$ can be an empty set. We define $n^{i}_{w-in}$ as the number of non-empty sets $\mathcal{N}^{i}_{m}$, where $m \in \{1,2,\dots,N_{zi}\}$. Similarly, $n^{i}_{w-in}$ is defined as the number of non-empty sets $\overline{\mathcal{N}}^{i}_{m}$, where $m \in \{1,2,\dots,N_{zi}\}$. Note that $n^{i}_{w-in} = N_{zi}$ because each zone is surrounded by at least one wall surface.

*STEP 2:* For each zone $m \in \{1,2,\dots,N_{zi}\}$, the temperatures of all wall surfaces in the set $\mathcal{N}^{i}_{m}$ and $\overline{\mathcal{N}}^{i}_{m}$ are aggregated into single states $\overline{T}^{i}_{w-in,m}$ and $\overline{T}^{i}_{w-out,n}$ respectively as shown (refer to nomenclature)

$$\overline{T}^{i}_{w-in,m} = \frac{\sum_{k\in\mathcal{N}^{i}_{m}} C^{i}_{w,k} T^{i}_{w,k}}{\sum_{k\in\mathcal{N}^{i}_{m}} C^{i}_{w,k}}, \quad (5.15)$$

$$\bar{T}^i_{w-out,m} = \frac{\sum_{k\in\bar{\mathcal{N}}^i_m} C^i_{w,k} T^i_{w,k}}{\sum_{k\in\bar{\mathcal{N}}^i_m} C^i_{w,k}} . \qquad (5.16)$$

*STEP 3:* Equivalent capacitances $\bar{C}^i_{w-in,m}$ and $\bar{C}^i_{w-out,m}$ associated with the states $\bar{T}^i_{w-in,m}$ and $\bar{T}^i_{w-out,m}$ respectively are computed as shown

$$\bar{C}^i_{w-in,m} = \sum_{k\in\mathcal{N}^i_m} C^i_{w,k} , \qquad (5.17)$$

$$\bar{C}^i_{w-out,m} = \sum_{k\in \bar{\mathcal{N}}^i_m} C^i_{w,k} . \qquad (5.18)$$

*STEP 4:* For each $m \in \{1,2,\dots,N_{zi}\}$, $\bar{R}^i_{w-in,w-out,m}$ is defined as the parallel equivalent of all resistances which connect an element in $\mathcal{N}^i_m$ with an element in $\bar{\mathcal{N}}^i_m$. The capacitors $\bar{C}^i_{w-in,m}$ and $\bar{C}^i_{w-out,m}$ are then connected using $\bar{R}^i_{w-in,w-out,m}$.

*STEP 5:* For each $m,n \in \{1,2,\dots,N_{zi}\}$, $\bar{R}^i_{w-in,z,m,n}$ is defined as the parallel equivalent of all resistances which connect an element in $\mathcal{N}^i_m$ with zone $n$. The capacitors $\bar{C}^i_{w-in,m}$ and $C^i_{z,n}$ (see nomenclature) are connected using $\bar{R}^i_{w-in,z,m,n}$.

*STEP 6:* For each $m,n \in \{1,2,\dots,N_{zi}\}$, $\bar{R}^i_{w-in,w-in,m,n}$ is defined as the parallel equivalent of all resistances which connect an element in $\mathcal{N}^i_m$ with an element in $\mathcal{N}^i_n$. The capacitors $\bar{C}^i_{w-in,m}$ and $\bar{C}^i_{w-in,n}$ are connected using $\bar{R}^i_{w-in,w-in,m,n}$.

*STEP 7:* For each $m \in \{1,2,\dots,N_{zi}\}$, $\bar{R}^i_{w-out,a,m}$ is defined as the parallel equivalent of all resistances which connect an element in $\bar{\mathcal{N}}^i_m$ with the ambient. The capacitor $\bar{C}^i_{w-out,m}$ and ambient are connected with $\bar{R}^i_{w-out,a,m}$. Similarly, $\bar{R}^i_{w-out,g,m}$ is defined for the ground instead of ambient.

*STEP 8:* For each cluster $j \in \{1,2,\dots p\}$, $j \neq i$ and for each zone $m \in \{1,2,\dots,N_{zi}\}$ in cluster $i$ and zone $n \in \{1,2,\dots,N_{zj}\}$ in cluster $j$, we define the resistance $\bar{R}^{i,j}_{w-out,z,m,n}$ as the parallel equivalent of all resistances which connect an element in $\bar{\mathcal{N}}_m^i$ with the zone $n$. The capacitor $\bar{C}^i_{w-out,m}$ is connected to the external zone $n$ (lying outside cluster $i$) using $\bar{R}^{i,j}_{w-out,z,m,n}$.

*STEP 9:* The reduced order RC network is constituted by the capacitances and resistances created in steps 3 – 8.

As an example, consider the 3-zone building shown in Figure 5.2 which is similar to the example in Figure 3.3, except that it is modeled using the 3R2C framework. The zones are marked as $z1 - z3$. The building has 20 wall surfaces which are marked as $w1 - w20$. The resistances representing heat transfer paths between the wall surfaces, zones, ambient and ground are also shown in the figure. The capacitances (not shown in the figure) corresponding to the wall surfaces and the zones are given by $C_{w1} - C_{w20}$ and $C_{z1} - C_{z3}$ respectively. Let the temperatures associated with these capacitances be represented by $T_{w1} - T_{w20}$ and $T_{z1} - T_{z3}$.

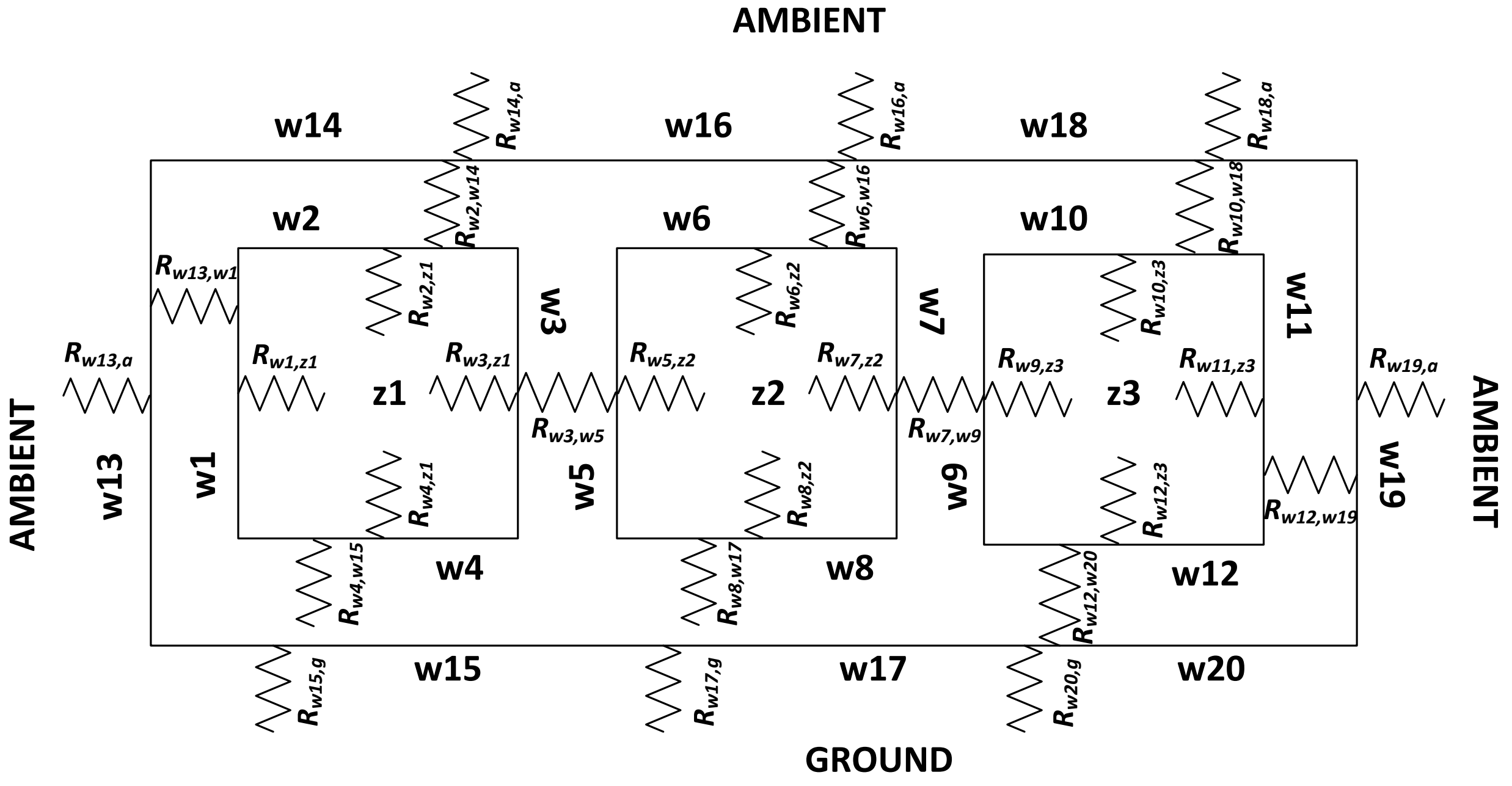

**Figure 5.2 Schematic of 3-zone building used for illustration of Algorithm 5.2**

Consider a 2-partition {{1,2},{3}} of the building. Application of each step of Algorithm 5.2 to the first cluster {1,2} is shown below.

STEP 1: $\mathcal{N}_1^1 = \{w1, w2, w3, w4\}, \bar{\mathcal{N}}_1^1 = \{w13, w14, w15\}, \mathcal{N}_2^1 = \{w5, w6, w7, w8\}$ and $\bar{\mathcal{N}}_2^1 = \{w16, w9, w17\}$.

STEP 2: $\bar{T}^1_{w-in,1} = \frac{C_{w1}T_{w1}+C_{w2}T_{w2}+C_{w3}T_{w3}+C_{w4}T_{w4}}{C_{w1}+C_{w2}+C_{w3}+C_{w4}}$, $\bar{T}^1_{w-out,1} = \frac{C_{w13}T_{w13}+C_{w14}T_{w14}+C_{w15}T_{w15}}{C_{w13}+C_{w14}+C_{w15}}$,

$\bar{T}^1_{w-in,2} = \frac{C_{w5}T_{w5}+C_{w6}T_{w6}+C_{w7}T_{w7}+C_{w8}T_{w8}}{C_{w5}+C_{w6}+C_{w7}+C_{w8}}$ and $\bar{T}^1_{w-out,2} = \frac{C_{w16}T_{w16}+C_{w9}T_{w9}+C_{w17}T_{w17}}{C_{w16}+C_{w9}+C_{w17}}$.

STEP 3: $\bar{C}^1_{w-in,1} = C_{w1} + C_{w2} + C_{w3} + C_{w4}$, $\bar{C}^1_{w-out,1} = C_{w13} + C_{w14} + C_{w15}$, $\bar{C}^1_{w-in,2} = C_{w5} + C_{w6} + C_{w7} + C_{w8}$ and $\bar{C}^1_{w-out,2} = C_{w16} + C_{w9} + C_{w17}$.

STEP 4: $\frac{1}{\bar{R}^1_{w-in,w-out,1}} = \frac{1}{R_{w2,w14}} + \frac{1}{R_{w13,w1}} + \frac{1}{R_{w4,w15}}$ and $\frac{1}{\bar{R}^1_{w-in,w-out,2}} = \frac{1}{R_{w6,w16}} + \frac{1}{R_{w7,w9}} + \frac{1}{R_{w8,w17}}$.

STEP 5: $\frac{1}{\bar{R}^1_{w-in,z,1,1}} = \frac{1}{R_{w1,z1}} + \frac{1}{R_{w2,z1}} + \frac{1}{R_{w3,z1}} + \frac{1}{R_{w4,z1}}$, $\frac{1}{\bar{R}^1_{w-in,z,1,2}} = 0$, $\frac{1}{\bar{R}^1_{w-in,z,2,1}} = 0$, and

$\frac{1}{\bar{R}^1_{w-in,z,2,2}} = \frac{1}{R_{w6,z2}} + \frac{1}{R_{w7,z2}} + \frac{1}{R_{w8,z2}} + \frac{1}{R_{w5,z2}}$.

STEP 6: $\frac{1}{\bar{R}^1_{w-in,w-in,1,2}} = \frac{1}{R_{w3,w5}}$.

STEP 7: $\frac{1}{\bar{R}^1_{w-out,a,1}} = \frac{1}{R_{w13,a}} + \frac{1}{R_{w14,a}}$, $\frac{1}{\bar{R}^1_{w-out,g,1}} = \frac{1}{R_{w15,g}}$, $\frac{1}{\bar{R}^1_{w-out,a,2}} = \frac{1}{R_{w16,a}}$ and $\frac{1}{\bar{R}^1_{w-out,g,2}} = \frac{1}{R_{w17,g}}$.

STEP 8 : $\frac{1}{\bar{R}^{1,2}_{w-out,z,1,1}} = 0$ and $\frac{1}{\bar{R}^{1,2}_{w-out,z,2,1}} = \frac{1}{R_{w9,z3}}$.

STEP 9*:* The reduced order RC network for the 3-zone building is shown in Figure 5.3. It has only 6 states compared to 23 states in the full order model.

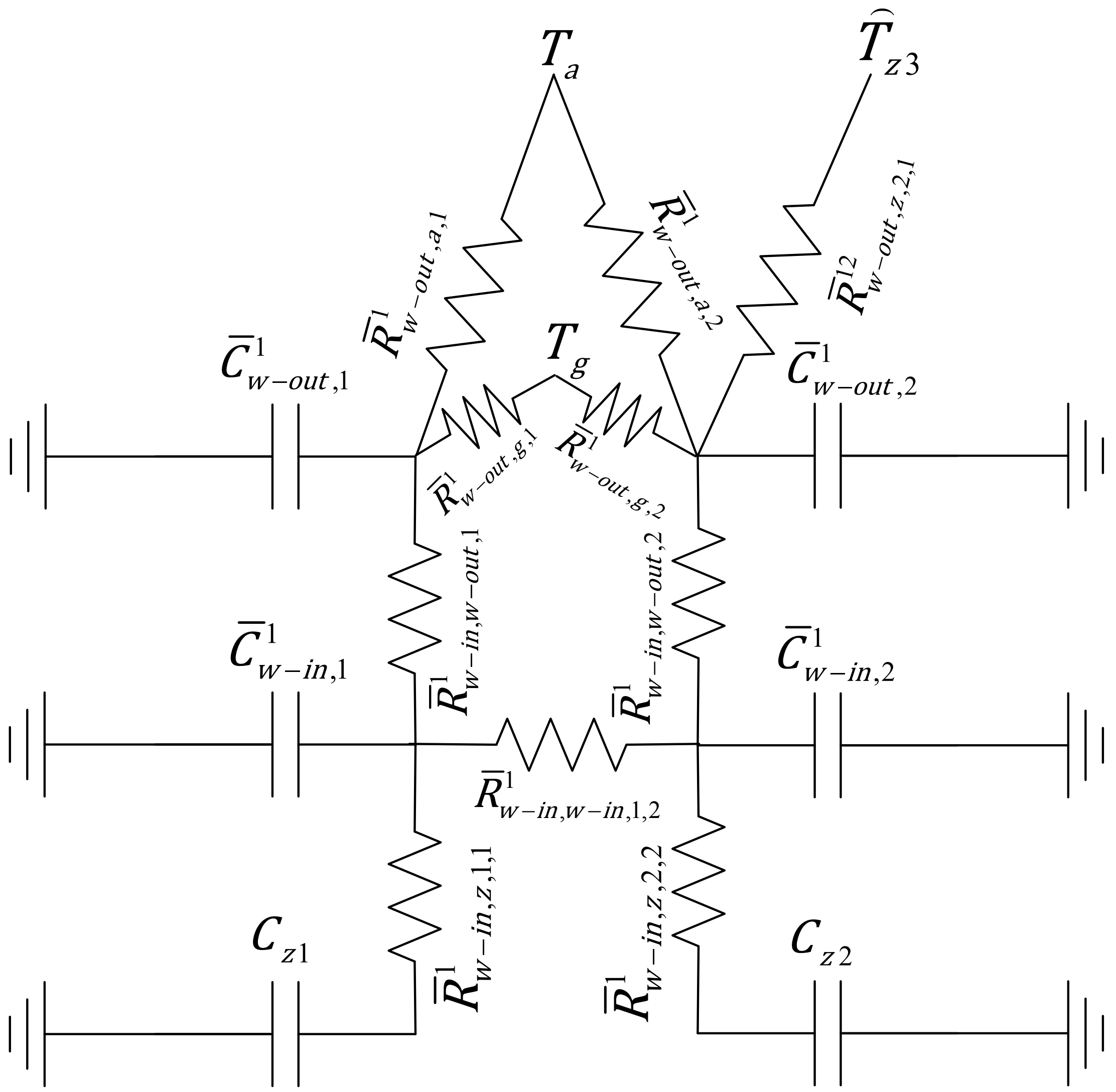

**Figure 5.3 Reduced order RC network obtained via Algorithm 5.2 for the cluster {1,2} of the 3-zone building in Figure 5.2**

### 5.2.4 State transformation

The states $\bar{\mathbf{T}}^{\mathbf{i}}_{\mathbf{w-in}}$ and $\bar{\mathbf{T}}^{\mathbf{i}}_{\mathbf{w-out}}$, and disturbances $\bar{\mathbf{d}}^{\mathbf{i}}_{\mathbf{w}}$ and $\bar{\mathbf{d}}^{\mathbf{i}}_{\mathbf{z}}$ in (5.12) are unknown. An extended state observer (ESO) [89, 90] can be designed to estimate them based on assumptions on the dynamics of the unknown disturbances. In this paper, we assume that the unknown disturbances are slowly time-varying quantities. Therefore, if the time window of interest for control design – such as prediction horizon in case of MPC – is sufficiently small, e.g. 30 minutes to an hour, we can assume that the disturbances are constant, that is $\dot{\bar{\mathbf{d}}}^{\mathbf{i}}_{\mathbf{w}} \approx \mathbf{0}$ and $\dot{\bar{\mathbf{d}}}^{\mathbf{i}}_{\mathbf{z}} \approx \mathbf{0}$. This is used to augment the dynamics in (5.12) as shown

$$\frac{d}{dt}\begin{bmatrix}\mathbf{x}^{\mathbf{i}}\\ \bar{\mathbf{d}}^{\mathbf{i}}_{\mathbf{w}}\\ \bar{\mathbf{d}}^{\mathbf{i}}_{\mathbf{z}}\end{bmatrix} = \begin{bmatrix}\bar{\mathbf{A}}^{\mathbf{i}} & \bar{\mathbf{B}}^{\mathbf{i}}_{\mathbf{dw}} & \bar{\mathbf{B}}^{\mathbf{i}}_{\mathbf{dz}}\\ \mathbf{0} & \mathbf{0} & \mathbf{0}\\ \mathbf{0} & \mathbf{0} & \mathbf{0}\end{bmatrix}\begin{bmatrix}\mathbf{x}^{\mathbf{i}}\\ \bar{\mathbf{d}}^{\mathbf{i}}_{\mathbf{w}}\\ \bar{\mathbf{d}}^{\mathbf{i}}_{\mathbf{z}}\end{bmatrix} + \begin{bmatrix}\bar{\mathbf{B}}^{\mathbf{i}}_{\mathbf{u}}\\ \mathbf{0}\\ \mathbf{0}\end{bmatrix}\mathbf{u}^{\mathbf{i}} + \begin{bmatrix}\bar{\mathbf{B}}^{\mathbf{i}}_{\mathbf{a}} & \bar{\mathbf{B}}^{\mathbf{i}}_{\mathbf{g}}\\ \mathbf{0} & \mathbf{0}\\ \mathbf{0} & \mathbf{0}\end{bmatrix}\begin{bmatrix}T_a\\ T_g\end{bmatrix} + \sum_{j\neq i}\begin{bmatrix}\bar{\mathbf{B}}^{\mathbf{ij}}_{\mathbf{z}}\\ \mathbf{0}\\ \mathbf{0}\end{bmatrix}\hat{\mathbf{T}}^{\mathbf{j}}_{\mathbf{z}}. \quad (5.19)$$

A limitation of representation (5.19) is that it does not guarantee observability when measurements are only available for the zone temperatures $\mathbf{T}^{\mathbf{i}}_{\mathbf{z}}$ in the cluster. To address this limitation, we define new states

$$\boldsymbol{\eta}^{\mathbf{i}}_{\mathbf{1}} = \bar{\mathbf{T}}^{\mathbf{i}}_{\mathbf{w}} + \left(\bar{\mathbf{A}}^{\mathbf{i}}_{\mathbf{w,w}}\right)^{-1}\bar{\mathbf{B}}^{\mathbf{i}}_{\mathbf{w,dw}}\bar{\mathbf{d}}^{\mathbf{i}}_{\mathbf{w}}, \quad (5.20)$$

$$\boldsymbol{\eta}^{\mathbf{i}}_{\mathbf{2}} = \bar{\mathbf{B}}^{\mathbf{i}}_{\mathbf{z,dz}}\bar{\mathbf{d}}^{\mathbf{i}}_{\mathbf{z}} - \bar{\mathbf{A}}^{\mathbf{i}}_{\mathbf{z,w}}\left(\bar{\mathbf{A}}^{\mathbf{i}}_{\mathbf{w,w}}\right)^{-1}\bar{\mathbf{B}}^{\mathbf{i}}_{\mathbf{w,dw}}\bar{\mathbf{d}}^{\mathbf{i}}_{\mathbf{w}}, \quad (5.21)$$

where,

$$\bar{\mathbf{T}}^{\mathbf{i}}_{\mathbf{w}} = \begin{bmatrix}\bar{\mathbf{T}}^{\mathbf{i}}_{\mathbf{w-in}}\\ \bar{\mathbf{T}}^{\mathbf{i}}_{\mathbf{w-out}}\end{bmatrix}, \bar{\mathbf{A}}^{\mathbf{i}}_{\mathbf{w,w}} = \begin{bmatrix}\bar{\mathbf{A}}^{\mathbf{i}}_{\mathbf{w-in,w-in}} & \bar{\mathbf{A}}^{\mathbf{i}}_{\mathbf{w-in,w-out}}\\ \bar{\mathbf{A}}^{\mathbf{i}}_{\mathbf{w-out,w-in}} & \bar{\mathbf{A}}^{\mathbf{i}}_{\mathbf{w-out,w-out}}\end{bmatrix}, \bar{\mathbf{B}}^{\mathbf{i}}_{\mathbf{w,dw}} = \begin{bmatrix}\bar{\mathbf{B}}^{\mathbf{i}}_{\mathbf{w-in,dw}}\\ \bar{\mathbf{B}}^{\mathbf{i}}_{\mathbf{w-out,dw}}\end{bmatrix},$$

$$\text{and } \bar{\mathbf{A}}^{\mathbf{i}}_{\mathbf{z,w}} = \begin{bmatrix}\bar{\mathbf{A}}^{\mathbf{i}}_{\mathbf{z,w-in}} & \mathbf{0}_{N_{zi}\times n^i_{w-out}}\end{bmatrix}.$$

The transformed state space model using the new states $\boldsymbol{\eta}^{\mathbf{i}}_{\mathbf{1}}$ and $\boldsymbol{\eta}^{\mathbf{i}}_{\mathbf{2}}$ is given by

$$\frac{d}{dt}\begin{bmatrix}\boldsymbol{\eta}^{\mathbf{i}}_{\mathbf{1}}\\ \mathbf{T}^{\mathbf{i}}_{\mathbf{z}}\\ \boldsymbol{\eta}^{\mathbf{i}}_{\mathbf{2}}\end{bmatrix} = \underbrace{\begin{bmatrix}\bar{\mathbf{A}}^{\mathbf{i}}_{\mathbf{w,w}} & \bar{\mathbf{A}}^{\mathbf{i}}_{\mathbf{w,z}} & \mathbf{0}\\ \bar{\mathbf{A}}^{\mathbf{i}}_{\mathbf{z,w}} & \bar{\mathbf{A}}^{\mathbf{i}}_{\mathbf{z,z}} & \mathbf{I}\\ \mathbf{0} & \mathbf{0} & \mathbf{0}\end{bmatrix}}_{\bar{\mathbf{A}}^{\mathbf{i}}_{\mathbf{tfo}}}\begin{bmatrix}\boldsymbol{\eta}^{\mathbf{i}}_{\mathbf{1}}\\ \mathbf{T}^{\mathbf{i}}_{\mathbf{z}}\\ \boldsymbol{\eta}^{\mathbf{i}}_{\mathbf{2}}\end{bmatrix} + \begin{bmatrix}\mathbf{0}\\ \bar{\mathbf{B}}^{\mathbf{i}}_{\mathbf{z,u}}\\ \mathbf{0}\end{bmatrix}\mathbf{u}^{\mathbf{i}} + \begin{bmatrix}\bar{\mathbf{B}}^{\mathbf{i}}_{\mathbf{w,a}} & \bar{\mathbf{B}}^{\mathbf{i}}_{\mathbf{w,g}}\\ \mathbf{0} & \mathbf{0}\\ \mathbf{0} & \mathbf{0}\end{bmatrix}\begin{bmatrix}T_a\\ T_g\end{bmatrix} + \sum_{j\neq i}\begin{bmatrix}\bar{\mathbf{B}}^{\mathbf{ij}}_{\mathbf{w,z}}\\ \mathbf{0}\\ \mathbf{0}\end{bmatrix}\hat{\mathbf{T}}^{\mathbf{j}}_{\mathbf{z}}, \quad (5.22)$$

where,

$$\bar{\mathbf{A}}^{\mathbf{i}}_{\mathbf{w,z}} = \begin{bmatrix}\bar{\mathbf{A}}^{\mathbf{i}}_{\mathbf{w-in,z}}\\ \mathbf{0}_{n^i_{w-out}\times N_z}\end{bmatrix}, \bar{\mathbf{B}}^{\mathbf{i}}_{\mathbf{w,a}} = \begin{bmatrix}\mathbf{0}_{n^i_{w-in}\times 1}\\ \bar{\mathbf{B}}^{\mathbf{i}}_{\mathbf{w-out,a}}\end{bmatrix}, \bar{\mathbf{B}}^{\mathbf{i}}_{\mathbf{w,g}} = \begin{bmatrix}\mathbf{0}_{n^i_{w-in}\times 1}\\ \bar{\mathbf{B}}^{\mathbf{i}}_{\mathbf{w-out,g}}\end{bmatrix}, \text{and } \bar{\mathbf{B}}^{\mathbf{ij}}_{\mathbf{w,z}} = \begin{bmatrix}\mathbf{0}_{n^i_{w-in}\times N_{zj}}\\ \bar{\mathbf{B}}^{\mathbf{ij}}_{\mathbf{w-out,z}}\end{bmatrix}.$$

The measurement model corresponding to (5.22) is given by

$$\mathbf{T^i_{z,meas}} = \underbrace{\begin{bmatrix} \mathbf{0}_{N_{zi}\times(n^i_{w-out}+n^i_{w-in})} & \mathbf{I}_{N_{zi}} & \mathbf{0}_{N_{zi}\times N_{zi}} \end{bmatrix}}_{\mathbf{C^i_{tfo}}} \begin{bmatrix} \boldsymbol{\eta}^i_1 \\ \mathbf{T^i_z} \\ \boldsymbol{\eta}^i_2 \end{bmatrix} \tag{5.23}$$

**Theorem 5.1**: *The pair* $\left(\mathbf{\bar{A}^i_{tfo}}, \mathbf{C^i_{tfo}}\right)$ *is observable.*

*Proof:* Using the definitions of $\mathbf{\bar{A}^i_{w,w}}$, $\mathbf{\bar{A}^i_{w,z}}$, $\mathbf{\bar{A}^i_{z,w}}$ and $\mathbf{\bar{A}^i_{z,z}}$ above, $\mathbf{\bar{A}^i_{tfo}}$ can be written as

$$\mathbf{\bar{A}^i_{tfo}} = \begin{bmatrix} \mathbf{\bar{A}^i_{w-in,w-in}} & \mathbf{\bar{A}^i_{w-in,w-out}} & \mathbf{\bar{A}^i_{w-in,z}} & \mathbf{0} \\ \mathbf{\bar{A}^i_{w-out,w-in}} & \mathbf{\bar{A}^i_{w-out,w-out}} & \mathbf{0} & \mathbf{0} \\ \mathbf{\bar{A}^i_{z,w-in}} & \mathbf{0} & \mathbf{\bar{A}^i_{z,z}} & \mathbf{I}_{N_{zi}} \\ \mathbf{0}_{N_{zi}\times n^i_{w-in}} & \mathbf{0}_{N_{zi}\times n^i_{w-out}} & \mathbf{0}_{N_{zi}\times N_{zi}} & \mathbf{0}_{N_{zi}\times N_{zi}} \end{bmatrix}. \tag{5.24}$$

For simplicity of notation, we define $\mathbf{A_{11}} = \mathbf{\bar{A}^i_{w-in,w-in}}$, $\mathbf{A_{12}} = \mathbf{\bar{A}^i_{w-in,w-out}}$, $\mathbf{A_{13}} = \mathbf{\bar{A}^i_{w-in,z}}$, $\mathbf{A_{21}} = \mathbf{\bar{A}^i_{w-out,w-in}}$, $\mathbf{A_{22}} = \mathbf{\bar{A}^i_{w-out,w-out}}$, $\mathbf{A_{31}} = \mathbf{\bar{A}^i_{z,w-in}}$ and $\mathbf{A_{33}} = \mathbf{\bar{A}^i_{z,z}}$. Furthermore, we define the matrix

$$\mathbf{X} = -\begin{bmatrix} \mathbf{A_{11}} & \mathbf{A_{12}} \\ \mathbf{A_{21}} & \mathbf{A_{22}} \end{bmatrix}. \tag{5.25}$$

A weighted graph is created with nodes corresponding to each of the aggregated wall temperature states in cluster $i$ ($\mathbf{\bar{T}^i_{w-in}}$ and $\mathbf{\bar{T}^i_{w-out}}$ in (5.12)), zone temperatures ($\mathbf{T^i_z}$), ambient temperature ($T_a$), ground temperatures ($T_g$), and zone temperature estimates ($\mathbf{\hat{T}^j_z}, j \neq i$ ) in all other clusters. These nodes are numbered in the order in which they are mentioned above. An edge is used to connect each pair of nodes which are connected by a resistance in the reduced order RC network (obtained from Algorithm 5.2). The weight such edges are set to be the inverse of the corresponding thermal resistances. The resulting weighted graph is denoted by $G^i = \left(V^i, E^i\right)$ along with the weight function $w: E^i \rightarrow \mathbb{R}^+$. Note that this procedure is an extension of the Algorithm 5.1 to a cluster in the building. We also define capacitance matrices $\mathbf{\bar{C}^i_{w-in}}$, $\mathbf{\bar{C}^i_{w-out}}$ and $\mathbf{C^i_z}$, which are diagonal matrices of the thermal capacitances associated with the state vectors $\mathbf{\bar{T}^i_{w-in}}$, $\mathbf{\bar{T}^i_{w-out}}$, and $\mathbf{T^i_z}$ respectively. The diagonal entries in these matrices are entered in the order of the corresponding node numbers in $G^i$. Let $\mathbf{L^i_G}$ denote the Laplacian matrix of $G^i$ and recall that $n_{w-in}^{i}$ and $n_{w-out}^{i}$ denote the lengths of the vectors $\overline{\mathbf{T}}_{\mathbf{w-in}}^{\mathbf{i}}$ and $\overline{\mathbf{T}}_{\mathbf{w-out}}^{\mathbf{i}}$ respectively. We now define the following sub-matrices of $\mathbf{L}_{\mathbf{G}}^{\mathbf{i}}$.

1. $\mathbf{L}_{\mathbf{11}}$ is the square sub-matrix of $\mathbf{L}_{\mathbf{G}}^{\mathbf{i}}$ which corresponds to its first $n_{w-in}^{i}$ rows and columns.

2. $\mathbf{L}_{\mathbf{X}}$ is the square sub-matrix of $\mathbf{L}_{\mathbf{G}}^{\mathbf{i}}$ which corresponds to its first $n_{w-in}^{i} + n_{w-out}^{i}$ rows and columns.

3. $\mathbf{L}_{\mathbf{31}}$ is the square sub-matrix of $\mathbf{L}_{\mathbf{G}}^{\mathbf{i}}$ which corresponds to rows $n_{w-in}^{i} + n_{w-out}^{i} + 1, \dots, n_{w-in}^{i} + n_{w-out}^{i} + N_{zi}$ and the first $n_{w-in}^{i}$ columns.

4. $\mathbf{L}_{\mathbf{12}}$ is defined as the sub-matrix of $\mathbf{L}_{\mathbf{G}}^{\mathbf{i}}$ which corresponds to its first $n_{w-in}^{i}$ rows and columns $n_{w-in}^{i} + 1, \dots, n_{w-in}^{i} + n_{w-out}^{i}$.

An illustration of the above matrix definitions is presented below. Consider a 2-partition {{1,2},{3}} of the building in Figure 5.2. The Laplacian matrix $\mathbf{L}_{\mathbf{G}}^{\mathbf{1}}$ for the first cluster {1,2} in this partition can be expressed as follows using its reduced order RC network representation in Figure 5.3.

$$\mathbf{L}_{\mathbf{G}}^{\mathbf{1}} = \begin{bmatrix} a_{1,1} & a_{1,2} & a_{1,3} & 0 & a_{1,5} & 0 & 0 & 0 & 0 \\ a_{2,1} & a_{2,2} & 0 & a_{2,4} & 0 & a_{2,6} & 0 & 0 & 0 \\ a_{3,1} & 0 & a_{3,3} & 0 & 0 & 0 & a_{3,7} & a_{3,8} & 0 \\ 0 & a_{4,2} & 0 & a_{4,4} & 0 & 0 & a_{4,7} & a_{4,8} & a_{4,9} \\ a_{5,1} & 0 & 0 & 0 & a_{5,5} & 0 & 0 & 0 & 0 \\ 0 & a_{6,2} & 0 & 0 & 0 & a_{6,6} & 0 & 0 & 0 \\ 0 & 0 & a_{7,3} & a_{7,4} & 0 & 0 & a_{7,7} & 0 & 0 \\ 0 & 0 & a_{8,3} & a_{8,4} & 0 & 0 & 0 & a_{8,8} & 0 \\ 0 & 0 & 0 & a_{9,4} & 0 & 0 & 0 & 0 & a_{9,9} \end{bmatrix},$$

where,

$$a_{1,2} = a_{2,1} = {}^{-1}\!\big/\bar{R}_{w-in,w-in,1,2}^{1}$$

$$a_{1,3} = a_{3,1} = {}^{-1}\!\big/\bar{R}_{w-in,w-out,1}^{1}$$

$$a_{1,5} = a_{5,1} = {}^{-1}\!\big/\bar{R}_{w-in,z,1,1}^{1}$$

$$a_{2,4} = a_{4,2} = {}^{-1}\!/_{\bar{R}^{1}_{w-in,w-out,2}}$$

$$a_{2,6} = a_{6,2} = {}^{-1}\!/_{\bar{R}^{1}_{w-in,z,2,2}}$$

$$a_{3,7} = a_{7,3} = {}^{-1}\!/_{\bar{R}^{1}_{w-out,a,1}}$$

$$a_{3,8} = a_{8,3} = {}^{-1}\!/_{\bar{R}^{1}_{w-out,g,1}}$$

$$a_{4,7} = a_{7,4} = {}^{-1}\!/_{\bar{R}^{1}_{w-out,a,2}}$$

$$a_{4,8} = a_{8,4} = {}^{-1}\!/_{\bar{R}^{1}_{w-out,g,2}}$$

$$a_{4,9} = a_{9,4} = {}^{-1}\!/_{\bar{R}^{1,2}_{w-out,z,2,1}}$$

$$a_{1,1} = -a_{1,2} - a_{1,3} - a_{1,5}$$

$$a_{2,2} = -a_{2,1} - a_{2,4} - a_{2,6}$$

$$a_{3,3} = -a_{3,1} - a_{3,7} - a_{3,8}$$

$$a_{4,4} = -a_{4,2} - a_{4,7} - a_{4,8} - a_{4,9}$$

$$a_{5,5} = -a_{5,1}$$

$$a_{6,6} = -a_{6,2}$$

$$a_{7,7} = -a_{7,3} - a_{7,4}$$

$$a_{8,8} = -a_{8,3} - a_{8,4}$$

$$a_{9,9} = -a_{9,4}\ .$$

The matrices $\mathbf{L_{11}}$, $\mathbf{L_X}$, $\mathbf{L_{31}}$ and $\mathbf{L_{12}}$ are then obtained as

$$\mathbf{L_{11}} = \begin{bmatrix} a_{1,1} & a_{1,2} \\ a_{2,1} & a_{2,2} \end{bmatrix},$$

$$\mathbf{L_X} = \begin{bmatrix} a_{1,1} & a_{1,2} & a_{1,3} & 0 \\ a_{2,1} & a_{2,2} & 0 & a_{2,4} \\ a_{3,1} & 0 & a_{3,3} & 0 \\ 0 & a_{4,2} & 0 & a_{4,4} \end{bmatrix},$$

$$\mathbf{L_{31}} = \begin{bmatrix} a_{5,1} & 0 \\ 0 & a_{6,2} \end{bmatrix},$$

$$\mathbf{L_{12}} = \begin{bmatrix} a_{1,3} & 0 \\ 0 & a_{2,4} \end{bmatrix}.$$

The following properties can be easily verified from the structure of graph $G^i$:

1. $\mathbf{L_{11}}$ and $\mathbf{L_X}$ are Hermitian, irreducible, diagonally dominant matrices with positive diagonal entries. Furthermore, at least one row of both $\mathbf{L_{11}}$ and $\mathbf{L_X}$ is strictly diagonally dominant.

2. $\mathbf{L_{31}}$ is a diagonal matrix with negative entries on diagonal.

3. $\mathbf{L_{12}}$ is a sparse matrix of nonpositive elements such that each column has exactly one nonzero entry. Furthermore, the nonzero entries in two different columns of $\mathbf{L_{12}}$ are in different row locations.

Now we state and prove the following Lemmas.

**Lemma 5.1:** $\mathbf{A_{31}}$ is a full rank matrix.

*Proof*: It can be shown that $\mathbf{A_{31}} = -(\mathbf{C_z^i})^{-1}\mathbf{L_{31}}$. Using property 2 of $\mathbf{L_{31}}$ above, we conclude that $\mathbf{A_{31}}$ is a diagonal matrix with positive entries on diagonal. Hence, $\mathbf{A_{31}}$ is full rank. □

**Lemma 5.2:** $\mathbf{A_{11}}$ is an invertible matrix.

*Proof*: $\mathbf{A_{11}}$ can be expressed as $-\left(\mathbf{\bar{C}^i_{w-in}}\right)^{-1}\mathbf{L_{11}}$. Since $\mathbf{L_{11}}$is an *irreducibly diagonally dominant* matrix from property 1 above, it is non-singular as a result of the *Levy-Desplanques* theorem [91]. Hence, $\mathbf{L_{11}^{-1}}$ exists. Therefore, $\mathbf{A_{11}^{-1}}$ exists and is given by $-\mathbf{L_{11}^{-1}\bar{C}^i_{w-in}}$. □

**Lemma 5.3:** **X** is a positive definite matrix.

*Proof*: **X** can be expressed as shown

$$\mathbf{X} = \begin{vmatrix} \mathbf{\bar{C}^i_{w-in}} & \mathbf{0} \\ \mathbf{0} & \mathbf{\bar{C}^i_{w-out}} \end{vmatrix}^{-1} \mathbf{L_X}. \qquad (5.26)$$

Since $\mathbf{L_X}$ is an *irreducibly diagonally dominant* matrix from property 1 above, it is non-singular as a result of the *Levy-Desplanques* theorem [91]. Also, since $\mathbf{L_X}$ is a Hermitian, diagonally dominant matrix with real non-negative entries on the diagonal, it is positive semi-definite. Since a non-singular positive semi-definite matrix can only be positive definite, we conclude that $\mathbf{L_X}$ is positive definite. Using the fact that $\bar{\mathbf{C}}^{\mathbf{i}}_{\mathbf{w-in}}$ and $\bar{\mathbf{C}}^{\mathbf{i}}_{\mathbf{w-out}}$ are diagonal matrices with positive entries, we establish using (5.26) that **X** is positive definite. □

**Lemma 5.4:** $\mathbf{A}^{\mathrm{T}}_{\mathbf{12}}\mathbf{A_{12}}$ *is a positive definite matrix.*

*Proof:* $\mathbf{A_{12}}$ can be written as $-\left(\bar{\mathbf{C}}^{\mathbf{i}}_{\mathbf{w-in}}\right)^{-1}\mathbf{L_{12}}$. Since $\bar{\mathbf{C}}^{\mathbf{i}}_{\mathbf{w-in}}$ is a diagonal matrix with positive entries, $\mathbf{A_{12}}$ share the properties of $\mathbf{L_{12}}$ above (property 3), except that its entries are nonnegative instead of nonpositive. Next,

$$\left(\mathbf{A}^{\mathrm{T}}_{\mathbf{12}}\mathbf{A_{12}}\right)_{p,q} = \sum_r \left(\mathbf{A}^{\mathrm{T}}_{\mathbf{12}}\right)_{p,r}(\mathbf{A_{12}})_{r,q} = \sum_r (\mathbf{A_{12}})_{r,p}(\mathbf{A_{12}})_{r,q}.$$

Since $(\mathbf{A_{12}})_{r,p}$ and $(\mathbf{A_{12}})_{r,q}$ are both nonzero simultaneously for some $r$ only if $p = q$, we conclude that:

$$\left(\mathbf{A}^{\mathrm{T}}_{\mathbf{12}}\mathbf{A_{12}}\right)_{p,q} \begin{cases} > 0 & \text{if } p = q \\ = 0 & \text{otherwise.} \end{cases}$$

Hence, $\mathbf{A}^{\mathrm{T}}_{\mathbf{12}}\mathbf{A_{12}}$ is a diagonal matrix with positive entries and is therefore positive definite. □

We rewrite $\mathbf{C}^{\mathbf{i}}_{\mathbf{tfo}}$ in the form (5.27 ) which is aligned with the expanded form of $\bar{\mathbf{A}}^{\mathbf{i}}_{\mathbf{tfo}}$ in (5.24).

$$\mathbf{C}^{\mathbf{i}}_{\mathbf{tfo}} = \begin{bmatrix} \mathbf{0}_{N_{zi} \times n^i_{w-in}} & \mathbf{0}_{N_{zi} \times n^i_{w-out}} & \mathbf{I}_{N_{zi}} & \mathbf{0}_{N_{zi}} \end{bmatrix}. \quad (5.27)$$

The resulting observability matrix for $\left(\bar{\mathbf{A}}^{\mathbf{i}}_{\mathbf{tfo}}, \mathbf{C}^{\mathbf{i}}_{\mathbf{tfo}}\right)$ is obtained as shown

$$\mathcal{O} = \begin{bmatrix} \mathbf{C}^{\mathbf{i}}_{\mathbf{tfo}} \\ \mathbf{C}^{\mathbf{i}}_{\mathbf{tfo}}\bar{\mathbf{A}}^{\mathbf{i}}_{\mathbf{tfo}} \\ \mathbf{C}^{\mathbf{i}}_{\mathbf{tfo}}\left(\bar{\mathbf{A}}^{\mathbf{i}}_{\mathbf{tfo}}\right)^2 \\ \vdots \\ \mathbf{C}^{\mathbf{i}}_{\mathbf{tfo}}\left(\bar{\mathbf{A}}^{\mathbf{i}}_{\mathbf{tfo}}\right)^{n^i_{w-in}+n^i_{w-out}+2N_{zi}} \end{bmatrix} = \begin{bmatrix} \mathbf{0} & \mathbf{0} & \mathbf{I} & \mathbf{0} \\ \mathbf{A_{31}} & \mathbf{0} & \mathbf{A_{33}} & \mathbf{I} \\ \mathbf{A_{31}}\mathbf{A_{11}} + \mathbf{A_{33}}\mathbf{A_{31}} & \mathbf{A_{31}}\mathbf{A_{12}} & \mathbf{A_{31}}\mathbf{A_{13}} + (\mathbf{A_{33}})^2 & \mathbf{A_{33}} \\ \mathbf{G_1} & \mathbf{G_2} & \mathbf{G_3} & \mathbf{G_4} \\ \vdots & \vdots & \vdots & \vdots \end{bmatrix}$$

( 5.28 )

Here, $\mathbf{G_1} = (\mathbf{A_{31}}\mathbf{A_{11}} + \mathbf{A_{33}}\mathbf{A_{31}})\mathbf{A_{11}} + \mathbf{A_{31}}\mathbf{A_{12}}\mathbf{A_{21}}\,[\mathbf{A_{31}}\mathbf{A_{13}} + (\mathbf{A_{33}})^2]\mathbf{A_{31}},$

$$\mathbf{G}_2 = (\mathbf{A}_{31}\mathbf{A}_{11} + \mathbf{A}_{33}\mathbf{A}_{31})\mathbf{A}_{12} + \mathbf{A}_{31}\mathbf{A}_{12}\mathbf{A}_{22},$$

$$\mathbf{G}_3 = (\mathbf{A}_{31}\mathbf{A}_{11} + \mathbf{A}_{33}\mathbf{A}_{31})\mathbf{A}_{13} + [\mathbf{A}_{31}\mathbf{A}_{13} + (\mathbf{A}_{33})^2]\mathbf{A}_{33},$$

$$\mathbf{G}_4 = \mathbf{A}_{31}\mathbf{A}_{13} + (\mathbf{A}_{33})^2.$$

Let us assume that the matrix $\boldsymbol{\mathcal{O}}$ does not have full rank. Hence, there exists a vector $\mathbf{v}$ whose elements are not all zero such that,

$$\boldsymbol{\mathcal{O}}\mathbf{v} = \mathbf{0} \tag{5.29}$$

We rewrite $\mathbf{v}$ in the form shown in below which is aligned with the expanded form of $\boldsymbol{\mathcal{O}}$ in (5.28)

$$\mathbf{v} = [\mathbf{v}_1 \quad \mathbf{v}_2 \quad \mathbf{v}_3 \quad \mathbf{v}_4]^T. \tag{5.30}$$

Using (5.28) and (5.30), (5.29) results in

$$\mathbf{v}_3 = \mathbf{0}, \tag{5.31}$$

$$\mathbf{A}_{31}\mathbf{v}_1 + \mathbf{A}_{33}\mathbf{v}_3 + \mathbf{v}_4 = \mathbf{0}, \tag{5.32}$$

$$(\mathbf{A}_{31}\mathbf{A}_{11} + \mathbf{A}_{33}\mathbf{A}_{31})\mathbf{v}_1 + \mathbf{A}_{31}\mathbf{A}_{12}\mathbf{v}_2 + [\mathbf{A}_{31}\mathbf{A}_{13} + (\mathbf{A}_{33})^2]\mathbf{v}_3 + \mathbf{A}_{33}\mathbf{v}_4 = \mathbf{0}, \tag{5.33}$$

$$\mathbf{G}_1\mathbf{v}_1 + \mathbf{G}_2\mathbf{v}_2 + \mathbf{G}_3\mathbf{v}_3 + \mathbf{G}_4\mathbf{v}_4 = \mathbf{0} \tag{5.34}$$

Using ( 5.31) in (5.32 ), we obtain

$$\mathbf{A}_{31}\mathbf{v}_1 + \mathbf{v}_4 = \mathbf{0}. \tag{5.35}$$

Using ( 5.31) and (5.35 ) in (5.33), we obtain

$$\mathbf{A}_{31}(\mathbf{A}_{11}\mathbf{v}_1 + \mathbf{A}_{12}\mathbf{v}_2) = \mathbf{0}. \tag{5.36}$$

Since $\mathbf{A}_{31}$ is full rank (Lemma 5.1), we obtain

$$\mathbf{A}_{11}\mathbf{v}_1 + \mathbf{A}_{12}\mathbf{v}_2 = \mathbf{0}. \tag{5.37}$$

Using ( 5.31), (5.35 ) and (5.37) in (5.34), and using the definitions of $\mathbf{G}_1$ to $\mathbf{G}_4$ results in

$$\mathbf{A}_{31}\mathbf{A}_{12}[-\mathbf{A}_{22} - \mathbf{A}_{21}(-\mathbf{A}_{11})^{-1}\mathbf{A}_{12}]\mathbf{v}_2 = \mathbf{0}. \tag{5.38}$$

Note that $(-\mathbf{A}_{11})^{-1}$ exists from Lemma 5.2. Since $\mathbf{A}_{31}$ is full rank (Lemma 5.1), we obtain

$$\mathbf{A}_{12}[-\mathbf{A}_{22} - \mathbf{A}_{21}(-\mathbf{A}_{11})^{-1}\mathbf{A}_{12}]\mathbf{v}_2 = \mathbf{0}. \tag{5.39}$$

We define $\mathbf{S} = -\mathbf{A_{22}} - \mathbf{A_{21}}(-\mathbf{A_{11}})^{-1}\mathbf{A_{12}}$. Note that $\mathbf{S}$ is the Schur complement of $\mathbf{X}$. Since $\mathbf{X}$ is positive definite (Lemma 5.3), using the Schur complement condition for positive definiteness, we conclude that $\mathbf{S}$ is positive definite. Next, (5.39) leads to

$$\mathbf{v_2^T S^T A_{12}^T A_{12} S v_2} = \mathbf{0}. \tag{5.40}$$

Since $\mathbf{S}$ and $\mathbf{A_{12}^T A_{12}}$ are positive definite matrices (Lemma 5.4), we conclude from (5.40) that $\mathbf{v_2} = \mathbf{0}$. Therefore, from (5.37), $\mathbf{v_1} = \mathbf{0}$ since $\mathbf{A_{11}}$ is full rank (Lemma 5.2). Using (5.35), this implies $\mathbf{v_4} = \mathbf{0}$. Hence, $\mathbf{v_1} = \mathbf{v_2} = \mathbf{v_3} = \mathbf{v_4} = \mathbf{0}$ implying from (5.30) that $\mathbf{v} = \mathbf{0}$. This contradicts the assumption that at least one element of $\mathbf{v}$ is nonzero. Therefore, $\boldsymbol{\mathcal{O}}$ is full rank and hence $\left(\mathbf{\overline{A}^i_{tfo}}, \mathbf{C^i_{tfo}}\right)$ is observable. □

# 5.3 Observer and controller design

The reduced order, transformed representation shown in (5.22) provides an observable state space representation of the cluster level dynamics. In this section, we use this property to design an observer at the cluster level to estimate the unknown states. A cluster level controller is then designed which utilizes these estimates to provide optimal control decisions that minimize an appropriate objective function subject to constraints.

## 5.3.1 Observer design

For any p-partition of a building, a family of discrete-time observers – one for each cluster *i* – can be designed to estimate the states $\boldsymbol{\eta}_1^{\mathbf{i}}$ and $\boldsymbol{\eta}_2^{\mathbf{i}}$ in the model (5.22) using the zone temperature measurements obtained in (5.23). From an implementation perspective, it is desired that the sampling rate of the observers should match that of the controllers. The temperature estimates $\mathbf{\widehat{T}_z^j}$ for the zones in the clusters other than $i$ appearing in (5.22) are treated as design parameters. For example, they can be set to the corresponding set-point temperatures $\mathbf{T^j_{z,ref}}$, based on the assumption that the controllers are able to accurately regulate the zone temperatures around the set-point.

## 5.3.2 Controller design

Here, we consider the design of output-feedback model predictive controllers which are decentralized at the cluster-level and use the state estimates provided by the observer in Section 5.3.1. The objective function to be minimized at any time step $k$ for the $i^{th}$cluster in a p-partition consists of a sum of appropriate performance and cost objectives (see (3.21)).

$$J_i = J_i^{perf} + J_i^{cost}, \tag{5.41}$$

where,

$$J_i^{perf} = \sum_{l=1}^{N_p} [\mathbf{e_i}(k+l)]^T \, diag(\boldsymbol{\beta}_\mathbf{i}) \mathbf{e_i}(k+l), \tag{5.42}$$

$$\mathbf{e_i}(k+l) = \mathbf{T_z^i}(k+l|k) - \mathbf{T_{z,ref}^i}(k), \tag{5.43}$$

$$J_i^{cost} = \sum_{l=0}^{N_u-1} [\mathbf{u^i}(k+l|k)]^T diag(\boldsymbol{\alpha}_\mathbf{i}) \, \mathbf{u^i}(k+l|k). \tag{5.44}$$

Here, $\boldsymbol{\alpha}_\mathbf{i}, \boldsymbol{\beta}_\mathbf{i} \in \mathbb{R}^{N_{zi}}$ are vector valued, component-wise positive weights on the performance (set-point tracking of zone temperatures) and cost (energy consumption) objectives; $N_p$ and $N_u$ represent the prediction and control horizon lengths respectively. A discrete-time version of the model (5.22) has the form

$$\begin{bmatrix} \boldsymbol{\eta}_1^\mathbf{i} \\ \mathbf{T_z^i} \\ \boldsymbol{\eta}_2^\mathbf{i} \end{bmatrix}(k+1) = \mathbf{\bar{A}_{tfo,D}^i} \begin{bmatrix} \boldsymbol{\eta}_1^\mathbf{i} \\ \mathbf{T_z^i} \\ \boldsymbol{\eta}_2^\mathbf{i} \end{bmatrix}(k) + \mathbf{\bar{B}_{u,tfo,D}^i u^i}(k) + \begin{bmatrix} \mathbf{\bar{B}_{a,tfo,D}^i} & \mathbf{\bar{B}_{g,tfo,D}^i} \end{bmatrix} \begin{bmatrix} T_a \\ T_g \end{bmatrix}(k)$$

$$+ \sum_{j \neq i} \mathbf{\bar{B}_{z,tfo,D}^{ij}} \, \mathbf{\hat{T}_z^j}(k). \tag{5.45}$$

The model ( 5.45) imposes the following constraints on the optimization:

1. For each $l \in \{1,2,..N_u\}$,

$$\begin{bmatrix} \boldsymbol{\eta}_1^\mathbf{i} \\ \mathbf{T_z^i} \\ \boldsymbol{\eta}_2^\mathbf{i} \end{bmatrix}(k+l|k) = \mathbf{\bar{A}_{tfo,D}^i} \begin{bmatrix} \boldsymbol{\eta}_1^\mathbf{i} \\ \mathbf{T_z^i} \\ \boldsymbol{\eta}_2^\mathbf{i} \end{bmatrix}(k+l-1|k) + \mathbf{\bar{B}_{u,tfo,D}^i u^i}(k+l-1|k)$$

$$+ \begin{bmatrix} \mathbf{\bar{B}_{a,tfo,D}^i} & \mathbf{\bar{B}_{g,tfo,D}^i} \end{bmatrix} \begin{bmatrix} T_a \\ T_g \end{bmatrix}(k) + \sum_{j \neq i} \mathbf{\bar{B}_{z,tfo,D}^{ij}} \, \mathbf{\hat{T}_z^j}(k), \tag{5.46}$$

2. For each $l \in \{N_u+1, \ldots, N_u + N_p\}$,

$$\begin{bmatrix} \boldsymbol{\eta}_1^\mathbf{i} \\ \mathbf{T_z^i} \\ \boldsymbol{\eta}_2^\mathbf{i} \end{bmatrix}(k+l|k) = \mathbf{\bar{A}_{tfo,D}^i} \begin{bmatrix} \boldsymbol{\eta}_1^\mathbf{i} \\ \mathbf{T_z^i} \\ \boldsymbol{\eta}_2^\mathbf{i} \end{bmatrix}(k+l-1|k) + \mathbf{\bar{B}_{u,tfo,D}^i u^i}(k+N_u-1|k)$$

$$+\begin{bmatrix}\bar{\mathbf{B}}^{\mathbf{i}}_{\mathbf{a,tfo,D}} & \bar{\mathbf{B}}^{\mathbf{i}}_{\mathbf{g,tfo,D}}\end{bmatrix}\begin{bmatrix}T_a\\T_g\end{bmatrix}(k)+\sum_{j\neq i}\bar{\mathbf{B}}^{\mathbf{ij}}_{\mathbf{z,tfo,D}}\,\hat{\mathbf{T}}^{\mathbf{j}}_{\mathbf{z}}(k). \qquad (5.47)$$

Additional constraints, such as heating and cooling capacity bounds are represented by

$$\mathbf{u^i}(k+l-1|k)\in\mathcal{U}^i,\ \text{for all}\ \ l\in\{1,2,..N_u\}. \qquad (5.48)$$

Here, $\mathcal{U}^i$ represents the set of feasible values of $\mathbf{u^i}(k+l-1|k)$. At each time instant $k$, the controller uses: (a) the state estimates from the observer $\hat{\boldsymbol{\eta}}^{\mathbf{i}}_{\mathbf{1}}(k)$ and $\hat{\boldsymbol{\eta}}^{\mathbf{i}}_{\mathbf{2}}(k)$, (b) the zone temperature measurements from thermostats $\mathbf{T}^{\mathbf{i}}_{\mathbf{z,meas}}(k)$, and (c) the signals $T_a(k)$, $T_g(k)$ and $\{\hat{\mathbf{T}}^{\mathbf{j}}_{\mathbf{z}}(k)\}_{j\neq i}$ to determine optimal values $\{\mathbf{u}_*{}^{\mathbf{i}}(k+l)\}_{l=0}^{N_u-1}$ of the control inputs $\{\mathbf{u^i}(k+l-1|k)\}_{l=1}^{N_u-1}$ which minimize the objective function (5.41) subject to the constraints (5.46) – (5.48). In accordance with the MPC methodology, the control input $\mathbf{u}_*{}^{\mathbf{i}}(k)$ corresponding to the current time instant is then applied to the plant. The observer and controller for the $i^{th}$ cluster are illustrated in Figure 5.4.

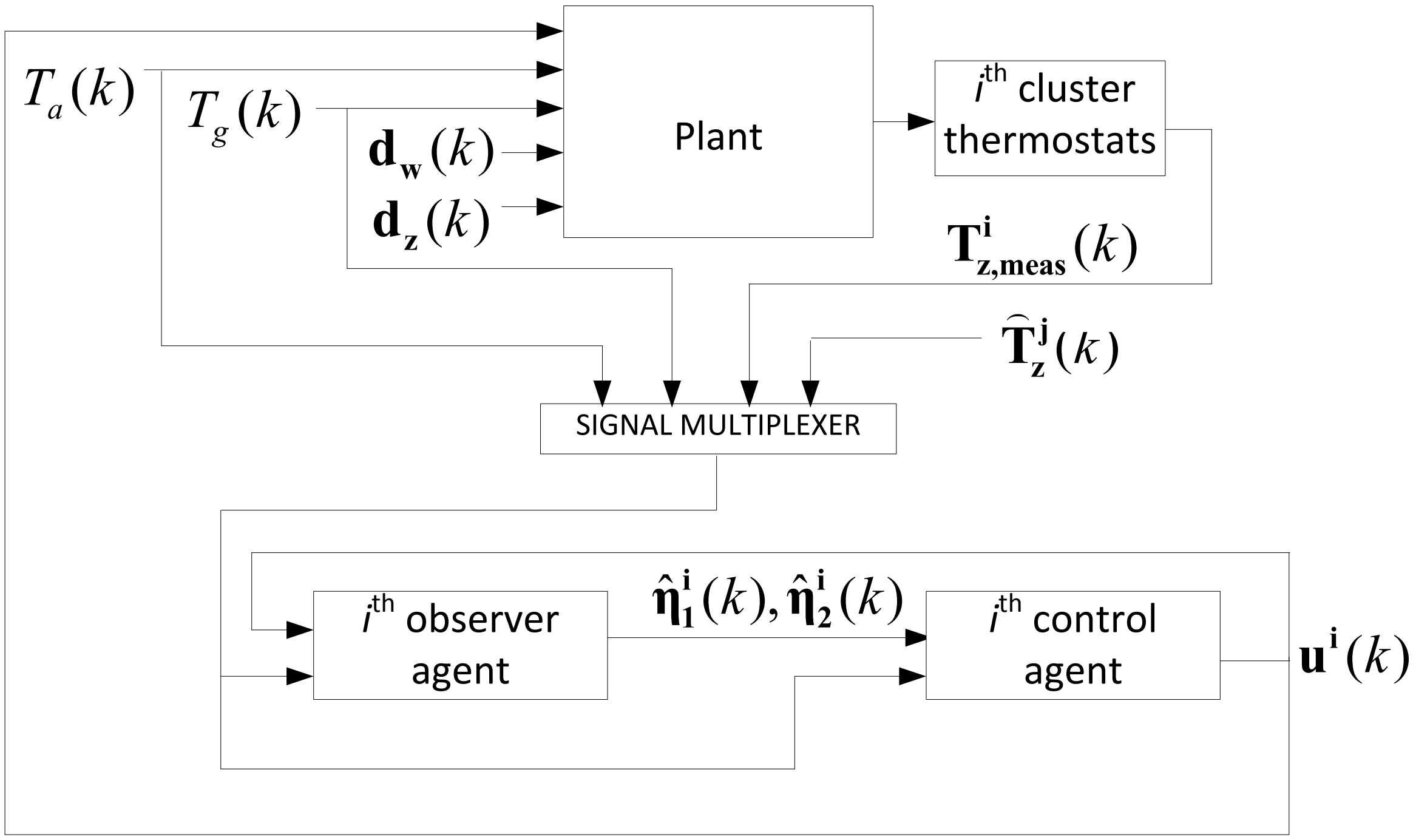

**Figure 5.4 Illustration of controller and observer for $i^{th}$ cluster.**

It should be noted that the design of cluster level decentralized controllers and observers discussed in this section leads to an output feedback control framework which only uses zone temperature measurements that are provided by thermostats. It also incorporates physical constraints such as heating or cooling capacity bounds associated with the HVAC system. Therefore, it builds upon the decentralized control framework that was presented in chapter 3 (Section 3.4) to make it practically implementable.

## 5.4 Optimization

Through a procedure analogous to that presented in Section 3.4.4, the optimization of the cost function $J_i$ in (5.41) for the $i^{th}$ control agent, in the presence of heating/cooling capacity constraints can be converted to a Quadratic Program (QP). The underlying procedure is shown in this section.

### 5.4.1 Re-statement of objective function

We augment the state space model in (5.45) by defining a new state $\mathbf{x_u^i}$ as shown below

$$\mathbf{x_u^i}(k) = \mathbf{u^i}(k-1). \tag{5.49}$$

Therefore,

$$\mathbf{x_u^i}(k+1) = \mathbf{x_u^i}(k) + \mathbf{\Delta u^i}(k), \tag{5.50}$$

Where $\mathbf{\Delta u^i}(k)$ is defined as

$$\mathbf{\Delta u^i}(k) = \mathbf{u^i}(k) - \mathbf{u^i}(k-1)\,. \tag{5.51}$$

Using (5.50) and (5.51), the augmented form of (5.45) is given by

$$\mathbf{x^i_{mpc}}(k+1) = \mathbf{A^i_{mpc}}\mathbf{x^i_{mpc}}(k) + \mathbf{B^i_{u,mpc}}\mathbf{\Delta u^i}(k) + \mathbf{B^i_{a,mpc}}T_a(k) + \mathbf{B^i_{g,mpc}}T_g(k)$$

$$+\textstyle\sum_{j\neq i}\mathbf{\bar{B}^{ij}_{mpc}}\,\mathbf{\hat{T}^j_z}(k), \tag{5.52}$$

where,

$$\mathbf{x^i_{mpc}} = \begin{bmatrix}\boldsymbol{\eta}^\mathbf{i}_1\\ \mathbf{T^i_z}\\ \boldsymbol{\eta}^\mathbf{i}_2\\ \mathbf{x^i_u}\end{bmatrix},\ \mathbf{A^i_{mpc}} = \begin{bmatrix}\mathbf{\bar{A}^i_{tfo,D}} & \mathbf{\bar{B}^i_{u,tfo,D}}\\ \mathbf{0} & \mathbf{I}_{N_{zi}}\end{bmatrix},\ \mathbf{B^i_{u,mpc}} = \begin{bmatrix}\mathbf{\bar{B}^i_{u,tfo,D}}\\ \mathbf{I}_{N_{zi}}\end{bmatrix},\ \mathbf{B^i_{a,mpc}} = \begin{bmatrix}\mathbf{\bar{B}^i_{a,tfo,D}}\\ \mathbf{0}_{N_{zi}\times 1}\end{bmatrix},$$

$$\mathbf{B}^{\mathbf{i}}_{\mathbf{g,mpc}} = \begin{bmatrix} \overline{\mathbf{B}}^{\mathbf{i}}_{\mathbf{g,tfo,D}} \\ \mathbf{0}_{N_{zi}\times 1} \end{bmatrix}, \overline{\mathbf{B}}^{\mathbf{ij}}_{\mathbf{mpc}} = \begin{bmatrix} \overline{\mathbf{B}}^{\mathbf{ij}}_{\mathbf{z,tfo,D}} \\ \mathbf{0}_{N_{zi}\times N_{zj}} \end{bmatrix}.$$

Based on (5.52), the constraints (5.46) and (5.47) can be written as the following.

1. For each $l \in \{1,2,\dots,N_u\}$:

$$\mathbf{x}^{\mathbf{i}}_{\mathbf{mpc}}(k+l|k) = \mathbf{A}^{\mathbf{i}}_{\mathbf{mpc}}\mathbf{x}^{\mathbf{i}}_{\mathbf{mpc}}(k+l-1|k) + \mathbf{B}^{\mathbf{i}}_{\mathbf{u,mpc}}\boldsymbol{\Delta}\mathbf{u}^{\mathbf{i}}(k+l-1|k)$$
$$+\mathbf{B}^{\mathbf{i}}_{\mathbf{a,mpc}}T_a(k+l-1|k) + \mathbf{B}^{\mathbf{i}}_{\mathbf{g,mpc}}T_g(k+l-1|k)$$
$$+ \textstyle\sum_{j\neq i}\overline{\mathbf{B}}^{\mathbf{ij}}_{\mathbf{mpc}}\,\widehat{\mathbf{T}}^{\mathbf{j}}_{\mathbf{z}}(k+l-1|k), \tag{5.53}$$

2. For each $l \in \{N_u+1,2,\dots,N_p\}$:

$$\mathbf{x}^{\mathbf{i}}_{\mathbf{mpc}}(k+l|k) = \mathbf{A}^{\mathbf{i}}_{\mathbf{mpc}}\mathbf{x}^{\mathbf{i}}_{\mathbf{mpc}}(k+l-1|k) + \mathbf{B}^{\mathbf{i}}_{\mathbf{a,mpc}}T_a(k+l-1|k) + \mathbf{B}^{\mathbf{i}}_{\mathbf{g,mpc}}T_g(k+l-1|k)$$
$$+ \textstyle\sum_{j\neq i}\overline{\mathbf{B}}^{\mathbf{ij}}_{\mathbf{mpc}}\,\widehat{\mathbf{T}}^{\mathbf{j}}_{\mathbf{z}}(k+l-1|k), \tag{5.54}$$

The derivation of (5.54) uses the fact that from definition (5.51), $\boldsymbol{\Delta}\mathbf{u}^{\mathbf{i}}(k+l-1|k) = 0$ outside the control horizon, i.e. when $l \in \{N_u+1,2,\dots,N_p\}$. Using (5.53) and (5.54), we obtain the following equations via iterations for each time step in the prediction horizon.

$$\mathbf{x}^{\mathbf{i}}_{\mathbf{mpc}}(k+1|k) = \mathbf{A}^{\mathbf{i}}_{\mathbf{mpc}}\mathbf{x}^{\mathbf{i}}_{\mathbf{mpc}}(k) + \mathbf{B}^{\mathbf{i}}_{\mathbf{u,mpc}}\boldsymbol{\Delta}\mathbf{u}^{\mathbf{i}}(k) + \mathbf{B}^{\mathbf{i}}_{\mathbf{a,mpc}}T_a(k) + \mathbf{B}^{\mathbf{i}}_{\mathbf{g,mpc}}T_g(k)$$
$$+\textstyle\sum_{j\neq i}\overline{\mathbf{B}}^{\mathbf{ij}}_{\mathbf{mpc}}\,\widehat{\mathbf{T}}^{\mathbf{j}}_{\mathbf{z}}(k).$$

$$\mathbf{x}^{\mathbf{i}}_{\mathbf{mpc}}(k+2|k) = \mathbf{A}^{\mathbf{i}}_{\mathbf{mpc}}\mathbf{x}^{\mathbf{i}}_{\mathbf{mpc}}(k+1|k) + \mathbf{B}^{\mathbf{i}}_{\mathbf{u,mpc}}\boldsymbol{\Delta}\mathbf{u}^{\mathbf{i}}(k+1|k) + \mathbf{B}^{\mathbf{i}}_{\mathbf{a,mpc}}T_a(k+1|k)$$
$$+\mathbf{B}^{\mathbf{i}}_{\mathbf{g,mpc}}T_g(k+1|k) + \textstyle\sum_{j\neq i}\overline{\mathbf{B}}^{\mathbf{ij}}_{\mathbf{mpc}}\,\widehat{\mathbf{T}}^{\mathbf{j}}_{\mathbf{z}}(k+1|k)$$
$$= \left(\mathbf{A}^{\mathbf{i}}_{\mathbf{mpc}}\right)^2\mathbf{x}^{\mathbf{i}}_{\mathbf{mpc}}(k) + \mathbf{A}^{\mathbf{i}}_{\mathbf{mpc}}\mathbf{B}^{\mathbf{i}}_{\mathbf{u,mpc}}\boldsymbol{\Delta}\mathbf{u}^{\mathbf{i}}(k) + \mathbf{B}^{\mathbf{i}}_{\mathbf{u,mpc}}\boldsymbol{\Delta}\mathbf{u}^{\mathbf{i}}(k+1|k)$$
$$+ \mathbf{A}^{\mathbf{i}}_{\mathbf{mpc}}\mathbf{B}^{\mathbf{i}}_{\mathbf{a,mpc}}T_a(k) + \mathbf{B}^{\mathbf{i}}_{\mathbf{a,mpc}}T_a(k+1|k) + \mathbf{A}^{\mathbf{i}}_{\mathbf{mpc}}\mathbf{B}^{\mathbf{i}}_{\mathbf{g,mpc}}T_g(k)$$
$$+ \mathbf{B}^{\mathbf{i}}_{\mathbf{g,mpc}}T_g(k+1|k) + \sum_{j\neq i}\left(\mathbf{A}^{\mathbf{i}}_{\mathbf{mpc}}\overline{\mathbf{B}}^{\mathbf{ij}}_{\mathbf{mpc}}\,\widehat{\mathbf{T}}^{\mathbf{j}}_{\mathbf{z}}(k) + \overline{\mathbf{B}}^{\mathbf{ij}}_{\mathbf{mpc}}\,\widehat{\mathbf{T}}^{\mathbf{j}}_{\mathbf{z}}(k+1|k)\right)$$

$$\vdots$$

$$\mathbf{x^i_{mpc}}(k+N_u|k) = \left(\mathbf{A^i_{mpc}}\right)^{N_u}\mathbf{x^i_{mpc}}(k) + \sum_{m=0}^{N_u-1}\left(\mathbf{A^i_{mpc}}\right)^{N_u-1-m}\Big(\mathbf{B^i_{u,mpc}}\,\mathbf{\Delta u^i}(k+m|k)$$

$$+\mathbf{B^i_{a,mpc}}T_a(k+m|k) + \mathbf{B^i_{g,mpc}}T_g(k+m|k)\Big) + \sum_{m=0}^{N_u-1}\sum_{j\neq i}\left(\mathbf{A^i_{mpc}}\right)^{N_u-1-m}\left(\overline{\mathbf{B}}^{\mathbf{ij}}_{\mathbf{mpc}}\,\widehat{\mathbf{T}}^{\mathbf{j}}_{\mathbf{z}}(k+m|k)\right)$$

$$\mathbf{x^i_{mpc}}(k+N_u+1|k) = \left(\mathbf{A^i_{mpc}}\right)^{N_u+1}\mathbf{x^i_{mpc}}(k) + \sum_{m=0}^{N_u-1}\left(\mathbf{A^i_{mpc}}\right)^{N_u-m}\mathbf{B^i_{u,mpc}}\mathbf{\Delta u^i}(k+m|k)$$

$$+\sum_{m=0}^{N_u}\left(\mathbf{A^i_{mpc}}\right)^{N_u-m}\left(\mathbf{B^i_{a,mpc}}T_a(k+m|k) + \mathbf{B^i_{g,mpc}}T_g(k+m|k)\right)$$

$$+\sum_{m=0}^{N_u}\sum_{j\neq i}\left(\mathbf{A^i_{mpc}}\right)^{N_u-m}\left(\overline{\mathbf{B}}^{\mathbf{ij}}_{\mathbf{mpc}}\,\widehat{\mathbf{T}}^{\mathbf{j}}_{\mathbf{z}}(k+m|k)\right)$$

$$\vdots$$

$$\mathbf{x^i_{mpc}}\left(k+N_p|k\right) = \left(\mathbf{A^i_{mpc}}\right)^{N_p}\mathbf{x^i_{mpc}}(k) + \sum_{m=0}^{N_u-1}\left(\mathbf{A^i_{mpc}}\right)^{N_p-1}\mathbf{B^i_{u,mpc}}\mathbf{\Delta u^i}(k+m|k)$$

$$+\sum_{m=0}^{N_p-1}\left(\mathbf{A^i_{mpc}}\right)^{N_p-1}\left(\mathbf{B^i_{a,mpc}}T_a(k+m|k) + \mathbf{B^i_{g,mpc}}T_g(k+m|k)\right)$$

$$+\sum_{m=0}^{N_p-1}\sum_{j\neq i}\left(\mathbf{A^i_{mpc}}\right)^{N_p-1}\left(\overline{\mathbf{B}}^{\mathbf{ij}}_{\mathbf{mpc}}\,\widehat{\mathbf{T}}^{\mathbf{j}}_{\mathbf{z}}(k+m|k)\right)$$

The above equations can be succinctly written as

$$\bar{\mathbf{x}}^{\mathbf{i}}_{\mathbf{mpc}} = \mathbf{M_i}\mathbf{x^i_{mpc}}(k) + \mathbf{S_{ui}}\overline{\mathbf{\Delta u^i}} + \mathbf{S_{ai}}\overline{\mathbf{T}}_{\mathbf{a}} + \mathbf{S_{gi}}\overline{\mathbf{T}}_{\mathbf{g}} + \textstyle\sum_{j\neq i}\mathbf{S_{ij}}\overline{\widehat{\mathbf{T}}}^{\mathbf{j}}_{\mathbf{z}}. \tag{5.55}$$

Here,

$$\bar{\mathbf{x}}^{\mathbf{i}}_{\mathbf{mpc}} = \begin{bmatrix}\mathbf{x^i_{mpc}}(k+1|k)\\ \mathbf{x^i_{mpc}}(k+2|k)\\ \vdots\\ \mathbf{x^i_{mpc}}(k+N_p|k)\end{bmatrix},\ \overline{\mathbf{\Delta u^i}} = \begin{bmatrix}\mathbf{\Delta u^i}(k|k)\\ \mathbf{\Delta u^i}(k+1|k)\\ \vdots\\ \mathbf{\Delta u^i}(k+N_u-1|k)\end{bmatrix},\ \overline{\mathbf{T}}_{\mathbf{a}} = \begin{bmatrix}T_a(k)\\ T_a(k+1|k)\\ \vdots\\ T_a(k+N_p-1|k)\end{bmatrix},$$

$$\overline{\mathbf{T}}_{\mathbf{g}} = \begin{bmatrix} T_g(k) \\ T_g(k+1|k) \\ \vdots \\ T_g(k+N_p-1|k) \end{bmatrix}, \overline{\widehat{\mathbf{T}}}_{\mathbf{z}}^{\mathbf{j}} = \begin{bmatrix} \widehat{\mathbf{T}}_{\mathbf{z}}^{\mathbf{j}}(k) \\ \widehat{\mathbf{T}}_{\mathbf{z}}^{\mathbf{j}}(k+1|k) \\ \vdots \\ \widehat{\mathbf{T}}_{\mathbf{z}}^{\mathbf{j}}(k+N_p-1|k) \end{bmatrix}, \mathbf{M}_{\mathbf{i}} = \begin{bmatrix} \mathbf{A}_{\mathbf{mpc}}^{\mathbf{i}} \\ \left(\mathbf{A}_{\mathbf{mpc}}^{\mathbf{i}}\right)^2 \\ \vdots \\ \left(\mathbf{A}_{\mathbf{mpc}}^{\mathbf{i}}\right)^{N_p} \end{bmatrix},$$

$$\mathbf{S}_{\mathbf{ui}} = \begin{bmatrix} \mathbf{B}_{\mathbf{u,mpc}}^{\mathbf{i}} & \mathbf{0} & \cdots & \mathbf{0} \\ \mathbf{A}_{\mathbf{mpc}}^{\mathbf{i}}\mathbf{B}_{\mathbf{u,mpc}}^{\mathbf{i}} & \ddots & \ddots & \vdots \\ \vdots & \ddots & \mathbf{B}_{\mathbf{u,mpc}}^{\mathbf{i}} & \mathbf{0} \\ \left(\mathbf{A}_{\mathbf{mpc}}^{\mathbf{i}}\right)^{N_u-1}\mathbf{B}_{\mathbf{u,mpc}}^{\mathbf{i}} & \cdots & \mathbf{A}_{\mathbf{mpc}}^{\mathbf{i}}\mathbf{B}_{\mathbf{u,mpc}}^{\mathbf{i}} & \mathbf{B}_{\mathbf{u,mpc}}^{\mathbf{i}} \\ \left(\mathbf{A}_{\mathbf{mpc}}^{\mathbf{i}}\right)^{N_u}\mathbf{B}_{\mathbf{u,mpc}}^{\mathbf{i}} & \cdots & \left(\mathbf{A}_{\mathbf{mpc}}^{\mathbf{i}}\right)^2\mathbf{B}_{\mathbf{u,mpc}}^{\mathbf{i}} & \mathbf{A}_{\mathbf{mpc}}^{\mathbf{i}}\mathbf{B}_{\mathbf{u,mpc}}^{\mathbf{i}} \\ \vdots & \ddots & \vdots & \vdots \\ \left(\mathbf{A}_{\mathbf{mpc}}^{\mathbf{i}}\right)^{N_p-1}\mathbf{B}_{\mathbf{u,mpc}}^{\mathbf{i}} & \cdots & \left(\mathbf{A}_{\mathbf{mpc}}^{\mathbf{i}}\right)^{N_p-N_u-1}\mathbf{B}_{\mathbf{u,mpc}}^{\mathbf{i}} & \left(\mathbf{A}_{\mathbf{mpc}}^{\mathbf{i}}\right)^{N_p-N_u}\mathbf{B}_{\mathbf{u,mpc}}^{\mathbf{i}} \end{bmatrix},$$

$$\mathbf{S}_{\mathbf{ai}} = \begin{bmatrix} \mathbf{B}_{\mathbf{a,mpc}}^{\mathbf{i}} & \mathbf{0} & \cdots & \mathbf{0} \\ \mathbf{A}_{\mathbf{mpc}}^{\mathbf{i}}\mathbf{B}_{\mathbf{a,mpc}}^{\mathbf{i}} & \ddots & \ddots & \vdots \\ \vdots & \ddots & \mathbf{B}_{\mathbf{a,mpc}}^{\mathbf{i}} & \mathbf{0} \\ \left(\mathbf{A}_{\mathbf{mpc}}^{\mathbf{i}}\right)^{N_p-1}\mathbf{B}_{\mathbf{a,mpc}}^{\mathbf{i}} & \cdots & \mathbf{A}_{\mathbf{i}}\mathbf{B}_{\mathbf{a,mpc}}^{\mathbf{i}} & \mathbf{B}_{\mathbf{a,mpc}}^{\mathbf{i}} \end{bmatrix},$$

$$\mathbf{S}_{\mathbf{gi}} = \begin{bmatrix} \mathbf{B}_{\mathbf{g,mpc}}^{\mathbf{i}} & \mathbf{0} & \cdots & \mathbf{0} \\ \mathbf{A}_{\mathbf{mpc}}^{\mathbf{i}}\mathbf{B}_{\mathbf{g,mpc}}^{\mathbf{i}} & \ddots & \ddots & \vdots \\ \vdots & \ddots & \mathbf{B}_{\mathbf{g,mpc}}^{\mathbf{i}} & \mathbf{0} \\ \left(\mathbf{A}_{\mathbf{mpc}}^{\mathbf{i}}\right)^{N_p-1}\mathbf{B}_{\mathbf{g,mpc}}^{\mathbf{i}} & \cdots & \mathbf{A}_{\mathbf{i}}\mathbf{B}_{\mathbf{g,mpc}}^{\mathbf{i}} & \mathbf{B}_{\mathbf{g,mpc}}^{\mathbf{i}} \end{bmatrix},$$

$$\mathbf{S}_{\mathbf{ij}} = \begin{bmatrix} \overline{\mathbf{B}}_{\mathbf{mpc}}^{\mathbf{ij}} & \mathbf{0} & \cdots & \mathbf{0} \\ \mathbf{A}_{\mathbf{mpc}}^{\mathbf{i}}\overline{\mathbf{B}}_{\mathbf{mpc}}^{\mathbf{ij}} & \ddots & \ddots & \vdots \\ \vdots & \ddots & \overline{\mathbf{B}}_{\mathbf{mpc}}^{\mathbf{ij}} & \mathbf{0} \\ \left(\mathbf{A}_{\mathbf{mpc}}^{\mathbf{i}}\right)^{N_p-1}\overline{\mathbf{B}}_{\mathbf{mpc}}^{\mathbf{ij}} & \cdots & \mathbf{A}_{\mathbf{i}}\overline{\mathbf{B}}_{\mathbf{mpc}}^{\mathbf{ij}} & \overline{\mathbf{B}}_{\mathbf{mpc}}^{\mathbf{ij}} \end{bmatrix}.$$

From (5.41), $J_i$ can be re-stated as

$$J_i = \left(\overline{\mathbf{T}}_\mathbf{z}^\mathbf{i} - \overline{\mathbf{T}}_\mathbf{z,ref}^\mathbf{i}\right)^T \mathbf{Q}_{\mathbf{1,i}}\left(\overline{\mathbf{T}}_\mathbf{z}^\mathbf{i} - \overline{\mathbf{T}}_\mathbf{z,ref}^\mathbf{i}\right) + \overline{\mathbf{u}}^{\mathbf{i}^T} \mathbf{Q}_{\mathbf{2,i}} \overline{\mathbf{u}}^\mathbf{i}$$

$$= \left(\begin{bmatrix} \overline{\mathbf{T}}_\mathbf{z}^\mathbf{i} \\ \overline{\mathbf{u}}^\mathbf{i} \end{bmatrix} - \begin{bmatrix} \overline{\mathbf{T}}_\mathbf{z,ref}^\mathbf{i} \\ \mathbf{0} \end{bmatrix}\right)^T \begin{bmatrix} \mathbf{Q}_{\mathbf{1,i}} & \mathbf{0} \\ \mathbf{0} & \mathbf{Q}_{\mathbf{2,i}} \end{bmatrix} \left(\begin{bmatrix} \overline{\mathbf{T}}_\mathbf{z}^\mathbf{i} \\ \overline{\mathbf{u}}^\mathbf{i} \end{bmatrix} - \begin{bmatrix} \overline{\mathbf{T}}_\mathbf{z,ref}^\mathbf{i} \\ \mathbf{0} \end{bmatrix}\right), \tag{5.56}$$

where,

$$\overline{\mathbf{T}}_\mathbf{z}^\mathbf{i} = \begin{bmatrix} \mathbf{T}_\mathbf{z}^\mathbf{i}(k+1|k) \\ \mathbf{T}_\mathbf{z}^\mathbf{i}(k+2|k) \\ \vdots \\ \mathbf{T}_\mathbf{z}^\mathbf{i}(k+N_p|k) \end{bmatrix}, \overline{\mathbf{T}}_\mathbf{zi,ref} = \begin{bmatrix} \mathbf{T}_\mathbf{z,ref}^\mathbf{i}(k) \\ \mathbf{T}_\mathbf{z,ref}^\mathbf{i}(k) \\ \vdots \\ \mathbf{T}_\mathbf{z,ref}^\mathbf{i}(k) \end{bmatrix}_{(N_p.N_{zi})\times 1},$$

$$\mathbf{Q}_{\mathbf{1},\mathrm{i}} = \begin{bmatrix} diag(\boldsymbol{\beta}_i) & \mathbf{0} & \dots & \mathbf{0} \\ \mathbf{0} & diag(\boldsymbol{\beta}_\mathbf{i}) & \dots & \mathbf{0} \\ \vdots & \vdots & \ddots & \vdots \\ \mathbf{0} & \mathbf{0} & \dots & diag(\boldsymbol{\beta}_i) \end{bmatrix}_{(N_{zi}.N_\mathrm{p})\times(N_{zi}.N_p)},$$

and $$\mathbf{Q}_{\mathbf{2,i}} = \begin{bmatrix} diag(\boldsymbol{\alpha}_i) & \mathbf{0} & \dots & \mathbf{0} \\ \mathbf{0} & diag(\boldsymbol{\alpha}_i) & \dots & \mathbf{0} \\ \vdots & \vdots & \ddots & \vdots \\ \mathbf{0} & \mathbf{0} & \dots & diag(\boldsymbol{\alpha}_\mathbf{i}) \end{bmatrix}_{(N_{zi}.N_u)\times(N_{zi}.N_\mathrm{u})}.$$

Let $$\mathbf{C}_\mathbf{mpc}^\mathbf{i} = \begin{bmatrix} \mathbf{0}_{N_{zi}\times N_x} & \mathbf{I}_{N_{zi}} & \mathbf{0}_{N_{zi}\times N_{zi}} & \mathbf{0}_{N_{zi}\times N_{zi}} \\ \mathbf{0}_{N_{zi}\times N_x} & \mathbf{0}_{N_{zi}\times N_{zi}} & \mathbf{0}_{N_{zi}\times N_{zi}} & \mathbf{I}_{N_{zi}} \end{bmatrix},$$

and $$\overline{\mathbf{C}}_\mathbf{mpc}^\mathbf{i} = \begin{bmatrix} \mathbf{C}_\mathbf{mpc}^\mathbf{i} & \mathbf{0} & \dots & \mathbf{0} \\ \mathbf{0} & \mathbf{C}_\mathbf{mpc}^\mathbf{i} & \dots & \mathbf{0} \\ \vdots & \vdots & \ddots & \vdots \\ \mathbf{0} & \mathbf{0} & \dots & \mathbf{C}_\mathbf{mpc}^\mathbf{i} \end{bmatrix}_{(2N_{zi}.N_p)\times((N_x+3N_{zi}).N_p)}.$$

Here, $N_x = n_{w-in}^i + n_{w-out}^i$.

Clearly, $$\begin{bmatrix} \overline{\mathbf{T}}_\mathbf{z}^\mathbf{i} \\ \overline{\mathbf{u}}^\mathbf{i} \end{bmatrix} = \overline{\mathbf{C}}_\mathbf{mpc}^\mathbf{i} \overline{\mathbf{x}}_\mathbf{mpc}^\mathbf{i}. \tag{5.57}$$

Using 5.55), (5.57) can be written as

$$\begin{bmatrix} \overline{\mathbf{T}}_\mathbf{z}^\mathbf{i} \\ \overline{\mathbf{u}}^\mathbf{i} \end{bmatrix} = \overline{\mathbf{C}}_\mathbf{mpc}^\mathbf{i}\left(\mathbf{M}_\mathbf{i}\mathbf{x}_\mathbf{mpc}^\mathbf{i}(k) + \mathbf{S}_\mathbf{ui}\overline{\boldsymbol{\Delta}\mathbf{u}^\mathbf{i}} + \mathbf{S}_\mathbf{ai}\overline{\mathbf{T}}_\mathbf{a} + \mathbf{S}_\mathbf{gi}\overline{\mathbf{T}}_\mathbf{g} + \textstyle\sum_{j\neq i}\mathbf{S}_\mathbf{ij}\widehat{\overline{\mathbf{T}}}_\mathbf{z}^\mathbf{j}\right). \tag{5.58}$$

We define the following:

$$\overline{\mathbf{T}}^{\mathbf{i}}_{\mathbf{ref}} = \begin{pmatrix} \overline{\mathbf{T}}^{\mathbf{i}}_{\mathbf{z,ref}} \\ \mathbf{0}_{(N_u.N_{zi})\times 1} \end{pmatrix}, \mathbf{Q_i} = \begin{bmatrix} \mathbf{Q_{1,i}} & \mathbf{0} \\ \mathbf{0} & \mathbf{Q_{2,i}} \end{bmatrix}. \tag{5.59}$$

Using (5.58) and (5.59) in (5.56), and ignoring terms which are independent of $\overline{\mathbf{\Delta u^i}}$ (the optimization variable), we obtain

$$J_i = (\overline{\mathbf{\Delta u^i}})^T \mathbf{H^i_{mpc}} \overline{\mathbf{\Delta u^i}} + \left(\mathbf{f^i_{mpc}}\right)^T \overline{\mathbf{\Delta u^i}}, \tag{5.60}$$

where,

$$\mathbf{H^i_{mpc}} = \mathbf{S}^T_{\mathbf{ui}} \left(\overline{\mathbf{C}}^{\mathbf{i}}_{\mathbf{mpc}}\right)^T \mathbf{Q_i} \overline{\mathbf{C}}^{\mathbf{i}}_{\mathbf{mpc}} \mathbf{S}^T_{\mathbf{ui}}, \tag{5.61}$$

and,

$$\mathbf{f_{dc,i}} = 2\mathbf{S}^T_{\mathbf{ui}} \left(\overline{\mathbf{C}}^{\mathbf{i}}_{\mathbf{mpc}}\right)^T \mathbf{Q_i} \left[\overline{\mathbf{C}}^{\mathbf{i}}_{\mathbf{mpc}} \left(\mathbf{M_i} \mathbf{x^i_{mpc}}(k) + \mathbf{S_{ui}} \overline{\mathbf{\Delta u^i}} + \mathbf{S_{ai}} \overline{\mathbf{T}}_{\mathbf{a}} + \mathbf{S_{gi}} \overline{\mathbf{T}}_{\mathbf{g}} + \textstyle\sum_{j\neq i} \mathbf{S_{ij}} \overline{\hat{\mathbf{T}}}^{\mathbf{j}}_{\mathbf{z}}\right) - \overline{\mathbf{T}}^{\mathbf{i}}_{\mathbf{ref}}\right]. \tag{5.62}$$

In the above framework, $\mathbf{x^i_{mpc}}(k)$ is constructed from the estimated and measured states. Therefore,

$$\mathbf{x^i_{mpc}}(k) = \begin{bmatrix} \hat{\boldsymbol{\eta}}^{\mathbf{i}}_{\mathbf{1}}(k) \\ \mathbf{T^i_{z,meas}}(k) \\ \hat{\boldsymbol{\eta}}^{\mathbf{i}}_{\mathbf{2}}(k) \end{bmatrix}. \tag{5.63}$$

## 5.4.2 Constraints

In this section, we consider constraints representing upper bounds on the heating and cooling provided to the zones by the HVAC system. These constraints translate into upper and lower bound inequality constraints on the control inputs. We assume that the HVAC system is a VAV system with reheating coils described in Chapter 2. For simplicity, we assume 100% recirculation of return air from each zone, i.e. there is no mixing of outside air in the air handling unit. Furthermore, we assume that the temperature of cold supply air is fixed; therefore the heating or cooling provided to the zones is manipulated only by changing the air mass flow rates via dampers in the VAV boxes and/or by varying the amount of heating provided by reheat coils (in case of heating only). In the remainder of this section, '≤' and '≥' when used to compare vectors, refer to inequalities taken component-wise.

#### 5.4.2.1 Cooling constraints

For each zone $n \in \{1,2,\dots,N_{zi}\}$ in the $i^{th}$ cluster, the cooling constraint due to the above assumptions is given by (refer to nomenclature):

$$u_n^i(k) \geq \dot{m}_{max,n}^i c_{pa}\left(T_{supp} - T_{z,n}^i(k)\right) \tag{5.64}$$

The combined form of (5.64) for all zones in the cluster is given by

$$\mathbf{u}^{\mathrm{i}}(k) \geq \mathbf{M}_{\mathbf{max}}^{\mathbf{i}} c_{pa}\left(\mathbf{T}_{\mathbf{supp}} - \mathbf{T}_{\mathbf{z}}^{\mathbf{i}}(k)\right) \tag{5.65}$$

Here,

$$\mathbf{M}_{\mathbf{max}}^{\mathbf{i}} = \begin{bmatrix} \dot{m}_{max,1}^i & 0 & \dots & 0 \\ 0 & \dot{m}_{max,2}^i & \dots & 0 \\ \vdots & \vdots & \ddots & \vdots \\ 0 & 0 & \dots & \dot{m}_{max,N_{zi}}^i \end{bmatrix}_{N_{zi} \times N_{zi}},$$

$$\mathbf{T}_{\mathbf{supp}} = \begin{bmatrix} T_{supp} \\ T_{supp} \\ \vdots \\ T_{supp} \end{bmatrix}_{N_{zi} \times 1}.$$

Equation (5.65) results in the following constraints over the control and prediction horizons.

1. For $l \in \{1,2,\dots,N_u - 1\}$,

$$\mathbf{u}^{\mathbf{i}}(k-1) + \mathbf{\Delta u}^{\mathbf{i}}(k) + \mathbf{\Delta u}^{\mathbf{i}}(k+1|k) + \dots + \mathbf{\Delta u}^{\mathbf{i}}(k+l|k) \geq \mathbf{M}_{\mathbf{max}}^{\mathbf{i}} c_{pa}\left(\mathbf{T}_{\mathbf{supp}} - \mathbf{T}_{\mathbf{z}}^{\mathbf{i}}(k+l|k)\right). \tag{5.66}$$

2. For $l \in \{N_u, N_u + 1, \dots, N_p\}$,

$$\mathbf{u}^{\mathbf{i}}(k-1) + \mathbf{\Delta u}^{\mathbf{i}}(k) + \mathbf{\Delta u}^{\mathbf{i}}(k+1|k) + \dots + \mathbf{\Delta u}^{\mathbf{i}}(k+N_u-1|k) \geq \mathbf{M}_{\mathbf{max}}^{\mathbf{i}} c_{pa}\left(\mathbf{T}_{\mathbf{supp}} - \mathbf{T}_{\mathbf{z}}^{\mathbf{i}}(k+l|k)\right). \tag{5.67}$$

Equations (5.66) and (5.67) can be succinctly written as

$$\mathbf{A}_{\mathbf{1}}^{\mathbf{i}} \overline{\mathbf{\Delta u}}^{\mathbf{i}} \geq c_{pa} \overline{\mathbf{M}}_{\mathbf{max}}^{\mathbf{i}} \left(\mathbf{A}_{\mathbf{2}}^{\mathbf{i}} \mathbf{T}_{\mathbf{supp}} - \begin{bmatrix} \mathbf{T}_{\mathbf{z}}^{\mathbf{i}}(k) \\ \overline{\mathbf{T}}_{\mathbf{z}}^{\mathbf{i}} \end{bmatrix}\right) - \mathbf{A}_{\mathbf{2}}^{\mathbf{i}} \mathbf{u}^{\mathbf{i}}(k-1), \tag{5.68}$$

where,

$$\mathbf{A_1^i} = \begin{bmatrix} \mathbf{A_{1,1}^i} \\ \mathbf{A_{1,2}^i} \end{bmatrix}, \mathbf{A_{1,1}^i} = \begin{bmatrix} \mathbf{I}_{N_{zi}} & \mathbf{0} & \cdots & \mathbf{0} \\ \mathbf{I}_{N_{zi}} & \mathbf{I}_{N_{zi}} & \cdots & \mathbf{0} \\ \vdots & \vdots & \ddots & \vdots \\ \mathbf{I}_{N_{zi}} & \mathbf{I}_{N_{zi}} & \cdots & \mathbf{I}_{N_{zi}} \end{bmatrix}_{(N_{zi}.N_u)\times(N_{zi}.N_u)},$$

$$\mathbf{A_{1,2}^i} = \begin{bmatrix} \mathbf{I}_{N_{zi}} & \mathbf{I}_{N_{zi}} & \cdots & \mathbf{I}_{N_{zi}} \\ \mathbf{I}_{N_{zi}} & \mathbf{I}_{N_{zi}} & \cdots & \mathbf{I}_{N_{zi}} \\ \vdots & \vdots & \ddots & \vdots \\ \mathbf{I}_{N_{zi}} & \mathbf{I}_{N_{zi}} & \cdots & \mathbf{I}_{N_{zi}} \end{bmatrix}_{\left(N_{zi}.(N_p-N_u+1)\right)\times(N_{zi}.N_u)}, \mathbf{A_2^i} = \begin{bmatrix} \mathbf{I}_{N_{zi}} \\ \mathbf{I}_{N_{zi}} \\ \vdots \\ \mathbf{I}_{N_{zi}} \end{bmatrix}_{\left(N_{zi}.(N_p+1)\right)\times N_{zi}},$$

$$\mathbf{\bar{M}_{max}^i} = \begin{bmatrix} \mathbf{M_{max}^i} & \mathbf{0} & \cdots & \mathbf{0} \\ \mathbf{0} & \mathbf{M_{max}^i} & \cdots & \mathbf{0} \\ \vdots & \vdots & \ddots & \vdots \\ \mathbf{0} & \mathbf{0} & \cdots & \mathbf{M_{max}^i} \end{bmatrix}_{\left(N_{zi}.(N_p+1)\right)\times\left(N_{zi}.(N_p+1)\right)}.$$

$\mathbf{\bar{T}_z^i}$ and $\overline{\mathbf{\Delta u^i}}$ were defined earlier as

$$\mathbf{\bar{T}_z^i} = \begin{bmatrix} \mathbf{T_z^i}(k+1|k) \\ \mathbf{T_z^i}(k+2|k) \\ \vdots \\ \mathbf{T_z^i}(k+N_p|k) \end{bmatrix}, \overline{\mathbf{\Delta u^i}} = \begin{bmatrix} \mathbf{\Delta u^i}(k|k) \\ \mathbf{\Delta u^i}(k+1|k) \\ \vdots \\ \mathbf{\Delta u^i}(k+N_u-1|k) \end{bmatrix}.$$

Let $\mathbf{C_{T_z}^i}$ and $\mathbf{\bar{C}_{T_z}^i}$ be defined as

$$\mathbf{C_{T_z}^i} = [\mathbf{0}_{N_{zi}\times N_x} \quad \mathbf{I}_{N_{zi}} \quad \mathbf{0}_{N_{zi}\times N_{zi}} \quad \mathbf{0}_{N_{zi}\times N_{zi}}],$$

$$\mathbf{\bar{C}_{T_z}^i} = \begin{bmatrix} \mathbf{C_{T_z}^i} & \mathbf{0} & \cdots & \mathbf{0} \\ \mathbf{0} & \mathbf{C_{T_z}^i} & \cdots & \mathbf{0} \\ \vdots & \vdots & \ddots & \vdots \\ \mathbf{0} & \mathbf{0} & \cdots & \mathbf{C_{T_z}^i} \end{bmatrix}_{(N_{zi}.N_p)\times((N_x+3N_{zi}).N_p)}.$$

Clearly,

$$\mathbf{\bar{T}_z^i} = \mathbf{\bar{C}_{T_z}^i}\mathbf{\bar{x}_{mpc}^i}. \tag{5.69}$$

Using (5.55), (5.69) can be written as

$$\bar{\mathbf{T}}_{\mathbf{z}}^{\mathbf{i}} = \bar{\mathbf{C}}_{\mathbf{T_z}}^{\mathbf{i}}\Big(\mathbf{M_i}\mathbf{x}_{\mathbf{mpc}}^{\mathbf{i}}(k) + \mathbf{S_{ui}}\overline{\boldsymbol{\Delta}\mathbf{u}^{\mathbf{i}}} + \mathbf{S_{ai}}\bar{\mathbf{T}}_{\mathbf{a}} + \mathbf{S_{gi}}\bar{\mathbf{T}}_{\mathbf{g}} + \textstyle\sum_{j\neq i}\mathbf{S_{ij}}\hat{\bar{\mathbf{T}}}_{\mathbf{z}}^{\mathbf{j}}\Big). \tag{5.70}$$

Substituting for $\bar{\mathbf{T}}_{\mathbf{z}}^{\mathbf{i}}$ from (5.70) in (5.68) we obtain:

$$\mathbf{A_1^i}\overline{\boldsymbol{\Delta}\mathbf{u}^{\mathbf{i}}} \geq c_{pa}\bar{\mathbf{M}}_{\mathbf{max}}^{\mathbf{i}}\left(\mathbf{A_2^i}\mathbf{T_{supp}} - \mathbf{A_3^i}\mathbf{x}_{\mathbf{mpc}}^{\mathbf{i}}(k) - \mathbf{A_4^i}\overline{\boldsymbol{\Delta}\mathbf{u}^{\mathbf{i}}} - \mathbf{A_5^i}\bar{\mathbf{T}}_{\mathbf{a}} - \mathbf{A_6^i}\bar{\mathbf{T}}_{\mathbf{g}} - \sum_{j\neq i}\mathbf{A_7^{ij}}\hat{\bar{\mathbf{T}}}_{\mathbf{z}}^{\mathbf{j}}\right)$$

$$-\mathbf{A_2^i}\mathbf{u^i}(k-1), \tag{5.71}$$

where,

$$\mathbf{A_3^i} = \begin{bmatrix}\mathbf{0}_{N_{zi}\times(N_x+3N_{zi})}\\ \bar{\mathbf{C}}_{\mathbf{T_z}}^{\mathbf{i}}\mathbf{M_i}\end{bmatrix} + \begin{bmatrix}\mathbf{C}_{\mathbf{T_z}}^{\mathbf{i}}\\ \mathbf{0}_{(N_p.N_{zi})\times(N_x+3N_{zi})}\end{bmatrix}, \mathbf{A_4^i} = \begin{bmatrix}\mathbf{0}_{N_{zi}\times(N_u.N_{zi})}\\ \bar{\mathbf{C}}_{\mathbf{T_z}}^{\mathbf{i}}\mathbf{S_{ui}}\end{bmatrix}, \mathbf{A_5^i} = \begin{bmatrix}\mathbf{0}_{N_{zi}\times N_p}\\ \bar{\mathbf{C}}_{\mathbf{T_z}}^{\mathbf{i}}\mathbf{S_{ai}}\end{bmatrix},$$

$$\mathbf{A_6^i} = \begin{bmatrix}\mathbf{0}_{N_{zi}\times N_p}\\ \bar{\mathbf{C}}_{\mathbf{T_z}}^{\mathbf{i}}\mathbf{S_{gi}}\end{bmatrix}, \mathbf{A_7^{ij}} = \begin{bmatrix}\mathbf{0}_{N_{zi}\times(N_u.N_{zj})}\\ \bar{\mathbf{C}}_{\mathbf{T_z}}^{\mathbf{i}}\mathbf{S_{ij}}\end{bmatrix}.$$

Note that in deriving (5.71), we used the fact that $\mathbf{T_z^i}(k) = \mathbf{C}_{\mathbf{T_z}}^{\mathbf{i}}\mathbf{x}_{\mathbf{mpc}}^{\mathbf{i}}(k)$. The inequality (5.71) can be re-stated as:

$$\left(\mathbf{A_1^i} + c_{pa}\bar{\mathbf{M}}_{\mathbf{max}}^{\mathbf{i}}\mathbf{A_4^i}\right)\overline{\boldsymbol{\Delta}\mathbf{u}^{\mathbf{i}}} \geq c_{pa}\bar{\mathbf{M}}_{\mathbf{max}}^{\mathbf{i}}\left(\mathbf{A_2^i}\mathbf{T_{supp}} - \mathbf{A_3^i}\mathbf{x}_{\mathbf{mpc}}^{\mathbf{i}}(k) - \mathbf{A_5^i}\bar{\mathbf{T}}_{\mathbf{a}} - \mathbf{A_6^i}\bar{\mathbf{T}}_{\mathbf{g}} - \sum_{j\neq i}\mathbf{A_7^{ij}}\hat{\bar{\mathbf{T}}}_{\mathbf{z}}^{\mathbf{j}}\right)$$

$$-\mathbf{A_2^i}\mathbf{u^i}(k-1). \tag{5.72}$$

#### 5.4.2.2 Heating constraints

For each zone $n \in \{1,2,\dots,N_{zi}\}$ in the $i^{th}$ cluster, the heating constraint due to the assumptions on the HVAC system mentioned earlier is given by (refer to nomenclature)

$$u_n^i(k) \leq Q_{RH-max,n}^i. \tag{5.73}$$

The combined form of (5.73) for all zones in the cluster is given by

$$\mathbf{u^i}(k) \leq \mathbf{Q_{RH-max}^i}. \tag{5.74}$$

Equation (5.74) results in the following constraints for $l \in \{1,2,\dots,N_u-1\}$ (the control horizon)

$$\mathbf{u^i}(k-1) + \mathbf{\Delta u^i}(k) + \mathbf{\Delta u^i}(k+1|k) + \cdots + \mathbf{\Delta u^i}(k+l|k) \leq \mathbf{Q^i_{RH-max}}. \quad (5.75)$$

Equation (5.75) can be succinctly written as

$$\mathbf{A^i_{1,1}}\overline{\mathbf{\Delta u^i}} \leq \overline{\mathbf{Q}}^\mathbf{i}_\mathbf{RH-max} - \mathbf{A^i_7 u^i}(k-1), \quad (5.76)$$

where, $\mathbf{A^i_{1,1}}$ was defined earlier as

$$\mathbf{A^i_{1,1}} = \begin{bmatrix} \mathbf{I}_{N_{zi}} & \mathbf{0} & \cdots & \mathbf{0} \\ \mathbf{I}_{N_{zi}} & \mathbf{I}_{N_{zi}} & \cdots & \mathbf{0} \\ \vdots & \vdots & \ddots & \vdots \\ \mathbf{I}_{N_{zi}} & \mathbf{I}_{N_{zi}} & \cdots & \mathbf{I}_{N_{zi}} \end{bmatrix}_{(N_{zi}.N_u)\times(N_{zi}.N_u)},$$

and,

$$\overline{\mathbf{Q}}^\mathbf{i}_\mathbf{RH-max} = \begin{bmatrix} \mathbf{Q^i_{RH-max}} \\ \mathbf{Q^i_{RH-max}} \\ \vdots \\ \mathbf{Q^i_{RH-max}} \end{bmatrix}_{(N_{zi}.N_u)\times 1}, \mathbf{A^i_7} = \begin{bmatrix} \mathbf{I}_{N_{zi}} \\ \mathbf{I}_{N_{zi}} \\ \vdots \\ \mathbf{I}_{N_{zi}} \end{bmatrix}_{(N_{zi}.N_u)\times N_{zi}}.$$

### 5.4.3 Quadratic program formulation

Using the results in Sections 5.4.1 and 5.4.2, the optimization problem corresponding to the decentralized controller for the $i^{th}$ cluster (see section 5.3.2) is expressed as the following quadratic program (QP)

$$\overline{\mathbf{\Delta u}_*}^\mathbf{i} = \arg\min_{\overline{\mathbf{\Delta u^i}}} \; g_i(\overline{\mathbf{\Delta u^i}}). \quad (5.77)$$

where, $$g_i(\overline{\mathbf{\Delta u^i}}) = (\overline{\mathbf{\Delta u^i}})^T \mathbf{H^i_{mpc}} \overline{\mathbf{\Delta u^i}} + \left(\mathbf{f^i_{mpc}}\right)^T \overline{\mathbf{\Delta u^i}} \quad (5.78)$$

subject to :

$$\left(\mathbf{A^i_1} + c_{pa}\overline{\mathbf{M}}^\mathbf{i}_\mathbf{max}\mathbf{A^i_4}\right)\overline{\mathbf{\Delta u^i}} \geq c_{pa}\overline{\mathbf{M}}^\mathbf{i}_\mathbf{max}\left(\mathbf{A^i_2 T_{supp}} - \mathbf{A^i_3 x^i_{mpc}}(k) - \mathbf{A^i_5}\overline{\mathbf{T}}_\mathbf{a} - \mathbf{A^i_6}\overline{\mathbf{T}}_\mathbf{g} - \sum_{j\neq i}\mathbf{A^{ij}_7}\widehat{\overline{\mathbf{T}}}^\mathbf{j}_\mathbf{z}\right)$$

$$-\mathbf{A^i_2 u^i}(k-1), \quad (5.79)$$

$$\mathbf{A^i_{1,1}}\overline{\mathbf{\Delta u^i}} \leq \overline{\mathbf{Q}}^\mathbf{i}_\mathbf{RH-max} - \mathbf{A^i_7 u^i}(k-1). \quad (5.80)$$

The above quadratic program can be solved at each time step $k$ using an appropriate tool such as the optimization toolbox in MATLAB [92]. The optimal control input, $\mathbf{u_*}^\mathbf{i}(k)$ at this time instant, is given by

$$\mathbf{u_*}^{\mathbf{i}}(k) \;= \mathbf{A_8^i}\overline{\mathbf{\Delta u}_*}^{\mathbf{i}} + \mathbf{u^i}(k-1), \quad (5.81)$$

where,

$$\mathbf{A_8^i} = \begin{bmatrix} \mathbf{I}_{N_{zi}} & \mathbf{0}_{N_{zi}\times((N_u-1).N_{zi})} \end{bmatrix}. \quad (5.82)$$

The control input $\mathbf{u_*}^{\mathbf{i}}(k)$ is applied to the $i^{th}$ cluster in the partition at the $k^{th}$ time step, and this procedure is then repeated for all clusters and all time steps.

# 5.5 Concluding remarks

A framework was presented in this chapter for the design of cluster level decentralized controllers corresponding to any partition of a building. The important elements of this framework were (i) a reduced order representation of the thermal dynamics associated with a cluster, (ii) a cluster level observer for estimation of unknown states and disturbances, and (iii) a cluster level controller based on an MPC framework which utilizes the reduced order model and the estimates provided by the observer.

Chapters 4 and 5 constitute a two-step decentralized control design process for the thermal control of a building. It consists of deciding a control architecture based on the tools developed in chapter 4, followed by design of decentralized controllers using the framework developed in this chapter. To demonstrate the use of these tools, the next chapter presents a real world building example, where this two-step procedure – involving control architecture selection and control design – was applied in simulation.

# Chapter 6
# Real World Building Simulation Study

## 6.1 Introduction

This chapter presents simulation studies using a real world building model to demonstrate the applicability of the tools developed in the previous chapters. A detailed model of the building is first developed in EnergyPlus [27] – a state-of-the-art modeling toolbox for buildings developed by the US Department of Energy (DOE). A linear representation of the thermal dynamics for the building is then obtained, which is used to partition it for decentralized control in accordance with the OLF-FPM method presented in Chapter 4. This is followed by the design of decentralized controllers using the methodology described in Chapter 5. Lastly, the optimality and robustness aspects associated with the decentralized controllers are investigated to conclude the chapter. The nomenclature used in this chapter is shown in Table 6.1

**Table 6.1 Nomenclature of symbols used in Chapter 6**

| **Symbol** | **Description** |
|---|---|
| $N_z$ | Number of zones in a building |
| $T_{z,j}$ | Temperature of zone $j$ |
| $Q_{int,j}$ | Internal thermal load generated in zone $j$ |
| $c_{p,air}$ | Specific heat capacity of air |
| $\dot{m}_{HVAC,j}$ | Mass flow rate of air supplied to zone $j$ by the HVAC system |
| $c_{zj}$ | Thermal capacitance of zone $j$ |
| $T_s$ | Sample time used for EnergyPlus simulations<br>(typically 1 minute in this work) |
| $\dot{m}_{inf,j}$ | Mass flow rate of infiltrated air from ambient to zone $j$ |
| $T_a$ | Ambient temperature |
| $T_{w,i}$ | Temperature of wall surface $i$ |
| $\mathrm{E}+$ | Superscript used to denote data obtained from EnergyPlus simulations |
| $*$ | Superscript used to denote optimal values or results obtained from an optimization |
| $\mathcal{N}_j$ | Set of wall surfaces which enclose zone $j$ |
| $h_{jl}$ | Heat transfer coefficient between zone $j$ and wall surface $l$ |
| $h_j$ | Heat transfer coefficient between zone $j$ and any enclosing wall surface |
| $A_i$ | Area of wall surface $i$ |
| $N$ | Length of time window (in samples) used in optimization |
| $d_{w-in,i}$ | Unmodeled thermal load acting on inside surface of wall $i$ |
| $T_{w-in,i}$ | Temperature of  inside surface of wall $i$ |
| $T_{int,i}$ | Temperature of interface (zone/ambient/ground) which thermally interacts with inside surface of wall $i$ |
| $T_{w-out,i}$ | Temperature of outside surface of wall $i$ |
| $d_{w-out,i}$ | Unmodeled thermal load acting on outside surface of wall $i$ |
| $T_{ext,i}$ | Temperature of interface (zone/ambient/ground) which thermally interacts with outside surface of wall $i$ |
| $N_w$ | Number of wall surfaces in building |
| $\mathbf{d_w}$ | Vector of unmodeled thermal loads acting on wall surfaces |
| $\mathbf{d_w}$ | Vector of unmodeled thermal loads acting on zones |
| $d_{w,i}$ | Thermal load acting on wall surface $i$ |
| $\sigma$ | Stefan-Boltzmann constant |
| $\alpha_i$ | Solar absorptance of wall surface $i$ |
| $q_{swr,i}$ | Incident solar radiation per unit area on wall surface $i$ |
| $\epsilon_i$ | Thermal absorptance of wall surface $i$ |
| $T_{gnd}$ | Temperature of ground outside building used for radiation calculations |
| $T_{sky}$ | Sky temperature used for radiation calculations |
| $d_{z,i}$ | Thermal load acting on zone $i$ |
| $\eta_{occ,i}$ | Actual occupancy in $i^{th}$ zone as a fraction of nominal occupancy |
| $N_{occ,i}$ | Nominal occupancy in $i^{th}$ zone |
| $Q_{occ}$ | Average rate of heat transfer from an occupant |
| $\eta_{light,i}$ | Actual lighting thermal load in $i^{th}$ zone as a fraction of nominal lighting thermal load |
| $Q_{light,i}$ | Nominal lighting thermal load in $i^{th}$ zone |
| $\eta_{appl,i}$ | Actual equipment thermal load in $i^{th}$ zone as a fraction of nominal equipment thermal load |
| $Q_{appl,i}$ | Nominal appliance thermal load in $i^{th}$ zone |
| $T_{z,i,\mathrm{ref}}$ | Set-point temperature for $i^{th}$ zone |

## 6.2 Test building and EnergyPlus model

The test building considered in this chapter is a multi-zone building, modeled based on the layout of the Siemens Corporate Research (SCR) building located in Princeton, New Jersey (NJ). A photograph of the actual building is shown in Figure 6.1. The various features of the building, which are used in the development of an EnergyPlus model, are explained this section.

### 6.2.1 Building Layout and Geometry

A Google SketchUp [94] illustration of the building is shown in Figure 6.2, whose plan view is shown in Figure 6.3. It consists of 5 blocks marked C, D, E, F and G (Figure 6.4) and has three floors including a basement (Figure 6.5). The building has 9 thermal zones, details of which are provided in Table 6.2. Note that the basement of the building has a section which is not thermally conditioned, indicated by NTCB in Table 6.2and shown in Figure 6.6. Each thermal zone is catered by its own air handling unit, except the zone labeled NTCB. For a background on thermal zones and air handling units in a building, the reader is directed to Chapter 2. As shown in Figure 6.2, the building also has external windows which are included in the development of an EnergyPlus model.

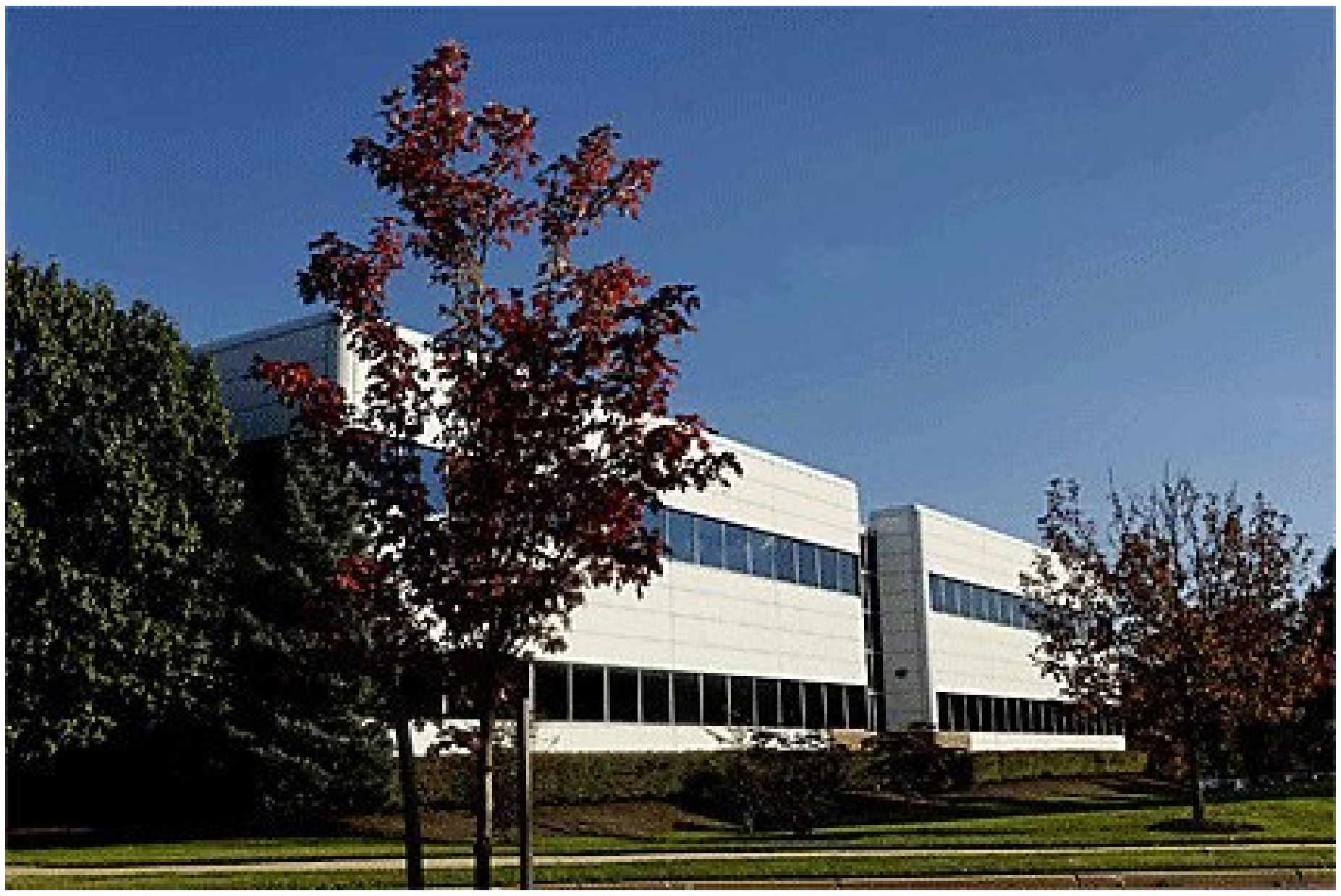

**Figure 6.1 Photograph of the SCR building (source: [93])**

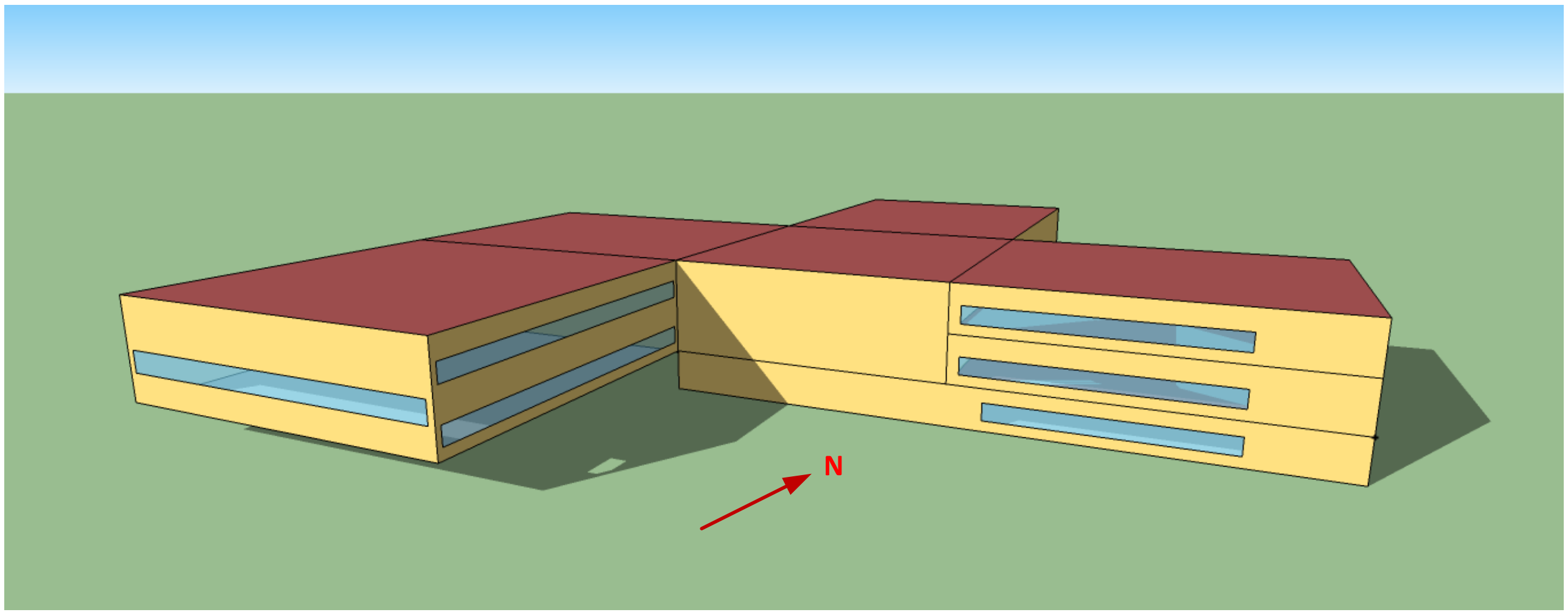

**Figure 6.2 Google SketchUp illustration of the test office building**

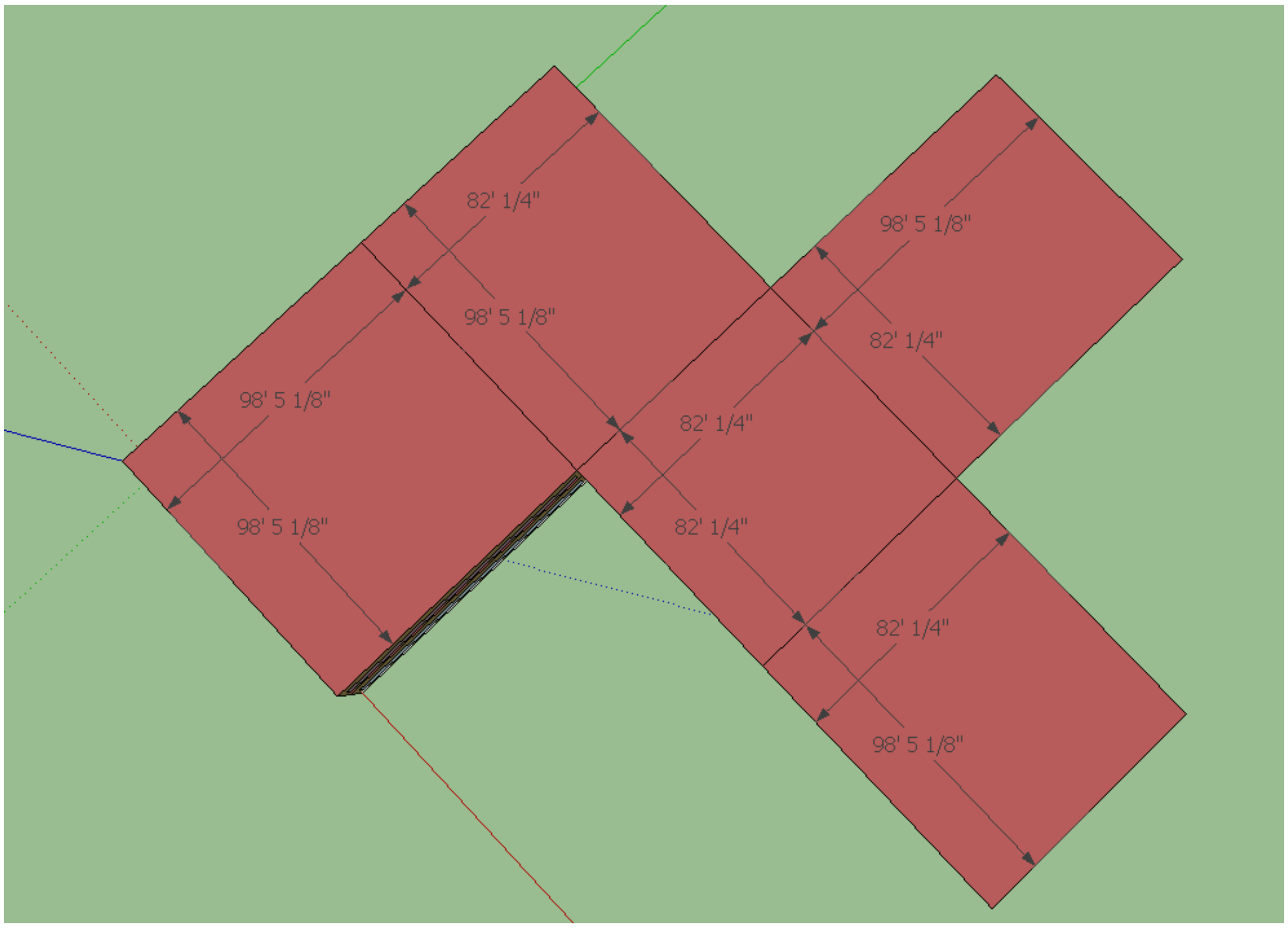

**Figure 6.3 Top (plan) view of test office building**

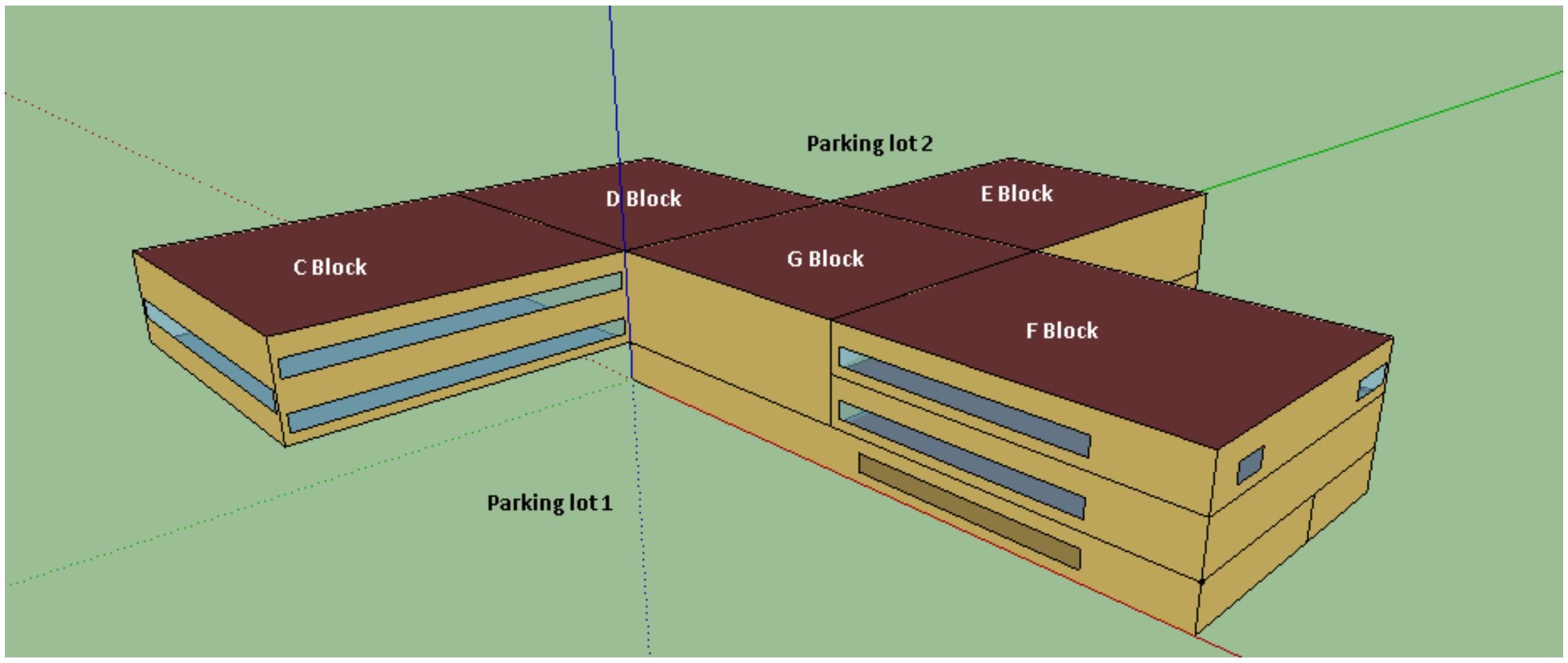

**Figure 6.4 Isometric view of the test office building**

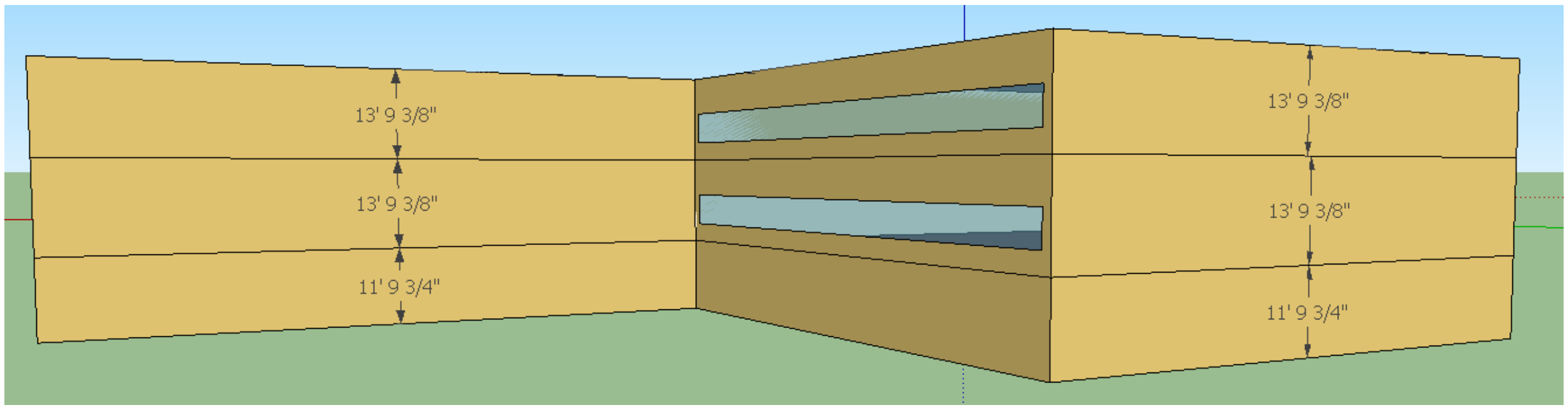

**Figure 6.5 Side view of test office building viewed from the North-East direction**

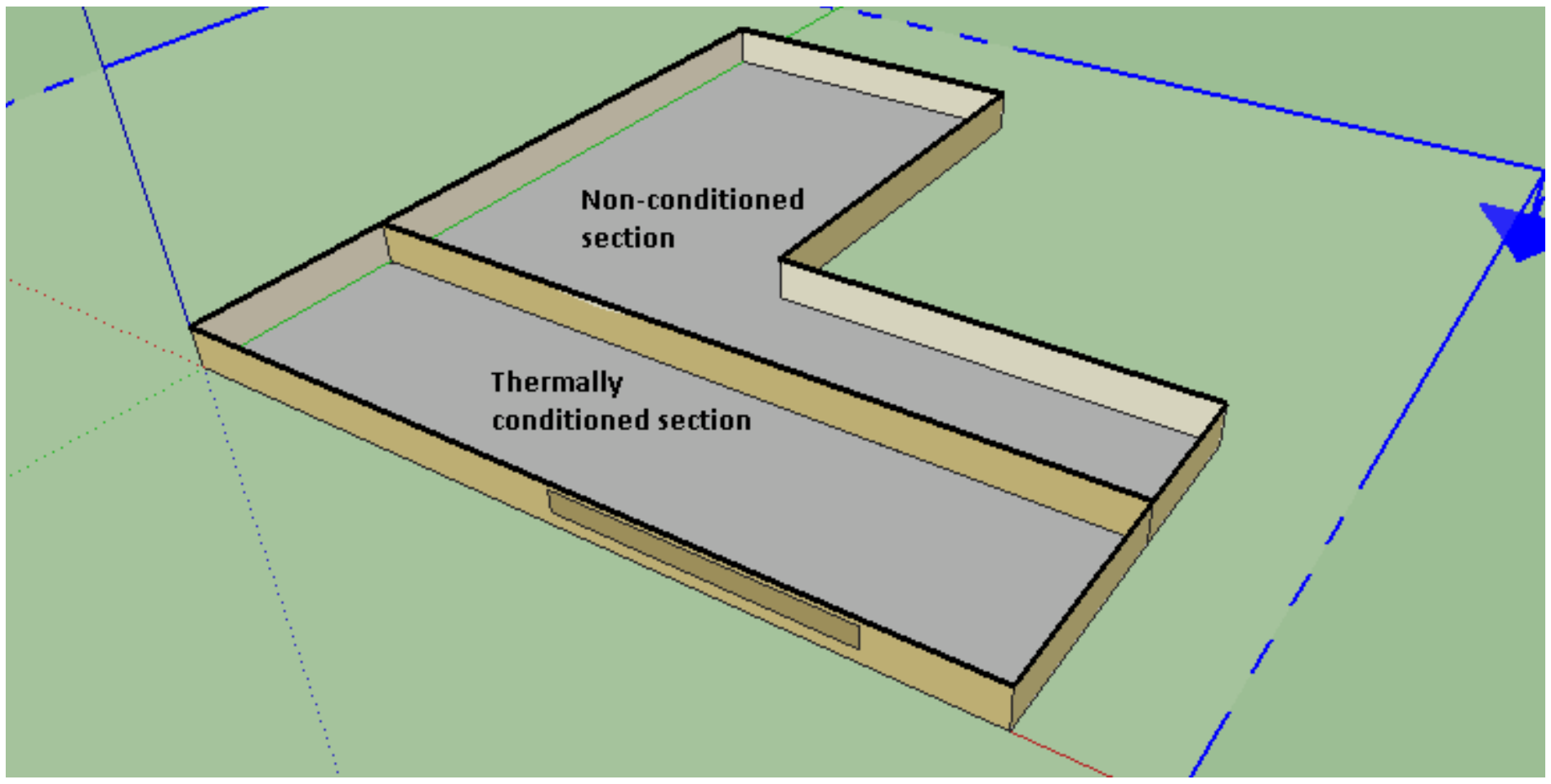

**Figure 6.6 Transverse cut of basement showing conditioned and non-conditioned sections of test office building**

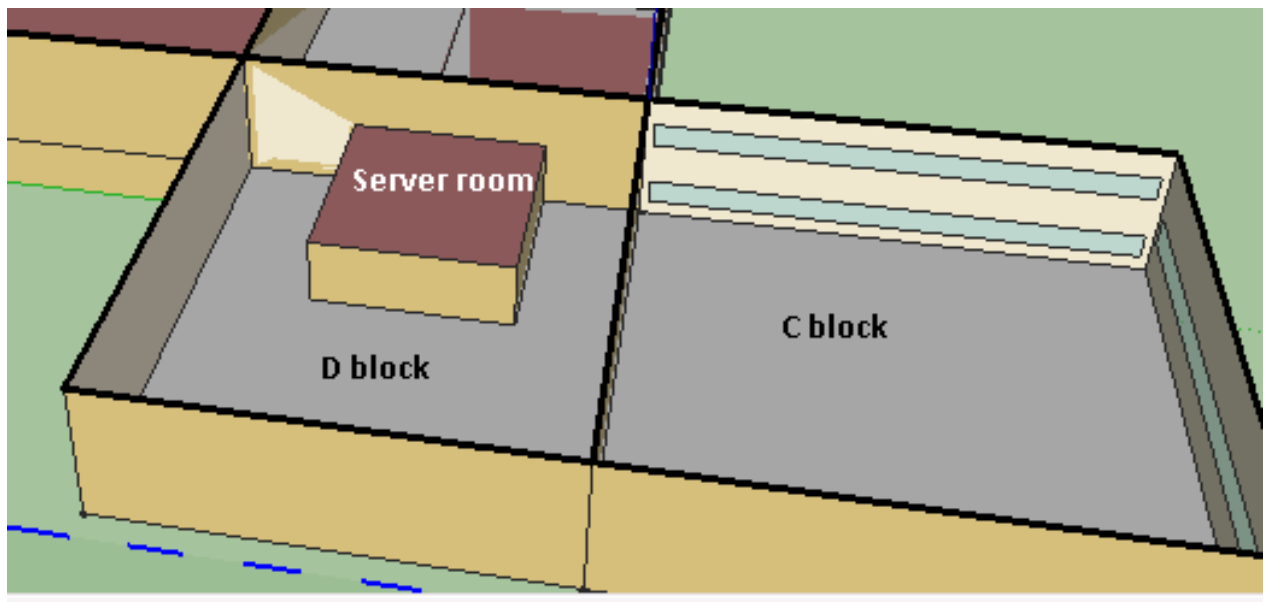

**Figure 6.7 Transverse cut of D block showing the server room of test office building**

**Table 6.2 Description of thermal zones in the test office building**

| Thermal zone number | Description | Alias |
|---|---|---|
| 1 | Entire G block (includes both 1st and 2nd floor sections) | G |
| 2 | Entire E block (includes both 1st and 2nd floor sections) | E |
| 3 | Entire C block (includes both 1st and 2nd floor sections) | C |
| 4 | Server room (Figure 6.7) located on 1st floor of D block | SR |
| 5 | 1st floor section of F block | F1 |
| 6 | Entire D block (includes both 1st and 2nd floor sections but excludes the server room) | D |
| 7 | 2nd floor section of F block | F2 |
| 8 | Thermally controlled section of the basement (Figure 6.6) | TCB |
| 9 | Non-thermally controlled section of the basement (Figure 6.6) | NTCB |

### 6.2.2 Construction data

The construction data for the building was based on a standard 'medium office building' construction template for ASHRAE climate zone 5 (NJ lies in that zone) provided in the OpenStudio tool [95] developed by the National Renewable Energy Laboratory (NREL). This sets appropriate properties for various construction elements in the building such as exterior and internal walls, floors and ceilings, and doors and glass windows. A description of the material layers associated with the various construction elements in the 'medium office building' construction template is shown in Table 6.3.

**Table 6.3 Details of construction template used for the test office building**

| Construction element in 'medium office building' template | Applicability | Material layers with thickness indicated in parentheses |
|---|---|---|
| ASHRAE_189.1-2009_ExtRoof_IEAD_ClimateZone 2-5 | Ceilings facing ambient | Roof membrane (0.0095 m), roof insulation (0.2105 m), metal decking (0.0015 m) |
| 000_Interior Wall | Interior walls | Gypsum board (0.019 m), air gap (0.15 m), gypsum board (0.019 m) |
| 000_Interior Door | Interior doors | Wood (0.0254 m) |
| ASHRAE_189.1-2009_ExtWall_Mass_ClimateZone 5 | Walls facing ambient | Stucco (0.0253 m), heavy-weight concrete (0.2033 m), wall insulation (0.0794 m), gypsum (0.0127 m) |
| 000_Interior Floor | Interior floors/ceilings | Acoustic tile (0.0191 m), air gap (0.18 m), light-weight concrete (0.1016 m) |
| 000_ExtSlabCarpet_4_in_ClimateZone 1-8 | Floors facing ground | Heavy-weight concrete (0.1016 m), carpet pad (0.2165 m) |
| ASHRAE_189.1-2009_ExtWindow_ClimateZone 4-5 | Windows facing ambient | Theoretical glass (0.003 m) |

### 6.2.3 Weather data

Weather information is required as an input for simulating EnergyPlus models. The weather information is provided to EnergyPlus through a weather file which contains information such as dry bulb temperature, wet bulb temperature, relative humidity, wind speed, etc. for any location. A library of weather files for various locations around the world is provided by the US DOE [96]. For the test building under consideration, a weather file specific to Princeton, NJ where the building is located was not available. Therefore, the weather file for the nearest location for which data is available (Trenton, NJ approximately 10 miles away) was used.

## 6.2.4 Internal loads

Internal loads are a key constituent of any EnergyPlus model. The calculation of internal thermal loads for each thermal zone in a building is performed in EnergyPlus based on information about occupancy, lighting and equipment corresponding to the zone.

### 6.2.4.1 Occupancy information

Occupancy information is specified by providing the nominal occupancy for the zones, occupancy schedules which determine what percentage of the nominal number of occupants is present at a given time, and activity schedules which determine the thermal contribution (in W) from each occupant at any time. The nominal occupancy can be entered in the form of total number of people, number of people per floor area or floor area per person. The occupancy and activity schedules can be provided as time-series data sampled hourly for each day in the week. Typically, three schedules – weekday schedule, Saturday schedule and Sunday schedule – are provided for occupancy and activity.

For the test building under consideration, the nominal occupancy values are shown in Table 6.4. Note that the table also shows nominal lighting and equipment loads which are explained later in this Section. The weekday and Saturday occupancy schedules are shown in Figure 6.8 and Figure 6.9 respectively. The Sunday occupancy is zero for the entire day. These occupancy schedules are based on a template called 'Medium_Office_Bldg_Occ' provided in the OpenStudio tool. The activity schedule for all zones at all times corresponds to a thermal contribution of 120 W per occupant. This value is based on the template "Medium_Office_Activity" provided in OpenStudio.

### 6.2.4.2 Lighting information

Similar to the occupancy information, the thermal load from lighting in each thermal zone in a building is specified by providing the nominal lighting thermal load (W) and lighting schedules which determine what percentage of the nominal lighting load is applicable at any given time. The nominal lighting thermal load can be entered in the form of an absolute load (W), load per floor area (W/ft$^2$) or load per person (W/person). The lighting schedules can be provided as time-series data sampled hourly for each day in the week.

For the test building under consideration, the nominal lighting thermal load values are shown in Table 6.4. The weekday and Saturday lighting schedules are shown in Figure 6.10 and Figure 6.11 respectively. Sunday's schedule for each zone is set to 5% of the corresponding nominal lighting thermal load at all times in the day. These lighting schedules are based on a template called 'Medium_Office_Bldg_Light' provided in the OpenStudio tool.

**Table 6.4 Nominal occupancy, lighting load and equipment load information used in EnergyPlus model of test building**

| Zone alias | Nominal occupancy | Nominal lighting thermal load (kW) | Nominal equipment thermal load (W per person) |
|---|---|---|---|
| G | 50 | 5.0 | 400 |
| E | 50 | 5.0 | 400 |
| C | 50 | 5.0 | 400 |
| SR | 0 | 5.0 | 400 |
| F1 | 75 | 5.0 | 400 |
| D | 50 | 5.0 | 400 |
| F2 | 75 | 5.0 | 400 |
| TCB | 10 | 5.0 | 400 |
| NTCB | 0 | 1.0 | 400 |

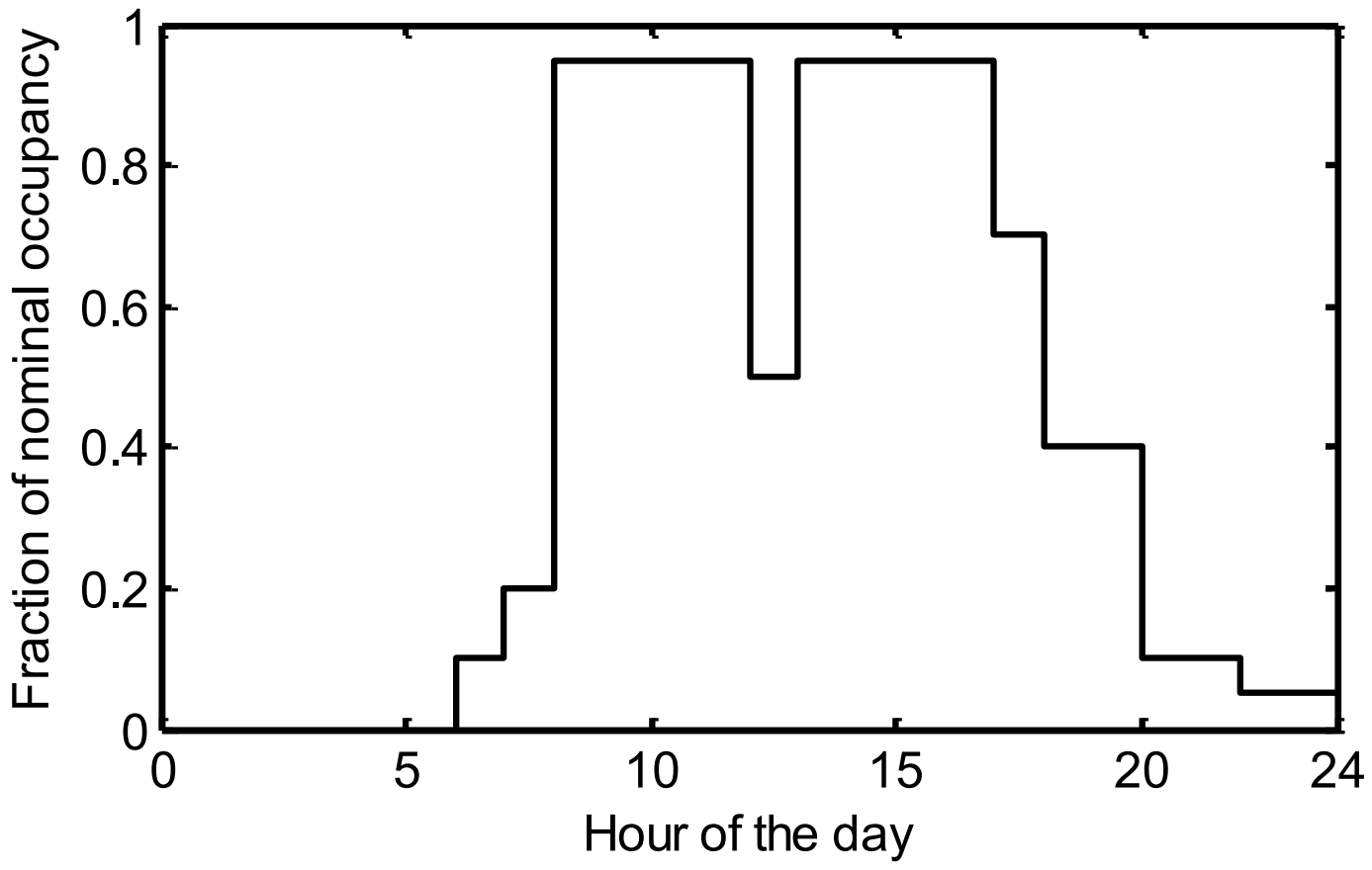

**Figure 6.8 Weekday occupancy schedule in each thermal zone in the test building**

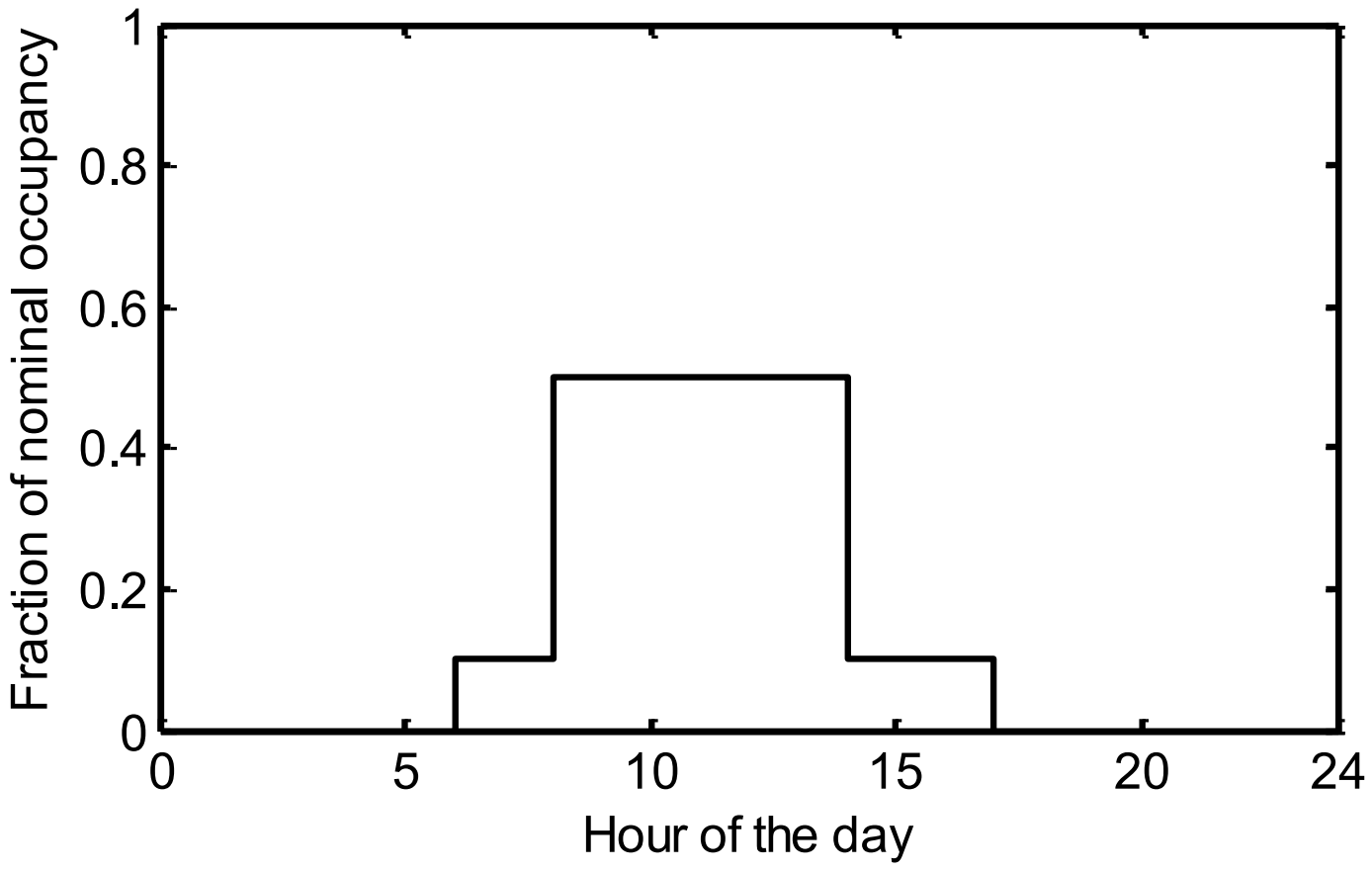

**Figure 6.9 Saturday's occupancy schedule in each thermal zone of the test building**

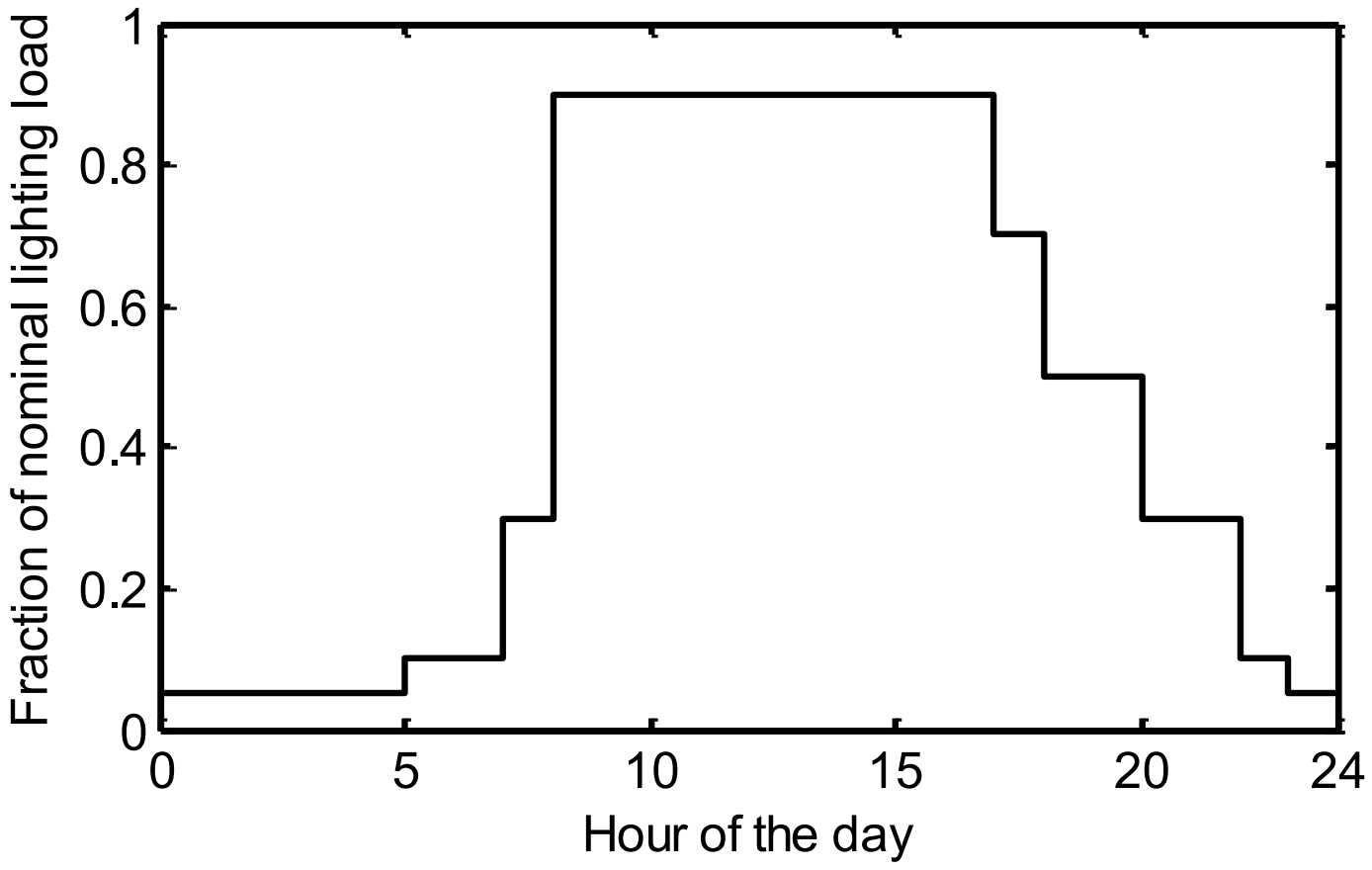

**Figure 6.10 Weekday lighting schedule in each thermal zone in the test building**

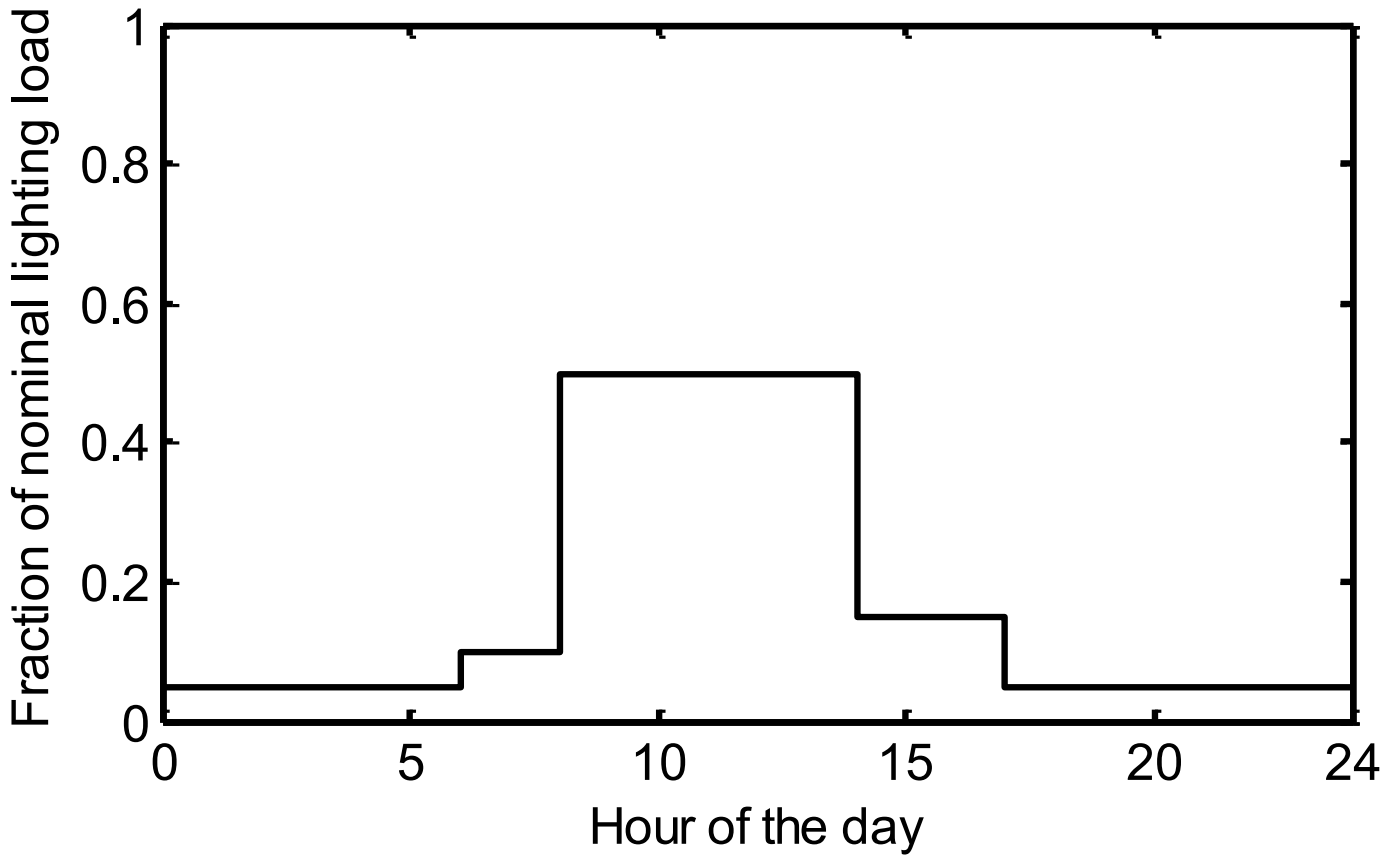

**Figure 6.11 Saturday's lighting schedule in each thermal zone in the test building**

### 6.2.5 Creation of EnergyPlus model

An EnergyPlus model for the test building was created using the OpenStudio tool [95] based on the layout, construction, weather and internal load information described above. The steps used for construction of the model are explained in Appendix E. The description includes a link to online tutorials on OpenStudio. For the interested reader, the Openstudio and EnergyPlus files that were generated for the test building under consideration are included in the media accompanying this thesis.

In the EnergyPlus model constructed above, the 'ideal air loads' option was turned on. This option provides the requisite amount of heating or cooling to each thermally controlled zone in order to achieve specified set-point temperatures, without the need to set up an HVAC system. Consideration of the HVAC system is beyond the scope of this thesis. However, OpenStudio provides the option of specifying an appropriate HVAC system, which can be auto-sized depending on design weather conditions and loads. More details are available in the tutorials referenced in Appendix E.

## 6.3 Generation of linearized model

In this section, we describe the development of a linearized model, based on the 3R2C framework (see Section 5.2.1) for the test building described in section 6.2. The purpose of the linearized model is to allow the use of the OLF-FPM method described in Chapter 4 to partition the building for decentralized control.

### 6.3.1 Overview of modeling framework

The identification of resistances and capacitances to construct a RC network model of a building is performed in two steps, which we refer to as 'zone level identification' and 'wall surface level identification'. In this chapter, the term wall is used to represent a general term to represent all surfaces constituting the building, i.e. ceilings, floors, vertical walls (both internal and external), windows and doors. The underlying details of these steps are described in Sections 6.3.2 and 6.3.3 below. The 3R2C modeling paradigm for walls which is used in these steps was illustrated in Figure 5.1 and is reproduced below in Figure 6.12.

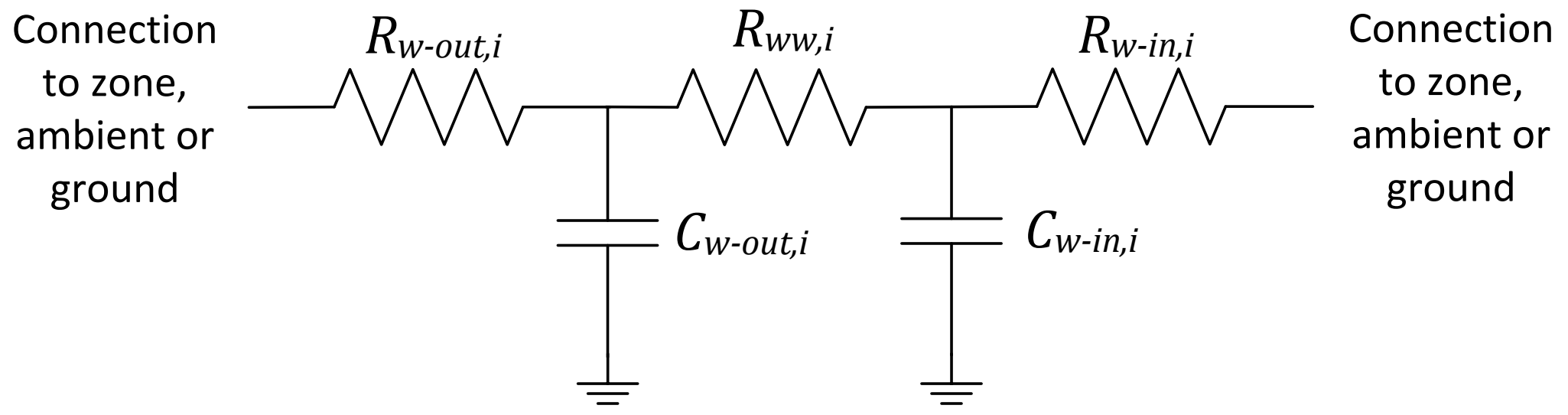

**Figure 6.12 Schematic of 3R2C modeling paradigm for wall *i*.**

The resistances and capacitances obtained through these steps are then used in Section 6.3.4 to create a linear time invariant (LTI) model of the building thermal dynamics.

### 6.3.2 Zone level identification

The zone level identification corresponds to the estimation of the following parameters

1. The thermal capacitance, $c_{zj}$ associated with each zone $j$ in the building.
2. The resistances $R_{w-in,i}$ and $R_{w-out,i}$ corresponding to each internal wall $i$ (flanked by zones on both sides) in the building as shown in Figure 6.12.
3. The resistance $R_{w-in,i}$ corresponding to each external wall $i$ (flanked by a zone on one side and ambient/ground on another) in the building as shown in Figure 6.12.

It is easy to verify that the above resistances are given by elements of the set $\{R_{jl} \mid j \in \{1,2,\ldots N_z\}, l \in \mathcal{N}_j\}$, where $R_{jl}$ represents the thermal resistance between zone $j$ and wall surface $l$, and $\mathcal{N}_j$ represents the set of wall surfaces which enclose zone $j$. EnergyPlus uses the following difference equation obtained using the Backward Euler method to represent the thermal dynamics for the $j^{th}$ zone (see nomenclature for details on notation)

$$T_{z,j}(k+1) = \frac{\dot{Q}_{int,j}(k+1)+c_{p,air}\dot{m}_{HVAC,j}(k+1)T_{HVAC,j}(k+1)+\frac{c_{zj}T_{z,j}(k)}{T_S}+c_{p,air}\dot{m}_{inf,j}(k)T_a(k)+\sum_{l\in\mathcal{N}_j}(UA)_{jl}(k)T_{w,l}(k)}{\frac{c_{zj}}{T_S}+c_{p,air}\dot{m}_{HVAC,j}(k+1)+c_{p,air}\dot{m}_{inf,j}(k)+\sum_{l\in\mathcal{N}_j}(UA)_{jl}(k)} \quad (6.1)$$

The above equation represents the conservation of energy (first law of Thermodynamics) for the zone. The interested reader is directed to [97] for details its derivation. Here, $(UA)_{jl}$ represents the overall heat transfer coefficient between the zone $j$ and wall surface $l$ multiplied by the area $A_l$ of surface $l$. This is related to $R_{jl}$ introduced earlier by

$$R_{jl} = \frac{1}{(UA)_{jl}}. \quad (6.2)$$

Equation (6.1) can be used to identify the unknown capacitance $c_{zj}$ and the thermal resistances $R_{jl}$ where $l \in \mathcal{N}_j$, based on data obtained from an EnergyPlus simulation. This can be formally expressed as a least-squares identification problem for zone $j$ given by the optimization below (see nomenclature for details on notation)

$$\left\{c_{zj}^*, \{(UA)_{jl}^*\}_{l\in\mathcal{N}_j}\right\} = \underset{\left\{c_{zj},\{(UA)_{jl}\}_{l\in\mathcal{N}_j}\right\}}{\arg\min} \sum_{k=0}^{N-1}\left(T_{z,j}(k+1) - T_{z,j}^{\mathrm{E+}}(k+1)\right)^2 \quad (6.3)$$

Subject to:

$$T_{z,j}(k+1) = \frac{\dot{Q}_{int,j}^{\mathrm{E+}}(k+1)+c_{p,air}\dot{m}_{HVAC,j}^{\mathrm{E+}}(k+1)T_{HVAC,j}^{\mathrm{E+}}(k+1)+\frac{c_{zj}T_{z,j}(k)}{T_S}+c_{p,air}\dot{m}_{inf,j}^{\mathrm{E+}}(k)T_a(k)+\sum_{l\in\mathcal{N}_j}(UA)_{jl}T_{w,l}^{\mathrm{E+}}(k)}{\frac{c_{zj}}{T_S}+c_{p,air}\dot{m}_{HVAC,j}^{\mathrm{E+}}(k+1)+c_{p,air}\dot{m}_{inf,j}^{\mathrm{E+}}(k)+\sum_{l\in\mathcal{N}_j}(UA)_{jl}}, \quad (6.4)$$

$$0 \le c_{zj} \le c_{zj,max}, \quad (6.5)$$

$$0 \le (UA)_{jl} \le (UA)_{jl,max} \quad \text{for all } l \in \mathcal{N}_j. \quad (6.6)$$

In the above optimization framework, the superscript "E+" is used to indicate data obtained from an EnergyPlus simulation and the superscript " * " represents optimal values or the result of the optimization. $N$ is the length of the time window used for simulation measured in terms of number of samples, where the sample period is $T_s$ seconds. Equation (6.4) is based on (6.1), whereas the constraints (6.5) and (6.6) are included to prevent the capacitance $c_{zj}$ and coefficients $(UA)_{jl}$ from becoming negative or unbounded. Solution of the above optimization problem for each zone $j \in \{1,2,\dots,N_z\}$ provides the required zone capacitances $c_{zj}$ and resistances $R_{jl}$ (computed from $(UA)_{jl}$ via (6.2)).

Data for use in the above optimization process is obtained by providing persistently excited zone set-point temperature signals in EnergyPlus in time-series format. In the simulation process, the 'fictitious' HVAC system corresponding to the 'ideal air loads' option heats or cools the zones, as needed, to achieve the specified set-point temperatures at each time instant.

The above optimization was performed for zone F2 in the test building using the 'fmincon' command in the MATLAB optimization toolbox [98]. The optimization time window was chosen to be 24 hours, from 12:00 AM on June 3 to 12:00 AM on June 4. The set-point temperature signal used to obtain EnergyPlus simulation data is shown in Figure 6.13, which is a pseudo-random binary signal (PRBS) generated using MATLAB. The relevant codes for performing this optimization are provided in Appendix F.

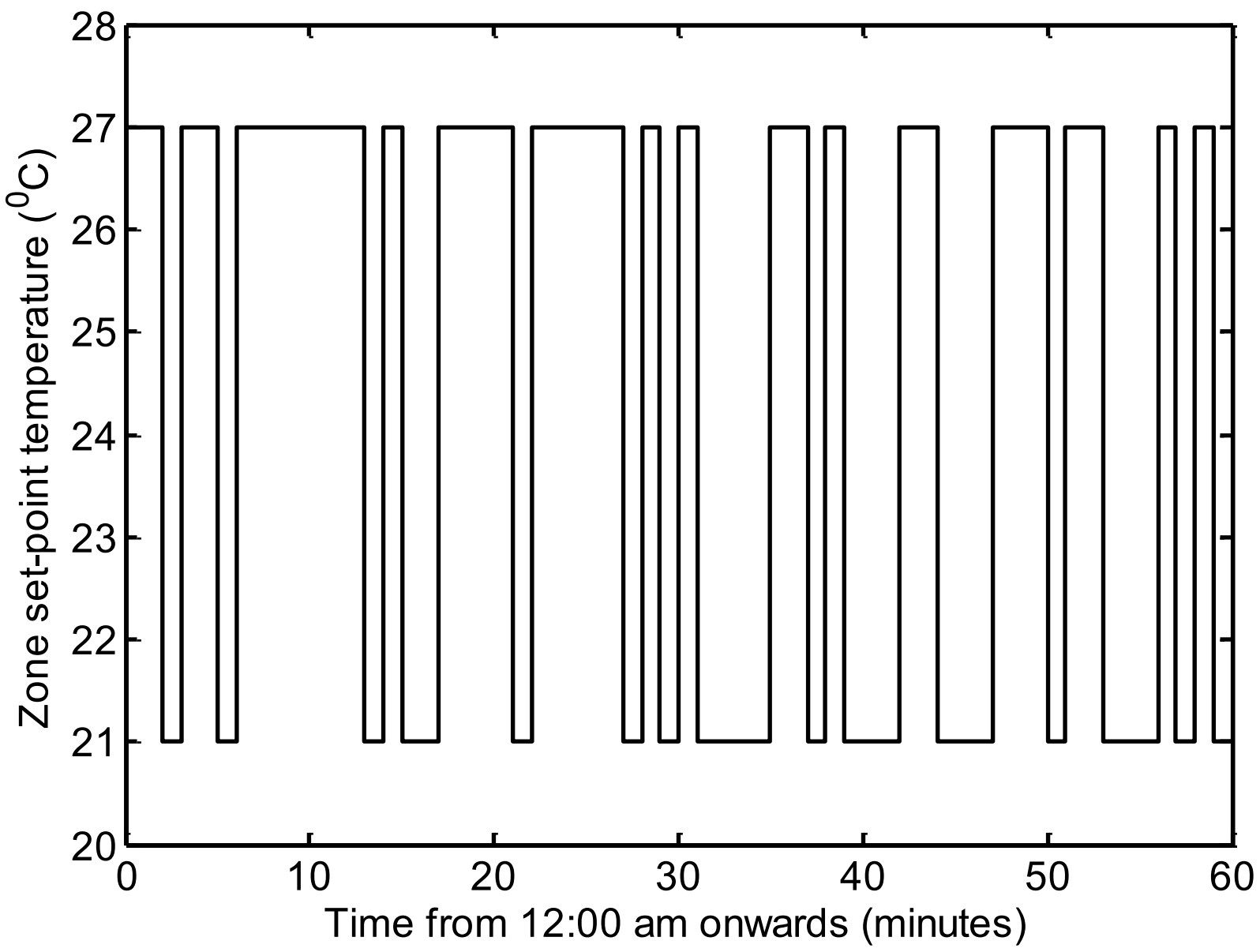

**Figure 6.13 Zone set point temperature signal from 12:00 AM to 1:00 AM used in EnergyPlus simulations for zone level identification**

Zone F2 is surrounded by 9 surfaces. The corresponding values of coefficients, $(UA)_{jl}$ in units of K/kW obtained from the optimization are 0, 0, 0, 0, 0, 0, 0, 0.617 and 3.857. A comparison of the zone temperature predicted using the identified model with the zone temperature obtained via EnergyPlus simulation is shown in Figure 6.14. It is observed that the zone temperature predicted using the identified model is very close to that obtained from EnergyPlus simulations all time instants. However, the identified parameter values are unrealistic. In particular, coefficients $(UA)_{jl}$ between zone F2 and seven of the surrounding surfaces are zero, which is physically untenable. A possible reason for this solution is the presence of too many unknown parameters (10 for zone F2 – 9 resistances and 1 capacitance) in the optimization framework. This may lead to a situation where the same total energy transfer to a zone from the surrounding surfaces can be achieved by multiple combinations of surface to zone heat transfer coefficients. Therefore, some of the heat transfer coefficients can be set to zero by augmenting the remaining heat transfer coefficients in such a way that the total energy transfer from the surfaces to the zone is the same as that corresponding to the EnergyPlus simulation.

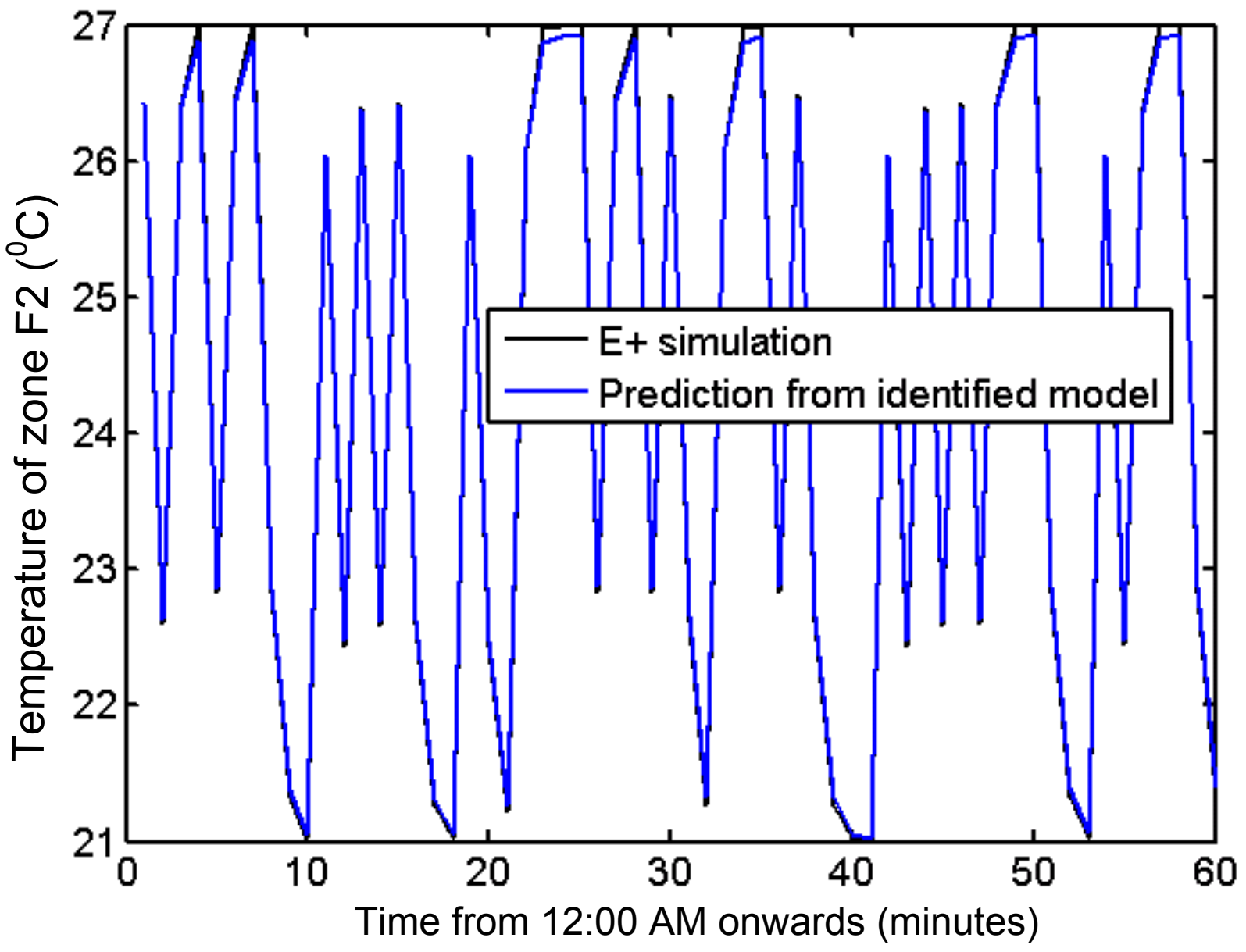

**Figure 6.14 Temperature of zone F2 from zone level least squares identification methodology compared with EnergyPlus data between 12:00 AM to 1:00 AM**

In an attempt to address the above problem, potentially arising due over-parameterization in the optimization framework, we propose a simplified framework which is based upon the following physical reasoning. Each coefficient $(UA)_{jl}$ (where $l \in \mathcal{N}_j$) can be written as the product of $h_{jl}$ and $A_j$, which denote the heat transfer coefficient between zone $j$ and surface $l$, and the area of surface $l$ respectively. The coefficient $h_{jl}$ corresponds to convection heat transfer which is primarily dependent on the properties of air and temperatures of zone $j$ and surface $l$. If we ignore the dependence of $h_{jl}$ on the temperatures of zone $j$ and surface $l$ in the range of operation of the building, we can assume that each of the heat transfer coefficients $h_{jl}$ are same for all $l \in \mathcal{N}_j$. This results in a simplified modeling framework which only uses a single heat transfer coefficient denoted by $h_j$ between zone $j$ and all surfaces $l$ enclosing it. Therefore, the following framework for least squares system identification is proposed.

$$\{c_{zj}^*, h_j^*\} = \underset{\{c_{zj}, h_j\}}{\arg\min} \quad \sum_{k=0}^{N-1} \left(T_{z,j}(k+1) - T_{z,j}^{\mathrm{E+}}(k+1)\right)^2 \tag{6.7}$$

Subject to:

$$T_{z,j}(k+1) = \frac{\dot{Q}_{int,j}^{\mathrm{E+}}(k+1) + c_{p,air}\dot{m}_{HVAC,j}^{\mathrm{E+}}(k+1)T_{HVAC,j}^{\mathrm{E+}}(k+1) + \frac{c_{zj}T_{z,j}(k)}{T_s} + c_{p,air}\dot{m}_{inf,j}^{\mathrm{E+}}(k)T_a(k) + h_j\sum_{l\in\mathcal{N}_j} A_l T_{w,l}^{\mathrm{E+}}(k)}{\frac{c_{zj}}{T_s} + c_{p,air}\dot{m}_{HVAC,j}^{\mathrm{E+}}(k+1) + c_{p,air}\dot{m}_{inf,j}^{\mathrm{E+}}(k) + h_j\sum_{l\in\mathcal{N}_j} A_l}, \tag{6.8}$$

$$0 \le c_{zj} \le c_{zj,max}, \tag{6.9}$$

$$0 \le h_j \le h_{j,max}. \tag{6.10}$$

Similar to the optimization framework given by (6.3) to (6.6), in the above equations, the superscript "E+" is used to indicate data obtained from an EnergyPlus simulation and the superscript " * " represents optimal values or the result of the optimization.

The above optimization was performed for all zones in the test building. The optimization time window was chosen to be 24 hours, from 12:00 AM on June 3 to 12:00 AM on June 4. The set-point temperature signal used to obtain EnergyPlus simulation data is shown in Figure 6.13 which is the same as that used for the optimization framework described earlier. The result of the optimization for each zone is shown in Table 6.5. A comparison of zone temperatures predicted using the identified models with zone temperatures obtained via EnergyPlus simulation is shown in Figure 6.15 to Figure 6.23. It is observed that the zone temperatures predicted using the identified models are very close to those obtained from EnergyPlus simulations all time instants. The relevant codes for performing this optimization are provided in Appendix G.

**Table 6.5 Results of zone level optimization for all zones in test building**

| **Thermal zone number** (*j*) | **Alias** | $c_{zj}^*$ (kJ/K) | $h_j^*$ (K/kW-m$^2$) |
|---|---|---|---|
| 1 | G | $6.182 \times 10^3$ | $2.005 \times 10^{-3}$ |
| 2 | E | $7.39 \times 10^3$ | $1.9 \times 10^{-3}$ |
| 3 | C | $8.89 \times 10^3$ | $1.795 \times 10^{-3}$ |
| 4 | SR | $7.06 \times 10^2$ | $1.924 \times 10^{-3}$ |
| 5 | F1 | $3.709 \times 10^3$ | $2.228 \times 10^{-3}$ |
| 6 | D | $7.391 \times 10^3$ | $1.921 \times 10^{-3}$ |
| 7 | F2 | $3.717 \times 10^3$ | $2.211 \times 10^{-3}$ |
| 8 | TCB | $3.449 \times 10^3$ | $1.651 \times 10^{-3}$ |
| 9 | NTCB | $6.499 \times 10^3$ | $8.85 \times 10^{-4}$ |

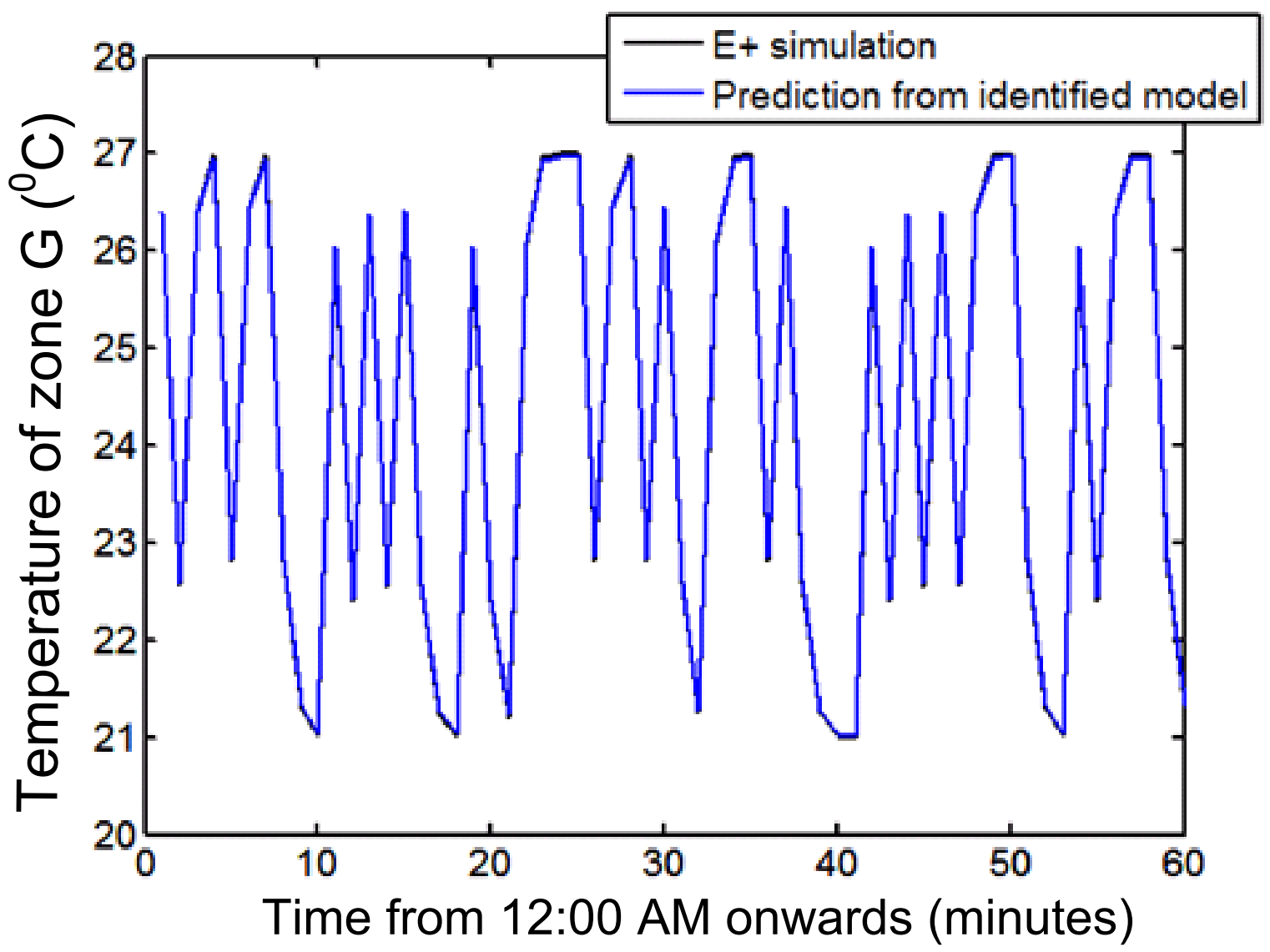

**Figure 6.15 Temperature of zone G from zone level least squares identification methodology compared with EnergyPlus data between 12:00 AM to 1:00 AM**

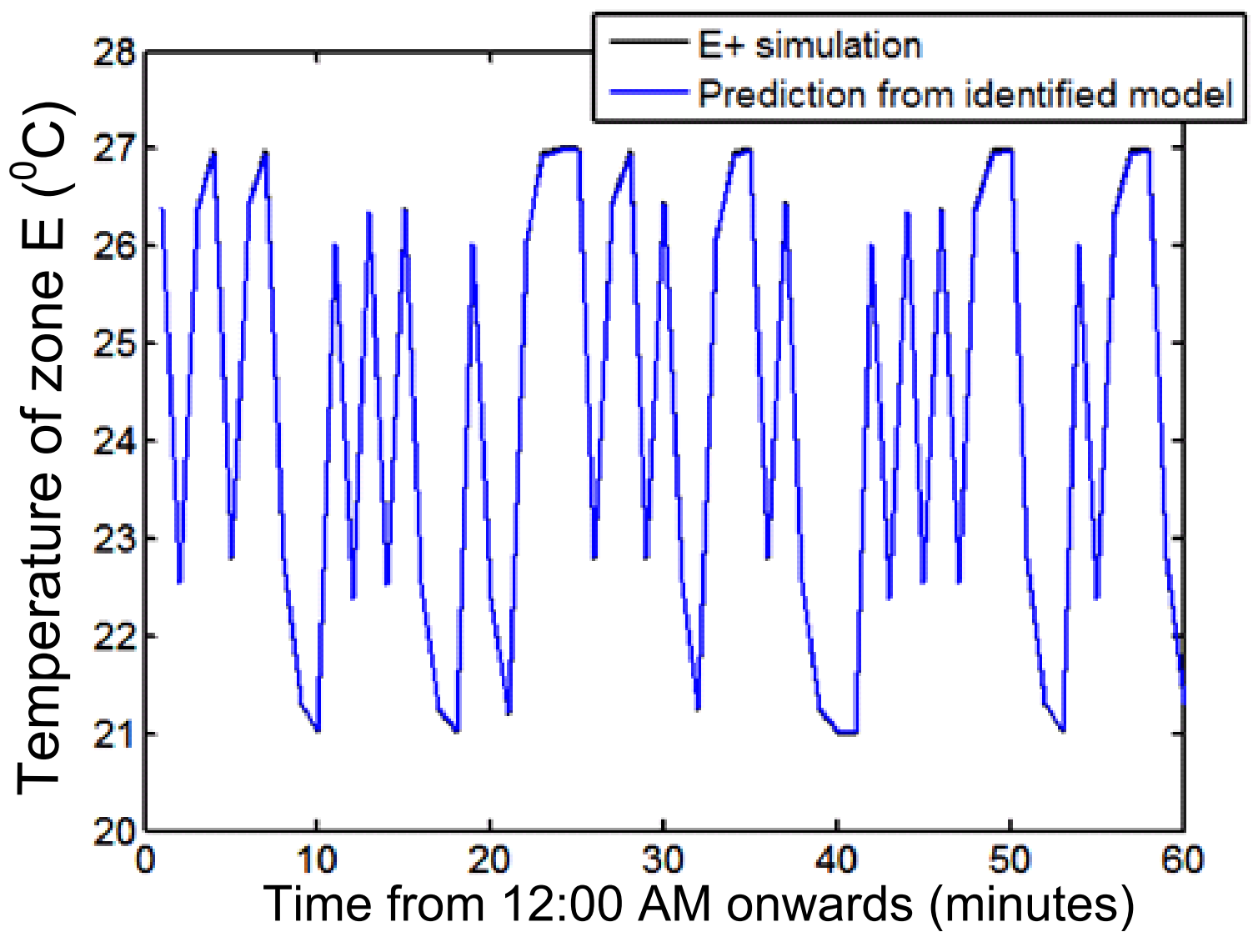

**Figure 6.16 Temperature of zone E from zone level least squares identification methodology compared with EnergyPlus data between 12:00 AM to 1:00 AM**

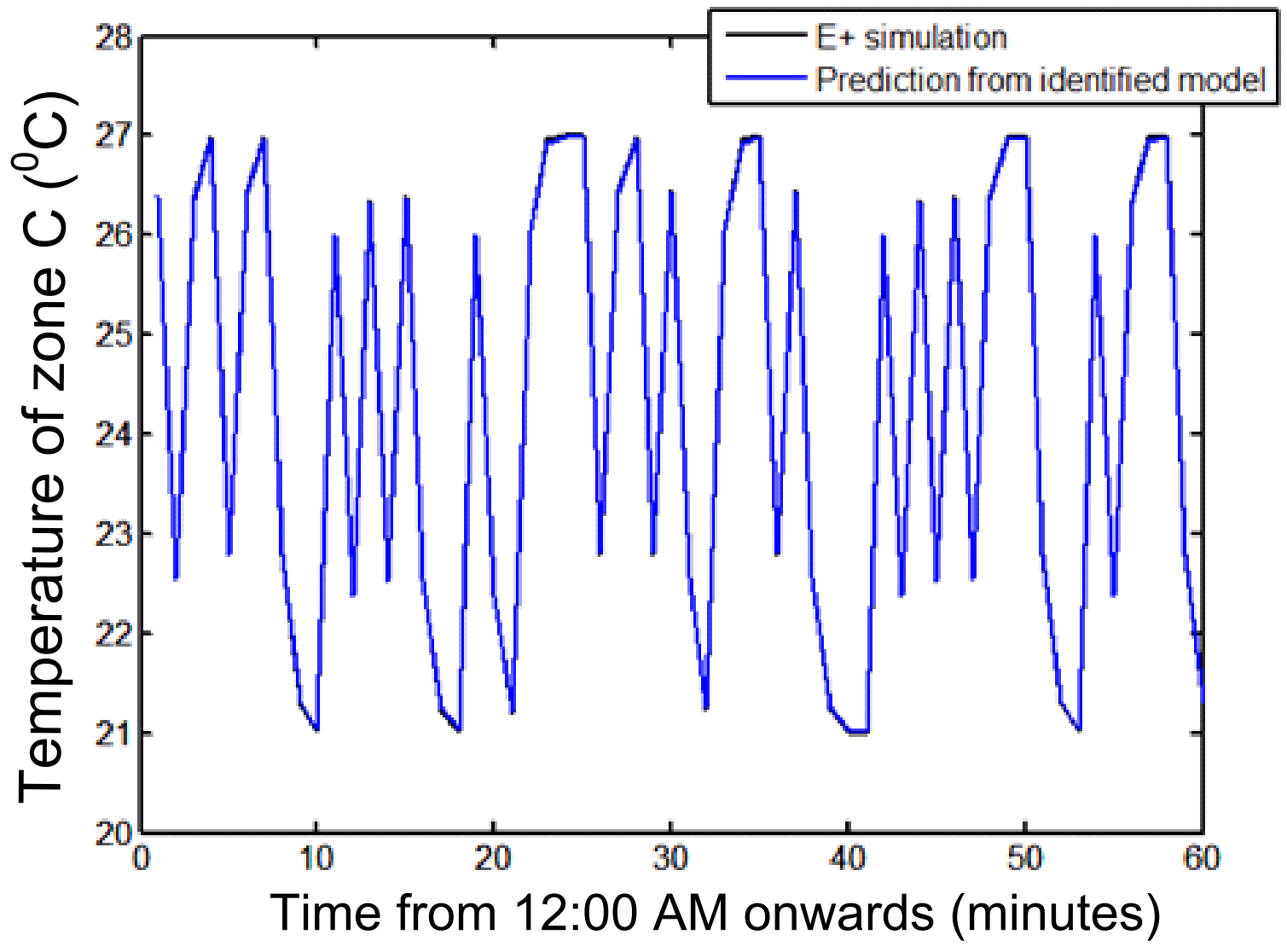

**Figure 6.17 Temperature of zone C from zone level least squares identification methodology compared with EnergyPlus data between 12:00 AM to 1:00 AM**

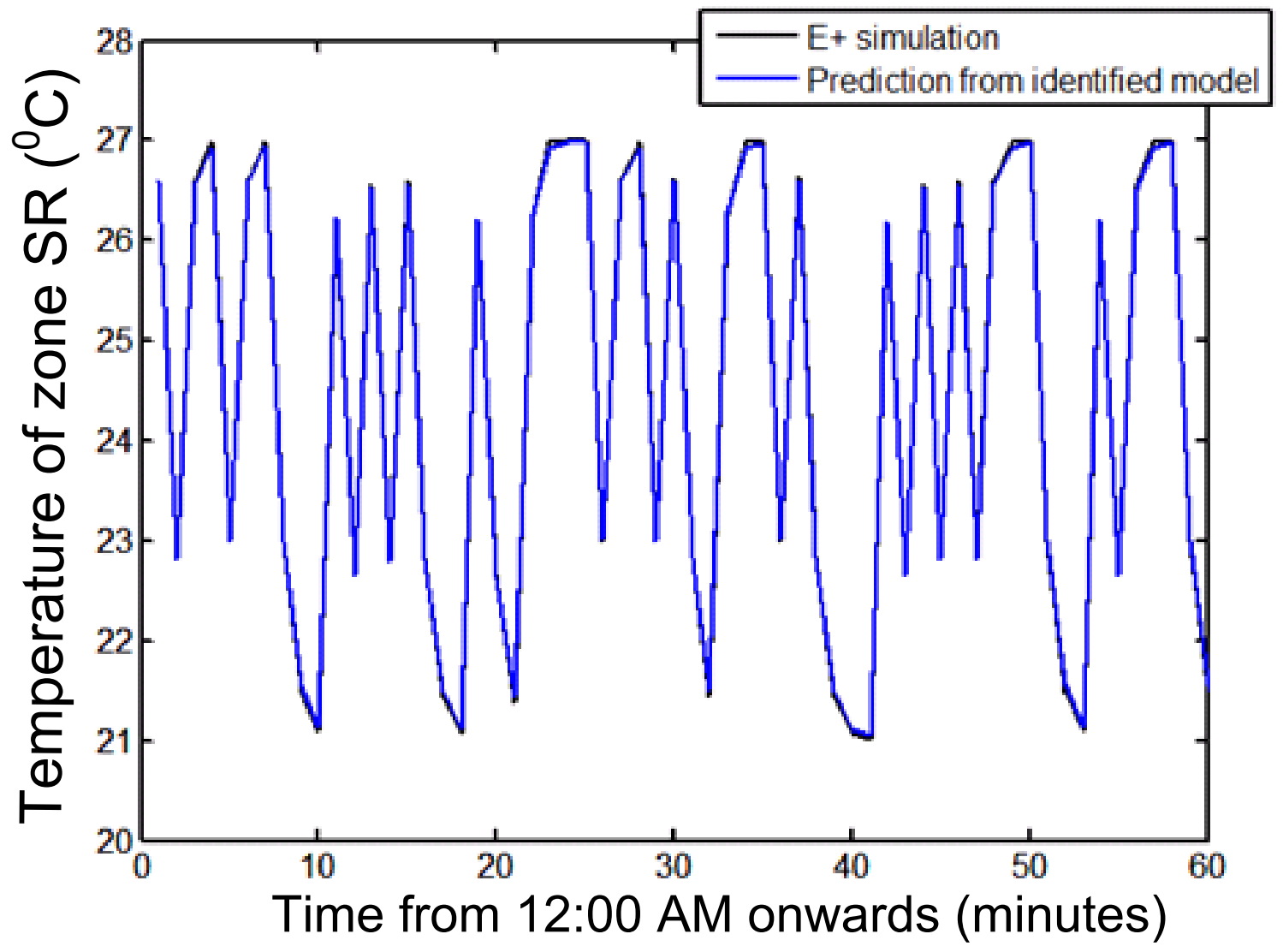

**Figure 6.18 Temperature of zone SR from zone level least squares identification methodology compared with EnergyPlus data between 12:00 AM to 1:00 AM**

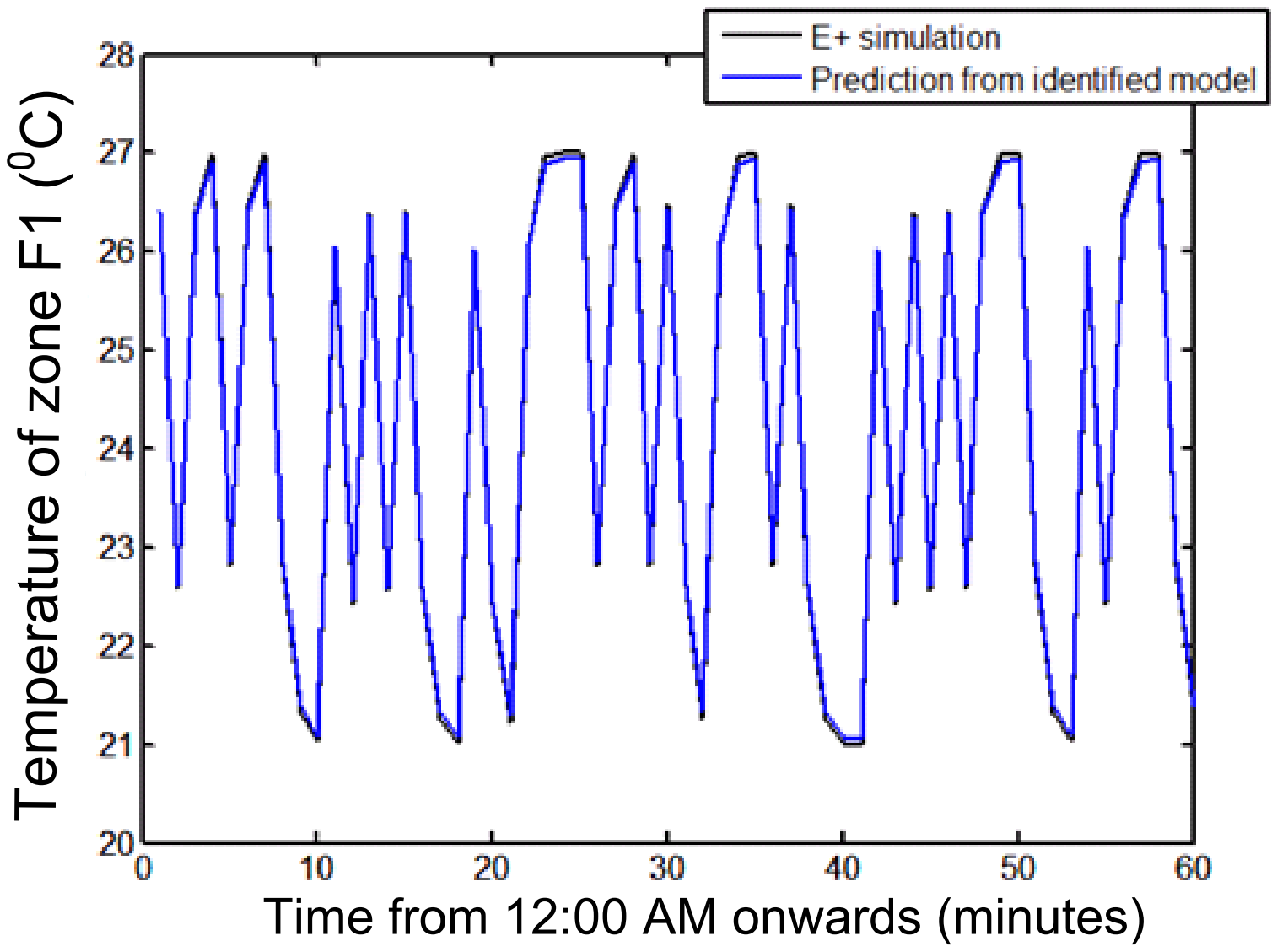

**Figure 6.19 Temperature of zone F1 from zone level least squares identification methodology compared with EnergyPlus data between 12:00 AM to 1:00 AM**

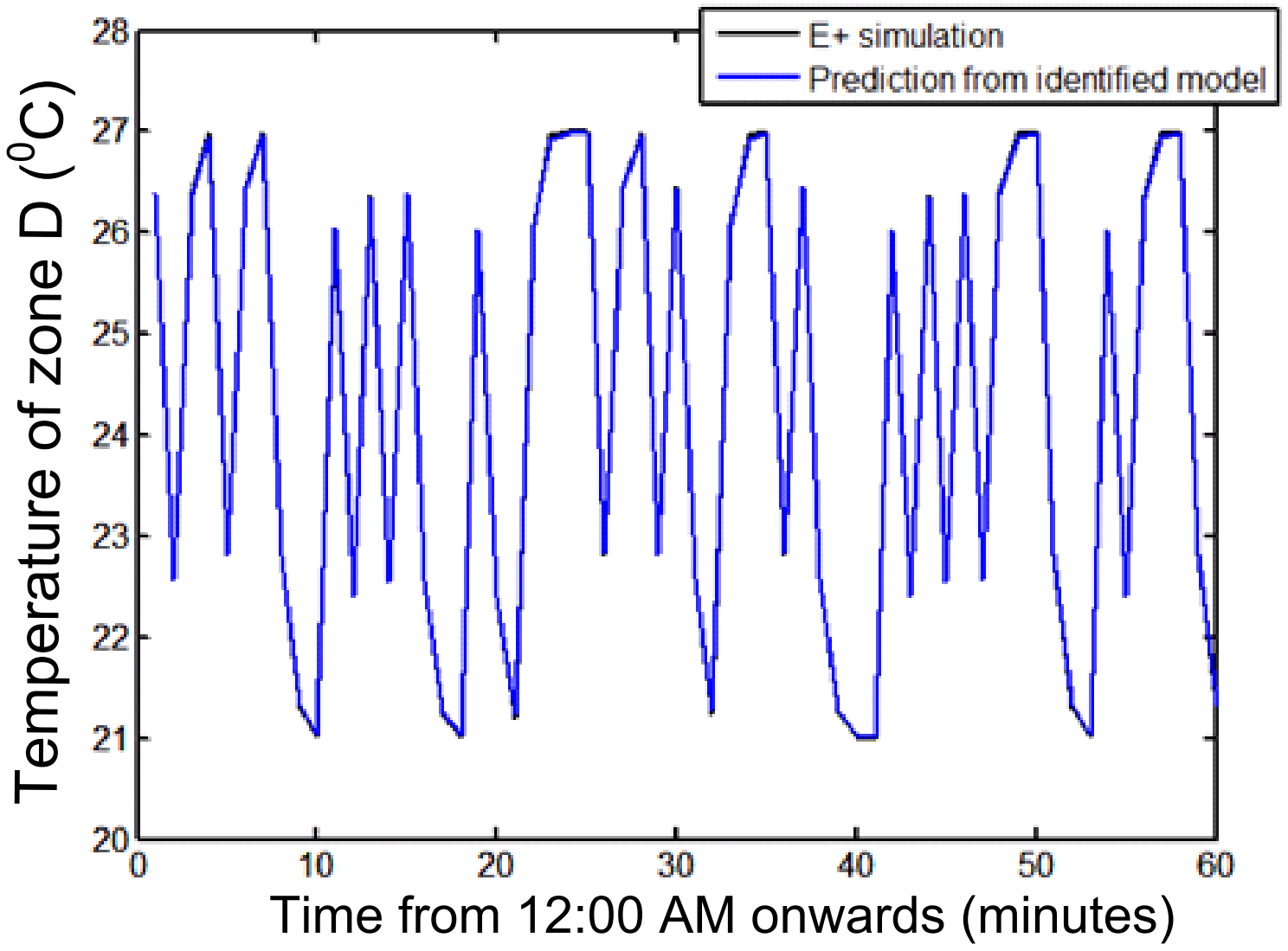

**Figure 6.20 Temperature of zone D from zone level least squares identification methodology compared with EnergyPlus data between 12:00 AM to 1:00 AM**

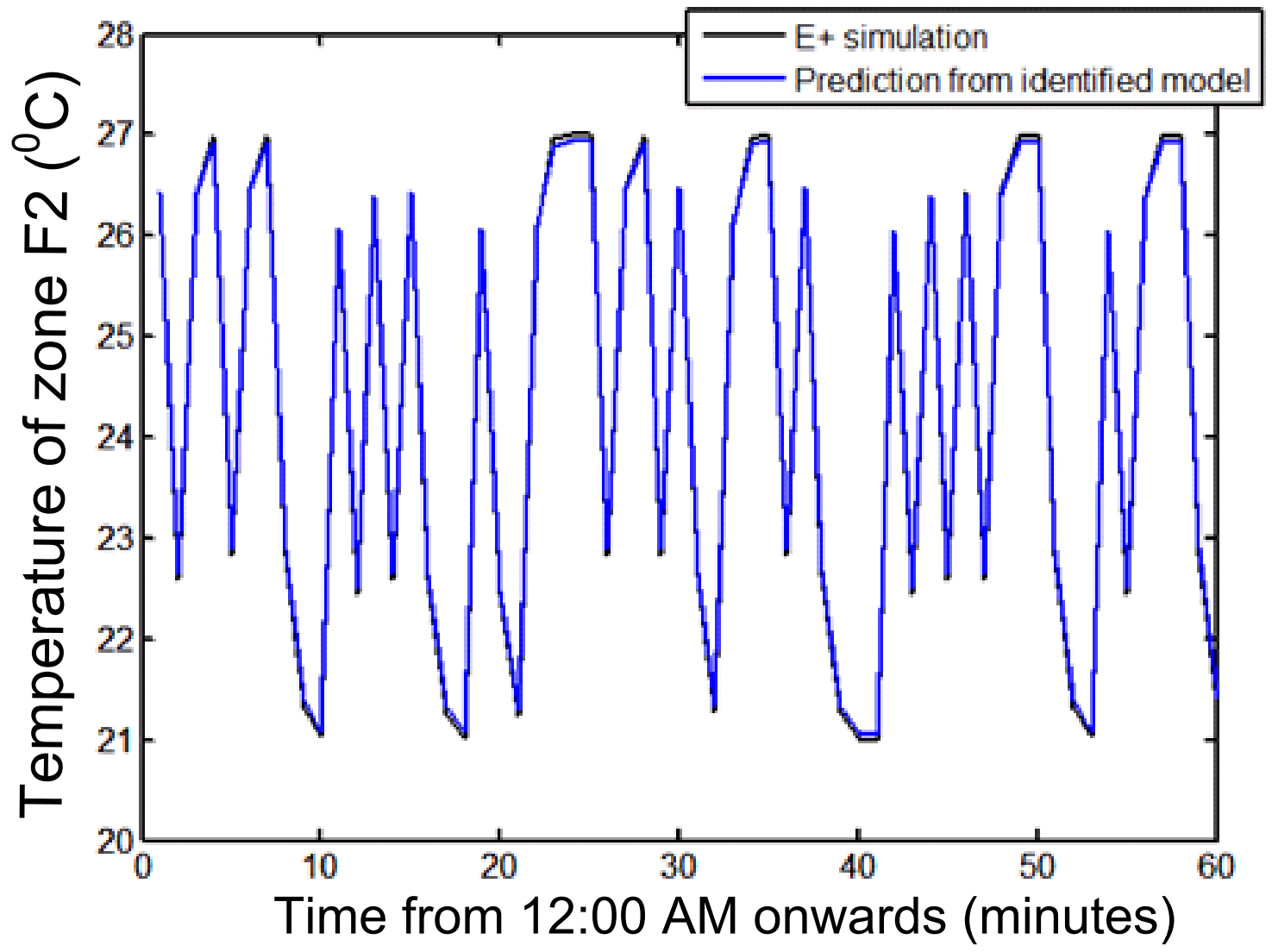

**Figure 6.21 Temperature of zone F2 from zone level least squares identification methodology compared with EnergyPlus data between 12:00 AM to 1:00 AM**

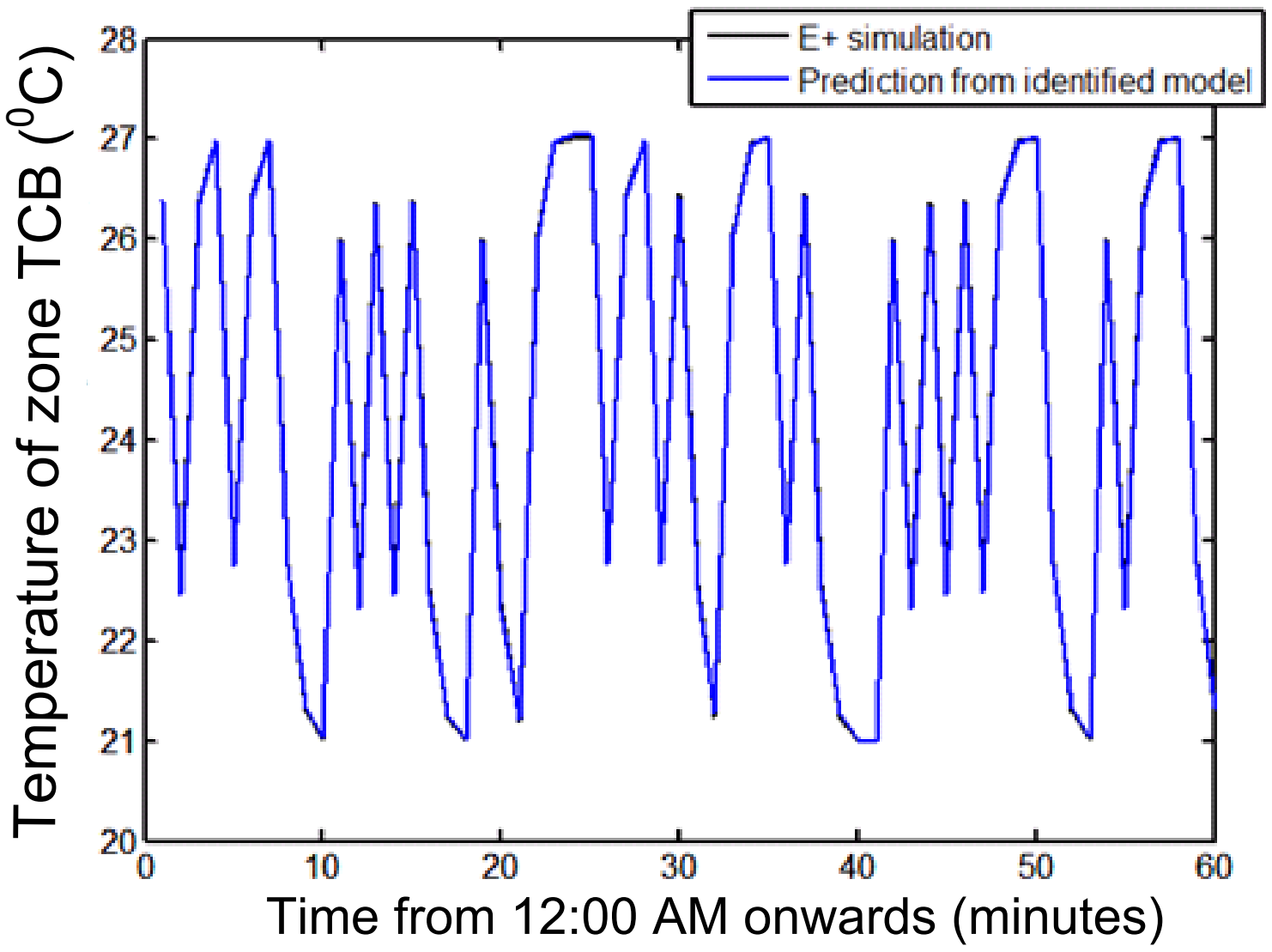

**Figure 6.22 Temperature of zone TCB from zone level least squares identification methodology compared with EnergyPlus data between 12:00 AM to 1:00 AM**

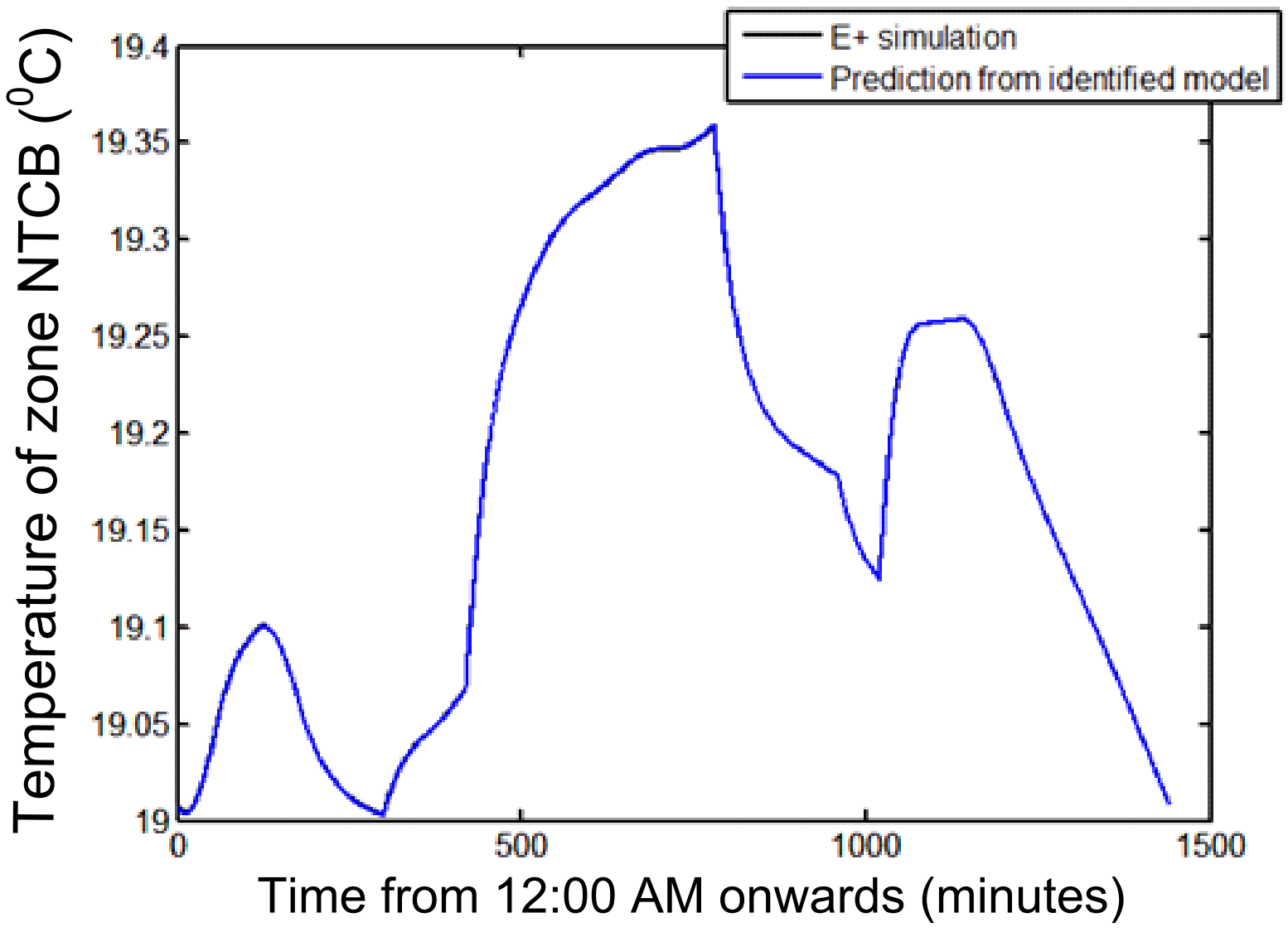

**Figure 6.23 Temperature of zone NTCB from zone level least squares identification methodology compared with EnergyPlus data between 12:00 AM to 1:00 AM**

The thermal resistance $R_{jl}$ between zone $j$ and its enclosing surfaces $l$ can be computed from the solution of the optimization problem given by (6.7) to (6.10) as shown below

$$R_{jl} = \frac{1}{(UA)_{jl}} = \frac{1}{h_j^* A_l}. \qquad (6.11)$$

As stated earlier, the resistances desired to be computed from the zone level identification are given by elements of the set $\{R_{jl} \mid j \in \{1,2, \dots N_z\}, l \in \mathcal{N}_j\}$. Therefore, the use of (6.11) allows the computation of all such resistances. The thermal capacitances associated with the zones appear as optimization variables in (6.7) and therefore are directly provided by the optimization. The calculation of resistances and capacitances in this way completes the zone level identification. For the test building under consideration, the capacitances obtained are shown in Table 6.5. The resistances are computed from the $h_j^*$ values shown in this table in a spreadsheet using (6.11) which is provided in the media accompanying this thesis.

### 6.3.3 Wall surface level identification

The wall surface level identification involves estimation of the following parameters with reference to Figure 6.12:

1. For internal walls (flanked by zones on both side): $R_{ww,i}$, $C_{w-in,i}$ and $C_{w-out,i}$
2. For external walls (flanked by a zone on one side and ambient/ground on another): $R_{w-out,i}$, $R_{ww,i}$, $C_{w-in,i}$, $C_{w-out,i}$

The above parameters, together with the parameters estimated using zone level identification in Section 6.3.2 provide the complete set of resistances and capacitances required to set up a Linear Time Invariant (LTI) model of the building thermal dynamics. Two variants of a least squares identification procedure – treating wall surface thermal loads as known and unknown respectively – were first implemented for the wall surface level identification. However, both these methods were found to have critical limitations and provided unsatisfactory results. Therefore, as an alternative, the wall surface level identification was performed by computing the values of the desired resistances and capacitances directly using material properties of the wall construction layers. The two least squares identification procedures mentioned above are presented below, followed by the procedure for direct computation of parameters.

#### 6.3.3.1 Least squares identification with unknown wall thermal loads

The least squares identification procedure described here treats the wall thermal loads ($d_{w-out,i}$ and $d_{w-in,i}$) as unknowns to be determined along with resistances and capacitances. The identification procedure involves solving a pair of optimization problems for each internal and external wall.

The pair of optimization problems corresponding to the $i^{th}$ internal wall is given by:

(i) $\left\{R^*_{ww,i}, C^*_{w-in,i}, \left\{d^*_{w-in,i}(l)\right\}_{l=0}^{N-1}\right\} =$

$$\underset{\left\{R_{ww,i}, C_{w-in,i}, \left\{d_{w-in,i}(l)\right\}_{l=0}^{N-1}\right\}}{\arg\min} \sum_{k=0}^{N-1} \left(T_{w-in,i}(k+1) - T^{\mathrm{E+}}_{w-in,i}(k+1)\right)^2 \quad (6.12)$$

Subject to:

$$C_{w-in,i} \frac{T_{w-in,i}(k+1) - T_{w-in,i}(k)}{T_s} = \frac{T^{\mathrm{E+}}_{int,i}(k) - T_{w-in,i}(k)}{R_{w-in,i}} + \frac{T^{\mathrm{E+}}_{w-out,i}(k) - T_{w-in,i}(k)}{R_{ww,i}} + d_{w-in,i}(k)$$

$$\text{for all } k \in \{0,1,\ldots,N-1\}, \quad (6.13)$$

$$0 \le R_{ww,i} \le R_{ww,i,max}, \quad (6.14)$$

$$0 \leq C_{w-in,i} \leq C_{w-in,i,max}, \quad (6.15)$$

$$d_{w-in,i,min} \leq d_{w-in,i}(k) \leq d_{w-in,i,max} \text{ for all } k \in \{0,1,\ldots,N-1\}. \quad (6.16)$$

(ii) $\left\{C^*_{w-out,i}, \left\{d^*_{w-out,i}(l)\right\}_{l=0}^{N-1}\right\} =$

$$\underset{\left\{C_{w-out,i},\left\{d_{w-out,i}(l)\right\}_{l=0}^{N-1}\right\}}{\arg\min} \sum_{k=0}^{N-1}\left(T_{w-out,i}(k+1) - T^{\mathrm{E+}}_{w-out,i}(k+1)\right)^2 \quad (6.17)$$

Subject to:

$$C_{w-out,i}\frac{T_{w-out,i}(k+1)-T_{w-out,i}(k)}{T_s} = \frac{T^{\mathrm{E+}}_{ext,i}(k)-T_{w-out,i}(k)}{R_{w-out,i}} + \frac{T^{\mathrm{E+}}_{w-in,i}(k)-T_{w-out,i}(k)}{R_{ww,i}} + d_{w-out,i}(k)$$

for all $k \in \{0,1,\ldots,N-1\}$, ( 6.18 )

$$0 \leq C_{w-out,i} \leq C_{w-out,i,max}, \quad (6.19)$$

$$d_{w-out,i,min} \leq d_{w-out,i}(k) \leq d_{w-out,i,max} \text{ for all } k \in \{0,1,\ldots,N-1\}. \quad (6.20)$$

The notations used in the above optimization problems are explained in the nomenclature (Table 6.1). Equations (6.13) and (6.18) represent energy conservation equations (first law of Thermodynamics) applied to the wall surfaces. Note that since $i$ is an internal wall, $T^{\mathrm{E+}}_{int,i}(k)$ and $T^{\mathrm{E+}}_{ext,i}(k)$ appearing above represent temperatures of zones that flank it on either side. The resistances $R_{w-in,i}$ and $R_{w-out,i}$ are assumed to be known since they can be determined from the zone level identification described in Section 6.3.2. The length of the optimization time-window $N$, and the bounds $R_{ww,i,max}$, $C_{w-in,i,max}$, $d_{w-in,i,min}$, $d_{w-in,i,max}$, , $C_{w-out,i,max}$, $d_{w-out,i,min}$ and $d_{w-out,i,max}$ are treated as design parameters which can be tuned if necessary. The pair of optimization problems corresponding to the $i^{th}$ external wall is given by:

(i) $\left\{R^*_{ww,i}, C^*_{w-in,i}, \left\{d^*_{w-in,i}(l)\right\}_{l=0}^{N-1}\right\} =$

$$\underset{\left\{R_{ww,i},C_{w-in,i},\left\{d_{w-in,i}(l)\right\}_{l=0}^{N-1}\right\}}{\arg\min} \sum_{k=0}^{N-1}\left(T_{w-in,i}(k+1) - T^{\mathrm{E+}}_{w-in,i}(k+1)\right)^2 \quad (6.21)$$

Subject to:

$$C_{w-in,i}\frac{T_{w-in,i}(k+1)-T_{w-in,i}(k)}{T_s} = \frac{T^{\mathrm{E+}}_{int,i}(k)-T_{w-in,i}(k)}{R_{w-in,i}} + \frac{T^{\mathrm{E+}}_{w-out,i}(k)-T_{w-in,i}(k)}{R_{ww,i}} + d_{w-in,i}(k)$$

for all $k \in \{0,1,\dots,N-1\}$, ( 6.22 )

$$0 \le R_{ww,i} \le R_{ww,i,max}, \quad (6.23)$$

$$0 \le C_{w-in,i} \le C_{w-in,i,max}, \quad (6.24)$$

$$d_{w-in,i,min} \le d_{w-in,i}(k) \le d_{w-in,i,max} \text{ for all } k \in \{0,1,\dots,N-1\}. \quad (6.25)$$

(ii) $\left\{R^*_{w-out,i}, C^*_{w-out,i}, \{d^*_{w-out,i}(l)\}_{l=0}^{N-1}\right\} =$

$$\underset{\left\{R_{w-out,i},C_{w-out,i},\{d_{w-out,i}(l)\}_{l=0}^{N-1}\right\}}{\arg\min} \sum_{k=0}^{N-1}\left(T_{w-out,i}(k+1) - T^{\mathrm{E+}}_{w-out,i}(k+1)\right)^2 \quad (6.26)$$

Subject to:

$$C_{w-out,i}\frac{T_{w-out,i}(k+1)-T_{w-out,i}(k)}{T_s} = \frac{T^{\mathrm{E+}}_{ext,i}(k)-T_{w-out,i}(k)}{R_{w-out,i}} + \frac{T^{\mathrm{E+}}_{w-in,i}(k)-T_{w-out,i}(k)}{R_{ww,i}} + d_{w-out,i}(k)$$

for all $k \in \{0,1,\dots,N-1\}$, ( 6.27 )

$$0 \le R_{w-out,i} \le R_{w-out,i,max}, \quad (6.28)$$

$$0 \le C_{w-out,i} \le C_{w-out,i,max}, \quad (6.29)$$

$$d_{w-out,i,min} \le d_{w-out,i}(k) \le d_{w-out,i,max} \text{ for all } k \in \{0,1,\dots,N-1\}. \quad (6.30)$$

The notations used in the above optimization problems are explained in the nomenclature (Table 6.1). Note that since $i$ is an external wall, $T^{\mathrm{E+}}_{int,i}(k)$ represents a zone temperature and $T^{\mathrm{E+}}_{ext,i}(k)$ represent ambient/ground temperature. The resistance $R_{w-in,i}$ is assumed to be known since it can be determined from the zone level identification described in Section 6.3.2. The length of the optimization time-window $N$, and the upper and lower bounds $R_{ww,i,max}$, $C_{w-in,i,max}$, $d_{w-in,i,min}$, $d_{w-in,i,max}$, $R_{w-out,i,max}$, $C_{w-out,i,max}$, $d_{w-out,i,min}$ and $d_{w-out,i,max}$ are treated as design parameters which can be tuned if necessary.

The above identification procedure was applied to the wall number 34 in the model, which is an external wall that faces zone F2 on one side and ambient on the other. A 10 hour long time window corresponding to 8 am – 6 pm was used for the optimization, which corresponds to the period when the building is occupied. The data used for the EnergyPlus simulation was the same as that generated in Section 6.3.2 using a PRBS set-point signal. The sample time, $T_s$ was set to 1 minute. The upper bound, $R_{ww,34,max}$, was chosen to be 100 K/kW. Figure 6.24 shows a comparison of predicted values of inside surface temperature $T_{w-in,34}$ obtained from the identified model with its values obtained from the EnergyPlus simulation. It is observed that the two plots match very well. However, an investigation of the identified parameters reveals that the identified resistance $R_{ww,34}$ has a value of 96 K/kW which is significantly large when compared to the theoretical value of 31 K/kW computed from material properties (see Section 6.3.3.3 for details on computation based on material properties). Similarly, on applying a more conservative upper bound of $R_{ww,34,max} = 40$ K/kW, the predictions from the identified model matched well with the EnergyPlus simulation as shown in Figure 6.25. The identified resistance $R_{ww,34}$ had a value of 40 K/kW in this case, which was close to its afore-mentioned theoretical value. Upon further lowering $R_{ww,34,max}$ to 10 K/kW, predictions from identified model were again close to EnergyPlus simulation data (Figure 6.26), but $R_{ww,34}$ was identified as 9 K/kW which was significantly smaller than its theoretical value. Note that for visual clarity, the results in Figure 6.24 to Figure 6.26 are shown only for the first hour of the optimization time window, i.e. from 8 am to 9 am. The relevant codes for performing these optimizations are provided in Appendix H.

These observations suggest that the least squares identification methodology presented above is not reliable since the identified resistance appears to be a function of the upper bound imposed on it. This behavior can be explained on the basis that since $d_{w-in,i}(k)$ is an independent variable in the above optimization framework, for each $k \in \{0,1,\ldots,N-1\}$, it can be chosen in such a way that $T_{w-in,i}$ matches $T^{E+}_{w-in,i}$, irrespective of the value of $R_{ww,i}$. This over-parameterization can be easily verified from the following constraint used in the optimization

$$C_{w-in,i}\frac{T_{w-in,i}(k+1)-T_{w-in,i}(k)}{T_s} = \frac{T^{\mathrm{E+}}_{int,i}(k)-T_{w-in,i}(k)}{R_{w-in,i}} + \frac{T^{\mathrm{E+}}_{w-out,i}(k)-T_{w-in,i}(k)}{R_{ww,i}} + d_{w-in,i}(k).$$

To address the above limitation of the least squares methodology presented, we seek a framework where $d_{w-in,i}(k)$ is not an independent variable for all values of $k \in \{0,1,\dots,N-1\}$, as proposed in the next section.

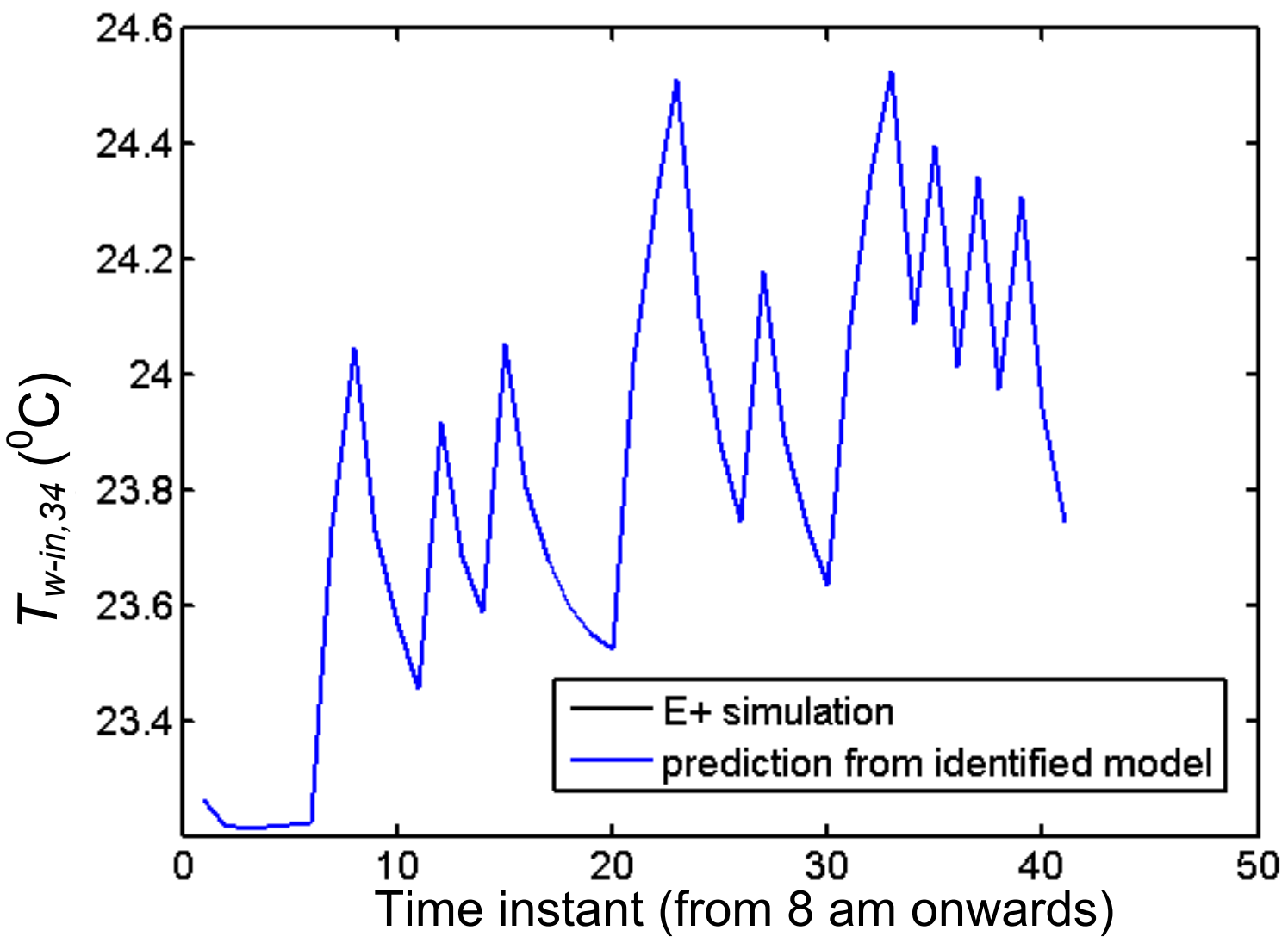

**Figure 6.24 Inside surface temperature of wall 34 from least squares identification methodology (section 6.3.3.1) compared with EnergyPlus data ($R_{ww,34,max}$ = 100 K/kW)**

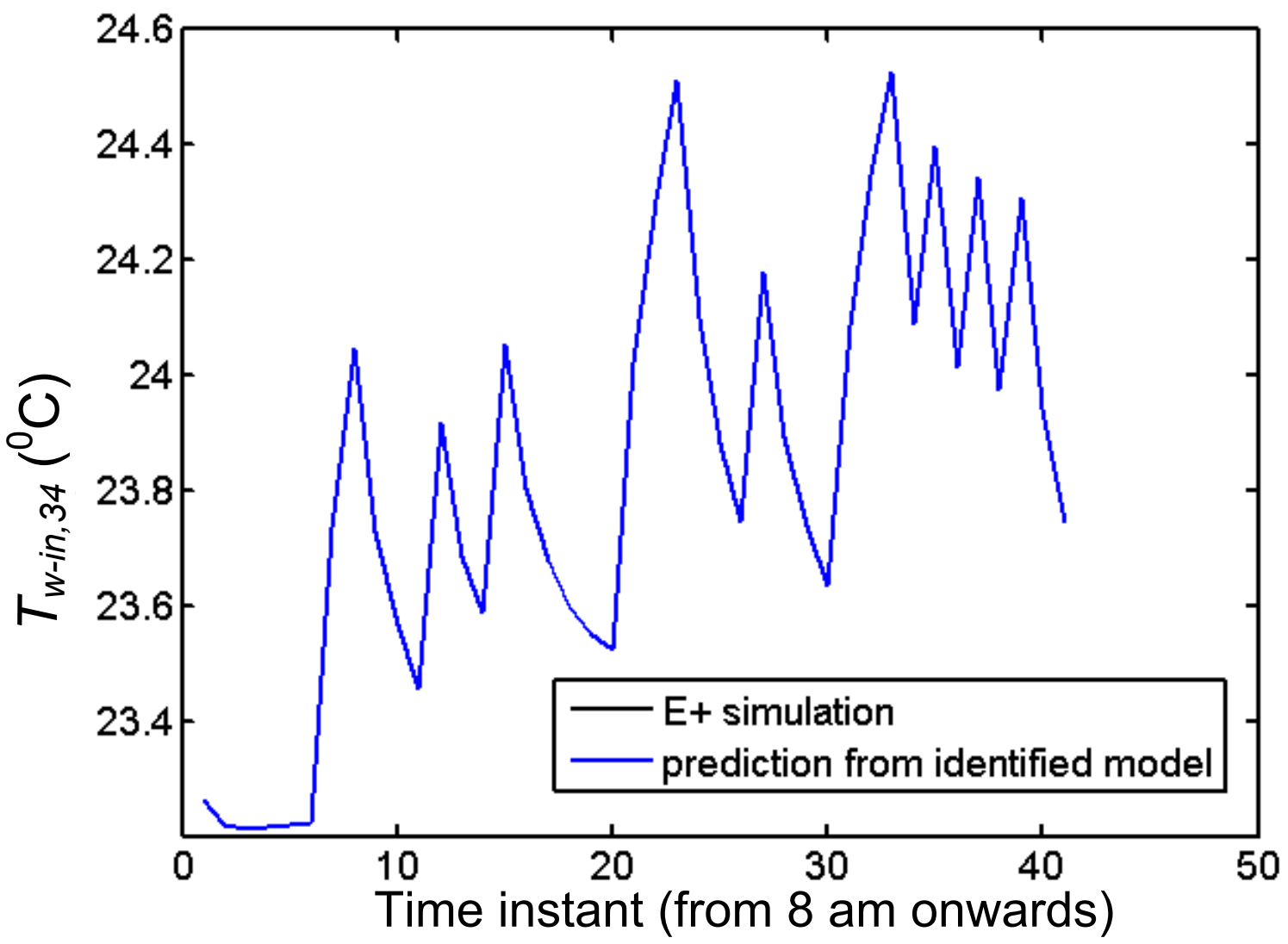

**Figure 6.25 Inside surface temperature of wall 34 from least squares identification methodology (section 6.3.3.1) compared with EnergyPlus data ($R_{ww,34,max}$ = 40 K/kW)**

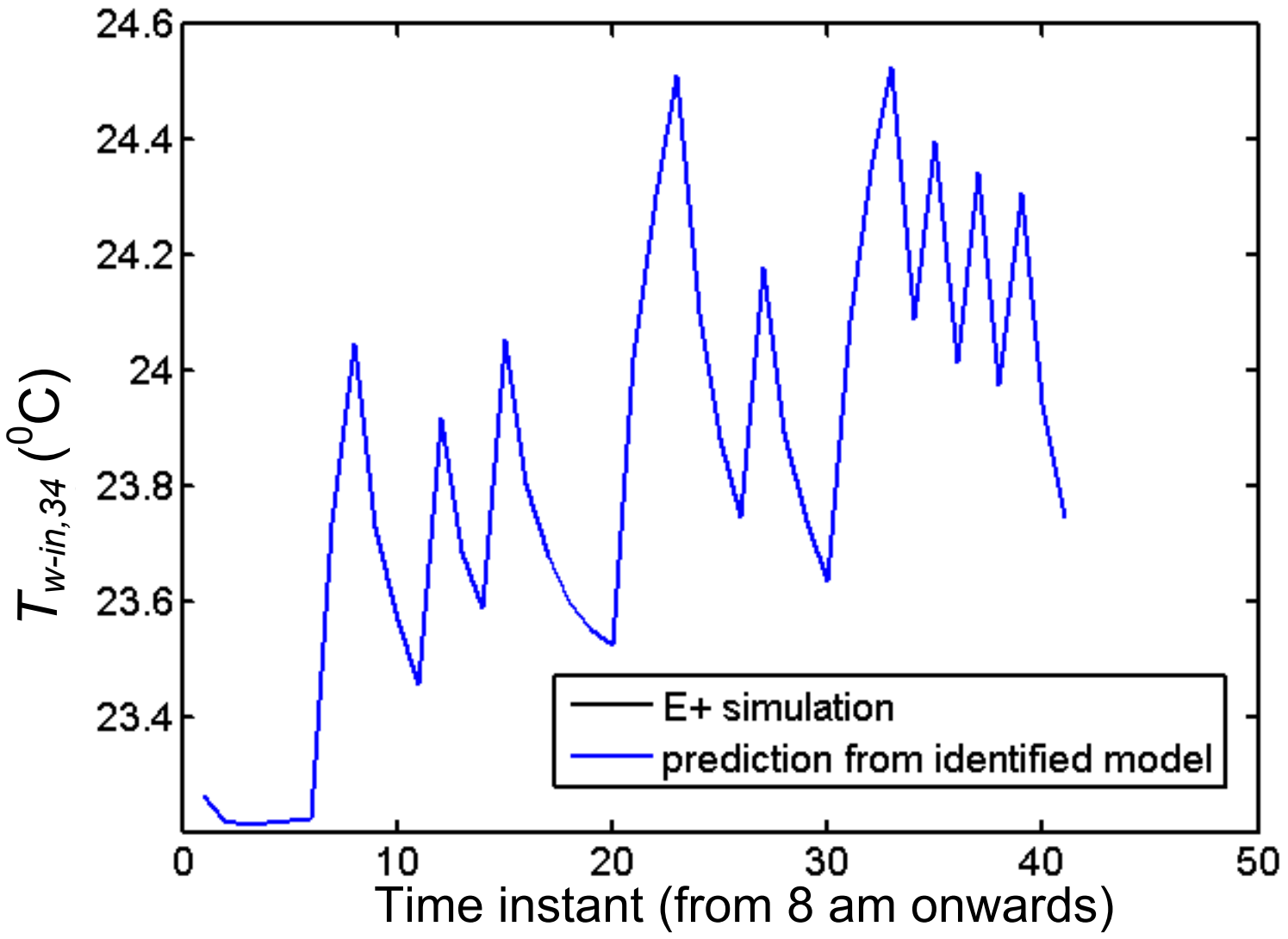

**Figure 6.26 Inside surface temperature of wall 34 from least squares identification methodology (section 6.3.3.1) compared with EnergyPlus data ($R_{ww,34,max}$ = 10 K/kW)**

### 6.3.3.2 Least squares identification with known wall thermal loads

The least squares identification procedure described here assumes that the wall thermal loads $d_{w-out,i}(k)$ and $d_{w-in,i}(k)$ are known at each time $k$. This is based on the assumption that these thermal loads are a result of radiation – both long-wave and short-wave – acting on the wall surfaces. Radiation heat transfer associated with each wall surface can be obtained from EnergyPlus simulations. The reader is directed to [97] for more details on procedure to obtain radiation data from EnergyPlus.

The identification procedure involves solving a pair of optimization problems for each internal and external wall, which are modifications of the optimization problems in Section 6.3.3.1. The pair of optimization problems corresponding to the $i^{th}$ internal wall is given by:

(i) $$\{R^*_{ww,i}, C^*_{w-in,i}\} = \underset{\{R_{ww,i}, C_{w-in,i}\}}{\arg\min} \sum_{k=0}^{N-1} \left(T_{w-in,i}(k+1) - T^{E+}_{w-in,i}(k+1)\right)^2 \quad (6.31)$$

Subject to:

$$C_{w-in,i} \frac{T_{w-in,i}(k+1) - T_{w-in,i}(k)}{T_s} = \frac{T^{E+}_{int,i}(k) - T_{w-in,i}(k)}{R_{w-in,i}} + \frac{T^{E+}_{w-out,i}(k) - T_{w-in,i}(k)}{R_{ww,i}} + d^{E+}_{w-in,i}(k)$$

for all $k \in \{0,1,\dots,N-1\}$, ( 6.32 )

$$0 \leq R_{ww,i} \leq R_{ww,i,max}, \quad (6.33)$$

$$0 \leq C_{w-in,i} \leq C_{w-in,i,max}. \quad (6.34)$$

(ii) $$C^{*}_{w-out,i} = \underset{C_{w-out,i}}{\arg\min} \sum_{k=0}^{N-1}\left(T_{w-out,i}(k+1) - T^{\mathrm{E+}}_{w-out,i}(k+1)\right)^{2} \quad (6.35)$$

Subject to:

$$C_{w-out,i}\frac{T_{w-out,i}(k+1)-T_{w-out,i}(k)}{T_s} = \frac{T^{\mathrm{E+}}_{ext,i}(k)-T_{w-out,i}(k)}{R_{w-out,i}} + \frac{T^{\mathrm{E+}}_{w-in,i}(k)-T_{w-out,i}(k)}{R_{ww,i}} + d^{\mathrm{E+}}_{w-out,i}(k)$$

for all $k \in \{0,1,\ldots,N-1\}$, ( 6.36 )

$$0 \leq C_{w-out,i} \leq C_{w-out,i,max}. \quad (6.37)$$

The pair of optimization problems corresponding to the $i^{th}$ external wall is given by:

(i) $$\left\{R^{*}_{ww,i}, C^{*}_{w-in,i}\right\} = \underset{\{R_{ww,i},C_{w-in,i}\}}{\arg\min} \sum_{k=0}^{N-1}\left(T_{w-in,i}(k+1) - T^{\mathrm{E+}}_{w-in,i}(k+1)\right)^{2} \quad (6.38)$$

Subject to:

$$C_{w-in,i}\frac{T_{w-in,i}(k+1)-T_{w-in,i}(k)}{T_s} = \frac{T^{\mathrm{E+}}_{int,i}(k)-T_{w-in,i}(k)}{R_{w-in,i}} + \frac{T^{\mathrm{E+}}_{w-out,i}(k)-T_{w-in,i}(k)}{R_{ww,i}} + d^{\mathrm{E+}}_{w-in,i}(k)$$

for all $k \in \{0,1,\ldots,N-1\}$, ( 6.39 )

$$0 \leq R_{ww,i} \leq R_{ww,i,max}, \quad (6.40)$$

$$0 \leq C_{w-in,i} \leq C_{w-in,i,max}. \quad (6.41)$$

(ii) $\left\{R^{*}_{w-out,i}, C^{*}_{w-out,i}\right\} =$

$$\underset{\{R_{w-out,i},C_{w-out,i}\}}{\arg\min} \sum_{k=0}^{N-1}\left(T_{w-out,i}(k+1) - T^{\mathrm{E+}}_{w-out,i}(k+1)\right)^{2} \quad (6.42)$$

Subject to:

$$C_{w-out,i}\frac{T_{w-out,i}(k+1)-T_{w-out,i}(k)}{T_s} = \frac{T^{\mathrm{E+}}_{ext,i}(k)-T_{w-out,i}(k)}{R_{w-out,i}} + \frac{T^{\mathrm{E+}}_{w-in,i}(k)-T_{w-out,i}(k)}{R_{ww,i}} + d^{\mathrm{E+}}_{w-out,i}(k)$$

for all $k \in \{0,1,\ldots,N-1\}$, ( 6.43 )

$$0 \leq R_{w-out,i} \leq R_{w-out,i,max}, \quad (6.44)$$

$$0 \leq C_{w-out,i} \leq C_{w-out,i,max}. \quad (6.45)$$

The notations used in the above optimization problems are explained in the nomenclature (Table 6.1). Similar to Section 6.3.3.1, the above identification procedure was applied to the inside surface of wall 34 in the model. A 10 hour long time window corresponding to 8 am – 6 pm was used for the optimization. The data used for the EnergyPlus simulation was the same as that generated in Section 6.3.2 using a PRBS set-point signal and the sample time, $T_s$ was set to 1 minute. Figure 6.27 shows the comparison of predicted values of surface temperature $T_{w-in.34}$ obtained using the identified model with its values obtained from the EnergyPlus simulation. It was observed that the two plots deviate significantly from one another. The relevant codes for performing this optimization are provided in Appendix I.

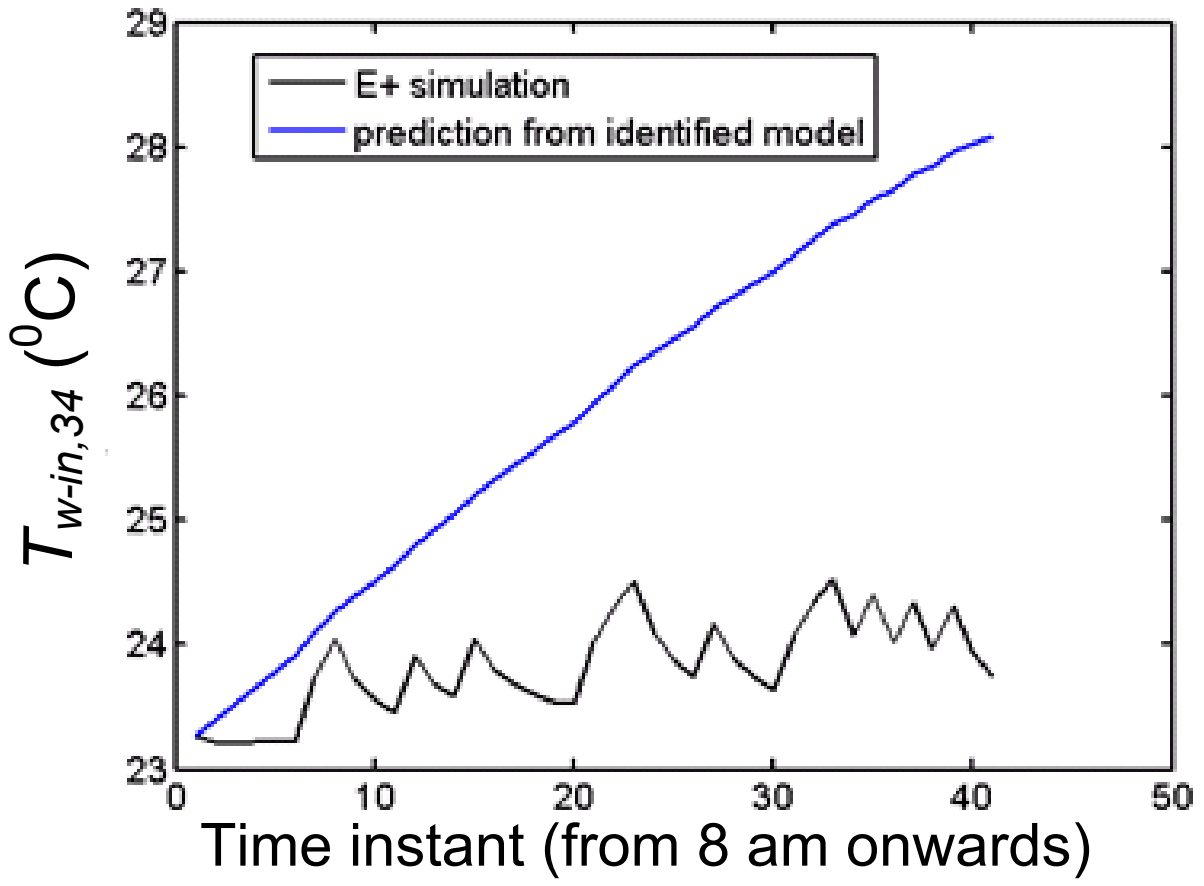

**Figure 6.27 Inside surface temperature of wall 34 from least squares identification methodology (section 6.3.3.2) compared with EnergyPlus data**

A possible reason for the above deviation is the difference in the order of the corresponding models. While the identified model for each wall is second because of its representation by two capacitances, EnergyPlus uses a model which is much higher in order, typically representing each construction layer by 6 – 18 capacitances connected by resistances [97].

The inability of the two identification approaches presented in Sections 6.3.3.1 and 6.3.3.2 to provide accurate and reliable parameter estimates suggests that data-based identification from EnergyPlus is not a suitable framework for obtaining 3R2C models for walls.

Therefore, we proceed to calculate the resistances and capacitances associated with the walls directly based on material properties of their corresponding construction layers, as described in the next section.

#### 6.3.3.3 Direct computation of parameters

The steps to compute the parameters associated with the walls from the material properties of the construction layers are as follows:

1. For each wall type in the building (Table 6.3), the quantities $\bar{C}$ and $\bar{R}$ are computed as follows:

$$\bar{C} = \sum_{j=1}^{n} \rho_j c_{pj} l_j \tag{6.46}$$

$$\bar{R} = \sum_{i=1}^{n} \frac{l_j}{k_j} \tag{6.47}$$

In the above equations, $n$ is the number of material layers, whereas $\rho_j$, $c_{pj}$, $l_j$ and $k_j$ represent the density, specific heat capacity, thickness and thermal conductivity, respectively, of layer $j$.

2. For each wall $i$, the total capacitance associated with it, $C_i$ is computed as

$$C_i = A_i \bar{C_i}, \tag{6.48}$$

where $A_i$ is the surface area of wall $i$ and $\bar{C_i}$ is the value of $\bar{C}$ for the corresponding wall type computed using (6.46). We assume that the capacitance $C_i$ is equally divided to represent the capacitances associated with the two surfaces of wall $i$. Therefore,

$$C_{w-in,i} = C_{w-out,i} = \frac{C_i}{2}. \tag{6.49}$$

3. For each wall $i$, the resistance $R_{ww,i}$ associated with heat transfer between its two surfaces, is computed as

$$R_{ww,i} = \frac{\bar{R}_i}{A_i}, \tag{6.50}$$

where, as before, $A_i$ is the surface area of wall $i$ and $\bar{R}_i$ is the value of $\bar{R}$ for the corresponding wall type computed using (6.47 ).

The above calculations together with the resistances calculated via the zone level identification described in section 6.3.2 provide all the parameters shown in Figure 6.12 except the resistances $R_{w-out,i}$ between external walls $i$ and ambient/ground. These resistances are computed as follows:

1. The convection heat transfer coefficient, $h_{w-out,i}$ between ambient and an external wall $i$ facing ambient can be directly obtained from EnergyPlus simulations. Let $\bar{h}_{w-out,i}$ denote the average value of $h_{w-out,i}$ over an appropriate simulation window such as 24 hours, i.e.

$$\bar{h}_{w-out,i} = \frac{\sum_{k=0}^{N-1} h_{w-out,i}(k)}{N}, \tag{6.51}$$

where $N$ denotes the number of samples in the simulation time window. The resistance $R_{w-out,i}$ is then computed as

$$R_{w-out,i} = \frac{1}{A_i \bar{h}_{w-out,i}}. \tag{6.52}$$

2. The resistance $R_{w-out,i}$ between ground and an external wall $i$ facing ground is set to 0, which is consistent with the modeling assumption in EnergyPlus that ground facing walls have the same temperature as ground. In other words, for such a wall, the temperature $T_{w-out,i}$ of the surface facing ground is not a state. Therefore, the dynamics for such a wall is first order represented only by the state $T_{w-in,i}$.

For the test building under consideration, the calculated values of $\bar{C}$ and $\bar{R}$ using (6.46) and (6.47 ) for the various wall types are shown in Table 1. Using the steps mentioned above, all resistances and capacitances corresponding to the wall surface level identification were computed using a spreadsheet which is provided in the media accompanying this thesis.

### 6.3.4 Construction of LTI model

The resistances and capacitances computed in sections 6.3.2 and 6.3.3 were used to construct an LTI model to represent the thermal dynamics for the test building. This was done in accordance with the graph based procedure described in Algorithm 5.1 The code used to obtain the model is provided in Appendix J. The resulting state space model has the structure shown in (5.1) and is of order 112. Of these, 103 states represent wall temperatures for the inside and outside surfaces and 9 states represent zone temperatures.

**Table 6.6 $\overline{C}$ and $\overline{R}$ values for various wall types in the test building**

| Wall type | $\overline{C}$ (kJ/K-m$^2$) | $\overline{R}$ (K/kW-m$^2$) |
|---|---|---|
| Ceilings facing ambient | $6.705 \times 10^1$ | $4.355 \times 10^3$ |
| Interior walls | $3.314 \times 10^1$ | $3.387 \times 10^2$ |
| Interior doors | $2.517 \times 10^1$ | $1.693 \times 10^2$ |
| Walls facing ambient | $4.271 \times 10^2$ | $2.071 \times 10^3$ |
| Interior floors/ceilings | $1.188 \times 10^3$ | $2.577 \times 10^3$ |
| Floors facing ground | $1.19 \times 10^2$ | $2.94 \times 10^2$ |
| Windows facing ambient | 6.3 | $2.256 \times 10^2$ |

# 6.4 Control architecture determination

In this section, the LTI model developed in Section 6.3.4 for the test building under investigation is used to determine appropriate control architectures which provide a satisfactory tradeoff between optimality and robustness. The OLF-FPM method described in Chapter 4 is used for this purpose. The reasons for using the OLF-FPM method over the CLF-MCS method were explained in Chapter 4.

## 6.4.1 Modifications to LTI model

The EnergyPlus model of the test building that was constructed in Section 6.2 was based on the assumption that the thermal zones are completely separated from one another by solid walls. However in the actual building being modeled, openings are present in the walls at several places to facilitate movement of people in the building as illustrated in Figure 6.28. From a thermodynamic point of view, these openings allow direct thermal interaction between the zones, in addition to the thermal interactions occurring through the walls which separate them. These additional thermal interactions can be modeled as resistances which directly couple the associated zone capacitances.

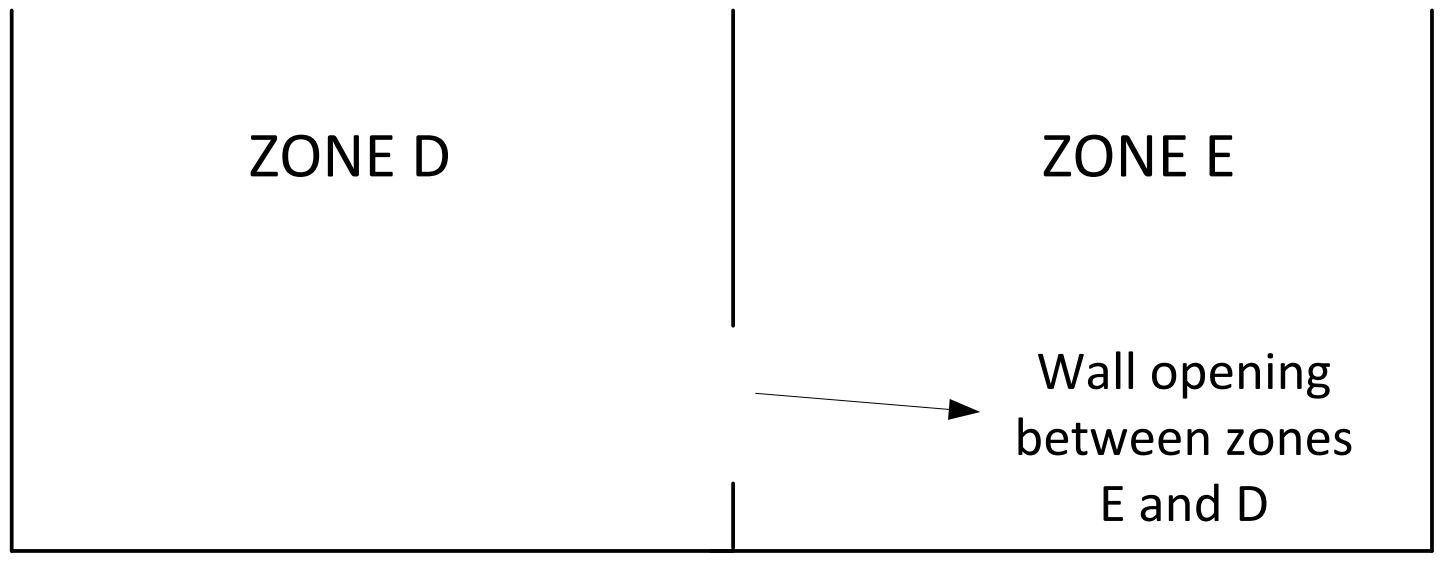

**Figure 6.28 Illustration of wall opening present between thermal zones E and D in the test building**

In the absence of accurate information about these openings, we assume that each pair of adjacent zones on a floor has a wall opening whose area is 5% of the surface area of the wall that separates these zones. The resistance $R_{ij}$ associated with the thermal interaction between each such pair of zones $\{i, j\}$ can be computed as

$$R_{ij} = \frac{1}{h_{opening}(0.05A_{ij})} . \quad (6.53)$$

Here, $h_{opening}$ is the coefficient for heat transfer between the pair of zones $\{i, j\}$ through the wall opening between them. Its value, based on the properties of the "air wall" element in OpenStudio is 0.06 kW/K-m$^2$. The values of these resistances for each adjacent pair of zones in each floor of the test building computed based on (6.53) are shown in Table 6.7

The LTI model obtained in Section 6.3.4 is modified by incorporating coupling terms in in its state transition matrix between the zone temperatures states. The code used to obtain the modified model is provided in Appendix J. Note that even after modification, the resulting LTI model has the structure given by (5.1). The LTI model was then discretized using a sample time of 60 seconds. The OLF-FPM partitioning procedure presented in Section 4.3 was applied based on this model. The OLF calculations were based on the parameters $N = 30$, $\alpha = 1\mathbf{e}_9$ and $\beta = 10^3\mathbf{e}_{11}$ (see nomenclature in Chapter 4). The p-partitions obtained via agglomerative clustering (Algorithm 4.1), are shown in Table 6.8. The resulting optimality robustness trade-off curve is shown in Figure 6.29. The codes used for discretization of the model and application of agglomerative clustering are included in Appendix J.

**Table 6.7 Thermal resistances due to wall opening between each pair of adjacent zones on each floor of the test building**

| Pair of zones | Thermal resistance due to wall opening (K/kW) |
|---|---|
| {C,D} | 1.318 |
| {G,D} | 2.079 |
| {SR,D} | 6.615 |
| {G,SR} | 6.615 |
| {G,E} | 1.582 |
| {G,F2} | 3.164 |
| {G,F1} | 3.164 |
| {TCB},{NTCB} | 1.683 |

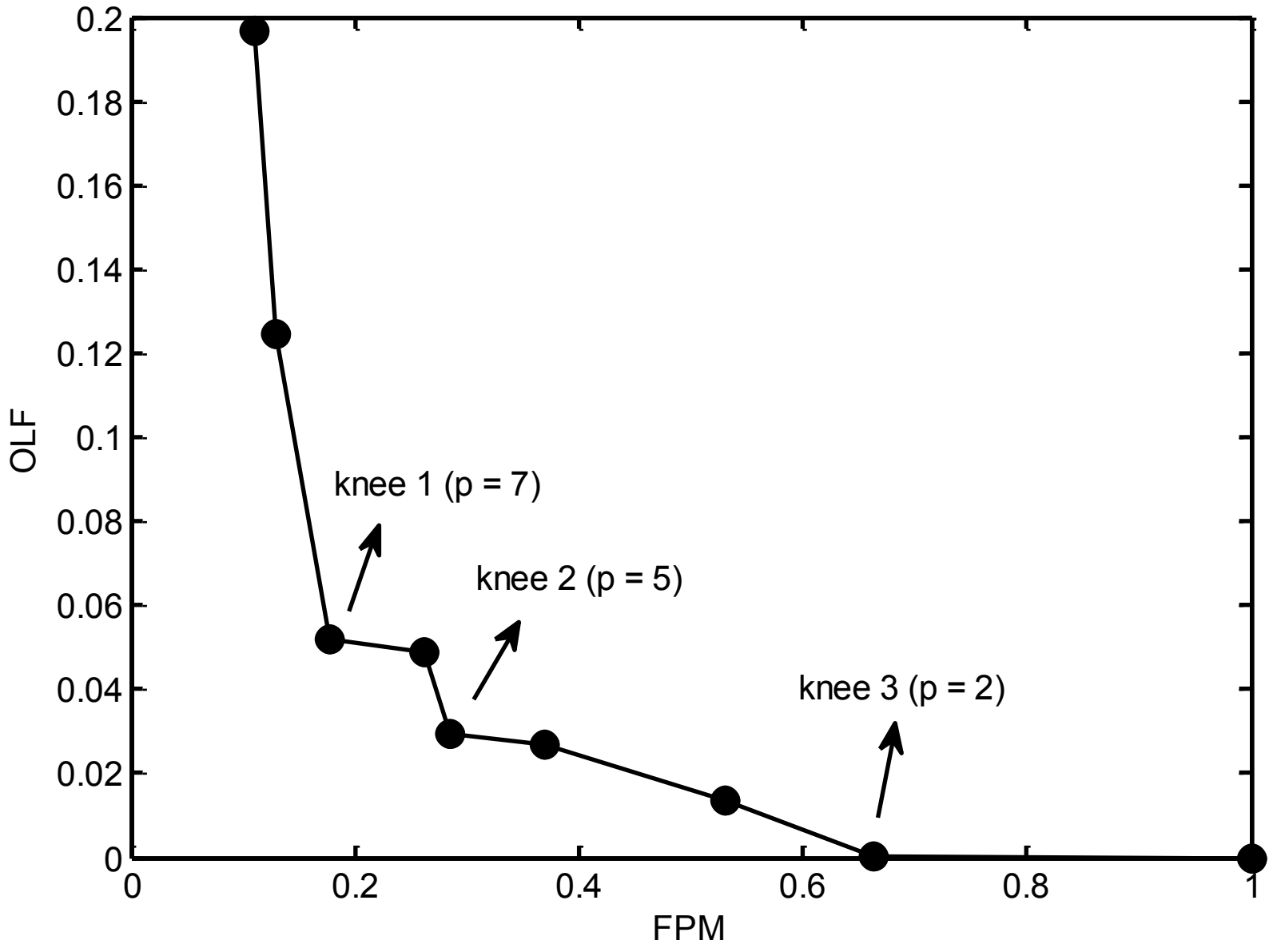

**Figure 6.29 Optimality-robustness tradeoff curve for test building (p denotes the number of clusters in a partition).**

**Table 6.8 Partitions using agglomeration for test building**

| ***p*** | **Partition from agglomeration** |
|---|---|
| 9 | {G}{E}{C}{SR}{F1}{D}{F2}{TCB}{NTCB} |
| 8 | {G}{E}{C}{SR,D}{F1}{F2}{TCB}{NTCB} |
| 7 (knee 1) | {G,SR,D}{E}{C}{F1}{F2}{TCB}{NTCB} |
| 6 | {G,SR,D,E}{C}{F1}{F2}{TCB}{NTCB} |
| 5 (knee 2, optimally decentralized partition) | {G,SR,D,E}{C}{F1}{F2}{TCB,NTCB} |
| 4 | {G,SR,D,E,F2}{C}{F1}{TCB,NTCB} |
| 3 | {G,SR,D,E,F2,C}{F1}{TCB,NTCB} |
| 2 (knee 3) | {G,SR,D,E,F2C,F1}{TCB,NTCB} |
| 1 | {G,SR,D,E,F2C,F1,TCB,NTCB} |

The following observations are made from these results.

1. From visual inspection, three knee points are observed as shown in Figure 6.29. These correspond to partitions of size 2, 5 and 7 clusters respectively.
2. Knee 2 is more centrally located than the other two knees and is therefore treated as the partition with the most appropriate balance between optimality and robustness. Therefore, we refer to the architecture corresponding to this partition as the *optimally decentralized* architecture.

A proposed explanation for the clusters appearing in the optimally decentralized architecture is presented below based on the building layout described in Section 6.2.1.

1. The zones G, D and E form a closely coupled triplet. This is because these zones are interconnected in a hub and spoke manner with zone G acting as the hub. The zone SR is fully contained within zone D and therefore is expected to be strongly coupled with it. The zones F1 and F2, which also act as spokes connected to the hub G, are not as strongly coupled to zone G as zones D and E. This is because the area of the wall connecting zones F1 or F2 to G is almost half of the area of the wall connecting zone E or D to G. Hence, it is expected that zones G, D, SR and E form a closely coupled quartet, verified from the fact that they constitute one cluster in the optimally decentralized architecture (see Table 6.8).

2. The zones F1 and F2 are expected to be weakly coupled because they lie on two different floors. The internal floor separating them has a significant amount of insulation due to air-gap in the construction layers (Table 6.3). Also, there is no wall opening because these zones are on different floors. Hence it is expected that zones F1 and F2 lie in different clusters in the optimally decentralized architecture, which is verified from Table 6.8.
3. Based on the layout shown in Figure 6.4, the zone C is expected to be weakly coupled to the rest of the building. This is because it is connected to the building through a single wall. Note that every other zone is connected to the rest of the building through at least two walls (for example, zone E is connected to zones G and NTCB). Hence, it is expected that zone C be put into a single cluster in the optimally decentralized architecture, which is verified from Table 6.8.
4. The zones TCB and NTCB are located entirely in the basement of the building and are therefore expected to be weakly coupled to the other zones in the building. However, the coupling between them is expected to be relatively strong because of a wall with a large surface area, and a proportionally large opening separating them. Hence, it is expected that zones TCB and NTCB constitute a cluster in the optimally decentralized architecture, which is verified from Table 6.8.

## 6.5 Control design and analysis

In this section, centralized and decentralized model predictive controllers are designed using the LTI model of the building thermal dynamics obtained in Section 6.3. The architectures for decentralized control are based on the results in Section 6.4. The control design is based on the framework presented in Chapter 5. Next, the performance of these controllers is evaluated in simulation on a nonlinear model of the building thermal dynamics, which is derived based on the EnergyPlus model developed in Section 6.2. Specifically, the effect of control architecture on optimality and robustness is investigated.

### 6.5.1 Control design

Centralized and decentralized output feedback model predictive controllers were designed using the control and observation framework described in Chapter 5. For the benefit of the reader, the underlying steps in the control design are summarized in Figure 6.30. The relevant sections in Chapter 5 are referenced. The codes used to implement each of the steps in Figure 6.30 for the test building are provided in Appendix K.

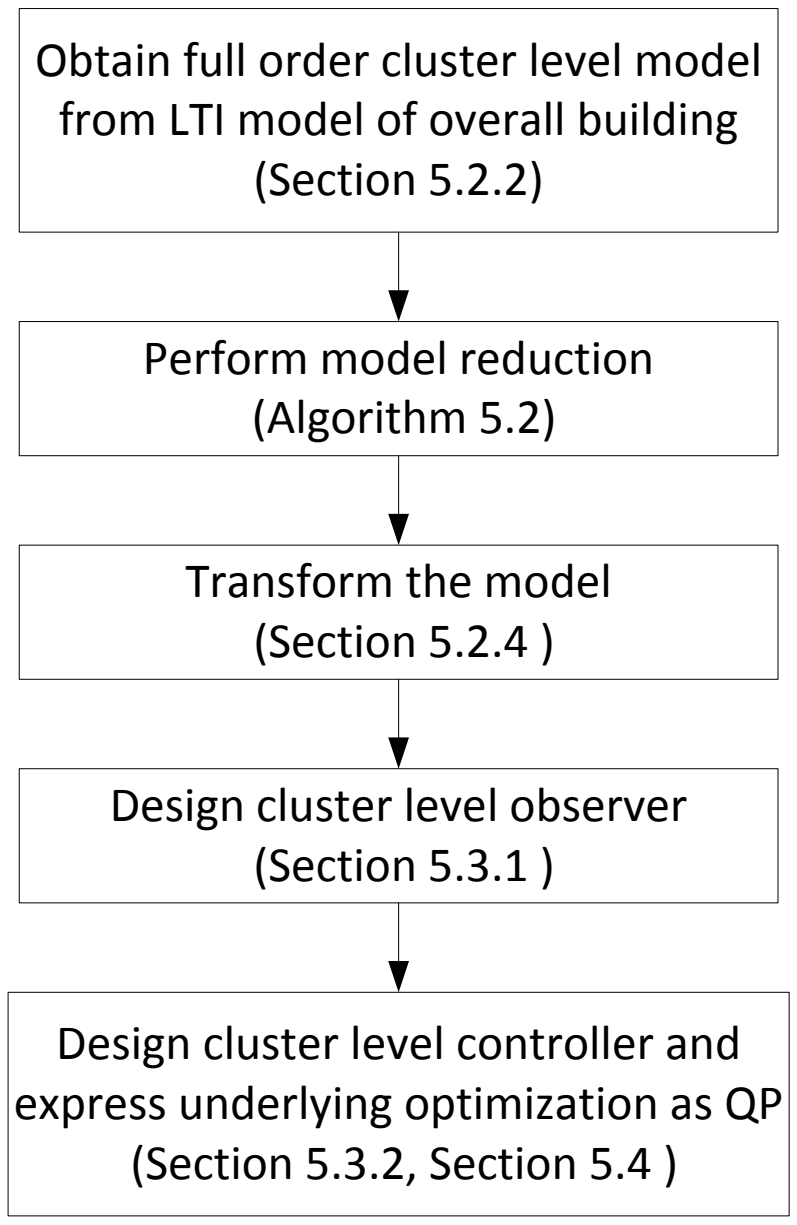

**Figure 6.30 Steps used for cluster level control design**

The decentralized controllers designed correspond to the partitions given by $p$ = 2, 5, 7 and 9 in Table 6.8. These architectures correspond to the three knee points in Figure 6.29 and also the fully decentralized case where each cluster is a zone. The centralized controller was designed using the same principles as for decentralized control. This is because a centralized architecture can be viewed as a decentralized architecture with only one cluster.

The observers were designed such that in the continuous time domain, their poles were 10 times further left on the real axis than the poles of the open loop model. As discussed in Section 5.3.1, for each cluster $i$, the temperature estimates $\mathbf{\hat{T}_z^j}$ for the zones in the clusters other than $i$ appearing in (5.11) are set to the corresponding set-point temperatures $\mathbf{T}^{\mathbf{j}}_{\mathbf{z,ref}}$. The other parameters used in the control design procedure are listed in Table 6.9 (refer to nomenclature in Chapter 5). The maximum supply air mass flow rates and reheating power which are required to set up the constraints in Section 5.4.2 are shown in Table 6.10.

**Table 6.9 Parameters used in designing controllers for each cluster $i$**

| Parameter | Description | Value |
|---|---|---|
| $N_p$ | Prediction horizon | 30 samples |
| $N_u$ | Control horizon | 15 samples |
| $\boldsymbol{\alpha}_{\mathbf{i}}$ | Penalty on performance term | $\mathbf{e}_{N_{zi}}$ |
| $\boldsymbol{\beta}_{\mathbf{i}}$ | Penalty on cost term | $10^3\mathbf{e}_{N_{zi}}$ |
| $T_{supp}$ | Supply air temperature to each zone | $12.8\ ^0\mathrm{C}$ |
| $c_{p,air}$ | Specific heat capacity of air | 1.005 kJ/kg-K |

**Table 6.10 Maximum supply air mass flow rate and reheat power available to each zone**

| Zone number (as per Table 6.2) | Maximum supply air mass flow rate (kg/s) | Maximum reheat power (kW) |
|---|---|---|
| 1 | 12.7 | 140 |
| 2 | 12.7 | 140 |
| 3 | 12.7 | 140 |
| 4 | 4.2 | 50 |
| 5 | 8.5 | 70 |
| 6 | 12.7 | 140 |
| 7 | 8.5 | 70 |
| 8 | 4.2 | 50 |
| 9 | 4.2 | 50 |

### 6.5.2 Closed loop performance assessment

#### 6.5.2.1 Nonlinear model for control evaluation

Reduced order LTI models were used for control design in Section 6.5.1, which were based on the assumption that the unmodeled thermal loads acting on the walls and zones were slowly time varying quantities. It is desired to evaluate the closed loop performance on a model which has higher fidelity when compared to the models used for control design. This is accomplished by developing a full order model which uses realistic nonlinear expressions for the thermal disturbances.

The afore-mentioned nonlinear model is obtained from the full order LTI model (5.1), by using the expressions (6.54) to (6.56) to model the thermal disturbances $\mathbf{d_w}$ and $\mathbf{d_z}$ for the walls and zones. These expressions emulate the models used by EnergyPlus to compute these disturbances [97]. Note that the temperature values in these equations should be in Kelvin. The reader is directed to the nomenclature for an explanation of the notations used.

For each internal wall surface $i \in \{1,2,\dots,N_w\}$,

$$d_{w,i}(t) = \underbrace{\sigma A_i \textstyle\sum_{j\in\mathcal{N}_i^w} \mathbf{F}_{i,j}\left(T_{w,j}(t)^4 - T_{w,i}(t)^4\right)}_{long\ wave\ radiation} + \underbrace{\alpha_i A_i q_{swr,i}(t)}_{short\ wave\ radiation}. \quad (6.54)$$

For each external wall surface $i$ ($i \in \{1,2,\dots,N_w\}$) facing the ambient,

$$d_{w,i}(t) = \underbrace{\sigma A_i \epsilon_i F_{gnd,i}\left(T_{gnd}(t)^4 - T_{w,i}(t)^4\right)}_{long\ wave\ radiation\ from\ ground} + \underbrace{\sigma A_i \epsilon_i F_{sky,i}\left(T_{sky}(t)^4 - T_{w,i}(t)^4\right)}_{long\ wave\ radiation\ from\ sky}$$

$$+ \underbrace{\sigma A_i \epsilon_i F_{air,i}\left(T_a(t)^4 - T_{w,i}(t)^4\right)}_{long\ wave\ radiation\ from\ air} + \underbrace{\alpha_i A_i q_{swr,i}(t).}_{short\ wave\ radiation} \quad (6.55)$$

There is no thermal disturbance acting on external wall surfaces facing the ground.

For each thermal zone, $i \in \{1,2,..,N_z\}$,

$$d_{z,i}(t) = \eta_{occ,i}(t) N_{occ,i} Q_{occ}(\mathrm{t}) + \eta_{light,i}(t) Q_{light,i} + \eta_{appl,i}(t) Q_{appl,i}. \quad (6.56)$$

In the above equations, $\mathbf{F}$ is a matrix of Script-F factors [97, 99], $\mathcal{N}_i^w$ is the set of all internal wall surfaces which share the same zone as internal wall surface $i$, and $F_{gnd,i}$, $F_{sky,i}$ and $F_{air,i}$ are view factors with respect to the ground, sky and air respectively for the external wall surface $i$. These quantities, along with $\epsilon_i$, $\alpha_i$, $A_i$, $T_{gnd}$, $T_{sky}$, $T_a$ and $q_{swr,i}$ can be directly obtained from the EnergyPlus model. A spreadsheet showing the values of these quantities is included in the media accompanying this thesis.

A 24-hour time window which starts at 12:00 AM on June 3 and ends at 12:00 AM on June 4 is used for simulation. The signal $T_a$ obtained from EnergyPlus for the simulation time is plotted in Figure 6.31. The quantities $\eta_{occ,i}, N_{occ,i}, \eta_{light,i}, Q_{light,i}, \eta_{appl,i}$ and $Q_{appl,i}$ in (6.56) correspond to nominal values and schedules related to occupancy, lighting and equipment, and their values used in the simulation are the same as described in Section 6.2.4. The temperature $T_g$ of the ground below the floor of the building at all times is set to be 18 $^0C$ which corresponds to the value used by EnergyPlus.

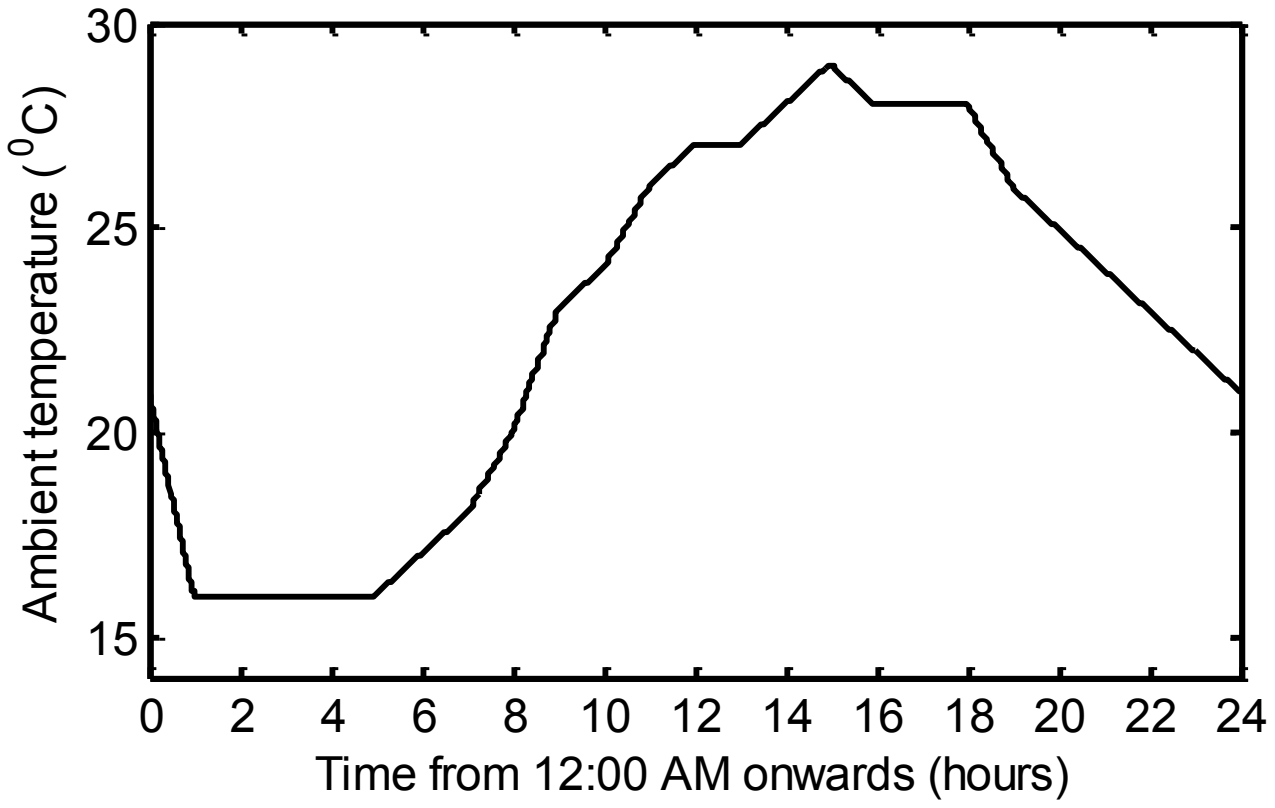

**Figure 6.31 Ambient temperature obtained from EnergyPlus weather file for the simulation time window**

#### 6.5.2.2 Optimality Analysis

The five controllers (1 centralized and 4 decentralized) designed in Section 6.5.1 were implemented on the nonlinear model obtained in Section 6.5.2.1. This was done in accordance with the block diagram shown in Figure 5.4. The associated MATLAB codes are provided in Appendix L and the SIMULINK models are enclosed in the media accompanying the thesis. A 24-hour long simulation was performed based on the settings described in Section 6.5.2.1. The desired set-point temperature for each zone is shown in Figure 6.32, which is obtained from the OpenStudio template called ‘Medium_Office_ClgSetp’. The initial temperature of all zones and walls is assumed to be 25 $^0C$. The ambient, ground and sky temperatures and short wave radiation data required to compute the disturbances in Section 6.5.2.1 were obtained by first simulating the EnergyPlus model for that day. This data is obtained from EnergyPlus in a spreadsheet which is provided in the media accompanying the thesis.

The zone temperature responses and associated control inputs (heating/cooling provided by HVAC system) corresponding to the various control architectures are shown in Figure 6.33 and Figure 6.42. From these figures, it can be observed that the temperature responses and control inputs signals vary depending on the control architecture.

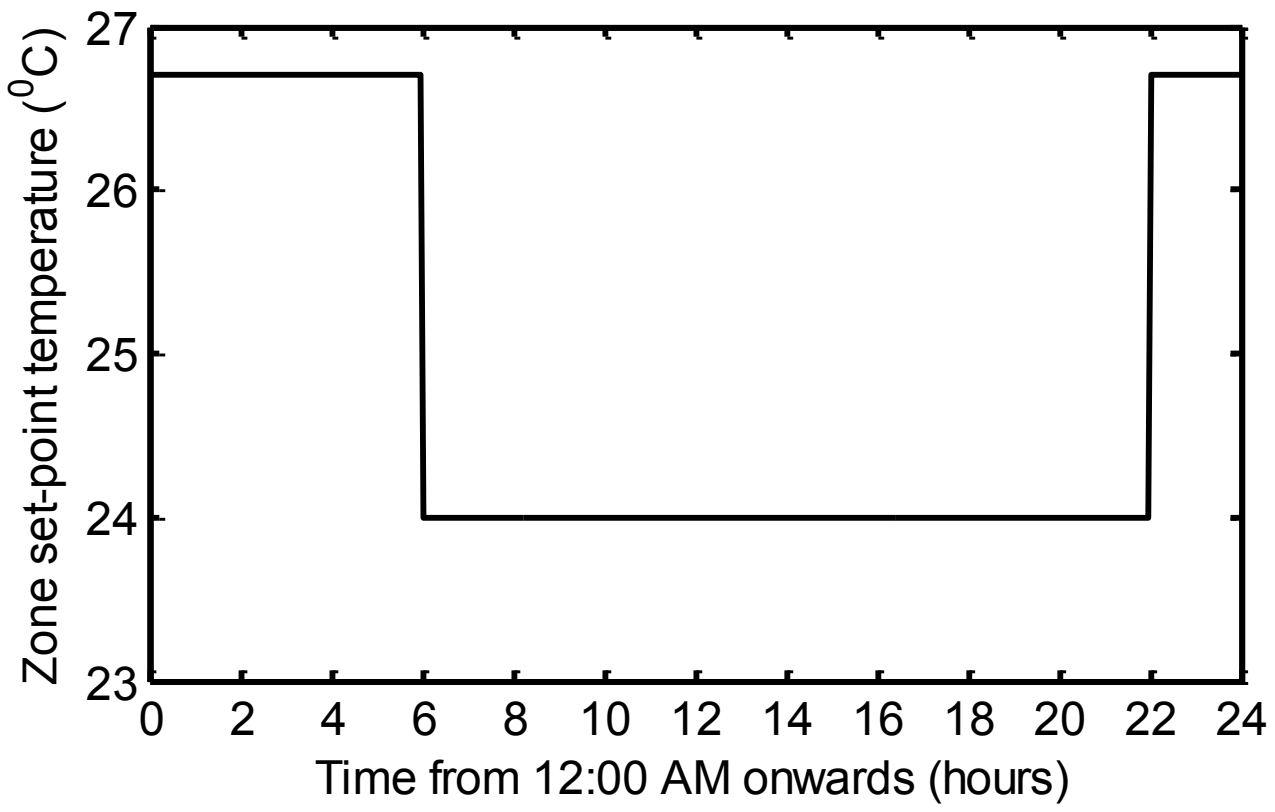

**Figure 6.32 Set-point temperatures for all zones during the simulation time window**

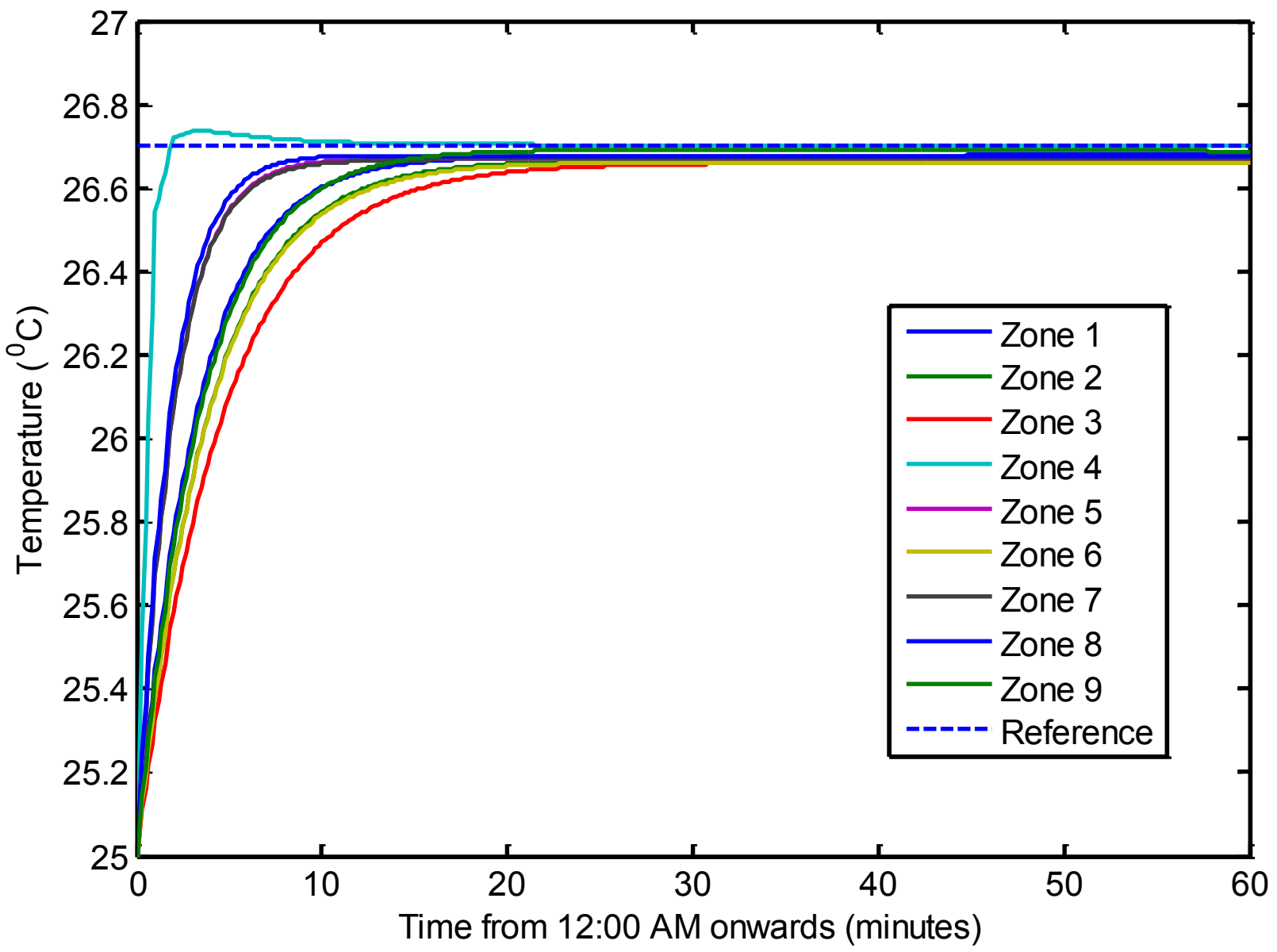

**Figure 6.33 Evolution of zone temperatures between 12:00 AM to 1:00 AM for centralized architecture**

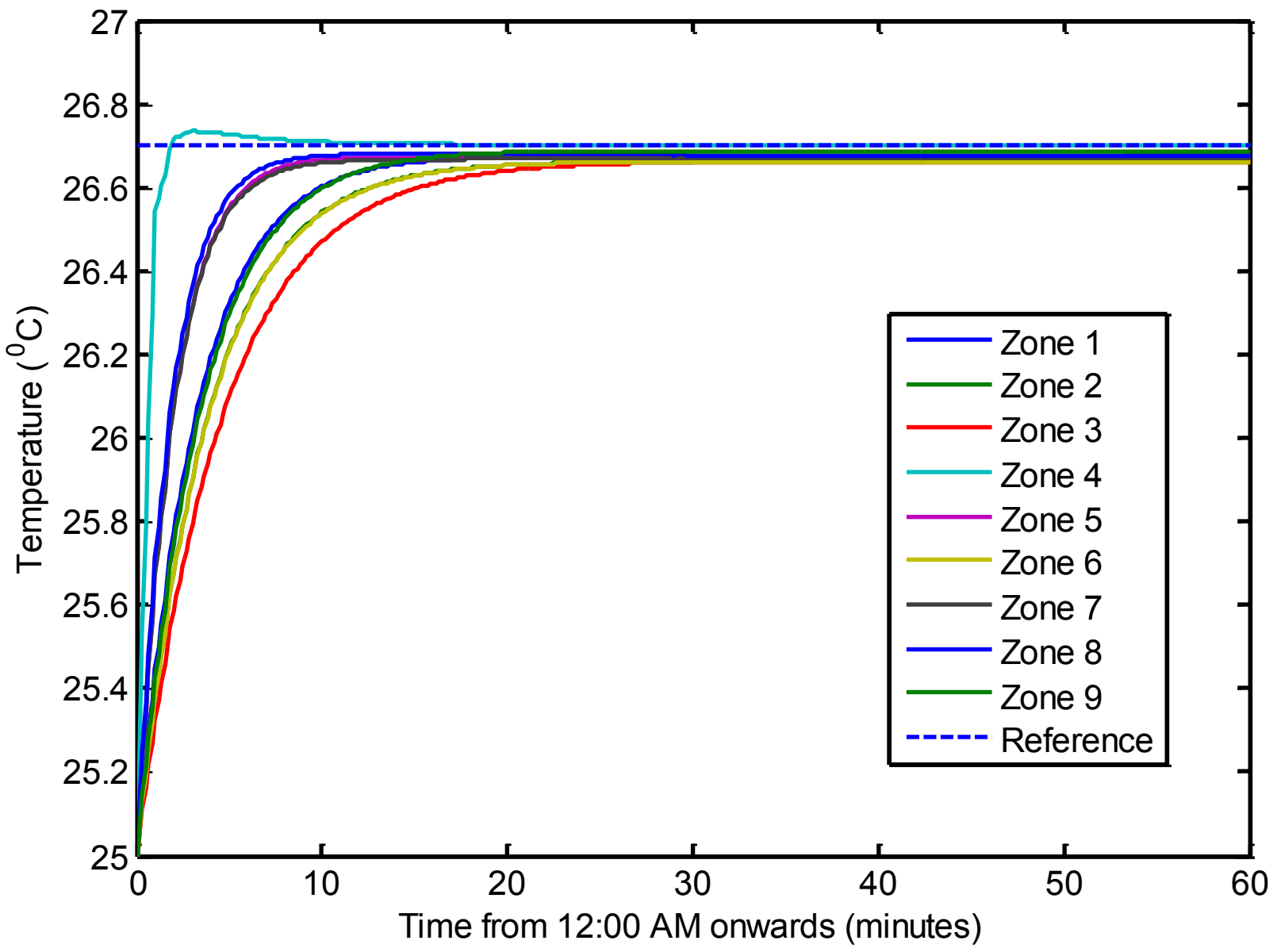

**Figure 6.34 Evolution of zone temperatures between 12:00 AM to 1:00 AM for decentralized architecture corresponding to knee 3**

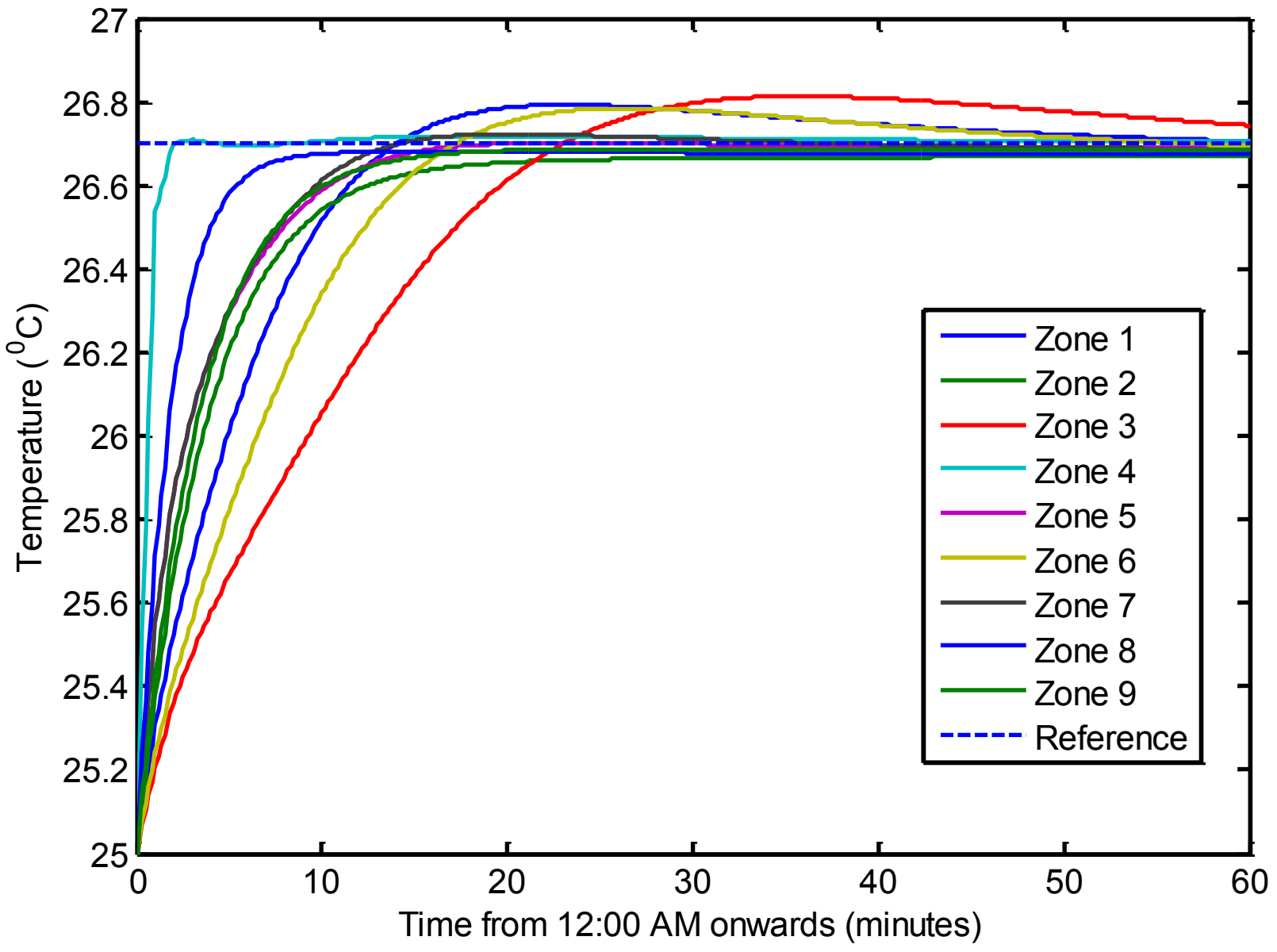

**Figure 6.35 Evolution of zone temperatures between 12:00 AM to 1:00 AM for decentralized architecture corresponding to knee 2**

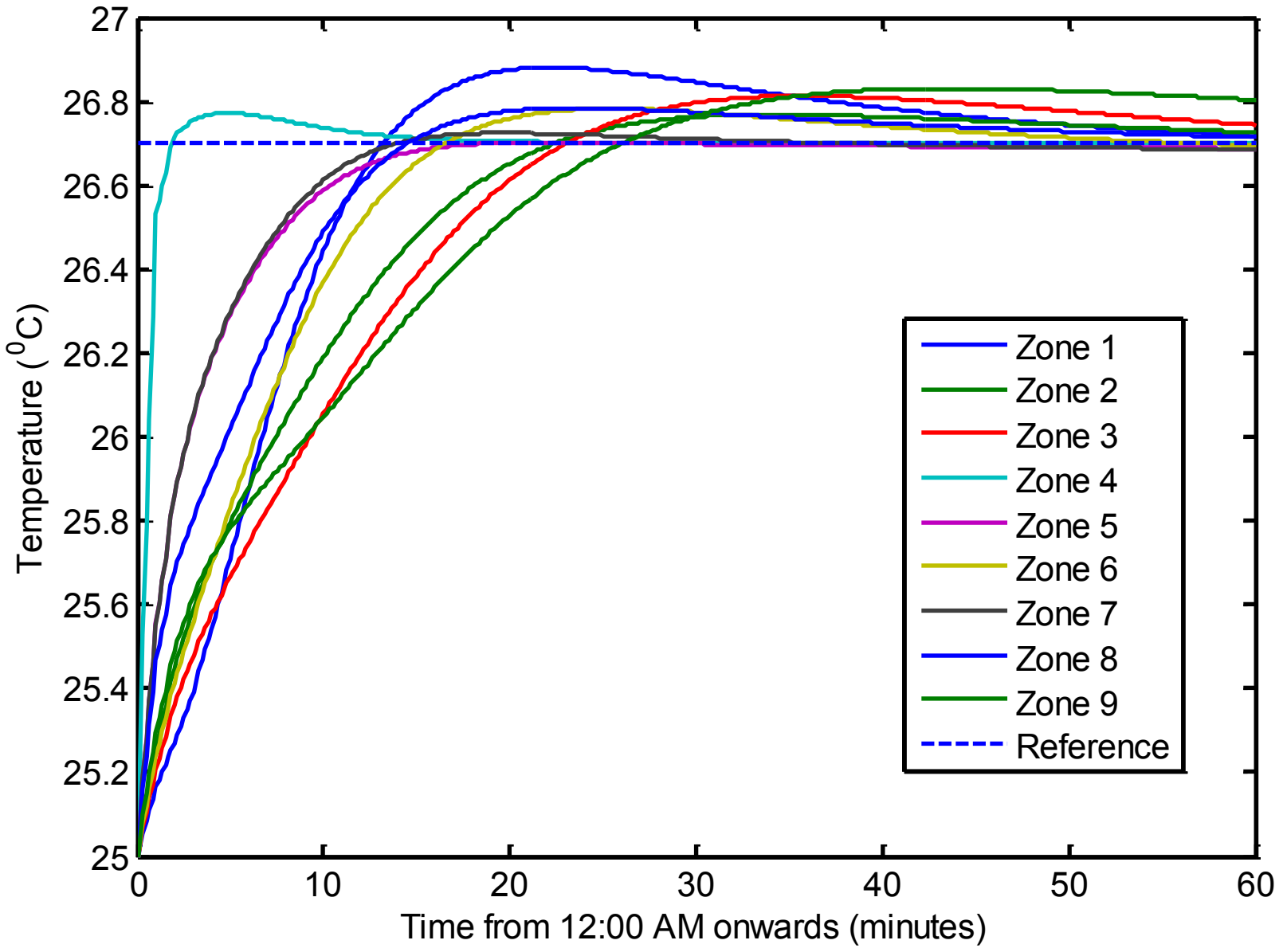

**Figure 6.36 Evolution of zone temperatures between 12:00 AM to 1:00 AM for decentralized architecture corresponding to knee 1**

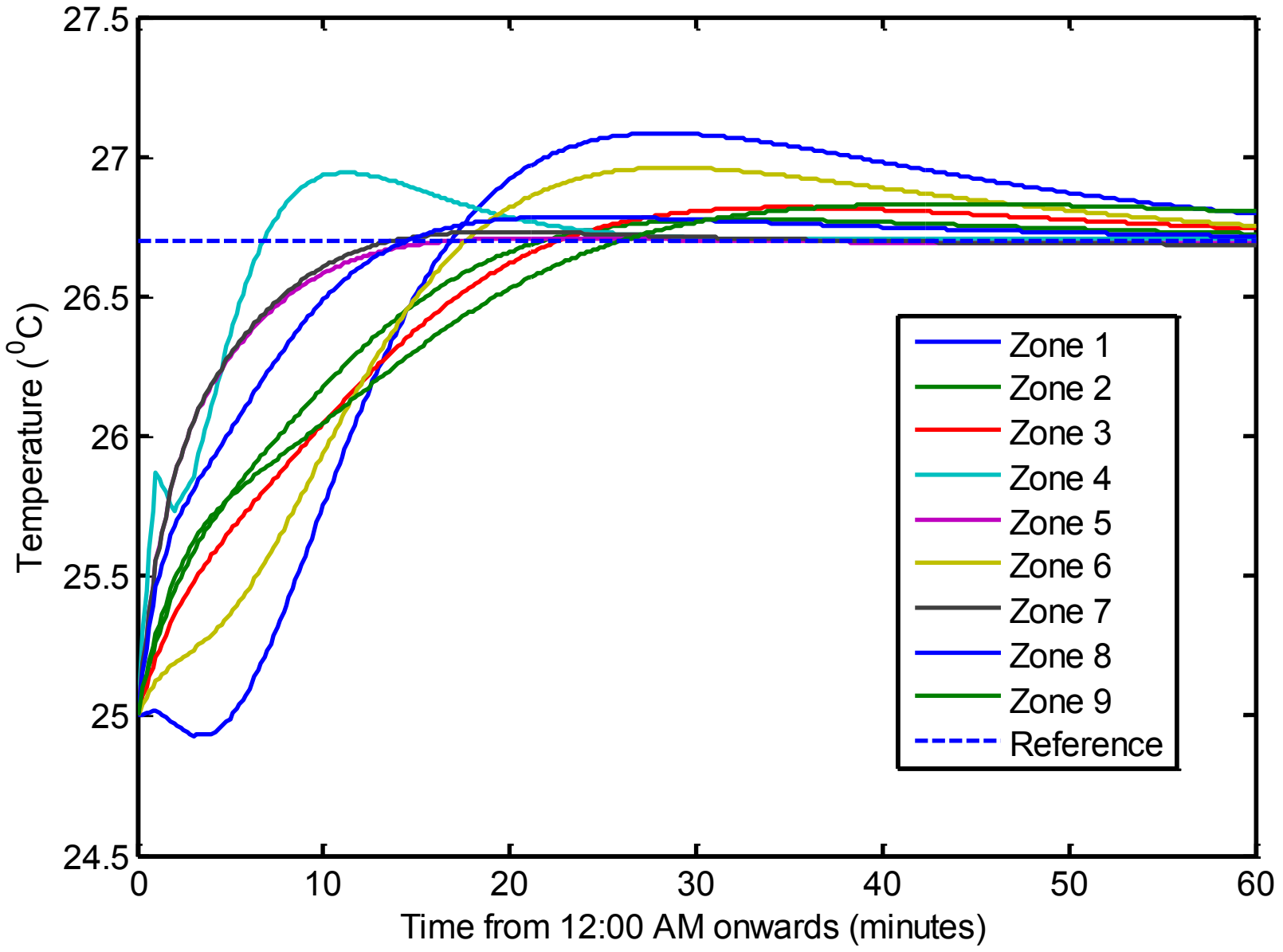

**Figure 6.37 Evolution of zone temperatures between 12:00 AM to 1:00 AM for fully decentralized architecture**

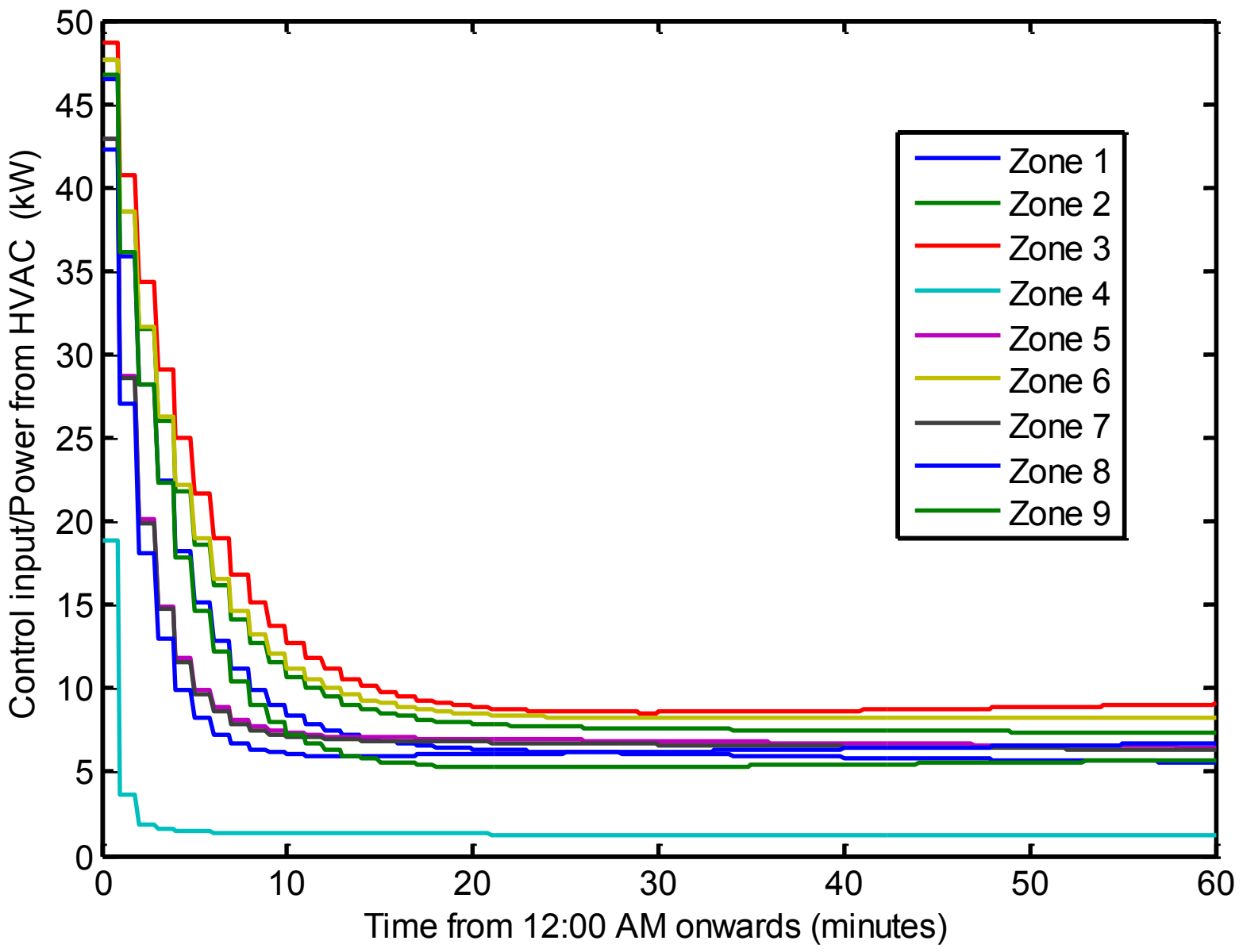

**Figure 6.38 Control inputs between 12:00 AM to 1:00 AM for centralized architecture**

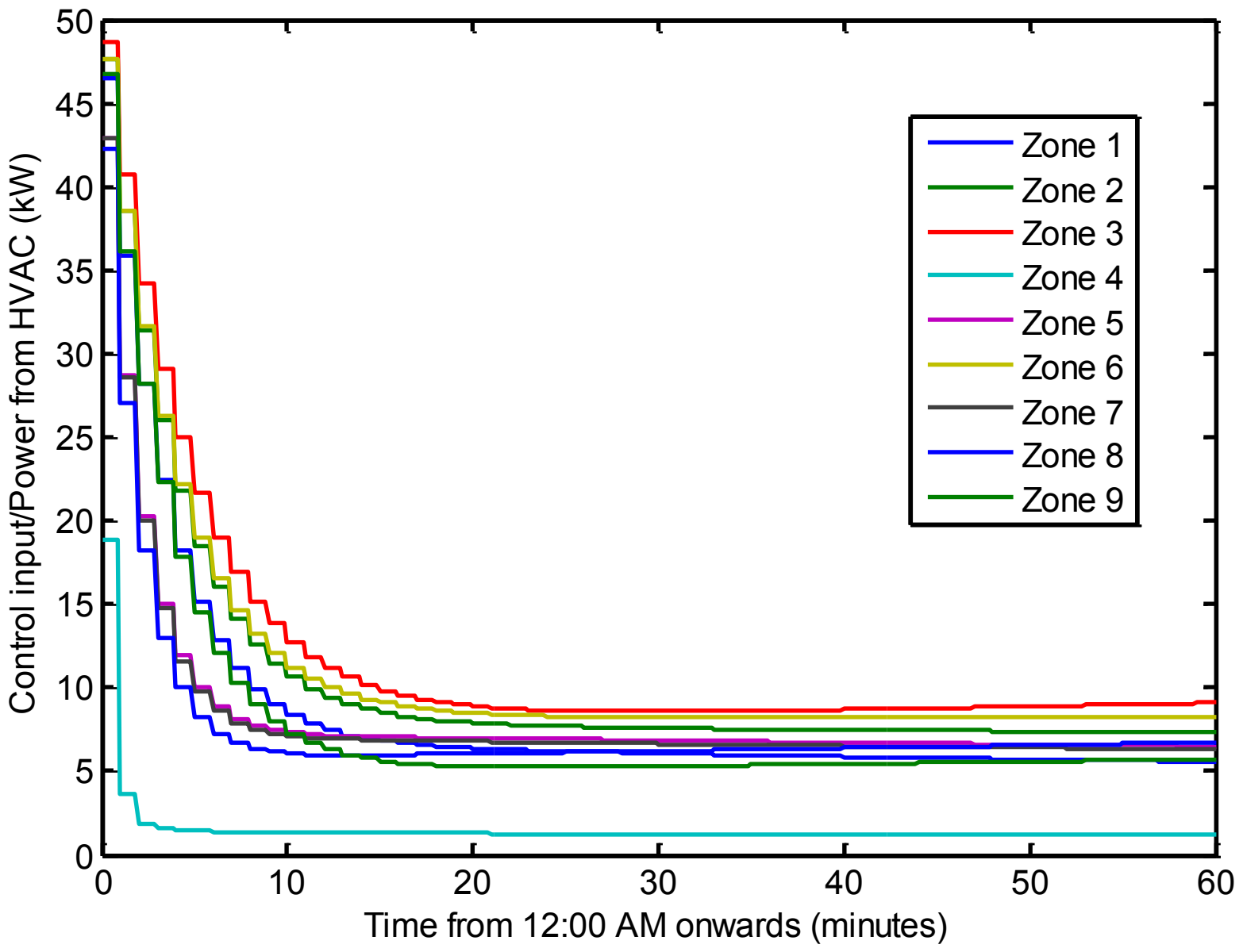

**Figure 6.39 Control inputs between 12:00 AM to 1:00 AM for decentralized architecture corresponding to knee 3**

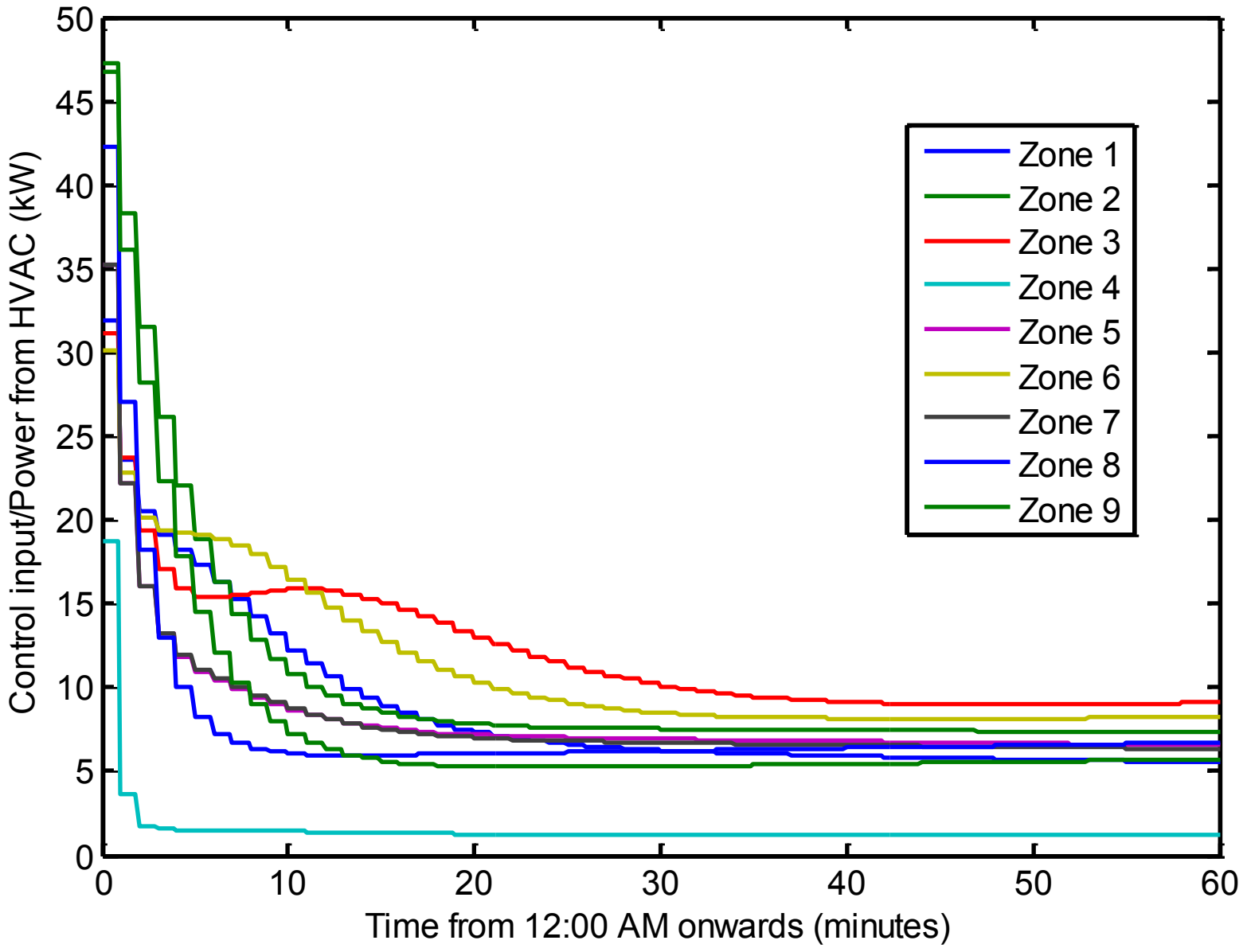

**Figure 6.40 Control inputs between 12:00 AM to 1:00 AM for decentralized architecture corresponding to knee 2**

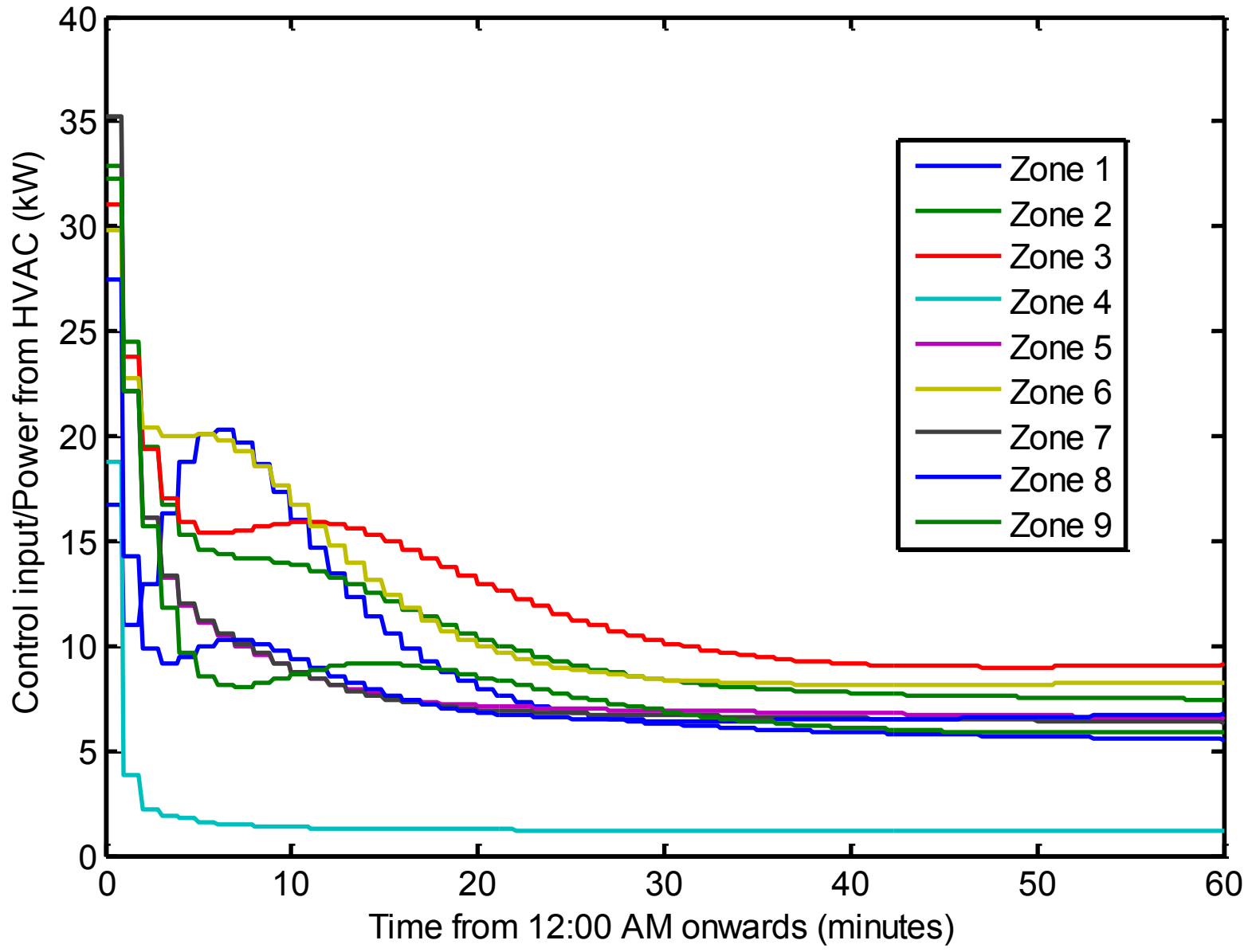

**Figure 6.41 Control inputs between 12:00 AM to 1:00 AM for decentralized architecture corresponding to knee 1**

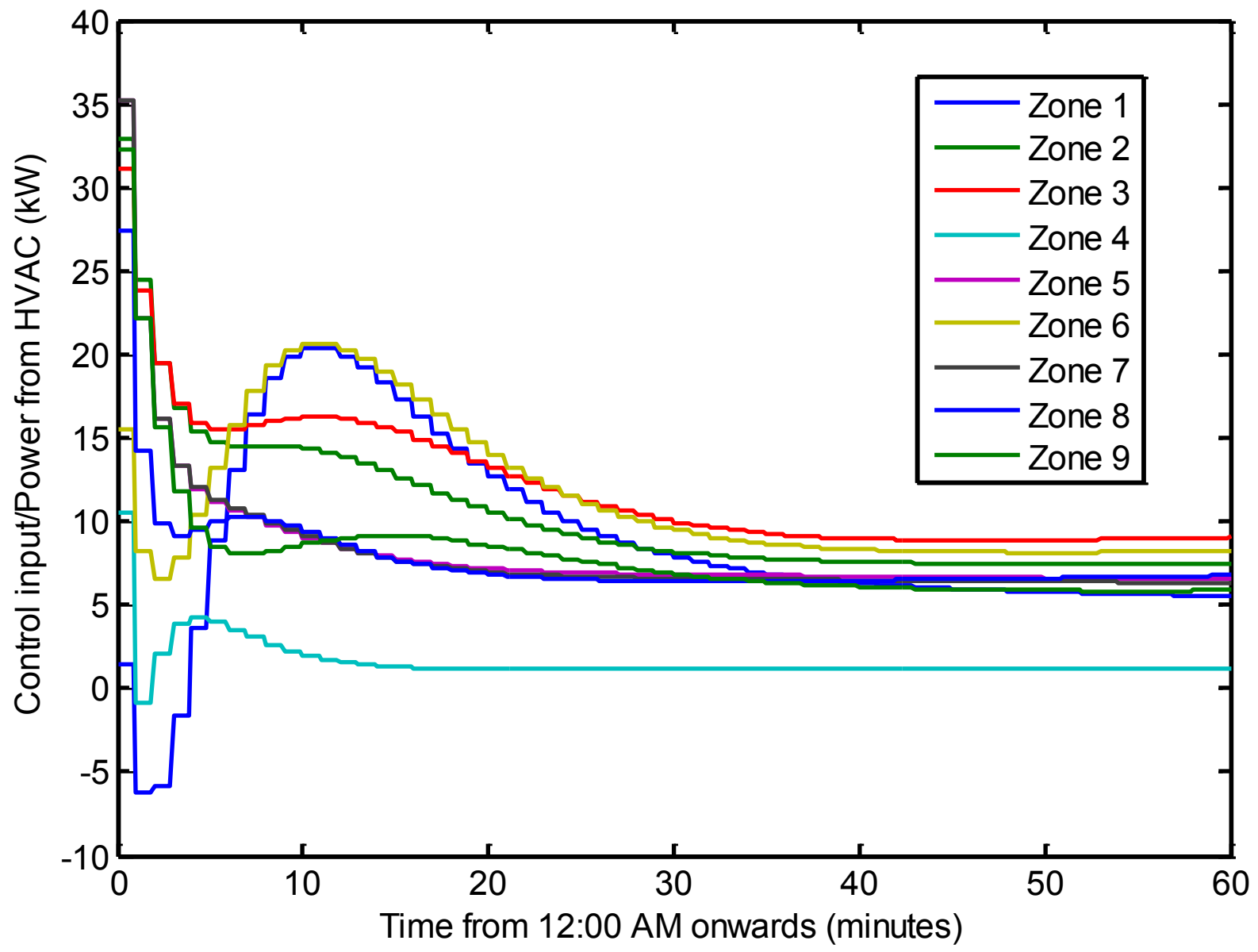

**Figure 6.42 Control inputs between 12:00 AM to 1:00 AM for fully decentralized architecture**

To compare the performance of various control architectures with regard to optimality, the integral $J_{opt}$ defined in (6.57) below is used, which is the continuous time analogue of the system-wide (centralized) objective function that the control was designed to minimize (see (5.41) and the parameters in Table 6.10).

$$J_{opt} = \int_{t=0}^{t=24\ hours} [\boldsymbol{u}(t)^T \mathbf{u}(t) + 10^3 \mathbf{e}(t)^T \mathbf{e}(t)\ ]\, dt, \qquad (6.57)$$

where, $\mathbf{e}(t) = \mathbf{T_z}(t) - \mathbf{T_{z,ref}}(t)$.

A comparison of the performance (optimality) for various control architectures is shown in Table 6.11. It is observed that the deviation in optimality increases with the level of decentralization, as expected.

#### 6.5.2.3 Robustness analysis

We set up a case study to examine the robustness of the controllers designed in Section 6.5.1 to a sensor failure event. It is assumed that the thermostat in zone G has developed a fault such that its temperature reading is 0℃ – the assumed lower limit of its sensing range – at all times. Zone G is chosen in this case study because it is connected to the most number of zones in the building (Figure 6.4). The simulation was re-run as described in Section 6.5.2.2 without changing any other setting. The associated MATLAB codes are provided in Appendix M and the SIMULINK models are enclosed in the media accompanying the thesis.

The zone temperature responses corresponding to the various control architectures are shown in Figure 6.43 to Figure 6.47. The resulting magnitudes of temperature deviations from set-points, i.e. $|T_{z,i}(t) - T_{z,i,\text{ref}}(t)|$ for all zones $i \in \{1,2, \dots 9\}$ in the building, at $t = 1$ hour, for the five control architectures are shown in Figure 6.48. Note that the zone numbers indicated correspond to Table 6.2.

**Table 6.11 Optimality analysis for test building under various control architectures**

| Control architecture | $J_{opt}$ | % deviation in $J_{opt}$ from centralized |
|---|---|---|
| Centralized | $6.29 \times 10^6$ | 0.00 |
| Knee 3 partition | $6.29 \times 10^6$ | 0.00 |
| Knee 2 partition | $6.85 \times 10^6$ | 8.90 |
| Knee 1 partition | $7.72 \times 10^6$ | 22.73 |
| Fully decentralized | $9.62 \times 10^6$ | 52.94 |

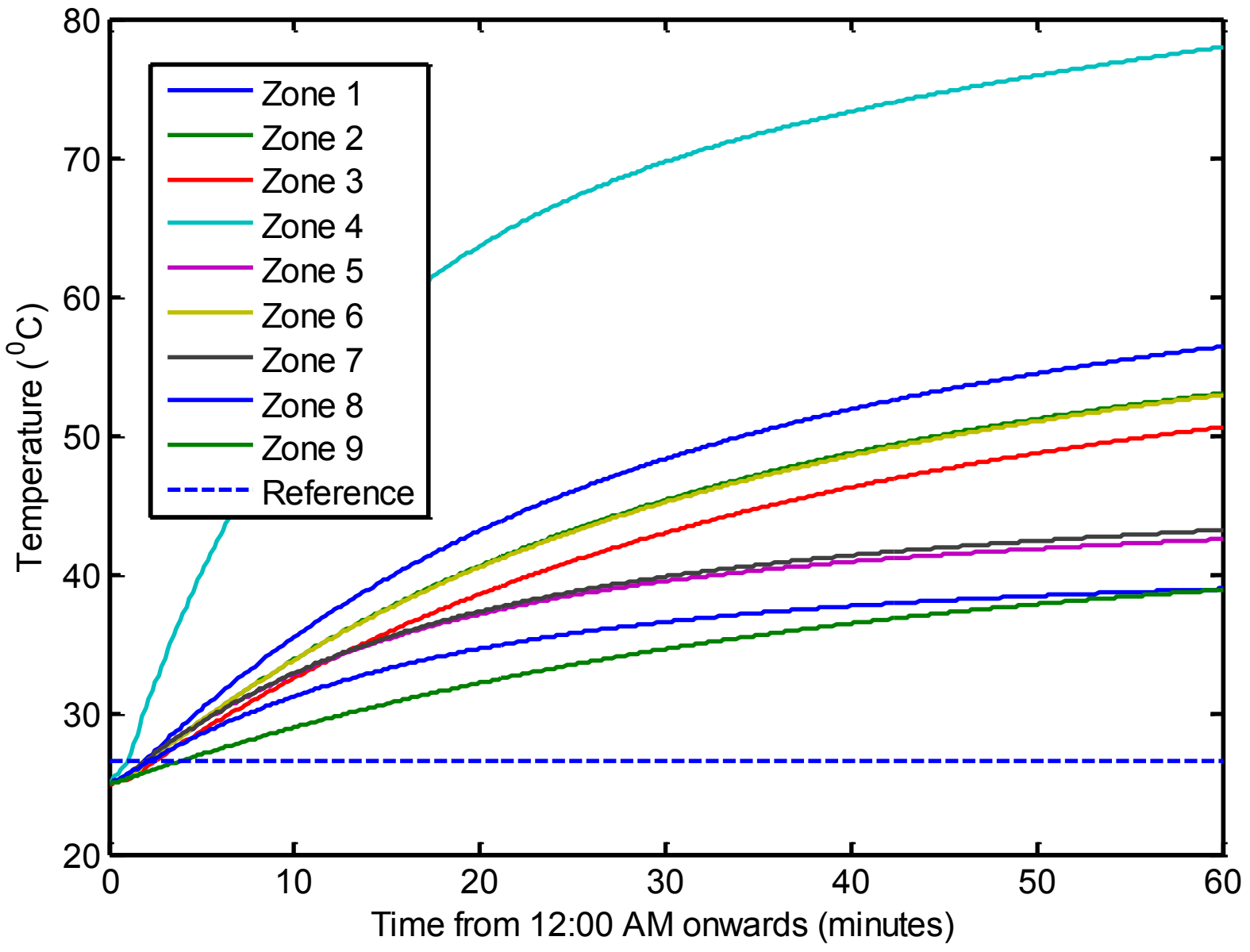

**Figure 6.43 Evolution of zone temperatures between 12:00 AM to 1:00 AM for centralized architecture in the event of sensor failure in zone 1**

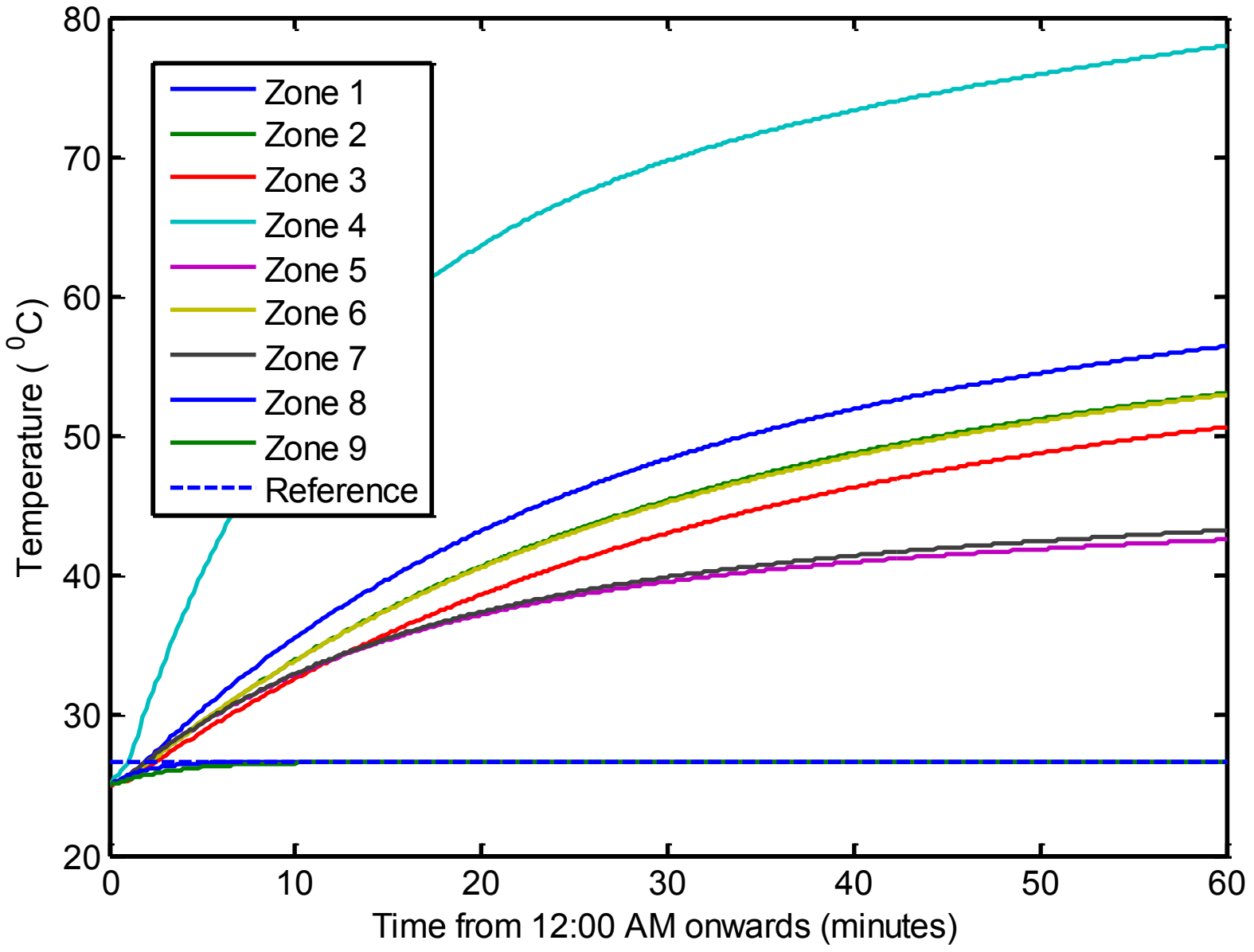

**Figure 6.44 Evolution of zone temperatures between 12:00 AM to 1:00 AM for decentralized architecture corresponding to knee 3 in the event of sensor failure in zone 1**

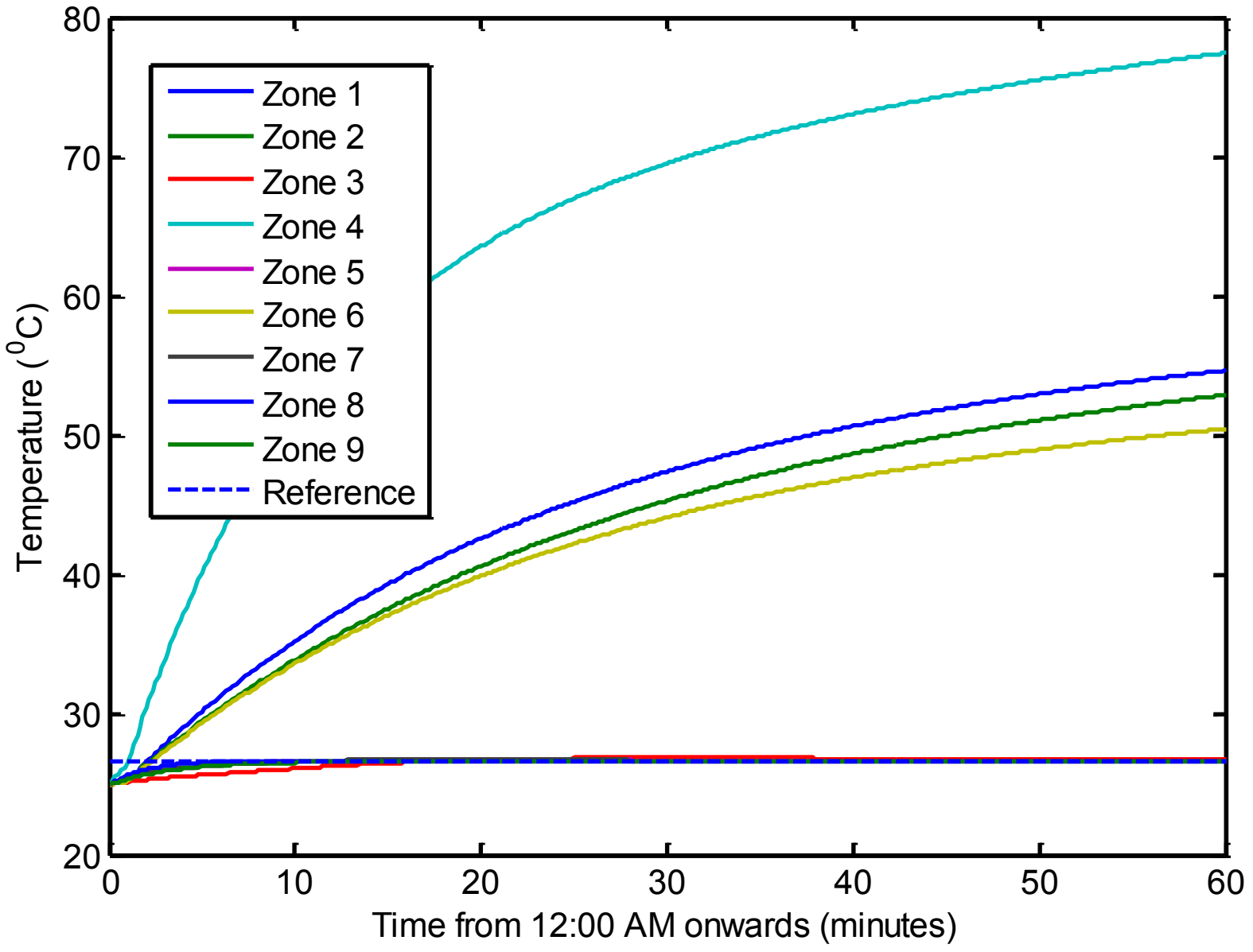

**Figure 6.45 Evolution of zone temperatures between 12:00 AM to 1:00 AM for decentralized architecture corresponding to knee 2 in the event of sensor failure in zone 1**

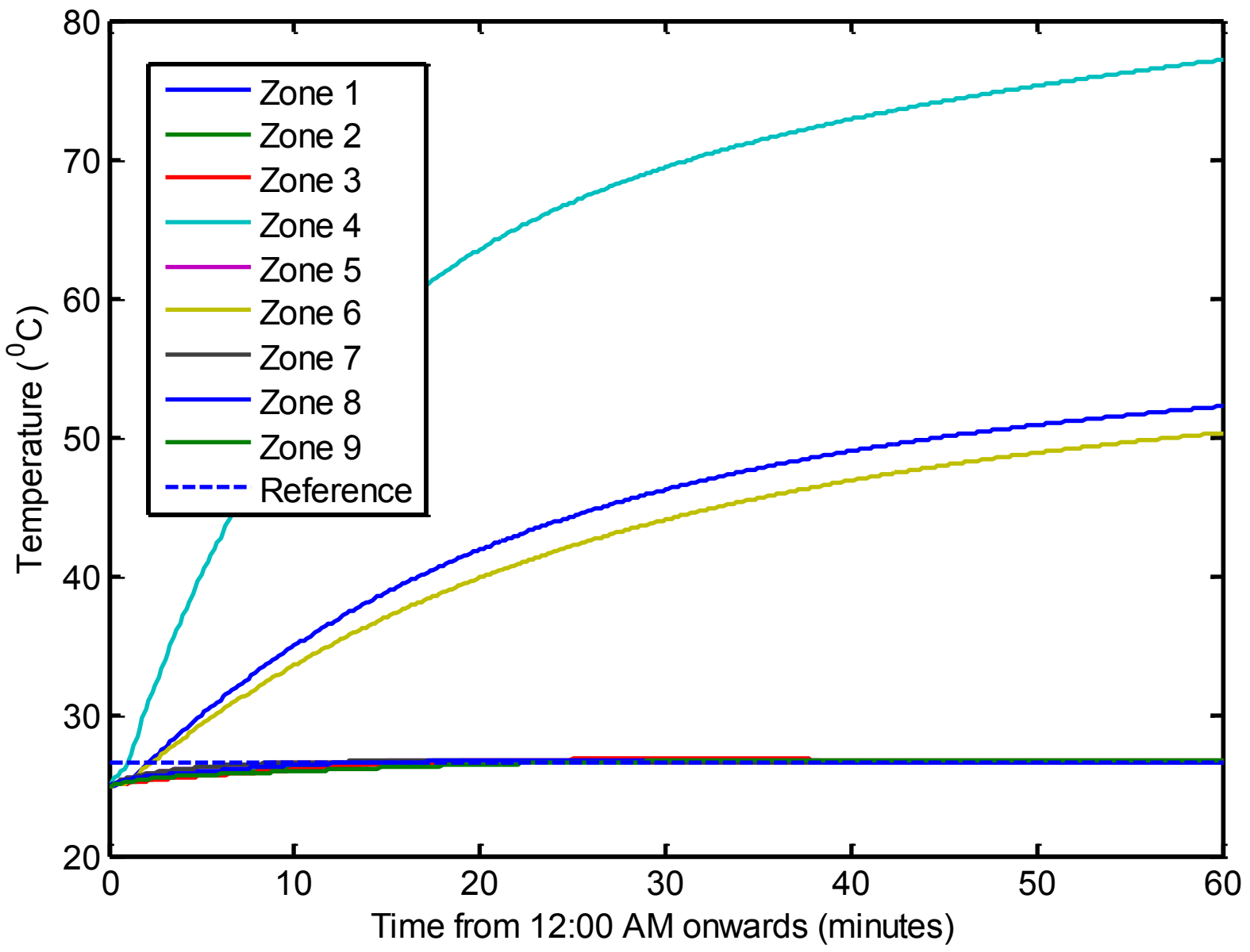

**Figure 6.46 Evolution of zone temperatures between 12:00 AM to 1:00 AM for decentralized architecture corresponding to knee 1 in the event of sensor failure in zone 1**

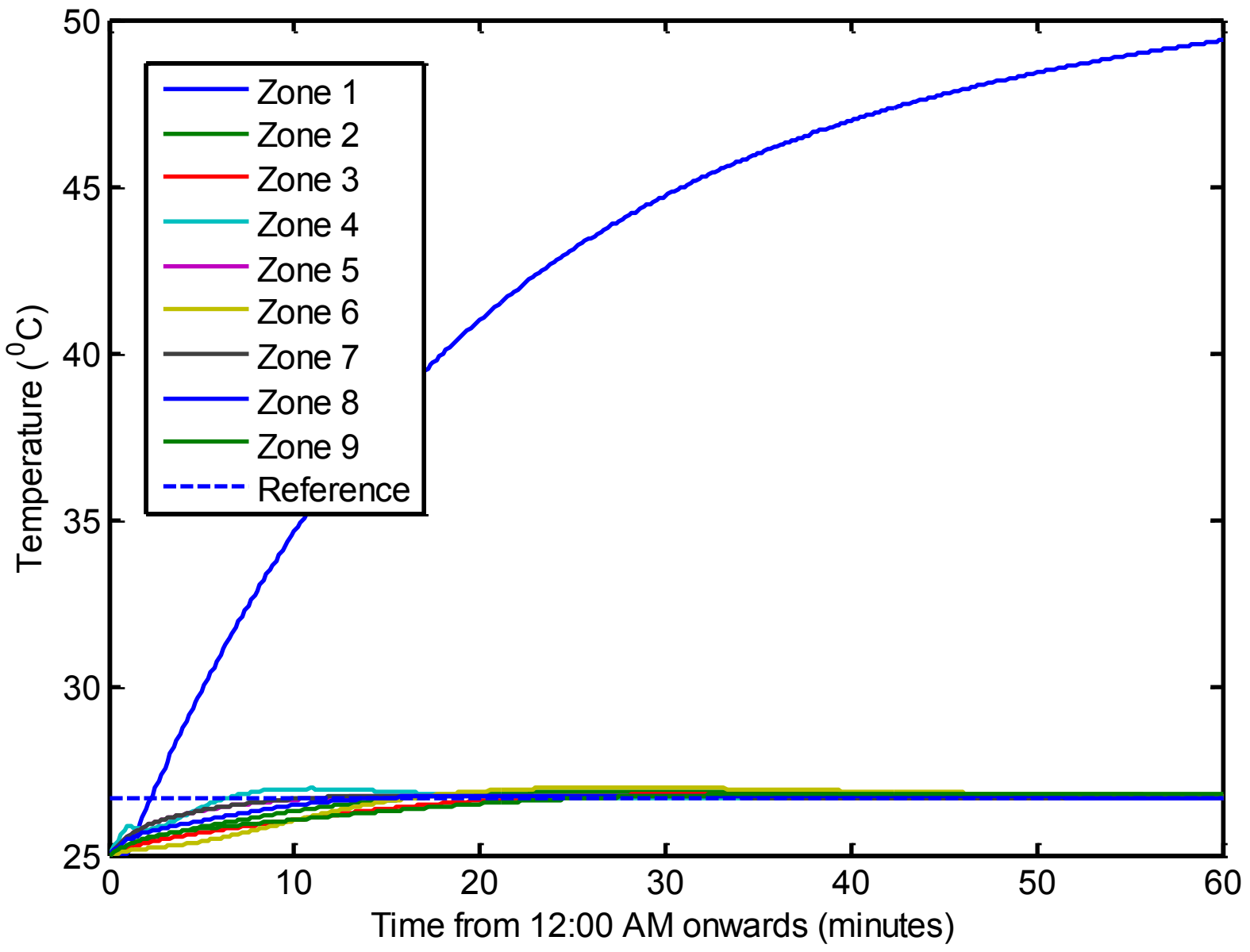

**Figure 6.47 Evolution of zone temperatures between 12:00 AM to 1:00 AM for fully decentralized architecture in the event of sensor failure in zone 1**

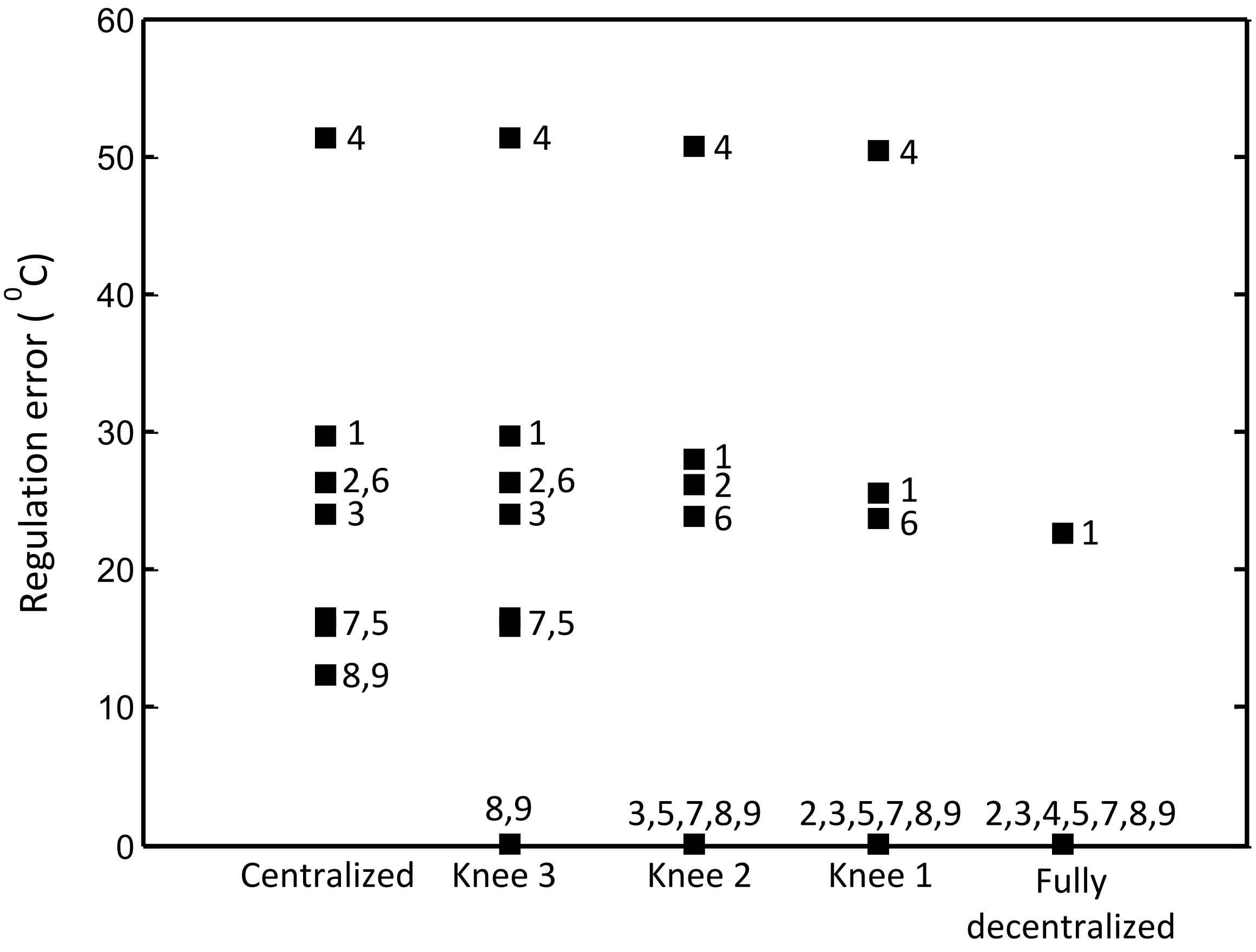

**Figure 6.48 Regulation errors evaluated at the end of 1 hour for all zones in the building under various control architectures in the event of sensor failure in zone 1 (zone numbers indicated correspond to Table 6.2)**

From Figure 6.43 to Figure 6.48, it is observed that large deviations from the set-point temperature result in zone 1 in each of the five control architectures. However, as seen in Figure 6.47, set-point temperatures are achieved in all other zones in the building in case of fully decentralized control. Table 6.12 shows the number of affected zones – where the temperatures do not achieve the set-point – and the corresponding fraction of building volume affected, obtained on the basis of data in Figure 6.48 for each of the control architectures considered. The spreadsheet used for computation of fraction of building volume affected is enclosed in the media accompanying the thesis. The clusters that constitute each control architecture, as obtained from Table 6.8, are also shown in Table 6.12 for reference. As expected, in each case it is observed from Table 6.12 that the number of affected zones matches with the size of the cluster containing zone 1 where the fault originates.

A thermodynamic explanation of the fault propagation phenomenon is as follows. Since the sensor in zone 1 records an incorrect value of 0°C, the controller corresponding to the cluster containing zone 1 dictates the HVAC system to overheat this zone at it maximum allowable heating capacity (Figure 6.49). This controller also overheats other zones in the cluster (see Figure 6.50 and Figure 6.51) – at their corresponding maximum allowable heating capacities - because the model for the thermal dynamics of the cluster incorrectly predicts a significant loss of thermal energy from these zones to zone 1 which is assumed to be at 0°C. Since the models used for other clusters do not use information from the sensors of this cluster, they are insulated from the sensor fault in zone 1 (verified from Figure 6.52).

**Table 6.12 Closed loop robustness analysis for test building for various control architectures in the event of sensor failure in zone 1**

| Control architecture | Number of affected zones | % building volume affected | Clusters |
|---|---|---|---|
| Centralized | 9 | 100.00 | {1,2,3,4,5,6,7,8,9} |
| Knee 3 | 6 | 79.25 | {1,2,3,4,5,6,7}{8,9} |
| Knee 2 | 4 | 45.30 | {1,2,4,6}{3}{5}{7}{8,9} |
| Knee 1 | 1 | 29.79 | {2}{3}{1,4,6}{5}{7}{8}{9} |
| Fully decentralized | 1 | 12.94 | {1}{2}{3}{4}{5}{6}{7}{8}{9} |

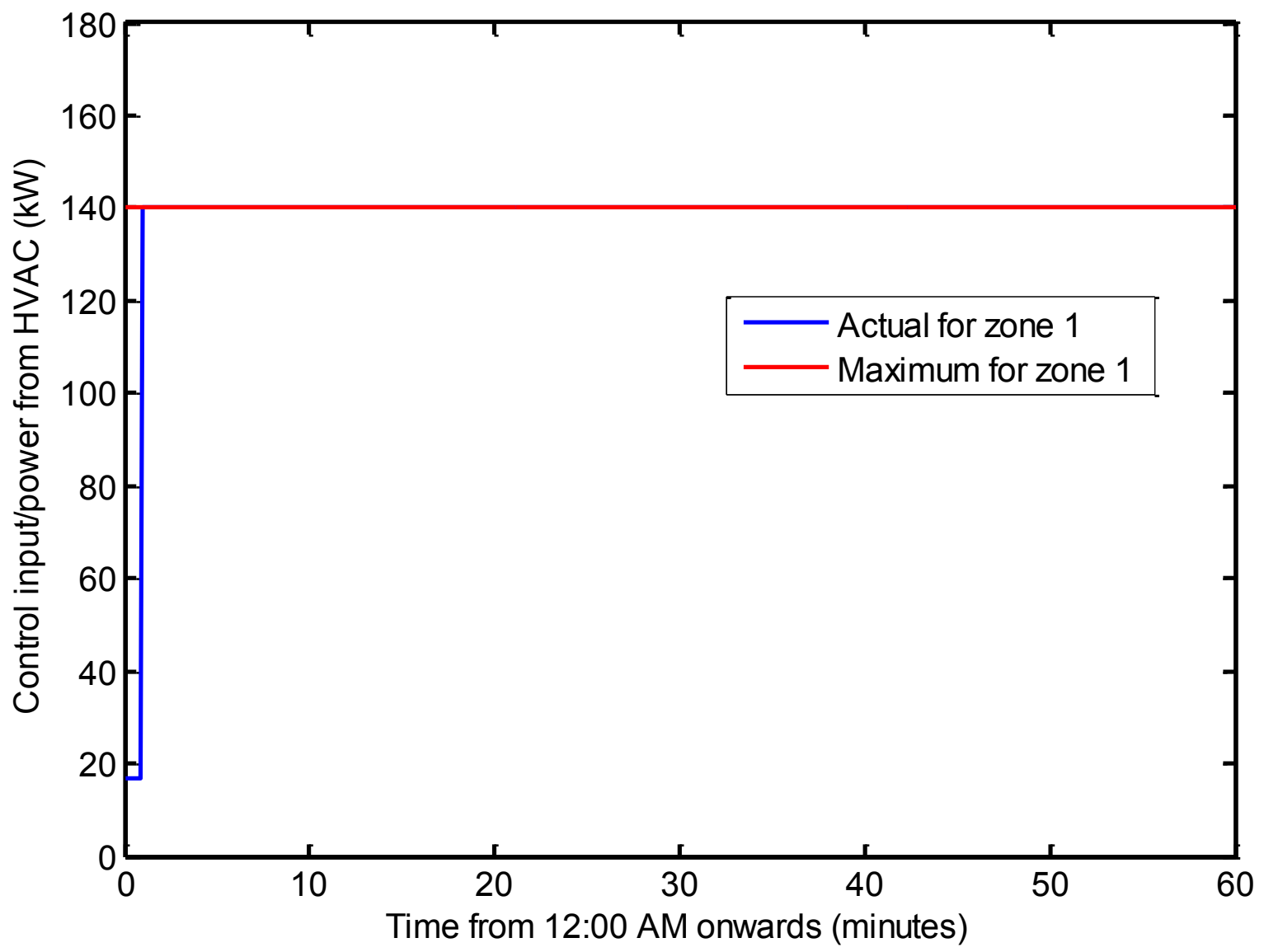

**Figure 6.49 Control input for zone 1 between 12:00 AM to 1:00 AM for decentralized architecture corresponding to knee 3 in the event of sensor failure in zone 1**

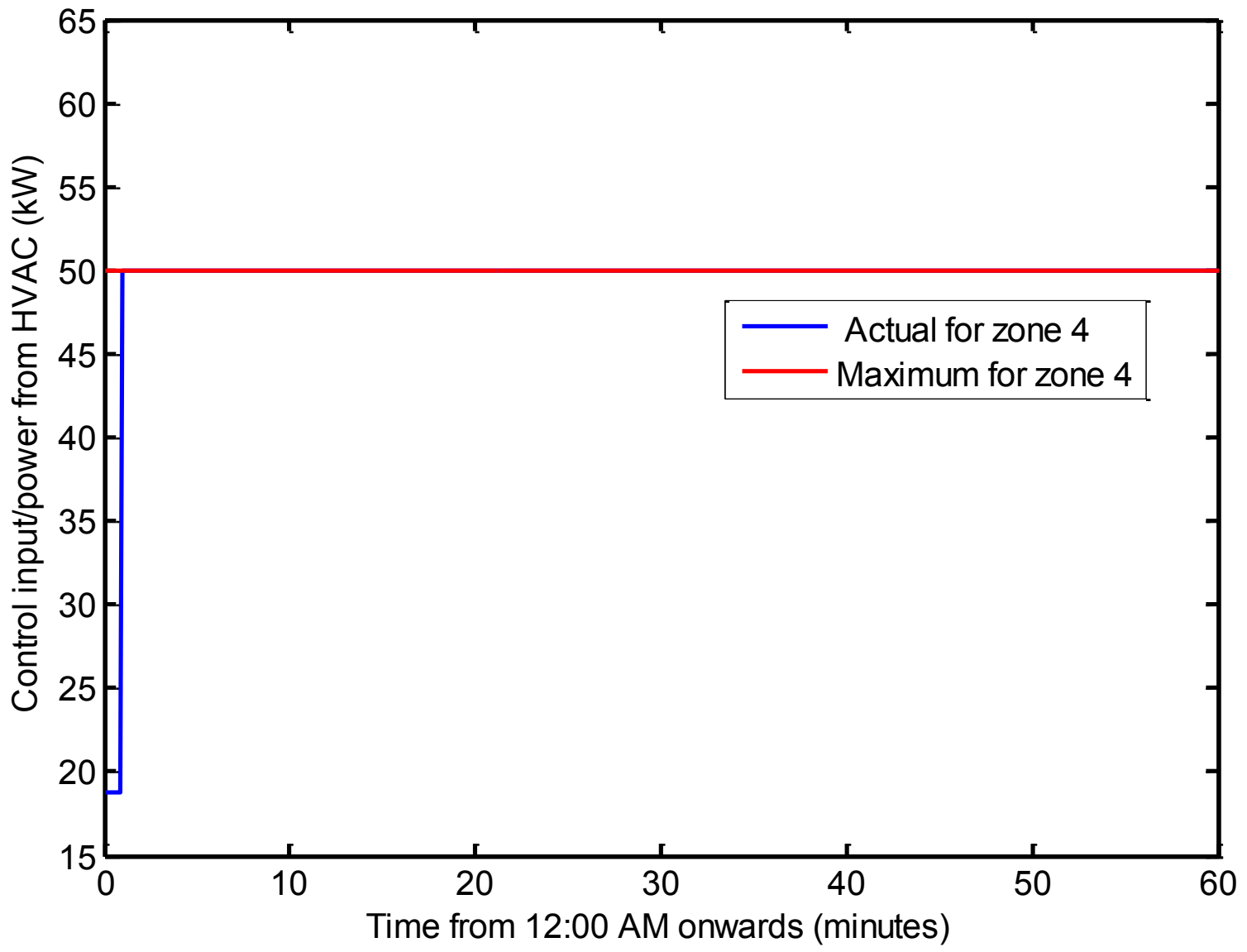

**Figure 6.50 Control input for zone 4 (lying in same cluster as zone 3) between 12:00 AM to 1:00 AM for decentralized architecture corresponding to knee 3 in the event of sensor failure in zone 1**

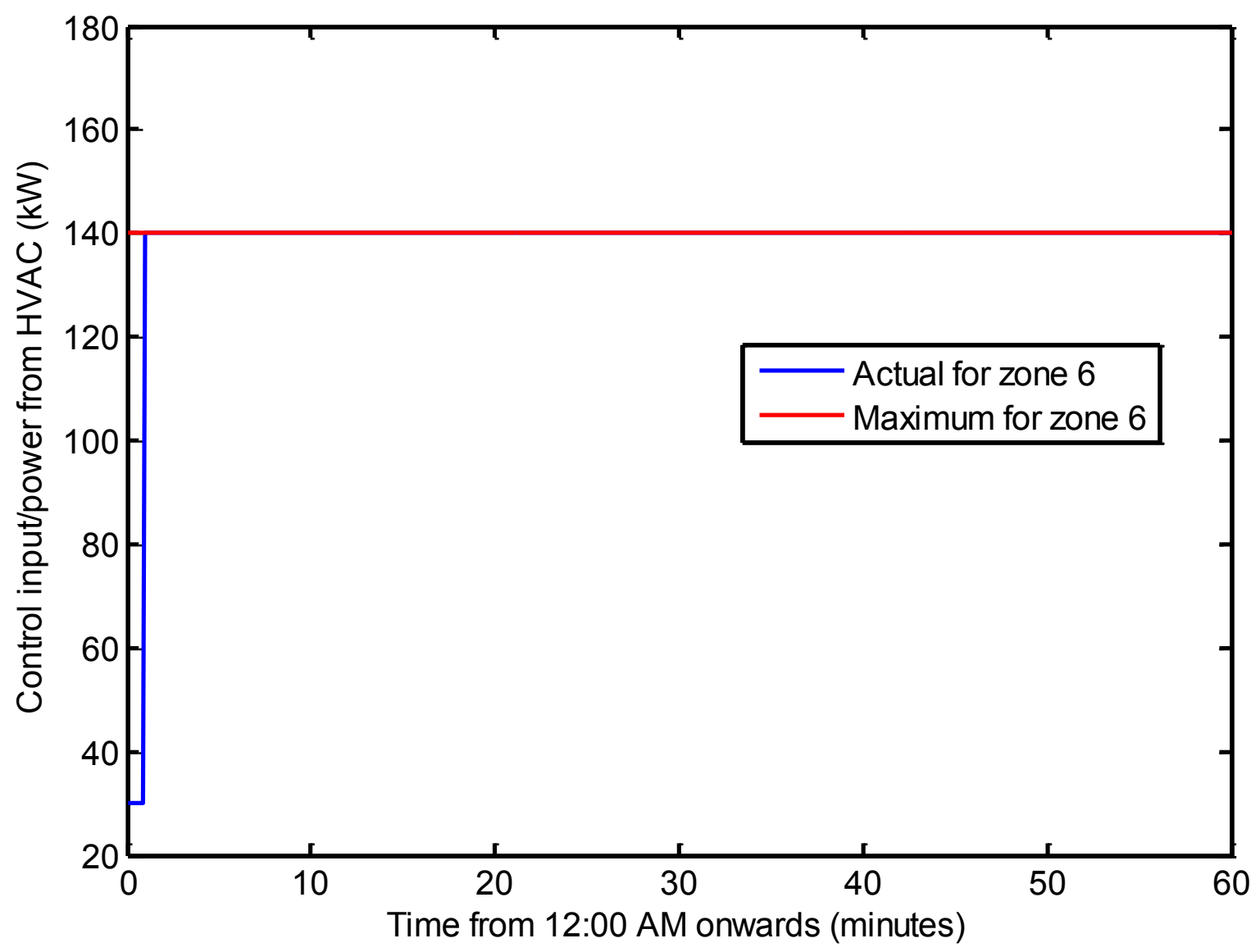

**Figure 6.51 Control input for zone 6 (lying in same cluster as zone 1) between 12:00 AM to 1:00 AM for decentralized architecture corresponding to knee 3 in the event of sensor failure in zone 1**

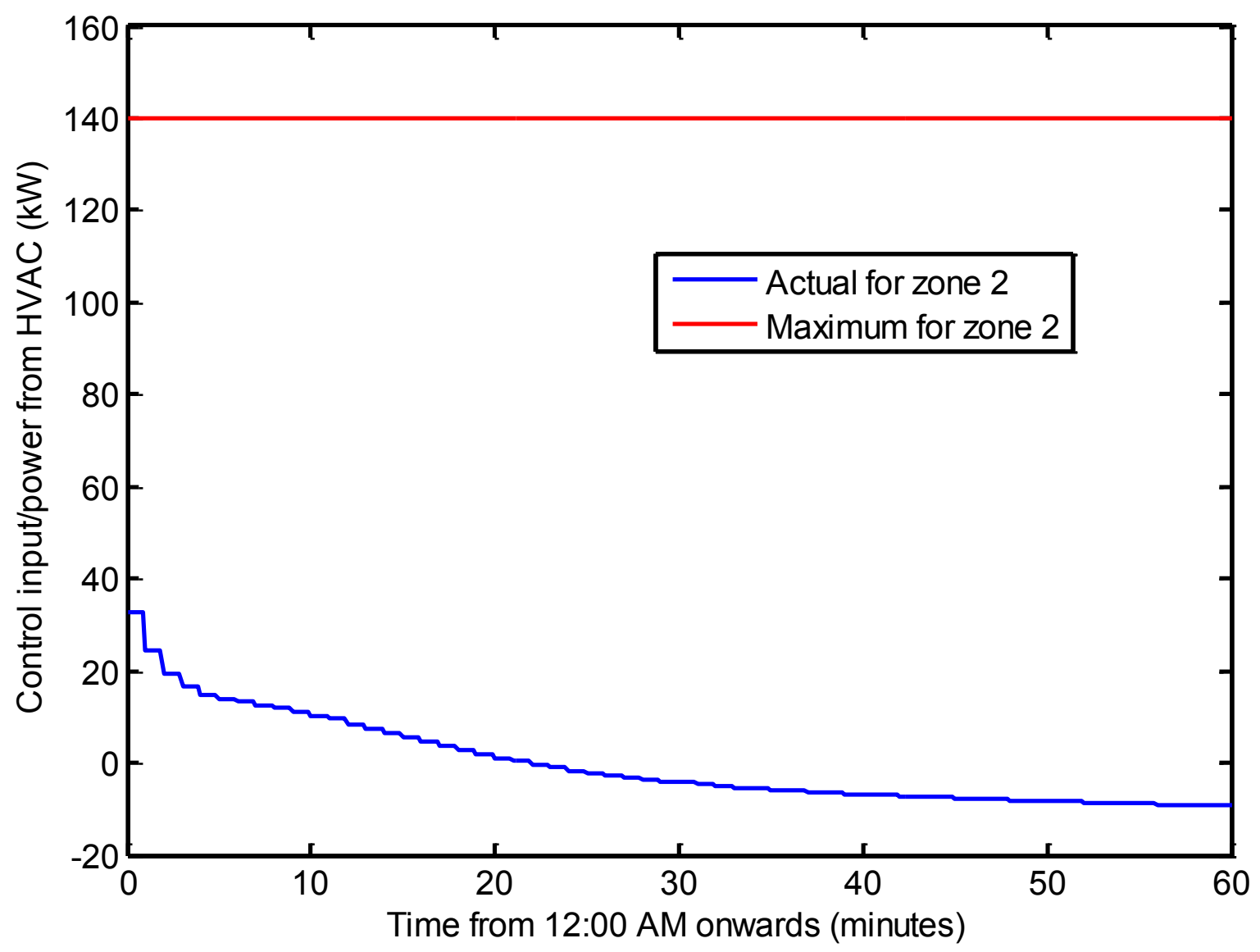

**Figure 6.52 Control input for zone 2 (in different cluster from zone 1) between 12:00 AM to 1:00 AM for decentralized architecture corresponding to knee 3 in the event of sensor failure in zone 1**

#### 6.5.2.4 Discussion

As seen from Table 6.12, less than half of the building volume is affected in case of the optimally decentralized partition (knee 2) in the event of sensor failure in zone 1. Also, as previously observed in Table 6.11, the deviation in optimality from centralized control in a situation without failures is less than 10%. Therefore, knee 2 appears to provide an appropriate trade-off between optimality and robustness objectives. This verifies the observation from Figure 6.29 that it is more centrally located on the optimality-robustness trade-off curve than other knees.

## 6.6 Concluding remarks

The two-step process of control architecture selection and control design presented in the previous chapters in the thesis was successfully applied in simulation on a real world building model. The optimality and robustness trends as were quantitatively investigated as a function of the degree of decentralization. Therefore this chapter demonstrates the applicability of the tools developed in the thesis and can be used by the interested reader to implement them on other building systems. The associated MATLAB codes, SIMULINK models and spreadsheets which are referenced in the chapter should be modified accordingly.

# Chapter 7
# Conclusions

In this final chapter, the work presented in this thesis is summarized in Section 7.1 followed by concluding remarks in Section 7.2. The main contributions of this work and future avenues of research are discussed in Sections 7.3 and 7.4.

## 7.1 Summary

A chapter-wise summary of this thesis is presented below:

1. Chapter 1 motivates the problem of building thermal control and describes the research objectives of this work. In particular, it motivates the need for decentralized thermal control of buildings. It also provides a literature survey of the area of building thermal control and the tools used in this work.
2. The physical aspects of building thermal control were presented in chapter 2. This included a description of underlying energy management systems, and sensing, actuation and control infrastructure.
3. Chapter 3 presented mathematical details of centralized and decentralized control architectures, which are used as preliminaries in the development of methodologies for partitioning a building for decentralized control. The objective functions, models and optimization frameworks corresponding to both centralized and decentralized architectures are described in detail.
4. Two methodologies – CLF-MCS and OLF-FPM – were developed in Chapter 4 to partitioning a building for decentralized control. In each of these methodologies, appropriate optimality and robustness metrics were developed and optimality-robustness trade-off curves were generated to decide a control architecture which provides a satisfactory balance between optimality and robustness. The metrics were based on the results of Chapter 3 and were tested on simple examples in simulation.

5. Chapter 5 considers the design of decentralized controllers for any partition of a building. Reduced order, observable models were developed for the thermal dynamics in a cluster. An observation framework which used these models to estimate both known states and disturbances was proposed. A state feedback decentralized model predictive framework was then developed which uses the estimates provided by the observer to minimize an objective function subject to physical constraints.
6. A real world simulation study was presented in Chapter 6 to demonstrate the applicability of the tools developed in this work for control architecture selection and control design. Both optimality and robustness analysis was performed to compare the closed loop performance of various control architectures.

## 7.2 Conclusions

The important conclusions from this work are as follows:

1. The thermal control of a complex interconnected system such as a building has multiple underlying objectives, most importantly occupant comfort, energy efficiency, robustness to faults and scalability.
2. The control architecture affects the extent to which these objectives are achieved. In particular, a fundamental tradeoff between optimality (occupant comfort and energy efficiency) and robustness (fault resilience) exists with respect to the degree of decentralization. Increase in the degree of decentralization results in improvements in robustness at the cost of optimality.
3. Two key challenges were identified with regard to the problem of determination of a control architecture that appropriately balances optimality and robustness requirements. Firstly, appropriate metrics are needed to quantify optimality and robustness. Secondly, the problem of partitioning a building into clusters for decentralized control is inherently computationally complex due to its combinatorial nature.

4. A CLF-MCS approach which used coupling loss factor (CLF) and mean cluster size (MCS) as heuristically defined optimality and robustness metrics was proposed. It used a divisive, stage-by-stage partitioning method to generate a family of partitions represented on an optimality-robustness trade-off curve. Its application to simulated examples revealed that the control architectures obtained were in sync with physical intuition.
5. An OLF-FPM approach was also proposed which used analytically derived optimality and robustness metrics – optimality loss factor (OLF) and fault propagation metric (FPM). It used an agglomerative clustering approach to generate a family of partitions represented on an optimality-robustness trade-off curve. Similar to the CLF-MCS approach, its application to simulated examples revealed that the control architectures obtained were in sync with physical intuition. It was concluded that the OLF-FPM approach was an improvement over the CLF-MCS approach because the metrics were analytically derived and the complexity of the partitioning procedure was only cubic in the number of zones, as opposed to the exponential complexity associated with the CLF-MCS approach.
6. The unavailability of measurements for wall temperature states and thermal disturbances was identified as a key challenge in the design of decentralized controllers, once a control architecture has been determined. A model reduction framework was proposed, which after a suitable state transformation resulted in an observable representation of the cluster level thermal dynamics. This allowed the design of extended state observers to estimate both unknown disturbances and states, which in turn allowed the design of output feedback decentralized controllers for thermal comfort. A model predictive framework with the ability to handle constraints was used for control design.
7. Identification of a simplified linear time invariant (LTI) model for a real world test building from its EnergyPlus model was investigated. It was observed that zone level identification using a standard least squares identification approach resulted in physically untenable parameters, potentially due to over-parameterization. This issue was addressed using a modified least squares identification framework which used fewer parameters. Use of least squares identification approaches for wall level identification resulted in inaccurate and unreliable parameter estimates, potentially due to over-parameterization and mismatch between the order of the identified model and the order of the model used in EnergyPlus. Therefore, direct computation of wall level parameters from construction layer properties was used as an alternative to least squares identification.

8. The application of the OLF-FPM approach on the LTI model of the real world test building resulted in a decentralized architecture which was physically explained on the basis of layout and construction properties of the building. Closed loop evaluation of this architecture on a nonlinear model of the building thermal dynamics verified that it provides a satisfactory tradeoff between optimality and robustness. The simulations also demonstrated and quantified the fundamental tradeoff between optimality and robustness that exists as a function of the degree of decentralization.

## 7.3 Contributions

Intelligent energy management in buildings is important due to the large scale impacts of the building sector on the economic and environmental aspects of energy. In this context, efficient thermal control is especially important because of the relatively significant contribution of space heating and cooling to the end use energy consumption of buildings. This thesis makes some important contributions to the problem of building thermal control, which are listed below.

1. The role of control architecture in achieving the objectives associated with the thermal control of buildings has not been properly investigated in literature. This work contributes to the area of building thermal control by motivating the incorporation of control architecture as an important dimension to be considered in control design. In this context, it specifically investigates the impact of decentralization of the control architecture on the attainment of optimality and robustness objectives.
2. A specific contribution of this work is the development of appropriate metrics to quantify the optimality and robustness attributes of any decentralized architecture for the thermal control of a building. Both heuristic (CLF and MCS) and analytically derived (OLF and FPM) metrics were presented.

3. Another contribution of this work is the adoption of existing tools and concepts from parallel technological fields, which have previously not been applied to area of building thermal control. In particular, the concept of agglomerative clustering was successfully employed to address the computational complexity concerns in the problem of control architecture determination.
4. In addition to providing methodologies for control architecture selection, this work also provides a methodology for control design based on the control architectures selected. It proposes a control design framework which addresses practical issues such as unavailability of certain states and disturbances and presence of physical constraints.
5. Lastly, this thesis complements the theoretical frameworks proposed for control architecture selection and control design by showing their applicability on a real world building example in simulation.

## 7.4 Future extensions

We identify the following areas of future research to build upon the work presented in this thesis.

### 7.4.1 Incorporation of HVAC system

The scope of this thesis was limited to consideration of the building side dynamics and control. The control variables considered correspond to the energy transfer rates (heating/cooling) provided by the HVAC system to the zones in the building. However, the HVAC itself was not considered.

To manipulate the heating or cooling provided to the zones for building thermal control, appropriate actuators in the HVAC system – such as dampers and reheaters in a VAV system (Chapter 2) – need to be adjusted. Therefore, for practical implementation of closed loop control, the HVAC system dynamics should be included. This can be achieved by modifying the control design framework in Chapter 5 by including appropriate actuator dynamics.

### 7.4.2 Experimental investigation

Implementation of the tools proposed in this work on a real building system can be undertaken. Such experimental studies would serve to complement the simulation studies presented in Chapter 6, and provide additional validation of the tools for control architecture selection and control design presented in this thesis.

As described in Section 7.4.1, the HVAC system dynamics would have to be incorporated in the control design framework for experimental implementation. This may necessitate the consideration of other practical aspects such as actuator limitations and slew rates in addition to the constraints described in Chapter 5.

Besides experimental validation, co-simulation approaches, which allow controllers based in MATLAB to interface directly with the higher fidelity EnergyPlus models can also be explored. This can be enabled by the use of appropriate platforms such as the Building Control Virtual Test Bed (BCVTB) [100].

### 7.4.3 Extension to other applications

Although the proposed tools in this thesis are developed specifically in the context of thermal control of buildings, they can also be potentially applied to other large scale energy management applications such as data center cooling, district heating and cooling for campuses, and distributed refrigeration systems for supermarkets. Moreover, the modularity of the decentralized framework allows for extensions to other energy efficiency domains apart from thermal. This includes electrical grid based systems having generation, distribution, consumption and recovery.

# Appendix A

# Codes for 12-zone building example in section 4.2.8

To obtain the results presented in this section, the following programs need to be run in the specified sequence:

STEP 1: Obtain weighted incidence matrix, and capacitance matrices (see Algorithm 3.1)

```
% *******************************************************************
clc;
clear all;
N_alpha = [];
Rinv_ext_h_in = 1/29.99;
Rinv_ext_h_out = 1/81.08;
Rinv_ext_v_in = 1/36.84;
Rinv_ext_v_out = 1/82.00;
Rinv_int_h_in = 1/21.32;
Rinv_int_h_out = 1/21.32;
Rinv_int_v_in = 1/21.32;
Rinv_int_v_out = 1/21.32;
% Factors rho_1, rho_2 and rho_3 being set to 1. These factors can be set to other values % to
run the various cases in section 4.2.8
rho_1 = 1;
rho_2 = 1;
rho_3 = 1;
% Set 1: Incidence matrix entries for horizontal external walls
count1 = 0;
vect_1 = [1,2,3,4,5,6,7,8,9,10,11,12];
vect_2 = [1,2,3,4,5,6,7,8,9,10,11,12];
for i = 1:length(vect_1)
   N_alpha(vect_1(i),vect_2(i)) = Rinv_ext_h_in;
   N_alpha(vect_1(i),13) = Rinv_ext_h_out;
   count1 = count1+1;
end
% Set 2: Incidence matrix entries for vertical external walls
count2 = 0;
vect_1 = [13,14,15,16,17,18,19,20,21,22,23,24];
vect_2 = [1,2,3,4,5,6,7,8,9,10,11,12];
for i = 1:length(vect_1)
   N_alpha(vect_1(i),vect_2(i)) = Rinv_ext_v_in;
   N_alpha(vect_1(i),13) = Rinv_ext_v_out;
   count2 = count2+1;
end
```

```
vect_1 = [37,38,39,40,41,42,43,44,45,46,47,48];
vect_2 = [1,2,5,6,11,12,7,8,9,10,11,12];
for i = 1:length(vect_1)
   N_alpha(vect_1(i),vect_2(i)) = Rinv_ext_v_in;
   N_alpha(vect_1(i),13) = Rinv_ext_v_out;
   count2 = count2+1;
end
% Set 3: Incidence matrix entries for horizontal internal walls
count3 = 0;
vect_1 = [25,26,27,28,29,30];
vect_2 = [3,4,7,8,9,10];
vect_3 = [5,6,9,10,11,12];
for i = 1:length(vect_1)
   N_alpha(vect_1(i),vect_2(i)) = Rinv_int_h_in/rho_1;
   N_alpha(vect_1(i),vect_3(i)) = Rinv_int_h_out/rho_1;
   count3 = count3+1;
end
% Set 4:  Incidence matrix entries for vertical internal walls
count4 = 0;
vect_1 = [31,32,33,34,35,36];
vect_2 = [1,3,5,7,9,11];
vect_3 = [2,4,6,8,10,12];
% Set 4, subset 1: Symmetrically splitting internal walls
for i = 1:length(vect_1)
   N_alpha(vect_1(i),vect_2(i)) = Rinv_int_V_in/rho_3;
   N_alpha(vect_1(i),vect_3(i)) = Rinv_int_V_out/rho_3;
   count4 = count4+1;
end
% Set 4, subset 2: Column separating internal walls
   N_alpha(37,1) = Rinv_int_V_in/rho_2;
   N_alpha(37,3) = Rinv_int_V_out/rho_2;
   N_alpha(38,2) = Rinv_int_V_in/rho_2;
   N_alpha(38,4) = Rinv_int_V_out/rho_2;
   N_alpha(39,3) = Rinv_int_V_in/rho_2;
   N_alpha(39,7) = Rinv_int_V_out/rho_2;
   N_alpha(40,4) = Rinv_int_V_in/rho_2;
   N_alpha(40,8) = Rinv_int_V_out/rho_2;
   N_alpha(41,5) = Rinv_int_V_in/rho_2;
   N_alpha(41,9) = Rinv_int_V_out/rho_2;
   N_alpha(42,6) = Rinv_int_V_in/rho_2;
   N_alpha(42,10) = Rinv_int_V_out/rho_2;
   count4 = count4+6;
% Capacitance matrices
C_w = blkdiag(8329.15*eye(count_1), 8329.15*eye(count_2), 4660*eye(count_3),
4660*eye(count_4));
```

```
C_z = blkdiag(250*eye(12));
C_cap = blkdiag(C_w,C_z);
% *****************************************************************
```

STEP 2: Obtaining matrices A, B, C and D for continuous time state space model (see Algorithm 3.1)

```
% *****************************************************************
nw = count_1+count_2+count_3+count_4;
nz = 12;
N_beta = -N_alpha;
N_alpha_z = N_alpha(:,1:nz);
N_beta_z = N_beta(:,1:nz);
S_r_beta = diag(sum(N_beta'));
S_c_alpha_z = diag(sum(N_alpha_z));
N_alpha_a = N_alpha(:,nz+1:nz+1);
inv_C_cap = inv(C_cap);
% Generation of A, B, C and D matrices denotes using the subscript "bldg."
A_bldg = inv_C_cap*[[S_r_beta,N_alpha_z];[-N_beta_z',-S_c_alpha_z]];
B_bldg = inv_C_cap*[[zeros(nw,nz),zeros(nw,nz),  N_alpha_a];[eye(nz), eye(nz), zeros(nz,1)]];
C_bldg = [zeros(nz,nw),eye(nz)];
D_bldg = zeros(nz,2*nz+1);
% *****************************************************************
```

STEP 3: Discretization of model and generation of Hessian matrix

```
% *****************************************************************
% Part 1: Discretization of continuous state space model, resulting matrices are A_fd, %B_fd, and C_fd
global N Hess Nu
sys = ss(A_bldg,B_bldg,C_bldg,D_bldg);
Ts = 600;
sysd = c2d(sys,Ts,'zoh');
A_fd = sysd.A;
B_f = sysd.B;
C_fd = sys.C;
B_fd = B_f(:,1:9);
% Part 2: Finding unswapped Hessian
N = 24; % Length of prediction horizon
Nu = 12;
for i = 0:N-1
   P(:,:,i+1) = C_fd*(A_fd)^i*B_fd;
end
Hess_us = [];
for i = 0:N-1
   blah = [];
   for j =0:N-1
      blah = [blah,(P(:,:,i+1))'*(P(:,:,j+1))];
```

```
   end
   Hess_us = [Hess_us;blah];
end
% Part 3: Swapping to create Hess which is the Hessian matrix in desired form
H_s = [];
for i = 1:Nu
   for j = 0:N-1
      H_s = [H_s;Hess_us(j*Nu+i,:)];
   end
end
Hess =[];
for i = 1:Nu
   for j = 0:N-1
      Hess = [Hess,H_s(:,j*Nu+i)];
   end
end
% ******************************************************************
```

STEP 4: Stage by stage combinatorial optimization (see section 4.2.3) and generation of optimality-robustness tradeoff curve

```
% ******************************************************************
global N Hess Nu
parentset = struct('values',{[1:1:Nu]});
rcs = 1;
k = 1;
clf_vect(k) = 0;
res_vect(k) = rcs;
while rcs > 1/Nu
[childset,clf,mcs] = comboptm(parentset); % call the function "comboptm" to calculate
% the child partition for a given parent partition (stage level combinatorial optimization)
childset.values % displays result (output child partition) of the combinatorial optimization
 parentset = childset;
 k = k + 1;
clf_vect(k) = clf;
 mcs_vect(k) = mcs;
end
plot(mcs_vect,clf_vect,'-') %Generates optimality-robustness tradeoff curve
    % ******************************************************************
```

MATLAB Function "comboptm" used in the above code is as shown:

```
% ******************************************************************
function [childset,clf1,mcs] = comboptm(parentset)
global N Hess Nu
no_parents = length(parentset);
```

```
minweight  = 99999;
for l = 1:no_parents
   list = parentset(l).values;
   n = length(list);
   for k = 1:n-1
      C = combnk(list,k); % Creates all possible intermediate cluster pairs
      nc = length(C(:,1));
% The following loop compares the intermediate cluster pairs to find the ones with
% smallest ILF
      for i = 1:nc
         innerlist = C(i,:);
         innerweight = weight_calculate(innerlist,list)/self_weight(list); %ILF (see (4.14))
         if(innerweight<minweight)
            minweight = innerweight;
            minlist = innerlist;
            minlist_comp = setdiff(list,minlist);
            minparent = l;
         end
      end
   end
end
% Create child clusters
childset = struct('values',{});
for l = 1:no_parents;
   if l<minparent
      childset(l).values = parentset(l).values;
   end
   if l == minparent
      childset(l).values = minlist;
      childset(l+1).values = minlist_comp;
   end
   if(l>minparent)
      childset(l+1).values = parentset(l).values;
   end
end
% The remainder of the program computes the CLF and MCS of child clusters
c = no_parents + 1;
tot_coupling_lost = [];
for i = 1:c-1
   tot_coupling_lost_i = [];
   for j = i+1:c
      coupling_i_j = findcoupling(childset(i).values,childset(j).values);
      if (i == j)
         coupling_i_j = 0*coupling_i_j;
      end
```

```
      tot_coupling_lost_i = [tot_coupling_lost_i;coupling_i_j];
   end
   tot_coupling_lost = [tot_coupling_lost;tot_coupling_lost_i];
end
clf1 = (norm(tot_coupling_lost))/norm(Hess);
mcs = 1/c;
% ********************************************************************
```

The MATLAB functions “weight_calculate”, “self_weight” and “findcoupling” used in the above function are shown below

```
% ********************************************************************
function result = weight_calculate(innerlist,list)
global N Hess Nu
cluster1 = innerlist;
cluster2 = setdiff(list,innerlist);
bout = [];
for i = 1:length(cluster1)
   bin = [];
   for j = 1:length(cluster2)
      bin = [bin,Hess((cluster1(i)-1)*N+1:cluster1(i)*N,(cluster2(j) - 1)*N+1:cluster2(j)*N)];
   end
   bout = [bout;bin];
end
result = norm(bout);
% ********************************************************************

% ********************************************************************
function result = self_weight(list)
global N Hess Nu
bout = [];
for i = 1:length(list)
   bin = [];
   for j = 1:length(list)
      bin = [bin,Hess((list(i)-1)*N+1:list(i)*N,(list(j)-1)*N+1:list(j)*N)];
   end
   bout = [bout;bin];
end
result = norm(bout);
% ********************************************************************

% ********************************************************************
function result = findcoupling(cluster1,cluster2)
global N Hess Nu
bout = [];
for i = 1:length(cluster1)
```

```
   bin = [];
   for j = 1:length(cluster2)
      bin = [bin,Hess((cluster1(i)-1)*N+1:cluster1(i)*N,(cluster2(j)-1)*N+1:cluster2(j)*N)];
   end
   bout = [bout;bin];
end
result = norm(bout);
% **********************************************************************
```

# Appendix B

# Codes for 9-zone building example in section 4.2.9

To obtain the results presented in this section, the following programs need to be run in the specified sequence:

STEP 1: Obtain weighted incidence matrix, and capacitance matrices (see Algorithm 3.1)

```
% *******************************************************************
% 9 zone building
clc
clear all
N_alpha = [];
rho = 1; % Factor rho is set here
rat1 = rho;
rat2 = 1/rho;
% SET 1: Incidence matrix entries for horizontal external walls
count_1 = 0;
for i = 1:3
    count_1 = count_1+1;
    currentrow = zeros(1,10);
    currentrow(i) = 1/29.99;
    currentrow(10) = 1/81.08;
    N_alpha = [N_alpha;currentrow];
end
for i = 7:9
    currentrow = zeros(1,10);
    currentrow(i) = 1/29.99;
    currentrow(10) = 1/81.08;
    count_1 = count_1+1;
    N_alpha = [N_alpha;currentrow];
end
% SET 2: Incidence matrix entries for vertical external walls
count_2 = 0;
for i = 1:3:7
    currentrow = zeros(1,10);
    currentrow(i) = 1/36.84;
    currentrow(10) = 1/82.00;
    count_2 = count_2+1;
    N_alpha = [N_alpha;currentrow];
end
for i = 1:9
    currentrow = zeros(1,10);
```

```
    currentrow(i) = 1/36.84;
    currentrow(10) = 1/82.00;
    count_2 = count_2+1;
   N_alpha = [N_alpha;currentrow];
end
for i = 3:3:9
    currentrow = zeros(1,10);
    currentrow(i) = 1/36.84;
    currentrow(10) = 1/82.00;
    count_2 = count_2+1;
    N_alpha = [N_alpha;currentrow];
end
for i = 1:9
    currentrow = zeros(1,10);
    currentrow(i) = 1/36.84;
    currentrow(10) = 1/82.00;
    count_2 = count_2+1;
    N_alpha = [N_alpha;currentrow];
end
%SET 3: Incidence matrix entries for horizontal internal walls
count_3 = 0;
for i = 1:6
currentrow = zeros(1,10);
    currentrow(i) = 1/(rat1*21.32);
    currentrow(i+3) = 1/(rat1*21.32);
    count_3 = count_3+1;
    N_alpha = [N_alpha;currentrow];
end
%SET 4: Incidence matrix entries for vertical internal walls
count_4 = 0;
for i = 1:2
    for j = i:3:i+6
        currentrow = zeros(1,10);
        currentrow(j) = 1/(rat2*21.32);
        currentrow(j+1) = 1/(rat2*21.32);
        count_4 = count_4+1;
        N_alpha = [N_alpha;currentrow];
    end
end
%Capacitance matrices
C_w = blkdiag(8329.15*eye(count_1), 8329.15*eye(count_2), 4660*eye(count_3),
4660*eye(count_4));
C_z = blkdiag(250*eye(9));
C_cap = blkdiag(C_w,C_z);
% ****************************************************************
```

STEP 2: Obtaining matrices A, B, C and D for continuous time state space model (see Algorithm 3.1)

```
% *******************************************************************
nw = count_1+count_2+count_3+count_4;
nz = 9;
N_beta = -N_alpha;
N_alpha_z = N_alpha(:,1:nz);
N_beta_z = N_beta(:,1:nz);
S_r_beta = diag(sum(N_beta'));
S_c_alpha_z = diag(sum(N_alpha_z));
N_alpha_a = N_alpha(:,nz+1:nz+1);
inv_C_cap = inv(C_cap);
% Generation of A, B, C and D matrices denotes using the subscript "bldg."
A_bldg = inv_C_cap*[[S_r_beta,N_alpha_z];[-N_beta_z',-S_c_alpha_z]];
B_bldg = inv_C_cap*[[zeros(nw,nz),zeros(nw,nz),N_alpha_a];[eye(nz),eye(nz),zeros(nz,1)]];
C_bldg = [zeros(nz,nw),eye(nz)];
D_bldg = zeros(nz,2*nz+1);
% *******************************************************************
```

STEP 3: Discretization of model and generation of Hessian matrix

```
% *******************************************************************
% Part 1: Discretization of continuous state space model, resulting matrices are A_fd, %B_fd, and C_fd
global N Hess Nu
sys = ss(A_bldg,B_bldg,C_bldg,D_bldg);
Ts = 600;
sysd = c2d(sys,Ts,'zoh');
A_fd = sysd.A;
B_f = sysd.B;
C_fd = sys.C;
B_fd = B_f(:,1:9);
N = 24; %Prediction horizon
Nu = 9;
%Finding unswapped Hessian
for i = 0:N-1
   P(:,:,i+1) = C_fd*(A_fd)^i*B_fd;
end
Hess_us = [];
for i = 0:N-1
   blah = [];
   for j =0:N-1
      blah = [blah,(P(:,:,i+1))'*(P(:,:,j+1))];
   end
   Hess_us = [Hess_us;blah];
end
```

```
%Finding swapped Hessian
H_s = [];
for i = 1:Nu
    for j = 0:N-1
        H_s = [H_s;Hess_us(j*Nu+i,:)];
    end
end
Hess =[];
for i = 1:Nu
    for j = 0:N-1
        Hess = [Hess,H_s(:,j*Nu+i)];
    end
end
% ********************************************************************
```

STEP 4: Stage by stage combinatorial optimization (see section 4.2.3) and generation of optimality-robustness tradeoff curve

```
% ********************************************************************
global N Hess Nu
parentset = struct('values',{[1:1:Nu]});
rcs = 1;
k = 1;
clf_vect(k) = 0;
res_vect(k) = rcs;
while rcs > 1/Nu
[childset,clf,mcs] = comboptm(parentset); % call the function "comboptm" to calculate
% the child partition for a given parent partition (stage level combinatorial optimization)
% To perform optimization using the MINCUT method instead, replace "comboptm" with
"comboptm_maxcut"
childset.values % displays result (output child partition) of the combinatorial optimization
 parentset = childset;
 k = k + 1;
clf_vect(k) = clf;
 mcs_vect(k) = mcs;
end
plot(mcs_vect,clf_vect,'-') %Generates optimality-robustness tradeoff curve
% ********************************************************************
```

The MATLAB functions "comboptm" and "comboptm_maxcut" used in the above codes are as shown:

```
% ********************************************************************
function [childset,clf1,mcs] = comboptm(parentset)
global N Hess Nu
no_parents = length(parentset);
```

```
minweight  = 99999;
for l = 1:no_parents
    list = parentset(l).values;
    n = length(list);
    for k = 1:n-1
        C = combnk(list,k); % Creates all possible intermediate cluster pairs
        nc = length(C(:,1));
% The following loop compares the intermediate cluster pairs to find the ones with
% smallest ILF
        for i = 1:nc
            innerlist = C(i,:);
            innerweight = weight_calculate(innerlist,list)/self_weight(list); %ILF (see (4.14))
            if(innerweight<minweight)
                minweight = innerweight;
                minlist = innerlist;
                minlist_comp = setdiff(list,minlist);
                minparent = l;
            end
        end
    end
end
% Create child clusters
childset = struct('values',{});
for l = 1:no_parents;
    if l<minparent
        childset(l).values = parentset(l).values;
    end
    if l == minparent
        childset(l).values = minlist;
        childset(l+1).values = minlist_comp;
    end
    if(l>minparent)
        childset(l+1).values = parentset(l).values;
    end
end
% The remainder of the program computes the CLF and MCS of child clusters
c = no_parents + 1;
tot_coupling_lost = [];
for i = 1:c-1
    tot_coupling_lost_i = [];
    for j = i+1:c
        coupling_i_j = findcoupling(childset(i).values,childset(j).values);
        if (i == j)
            coupling_i_j = 0*coupling_i_j;
        end
```

```
      tot_coupling_lost_i = [tot_coupling_lost_i;coupling_i_j];
   end
   tot_coupling_lost = [tot_coupling_lost;tot_coupling_lost_i];
end
clf1 = (norm(tot_coupling_lost))/norm(Hess);
mcs = 1/c;
% **********************************************************************

% **********************************************************************
function [childset,clf1,mcs] = comboptm_maxcut(parentset)
global N Hess Nu
no_parents = length(parentset);
minweight  = 99999;
for l = 1:no_parents
   list = parentset(l).values;
   n = length(list);
   if(n>1)
   %Create Hp
   Hp = [];
   for i = 1:n
      Hp_row = [];
      for j = 1:n
         Hp_row =[Hp_row,Hess((list(i)-1)*N+1:list(i)*N,(list(j)-1)*N+1:list(j)*N)];
      end
      Hp = [Hp;Hp_row];
   end
   for i = 1:n
   Hp((i-1)*N+1:i*N,(i-1)*N+1:i*N) = zeros(N,N);
   end
      %Create Q
   Q = zeros(N*n,n);
   for i = 1:n
      Q((i-1)*N+1:i*N,i) = ones(N,1);
   end
   %Create weighting matrix and run MINCUT
   W = Q'*(Hp.*Hp)*Q;
   [MinCutGroupsList, MinCutWeight] = MinCut([1],W);
   innerlist_indices = MinCutGroupsList(1,:);
   innerlist_indices = innerlist_indices(innerlist_indices ~= 0);
   innerlist = zeros(1,length(innerlist_indices));
   for i = 1:length(innerlist_indices)
      innerlist(i) = list(innerlist_indices(i));
   end
   %Compare across all parents
   innerweight = weight_calculate(innerlist,list)/self_weight(list);
```

```
    if(innerweight<minweight)
        minweight = innerweight;
        minlist = innerlist;
        minlist_comp = setdiff(list,minlist);
        minparent = l;
    end
    end
end
childset = struct('values',{});
for l = 1:no_parents;
    if l<minparent
        childset(l).values = parentset(l).values;
    end
    if l == minparent
        childset(l).values = minlist;
        childset(l+1).values = minlist_comp;
    end
    if(l>minparent)
        childset(l+1).values = parentset(l).values;
    end
end
c = no_parents + 1;
tot_coupling_lost = [];
for i = 1:c-1
    tot_coupling_lost_i = [];
    for j = i+1:c
        coupling_i_j = findcoupling(childset(i).values,childset(j).values);
        if (i == j)
            coupling_i_j = 0*coupling_i_j;
        end
         tot_coupling_lost_i = [tot_coupling_lost_i;coupling_i_j];
    end
    tot_coupling_lost = [tot_coupling_lost;tot_coupling_lost_i];
end
clf1 = (norm(tot_coupling_lost))/norm(Hess);
mcs = 1/c;
% *****************************************************************
```

The MATLAB functions “weight_calculate”, “self_weight” and “findcoupling” used in the above function are shown below

```
% *****************************************************************
function result = weight_calculate(innerlist,list)
global N Hess Nu
cluster1 = innerlist;
cluster2 = setdiff(list,innerlist);
```

```
bout = [];
for i = 1:length(cluster1)
   bin = [];
   for j = 1:length(cluster2)
      bin = [bin,Hess((cluster1(i)-1)*N+1:cluster1(i)*N,(cluster2(j) - 1)*N+1:cluster2(j)*N)];
   end
   bout = [bout;bin];
end
result = norm(bout);
% ********************************************************************

% ********************************************************************
function result = self_weight(list)
global N Hess Nu
bout = [];
for i = 1:length(list)
   bin = [];
   for j = 1:length(list)
      bin = [bin,Hess((list(i)-1)*N+1:list(i)*N,(list(j)-1)*N+1:list(j)*N)];
   end
   bout = [bout;bin];
end
result = norm(bout);
% ********************************************************************

% ********************************************************************
function result = findcoupling(cluster1,cluster2)
global N Hess Nu
bout = [];
for i = 1:length(cluster1)
   bin = [];
   for j = 1:length(cluster2)
      bin = [bin,Hess((cluster1(i)-1)*N+1:cluster1(i)*N,(cluster2(j)-1)*N+1:cluster2(j)*N)];
   end
   bout = [bout;bin];
end
result = norm(bout);
% ********************************************************************

% ********************************************************************
function [MinCutGroupsList, MinCutWeight] = MinCut(SourceVertices, WeightedGraph)
%%% performs Min Cut algorithm described in "A Simple Min Cut Algorithm" by
%%% M. Stoer and F. Wagner
%%%[source: http://www.mathworks.com/matlabcentral/fileexchange/13892-a-simple-min-cut-
algorithm]
```

```
%%% input -
%%%    SourceVertices - a list of vertices that are forced to be kept in one side of the cut.
%%%    WeightedGraph - symetric matrix of edge weights. Wi,j is the edge connecting
%vertices i,j
%%%                  use Wi,j=0 or Wi,j == inf to indicate unconnected vertices.
%%% output -
%%%   MinCutGroupsList - two lists of verices, SECOND one contains the sourve vertives
%%%   MinCutWeight - sum of weight of edges alosng the cut
%%% (C) Yohai Devir 2006
%%% <my first name> AT WHOEVER D0T COM
   GraphDim = size(WeightedGraph,1);
   SourceVertices = SourceVertices(SourceVertices ~= 0); %remove zero vertices
   %%% remove self edges and ZEROed ones
   WeightedGraph = WeightedGraph+diag(inf(1,GraphDim));
   % for ii = 1:GraphDim
   %    WeightedGraph(ii,ii) = inf;
   % end
   WeightedGraph(WeightedGraph == 0) = inf;
   %%%Merge all Source Vrtices to one, so they'll be unbreakable, descending order is
%VITAL!!!
   SourceVertices = sort(SourceVertices);
   GroupsList = zeros(GraphDim);   %each row are the vertices melted into one vertex in %the
table.
   GroupsList(:,1) = 1:GraphDim;
   for ii=length(SourceVertices):-1:2;
      [WeightedGraph,GroupsList] =
MeltTwoVertices(SourceVertices(1),SourceVertices(ii),WeightedGraph,GroupsList);
   end
   Split = GroupsList(:,1);
   %%% By now I have a weighted graph in which all seed vertices are
   %%% merged into one vertex. Run Mincut algrithm on this graph
   [MinCutGroupsList_L, MinCutWeight] = MinCutNoSeed(WeightedGraph);
   %%% Convert Data so the seed vertices will be reconsidered as different
   %%% vertices and not one vertex.
   for ii = 1:2
      MinCutGroupsList(ii,:) = Local2GlobalIndices(MinCutGroupsList_L(ii,:), Split);
   end
   if (length(find(MinCutGroupsList(1,:) == SourceVertices(1))) == 1)
      SeedLocation = 1;
   else
      SeedLocation = 2;
   end
   MinCutGroupsList_withSeed =
[MinCutGroupsList(SeedLocation,(MinCutGroupsList(SeedLocation,:)~=0))
SourceVertices(2:length(SourceVertices))];
```

```
   MinCutGroupsList_withSeed = sort(MinCutGroupsList_withSeed);
   MinCutGroupsList_withSeed = [MinCutGroupsList_withSeed zeros(1,GraphDim -
length(MinCutGroupsList_withSeed))];
   MinCutGroupsList_NoSeed = MinCutGroupsList(3 - SeedLocation,(MinCutGroupsList(3 -
SeedLocation,:)~=0));
   MinCutGroupsList_NoSeed = sort(MinCutGroupsList_NoSeed);
   MinCutGroupsList_NoSeed = [MinCutGroupsList_NoSeed zeros(1,GraphDim -
length(MinCutGroupsList_NoSeed))];
   MinCutGroupsList = [MinCutGroupsList_NoSeed ; MinCutGroupsList_withSeed];
return
%%%%%%%%%%%%%%%%%%%%%%%%%%%%%%%%%%%%%%%%%%%%%%%%%%%%%%%%%%%%%%%%%%%%%%%%%%%%%%%%
%%%%%%%%%%%%%%%%%%%%%%%%%%%%%%%%%%%%%%%%%%%%%%%%
%%% Perform ordinary Stoer & Wagner algorithm Min Cut algorithm
%%%%%%%%%%%%%%%%%%%%%%%%%%%%%%%%%%%%%%%%%%%%%%%%%%%%%%%%%%%%%%%%%%%%%%%%%%%%%%%%
%%%%%%%%%%%%%%%%%%%%%%%%%%%%%%%%%%%%%%%%%%%%%%%%
function [MinCutGroupsList, MinCutWeight] = MinCutNoSeed(WeightedGraph)
   GraphDim = size(WeightedGraph,1);
   GroupsList = zeros(GraphDim);
   GroupsList(:,1) = 1:GraphDim;
   MinCutWeight = inf;
   MinCutGroup = [];
   for ii = 1:GraphDim-1
      [OneBefore, LastVertex] = MinimumCutPhase(WeightedGraph);
      if OneBefore == -1 %Graph is not connected. LastVertex is a group of vertices not
connected to Vertex 1
         MinCutGroup_L = LastVertex(LastVertex~=0); clear LastVertex; %it's not the last
vertex
         MinCutGroupsList = [];
         for jj = 1:length(MinCutGroup_L);
            MinCutGroup_temp = GroupsList(MinCutGroup_L(jj));
            MinCutGroup_temp = MinCutGroup_temp(MinCutGroup_temp~=0);
            MinCutGroupsList = [MinCutGroupsList MinCutGroup_temp];
         end
         MinCutGroupsList = [MinCutGroupsList zeros(1,GraphDim -
length(MinCutGroupsList))];
         jj = 1;
         for kk=1:GraphDim
            if (find(MinCutGroupsList(1,:) == kk))
               MinCutGroupsList(2 ,jj) = kk;
               jj = jj + 1;
            end
         end
         MinCutWeight = 0;
         return
      end %of: If graph is not connected
```

```
        Edges = WeightedGraph(LastVertex,:);
        Edges = Edges(isfinite(Edges));
        MinCutPhaseWeight = sum(Edges);
        if MinCutPhaseWeight < MinCutWeight
            MinCutWeight = MinCutPhaseWeight;
            MinCutGroup = GroupsList(LastVertex,:);
            MinCutGroup = MinCutGroup(MinCutGroup~=0);
        end
        [WeightedGraph,GroupsList] =
MeltTwoVertices(LastVertex,OneBefore,WeightedGraph,GroupsList);
    end
    MinCutGroup = sort(MinCutGroup);
    MinCutGroupsList = [MinCutGroup zeros(1,GraphDim - length(MinCutGroup))];
    jj = 1;
    for kk=1:GraphDim
        if isempty(find(MinCutGroup(1,:) == kk,1))
            MinCutGroupsList(2 ,jj) = kk;
            jj = jj + 1;
        end
    end
return
%%% This function takes V1 and V2 as vertexes in the given graph and MERGES
%%% THEM INTO V1 !!
%%% The output is the UpdatedGraph inwhich both vertices are considered
%%% one, and updated GroupsList to reflect the change.
function        [UpdatedGraph,GroupsList] =
MeltTwoVertices(V1,V2,WeightedGraph,GroupsList)
    t = min(V1,V2);
    V2 = max(V1,V2);
    V1 = t;
    GraphDim = size(WeightedGraph,1);
    UpdatedGraph = WeightedGraph;
    Mask1 = isinf(WeightedGraph(V1,:) );
    Mask2 = isinf(WeightedGraph(V2,:) );
    UpdatedGraph(V1,Mask1) = 0;
    UpdatedGraph(V2,Mask2) = 0;
    infMask = zeros(1,size(Mask1,2));
%    infMask(find(Mask1 & Mask2)) = inf;
    infMask((Mask1 & Mask2)) = inf;
    UpdatedGraph(V1,:)  =  UpdatedGraph(V1,:)  + UpdatedGraph(V2,:) + infMask;
    UpdatedGraph(:,V1) = UpdatedGraph(V1,:)';
    selectVec = true(1,GraphDim); selectVec(V2) = false;
%    UpdatedGraph = [UpdatedGraph(1:V2-1,:) ; UpdatedGraph(V2+1:GraphDim,:)]; %remove
second vertex from graph
%    UpdatedGraph = [UpdatedGraph(:,1:V2-1)  UpdatedGraph(:,V2+1:GraphDim)];
```

```
   UpdatedGraph = UpdatedGraph(selectVec,selectVec);
   UpdatedGraph(V1,V1) = inf;                                    % group-group distance
   V1list = GroupsList(V1,( GroupsList(V1,:)~=0 ) );
   V2list = GroupsList(V2,( GroupsList(V2,:)~=0 ) );
   GroupsList(V1,:) = [V1list V2list zeros(1,size( GroupsList,2)- length(V1list) - length(V2list))
]; %shorten grouplist
%     GroupsList = [GroupsList(1:V2-1,:) ;GroupsList(V2+1:GraphDim,:) ];
   GroupsList = GroupsList(selectVec,:);
return
%%%%%%%%%%%%%%%%%%%%%%%%%%%%%%%%%%%%%%%%%%%%%%
%%%%%%%%%%%%%%%%%%%%%%%%%%%%%%%%
%%% perform one phase of the algorithm
%%%
%%% return [-1, B  ] in case of Unconnected Graph when B is a subgraph(s)
%%% that are not connected to Vertex 1
%%%%%%%%%%%%%%%%%%%%%%%%%%%%%%%%%%%%%%%%%%%%%%
%%%%%%%%%%%%%%%%%%%%%%%%%%%%%%%%
function [OneBefore, LastVertex] = MinimumCutPhase(WeightedGraph)
   GraphDim = size(WeightedGraph,1);
   GroupsList = zeros(GraphDim);
   GroupsList(:,1) = 1:GraphDim;
   if size(WeightedGraph,1) > 2
      FarestVertexGroup = 0;
      for ii = 1:GraphDim-1
         OneBefore        = FarestVertexGroup(1);
         PossibleVertices = WeightedGraph(1,1:size(WeightedGraph,2));
         PossibleVertices(isinf(PossibleVertices)) = 0;
         FarestVertex = find(PossibleVertices == max(PossibleVertices),1,'first');
         if FarestVertex == 1 %In case the Graph is not connected
            OneBefore = -1;
            LastVertex = GroupsList(1,:);
            return
         end
         FarestVertexGroup = GroupsList(FarestVertex,:);
         [WeightedGraph,GroupsList] =
MeltTwoVertices(1,FarestVertex,WeightedGraph,GroupsList);
      end
      LastVertex = FarestVertexGroup(1);
   else
      OneBefore  = 1;
      LastVertex = 2;
   end
return
%%% Having a local list of indices in a global list and sublist of the
%%% local list, find the corresponding global indices.
```

```
function GlobalIndices = Local2GlobalIndices(LocalIndices, Split)
   if max(LocalIndices) > length(Split)
      error('Local indices are bigger than local split\n');
   end
   GlobalIndices = nan(length(LocalIndices),1);
   for ii=1:length(LocalIndices)
      if LocalIndices(ii) == 0
         GlobalIndices(ii) = 0;
      else
         GlobalIndices(ii) = Split(LocalIndices(ii));
      end
   end
return
% **********************************************************************
```

# Appendix C

# Codes for 9-zone building example in section 4.3.10

To obtain the results presented in this section, the following programs need to be run in the specified sequence:

STEP 1: Obtain weighted incidence matrix, capacitance matrices, and state space matrices (see Algorithm 3.1)

```
% ******************************************************************
clc
clear all;
% Part 1: Generation of Incidence matrix (similar to code used for 9-zone building   % model in section 4.2.8)
clc
clear all
N_alpha = [];
Incid = [];
overall_rat = 0.06;
rat1 = 1;
rat2 = 1/3;
% Set 1: Horizontal external walls
count_1 = 0;
for i = 1:3
   count_1 = count_1+1;
   currentrow = zeros(1,10);
   booleanrow = zeros(1,10);
   currentrow(i) = 1/(29.99*overall_rat);
   booleanrow(i) = 1;
   currentrow(10) = 1/(81.08*overall_rat);
   booleanrow(10) = 1;
   N_alpha = [N_alpha;currentrow];
   Incid = [Incid;booleanrow];
end
for i = 7:9
   currentrow = zeros(1,10);
   booleanrow = zeros(1,10);
   currentrow(i) = 1/(29.99*overall_rat);
```

```
    booleanrow(i) = 1;
    currentrow(10) = 1/(81.08*overall_rat);
    booleanrow(10) = 1;
    count_1 = count_1+1;
    N_alpha = [N_alpha;currentrow];
    Incid = [Incid;booleanrow];
end
% Set 2: Vertical external walls
count_2 = 0;
for i = 1:3:7
    currentrow = zeros(1,10);
    booleanrow = zeros(1,10);
    currentrow(i) = 1/(36.84*overall_rat);
    booleanrow(i) = 1;
    currentrow(10) = 1/(82.00*overall_rat);
    booleanrow(10) = 1;
    count_2 = count_2+1;
    N_alpha = [N_alpha;currentrow];
    Incid = [Incid;booleanrow];
end
for i = 1:9
    currentrow = zeros(1,10);
    booleanrow = zeros(1,10);
    currentrow(i) = 1/(36.84*overall_rat);
    booleanrow(i) = 1;
    currentrow(10) = 1/(82.00*overall_rat);
    booleanrow(10) = 1;
    count_2 = count_2+1;
    N_alpha = [N_alpha;currentrow];
    Incid = [Incid;booleanrow];
end
for i = 3:3:9
    currentrow = zeros(1,10);
    booleanrow = zeros(1,10);
    currentrow(i) = 1/(36.84*overall_rat);
    booleanrow(i) = 1;
    currentrow(10) = 1/(82.00*overall_rat);
    booleanrow(10) = 1;
    count_2 = count_2+1;
```

```
    N_alpha = [N_alpha;currentrow];
    Incid = [Incid;booleanrow];
end
for i = 1:9
    currentrow = zeros(1,10);
    booleanrow = zeros(1,10);
    currentrow(i) = 1/(36.84*overall_rat);
    booleanrow(i) = 1;
    currentrow(10) = 1/(82.00*overall_rat);
     booleanrow(10) = 1;
    count_2 = count_2+1;
    N_alpha = [N_alpha;currentrow];
    Incid = [Incid;booleanrow];
end
% Set 3: Horizontal internal walls
count_3 = 0;
for i = 1:6
currentrow = zeros(1,10);
booleanrow = zeros(1,10);
    currentrow(i) = 1/(rat1*21.32*overall_rat);
    booleanrow(i) = 1;
    currentrow(i+3) = 1/(rat1*21.32*overall_rat);
    booleanrow(i+3) = 1;
    count_3 = count_3+1;
    N_alpha = [N_alpha;currentrow];
    Incid = [Incid;booleanrow];
end
% Set 4: Vertical internal walls
count_4 = 0;
for i = 1:2
    for j = i:3:i+6
        currentrow = zeros(1,10);
        booleanrow = zeros(1,10);
        currentrow(j) = 1/(rat2*21.32*overall_rat);
         booleanrow(j) = 1;
        currentrow(j+1) = 1/(rat2*21.32*overall_rat);
        booleanrow(j+1) = 1;
        count_4 = count_4+1;
        N_alpha = [N_alpha;currentrow];
```

```
    Incid = [Incid;booleanrow];
  end
end
% Part 2: Generation of capacitance matrices
C_w =
0.1*blkdiag(8329.15*eye(count_1),8329.15*eye(count_2),4660*eye(count_3),4660*eye(count_
4));
c_z = 1*250*ones(9,1);
C_z = blkdiag(250*eye(9));
C_cap = blkdiag(C_w,C_z);
% Part 3: Generation of state space matrices
nw = count_1+count_2+count_3+count_4;
nz = 9;
nx = nw+nz;
nu = 9;
N_beta = -N_alpha;
N_alpha_z = N_alpha(:,1:nz);
N_beta_z = N_beta(:,1:nz);
S_r_beta = diag(sum(N_beta'));
S_c_alpha_z = diag(sum(N_alpha_z));
N_alpha_a = N_alpha(:,nz+1:nz+1);
inv_C_cap = inv(C_cap);
A_bldg = inv_C_cap*[[S_r_beta,N_alpha_z];[-N_beta_z',-S_c_alpha_z]];
B_bldg = inv_C_cap*[[zeros(nw,nz),N_alpha_a];[eye(nz),zeros(nz,1)]];
C_bldg = [zeros(nz,nw),eye(nz)];
D_bldg = zeros(nz,nz+1);
% Part 4: State space matrices in desired form, and zone-wall matrix
A_ww = A_bldg(1:nw,1:nw);
A_wz = A_bldg(1:nw,nw+1:nw+nz);
A_zw = A_bldg(nw+1:nw+nz,1:nw);
A_zz = A_bldg(nw+1:nw+nz,nw+1:nw+nz);
B_zd = B_bldg(nw+1:nw+nz,1:nz);
ZW = (Incid(:,1:nz))';
Nw = nw;
Nz = nz;
% Part 5: MPC Parameters
Ts = 60;
alphaa = 0.1;
N = 5;
```

```
% Part 6: Discretization of model
Ad_ww = eye(Nw) + Ts*A_ww;
Ad_wz = Ts*A_wz;
Ad_zw = Ts*A_zw;
Ad_zz = eye(Nz) + Ts*A_zz;
Bd_zd = Ts*B_zd;
    save nine_zone_model
% ***********************************************************************
```

STEP 2: Obtain true optimal partitions when the number of clusters is *i* (program has to be rerun for $1 \leq i \leq 9$)

```
% ***********************************************************************
i = 1;
min_olf = 99999999999;
   res = SetPartition(Nz,i); %Obtains set partitions
   n_part_i = length(res);
   local_part = struct;
   for k = 1:n_part_ilo
      for j = 1:i
         local_part(j).value = (res{k}{j})';
      end
      local_olf(k) = find_new_olf(i,local_part);
      hold on
      if(local_olf(k) < min_olf)
         min_olf = local_olf(k);
         min_part = local_part;
      end
   end
   hold on;
    scatter(i*ones(n_part_i,1),local_olf,'k','filled'); %Plots trade-off curve
   optimal_olf(i) = min_olf;
   optimal_part = min_part;
% ***********************************************************************
```

The function “SetPartition” appearing in the above program is part of a package downloaded from MATLAB Central. It is included in the media accompanying this thesis.

STEP 3: Obtain optimal partitions via agglomerative clustering (Algorithm 4.1)

```
% ***********************************************************************
n_parent = 9;
s_dc_parent = struct;
s_dc_parent(1).value = [1]';
s_dc_parent(2).value = [2]';
s_dc_parent(3).value = [3]';
s_dc_parent(4).value = [4]';
s_dc_parent(5).value = [5]';
```

```
s_dc_parent(6).value = [6]';
s_dc_parent(7).value = [7]';
s_dc_parent(8).value = [8]';
s_dc_parent(9).value = [9]';
olf = zeros(Nz,1);
olf(Nz) = find_new_olf(n_parent,s_dc_parent);
fpm(Nz) = find_fpm(s_dc_parent);
for p = Nz-1:-1:1
olf_min = 9999999999999999999999;
s_dc_child_min = struct;
for i = 1:n_parent-1
   for j = i+1:n_parent
      n_child = n_parent-1;
      s_dc_child = struct;
        for k = 1:j-1
        s_dc_child(k).value = s_dc_parent(k).value;
        end
        s_dc_child(i).value = [s_dc_parent(i).value;s_dc_parent(j).value];
      if(j<n_parent)
         for k = j:n_parent-1
            s_dc_child(k).value = s_dc_parent(k+1).value;
         end
      end
      olf_child = find_new_olf(n_child,s_dc_child);
      if(olf_child <= olf_min)
         olf_min = olf_child;
         l_max = length(s_dc_child);
         for l = 1:l_max
         s_dc_child_min(l).value = s_dc_child(l).value;
         end
      end
   end
end
n_parent = n_child;
l_max = length(s_dc_child_min);
for l = 1:l_max
s_dc_parent(l).value = s_dc_child_min(l).value;
end
olf(p) = olf_min;
fpm(p) = find_fpm(s_dc_child_min);
disp('***************************************');
disp('The Optimal child partition for p = ');
disp(p);
for l = 1:p
   disp(s_dc_child_min(l).value);
```

```
end
end
% *********************************************************************
```

The functions “find_new_olf” and “find_fpm” used in the above code are shown below.

```
% *********************************************************************
function OLF = find_new_olf(n,s_dc)
load nine_zone_model
%********************************************************************
%*****************STRUCTURE CONVERSION PART********************
%********************************************************************
%Decentralized Boolean matrices
% Generate r_dc data structure (find walls for each cluster)
r_dc = struct;
N_dc = zeros(n,1);
N_dc_w = zeros(n,1);
for i = 1:n
   N_dc(i) = length(s_dc(i).value);
   current_wall_set = [];
   zone_set = s_dc(i).value;
   for l = 1:N_dc(i)
      wall_choice = [];
      for j = 1:Nw
         if(ZW(zone_set(l),j)==1)
            wall_choice = union(wall_choice,j);
         end
      end
     current_wall_set = union(current_wall_set,wall_choice);
   end
   r_dc(i).value = current_wall_set;
   N_dc_w(i) = length(current_wall_set);
end
%generate Pi and Qi
P_dc = struct;
Q_dc = struct;
for i = 1:n
   mat_P = zeros(N_dc(i),Nz);
   si = s_dc(i).value;
   for l = 1:N_dc(i)
      mat_P(l,si(l)) = 1;
   end
   P_dc(i).value = mat_P;
   mat_Q = zeros(N_dc_w(i),Nw);
   ri = r_dc(i).value;
```

```
    for l = 1:N_dc_w(i)
        mat_Q(l,ri(l)) = 1;
    end
    Q_dc(i).value = mat_Q;
end
%Generation of combined and lifted matrices
P_l_dc = struct;
P = zeros(Nz,Nz);
P_l = zeros(N*Nz,N*Nz);
for i = 1:n
    local_mat = zeros(N*N_dc(i),N*N_dc(i));
    for j = 1:N
        local_mat(N_dc(i)*(j-1)+1:N_dc(i)*j,Nz*(j-1)+1:Nz*j) = P_dc(i).value;
    end
    P_l_dc(i).value = local_mat;
end
rowpos = 0;
for i = 1:n
    P(rowpos+1:rowpos+N_dc(i),:) = P_dc(i).value;
    rowpos = rowpos + N_dc(i);
end
rowpos = 0;
for i = 1:n
    P_l(rowpos+1:rowpos+N_dc(i)*N,:) = P_l_dc(i).value;
    rowpos = rowpos+N_dc(i)*N;
end
%Decentralized MPC simplified model
A_dc = struct;
B_dc = struct;
for i = 1:n
    for j = 1:n
        if (i == j)
            A_dc(i,j).value = [(Q_dc(i).value)*Ad_ww*(Q_dc(i).value)'
(Q_dc(i).value)*Ad_wz*(P_dc(i).value)';(P_dc(i).value)*Ad_zw*(Q_dc(i).value)'
(P_dc(i).value)*Ad_zz*(P_dc(i).value)'];
        else
            A_dc(i,j).value = [(Q_dc(i).value)*Ad_wz*(P_dc(j).value)';zeros(N_dc(i),N_dc(j))];
        end
    end
    B_dc(i).value = [zeros(N_dc_w(i),N_dc(i));(P_dc(i).value)*Bd_zd*(P_dc(i).value)'];
end
%*****************************************************************
%*****************DECENTRALIZED MATRICES PART*****************
%*****************************************************************
%M_dc data structure
```

```
M_dc = struct;
for i = 1:n
   for k = 1:N
      temp_mat = zeros(N_dc(i)+N_dc_w(i),(N_dc(i)+N_dc_w(i))*N);
      for l = 1:k
         temp_mat(:,(N_dc(i)+N_dc_w(i))*(l-1)+1:(N_dc(i)+N_dc_w(i))*l) =
(A_dc(i,i).value)^(k-l);
      end
      temp_mat(:,(N_dc(i)+N_dc_w(i))*k+1:(N_dc(i)+N_dc_w(i))*N) =
zeros((N_dc(i)+N_dc_w(i)),(N-k)*(N_dc(i)+N_dc_w(i)));
      M_dc(i,k).value = temp_mat;
   end
end
%B_diag data structure
B_diag = struct;
for i = 1:n
   local_mat = zeros((N_dc(i)+N_dc_w(i))*N,N_dc(i)*N);
   for k = 1:N
      local_mat((N_dc(i)+N_dc_w(i))*(k-1)+1:(N_dc(i)+N_dc_w(i))*k,N_dc(i)*(k-
1)+1:N_dc(i)*k) = B_dc(i).value;
   end
   B_diag(i).value = local_mat;
end
%R_dc data structure
R_dc = struct;
for i = 1:n
   mat_sum = zeros((N_dc(i)+N_dc_w(i))*N,(N_dc(i)+N_dc_w(i)));
   for k = 1:N
      mat_sum = mat_sum + (M_dc(i,k).value)';
   end
   R_dc(i).value = 2*(B_diag(i).value)'*mat_sum;
end
%S_dc data structure
S_dc = struct;
for i = 1:n
   for j = 1:n
      if (i==j)
         S_dc(i,j).value = zeros(N*N_dc(i),N*N_dc(j));
      else
      mat_sum = zeros((N_dc(i)+N_dc_w(i))*N,(N_dc(i)+N_dc_w(i))*N);
      blk_mat = zeros(N*(N_dc(i)+N_dc_w(i)),N*N_dc(j));
      for k = 1:N
         mat_sum = mat_sum + (M_dc(i,k).value)'*(M_dc(i,k).value);
         blk_mat((N_dc(i)+N_dc_w(i))*(k-1)+1:(N_dc(i)+N_dc_w(i))*k,N_dc(j)*(k-
1)+1:N_dc(j)*k)=A_dc(i,j).value;
```

```
        end
        S_dc(i,j).value = 2*(B_diag(i).value)'*mat_sum*blk_mat;
        end
    end
end
%H_dc data structure
H_dc = struct;
for i = 1:n
    mat_sum = zeros((N_dc(i)+N_dc_w(i))*N,(N_dc(i)+N_dc_w(i))*N);
    for k = 1:N
        mat_sum = mat_sum + (M_dc(i,k).value)'*(M_dc(i,k).value);
    end
    H_dc(i).value = alphaa*eye(N*N_dc(i))+(B_diag(i).value)'*mat_sum*B_diag(i).value;
end
%Combination Matrices
%Hdc
Hdc = zeros(N*Nz,N*Nz);
pos = 0;
for i = 1:n
    Hdc(pos+1:pos+N*N_dc(i),pos+1:pos+N*N_dc(i)) = H_dc(i).value;
    pos = pos+N*N_dc(i);
end
%Sdc
Sdc = zeros(N*Nz,N*Nz);
rowpos = 0;
for i = 1:n
    colpos = 0;
    for j = 1:n
    Sdc(rowpos+1:rowpos+N*N_dc(i),colpos+1:colpos+N*N_dc(j)) = S_dc(i,j).value;
    colpos = colpos+N*N_dc(j);
    end
    rowpos = rowpos + N*N_dc(i);
end
%Rdc
Rdc = zeros(N*Nz,Nz+Nw);
rowpos = 0;
colpos = 0;
for i = 1:n
    Rdc(rowpos+1:rowpos+N*N_dc(i),colpos+1:colpos+N_dc(i)+N_dc_w(i)) = R_dc(i).value;
    rowpos = rowpos + N*N_dc(i);
    colpos = colpos + N_dc(i)+N_dc_w(i);
end
%*********************************************************************
%*****************H_c CALCULATION PART****************************
%*********************************************************************
```

```
%Centralized model matrices
A_c = [Ad_ww,Ad_wz;Ad_zw,Ad_zz];
B_c = [zeros(Nw,Nz);Bd_zd];
%M_c data structure
M_c = struct;
   for k = 1:N
      temp_mat = zeros(Nz+Nw,(Nz+Nw)*N);
      for l = 1:k
         temp_mat(:,(Nz+Nw)*(l-1)+1:(Nz+Nw)*l) = (A_c)^(k-l);
      end
      temp_mat(:,(Nz+Nw)*k+1:(Nz+Nw)*N) = zeros((Nz+Nw),(N-k)*(Nz+Nw));
      M_c(k).value = temp_mat;
   end
%B_diag_c matrix
   B_diag_c = zeros((Nz+Nw)*N,Nz*N);
   for k = 1:N
      B_diag_c((Nz+Nw)*(k-1)+1:(Nz+Nw)*k,Nz*(k-1)+1:Nz*k) = B_c;
   end
%H_c matrix
   mat_sum = zeros((Nz+Nw)*N,(Nz+Nw)*N);
   for k = 1:N
      mat_sum = mat_sum + (M_c(k).value)'*(M_c(k).value);
   end
   H_c = alphaa*eye(N*Nz)+(B_diag_c)'*mat_sum*B_diag_c;
%*****************************************************************
%*****************OLF CALCULATION PART**************************
%*****************************************************************
M = inv(P_l)*inv(Hdc)*Sdc*P_l;
OLF_mat = M'*H_c*M;
OLF = max(eig(OLF_mat));
%OLF = norm(M,inf);
End
% ****************************************************************

% ****************************************************************
function FPM = find_fpm(s_dc)
load nine_zone_model
p = length(s_dc);
N_dc = zeros(p,1);
summ = 0;
for i = 1:p
   cluster_i = s_dc(i).value;
   N_dc_i = length(cluster_i);
   cap_summ_i = 0;
   for j = 1:N_dc_i
```

```
        cap_summ_i = cap_summ_i + c_z(cluster_i(j));
    end
    summ = summ+ cap_summ_i*N_dc_i;
    %summ = summ+ (N_dc_i)^2;
end
FPM = summ/(Nz*sum(c_z));
end
% *********************************************************************
```

# Appendix D

# Codes for 11-zone building example in section 4.3.11

To perform the agglomerative clustering analysis presented in this section, the following programs need to be run in the specified sequence:

STEP 1: Obtain weighted incidence matrix, capacitance matrices, and state space matrices (see Algorithm 3.1)

```
% ***********************************************************************
%Model Inputs and MPC parameters for 11 zone circular bldg
clc
clear all;
% Part 1: Generation of Incidence matrix
clc
clear all
N_alpha = [];
Incid = [];
rat1 = 50;
rat2 = 1;
% Set 1: Horizontal external walls
count_1 = 0;
for i = 1:11
   count_1 = count_1+1;
   currentrow = zeros(1,12);
   booleanrow = zeros(1,12);
   if(i==1)
      currentrow(i) = 1/(0.5*18.356);
      booleanrow(i) = 1;
      currentrow(12) = 1/(0.5*18.356);
      booleanrow(12) = 1;
      N_alpha = [N_alpha;currentrow];
      Incid = [Incid;booleanrow];
   elseif((i>=2)&&(i<=6))
      currentrow(i) = 1/(0.5*30.594);
      booleanrow(i) = 1;
      currentrow(12) = 1/(0.5*30.594);
      booleanrow(12) = 1;
      N_alpha = [N_alpha;currentrow];
      Incid = [Incid;booleanrow];
```

```
    else
        currentrow(i) = 1/(0.5*16.6874);
        booleanrow(i) = 1;
        currentrow(12) = 1/(0.5*16.6874);
        booleanrow(12) = 1;
        N_alpha = [N_alpha;currentrow];
        Incid = [Incid;booleanrow];
    end
end
for i = 1:11
    count_1 = count_1+1;
    currentrow = zeros(1,12);
    booleanrow = zeros(1,12);
    if(i==1)
        currentrow(i) = 1/(0.5*18.356);
        booleanrow(i) = 1;
        currentrow(12) = 1/(0.5*18.356);
        booleanrow(12) = 1;
        N_alpha = [N_alpha;currentrow];
        Incid = [Incid;booleanrow];
    elseif((i>=2)&&(i<=6))
        currentrow(i) = 1/(0.5*30.594);
        booleanrow(i) = 1;
        currentrow(12) = 1/(0.5*30.594);
        booleanrow(12) = 1;
        N_alpha = [N_alpha;currentrow];
        Incid = [Incid;booleanrow];
    else
        currentrow(i) = 1/(0.5*16.6874);
        booleanrow(i) = 1;
        currentrow(12) = 1/(0.5*16.6874);
        booleanrow(12) = 1;
        N_alpha = [N_alpha;currentrow];
        Incid = [Incid;booleanrow];
    end
end
% Set 2: Vertical external walls
count_2 = 0;
for i = 7:11
    currentrow = zeros(1,12);
    booleanrow = zeros(1,12);
    currentrow(i) = 1/(0.5*18.827);
    booleanrow(i) = 1;
    currentrow(12) = 1/(0.5*18.827);
    booleanrow(12) = 1;
```

```
    count_2 = count_2+1;
    N_alpha = [N_alpha;currentrow];
    Incid = [Incid;booleanrow];
end
%Set 3: Horizontal internal walls – no such walls since single floor building
count_3 = 0;
%Set 4: Vertical internal walls
count_4 = 0;
for i = 2:6
    currentrow = zeros(1,12);
    booleanrow = zeros(1,12);
    currentrow(1) = 1/(0.5*18.32);
    booleanrow(1) = 1;
    currentrow(i) = 1/(0.5*18.32);
    booleanrow(i) = 1;
    count_4 = count_4+1;
    N_alpha = [N_alpha;currentrow];
    Incid = [Incid;booleanrow];
end
for i = 2:5
    currentrow = zeros(1,12);
    booleanrow = zeros(1,12);
    currentrow(i) = 1/(0.5*23.03);
    booleanrow(i) = 1;
    currentrow(i+1) = 1/(0.5*23.03);
    booleanrow(i+1) = 1;
    count_4 = count_4+1;
    N_alpha = [N_alpha;currentrow];
    Incid = [Incid;booleanrow];
end
    currentrow = zeros(1,12);
    booleanrow = zeros(1,12);
    currentrow(2) = 1/(0.5*46.06*rat1);
    booleanrow(2) = 1;
    currentrow(6) = 1/(0.5*46.06*rat1);
    booleanrow(6) = 1;
    count_4 = count_4+1;
    N_alpha = [N_alpha;currentrow];
    Incid = [Incid;booleanrow];
for i = 2:6
    currentrow = zeros(1,12);
    booleanrow = zeros(1,12);
    currentrow(i) = 1/(0.5*17.25*rat1);
    booleanrow(i) = 1;
    currentrow(i+5) = 1/(0.5*17.25*rat1);
```

```
    booleanrow(i+5) = 1;
    count_4 = count_4+1;
    N_alpha = [N_alpha;currentrow];
    Incid = [Incid;booleanrow];
end
for i = 7:10
    currentrow = zeros(1,12);
    booleanrow = zeros(1,12);
    currentrow(i) = 1/(0.5*23.03);
    booleanrow(i) = 1;
    currentrow(i+1) = 1/(0.5*23.03);
    booleanrow(i+1) = 1;
    count_4 = count_4+1;
    N_alpha = [N_alpha;currentrow];
    Incid = [Incid;booleanrow];
end
    currentrow = zeros(1,12);
    booleanrow = zeros(1,12);
    currentrow(7) = 1/(0.5*46.06*rat1);
    booleanrow(7) = 1;
    currentrow(11) = 1/(0.5*46.06*rat1);
    booleanrow(11) = 1;
    count_4 = count_4+1;
    N_alpha = [N_alpha;currentrow];
    Incid = [Incid;booleanrow];
% Part 2: Generation of capacitance matrices
c_w =
[4.27*10^4;ones(5,1)*2.56*10^4;ones(5,1)*3.36*10^4;4.27*10^4;ones(5,1)*2.56*10^4;ones(5,1
)*3.36*10^4;ones(5,1)*4.17*10^4;ones(5,1)*6.53*10^3;ones(4,1)*5.2*10^3;1.04*10^4;ones(5,1
)*2.77*10^4;ones(4,1)*5.2*10^3;1.04*10^4];
C_w = diag(c_w);
c_z = [643.72;ones(5,1)*386.23;ones(5,1)*708.09];
C_z = diag(c_z);
C_cap = blkdiag(C_w,C_z);
% Part 3: Generation of state space matrices
nw = count_1+count_2+count_3+count_4;
nz = 11;
N_beta = -N_alpha;
N_alpha_z = N_alpha(:,1:nz);
N_beta_z = N_beta(:,1:nz);
S_r_beta = diag(sum(N_beta'));
S_c_alpha_z = diag(sum(N_alpha_z));
N_alpha_a = N_alpha(:,nz+1:nz+1);
inv_C_cap = inv(C_cap);
A_bldg = inv_C_cap*[[S_r_beta,N_alpha_z];[-N_beta_z',-S_c_alpha_z]];
```

```
%B_bldg = inv_C_cap*[[zeros(nw,nz),zeros(nw,nz),N_alpha_a];[eye(nz),eye(nz),zeros(nz,1)]];
B_bldg = inv_C_cap*[[zeros(nw,nz),N_alpha_a];[eye(nz),zeros(nz,1)]];
C_bldg = [zeros(nz,nw),eye(nz)];
D_bldg = zeros(nz,nz+1);
% Part 4: State space matrices in desired form, and zone-wall matrix
A_ww = A_bldg(1:nw,1:nw);
A_wz = A_bldg(1:nw,nw+1:nw+nz);
A_zw = A_bldg(nw+1:nw+nz,1:nw);
A_zz = A_bldg(nw+1:nw+nz,nw+1:nw+nz);
B_normal_w = B_bldg(1:nw,:);
B_normal_z = B_bldg(nw+1:nw+nz,:);
B_zd = B_bldg(nw+1:nw+nz,1:nz);
ZW = (Incid(:,1:nz))';
Nw = nw;
Nz = nz;
%Part 5: MPC Parameters
Ts = 600;
alphaa = 0.1;
N = 5;
%Part 6: Discretization of model
Ad_ww = eye(Nw) + Ts*A_ww;
Ad_wz = Ts*A_wz;
Ad_zw = Ts*A_zw;
Ad_zz = eye(Nz) + Ts*A_zz;
Bd_zd = Ts*B_zd;
%Part 7: Simulation amplitudes – needed for open loop simulations later in this section
a1 = 0.5;
a2 = 0.5;
b1 = 10;
b2 = 20;
save circular_model
% ****************************************************************
```

STEP 2: To run the agglomerative clustering, run the program in Step 3 of Appendix C with the following changes:

(i) Replace “n_parent = 9;” with “n_parent = 11;”

(ii) Include the lines “s_dc_parent(10).value = [10]';” and “s_dc_parent(11).value = [11]';” after the line “s_dc_parent(9).value = [9]';”

(iii) In the functions “find_new_olf” and “find_fpm”, replace “load nine_zone_model” with “load circular_model”

To obtain the plot in Fig. 4.21 corresponding to the scalarized analysis, the following code was used. Note that this code provides the plot only for the case when $\lambda = 0.5$. It can be re-run with the appropriate modification “lambda = 0.85;” for the case when $\lambda = 0.85$.

```
% ******************************************************************
J = zeros(Nz,1);
lambda = 0.5;
for i = 1:Nz
   J(i) = lambda*olf(i)/max(olf) + (1-lambda)*(fpm(i))/(max(fpm));
end
figure;
plot(J,'k');
% ******************************************************************
```

To obtain the open loop simulation results, cluster level state-space models can be obtained using the following code. The models hence obtained are used in Simulink models to perform the necessary simulations. These Simulink models (for centralized, knee2 and fully decentralized cases) are included in the media accompanying this thesis.

```
% ******************************************************************
zone_set = [1,2,3,4,5,6]; % Zone numbers corresponding to the cluster go here. In this
 % example, we use the first cluster for knee 2 in Fig. 4.20
n_zone_set = length(zone_set);
wall_set = [];
for i = 1:n_zone_set
   w_i = [];
   for j = 1:nw
      if(ZW(zone_set(i),j)==1)
         w_i = [w_i;j];
      end
   end
   wall_set = union(wall_set,w_i);
end
n_wall_set = length(wall_set);
Adecen_ww = zeros(n_wall_set,n_wall_set);
Adecen_wz = zeros(n_wall_set,n_zone_set);
Adecen_zw = zeros(n_zone_set,n_wall_set);
Adecen_zz = zeros(n_zone_set,n_zone_set);
Bdecen_normal_w = zeros(n_wall_set,nz+1);
Bdecen_normal_z = zeros(n_zone_set,nz+1);
Bdecen_est_w = zeros(n_wall_set,nz);
Bdecen_est_z = zeros(n_zone_set,nz);
for i = 1:n_wall_set
   Adecen_ww(i,i) = A_ww(wall_set(i),wall_set(i));
end
for i = 1:n_wall_set
   for j = 1:n_zone_set
      Adecen_wz(i,j) = A_wz(wall_set(i),zone_set(j));
   end
```

```
end
for i = 1:n_zone_set
   for j = 1:n_wall_set
      Adecen_zw(i,j) = A_zw(zone_set(i),wall_set(j));
   end
end
for i = 1:n_zone_set
   Adecen_zz(i,i) = A_zz(zone_set(i),zone_set(i));
end
for i = 1:n_wall_set
   for j = 1:nz+1
   Bdecen_normal_w(i,j) = B_normal_w(wall_set(i),j);
   end
end
for i = 1:n_zone_set
   for j = 1:nz+1
   Bdecen_normal_z(i,j) = B_normal_z(zone_set(i),j);
   end
end
for i = 1:n_wall_set
   for j = 1:nz
      Bdecen_est_w(i,j) = A_wz(wall_set(i),j);
   end
end
for i = 1:n_wall_set
   for j = 1:n_zone_set
      Bdecen_est_w(i,zone_set(j)) = 0;
   end
end
Adecen = [Adecen_ww,Adecen_wz;Adecen_zw,Adecen_zz];
Bdecen = [Bdecen_normal_w,Bdecen_est_w;Bdecen_normal_z,Bdecen_est_z];
Cdecen = [zeros(n_zone_set,n_wall_set),eye(n_zone_set)];
Ddecen = zeros(n_zone_set,2*nz+1);
% The use of the number "1" in the variables "nwpart1" to "Dpart1" below corresponds to
% the variable names that are used to refer to the state space model corresponding to this
% cluster in the SIMULINK model
nwpart1 = n_wall_set;
nzpart1 = n_zone_set;
Apart1= Adecen;
Bpart1= Bdecen;
Cpart1= Cdecen;
Dpart1 = Ddecen;
% *******************************************************************
```

# Appendix E

# Steps used in creating an EnergyPlus model using OpenStudio

The following video tutorials provide information on creation of an EnergyPlus model using OpenStudio: http://www.youtube.com/user/NRELOpenStudio. Based on these tutorials, the procedure involves the following basic steps:

Step 1: Choose the construction template using the OpenStudio plugin in Google SketchUp
Step 2: Draw the building in Google SketchUp. This involves the following substeps:

- (i) Create the plan drawing of the building using SketchUp drawing tools
- (ii) Use the "spaces from diagram" tool in the plugin to create floors based on the above plan
- (iii) Draw fenestration surfaces using the SketchUp drawing tools
- (iv) Use the "project loose geometry" tool in the plugin to project the fenestration surfaces on appropriate building surfaces
- (v) Additional geometry if needed can be created using the "shading surface tool", "internal partition surface tool, etc.

Step 3: Use "Surface matching tool" in plugin to set appropriate boundary conditions for each surface.

Step 4: Use "Space attributes tool" in plugin to assign stories, thermal zones and other attributes to each of the spaces in the building. "Render modes" can be used to check the successful application of these attributes.

Step 5: Launch the OpenStudio application from the plugin to open the .osm file.

Step 6: In the .osm file, set the paths for weather and design day files.

Step 7: Examine the schedules via the schedules tab. The relevant schedules should already have been set to default values based on the construction template chosen in Step 1 above. If necessary, one or more schedules can be changed at this point by dragging and dropping schedules from the 'my model' or 'library' tabs.

Step 8: Examine the constructions via the constructions tab. Default construction sets should already be present based on the construction template chosen in Step 1 above. For each construction set, the relevant applicability –the entire building, a story, a space type or a space – should also be assigned by default. This information can be changed if needed. Also, each construction set consists of constructions, which in turn consist of a set of materials. The constructions and/or materials constituting any such construction set can be changed at this point using the 'my model' or 'library' tabs.

Step 9: Examine the load objects and the underlying attributes. Default values can be changed at this point.

Step 10: Examine the space types using the "space types" tab. Construction sets, schedule sets and loads are assigned to each space type. These assignments can be changed for existing space types. Also new space types can be created with appropriate construction set, schedule set and load assignments.

Step 11: Examine the stories using the "stories" tab. For each story, the underlying construction and schedule sets can be changed. Also new stories can be created by assigning appropriate construction and schedule sets.

Step 12: Click on the "facilities" tab. Choose appropriate view type – view by story/ space type/ thermal zone etc. Check the underlying assignments and ensure that the sets "unassigned space type", "unassigned thermal zone" etc. are empty. The facilities tab can be used to assign attributes such as loads at the building/space type/story level.

Step 13: Click on the "thermal zones" tab. The "ideal air loads" option can be turned on/off at this stage. If the "ideal air loads" option is turned off, zone equipment can be assigned at this stage. Also, thermostats can be set at this point to each thermal zone.

Step 14: The "HVAC systems tab" can be used to create HVAC system loops and specify the thermal zones which are serviced by the thermal "sink" element (e.g. Air handling Unit) in the loops. It can be used to modify the existing loops by adding/removing sink/source components, valves, terminal units etc.

Step 15: The variables which are desired to be included the SQL file generated at the end of the EnergyPlus simulation can be selected, along with the reporting frequency using "output variables" tab.

Step 16: Run the simulation using the "run simulation" tab. The directory containing the EnergyPlus input data file (idf) can be accessed using the "tree" sub-tab.

Step 17: The "EP-Launch" EnergyPlus application can be used to perform an EnergyPlus simulation via the idf file generated above. The results of the simulation are stored in a .csv file which can be accessed using Microsoft Excel. The columns of this file provide the values of the output variables which were selected in Step 15 above.

# Appendix F

# Codes for performing zone level identification as per optimization framework presented in (6.3) – (6.6)

To obtain the results presented in section 6.3.2 for the optimization framework corresponding to (6.3) – (6.6), the following steps are required. For the purposes of illustration, these steps are explained for Zone F2. For the other zones, the steps are the same and the appropriate codes are provided in the media accompanying this thesis.

STEP 1: In the EnergyPlus model, PRBS set-points are prescribed for all zones in the building. The EnergyPlus model (.idf file), hence modified, is included in the media accompanying this thesis.
STEP 2: A day-long simulation is run using the above EnergyPlus model and the generic spreadsheet containing the output variables is obtained.
STEP 3: From the above generic spreadsheet, columns containing the variables needed for the zone level identification are used to create a new spreadsheet called "sim_data_zone_F2". This file is also included in the media accompanying this thesis.
STEP 4: Simulation data contained in "sim_data_zone_F2" is exported to the MATLAB workspace and saved in a MATLAB file called "zone F2 data" using the following MATLAB code:

```
% ********************************************************************
% Import data for zone F2 in MATLAB
 clc;
 clear all;
 Q_dot_in = 0.001*xlsread('sim_data_zone_F2.xls','F2_data','a2:a1441');
 m_dot_supp = xlsread('sim_data_zone_F2.xls','F2_data','b2:b1441');
 T_supp = xlsread('sim_data_zone_F2.xls','F2_data','c2:c1441');
 T_z = xlsread('sim_data_zone_F2.xls','F2_data','d2:d1441');
 T_w1 = xlsread('sim_data_zone_F2.xls','F2_data','e2:e1441');
 T_w2 = xlsread('sim_data_zone_F2.xls','F2_data','f2:f1441');
 T_w3 = xlsread('sim_data_zone_F2.xls','F2_data','g2:g1441');
 T_w3_sub1 = xlsread('sim_data_zone_F2.xls','F2_data','h2:h1441');
 T_w3_sub2 = xlsread('sim_data_zone_F2.xls','F2_data','i2:i1441');
 T_w4 = xlsread('sim_data_zone_F2.xls','F2_data','j2:j1441');
 T_w5 = xlsread('sim_data_zone_F2.xls','F2_data','k2:k1441');
 T_w5_sub1 = xlsread('sim_data_zone_F2.xls','F2_data','l2:l1441');
```

```
 T_w6 = xlsread('sim_data_zone_F2.xls','F2_data','m2:m1441');
 V_inf = xlsread('sim_data_zone_F2.xls','F2_data','n2:n1441');
 T_amb = xlsread('sim_data_zone_F2.xls','F2_data','o2:o1441');
 save zone_F2_data
% ***************************************************************
```

STEP 5: The following MATLAB code can then be run to perform the zone level optimization.

```
% ***************************************************************
clc
clear all
% Part 1: Generation of optimal parameter values
area1 = 98.4271*82.0208*0.3048^2;
area2 = 98.4271*13.7812*0.3048^2;
area4 = (72.1771-62.3333)*(7.875-2.95312)*0.3048^2;
area5 = (16.4062-1.64062)*(7.875-2.95312)*0.3048^2;
area3 = 82.0208*13.7812*0.3048^2 - area4-area5;
area6 = 82.0208*13.7812*0.3048^2;
area8 = (72.2344-3.28125)*(7.875-2.95312)*0.3048^2;
area7 = 98.4271*13.7812*0.3048^2 - area8;
area9 = 98.4271*82.0208*0.3048^2;
areas = [area1 area2 area3 area4 area5 area6 area7 area8 area9]';
matlabpool open 4
options =
optimset('Algorithm','sqp','MaxFunEvals',100000,'MaxIter',100000,'UseParallel','always','TolFun
', 1e-4, 'TolX', 1e-12, 'TolCon', 1e-12 );
init_param = [1*0.01*800*ones(9,1);5000]; % initial parameter valued
UB = [0.025*800*ones(9,1);10000]; % upper bounds on parameter values
LB = [0*ones(9,1);0]; % lower bounds on parameter values
optim_param = fmincon(@zone_F2_objective_fn_case1,init_param,[],[],[],[],LB,UB,[],options);
% call to
% fmincon to obtain optimal parameter values using least squares optimization
% Part 2: Comparison of zone temperature from least squares identification with EnergyPlus
simulation data
UA = optim_param(1:9);
C = optim_param(10);
load zone_F2_data
Tw_meas = [T_w1 T_w2 T_w3 T_w3_sub1 T_w3_sub2 T_w4 T_w5 T_w5_sub1 T_w6];
startt_index = 0;
startt = 60*startt_index;
Tz_pred(startt+1,1) = T_z(startt+1);
for k = startt+2:60*(startt_index+24)
	Tz_pred(k,1) = (Q_dot_in(k) + m_dot_supp(k)*1.012*T_supp(k) + C*T_z(k-1)/60 +
(Tw_meas(k-1,:))*(UA) + 1.2041*V_inf(k-1)*1.012*T_amb(k-1))/(C/60 + sum(UA) +
1.2041*V_inf(k-1)*1.012 + m_dot_supp(k)*1.012);
end
```

```
figure
plot(T_z(startt+1:60*(startt_index+1),1),'k');
hold on
plot(Tz_pred(startt+1:60*(startt_index+1)));
% *******************************************************************
```

The MATLAB function “zone_F2_objective_fn_case1” used in the above code is as follows:

```
% *******************************************************************
function cost = zone_F2_objective_fn_case1 (param)
% Part 1: Extract optimization variables
UA = param(1:9,1);
C = param(10);
% Part 2: Load measurement data
load zone_F2_data
Tw_meas = [T_w1 T_w2 T_w3 T_w3_sub1 T_w3_sub2 T_w4 T_w5 T_w5_sub1 T_w6];

% Part 3: Evolve the dynamics
area1 = 98.4271*82.0208*0.3048^2;
area2 = 98.4271*13.7812*0.3048^2;
area4 = (72.1771-62.3333)*(7.875-2.95312)*0.3048^2;
area5 = (16.4062-1.64062)*(7.875-2.95312)*0.3048^2;
area3 = 82.0208*13.7812*0.3048^2 - area4-area5;
area6 = 82.0208*13.7812*0.3048^2;
area8 = (72.2344-3.28125)*(7.875-2.95312)*0.3048^2;
area7 = 98.4271*13.7812*0.3048^2 - area8;
area9 = 98.4271*82.0208*0.3048^2;
areas = [area1 area2 area3 area4 area5 area6 area7 area8 area9]';
startt_index = 0;
startt = 60*startt_index;
Tz_pred(startt+1,1) = T_z(startt+1);
for k = startt+2:60*(startt_index+24)
        Tz_pred(k,1) = (Q_dot_in(k) + m_dot_supp(k)*1.012*T_supp(k) + C*T_z(k-1)/60 +
(Tw_meas(k-1,:))*(UA) + 1.2041*V_inf(k-1)*1.012*T_amb(k-1))/(C/60 + sum(UA) +
1.2041*V_inf(k-1)*1.012 + m_dot_supp(k)*1.012);
end
% Part 4: Calculate cost
deviation = Tz_pred(startt+1:60*(startt_index+24),1) - T_z(startt+1:60*(startt_index+24),1);
cost = norm(deviation);
% *******************************************************************
```

# Appendix G

# Codes for performing zone level identification as per optimization framework presented in (6.7) – (6.10)

To obtain the results presented in section 6.3.2 for the optimization framework corresponding to (6.7) – (6.10), the following steps are required. For the purposes of illustration, these steps are explained for Zone F2. For the other zones, the steps are the same and the appropriate codes are provided in the media accompanying this thesis.

STEPS 1- 4: Same as Steps 1-4 in Appendix F.
STEP 5: The following MATLAB code can then be run to perform the zone level optimization.

```
% *****************************************************************
clc
clear all
% Part 1: Generation of optimal parameter values
area1 = 98.4271*82.0208*0.3048^2;
area2 = 98.4271*13.7812*0.3048^2;
area4 = (72.1771-62.3333)*(7.875-2.95312)*0.3048^2;
area5 = (16.4062-1.64062)*(7.875-2.95312)*0.3048^2;
area3 = 82.0208*13.7812*0.3048^2 - area4-area5;
area6 = 82.0208*13.7812*0.3048^2;
area8 = (72.2344-3.28125)*(7.875-2.95312)*0.3048^2;
area7 = 98.4271*13.7812*0.3048^2 - area8;
area9 = 98.4271*82.0208*0.3048^2;
areas = [area1 area2 area3 area4 area5 area6 area7 area8 area9]';
%matlabpool open 4
options =
optimset('Algorithm','sqp','MaxFunEvals',100000,'MaxIter',100000,'UseParallel','always','TolFun
', 1e-4, 'TolX', 1e-12, 'TolCon', 1e-12 );
init_param = [0.001;5000];
UB = [0.01;10000];
LB = [0;0];
optim_param = fmincon(@zone_F2_objective_fn_case2,init_param,[],[],[],[],LB,UB,[],options);
% call to
% fmincon to obtain optimal parameter values using least squares optimization
% Part 2: Comparison of zone temperature from least squares identification with EnergyPlus
```

```
simulation data
h = optim_param(1);
C = optim_param(2);
load zone_F2_data
Tw_meas = [T_w1 T_w2 T_w3 T_w3_sub1 T_w3_sub2 T_w4 T_w5 T_w5_sub1 T_w6];
startt_index = 0;
startt = 60*startt_index;
Tz_pred(startt+1,1) = T_z(startt+1);
for k = startt+2:60*(startt_index+24)
    Tz_pred(k,1) = (Q_dot_in(k) + m_dot_supp(k)*1.012*T_supp(k) + C*T_z(k-1)/60 +
(Tw_meas(k-1,:))*(h*areas) + 1.2041*V_inf(k-1)*1.012*T_amb(k-1))/(C/60 + sum(h*areas) +
1.2041*V_inf(k-1)*1.012 + m_dot_supp(k)*1.012);
end
figure
plot(T_z(startt+1:60*(startt_index+1),1),'k');
hold on
plot(Tz_pred(startt+1:60*(startt_index+1)));
% ********************************************************************
```

The MATLAB function “zone_F2_objective_fn_case2” used in the above code is as follows:

```
% ********************************************************************
function cost = zone_F2_objective_fn_case2 (param)
% Part 1: Extract optimization variables
h = param(1);
C = param(2);
% Part 2: Load measurement data
load zone_F2_data
Tw_meas = [T_w1 T_w2 T_w3 T_w3_sub1 T_w3_sub2 T_w4 T_w5 T_w5_sub1 T_w6];
% Part 3: Evolve the dynamics
area1 = 98.4271*82.0208*0.3048^2;
area2 = 98.4271*13.7812*0.3048^2;
area4 = (72.1771-62.3333)*(7.875-2.95312)*0.3048^2;
area5 = (16.4062-1.64062)*(7.875-2.95312)*0.3048^2;
area3 = 82.0208*13.7812*0.3048^2 - area4-area5;
area6 = 82.0208*13.7812*0.3048^2;
area8 = (72.2344-3.28125)*(7.875-2.95312)*0.3048^2;
area7 = 98.4271*13.7812*0.3048^2 - area8;
area9 = 98.4271*82.0208*0.3048^2;
areas = [area1 area2 area3 area4 area5 area6 area7 area8 area9]';
startt_index = 0;
startt = 60*startt_index;
Tz_pred(startt+1,1) = T_z(startt+1);
for k = startt+2:60*(startt_index+24)
    Tz_pred(k,1) = (Q_dot_in(k) + m_dot_supp(k)*1.012*T_supp(k) + C*T_z(k-1)/60 +
(Tw_meas(k-1,:))*(h*areas) + 1.2041*V_inf(k-1)*1.012*T_amb(k-1))/(C/60 + sum(h*areas) +
```

```
1.2041*V_inf(k-1)*1.012 + m_dot_supp(k)*1.012);
end
% Part 4: Calculate cost
deviation = Tz_pred(startt+1:60*(startt_index+24),1) - T_z(startt+1:60*(startt_index+24),1);
cost = norm(deviation);
% **********************************************************
```

# Appendix H

# Codes for performing wall level identification as per optimization framework presented in section 6.3.3.1

To obtain the results presented in section 6.3.3.1 for the optimization framework corresponding to (6.12) – (6.20), the following steps are required. Note that these results are for the internal (zone F2 facing) surface of wall 34.

STEPS 1- 2: Same as Steps 1-2 in Appendix F.
STEP 3: From the generic spreadsheet obtained in Step 2, columns containing the variables needed for the wall level identification are used to create a new spreadsheet called "sim_data_surf34_internal". This file is included in the media accompanying this thesis.
STEP 4: Simulation data contained in "sim_data_surf34_internal" is exported to the MATLAB workspace and saved in a MATLAB file called "surf34_internal_data" using the following MATLAB code:

```
% *****************************************************************
% Import data for inside surface of wall 34 in MATLAB
 clc;
 clear all;
 T_in = xlsread('sim_data_surf34_internal.xls','surf34_internal_data','a2:a1441');
 T_out = xlsread('sim_data_surf34_internal.xls','surf34_internal_data','b2:b1441');
 T_z = xlsread('sim_data_surf34_internal.xls','surf34_internal_data','c2:c1441');
 T_surf = xlsread('sim_data_surf34_internal.xls','surf34_internal_data','d2:k1441');
 save surf34_internal_data
% *****************************************************************
```

STEP 5: The following MATLAB code can then be run to perform the wall level optimization.

```
% *****************************************************************
clc
clear all
% Part 1: Generation of optimal parameter values
matlabpool open 4
options =
optimset('Algorithm','sqp','MaxFunEvals',10000000,'MaxIter',100000,'UseParallel','always','TolF
un', 1e-4, 'TolX', 1e-12, 'TolCon', 1e-12 );
window_length = 24;
init_param = [1e2;1;1*ones(24*60,1)];
```

```
UB = [1e4;100;5*ones(24*60,1)]; % the upper bound on the resistance is set to 100 here. It can be changed to 40 and 10 here for the other two cases presented in section 6.3.3.1.
LB = [0;0;-5*ones(24*60,1)];
optim_param = fmincon(@id_surf34_internal_case1,init_param,[],[],[],[],LB,UB,[],options)% call to
% fmincon to obtain optimal parameter values using least squares optimization
% Part 2: Comparison of wall temperature from least squares identification with EnergyPlus simulation data
C_in = optim_param(1);
R_cond = optim_param(2);
d_in = optim_param(2+1:2+24*60);
R_in = 1/0.207;
load surf34_internal_data
T_in_pred(1,1) = T_in(1);
for k = 2:60*(window_length)
    T_in_pred(k,1) = T_in_pred(k-1) + (60/C_in)*((T_z(k-1)-T_in_pred(k-1))/R_in + (T_out(k-1)-T_in_pred(k-1))/R_cond + d_in(k-1));
end
figure
plot(T_in(1:60*window_length,1),'k');
hold on
plot(T_in_pred(1:60*window_length,1));
% *****************************************************************
```

The MATLAB function "id_surf34_internal_case1" used in the above code is as follows:

```
% *****************************************************************
function cost = id_surf34_internal_case1 (param)
% Part 1: Extract optimization variables
window_length = 24;
C_in = param(1);
R_cond = param(2);
d_in = param(2+1:2+24*60);
R_in = 1/0.207;
% Part 2: Load measurement data
load surf34_internal_data
% Part 3: Evolve the dynamics
T_in_pred(1,1) = T_in(1);
for k = 2:60*(window_length)
    T_in_pred(k,1) = T_in_pred(k-1) + (60/C_in)*((T_z(k-1)-T_in_pred(k-1))/R_in + (T_out(k-1)-T_in_pred(k-1))/R_cond + d_in(k-1));
end
% Part 4: Calculate cost
deviation = T_in_pred(1:60*window_length,1) - T_in(1:60*window_length,1);
cost = norm(deviation);
% *****************************************************************
```

# Appendix I

# Codes for performing wall level identification as per optimization framework presented in section 6.3.3.2

To obtain the results presented in section 6.3.3.2 for the optimization framework corresponding to (6.31) – (6.45), the following steps are required. Note that these results are for the internal (zone F2 facing) surface of wall 34.

STEP 1: In the EnergyPlus model, PRBS set-points are prescribed for all zones in the building. Also, the short-wave radiation incident on each wall surface is specified as an output variable. The EnergyPlus model (.idf file), hence modified, is included in the media accompanying this thesis.
STEP 2: A day-long simulation is run using the above EnergyPlus model and the generic spreadsheet containing the output variables is obtained.
STEP 3: From the above generic spreadsheet, columns containing the variables needed for the wall level identification for the internal surface of wall 34 are used to create a new spreadsheet called "sim_data_surf34_internal_with_loads". In this spreadsheet, the long-wave radiation incident on the wall surface is computed using the ScriptF factors obtained from EnergyPlus and the wall surface temperatures. The long-wave radiation and short-wave radiation values are added to give the total thermal load acting on the wall surface. This spreadsheet is also included in the media accompanying this thesis.
STEP 4: Simulation data contained in "sim_data_surf34_internal_with_loads" is exported to the MATLAB workspace and saved in a MATLAB file called "surf34_internal_data_with_loads" using the following MATLAB code:

```
% *******************************************************************
clc;
 clear all;
 T_in = xlsread('sim_data_surf34_internal_with_loads.xls','a33:a73');
 T_out = xlsread('sim_data_surf34_internal_with_loads.xls','b33:b73');
 T_z = xlsread('sim_data_surf34_internal_with_loads.xls','c33:c73');
 d_in = 0.001*xlsread('sim_data_surf34_internal_with_loads.xls','o33:o73');
 save surf34_internal_data_with_loads
% *******************************************************************
```

STEP 5: The following MATLAB code can then be run to perform the wall level optimization.

```
% *******************************************************************
clc
```

```
clear all
% Part 1: Generation of optimal parameter values
matlabpool open 4
options =
optimset('Algorithm','sqp','MaxFunEvals',100000,'MaxIter',100000,'UseParallel','always','TolFun
', 1e-4, 'TolX', 1e-12, 'TolCon', 1e-12 );
window_length = 23;
init_param = [1.1e3;1;1];
UB = [2e3;10;10];
LB = [0;0;0];
optim_param = fmincon(@id_surf34_internal_case2,init_param,[],[],[],[],LB,UB,[],options); %
call to fmincon to obtain optimal parameter values using least squares optimization
% Part 2: Comparison of wall temperature from least squares identification with EnergyPlus
simulation data
C_in = optim_param(1);
R_cond = optim_param(2);
R_rad = optim_param(3);
R_in = 1/0.207;
load surf34_internal_data
T_in_pred(1,1) = T_in(1);
for k = 2:60*(window_length)
    d_in(k-1) = (1/R_rad)*(sum(T_surf(k-1,:))-8*T_in_pred(k-1));
    T_in_pred(k,1) = T_in_pred(k-1) + (60/C_in)*((T_z(k-1)-T_in_pred(k-1))/R_in + (T_out(k-
1)-T_in_pred(k-1))/R_cond + d_in(k-1));
end
figure
plot(T_in(1:60*window_length,1),'k');
hold on
plot(T_in_pred(1:60*window_length,1));
% ********************************************************************
```

The MATLAB function “id_surf34_internal_case2” used in the above code is as follows:

```
% ********************************************************************
function cost = id_surf34_internal_case2 (param)
% Part 1: Extract optimization variables
window_length = 24;
C_in = param(1);
R_cond = param(2);
R_rad = param(3);
R_in = 1/0.207;
% Part 2: Load measurement data
load surf34_internal_data_with_loads
% Part 3: Evolve the dynamics
T_in_pred(1,1) = T_in(1);
for k = 2:60*(window_length)
```

```
    d_in(k-1) = (1/R_rad)*(sum(T_surf(k-1,:))-8*T_in_pred(k-1));
    T_in_pred(k,1) = T_in_pred(k-1) + (60/C_in)*((T_z(k-1)-T_in_pred(k-1))/R_in + (T_out(k-1)-T_in_pred(k-1))/R_cond + d_in(k-1));
end
% Part 4: Calculate cost
deviation = T_in_pred(1:60*window_length,1) - T_in(1:60*window_length,1);
cost = norm(deviation);
% *****************************************************************
```

# Appendix J

# Codes to obtain LTI model of SCR building and perform agglomerative clustering (sections 6.3.4 – 6.4)

To perform the agglomerative clustering analysis presented in this section, the following programs need to be run in the specified sequence:

STEP 1: Obtain weighted incidence matrix, capacitance matrices, and state space matrices (see Algorithm 5.1). The spreadsheet "identified_parameters_mod.xlsx" contains the data obtained from the zone and wall level identification exercises performed in section 6.3. It is provided in the media accompanying this thesis

```
% ******************************************************************
% Part 1: Generation of A_G (see Algorithm 5.1)
clc
clear all;
Nw = 103;
Nz = 9;
Nw_int = 59;
resistance_matrix = xlsread('identified_parameters_mod','consolidated','B2:BH11');
no_col = size(resistance_matrix,2);
inside_interface = resistance_matrix(1,:);
outside_interface = resistance_matrix(2,:);
internal_surface = resistance_matrix(3,:);
external_surface = resistance_matrix(4,:);
R_cond = resistance_matrix(6,:);
R_in = resistance_matrix(9,:);
R_out = resistance_matrix(10,:);
A_G = zeros(114,114);
for i = 1:no_col
   A_G(inside_interface(i),internal_surface(i))=1/R_in(i);
   A_G(internal_surface(i),inside_interface(i))=1/R_in(i);
   A_G(internal_surface(i),external_surface(i))=1/R_cond(i);
   A_G(external_surface(i),internal_surface(i))=1/R_cond(i);
   if (external_surface(i)~=114)
      A_G(external_surface(i),outside_interface(i)) = 1/R_out(i);
      A_G(outside_interface(i),external_surface(i)) = 1/R_out(i);
   end
```

```
end
inv_R_inf =
1*(0.06*0.05)*[252.859;160.319;50.392;50.392;210.712;105.356;105.356;198.027];
% Implementation of modifications as per section 6.4.1.
A_G(103+3,103+6) = inv_R_inf(1);
A_G(103+6,103+3) = inv_R_inf(1);
A_G(103+1,103+6) = inv_R_inf(2);
A_G(103+6,103+1) = inv_R_inf(2);
A_G(103+6,103+4) = inv_R_inf(3);
A_G(103+4,103+6) = inv_R_inf(3);
A_G(103+1,103+4) = inv_R_inf(4);
A_G(103+4,103+1) = inv_R_inf(4);
A_G(103+1,103+2) = inv_R_inf(5);
A_G(103+2,103+1) = inv_R_inf(5);
A_G(103+1,103+7) = inv_R_inf(6);
A_G(103+7,103+1) = inv_R_inf(6);
A_G(103+1,103+5) = inv_R_inf(7);
A_G(103+5,103+1) = inv_R_inf(7);
A_G(103+8,103+9) = inv_R_inf(8);
A_G(103+9,103+8) = inv_R_inf(8);
% Part 2: Generation of L_Gx
D_G = diag(sum(A_G,2));
L_G = A_G-D_G;
L_Gx = L_G(1:Nw+Nz,1:Nw+Nz);
% Part 3: Generation of C_w
C_in = resistance_matrix(7,:);
C_out = resistance_matrix(8,:);
c_w = zeros(Nw,1);
for i = 1:no_col
   c_w(internal_surface(i)) = C_in(i);
   if (external_surface(i)~=114)
      c_w(external_surface(i)) = C_out(i);
   end
end
C_w = diag(c_w);
% Part 4: Generation of C_z
c_z = zeros(Nz,1);
c_z(1) = 6181.45; %zone G
c_z(2) = 7389.79; %zone E
c_z(3) = 8889.79; %zone C
c_z(4) = 706.02; %zone comp_room
c_z(5) = 3709.49; %zone F1
c_z(6) = 7390.85; %zone D
c_z(7) = 3716.90; %zone F2
c_z(8) = 3449.18; %zone basement_TC
```

```
c_z(9) = 6498.85; %zone basement_NTC
C_z = diag(c_z);
% Part 5: Generation of L_a
L_a = L_G(1:Nw,Nw+Nz+1);
% Part 6: Generation of L_Gg
L_g = L_G(1:Nw,end);
% Part 7: Generation of State Space matrices(continuous time)
A_cont = (blkdiag(C_w,C_z))\L_Gx;
B_a_cont = [C_w\L_a;zeros(Nz,1)];
B_g_cont = [C_w\L_g;zeros(Nz,1)];
B_u_cont = [zeros(Nw,Nz);inv(C_z)];
B_dw_cont = [inv(C_w);zeros(Nz,Nw)];
B_dz_cont = B_u_cont;
% Part 8: State space matrices in desired form and zone-wall matrix
A_ww = A_cont(1:Nw,1:Nw);
A_wz = A_cont(1:Nw,Nw+1:Nw+Nz);
A_zw = A_cont(Nw+1:Nw+Nz,1:Nw);
A_zz = A_cont(Nw+1:Nw+Nz,Nw+1:Nw+Nz);
B_zd = B_u_cont(Nw+1:Nw+Nz,1:Nz);
ZW = zeros(Nz,Nw);
for i = 1:Nz
   for j = 1:Nw
      if(A_G(Nw+i,j)~=0)
         ZW(i,j) = 1;
         for k = 1:Nw
            if(A_G(j,k)~=0)
               ZW(i,k) = 1;
            end
         end
      end
   end
end
% Part 9: MPC Parameters
Ts = 60;
alphaa = 0.001;
N = 30;
% Part 10: Discretization
Ad_ww = eye(Nw) + Ts*A_ww;
Ad_wz = Ts*A_wz;
Ad_zw = Ts*A_zw;
Ad_zz = eye(Nz) + Ts*A_zz;
Bd_zd = Ts*B_zd;
save scr_model
% ****************************************************************
```

STEP 2: To run the agglomerative clustering, run the program in Step 3 of Appendix C with the following change: in the functions “find_new_olf” and “find_fpm”, replace “load nine_zone_model” with “load scr_model”.

# Appendix K

# Codes for cluster level control design (Fig. 6.30)

To perform the steps used in the cluster level control design shown in Figure 6.30, the following programs are used. It is recommended that this Appendix be read along with Appendices L and M.

STEP 1: Obtain weighted incidence matrix, capacitance matrices, and state space matrices (see Algorithm 5.1). The spreadsheet "identified_parameters_mod.xlsx" contains the data obtained from the zone and wall level identification exercises performed in section 6.3. It is provided in the media accompanying this thesis.

```
% ********************************************************************
% Part 1: Generation of A_G (see Algorithm 5.1)
clc
clear all;
Nw = 103;
Nz = 9;
Nw_int = 59;
resistance_matrix = xlsread('identified_parameters_mod','consolidated','B2:BH11');
no_col = size(resistance_matrix,2);
inside_interface = resistance_matrix(1,:);
outside_interface = resistance_matrix(2,:);
internal_surface = resistance_matrix(3,:);
external_surface = resistance_matrix(4,:);
R_cond = resistance_matrix(6,:);
R_in = resistance_matrix(9,:);
R_out = resistance_matrix(10,:);
A_G = zeros(114,114);
for i = 1:no_col
   A_G(inside_interface(i),internal_surface(i))=1/R_in(i);
   A_G(internal_surface(i),inside_interface(i))=1/R_in(i);
   A_G(internal_surface(i),external_surface(i))=1/R_cond(i);
   A_G(external_surface(i),internal_surface(i))=1/R_cond(i);
   if (external_surface(i)~=114)
      A_G(external_surface(i),outside_interface(i)) = 1/R_out(i);
      A_G(outside_interface(i),external_surface(i)) = 1/R_out(i);
```

```
    end
end
% Implementation of modifications as per section 6.4.1.
inv_R_inf =
1*(0.06*0.05)*[252.859;160.319;50.392;50.392;210.712;105.356;105.356;198.027];
A_G(103+3,103+6) = inv_R_inf(1);
A_G(103+6,103+3) = inv_R_inf(1);
A_G(103+1,103+6) = inv_R_inf(2);
A_G(103+6,103+1) = inv_R_inf(2);
A_G(103+6,103+4) = inv_R_inf(3);
A_G(103+4,103+6) = inv_R_inf(3);
A_G(103+1,103+4) = inv_R_inf(4);
A_G(103+4,103+1) = inv_R_inf(4);
A_G(103+1,103+2) = inv_R_inf(5);
A_G(103+2,103+1) = inv_R_inf(5);
A_G(103+1,103+7) = inv_R_inf(6);
A_G(103+7,103+1) = inv_R_inf(6);
A_G(103+1,103+5) = inv_R_inf(7);
A_G(103+5,103+1) = inv_R_inf(7);
A_G(103+8,103+9) = inv_R_inf(8);
A_G(103+9,103+8) = inv_R_inf(8);
% Part 2: Generation of L_Gx
D_G = diag(sum(A_G,2));
L_G = A_G-D_G;
L_Gx = L_G(1:Nw+Nz,1:Nw+Nz);
%Generation of C_w
C_in = resistance_matrix(7,:);
C_out = resistance_matrix(8,:);
c_w = zeros(Nw,1);
for i = 1:no_col
    c_w(internal_surface(i)) = C_in(i);
    if (external_surface(i)~=114)
        c_w(external_surface(i)) = C_out(i);
    end
end
C_w = diag(c_w);
%Generation of C_z
c_z = zeros(Nz,1);
c_z(1) = 6181.45; %zone G
```

```
c_z(2) = 7389.79; %zone E
c_z(3) = 8889.79; %zone C
c_z(4) = 706.02; %zone comp_room
c_z(5) = 3709.49; %zone F1
c_z(6) = 7390.85; %zone D
c_z(7) = 3716.90; %zone F2
c_z(8) = 3449.18; %zone basement_TC
c_z(9) = 6498.85; %zone basement_NTC
C_z = diag(c_z);
% Part 3: Generation of L_Ga
L_Ga = L_G(1:Nw,Nw+Nz+1);
% Part 4: Generation of L_Gg
L_Gw = L_G(1:Nw,end);
% Part 5: Generation of State Space matrices(continuous time)
A_cont = (blkdiag(C_w,C_z))\L_Gx;
B_a_cont = [C_w\L_Ga;zeros(Nz,1)];
B_g_cont = [C_w\L_Gw;zeros(Nz,1)];
B_u_cont = [zeros(Nw,Nz);inv(C_z)];
B_dw_cont = [inv(C_w);zeros(Nz,Nw)];
B_dz_cont = B_u_cont;
ZW = zeros(Nz,Nw);
for i = 1:Nz
  for j = 1:Nw
    if(A_G(Nw+i,j)~=0)
      ZW(i,j) = 1;
      for k = 1:Nw
        if(A_G(j,k)~=0)
          ZW(i,k) = 1;
        end
      end
    end
  end
end
save fo_model
% *******************************************************************
```

STEP 2: Obtain the parameters for design of controller and observer at the cluster level corresponding to any decentralized architecture

```
% *******************************************************************
% Part 1: Input cluster information for fully decentralized case. For other architectures, the
```

```
% code can be modified appropriately
n_clust = 9;
for i = 1:9
    struct_clust_zones(i).value = i;
end
% Part 2: Compute parameters for implementation of controller and observer for each cluster %
and store them as fields of appropriate data structures
for i = 1:n_clust
[struct_tfo_aggreg_A_cont(i).value,struct_tfo_aggreg_B_u_cont(i).value,struct_tfo_aggreg_B_a
_cont(i).value,struct_tfo_aggreg_B_g_cont(i).value,struct_tfo_aggreg_B_hat_cont(i).value,struct
_aggregates(i).value,struct_clust_walls(i).value,struct_ext_zones(i).value,Np,Nu,struct_S(i).valu
e,struct_C_bar(i).value,struct_Q1(i).value,struct_T(i).value,struct_W1(i).value,struct_W2(i).valu
e,struct_W3(i).value,struct_H(i).value,Ts,struct_A1(i).value,struct_A2(i).value,struct_A3(i).valu
e,struct_A4(i).value,struct_A5(i).value,struct_A6(i).value,struct_A7(i).value,rho_a,cp_a,struct_T
_supp(i).value,struct_M_bar_max(i).value,struct_C_bar_temp(i).value,struct_Q_max(i).value,str
uct_L(i).value,struct_init_est(i).value] =
find_model_and_mpc_param_with_dist(struct_clust_zones(i).value);
end
% Part 3: Save the data structures generated above in the Matlab file “clust_info”
save clust_info struct_tfo_aggreg_A_cont struct_tfo_aggreg_B_u_cont
struct_tfo_aggreg_B_a_cont struct_tfo_aggreg_B_g_cont struct_tfo_aggreg_B_hat_cont
struct_Q1 struct_clust_walls Ts Np Nu struct_aggregates c_w struct_clust_zones struct_S
struct_C_bar struct_T struct_W1 struct_W2 struct_W3 struct_ext_zones struct_H Nw Nz
struct_A1 struct_A2 struct_A3 struct_A4 struct_A5 struct_A6 struct_A7 rho_a cp_a
struct_T_supp struct_M_bar_max struct_C_bar_temp struct_Q_max struct_L struct_init_est
u_prev = zeros(Nz,1);
x_init_est = [];
for i = 1:n_clust
x_init_est = [x_init_est;struct_init_est(i).value]; % create a vector using initial state estimates %
computed for each cluster
end
save u_values u_prev % these parameters are required later in the implementation of MPC   %
based decentralized controllers
% *********************************************************************
```

The function “find_model_and_mpc_param_with_dist” used in the above code is shown below:

```
% *********************************************************************
function[tfo_aggreg_A_cont,tfo_aggreg_B_u_cont,tfo_aggreg_B_a_cont,tfo_aggreg_B_g_cont,t
fo_aggreg_B_hat_cont,aggregates,clust_walls,ext_zones,Np,Nu,S,C_bar,Q1,T,W1,W2,W3,H,Ts
```

```
,A1,A2,A3,A4,A5,A6,A7,rho_a,cp_a,T_supp,M_bar_max,C_bar_temp,Q_max,L_tfo_aggreg,ini
t_est] = find_model_and_mpc_param_with_dist(clust_zones)
%%%%%%%%%%%%%%%%%%%%%%%%%%%%%%%%%%%%%%%%%%%%
%SECTION 1: FINDING CLUSTER LEVEL MODEL (FULL ORDER)
%%%%%%%%%%%%%%%%%%%%%%%%%%%%%%%%%%%%%%%%%%%%
load fo_model
n_clust_zones = length(clust_zones);
% Part 1: Determine walls of cluster
clust_walls = [];
for i = 1:n_clust_zones
   local_walls = find(ZW(clust_zones(i),:)~=0);
   clust_walls = union(clust_walls,local_walls);
end
n_clust_walls = length(clust_walls);

% Part 2: Determine zones external to cluster
ext_zones = [];
for i = 1:n_clust_walls
   local_zones = find(ZW(:,clust_walls(i))~=0);
   ext_zones = union(ext_zones,local_zones);
end
ext_zones = setdiff(ext_zones,clust_zones);
n_clust_ext_zones = length(ext_zones);
% Part 3: Determine connectivity among walls and zones inside cluster
clust_A_G = zeros(n_clust_walls + n_clust_zones,n_clust_walls + n_clust_zones);
for i = 1:n_clust_walls
   for j = 1:n_clust_walls
clust_A_G(i,j) = A_G(clust_walls(i),clust_walls(j));
   end
   for k = n_clust_walls+1:n_clust_walls+n_clust_zones
      clust_A_G(i,k) = A_G(clust_walls(i),Nw+clust_zones(k-n_clust_walls));
      clust_A_G(k,i) = A_G(Nw+clust_zones(k-n_clust_walls),clust_walls(i));
   end
end
for i = n_clust_walls+1:n_clust_walls+n_clust_zones
   for j = n_clust_walls+1:n_clust_walls+n_clust_zones
   clust_A_G(i,j) = A_G(Nw+clust_zones(i-n_clust_walls),Nw+clust_zones(j-n_clust_walls));
   end
end
```

```
% Part 4: Determine connectivity among walls and zones inside cluster and outside zones
clust_A_hat = zeros(n_clust_walls,n_clust_ext_zones);
for i = 1:n_clust_walls
   for j = 1:n_clust_ext_zones
      clust_A_hat(i,j) = A_G(clust_walls(i),Nw+ext_zones(j));
   end
end
clust_Az_hat = zeros(n_clust_zones,n_clust_ext_zones);
for i = 1:n_clust_zones
    for j = 1:n_clust_ext_zones
    clust_Az_hat(i,j) = A_G(Nw+clust_zones(i),Nw+ext_zones(j));
    end
end
% Part 5: Determine connectivity of walls with ambient and ground
clust_B_w_a = zeros(n_clust_walls,1);
for i = 1:n_clust_walls
   clust_B_w_a(i,1) = A_G(clust_walls(i),Nw+Nz+1);
end
clust_B_w_g = zeros(n_clust_walls,1);
for i = 1:n_clust_walls
   clust_B_w_g(i,1) = A_G(clust_walls(i),Nw+Nz+2);
end
% Part 6: Determine capacitance matrices
clust_c_w = zeros(n_clust_walls,1);
for i = 1:n_clust_walls
   clust_c_w(i) = c_w(clust_walls(i),1);
end
clust_c_z = zeros(n_clust_zones,1);
for i = 1:n_clust_zones
   clust_c_z(i) = c_z(clust_zones(i),1);
end
clust_C_w = diag(clust_c_w);
clust_C_z = diag(clust_c_z);
%%%%%%%%%%%%%%%%%%%%%%%%%%%%%%%%%%%%%%%%%%%%%%
%SECTION 2: OBTAIN REDUCED ORDER CLUSTER LEVEL MODEL
%%%%%%%%%%%%%%%%%%%%%%%%%%%%%%%%%%%%%%%%%%%%%%
% Part 1: Identify internal wall aggregations
aggregates = struct;
for i = n_clust_walls+1:n_clust_walls+n_clust_zones
```

```
    local_int_aggregation = find(clust_A_G(i,1:n_clust_walls)~=0);
    aggregates(i-n_clust_walls).value = local_int_aggregation;
end
% Part 2: Identify external wall aggregations
cnt = n_clust_zones;
for i = n_clust_walls+1:n_clust_walls+n_clust_zones
    local_ext_aggregation = [];
    local_int_aggregation = aggregates(i-n_clust_walls).value;
    n_local_int_aggregation = length(local_int_aggregation);
    for j = 1:n_local_int_aggregation
        temp = find(clust_A_G(local_int_aggregation(j),1:n_clust_walls)~=0);
        if(norm(clust_A_G(temp,n_clust_walls+1:n_clust_walls+n_clust_zones))==0)
        local_ext_aggregation = [local_ext_aggregation,temp];
        end
    end
    if(isempty(local_ext_aggregation) == 0)
         cnt = cnt + 1;
        aggregates(cnt).value = local_ext_aggregation;
    end
end
% Part 3: Determine aggregated capacitances
n_aggregations = length(aggregates);
aggreg_c_w = zeros(n_aggregations,1);
for i = 1:n_aggregations
    aggreg_c_w(i) = sum(clust_c_w(aggregates(i).value));
end
aggreg_C_w = diag(aggreg_c_w);
% Part 4: Determine equivalent resistances between aggregated walls and cluster zones
aggreg_A_G = zeros(n_aggregations+n_clust_zones,n_aggregations+n_clust_zones);
for i = 1:n_aggregations
    local_i_aggregation = aggregates(i).value;
    for j = 1:n_aggregations
        local_j_aggregation = aggregates(j).value;
        summ = 0;
        for m = 1:length(local_i_aggregation)
            for n = 1:length(local_j_aggregation)
                summ = summ + clust_A_G(local_i_aggregation(m),local_j_aggregation(n));
            end
        end
```

```
        aggreg_A_G(i,j) = summ;
    end
    for j = 1:n_clust_zones
        summ = 0;
        for k = 1:length(local_i_aggregation)
            summ = summ + clust_A_G(local_i_aggregation(k),n_clust_walls+j);
        end
        aggreg_A_G(i,n_aggregations+j) = summ;
        aggreg_A_G(n_aggregations+j,i) = summ;
    end
end
for i = 1:n_clust_zones
    for j = 1:n_clust_zones
        aggreg_A_G(n_aggregations+i,n_aggregations+j) =
clust_A_G(n_clust_walls+i,n_clust_walls+j);
    end
end
% Part 5: Determine equivalent resistances between aggregated walls and cluster zones and %
external zones
aggreg_A_hat = zeros(n_aggregations,n_clust_ext_zones);
for i = 1:n_aggregations
    local_aggregation = aggregates(i).value;
    for j = 1:n_clust_ext_zones
        summ = 0;
        for m = 1:length(local_aggregation)
            summ = summ + clust_A_hat(local_aggregation(m),j);
        end
        aggreg_A_hat(i,j) = summ;
    end
end
aggreg_Az_hat = clust_Az_hat;
% Part 6: Determine equivalent resistances between aggregated walls and ambient and        %
ground
aggreg_B_w_a = zeros(n_aggregations,1);
aggreg_B_w_g = zeros(n_aggregations,1);
for i = 1:n_aggregations
    local_aggregation = aggregates(i).value;
    summ1 = 0;
    summ2 = 0;
```

```
    for m = 1:length(local_aggregation)
        summ1 = summ1 + clust_B_w_a(local_aggregation(m));
        summ2 = summ2 + clust_B_w_g(local_aggregation(m));
    end
    aggreg_B_w_a(i,1) = summ1;
    aggreg_B_w_g(i,1) = summ2;
end
% Part 7: Obtain continuous-time, reduced order state space model
aggreg_D_G =
sum([aggreg_A_G,[aggreg_A_hat;zeros(n_clust_zones,n_clust_ext_zones)],[aggreg_B_w_a;zer
os(n_clust_zones,1)],[aggreg_B_w_g;zeros(n_clust_zones,1)]],2);
aggreg_A_cont = (blkdiag(aggreg_C_w,clust_C_z))\(aggreg_A_G-diag(aggreg_D_G));
aggreg_B_u_cont = [zeros(n_aggregations,n_clust_zones);inv(clust_C_z)];
aggreg_B_a_cont = [aggreg_C_w\aggreg_B_w_a;zeros(n_clust_zones,1)];
aggreg_B_g_cont = [aggreg_C_w\aggreg_B_w_g;zeros(n_clust_zones,1)];
aggreg_B_hat_cont = [aggreg_C_w\aggreg_A_hat;clust_C_z\aggreg_Az_hat];
% Part 8: Obtain continuous-time, reduced order, transformed, state space model
tfo_aggreg_A_cont =
[aggreg_A_cont,[zeros(n_aggregations,n_clust_zones);inv(clust_C_z)];zeros(n_clust_zones,n_a
ggregations+2*n_clust_zones)];
tfo_aggreg_B_u_cont = [aggreg_B_u_cont;zeros(n_clust_zones)];
tfo_aggreg_B_a_cont = [aggreg_B_a_cont;zeros(n_clust_zones,1)];
tfo_aggreg_B_g_cont = [aggreg_B_g_cont;zeros(n_clust_zones,1)];
tfo_aggreg_B_hat_cont = [aggreg_B_hat_cont;zeros(n_clust_zones,n_clust_ext_zones)];
%%%%%%%%%%%%%%%%%%%%%%%%%%%%%%%%%%%%%%%%%%%%%%%%%%
%SECTION 3: FINDING CLUSTER LEVEL MPC PARAMETERS
%%%%%%%%%%%%%%%%%%%%%%%%%%%%%%%%%%%%%%%%%%%%%%%%%%
% Part 1: specify control and prediction horizons, sample time, penalties, upper and lower
% bounds
Np = 30;
Nu = Np/3;
Ts = 60;
gammaa_temp = 1000*ones(n_clust_zones,1);
gammaa_contrl =1*ones(n_clust_zones,1);
sigmaa = 0*ones(n_clust_zones,1); % this parameter is not used
cp_a = 1.05;
rho_a = 1.02;
m_max_full = 1*[12.7426;12.7426;12.7426;4.2475;8.4951;12.7426;8.4951;4.2475;4.2475];
T_supp_full = 12.8*ones(9,1);
```

```
Q_max_full = 1*[140;140;140;50;70;140;70;50;50];
% Part 2: Model discretization
sysc =
ss(tfo_aggreg_A_cont,[tfo_aggreg_B_u_cont,tfo_aggreg_B_a_cont,tfo_aggreg_B_g_cont,tfo_ag
greg_B_hat_cont],eye(n_aggregations+2*n_clust_zones),zeros(n_aggregations+2*n_clust_zones
,n_clust_zones+2+n_clust_ext_zones));
sysd = c2d(sysc,Ts,'zoh');
tfo_aggreg_A_disc = sysd.A;
tfo_aggreg_B_disc = sysd.B;
tfo_aggreg_B_u_disc = tfo_aggreg_B_disc(:,1:n_clust_zones);
tfo_aggreg_B_a_disc = tfo_aggreg_B_disc(:,n_clust_zones+1);
tfo_aggreg_B_g_disc = tfo_aggreg_B_disc(:,n_clust_zones+2);
tfo_aggreg_B_hat_disc =
tfo_aggreg_B_disc(:,n_clust_zones+2+1:n_clust_zones+2+n_clust_ext_zones);
% Part 3: Model augmentation
A_mpc =
[tfo_aggreg_A_disc,tfo_aggreg_B_u_disc;zeros(n_clust_zones,2*n_clust_zones+n_aggregations
),eye(n_clust_zones)];
B_mpc = [tfo_aggreg_B_u_disc;eye(n_clust_zones)];
Ba_mpc = [tfo_aggreg_B_a_disc;zeros(n_clust_zones,1)];
Bg_mpc = [tfo_aggreg_B_g_disc;zeros(n_clust_zones,1)];
B_hat_mpc = [tfo_aggreg_B_hat_disc;zeros(n_clust_zones,n_clust_ext_zones)];
C_mpc=
[zeros(n_clust_zones,n_aggregations),eye(n_clust_zones),zeros(n_clust_zones),zeros(n_clust_zo
nes,n_clust_zones);zeros(n_clust_zones,n_aggregations),zeros(n_clust_zones,n_clust_zones),zer
os(n_clust_zones),eye(n_clust_zones)];
% Part 4: T,S,W,C_bar,Q1 and Q2 matrices
T = zeros((n_aggregations+3*n_clust_zones)*Np,n_aggregations + 3*n_clust_zones);
for i = 1:Np
   T((i-1)*(n_aggregations+3*n_clust_zones)+1:i*(n_aggregations+3*n_clust_zones),:) =
(A_mpc)^i;
end
S = zeros((n_aggregations+3*n_clust_zones)*Np,n_clust_zones*Nu);
for i = 1:Np
   for j = 1:Nu
      if i-j < 0
         S((i-1)*(n_aggregations+3*n_clust_zones)+1:i*(n_aggregations+3*n_clust_zones),(j-
1)*n_clust_zones+1:j*n_clust_zones) =
zeros((n_aggregations+3*n_clust_zones),n_clust_zones);
```

```
        elseif i-j == 0
            S((i-1)*(n_aggregations+3*n_clust_zones)+1:i*(n_aggregations+3*n_clust_zones),(j-
1)*n_clust_zones+1:j*n_clust_zones) = B_mpc;
        elseif i-j > 0
            S((i-1)*(n_aggregations+3*n_clust_zones)+1:i*(n_aggregations+3*n_clust_zones),(j-
1)*n_clust_zones+1:j*n_clust_zones) = A_mpc^(i-j)*B_mpc;
        end
    end
end
W1 = zeros((n_aggregations+3*n_clust_zones)*Np,Np);
for i = 1:Np
    for j = 1:Np
        if i-j < 0
            W1((i-1)*(n_aggregations+3*n_clust_zones)+1:i*(n_aggregations+3*n_clust_zones),j) =
zeros((n_aggregations+3*n_clust_zones),1);
        elseif i-j == 0
            W1((i-1)*(n_aggregations+3*n_clust_zones)+1:i*(n_aggregations+3*n_clust_zones),j) =
Ba_mpc;
        elseif i-j > 0
            W1((i-1)*(n_aggregations+3*n_clust_zones)+1:i*(n_aggregations+3*n_clust_zones),j) =
A_mpc^(i-j)*Ba_mpc;
        end
    end
end
W2 = zeros((n_aggregations+3*n_clust_zones)*Np,Np);
for i = 1:Np
    for j = 1:Np
        if i-j < 0
            W2((i-1)*(n_aggregations+3*n_clust_zones)+1:i*(n_aggregations+3*n_clust_zones),j) =
zeros((n_aggregations+3*n_clust_zones),1);
        elseif i-j == 0
            W2((i-1)*(n_aggregations+3*n_clust_zones)+1:i*(n_aggregations+3*n_clust_zones),j) =
Bg_mpc;
        elseif i-j > 0
            W2((i-1)*(n_aggregations+3*n_clust_zones)+1:i*(n_aggregations+3*n_clust_zones),j) =
A_mpc^(i-j)*Bg_mpc;
        end
    end
end
```

```
W3 = zeros((n_aggregations+3*n_clust_zones)*Np,n_clust_ext_zones*Np);
for i = 1:Np
   for j = 1:Np
      if i-j < 0
         W3((i-1)*(n_aggregations+3*n_clust_zones)+1:i*(n_aggregations+3*n_clust_zones),(j-
1)*n_clust_ext_zones+1:j*n_clust_ext_zones) =
zeros((n_aggregations+3*n_clust_zones),n_clust_ext_zones);
      elseif i-j == 0
         W3((i-1)*(n_aggregations+3*n_clust_zones)+1:i*(n_aggregations+3*n_clust_zones),(j-
1)*n_clust_ext_zones+1:j*n_clust_ext_zones) = B_hat_mpc;
      elseif i-j > 0
         W3((i-1)*(n_aggregations+3*n_clust_zones)+1:i*(n_aggregations+3*n_clust_zones),(j-
1)*n_clust_ext_zones+1:j*n_clust_ext_zones) = A_mpc^(i-j)*B_hat_mpc;
      end
   end
end
C_bar = zeros(Np*2*n_clust_zones,Np*(n_aggregations+3*n_clust_zones));
for i = 1:Np
   C_bar((i-1)*2*n_clust_zones+1:i*2*n_clust_zones,(i-
1)*(n_aggregations+3*n_clust_zones)+1:i*(n_aggregations+3*n_clust_zones)) = C_mpc;
end
gammaa = [gammaa_temp;gammaa_contrl];
Q1 = zeros(Np*2*n_clust_zones,Np*2*n_clust_zones);
for i = 1:Np
   Q1((i-1)*2*n_clust_zones+1:i*2*n_clust_zones,(i-1)*2*n_clust_zones+1:i*2*n_clust_zones)
= diag(gammaa);
end
Q2 = zeros(Nu*n_clust_zones,Nu*n_clust_zones);
for i = 1:Nu
   Q2((i-1)*n_clust_zones+1:i*n_clust_zones,(i-1)*n_clust_zones+1:i*n_clust_zones) =
diag(sigmaa);
end
% Part 5: H matrix
H1 = S'*C_bar'*Q1*C_bar*S + Q2;
H = (H1+H1')/2;
% Part 6: Constraint matrices
A1 = zeros(n_clust_zones*(Np+1),n_clust_zones*Nu);
for i = 1:Np+1
   for j = 1:Nu
```

```
        if i-j >= 0
            A1((i-1)*n_clust_zones+1:i*n_clust_zones,(j-
1)*n_clust_zones+1:j*n_clust_zones)=eye(n_clust_zones);
        else
            A1((i-1)*n_clust_zones+1:i*n_clust_zones,(j-1)*n_clust_zones+1:j*n_clust_zones)=
zeros(n_clust_zones);
        end
    end
end
m_max = m_max_full(clust_zones);
Q_max = Q_max_full(clust_zones);
T_supp = T_supp_full(clust_zones);
M_max = diag(m_max);
for i = 1:Np+1
    M_bar_max((i-1)*n_clust_zones+1:i*n_clust_zones,(i-1)*n_clust_zones+1:i*n_clust_zones)
= M_max;
end
A4 = [zeros(n_aggregations+3*n_clust_zones,Nu*n_clust_zones);S];
A2 = zeros((Np+1)*n_clust_zones,n_clust_zones);
for i = 1:Np+1
    A2((i-1)*n_clust_zones+1:i*n_clust_zones,:) = eye(n_clust_zones);
end
A3 = [eye(n_aggregations+3*n_clust_zones);T];
c_temp =
[zeros(n_clust_zones,n_aggregations),eye(n_clust_zones),zeros(n_clust_zones,2*n_clust_zones)
];
C_bar_temp = zeros((Np+1)*n_clust_zones,(Np+1)*(n_aggregations+3*n_clust_zones));
for i = 1:Np+1
    C_bar_temp((i-1)*n_clust_zones+1:i*n_clust_zones,(i-
1)*(n_aggregations+3*n_clust_zones)+1:i*(n_aggregations+3*n_clust_zones)) = c_temp;
end
A5 = [zeros(n_aggregations+3*n_clust_zones,Np);W1];
A6 = [zeros(n_aggregations+3*n_clust_zones,Np);W2];
A7 = [zeros(n_aggregations+3*n_clust_zones,n_clust_ext_zones*Np);W3];
%%%%%%%%%%%%%%%%%%%%%%%%%%%%%%%%%%%%%%%%%%%%%%%%%%%%
%SECTION 4: FINDING OBSERVER GAINS and INITIAL CONDITIONS
%%%%%%%%%%%%%%%%%%%%%%%%%%%%%%%%%%%%%%%%%%%%%%%%%%%%
gain_mat = 0.5*(H\(S'*C_bar'*Q1'*C_bar*T));
A_cl = tfo_aggreg_A_cont-tfo_aggreg_B_u_cont*gain_mat(1:n_clust_zones,1:end-
```

```
n_clust_zones);
tfo_aggreg_C =
[zeros(n_clust_zones,n_aggregations),eye(n_clust_zones),zeros(n_clust_zones,n_clust_zones)];
eigen = eig(A_cl);
des_obsv_poles =
[10*eigen(1:n_aggregations+n_clust_zones);10*min(eigen)*(linspace(1.1,1.2,n_clust_zones))'];
L_tfo_aggreg = (place(tfo_aggreg_A_cont',tfo_aggreg_C',des_obsv_poles))';
init_est = [25*ones(n_aggregations+n_clust_zones,1);zeros(n_clust_zones,1)];
end
% ****************************************************************
```

STEP 3: Implement the decentralized controllers and observers whose parameters were computed in Step 2. This is done using appropriate MATLAB functions invoked real-time from the SIMULINK model. Instead of providing these codes here, they are provided in Appendix L where the Simulation framework used to perform the optimality analysis in chapter 6 is described.

# Appendix L

# Codes for performing the optimality analysis (section 6.5.2.2)

The steps for performing the optimality analysis are listed below:

1. Execute the code shown in Step 1 in Appendix K.
2. Execute the code shown in Step 2 in Appendix K.. Note that this code must be modified to reflect the appropriate control architecture (centralized/knee3/knee2/knee1/fully decentralized) for which the results are desired.
3. Execute the program "disturbance_param.m" provided later in this section. This program sets the parameters used to speify the disturbance vectors "d_z" and "d_w" and the ambient temperature as per section 6.5.2.1.
4. Run the SIMULINK model "output_feedback_decentralized_mpc" provided in the media accompanying this thesis. This model invokes the following MATLAB functions in real-time:
    a. "find_amb": Provides the ambient temperature at each time instant
    b. "disturbance_function": Creates the vectors "d_w" and "d_z" at each time instant as per section 6.5.2.1.
    c. "temp_sensor": Introduces a fault in the thermostat of a desired zone. Make sure that the line "meas_temp(1) = 0" is commented out.
    d. "decen_obsv": Implements the observers designed in Appendix K for each cluster in the architecture
    e. "extract_eta1": Extract the vector of estimates of "eta_hat_1" for each cluster lumped into a giant vector
    f. "extract_eta2": Extract the of estimates of "eta_hat_2" for each cluster lumped into a giant vector
    g. "extract_zone_temp": Extract the vector of zone temperature estimates
    h. "find_ref": Provide the zone set-point temperature at each time instant
    i. "decen_mpc_with_dist": Implements the model predictive controllers designed in Appendix K for each cluster in the architecture
    j. "actuator": Provides the option of introducing actuator faults. This option was not used in this thesis

The MATLAB programs and functions referenced above are shown below:

1. Program "disturbance_param.m"

```
%*******************************************************************
%matrix F
F = zeros(Nw);
mat_G = xlsread('identified_parameters_mod','int_wall_dist','B1:K10');
for i = 2:size(mat_G,1)
   for j = 2:size(mat_G,2)
      F(mat_G(i,1),mat_G(1,j)) = mat_G(i,j);
   end
end
mat_D = xlsread('identified_parameters_mod','int_wall_dist','B12:J20');
for i = 2:size(mat_D,1)
   for j = 2:size(mat_D,2)
      F(mat_D(i,1),mat_D(1,j)) = mat_D(i,j);
   end
end
mat_E = xlsread('identified_parameters_mod','int_wall_dist','B22:H28');
for i = 2:size(mat_E,1)
   for j = 2:size(mat_E,2)
      F(mat_E(i,1),mat_E(1,j)) = mat_E(i,j);
   end
end
mat_TC = xlsread('identified_parameters_mod','int_wall_dist','B30:J38');
for i = 2:size(mat_TC,1)
   for j = 2:size(mat_TC,2)
      F(mat_TC(i,1),mat_TC(1,j)) = mat_TC(i,j);
   end
end
mat_C = xlsread('identified_parameters_mod','int_wall_dist','B40:K49');
for i = 2:size(mat_C,1)
   for j = 2:size(mat_C,2)
      F(mat_C(i,1),mat_C(1,j)) = mat_C(i,j);
   end
end
mat_F1 = xlsread('identified_parameters_mod','int_wall_dist','B51:J59');
for i = 2:size(mat_F1,1)
   for j = 2:size(mat_F1,2)
```

```
        F(mat_F1(i,1),mat_F1(1,j)) = mat_F1(i,j);
    end
end
mat_F2 = xlsread('identified_parameters_mod','int_wall_dist','B61:K70');
for i = 2:size(mat_F2,1)
    for j = 2:size(mat_F2,2)
        F(mat_F2(i,1),mat_F2(1,j)) = mat_F2(i,j);
    end
end
mat_SR = xlsread('identified_parameters_mod','int_wall_dist','B72:H78');
for i = 2:size(mat_SR,1)
    for j = 2:size(mat_SR,2)
        F(mat_SR(i,1),mat_SR(1,j)) = mat_SR(i,j);
    end
end
mat_NTC = xlsread('identified_parameters_mod','int_wall_dist','B80:L90');
for i = 2:size(mat_NTC,1)
    for j = 2:size(mat_NTC,2)
        F(mat_NTC(i,1),mat_NTC(1,j)) = mat_NTC(i,j);
    end
end
%areas
all_walls =
[xlsread('identified_parameters_mod','consolidated','B4:BH4'),xlsread('identified_parameters
_mod','consolidated','B5:BH5')];
all_areas =
[xlsread('identified_parameters_mod','consolidated','B6:BH6'),xlsread('identified_parameters
_mod','consolidated','B6:BH6')];
for i = 1:length(all_walls)
    areas(all_walls(i)) = all_areas(i);
end
areas = areas(1:Nw);
%stephen boltzmann constant
sb_const = 5.67e-8;
%ground,sky and ambient temp
T_gnd = xlsread('identified_parameters_mod','rad_signals','B2:B7201');
T_sky = xlsread('identified_parameters_mod','rad_signals','C2:C7201');
T_air = T_gnd;
%view factors
```

```
temp_mat1 = xlsread('identified_parameters_mod','ext_wall_dist','C2:F119');
for i = 1:118
   F_gnd(temp_mat1(i,1)) = temp_mat1(i,2);
   F_sky(temp_mat1(i,1)) = temp_mat1(i,3);
   F_air(temp_mat1(i,1)) = temp_mat1(i,4);
end
F_gnd = F_gnd(1:Nw);
F_sky = F_gnd(1:Nw);
F_air = F_gnd(1:Nw);
%short wave radiations
q_swr = zeros(7200,Nw);
temp_mat2 = xlsread('identified_parameters_mod','rad_signals','D1:AD7201');
for i = 1:size(temp_mat2,2)
   q_swr(:,temp_mat2(1,i)) = temp_mat2(2:end,i);
end
%schedules
eta_occ = [0 0 0 0 0 0 0.1 0.2 0.95 0.95 0.95 0.95 0.5 0.95 0.95 0.95 0.95 0.7 0.4 0.4 0.1 0.1
0.05 0.05];
eta_light = [0.05 0.05 0.05 0.05 0.05 0.1 0.1 0.3 0.9 0.9 0.9 0.9 0.9 0.9 0.9 0.9 0.9 0.7 0.5 0.5
0.3 0.3 0.1 0.05];
%eta_appl = [0.4 0.4 0.4 0.4 0.4 0.4 0.4 0.4 0.9 0.9 0.9 0.9 0.8 0.9 0.9 0.9 0.9 0.8 0.6 0.6 0.5
0.5 0.4 0.4];
eta_appl = ones(1,24);
%Nominal loads
N_occ = xlsread('identified_parameters_mod','zone_dist','B2:B10');
W_light = xlsread('identified_parameters_mod','zone_dist','C2:C10');
W_appl = 0.5*xlsread('identified_parameters_mod','zone_dist','D2:D10');
save dist_param F areas sb_const T_gnd T_sky T_air F_gnd F_sky F_air q_swr eta_occ
eta_light eta_appl N_occ W_light W_appl
%*****************************************************************
```

2. Function “find_amb”

```
%*****************************************************************
function Ta = find_amb(t)
load dist_param
Ta = T_air(floor(t/60)+1);
end
%*****************************************************************
```

3. Function “disturbance_function”

```
%*****************************************************************
function dist = disturbance_function(u)
load dist_param
Nw = 103;
Nz = 9;
wall_temp = u(1:Nw);
zone_temp = u(Nw+1:Nw+Nz);
curr_time = u(Nw+Nz+1);
dw = zeros(Nw,1);
dz = zeros(Nz,1);
%Internal surfaces
for i = 1:Nw
   summ = 0;
   for j = 1:Nw
      summ = summ + F(i,j)*((273.15+wall_temp(j))^4 - (273.15+wall_temp(i))^4);
   end
   dw_in(i) = sb_const*areas(i)*summ;
end
%External surfaces
for i = 1:Nw
   if(F_sky(i)==1)
      factor_ab = 0.7;
   else
      factor_ab = 0.9;
   end
   dw_out(i) = sb_const*areas(i)*(F_gnd(i)*((273.15+T_gnd(floor(curr_time/60)+1))^4 -
(273.15+wall_temp(i))^4) + F_sky(i)*((273.15+T_sky(floor(curr_time/60)+1))^4 -
(273.15+wall_temp(i))^4) + F_air(i)*((273.15+T_air(floor(curr_time/60)+1))^4 -
(273.15+wall_temp(i))^4))+areas(i)*factor_ab*q_swr(floor(curr_time/60)+1);
end
%Determine hour in which time lies
dayy = floor(curr_time/(24*3600))+1;
t_bar = curr_time-(dayy-1)*24*3600;
hourr = floor(t_bar/3600)+1;
%Zones
for i = 1:Nz
```

```
  %dz(i) = 100*eta_occ(hourr)*N_occ(i) + 1000*eta_light(hourr)*W_light(i) + 
1000*eta_appl(hourr)*W_appl(i);
  dz(i) = 100*eta_occ(hourr)*N_occ(i) + 1000*eta_light(hourr)*W_light(i) + 
400*eta_appl(hourr)*eta_occ(hourr)*N_occ(i);
end
dw = (dw_in + dw_out)/1000;
dz = dz/1000;
dist = [dw';dz];
end
%*******************************************************************
```

4. Function "temp_sensor":

```
%*******************************************************************
function meas_temp = temp_sensor(u)
act_temp = zeros(9,1);
act_temp(1:9,1) = u(1:9);
curr_time = u(10);
meas_temp = act_temp;
%meas_temp(1) = 0;
%*******************************************************************
```

5. Function "decen_obsv":

```
%*******************************************************************
function der_estimates = decen_obsv(u)
load clust_info
%Obtain inputs
n_clust = length(struct_clust_zones);
summ = 0;
for j = 1:n_clust
  summ = summ + length(struct_aggregates(j).value)+2*length(struct_clust_zones(j).value);
end
n_all_states = summ;
all_states = zeros(n_all_states,1);
all_states(1:n_all_states,1) = u(1:n_all_states);
all_inputs = zeros(Nz,1);
all_inputs(1:Nz,1) = u(n_all_states+1:n_all_states+Nz);
Ta_meas = u(n_all_states+Nz+1);
Tg_meas = u(n_all_states+Nz+2);
all_err = zeros(Nz,1);
all_err(1:Nz,1) = u(n_all_states+Nz+2+1:n_all_states+Nz+2+Nz);
```

```
overall_T_hat = zeros(Nz,1);
overall_T_hat = u(n_all_states+Nz+2+Nz+1:n_all_states+Nz+2+Nz+Nz);
n_clust = length(struct_clust_zones);
summ = 0;
der_estimates = [];
 for j = 1:n_clust
        aggregates = struct_aggregates(j).value;
        n_aggregations = length(aggregates);
        clust_zones = struct_clust_zones(j).value;
        n_clust_zones = length(clust_zones);
        x =
all_states(summ+1:summ+length(struct_aggregates(j).value)+2*length(struct_clust_zones(j).
value));
        summ = summ +
length(struct_aggregates(j).value)+2*length(struct_clust_zones(j).value);
        u = zeros(n_clust_zones,1);
        u(1:n_clust_zones,1) = all_inputs(clust_zones);
        local_err = zeros(n_clust_zones,1);
        local_err(1:n_clust_zones,1) = all_err(clust_zones);
        ext_zones = struct_ext_zones(j).value;
        T_hat = overall_T_hat(ext_zones);
        der_x = struct_tfo_aggreg_A_cont(j).value*x + struct_tfo_aggreg_B_u_cont(j).value*u
+ struct_tfo_aggreg_B_a_cont(j).value*Ta_meas +
struct_tfo_aggreg_B_g_cont(j).value*Tg_meas +
struct_tfo_aggreg_B_hat_cont(j).value*T_hat + struct_L(j).value*local_err;
        der_estimates = [der_estimates;der_x];
 end
%*****************************************************************
```

6. Function “extract_eta1”:

```
%*****************************************************************
function all_eta = extract_eta1(u)
load clust_info
all_eta = [];
n_clust = length(struct_clust_zones);
summ = 0;
for j = 1:n_clust
        aggregates = struct_aggregates(j).value;
        n_aggregations = length(aggregates);
```

```
    x =
u(summ+1:summ+length(struct_aggregates(j).value)+2*length(struct_clust_zones(j).value));
    summ = summ +
length(struct_aggregates(j).value)+2*length(struct_clust_zones(j).value);
    local_eta = zeros(n_aggregations,1);
    local_eta(1:n_aggregations,1) = x(1:n_aggregations);
    all_eta = [all_eta;local_eta];
end
%******************************************************************
```

7. Function “extract eta2”:

```
%******************************************************************
function zone_dist = extract_eta2(u)
load clust_info
zone_dist = zeros(Nz,1);
n_clust = length(struct_clust_zones);
summ = 0;
for j = 1:n_clust
    aggregates = struct_aggregates(j).value;
    n_aggregations = length(aggregates);
    clust_zones = struct_clust_zones(j).value;
    n_clust_zones = length(clust_zones);
    x =
u(summ+1:summ+length(struct_aggregates(j).value)+2*length(struct_clust_zones(j).value));
    summ = summ +
length(struct_aggregates(j).value)+2*length(struct_clust_zones(j).value);
    local_dist = zeros(n_clust_zones,1);
    local_dist(1:n_clust_zones,1) =
x(n_aggregations+n_clust_zones+1:n_aggregations+n_clust_zones+n_clust_zones);
    zone_dist(clust_zones) = local_dist;
end
%******************************************************************
```

8. Function “extract_zone_temp”

```
%******************************************************************
function zone_temp = extract_zone_temp(u)
load clust_info
zone_temp = zeros(Nz,1);
n_clust = length(struct_clust_zones);
summ = 0;
```

```
for j = 1:n_clust
        aggregates = struct_aggregates(j).value;
        n_aggregations = length(aggregates);
        clust_zones = struct_clust_zones(j).value;
        n_clust_zones = length(clust_zones);
        x =
u(summ+1:summ+length(struct_aggregates(j).value)+2*length(struct_clust_zones(j).value));
        summ = summ +
length(struct_aggregates(j).value)+2*length(struct_clust_zones(j).value);
        local_zone_temp = zeros(n_clust_zones,1);
        local_zone_temp(1:n_clust_zones,1) =
x(n_aggregations+1:n_aggregations+n_clust_zones);
        zone_temp(clust_zones) = local_zone_temp;
end
%*****************************************************************
```

9. Function “find_ref”

```
%*****************************************************************
function ref = find_ref(t)
sp = [26.7*ones(1,6),24*ones(1,16),26.7*ones(1,2)];
%Determine hour in which time lies
dayy = floor(t/(24*3600))+1;
t_bar = t-(dayy-1)*24*3600;
hourr = floor(t_bar/3600)+1;
ref = sp(hourr);
%*****************************************************************
```

10. Function “decen_mpc_with_dist”:

```
%*****************************************************************
function u_star = decen_mpc_with_dist(u)
load clust_info
load u_values
%Obtain parameters
n_clust = length(struct_clust_zones);
summ = 0;
for j = 1:n_clust
    summ = summ +
length(struct_aggregates(j).value)+2*length(struct_clust_zones(j).value);
end
n_all_states = summ;
```

```
all_states = zeros(n_all_states,1);
all_states(1:n_all_states,1) = u(1:n_all_states);
all_inputs = zeros(Nz,1);
all_inputs(1:Nz,1) = u(n_all_states+1:n_all_states+Nz);
Ta_meas = u(n_all_states+Nz+1);
Tg_meas = u(n_all_states+Nz+2);
all_ref = zeros(Nz,1);
all_ref(1:Nz,1) = u(n_all_states+Nz+1+1+1:n_all_states+Nz+1+1+Nz);
current_time = u(n_all_states+Nz+1+1+Nz+1);
overall_T_hat = all_ref;
if(mod(current_time,Ts) == 0)
    summ = 0;
     for j = 1:n_clust
        %assignments for each cluster
        aggregates = struct_aggregates(j).value;
        n_aggregations = length(aggregates);
        clust_zones = struct_clust_zones(j).value;
        n_clust_zones = length(clust_zones);
        x0 =
all_states(summ+1:summ+length(struct_aggregates(j).value)+2*length(struct_clust_zones(j)
.value));
        summ = summ +
length(struct_aggregates(j).value)+2*length(struct_clust_zones(j).value);
        u0 = zeros(n_clust_zones,1);
        u0(1:n_clust_zones,1) = all_inputs(clust_zones);
        r_temp = zeros(n_clust_zones,1);
        r_temp(1:n_clust_zones,1) = all_ref(clust_zones);
        S = struct_S(j).value;
        C_bar = struct_C_bar(j).value;
        Q1 = struct_Q1(j).value;
        T = struct_T(j).value;
        W1 = struct_W1(j).value;
        W2 = struct_W2(j).value;
        W3 = struct_W3(j).value;
        ext_zones = struct_ext_zones(j).value;
        n_clust_ext_zones = length(ext_zones);
        T_hat = overall_T_hat(ext_zones);
        H = struct_H(j).value;
        A1 = struct_A1(j).value;
```

```
        A2 = struct_A2(j).value;
        A3 = struct_A3(j).value;
        A4 = struct_A4(j).value;
        A5 = struct_A5(j).value;
        A6 = struct_A6(j).value;
        A7 = struct_A7(j).value;
        T_supp = struct_T_supp(j).value;
        M_bar_max = struct_M_bar_max(j).value;
        Q_max = struct_Q_max(j).value;
        C_bar_temp = struct_C_bar_temp(j).value;
        %initial conditions, lifted disturbances and references
        x_bar0 = [x0;u0];
        Ta_bar = Ta_meas*ones(Np,1);
        Tg_bar = Tg_meas*ones(Np,1);
        T_hat_bar = zeros(n_clust_ext_zones*Np,1);
        for i = 1:Np
           T_hat_bar((i-1)*n_clust_ext_zones+1:i*n_clust_ext_zones,1) = T_hat;
        end
        r = [r_temp;zeros(n_clust_zones,1)];
        %R matrix
        R = zeros(Np*2*n_clust_zones,1);
        for i = 1:Np
           R((i-1)*2*n_clust_zones+1:i*2*n_clust_zones,1) = r;
        end
        %F matrix
        F = S'*C_bar'*Q1'*(C_bar*T*x_bar0 + C_bar*W1*Ta_bar + C_bar*W2*Tg_bar +
C_bar*W3*T_hat_bar - R);
A_constraint = [-
(A1+rho_a*cp_a*M_bar_max*C_bar_temp*A4);(A1+0*rho_a*cp_a*M_bar_max*C_bar_t
emp*A4)];
b_constraint = [-rho_a*cp_a*M_bar_max*(A2*T_supp -
C_bar_temp*(A3*x_bar0+A5*Ta_bar+A6*Tg_bar+A7*T_hat_bar)) + A2*u0 ;
0*rho_a*cp_a*M_bar_max*(A2*T_supp -
C_bar_temp*(A3*x_bar0+A5*Ta_bar+A6*Tg_bar+A7*T_hat_bar)) - A2*u0 +
A2*Q_max];
        %Optimization
        options = optimset('display', 'off', 'Algorithm', 'active-
set','MaxFunEvals',100000,'MaxIter',100000,'UseParallel','always','TolFun', 1e-4, 'TolX', 1e-
12, 'TolCon', 1e-12);
```

```
        init_point = zeros(Nu*n_clust_zones,1);
        D_u_star_lifted = quadprog(H,F,A_constraint,b_constraint,[],[],[],[],init_point,options);
        %D_u_star_lifted = quadprog(H,F,[],[],[],[],[],[],init_point,options);
        %Extraction of optimal control input
        D_u_star = D_u_star_lifted(1:n_clust_zones);
        u_star(clust_zones) = D_u_star + u0;
        u_prev(clust_zones) = u_star(clust_zones);
    end
    save u_values u_prev
else
    u_star = u_prev;
end
%u_star(1) = 140;
end
%*******************************************************************
```

11. Function "actuator":

```
%*******************************************************************
function u_z_supp = actuator(u)
u_act = zeros(9,1);
u_max = zeros(9,1);
u_min = zeros(9,1);
u_act(1:9,1) = u(1:9);
u_max(1:9,1) = u(10:18);
u_min(1:9,1) = u(19:27);
u_z_supp = u_act;
%*******************************************************************
```

# Appendix M

# Codes for performing the robustness analysis (section 6.5.2.3)

The steps to perform the robustness analysis are the same as that for optimality analysis (Appendix L), the only difference being that the line "meas_temp(1) = 0" in the function "temp_sensor" should not be commented out.